\documentclass[twocolumn]{aastex63}

\newcommand{\herschel}{\textit{Herschel}}
\newcommand{\spitzer}{\textit{Spitzer}}
\newcommand{\hst}{\textit{HST}}

\usepackage{savesym}
\savesymbol{tablenum}

\usepackage[binary-units]{siunitx}
\restoresymbol{SIX}{tablenum}

\renewcommand\micron{\mbox{\si{\micro\metre}}}

\renewcommand\arcdeg{\mbox{$^\circ$}}%
\renewcommand\arcmin{\mbox{$^\prime$}}%
\renewcommand\arcsec{\mbox{$^{\prime\prime}$}}%

\newcommand\lir{\mbox{$L_\mathrm{IR}$}}
\newcommand\msun{\mbox{\si{M_\odot}}}
\newcommand\lsun{\mbox{\si{L_\odot}}}
\newcommand\smpy{\mbox{\si{M_\odot. yr^{-1}}}}

\newcommand{\cii}{[\ion{C}{2}]}

\newcommand{\zph}{\mbox{$z_\mathrm{phot}$}}
\newcommand{\zsp}{\mbox{$z_\mathrm{spec}$}}

\newcommand{\rmsp}{\mbox{$\mathrm{RMS}_\mathrm{prior}$}}
\newcommand{\rmsm}{\mbox{$\mathrm{RMS}_\mathrm{map}$}}
\defcitealias{rawle16}{R16}

\graphicspath{{./}{figures/}}

\received{October 29, 2021}
\revised{April 22, 2022}
\accepted{May 3, 2022}
\submitjournal{ApJ}

\shorttitle{ALCS-\herschel: Lensed Dusty Star-Forming Galaxies across $z\simeq0.5-6$}
\shortauthors{Sun et al.}

\begin{document}


\title{ALMA Lensing Cluster Survey: \\
ALMA-\textit{Herschel} Joint Study of Lensed Dusty Star-Forming Galaxies across $z\simeq0.5-6$
}



\correspondingauthor{Fengwu Sun}
\email{fengwusun@email.arizona.edu}

\author[0000-0002-4622-6617]{Fengwu Sun}
\affiliation{Steward Observatory, University of Arizona, 933 N. Cherry Avenue, Tucson, 85721, USA}

\author[0000-0003-1344-9475]{Eiichi Egami}
\affiliation{Steward Observatory, University of Arizona, 933 N. Cherry Avenue, Tucson, 85721, USA}

\author[0000-0001-7201-5066]{Seiji Fujimoto}
\affiliation{Cosmic Dawn Center (DAWN), Jagtvej 128, DK2200 Copenhagen N, Denmark}
\affiliation{Niels Bohr Institute, University of Copenhagen, Lyngbyvej 2, DK2100 Copenhagen {\O}, Denmark}

\author[0000-0002-7028-5588]{Timothy Rawle}
\affiliation{European Space Agency (ESA), ESA Office, Space Telescope Science Institute, 3700 San Martin Drive, Baltimore, MD 21218, USA}

\author[0000-0002-8686-8737]{Franz E. Bauer}
\affiliation{Instituto de Astrofısica, Facultad de Fısica, Pontificia Universidad Catolica de Chile Av. Vicuna Mackenna 4860, 782-0436
Macul, Santiago, Chile}
\affiliation{Millennium Institute of Astrophysics, Nuncio Monse{\~n}or S\'otero Sanz 100, Providencia, Santiago, Chile}

\author[0000-0002-4052-2394]{Kotaro Kohno}
\affiliation{Institute of Astronomy, Graduate School of Science, The University of Tokyo, 2-21-1 Osawa, Mitaka, Tokyo 181-0015, Japan}
\affiliation{Research Center for the Early Universe, School of Science, The University of Tokyo, 7-3-1 Hongo, Bunkyo-ku, Tokyo 113-0033, Japan}

\author[0000-0003-3037-257X]{Ian Smail}
\affiliation{Centre for Extragalactic Astronomy, Department of Physics, Durham University, South Road, Durham, DH1 3LE, UK}

\author[0000-0003-4528-5639]{Pablo G.\ P\'erez-Gonz\'alez}
\affiliation{Centro de Astrobiolog\'ia, Departamento de Astrof\'isica, CSIC-INTA, Cra.\ de Ajalvir km.4 E-28850--Torrej\'on de Ardoz, Madrid, Spain}

\author[0000-0003-3139-2724]{Yiping Ao}
\affiliation{Purple Mountain Observatory \& Key Laboratory for Radio Astronomy, Chinese Academy of Sciences, Nanjing, China}
\affiliation{School of Astronomy and Space Science, University of Science and Technology of China, Hefei, Anhui, China}

\author{Scott C. Chapman}
\affiliation{Eureka Scientific, Inc. 2452 Delmer Street Suite 100, Oakland, CA 94602-3017, USA}
\affiliation{Dalhousie University Dept. of Physics and Atmospheric Science Coburg Road Halifax, B3H1A6, Canada}

\author[0000-0003-2658-7893]{Francoise Combes}
\affiliation{Sorbonne Universit\'e, Observatoire de Paris, Universit\'e PSL, CNRS, LERMA, 75014 Paris, France}
\affiliation{Coll\`ege de France, 11 Place Marcelin Berthelot, 75231 Paris, France}

\author[0000-0003-0348-2917]{Miroslava Dessauges-Zavadsky}
\affiliation{Observatoire de Gen\`{e}ve, Universit\'{e} de Gen\`{e}ve, 51, Ch.\ des Maillettes, 1290 Versoix, Switzerland}

\author[0000-0002-8726-7685]{Daniel Espada}
\affiliation{SKA Organisation, Lower Withington, Macclesfield, Cheshire SK11 9DL, UK}
\affiliation{Departamento de F\'isica Te\'orica y del Cosmos, Campus de Fuentenueva, Universidad de Granada, E18071-Granada, Spain}

\author[0000-0003-3926-1411]{Jorge Gonz\'alez-L\'opez}
\affil{Las Campanas Observatory, Carnegie Institution of Washington, Casilla 601, La Serena, Chile}
\affil{N\'ucleo de Astronom\'ia de la Facultad de Ingenier\'ia y Ciencias, Universidad Diego Portales, Av. Ej\'ercito Libertador 441, Santiago, Chile}

\author[0000-0002-6610-2048]{Anton M. Koekemoer}
\affiliation{Space Telescope Science Institute, 3700 San Martin Dr., Baltimore, MD 21218, USA}

\author[0000-0002-5588-9156]{Vasily Kokorev}
\affiliation{Cosmic Dawn Center (DAWN), Jagtvej 128, DK2200 Copenhagen N, Denmark}
\affiliation{Niels Bohr Institute, University of Copenhagen, Lyngbyvej 2, DK2100 Copenhagen {\O}, Denmark}

\author[0000-0002-2419-3068]{Minju M. Lee}
\affiliation{Max-Planck-Institut fur Extraterrestrische Physik (MPE), Giessenbachstr. 1, D-85748 Garching, Germnay}

\author[0000-0003-3932-0952]{Kana Morokuma-Matsui}
\affiliation{Institute of Astronomy, Graduate School of Science, The University of Tokyo, 2-21-1 Osawa, Mitaka, Tokyo 181-0015, Japan}
\affiliation{Institute of Space and Astronautical Science, Japan Aerospace Exploration Agency, 3-1-1 Yoshinodai, Chuo-ku, Sagamihara, Kanagawa
252-5210, Japan}
\affiliation{Chile Observatory, National Astronomical Observatory of Japan, 2-21-1 Osawa, Mitaka-shi, Tokyo 181-8588, Japan}

\author[0000-0002-8722-516X]{Alejandra M. Mu{\~n}oz Arancibia}
\affiliation{Millennium Institute of Astrophysics, Nuncio Monse{\~n}or S\'otero Sanz 100, Providencia, Santiago, Chile}
\affiliation{Center for Mathematical Modeling, University of Chile, AFB170001, Chile}

\author[0000-0003-3484-399X]{Masamune Oguri}
\affiliation{Research Center for the Early Universe, Graduate School of Science, The University of Tokyo, 7-3-1 Hongo, Bunkyo-ku, Tokyo 113-0033, Japan}
\affiliation{{Center for Frontier Science, Chiba University, 1-33 Yayoi-cho, Inage-ku, Chiba 263-8522, Japan}}
\affiliation{Kavli Institute for the Physics and Mathematics of the Universe (WPI), The University of Tokyo, 5-1-5 Kashiwanoha, Kashiwa-shi, Chiba, 277-8583, Japan}

\author[0000-0003-0858-6109]{Roser Pell\'o}
\affiliation{Aix Marseille Universit\'e, CNRS, CNES, LAM (Laboratoire d’Astrophysique de Marseille), UMR 7326, 13388, Marseille, France}

\author[0000-0001-7821-6715]{Yoshihiro Ueda}
\affiliation{{Department of Astronomy, Kyoto University, Kyoto 606-8502, Japan}}

\author[0000-0001-6653-779X]{Ryosuke Uematsu}
\affiliation{{Department of Astronomy, Kyoto University, Kyoto 606-8502, Japan}}

\author[0000-0001-6477-4011]{Francesco Valentino}
\affiliation{Cosmic Dawn Center (DAWN), Jagtvej 128, DK2200 Copenhagen N, Denmark}
\affiliation{Niels Bohr Institute, University of Copenhagen, Lyngbyvej 2, DK2100 Copenhagen {\O}, Denmark}

\author[0000-0001-5434-5942]{Paul Van der Werf}
\affiliation{Leiden Observatory, Leiden University, P.O. Box 9513, NL-2300 RA Leiden, The Netherlands}

\author[0000-0002-6313-6808]{Gregory L. Walth}
\affiliation{The Observatories of the Carnegie Institution for Science, 813 Santa Barbara Street, Pasadena, CA 91101, USA}

\author[0000-0001-8253-1451]{Michael Zemcov}
\affiliation{Rochester Institute of Technology, 1 Lomb Memorial Drive, Rochester, NY 14623, USA}
\affiliation{Jet Propulsion Laboratory, California Institute of Technology, Pasadena, CA 91109, USA}

\author[0000-0002-0350-4488]{Adi Zitrin}
\affiliation{Physics Department, Ben-Gurion University of the Negev, P.O. Box 653, Be'er-Sheva 84105, Israel}


\begin{abstract}
We present an ALMA-\textit{Herschel} joint analysis of sources detected by the ALMA Lensing Cluster Survey (ALCS) {at 1.15\,mm}.
\herschel/PACS and SPIRE data at 100--500\,\micron\ are deblended for 180 ALMA sources in 33 lensing cluster fields that are either detected securely (141 sources; {in our} main sample) or tentatively at S/N$\geq$4 with cross-matched \hst/\spitzer\ counterparts, down to a delensed 1.15\,mm flux density of $\sim0.02$\,mJy.
We performed far-infrared spectral energy distribution modeling and derived the physical properties of dusty star formation for 125 sources (109 independently) that are detected at $>2\sigma$ in at least one \herschel\ band. 
27 secure ALCS sources are not detected in any \herschel\ bands, including 17 optical/near-IR-dark sources that likely reside at $z=4.2\pm1.2$.
The 16--50--84 percentiles of {the} redshift distribution are {1.15--2.08--3.59} for ALCS sources in the main sample, suggesting an increasing fraction of $z\simeq1-2$ galaxies among fainter millimeter sources ($f_{1150}\sim$\,0.1\,mJy). 
With a median lensing magnification factor of $\mu = 2.6_{-0.8}^{+2.6}$, ALCS sources in the main sample exhibit a median intrinsic star-formation rate of $94_{-54}^{+84}$\,\smpy, 
lower than that of conventional submillimeter galaxies at similar redshifts by a factor of $\sim$\,3.
Our study suggests weak or no redshift evolution of dust temperature with $L_\mathrm{IR}<10^{12}$\,\lsun\ galaxies {within our sample} at $z \simeq 0 - 2$.
At $L_\mathrm{IR}>10^{12}$\,\lsun, the dust temperatures show no evolution across $z \simeq 1 -4$ while being lower than those in the local Universe.
For the highest-redshift source in our sample ($z=6.07$), we can rule out an extreme dust temperature ($>$80\,K) that was reported for MACS0416\,Y1 at $z=8.31$.
\end{abstract}




\section{Introduction}
\label{sec:01_intro}

In massive star-forming galaxies at cosmological distances, a large fraction of star formation is found to be obscured by dust (e.g., \citealt{ivison98,heinis14, whitaker17,fudamoto20}).
Observations at far-infrared (far-IR) wavelengths directly sample the thermal continuum emission from dust grains in the interstellar medium (ISM), a reliable tracer of recent star formation activity \citep[e.g.,][]{ke12}.
With the high sensitivity and spatial resolution of  Atacama Large Millimeter/submillimeter Array (ALMA), submillimeter galaxies (SMGs; or dusty star-forming galaxies, DSFGs as they are often called) have been studied up to a redshift of 6.9 \citep{strandet17, marrone18}, and dust continuum emission from Lyman-break galaxies (LBGs) have also been revealed up to a redshift of 8.3 \citep[e.g.,][]{tamura19,bakx2020}.
ALMA studies of dust obscured star formation, combined with observations obtained at rest-frame UV/optical wavelengths (e.g., with \textit{Hubble Space Telescope}, \hst), provide a comprehensive picture on galaxy formation and evolution across the past 13\,Gyrs (e.g., \citealt{bouwens20}; see a recent review by \citealt{hodge20}).

In order to discover and study the physical properties of SMGs that are intrinsically faint (0.01--1\,mJy around 1\,mm wavelength), {the effect of} gravitational lensing has been widely utilized, which allowed the first detection of SMGs \citep{smail97}.
The ALMA Lensing Cluster Survey (ALCS) is an ALMA Cycle-6 large programs (PI: Kohno; \citealt{kohno19conf, kkprep}) dedicated to survey intrinsically faint continuum sources and line emitters with the assistance of gravitational lensing.
By surveying a total {image-plane} sky area of $\sim$\,134\,\si{arcmin^2} (primary beam response greater than 0.3) down to a depth of 0.07\,mJy\,beam$^{-1}$ ($1\sigma$), ALCS {aimed to} detect $>100$ continuum sources at $\geq 5\sigma$ significance at 1.15\,mm. 
{ALCS has an effective survey area of $\sim$\,10\,arcmin$^2$ for sources brighter than 0.1\,mJy at 1.15\,mm ($>5\sigma$; lensing corrected). 
This is the largest survey obtained with ALMA Band 6 at comparable depth so far.
The detected continuum sources} 
can then be used to examine the origin of cosmic infrared background (CIB), measure the \cii\ luminosity function in the Epoch of Reionization (EoR; \citealt{fujimoto21}), and constrain the evolution of gas and dust content of galaxies around the peak of cosmic star-formation history (CSFH).

All the 33 cluster fields were selected from the best studied clusters primarily from the Abell \citep{abell58,abell89} and MACS \citep{ebeling01} catalogs that have been observed with \hst\ Treasury Programs including Cluster Lensing And Supernova survey with \textit{Hubble} \citep[CLASH, 12 clusters; PI: Postman;][]{postman12}, \textit{Hubble} Frontier Fields \citep[HFF, 5 clusters; PI: Lotz;][]{lotz17} and Reionization Lensing Cluster Survey \citep[RELICS, 16 clusters; PI: Coe;][]{coe19}.
These survey programs utilized both the Advanced Camera for Surveys (ACS) and Wide Field Camera 3 (WFC3) to obtain deep and high-resolution images of massive galaxy clusters at $z = 0.2 - 0.9$ from the optical to near-infrared (near-IR).
Combined with \spitzer/IRAC coverage of at least at medium depth {(the median $5\sigma$ depth at 4.5\,\micron\ is $23.3\pm0.1$\,AB mag; \citealt{sun21b})}, these \hst\ data provide direct constraints on unobscured or mildly obscured stellar components in both the environment of the massive galaxy cluster as well as in gravitationally lensed galaxies in the distant Universe.


The ALCS fields were also observed by the \herschel\ \textit{Space Observatory} \citep{herschel10} in the wavelength range from 100 to 500\,\micron.
Launched in 2009 and retired in 2013, \herschel\ was designed to study the dust-obscured universe at submillimeter wavelengths.
\herschel\ data are critical for the interpretation of the ALCS fields due to its unique wavelength coverage of thermal dust continuum of high-redshift ($z\gtrsim 1$) galaxies.
{\herschel\ and ALMA observations both detect thermal emissions from dust heated by the UV radiation from young stars.}
With a good sampling of far-IR spectral energy distributions (SEDs) with 4--6 bands in total, ALMA and \herschel\ data provide critical constraints on photometric redshifts, thermal dust temperature and dust mass of ALCS-selected galaxies.

\herschel\ data of the ALCS clusters were obtained through various programs.
Among them, the \herschel\ Lensing Survey (HLS; \citealt{egami10}; \citealt{sun21a}) is the largest program {imaging the fields of massive galaxy clusters to study cluster-lensed high-redshift galaxies}.
Deep \herschel\ observations in blank fields are often subject to confusion noise \citep{nguyen10}. 
This restricts the detection of typical ultra luminous infrared galaxies (ULIRGs; $L_\mathrm{IR}\geq 10^{12}$\,\lsun) beyond $z\sim 2$ \citep[c.f.][]{rawle16}.  
However, with the lensing magnification by massive clusters, which typically do not contain far-IR-bright galaxies in the cluster core \citep[e.g.,][]{rawle12}, we are able to break the confusion limit and discover intrinsically faint sources \citep{smail97}. 
This has been demonstrated by the \herschel\ detection of the $z=2.8$ LIRG ($L_\mathrm{IR}=10^{11}- 10^{12}$\,\lsun) behind the Bullet cluster with a lensing magnification of $\mu\sim 75$ \citep{rex10}. 
With cluster lensing, \citet{sklias14} and \citet{mdz15} explored the star formation history, dust extinction and molecular gas content of LIRGs at $z\simeq 1.5 - 3$, and recent ALMA observations of cluster-lensed \herschel\ sources revealed the existence of low-surface-brightness SMGs with extended dust continua \citep{sun21a}.

In this work, we present the ALMA-\herschel\ joint analysis of the dusty star-forming galaxies detected by ALCS. 
Similar work has been presented in \citet[hereafter \citetalias{rawle16}]{rawle16} for the six HFF clusters, and here we expand the sample to 28 more cluster fields and use high-resolution ($\sim$1\arcsec) ALMA continuum maps as priors for source extraction, in contrast to the mid-IR priors (\spitzer\ and \textit{WISE}) used in \citetalias{rawle16}.
Because of the well known negative $K$-correction, the selection function of SMGs at millimeter wavelengths is nearly constant  {in terms of cold dust mass} across $z\simeq 1 - 6$. 
Therefore, compared with \citetalias{rawle16}, the use of ALMA priors allows a more extensive and accurate measurements of \herschel\ flux densities of sources at higher redshifts ($z>2$), constraining the redshift distribution, dust temperatures and star-formation rates (SFRs) of millimeter sources towards the faint end {($f_{1150}\sim0.02$\,mJy)}.

This paper is arranged as follows:
In Section~\ref{sec:02_obs}, we introduce the sample discovered by the ALCS \citep[will be described in greater detail by][]{sfprep} and the obtained \herschel\ data together with the data reduction techniques.
Section~\ref{sec:03_res} presents the procedure for source extraction using the \herschel\ data.
Section~\ref{sec:04_sed} presents the far-IR SED fitting and photometric redshift estimate.
Section~\ref{sec:05_lens} presents the analysis of lensing magnification and uncertainty.
In Section~\ref{sec:06_dis}, we discuss the statistical results of galaxy properties and their implications.
The summary can be found in Section~\ref{sec:07_sum}.
Throughout this work, we assume a flat $\Lambda$CDM cosmology with $h=0.7$ and $\Omega_m = 0.3$. 
We define the IR luminosity ($L_\mathrm{IR}$) as the integrated luminosity over a rest-frame wavelength range from 8 to 1000\,\micron.


\section{Observations and Data Reduction}
\label{sec:02_obs}

\subsection{ALMA Data and the Sample}
\label{ss:02a_sample}

All of the sources in this work are selected with the ALCS, which will be detailed in \citet{sfprep}.
ALMA Band-6 observations for the 33 clusters were conducted through Program 2018.1.00035.L (ALCS; PI: Kohno), and we also combined archival data from Programs 2013.1.00999.S and 2015.1.01425.S (ALMA Frontier Fields; PI: Bauer; \citealt{glopez17}).
The list of the ALCS clusters with their coordinates, short names (e.g., M0553 for MACSJ0553.4--3342) and \hst\ program names is presented in Table~\ref{tab:01_obs}.
The observations were obtained at a central wavelength of 1.15\,mm with a 15\,GHz total bandwidth (i.e., two tunings of dual polarization; 250.0--257.5\,GHz and 265.0–-272.5\,GHz).
{The use of two tunings instead of one allows us to search for line-emitting galaxies over a larger volume, which is another important science goal of ALCS. This, for example, led to the serendipitous discovery of a \cii\ emitter at $z=6.072$ \citep{fujimoto21}.
}
All the ALMA data were reduced with \textsc{casa} \citep{casa} with different pipelines versions for observations obtained in different cycles (e.g., v5.4.0 for 26 clusters observed in Cycle 6 and v5.6.1 for the remaining clusters in Cycle 7).
Natural-weighting continuum imaging was performed at both the native (full width at half maximum, FWHM\,$\sim 1$\arcsec) and \textit{uv}-tapered ($\sim2$\arcsec) resolutions with the \textsc{casa} \textsc{tclean} algorithm.


Through a peak pixel identification routine of \textsc{SExtractor} \citep{sextractor} with the ALMA maps at both the native and 2\arcsec-tapered resolutions (before primary beam correction), we securely detected 141 sources that are either at (\romannumeral1) signal-to-noise ratio $\mathrm{S/N}_\mathrm{nat} \geq 5$ in the native-resolution maps, or (\romannumeral2) $\mathrm{S/N}_\mathrm{tap} \geq 4.5$ in the 2\arcsec-tapered maps, over an area of $\sim$134\,arcmin$^2$ with primary beam response greater than 0.3 \citep{sfprep}. 
Based on the number count of negative peaks, the number of spurious source above these S/N cuts is expected to be around one.
We further refer to these 141 secure ALCS sources as the \textit{Main Sample}.

258 sources were tentatively detected at (\romannumeral1) $\mathrm{S/N}_\mathrm{nat} = 4 - 5$ in the native-resolution maps and (\romannumeral2) $\mathrm{S/N}_\mathrm{tap} < 4.5$ in the 2\arcsec-tapered maps, down to a minimal flux density of $\sim 0.2$\,mJy at 1.15\,mm.
Based on \hst\ and \spitzer/IRAC images, we identified 39 of these sources with near/mid-IR counterparts within a separation of 1\arcsec.
Given the high source densities in cluster fields ($\sim$\,0.06\,arcsec$^{-2}$ in the \hst/F160W band; \citealt{sun21b}), we expect $7\pm3$ pairs of random associations among these tentative ALCS sources and cross-matched near/mid-IR counterparts. 
Assuming that the majority of these \hst/\spitzer-matched sources are real, we note that the number found is consistent with the number difference between positive and negative peaks in the ALMA maps in this S/N range (37 in total). 
We further refer to these 39 tentative ALCS sources as the \textit{Secondary Sample}, {but warn that $18\pm8$\% of these sources are likely to be false matched}. 

We note that the continuum S/N of ALCS sources could be boosted by serendipitous emission line detections, for example, CO(5--4) line at $z=1.11-1.17$ and $1.24-1.31$ (e.g., M0553-ID133/190/249 at $z=1.142$, \citealt{ebeling17,sun21a}).
All the line emitters will be reported by a future paper of the collaboration.
However, because of a large bandwidth (15\,GHz), the CO line contamination to continuum S/N and flux density is limited to $\lesssim$\,1--10\%.
In addition, ALCS can only sample faint high-$J$ CO lines (upper $J$ number at $\geq$\,7) for sources at $z\gtrsim2$. 
{According to the CO spectral line energy distribution (SLED) of high-redshift SMGs reported in the literature \citep[e.g.,][]{greve14,bethermin16,yang17,birkin21}, even the high-$J$ CO SLED is as flat as those reported for local active-galactic-nucleus-host (AGN-host) galaxies \citep[e.g.,][]{rosenberg15}
, the CO contamination will be $\sim1-10$\%\ at most.}
{Only one \cii\ emitter was found among all continuum sources \citep{fujimoto21,laporte21}, and the continuum flux density of this source was measured on line-subtracted continuum image.
}

\subsection{\herschel/PACS}
\label{ss:02b_pacs}

\begin{figure*}[!t]
\centering
\includegraphics[width=\linewidth]{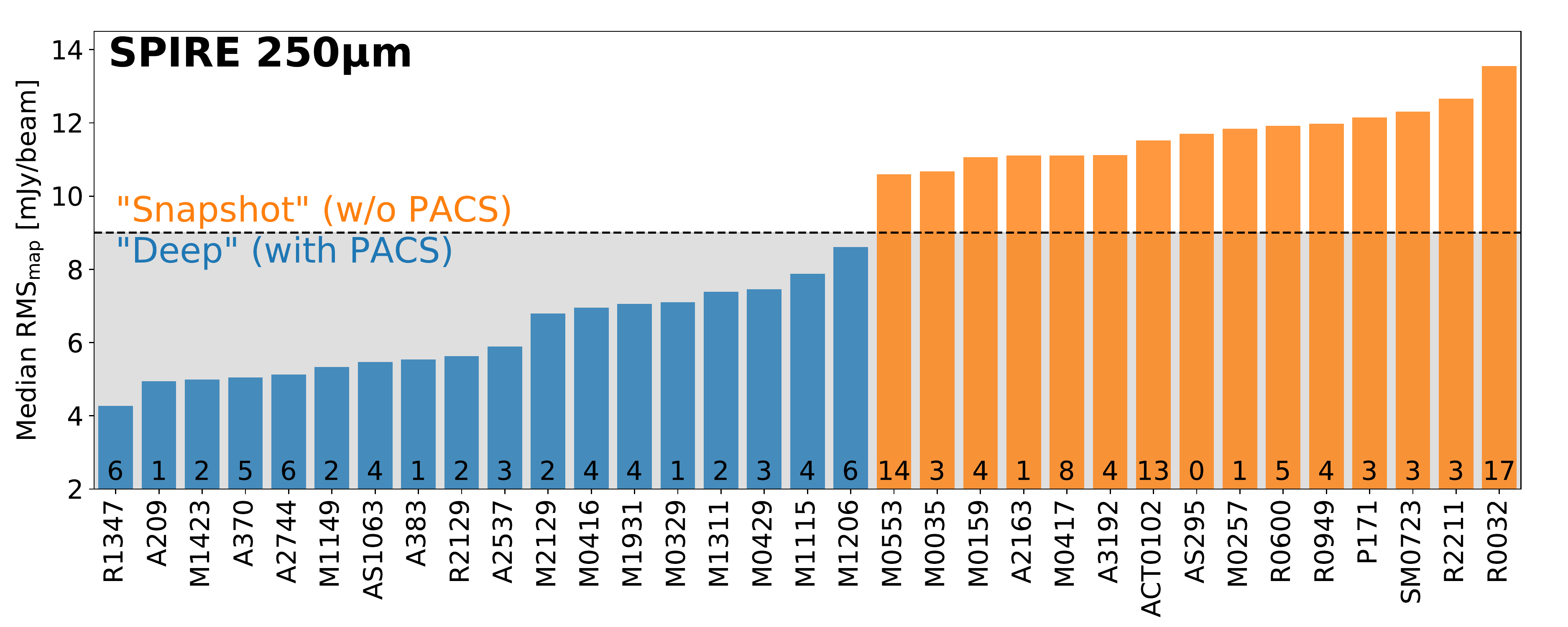}
\caption{Median RMS noise of SPIRE pixels ($\mathrm{RMS}_\mathrm{map}$; measured as the noise of sky background after sources being filtered out) at 250\,\micron\ within the footprints of 33 ALCS clusters.
18 clusters were observed with \herschel\ in the ``deep'' mode (in blue), and they have deep SPIRE images ($\mathrm{RMS}_\mathrm{map}<9$\,\si{mJy.beam^{-1}}) as well as PACS ones.
The remaining 15 clusters were observed in the ``snapshot'' mode (in orange), and they have shallower SPIRE data ($\mathrm{RMS}_\mathrm{map}>9$\,\si{mJy.beam^{-1}}; see Section~\ref{ss:02c_spire}) with no PACS coverage.
The number of main-sample ALCS sources in each cluster field is noted at the bottom of each bar.
}
\label{fig:0_unc_bar}
\end{figure*}

The Photodetector Array Camera and Spectrometer (PACS; \citealt{pacs}) on \herschel\ {enabled simultaneous observations} at 160\,\micron\ (red channel) with the long-wavelength camera and either at 70 or 100\,\micron\ (blue/green channel) with the short-wavelength camera.
Eighteen out of the total 33 clusters were imaged with PACS at both 100 and 160\,\micron, and two clusters were also observed with PACS at 70\,\micron\ (M1149 and AS1063). 
The analysis of PACS 70\,\micron\ data in these cluster fields has been presented by \citetalias{rawle16}, and the only two matched sources are AS1063-ID17 ($z=1.44$, $f_{70} = 7.3	\pm 0.9$\,mJy) and AS1063-ID147 ($z = 0.610$, $f_{70} = 28.8 \pm 2.3$\,mJy; analyzed in detail by \citealt{walth19}).

16 out of the 18 clusters were observed by PACS as part of the HLS \citep{egami10,sun21a}, which combines an Open-Time Key Program (KPOT; program ID: \texttt{KPOT\_eegami\_1}; nine clusters) and an Open-Time Cycle 2 (OT2; program ID: \texttt{OT2\_eegami\_5}; seven clusters) Program (both PI: Egami).
The remaining two clusters, namely Abell370 and RXJ1347--1145, were observed as part of the PACS Evolutionary Probe \citep[PEP; program ID: \texttt{KPGT\_dlutz\_1}, PI: Lutz;][]{lutz11}.
All of the PACS 100 and 160\,\micron\ observations consist of two orthogonal scan maps,
each comprising 18--22 repetitions of 13 parallel 4-arcmin scan legs.
The summary of the PACS observations, including the observation IDs and total scan time for each cluster, is presented in Table~\ref{tab:01_obs}. 

We followed the same data reduction procedure as detailed in \citetalias{rawle16} for the HFF clusters.
The PACS images were generated with
\textsc{UniMap} \citep{piazzo15} with a pixel scale of 1\farcs0 at 100\,\micron\ and 2\farcs0 at 160\,\micron.
The final PACS image products have a typical field of view (FoV) with a radius of $\sim4$\arcmin, covering the full ALMA footprints of the 18 clusters.
The typical beam sizes are 7\farcs4 and 11\farcs4\ at 100 and 160\,\micron, and the depths of the PACS data at the cluster center are presented in Table~\ref{tab:02_unc}.

\subsection{\herschel/SPIRE}
\label{ss:02c_spire}

The Spectral and Photometric Imaging Receiver \citep[SPIRE;][]{spire} on \herschel\ worked simultaneously at 250, 350 and 500\,\micron. 
All of the 33 clusters were imaged with SPIRE in two observing modes with different depths.
The 18 clusters also observed with PACS were scanned with SPIRE in the ``deep" mode down to confusion-limited depths ($\mathrm{RMS}_\mathrm{map}\sim6$\,mJy\,beam$^{-1}$ at 250\,\micron; measured as the noise of sky background after sources being filtered out), and the remaining 15 clusters were observed in the ``snapshot" mode with a shorter scan duration and thus at shallower depths ($\mathrm{RMS}_\mathrm{map}\sim11$\,mJy\,beam$^{-1}$; as visualized in Figure~\ref{fig:0_unc_bar}). 

Among the total 18 clusters in the ``deep'' mode, 16 of them were observed as part of the HLS.
Observations of the nine clusters through \texttt{KPOT\_eegami\_1} consisted of 20 repetitions in the large scan map mode, each with two 4\arcmin\ scans and cross-scans (total scan time as $t_\mathrm{scan} \sim 1.7$\,h per cluster).
The other seven clusters observed through Open-Time Cycle 1/2 Programs (\texttt{OT1\_eegami\_4} and \texttt{OT2\_eegami\_5}; both PI: Egami) were imaged through 11-repetition small scan maps (one in OT1 and ten in OT2), and each repetition consisted of one scan and one cross-scan of 4\arcmin\ length ($t_\mathrm{scan}\sim 0.4$\,h per cluster).
The remaining two clusters, A370 and R1347, were observed as part of the \textit{Herschel} Multi-tiered Extragalactic Survey \citep[HerMES; program ID: \texttt{KPGT\_soliver\_1}; PI: Oliver;][]{oliver12}. 
Both of these clusters were observed with eight small scan maps (six repetitions per each covering the cluster core) and three large scan maps (one repetition per each with 38\arcmin\ length covering a wider area), and the total scan time is 3.5\,h per cluster.
The final SPIRE map products of these two clusters have a wider FoV, but the central depths are comparable to those of the HLS data.

All of the 15 clusters in the ``snapshot'' mode were observed as part of the HLS through single-repetition small scan maps, and each repetition consisted of one scan and one cross-scan of a 4\arcmin\ length ($t_\mathrm{scan}\sim 3$\,min per cluster).
Fourteen of them were observed by the OT1 program \texttt{OT1\_eegami\_4}, and the remaining one was observed by the OT2 program \texttt{OT2\_eegami\_6} (PI: Egami).

Table~\ref{tab:01_obs} summarizes the IDs and total scan times of all the SPIRE observations.
All of the SPIRE data were processed by the standard reduction pipeline in HIPE v12.2 \citep{ott10} which is also detailed in \citetalias{rawle16}.
The output pixel sizes of the final image products are 6\arcsec, 9\arcsec\ and 12\arcsec\ at 250, 350 and 500\,\micron.
The typical radii of the SPIRE FoVs are 7\arcmin\ for the 15 clusters observed in the ``snapshot'' mode,  8\arcmin\ for the seven clusters observed in OT1/2, 11\arcmin\ for the nine clusters observed in KPOT, and 33\arcmin\ for A370 and R1347.
The full survey area of the ALCS was covered by these SPIRE images.
The typical beam sizes are 18\arcsec, 24\arcsec\ and 35\arcsec\ in these three bands, and the depths of the SPIRE data at the cluster center are presented in Table~\ref{tab:02_unc}.

\subsection{Ancillary \hst\ and \spitzer\ Data}
\label{ss:02d_other}

For the purpose of enhancing the astrometric accuracy of \herschel\ data, we include the \spitzer/IRAC data of these 33 cluster fields in our analysis obtained from the NASA/IPAC Infrared Science Archive (IRSA\footnote{https://irsa.ipac.caltech.edu/}). 
We also include the \hst\ data of all the cluster fields but only for comparing the positions of dust continuum sources with the stellar components. 
We defer the study of optical/near-IR counterparts and panchromatic SED modeling of ALCS sources to another paper from the collaboration. 

\subsection{Redshift Catalogs}
\label{ss:02e_redshift}

To supply accurate redshifts for far-IR SED modeling (Section~\ref{sec:04_sed}), we cross-matched the ALCS source sample with the spectroscopic redshift (\zsp) catalogs made available by the CLASH-VLT spectroscopic survey \citep{biviano13}, Grism Lens-Amplified Survey from Space (GLASS; \citealt{schmidt14,treu15,wang15}) and recent VLT/MUSE surveys of massive cluster fields by \citet{caminha19} and \citet{richard21}. 
A maximum separation of 1\farcs5 is allowed for cross-matching, which is comparable to the FWHM of IRAC PSF.
We also include redshifts for a few sources reported by various studies in the literature \citep[e.g., M0553 triply lensed system at $z=1.14$;][]{ebeling17} or private communication (e.g., M0417-ID46/58/121, an \hst\ $H$-faint triply lensed system at $z=3.65$; \citealt{kkprep}).
In addition, we also include two ALMA-HFF sources reported by \citet{laporte17} with their \zsp's derived from the GLASS detection of the 4000\,\AA\ break, and a triply lensed ALCS source system that belongs to a MUSE-confirmed galaxy group at $z=4.32$ \citep{caputi21}. 
Spectroscopic redshifts are available for 60 ALCS sources in both the main and secondary samples.

We also utilized the \hst\ photometric redshift (\zph) catalogs of optical/near-IR sources tabulated by CLASH \citep{molino17}, HFF \citep{shipley18} and RELICS \citep{coe19} groups.
Sources are cross-matched by their coordinates and a maximum separation of 1\farcs5 is allowed. 
\citet{fujimoto16} reported a median offset of 0\farcs25 between the \hst\ and ALMA centroids of ALMA sources, and such an observed offset could be larger in cluster fields because of the lensing magnification.
We also apply visual inspections of the \hst\ F814W, F105W and F160W images to remove any conspicuous mismatch. 
We identified catalogued \hst\ $z_\mathrm{phot}$ measurements for {125} ALCS sources in both the main and secondary samples,
{including 49 sources with additional spectroscopic redshifts.}

\section{Herschel Source Extraction}
\label{sec:03_res}



\subsection{Preparation}

\subsubsection{Image Alignment}
\label{ss:03a_aligh}

Following \citetalias{rawle16}, we first aligned all the \herschel\ images to the IRAC Channel 1 (3.6\,\micron) images before the actual source extraction at 100--500\,\micron.
We cross-matched the IRAC 3.6\,\micron\ source catalog in each field with the $\sim$10--20 brightest sources detected in \herschel\ bands using \textsc{DAOFIND} \citep{daofind}. 
We then computed the median RA and Dec offsets of the matched sources in the \herschel\ and IRAC bands, and corrected these for the \herschel\ data. 
We only calculated the offsets independently for the PACS 100\,\micron\ and SPIRE 250\,\micron\ data and applied the same astrometric shift to other bands of the same instrument.
This is because the offsets between different bands of the same \herschel\ instrument have been well calibrated. 
The median corrected offsets are 0\farcs9 and 1\farcs3 for PACS and SPIRE images, consistent with those reported in \citetalias{rawle16}.

\subsubsection{Input ALCS Source Catalog}
\label{ss:03b_input}

We constructed the input catalog for \herschel\ source extraction using 180 ALCS sources at $\mathrm{S/N} \geq 4$ (Section~\ref{ss:02a_sample}).
Among them, 141 secure sources (main sample) were extracted in the first two iterations, and 39 tentative sources with matched \hst/\spitzer\ counterparts (secondary sample) were then extracted on the residual images.
We note that 85\% of tentative ALCS sources at $\mathrm{S/N}_\mathrm{nat} = 4-5$ and $\mathrm{S/N}_\mathrm{tap} < 4.5$ do not show any near/mid-IR counterpart.
These sources are expected to mostly be spurious and not included for \herschel\ photometry. 
However, it is possible that a few of them represent highly obscured high-redshift galaxies ($A_V\gtrsim5$, $z \gtrsim 4$), which are missed by this study.
The coordinates of ALMA sources were used as positional priors,
and the S/N's of ALMA detections were later used to rank the priority of extraction in Section~\ref{ss:03c_iter}.

\begin{figure*}[!htp]
\centering
\includegraphics[width=0.49\linewidth]{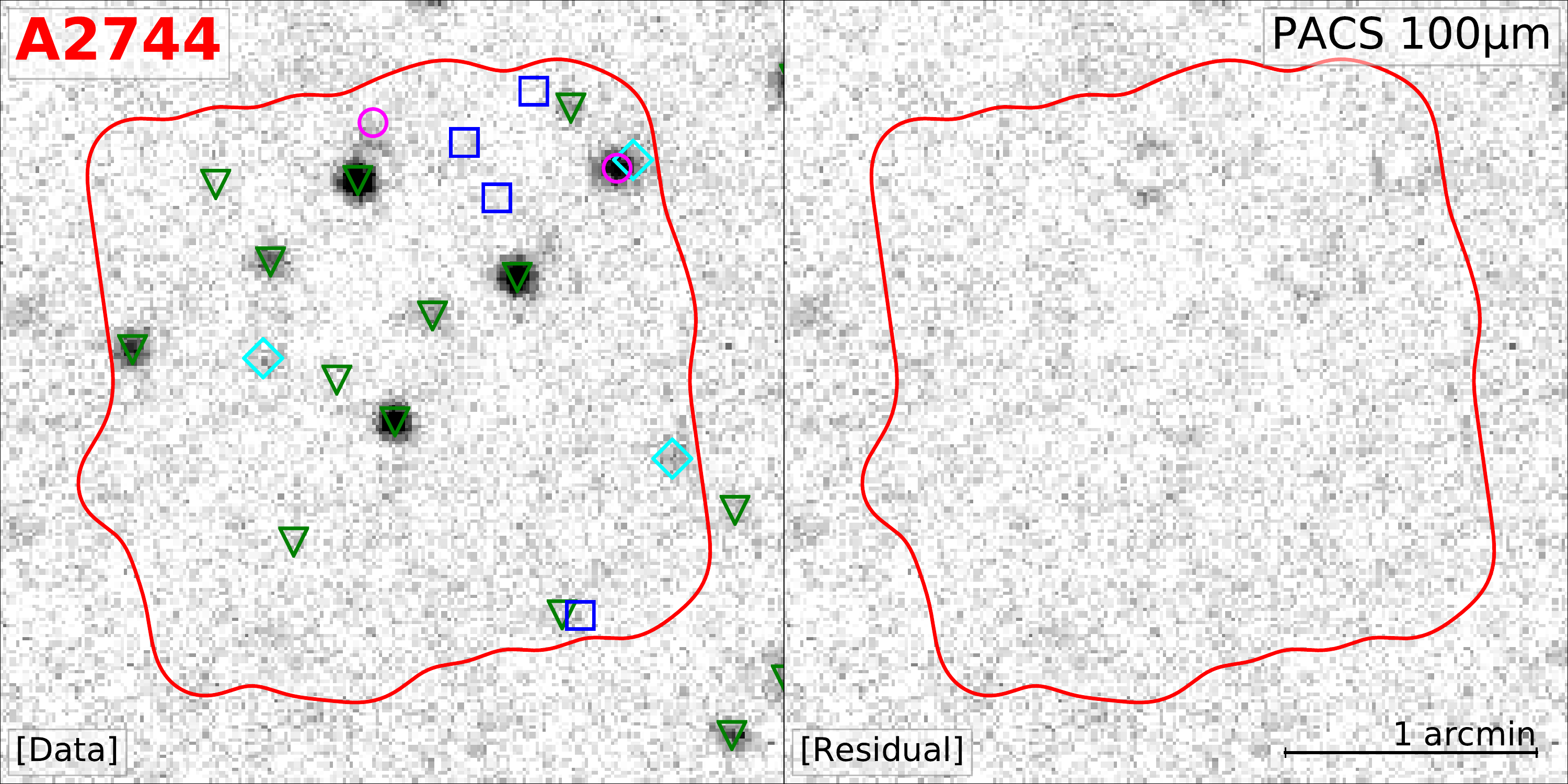}
\includegraphics[width=0.49\linewidth]{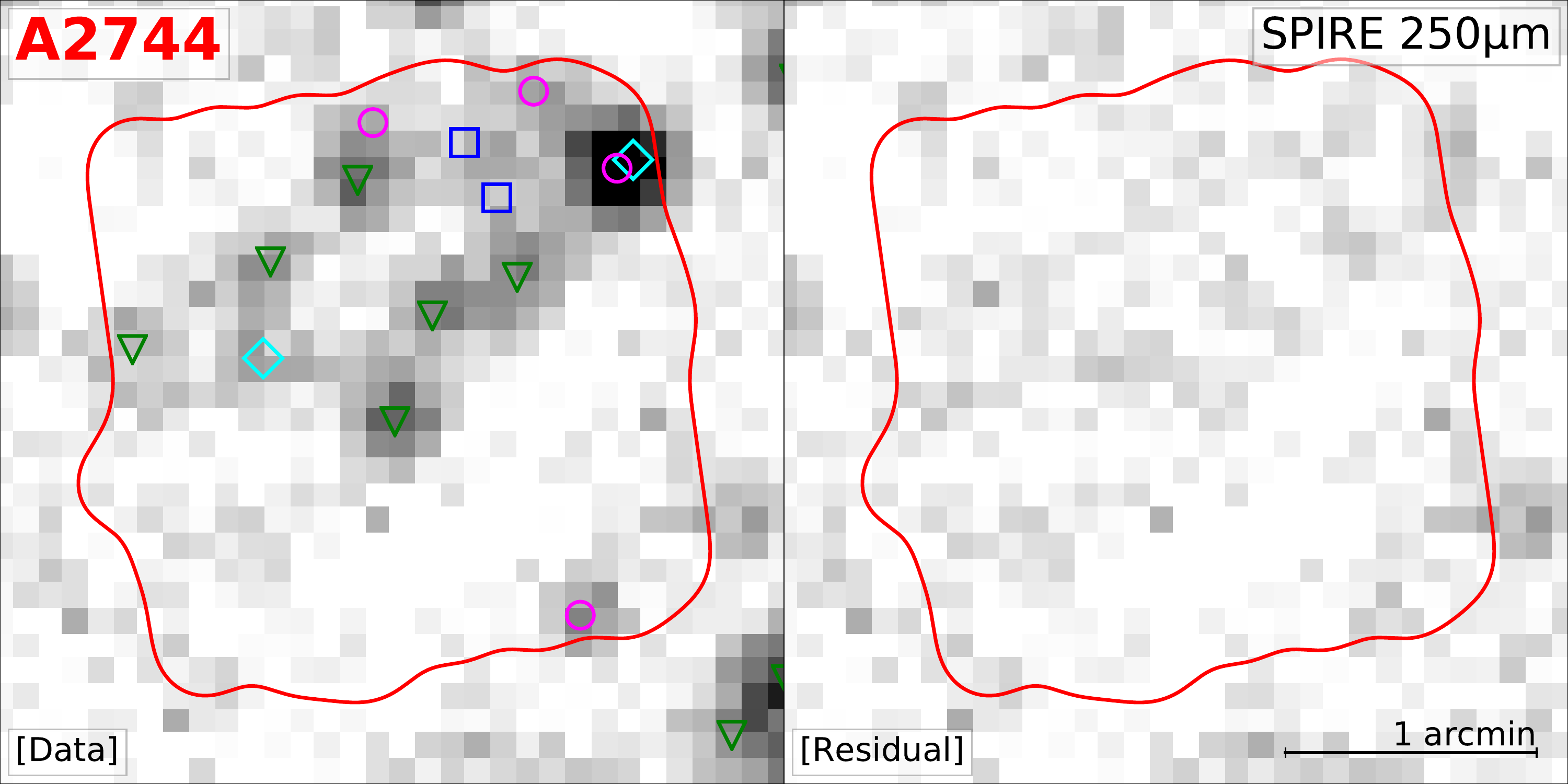}
\includegraphics[width=0.49\linewidth]{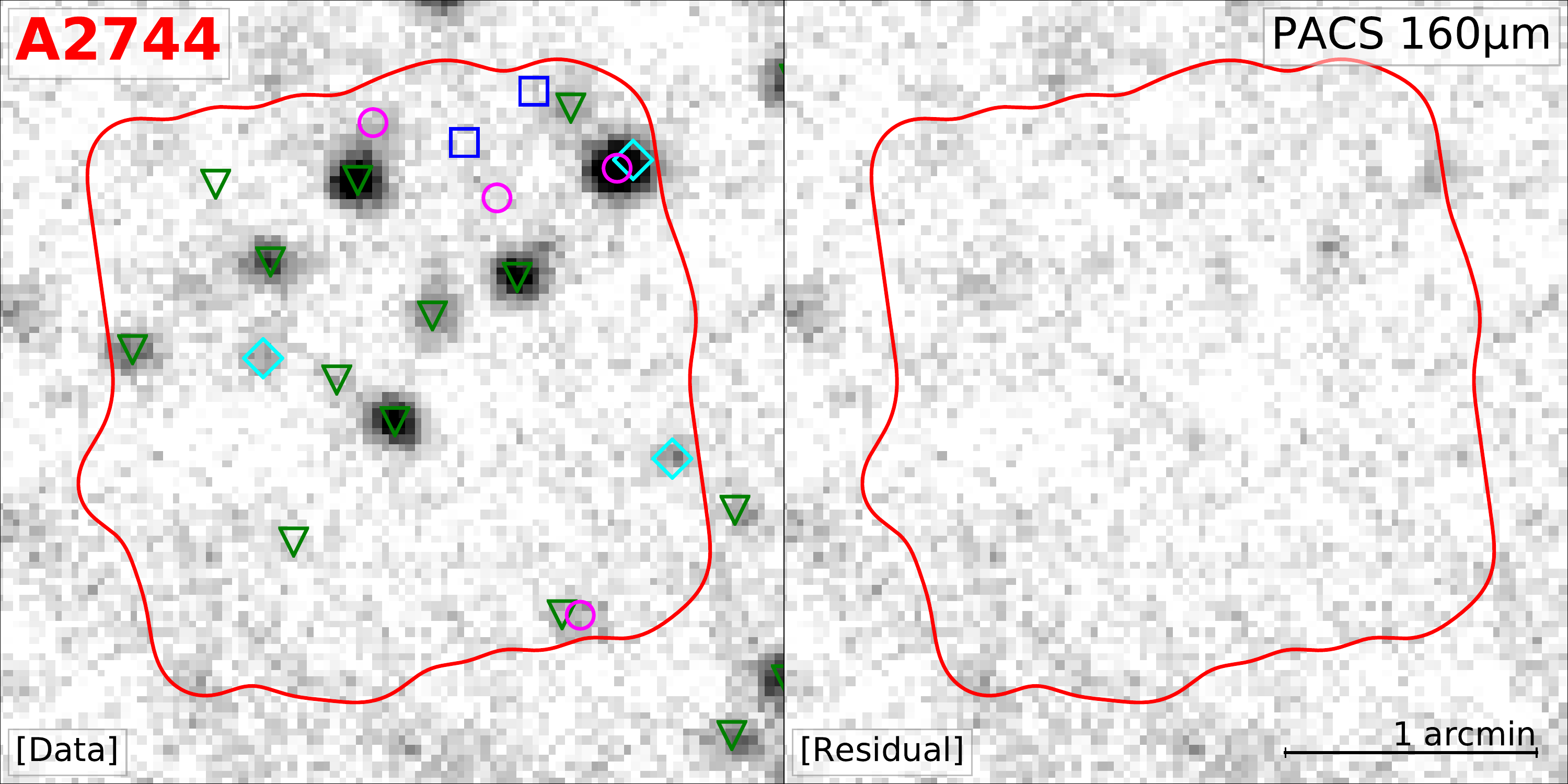}
\includegraphics[width=0.49\linewidth]{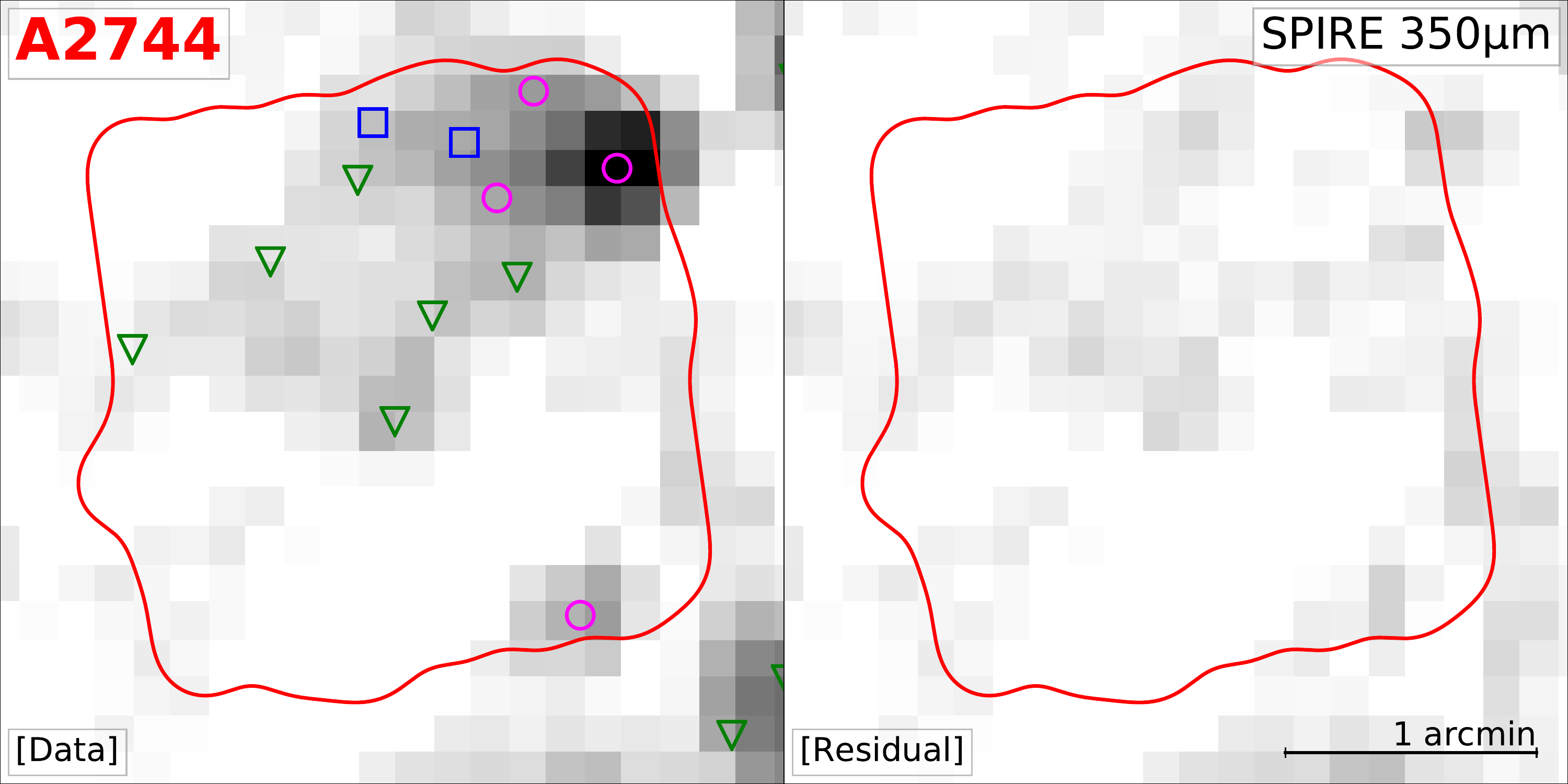}
\includegraphics[width=0.49\linewidth]{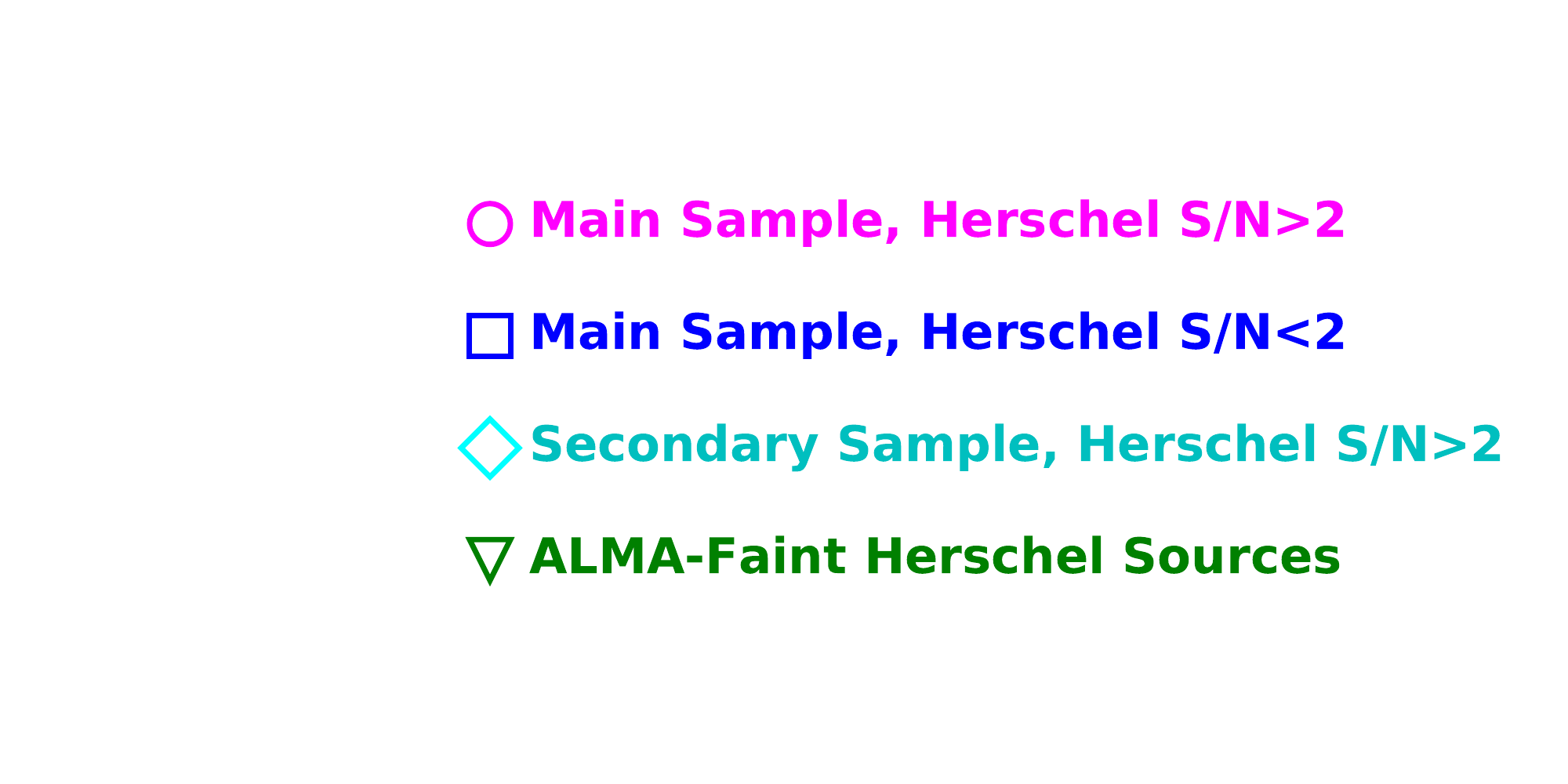}
\includegraphics[width=0.49\linewidth]{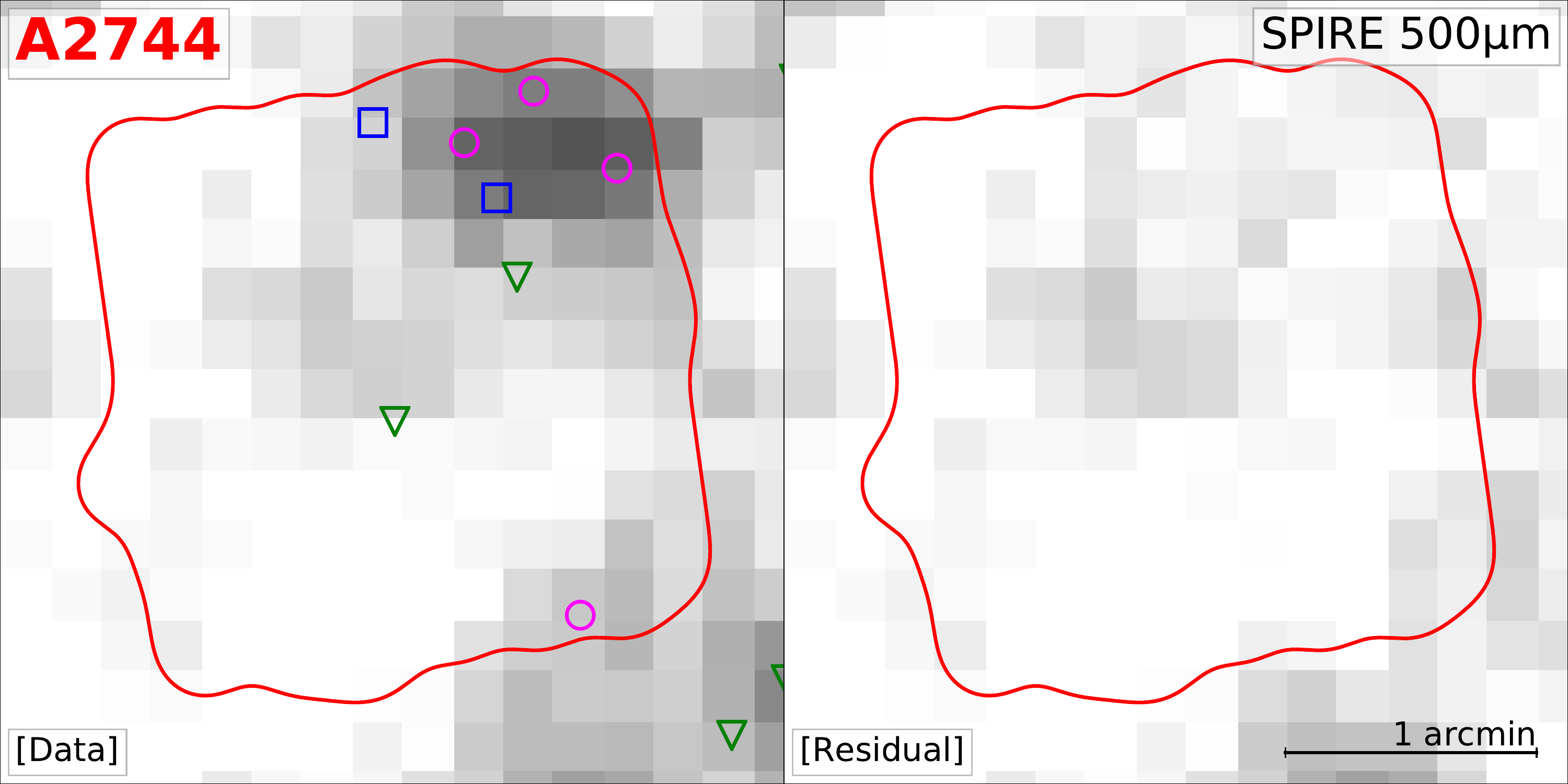}
\caption{\herschel\ images of Abell2744 (A2744) observed in the ``deep'' mode.
PACS 100/160\,\micron\ images are shown on the left and SPIRE 250/350/500\,\micron\ images are on the right.
Each panel consists of the scientific image (left) and residual image after PSF source extraction (right).
In each panel, the red patch represents the ALCS footprint (primary beam response cut at 0.25), and the magenta {circles (blue squares)} represent the secure ALCS sources (main sample) extracted at S/N$>$2 (S/N$<$2) in each \herschel\ band.
The cyan {diamonds} are tentative ALCS sources (secondary sample) extracted at S/N$>$2 in the \herschel\ bands.
The green {triangles} are ALMA-faint \herschel\ sources (see Section~\ref{ss:03c_iter}), which we also included for source extraction. 
One-arcmin scale bar is shown at the lower-right corner of each panel. 
}
\label{fig:01_deep}
\end{figure*}

\subsubsection{Background Subtraction}

We estimated and subtracted the 2D sky background of \herschel\ images using sigma-clipped statistics in each mesh of a grid that covers the whole input data frame.
This is the same algorithm used by \textsc{SExtractor} \citep{sextractor}.
The typical box size of the mesh was 21$\times$21/16$\times$16 pixels for PACS/SPIRE bands, corresponding to an area of 10/20 beams.
The size of median filter, which was applied to suppress possible overestimate of background due to bright sources, was 7$\times$7/5$\times$5 pixels for PACS/SPIRE bands (corresponding to an area of 1/2 beams).
In the cluster fields with strong extended emissions seen in the SPIRE bands (R0032, M2129, P171 and A2163), we slightly reduced the sizes of mesh and filter for a better removal of foreground large-scale emissions. 
The 2D RMS map was then created as a by-product of this background subtraction process.

\subsubsection{Neighborhood Examination}

Our \herschel\ source extraction process started from the bluest band for each cluster, i.e., 100\,\micron\ for the 18 cluster observed with PACS in the ``deep'' mode, and 250\,\micron\ for the other 15 clusters observed in the ``snapshot'' mode.
Therefore, in the bluest band, we first extracted a source catalog using the \textsc{DAOFIND} algorithm, which contains sources detected or undetected by the ALCS.
We then matched and removed the ALCS sources in this catalog within a maximum separation of 1/3 the point-spread function (PSF) FWHM (i.e., 2\arcsec\ at 100\,\micron\ and 6\arcsec\ at 250\,\micron).
Therefore, this catalog represents the 1.15\,mm-undetected \herschel\ sources that may blend with ALMA-detected sources in \herschel\ data at longer wavelengths (e.g., 500\,\micron). 
Compared with ALCS sources, these sources are likely at a lower redshift, and thus their SEDs peak at shorter wavelength and drop rapidly at longer wavelength.
We only included ALMA-faint \herschel\ sources within a separation of 30\arcsec\ (100\,\micron) or 36\arcsec\ (250\,\micron) from the ALCS sources.
This is because a larger searching area would not further increase the quality of source deblending for the ALCS sources.


To enhance the accuracy of flux extraction, we also manually added or adjusted the positions of several ALMA-faint \herschel\ sources according to the coordinates of their IRAC counterparts. 
This affects 5\% of all ALMA-faint \herschel\ sources.
For \herschel\ bands at longer wavelengths, we directly used this list of ALMA-faint \herschel\ sources in the bluest band. 
However, we find that if an ALMA-faint \herschel\ source cannot be extracted at $>$10\,mJy for ``deep''-mode clusters or $>$18\,mJy for ``snapshot''-mode clusters in a given SPIRE band, this source will be very unlikely to be detected above 2$\sigma$ at any redder band. 
Therefore, such sources will be removed from fitting at longer wavelengths.

\subsection{Iterative Herschel Photometry}
\label{ss:03c_iter}

\begin{figure}[!tb]
\centering
\includegraphics[width=\linewidth]{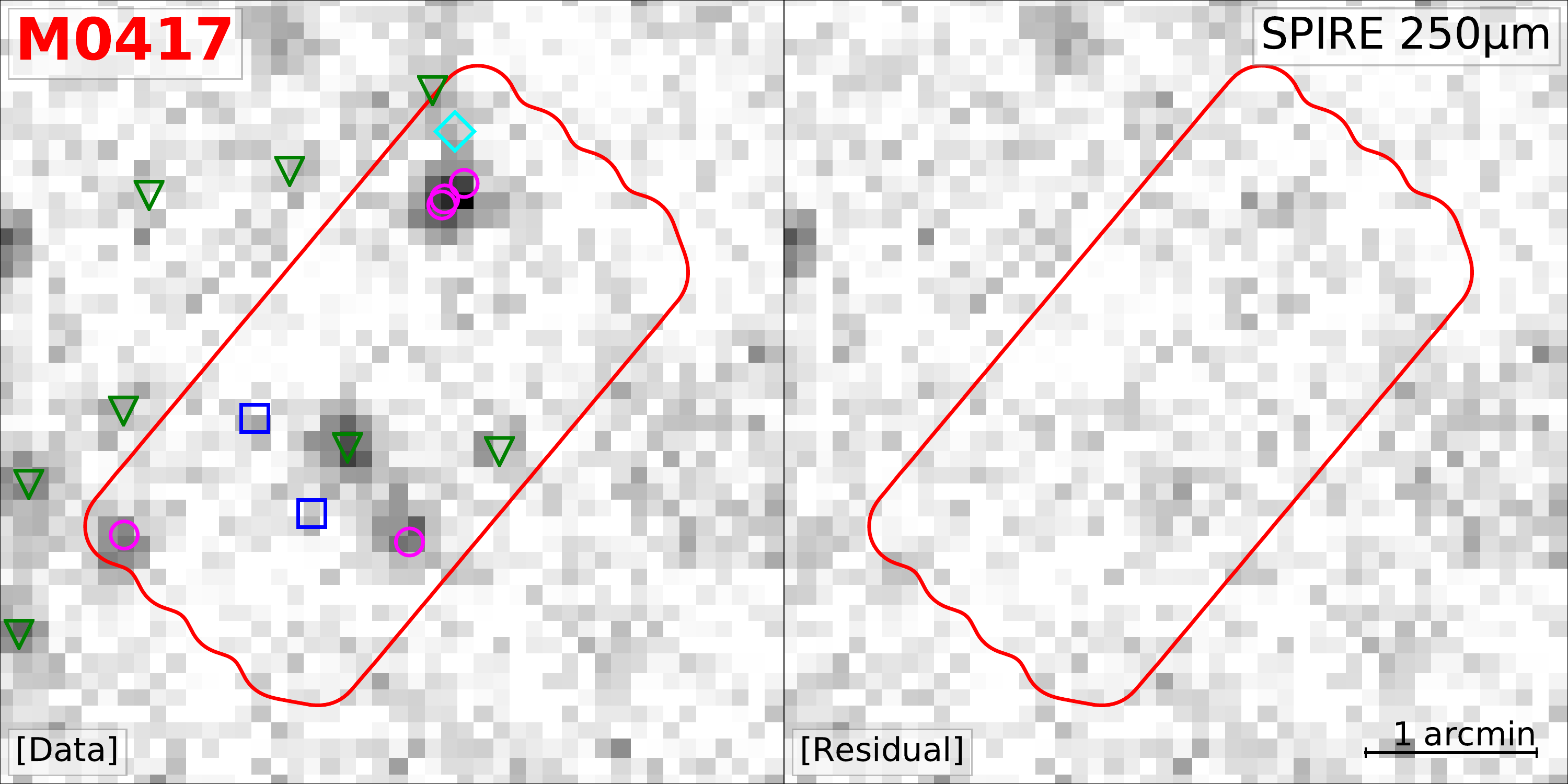}
\includegraphics[width=\linewidth]{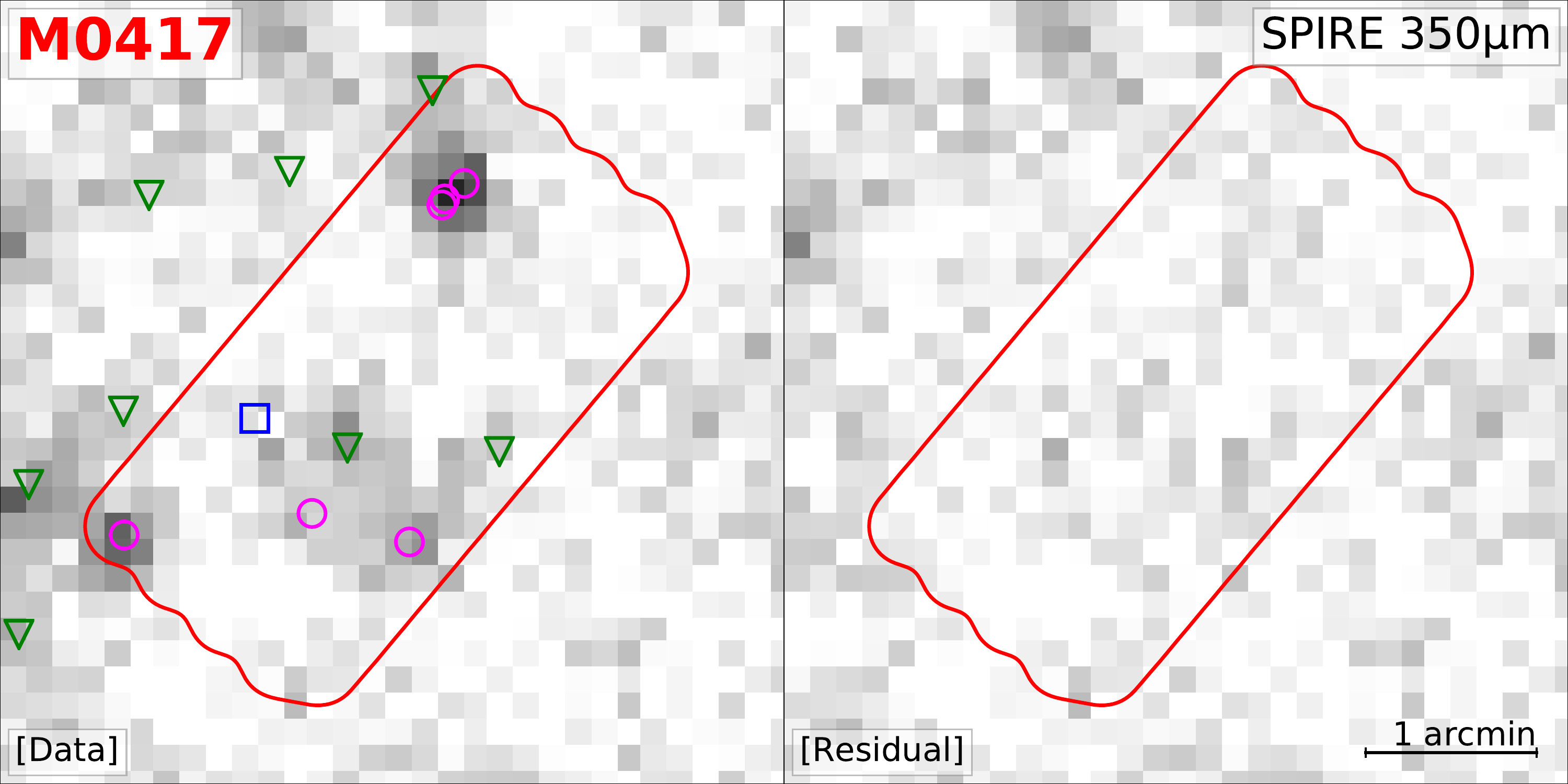}
\includegraphics[width=\linewidth]{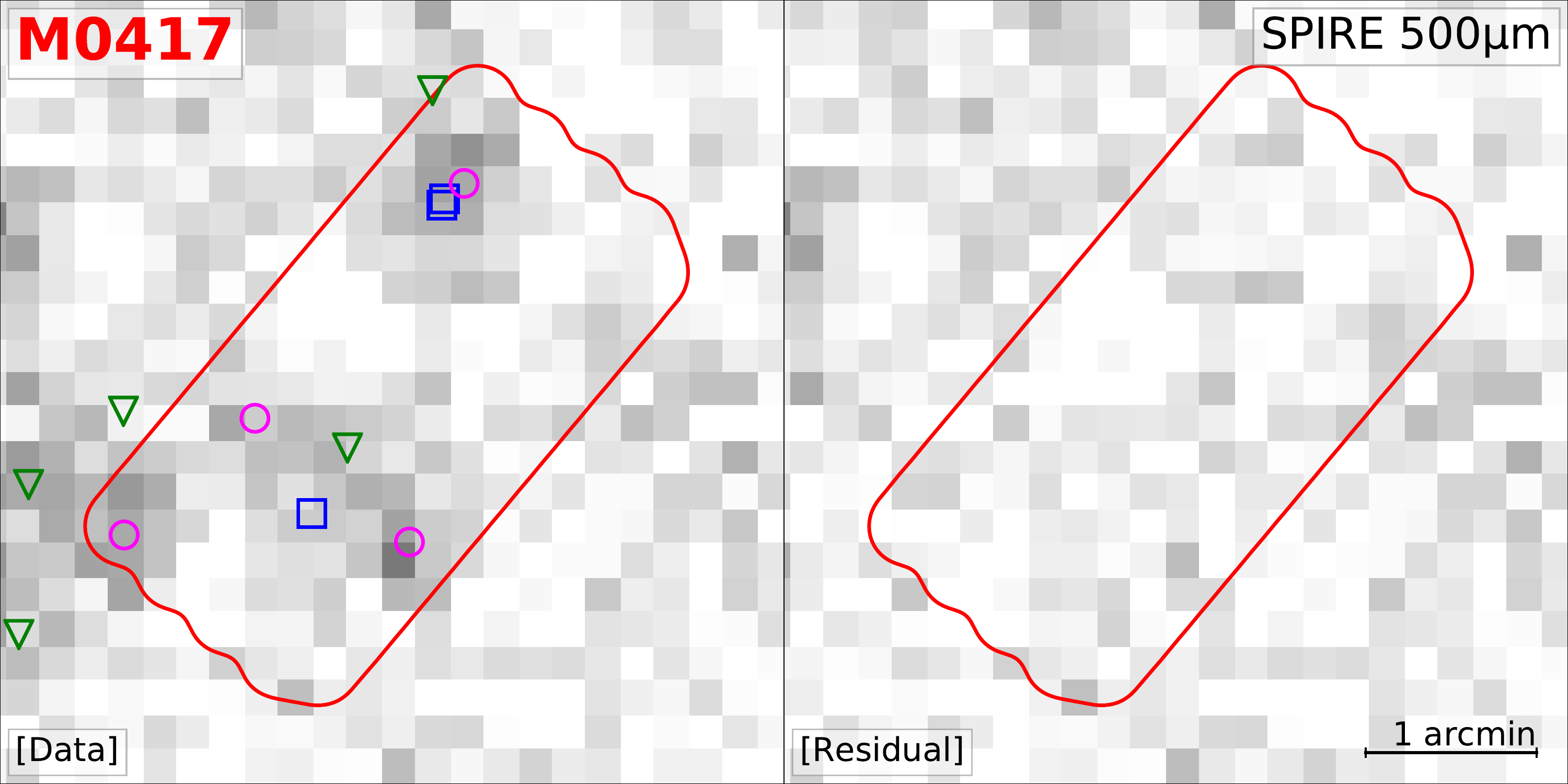}
\caption{\herschel/SPIRE images of MACSJ0417.5-1154 (M0417) observed in the ``snapshot'' mode.
The layout and symbols are the same as those in Figure~\ref{fig:01_deep}.
}
\label{fig:02_snap}
\end{figure}

\herschel\ source extraction was performed with an iterative PSF photometry approach in an increasing order of wavelength using \textsc{photutils} \citep{photutils}.
We adopted the PSF models of PACS and SPIRE from the \herschel\ Science Archive (HSA)\footnote{https://www.cosmos.esa.int/web/herschel/ancillary-data-products}. 
We also applied the spacecraft orientation angle to calculate the realistic PSF for the data taken in each cluster field. 

\subsubsection{Initial Guess of Flux Densities}

To provide initial guesses of flux densities for PSF photometry, we performed circular aperture photometry for all the ALCS sources and ALMA-faint \herschel\ sources.
The radii of the apertures were 5\arcsec, 8\arcsec, 12\arcsec, 15\arcsec, 18\arcsec\ from 100 to 500\,\micron, and the aperture correction factors (1.92, 1.90, 1.66, 1.79, 2.24) were computed based on the PSF models.
In the PACS bands, these initial guesses of flux densities were used for all the sources in the main sample and ALMA-faint \herschel\ sources.
In the SPIRE bands, these initial guesses were only used for all the $\mathrm{S/N}_\mathrm{ALMA}\equiv\max(\mathrm{S/N}_\mathrm{nat},\, \mathrm{S/N}_\mathrm{tap})\geq10$ sources and ALMA-faint \herschel\ sources due to a stronger source blending issue.

\subsubsection{Iterations of PSF Photometry}

The PSF photometry was performed in three rounds of iterations in an decreasing order of the significance of detection. 
Because most of the ALCS sources are compact in spatial extent (FWHM$\lesssim$1\arcsec), we assumed point-like profile for all of the sources to be extracted.
In each iteration, we only kept the results of those sources with positive extracted flux densities.
The uncertainties of extracted flux densities were estimated from the covariance matrix of least squares fitting.

In the first iteration, we tentatively extract sources that were (\romannumeral1) ALCS sources at $\mathrm{S/N}_\mathrm{ALMA}\geq10$, or (\romannumeral2) ALMA-faint \herschel\ sources described above.
These two types of objects should be the brightest sources seen in a given \herschel\ map. 
Therefore, an accurate flux density modeling of these sources will provide helpful guesses for the final combined source models in the whole data frame. 
We applied the rotated PSFs and modeled the flux densities at given source positions. 
The \textsc{DAOGROUP} algorithm \citep{daofind} was adopted to group sources within a separation of one beam FWHM. 
The best-fit model was then stored and used as prior information for the next iteration.

In the second iteration, we extracted sources that were (\romannumeral1) ALCS sources in the main sample, or (\romannumeral2) ALMA-faint \herschel\ sources. 
These two types of objects should be secure sources and thus their flux densities should be positive in any \herschel\ band. 
The flux priors were given by the first iteration or aperture photometry if the sources were not modeled previously. 
In SPIRE bands, the initial flux guesses of sources at $ \mathrm{S/N}_\mathrm{ALMA} < 10$ were assumed as $\mathrm{RMS}_\mathrm{map}$. 
With a similar PSF photometry routine, we modeled and subtracted sources with positive best-fit flux densities. 
In this intermediate step, most of the real sources in the current \herschel\ band have been extracted. 
The best-fit model and residual maps were recorded for the next iteration.

In the last iteration, we extracted tentative ALCS sources in the secondary sample. 
Source extraction was performed on the residual map of the second iteration, and the initial guesses of the flux densities were set to zero.
Only sources modeled with positive flux densities were kept. 
The best-fit model and residual was then recorded as part of the final products.

Figure~\ref{fig:01_deep} and \ref{fig:02_snap} display the \herschel\ scientific and residual images (i.e., before and after the source extraction) of two cluster fields observed in both the ``deep'' and ``snapshot'' modes, namely A2744 (Figure~\ref{fig:01_deep}; ``deep'') and M0417 (Figure~\ref{fig:02_snap}; ``snapshot'').
Secure, tentative ALCS sources and ALMA-faint \herschel\ sources extracted at $\mathrm{S/N}>2$ are shown as open magenta, cyan and green circles, respectively. 
No significant residual in SPIRE bands can be found within the ALCS footprint shown as the region enclosed by red solid line, although weak ring-shape residuals can be identified in PACS bands for a few very bright sources ($f\gtrsim50$\,mJy). 
This could be caused by the invalidity of point-source assumption or mismatch of PSF models, but our examination in Appendix~\ref{apd:01_xid} suggests no gain or loss of PACS flux densities through this PSF photometry routine.

\subsection{Non-detections}
\label{ss:03d_und}

\begin{figure}[!t]
\centering
\includegraphics[width=\linewidth]{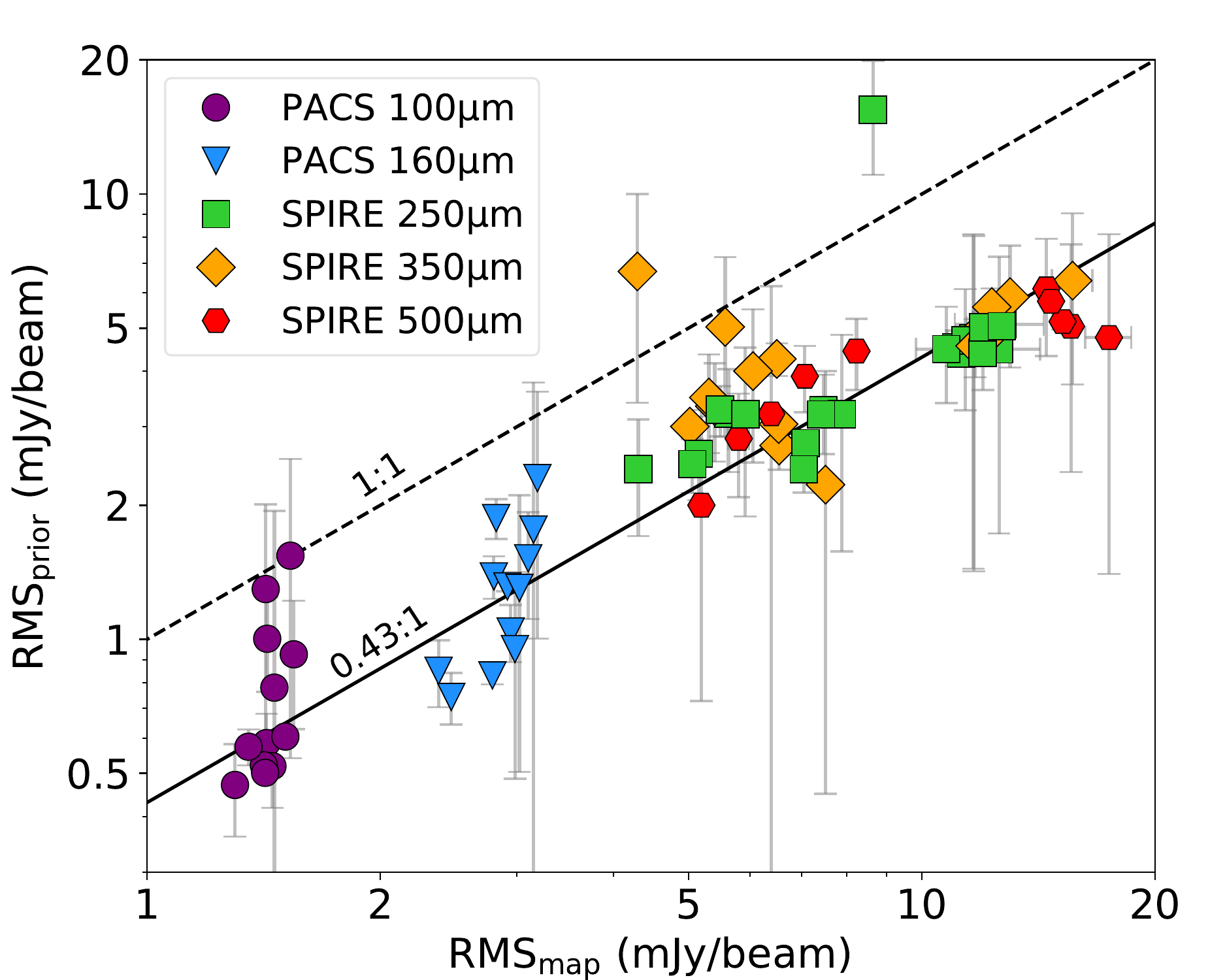}
\caption{RMS noise of \herschel\ images measured from the prior-based catalogs (\rmsp; see Section~\ref{ss:03d_und}) versus the 2D uncertainty maps (\rmsm).
Measurements obtained on the maps from 100 to 500\,\micron\ are shown as symbols with different colors as labeled in the upper-left legend.
The dashed line indicates the case in which the two RMS noises are identical, and the solid line represents the median ratio of \rmsp\ to \rmsm\ as $0.43\pm0.01$.
}
\label{fig:05_hers_unc}
\end{figure}

As a consequence of the limited depths of the \herschel\ data when compared with the deep ALCS data, a significant number of sources were not successfully extracted in the \herschel\ bands (see statistics in Section~\ref{ss:03f_stats}). 
Therefore, we only provide $3\sigma$ upper limits of their flux densities.
As pointed out by \citetalias{rawle16} and other works, because the source positions are known from the ALMA data, the actual \herschel\ flux limit of non-detections are lower than the nominal confusion noise limit \citep[e.g.,][]{nguyen10}.

Based on the flux densities and their uncertainties modeled with PSF photometry, we calculated the median flux density uncertainty of the extracted secure sources in each band and each cluster field (main ALCS sample, \herschel\ $\mathrm{S/N} > 2$).
These uncertainties were obtained based on positional priors using the covariance matrix of least squares PSF modeling, and thus we define them as \rmsp.
Table~\ref{tab:02_unc} presents all the measured \rmsp\ along with the \rmsm\ which is directly measured from the 2D uncertainty map within the ALCS footprint.
We also compare the \rmsp\ with \rmsm\ measured for all our \herschel\ data in Figure~\ref{fig:05_hers_unc}, and we find a median ratio of \rmsp\ to \rmsm\ as $0.43\pm0.01$ in the \herschel/SPIRE bands.
Similar value can also be found for the PACS bands.
This means that with prior knowledge of source positions, the actual $3\sigma$ limit of non-detection is around 1.3 times of the local background \rmsm.
Such an upper limit is adopted for all the \herschel\ non-detections in this work.

The median $3\sigma$ depths derived for the 18 clusters observed in the ``deep'' mode are 7.5, 7.6 and 8.2\,mJy at 250, 350 and 500\,\micron.
These are consistent with the $3\sigma$ RMS of deblended flux densities using the cross identification procedure reported by \citet[for HerMES fields]{roseboom10}, and only slightly lower than the reported value in \citet{swinbank14} at 500\,\micron, which also included ALMA positional priors for deblending.

\subsection{Special Sources}
\label{ss:03e_special}

\begin{figure*}[!ht]
\centering
\includegraphics[width=0.49\linewidth]{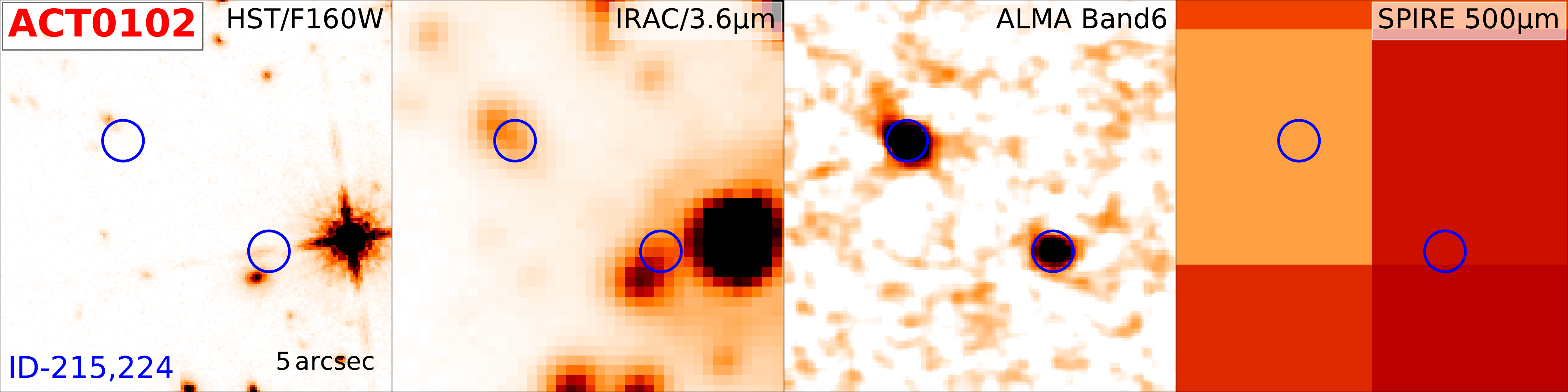}
\includegraphics[width=0.49\linewidth]{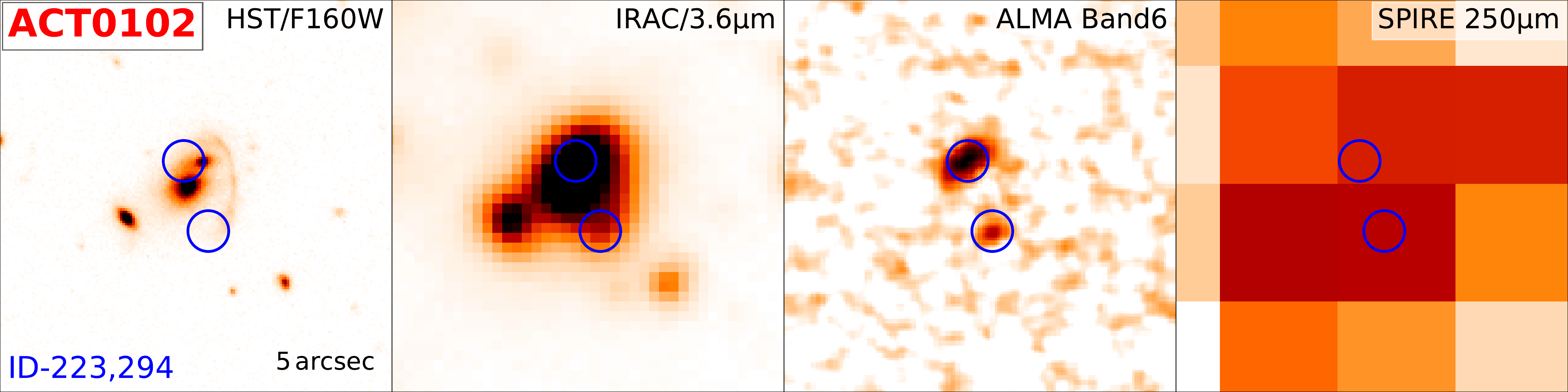}
\includegraphics[width=0.49\linewidth]{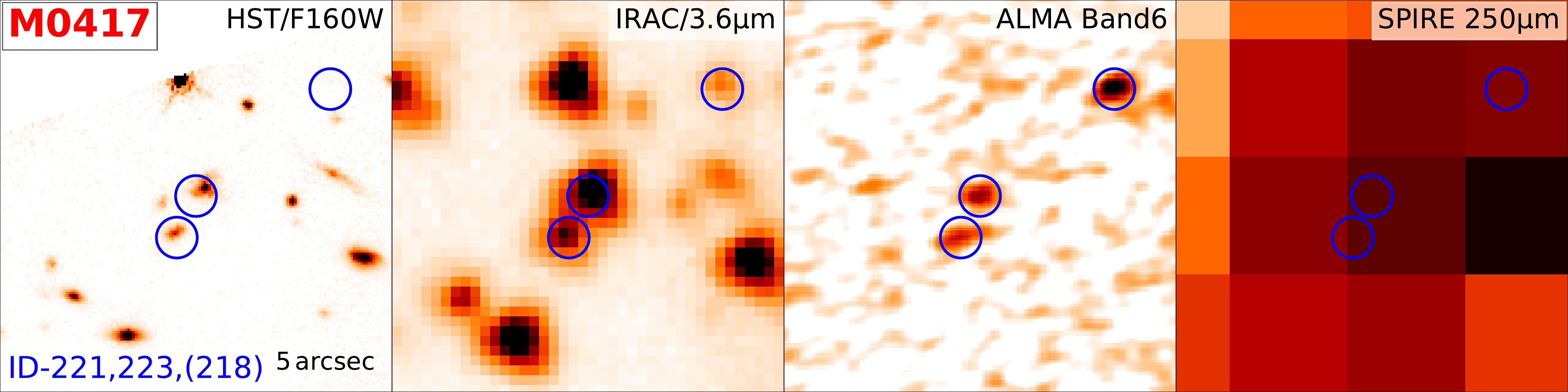}
\includegraphics[width=0.49\linewidth]{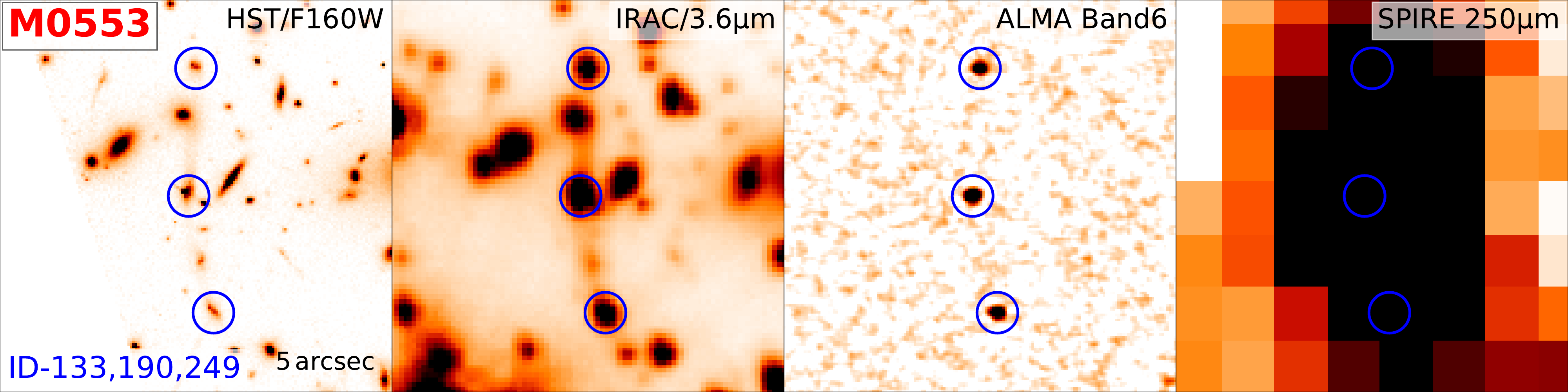}
\includegraphics[width=0.49\linewidth]{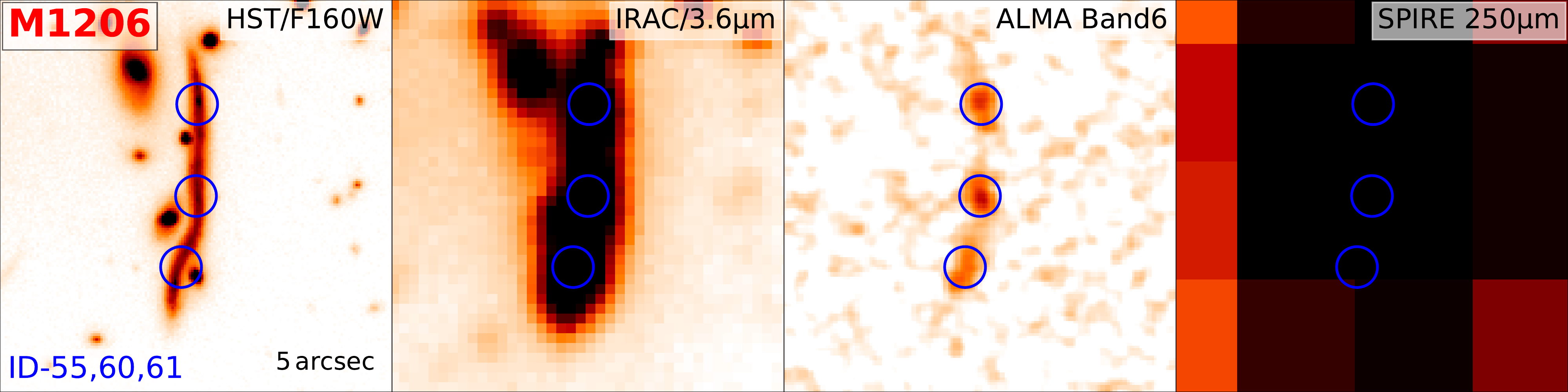}
\includegraphics[width=0.49\linewidth]{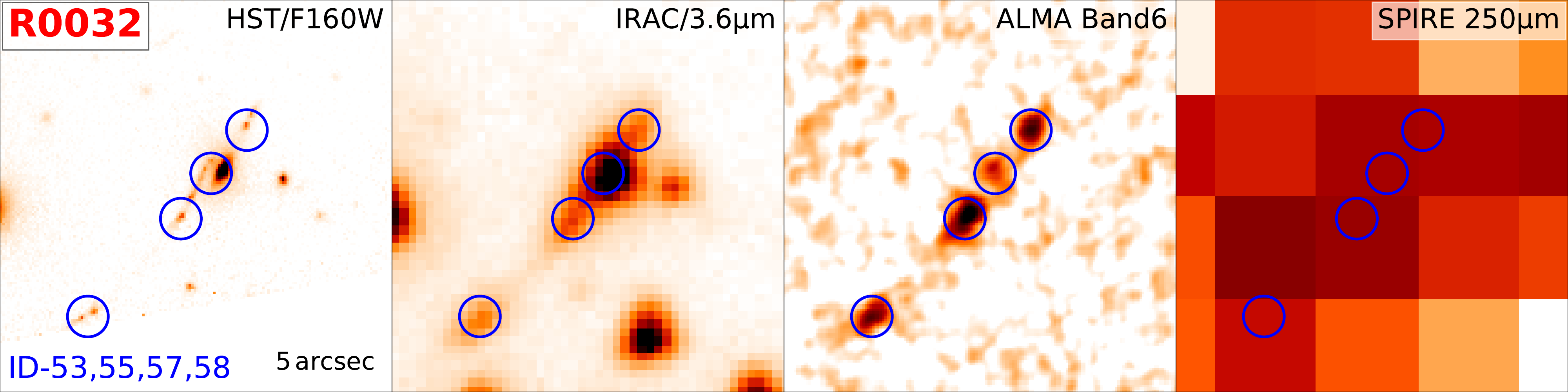}
\includegraphics[width=0.49\linewidth]{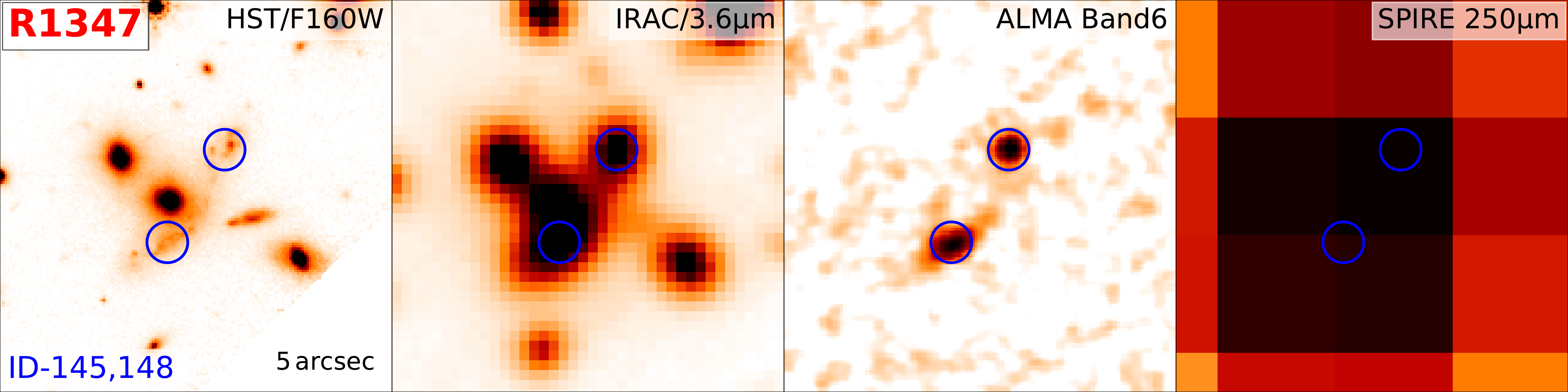}
\includegraphics[width=0.49\linewidth]{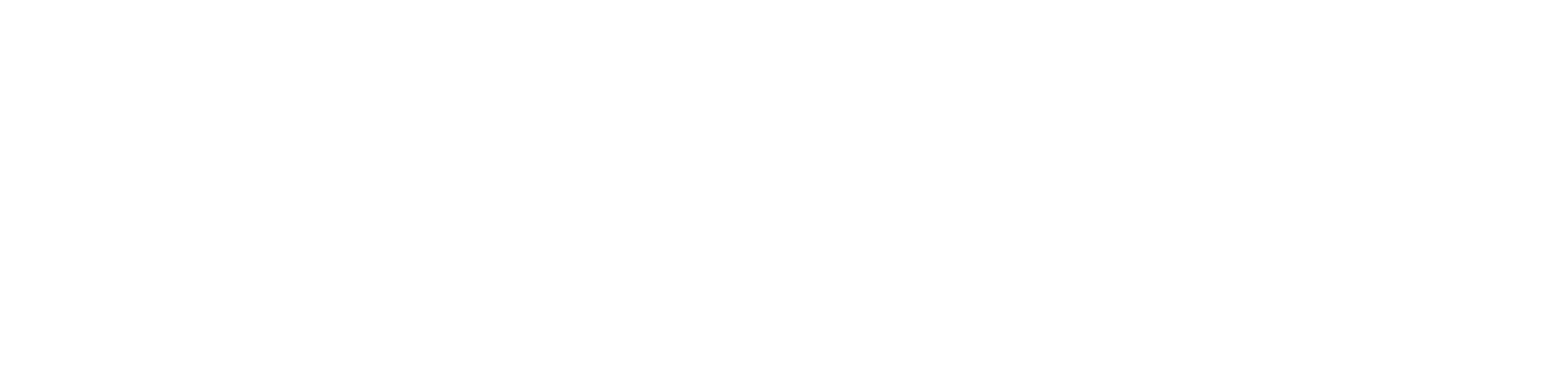}
\caption{The seven special cases of sources with close companions that we redistributed the \herschel\ flux densities based on the ALMA flux density ratios (see Section~\ref{ss:03e_special}). 
In each postage stamp image, we show the \hst\ WFC3-IR/F160W image, {\spitzer\ IRAC 3.6\,\micron\ image}, ALMA Band-6 image at native resolution and SPIRE 250\,\micron\ image (500\,\micron\ for ACT0102-ID215/224) from the left to right.
ALCS sources are labeled with open blue circles, and cluster names, ALMA source IDs and scale bars are shown in the corners of the F814W images.
}
\label{fig:a1_sp}
\end{figure*}

Several ALCS sources show secure close companions (i.e., angular separation less than 6\arcsec, which is one third of the beam FWHM at 250\,\micron) at $\mathrm{S/N} \geq 5$ in the ALMA maps.
Five of these seven systems have already been confirmed as lensed arcs or multiply lensed systems with \hst\ or ALMA data.
Due to the coarse resolution of the \herschel\ data, especially those of SPIRE, the flux density ratios among these source groups might be incorrectly modeled in Section~\ref{ss:03c_iter}. 

In this step, we re-distributed the \herschel\ flux densities of these source groups according to their ALMA flux density ratios.
If a source was resolved on the native-resolution ALMA map with a major-axis FWHM less than 3\arcsec\ (morphological parameters modeled with \textsc{casa imfit}; \citealt{sfprep}), we adopted the ALMA flux density measured with a circular aperture of $r = 2\arcsec$.
For sources with larger FWHMs, we adopted the best-fit ALMA flux densities derived from surface brightness profile modeling (assuming 2D Gaussian profile with \textsc{imfit}).
For unresolved sources, we used the peak flux densities per beam measured on the 2\arcsec-tapered maps.
{We note that the redistribution of \herschel\ flux densities assumes a fixed far-IR SED shape among blended sources in each group. 
Only one source from each blended group is considered for the discussions of dust temperature in Section~\ref{ss:06d_temp}.
}

We redistributed the \herschel\ fluxes for all the ALCS sources in the main sample and within a separation of 6\arcsec.
This includes:
(\romannumeral1) ACT0102-ID215/224 \citep[lensed galaxy pair known as ``la Flaca'';][]{linder15, wu18, caputi21},
(\romannumeral2) ACT0102-ID223/294 \citep[lensed galaxy pair;][]{wu18},
(\romannumeral3) M0417-ID221/223 (two ALMA sources at $\mathrm{S/N}\gtrsim 6$ with a separation of 2.3\arcsec; \citealt{kkprep}),
(\romannumeral4) M0553-ID133/190/249 \citep[triply lensed arc;][]{ebeling17,sun21a},
(\romannumeral5) M1206-ID55/60/61 \citep[known as the ``Cosmic Snake'';][]{ebeling09,cava18,mdz19},
(\romannumeral6) R0032-ID53/55/57/58 (lensed arc; \citealt{mdz17}),
(\romannumeral7) R1347-ID145/148 (IR-bright lensed arc in R1347).
In Figure~\ref{fig:a1_sp} we display the \hst, ALMA and SPIRE images of all these source groups. 
Note that we only redistributed the 500\,\micron\ flux densities of ACT0102-ID215/224 because of a moderate separation (9\farcs3).



\subsection{Statistics of \herschel\ Detections}
\label{ss:03f_stats}

\begin{figure*}[!thb]
\centering
\includegraphics[width=0.9\linewidth]{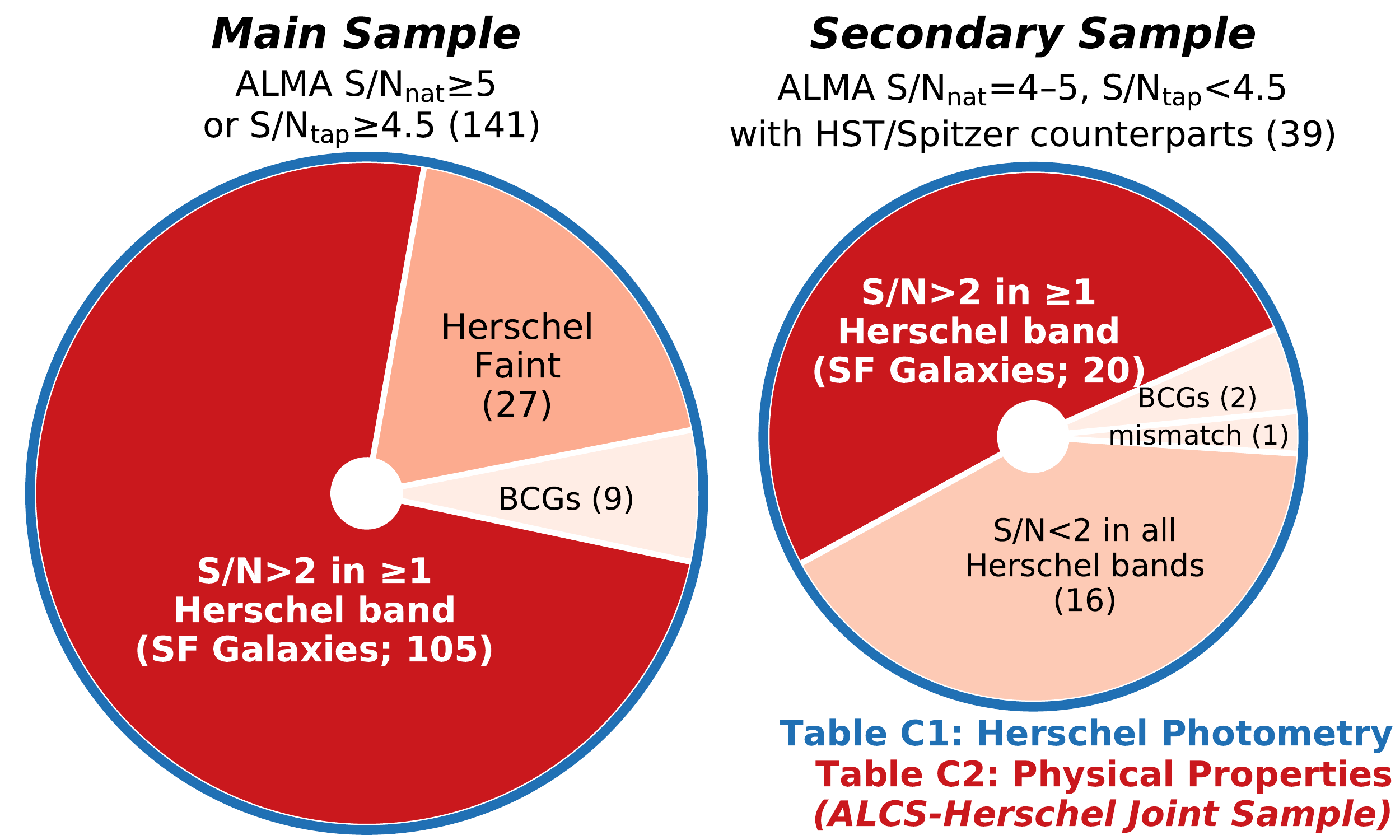}
\caption{Summary of ALMA and \herschel\ detections and sample definition. 
The \textit{Main Sample} includes all 141 secure ALCS sources detected at either $\mathrm{S/N}_\mathrm{nat} \geq 5$ in the native-resolution maps or $\mathrm{S/N}_\mathrm{tap} \geq 4.5$ in the 2\arcsec-tapered maps.
The \textit{Secondary Sample} includes 39 tentative ALCS sources ($\mathrm{S/N}_\mathrm{nat} = 4 - 5$ and $\mathrm{S/N}_\mathrm{tap} < 4.5$) showing \hst/\spitzer\ counterparts within 1\arcsec\ offset.
Table~\ref{tab:03_phot} presents the \herschel\ photometric catalog of these 180 sources (shown as blue circles).
Among 141 sources in the main sample, 105 of them are star-forming galaxies which are detected at $\mathrm{S/N} > 2$ in at least one \herschel\ band.
Together with 20 star-forming galaxies in the secondary sample above the same \herschel\ detection threshold, these galaxies are presented in Table~\ref{tab:04_sed} for their physical properties (shown as red filled regions).
43 sources that are undetected in all \herschel\ bands (including 27 ``\herschel-faint'' galaxies in the main sample; see Section~\ref{ss:03f_stats} and Section~\ref{ss:04c_dropout}), 11 BCGs and one mismatched source (A2537-ID06) are not included in Table~\ref{tab:04_sed} (shown as shallow filled regions).
 }
\label{fig:catalog}
\end{figure*}

\begin{figure}[!ht]
\centering
\includegraphics[width=\linewidth]{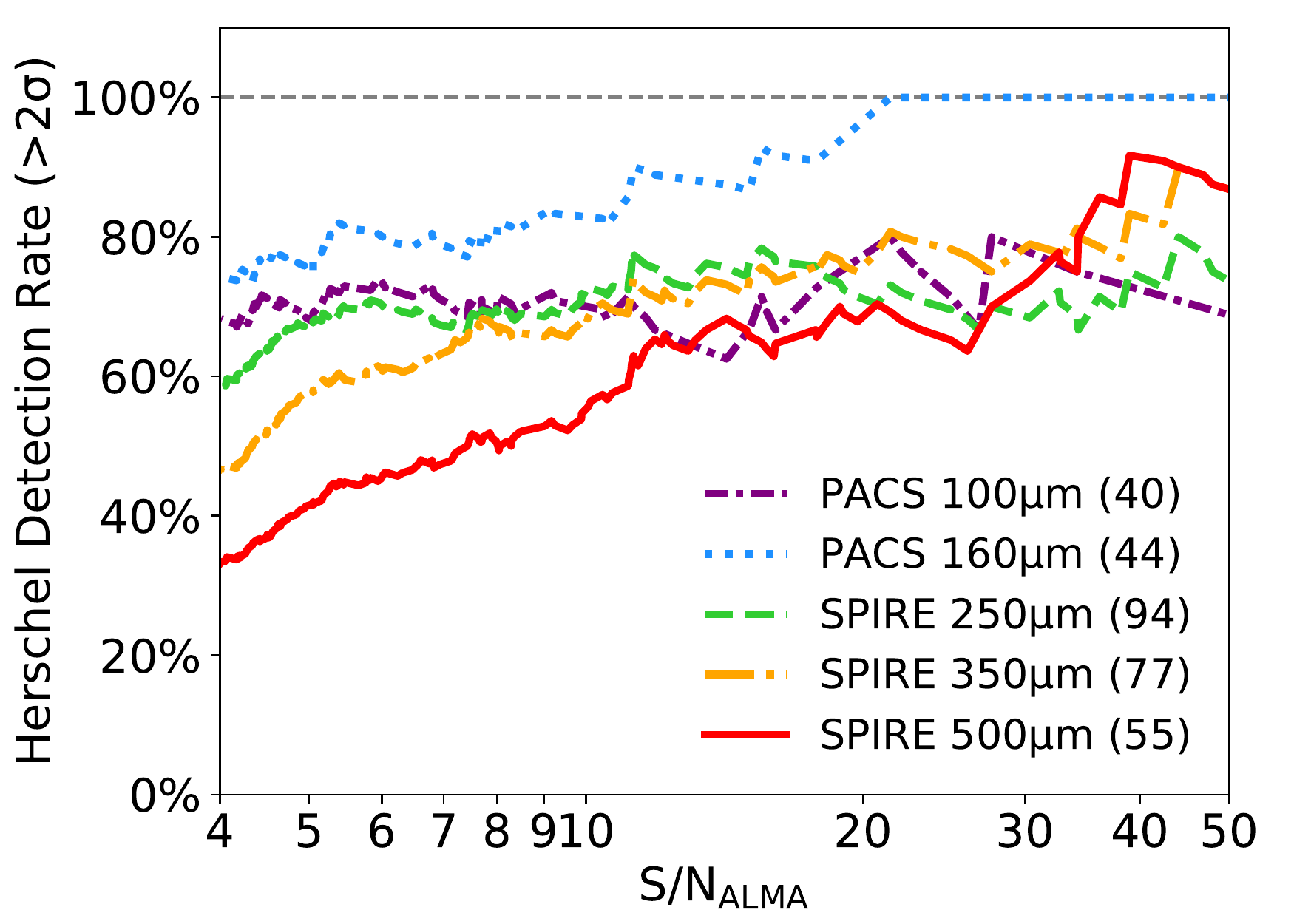}
\caption{Detection ($>2\sigma$) rates of ALCS sources as functions of ALMA S/N cut in all five \herschel\ bands.
Line color and style of each band are shown in the lower-right legend, and values in parentheses indicate numbers of $>2\sigma$ \herschel\ detections among secure ALCS sources ($\mathrm{S/N}_\mathrm{ALMA} \geq 5$). 
}
\label{fig:03_detect}
\end{figure}


By comparing our PSF photometric results with those derived with aperture photometry, PSF photometry using mid-IR priors \citepalias{rawle16} and different software (\textsc{xid+}; \citealt{hurley17}) as presented in Appendix~\ref{apd:01_xid}, we confirmed the quality of our \herschel\ flux density measurements.
In Table~\ref{tab:03_phot} we present the \herschel\ photometric catalog of 180 ALCS sources in both the main and secondary sample (Section~\ref{ss:02a_sample}).
The definition of these samples is also illustrated in Figure~\ref{fig:catalog}.

Figure~\ref{fig:03_detect} shows the \herschel\ detection rates as functions of ALMA S/N cut from 100 to 500\,\micron.
The rates of 100\,\micron\ and 250\,\micron\ detection ($>2\sigma$) are almost constant at $\mathrm{S/N}_\mathrm{ALMA} \geq 5$.
However, the detection rates at long wavelength (350 and 500\,\micron) are clearly correlated with the significance of ALMA sources.
This is likely the consequence of (\romannumeral1) a larger beam size and stronger blending effect towards longer wavelength, and (\romannumeral2) a decreasing fraction of high-redshift sources ($z>3$) towards lower 1.15\,mm flux density \citep[e.g.,][]{bethermin15,casey18a,popping20}.
{We also note one caveat that certain extragalactic ALMA surveys of rest-frame UV/optical-selected galaxies may have a selection bias against highly dust-obscured sources at $z\gtrsim3$.
For such surveys, the most accurate measurement of the redshift distribution can be obtained after the sample is spectroscopically complete  \citep[cf.,][]{reuter20,chen22}.
}

Among the total of 141 secure ALCS sources at $\mathrm{S/N}_\mathrm{nat} \geq 5$ or $\mathrm{S/N}_\mathrm{tap} \geq 4.5$ (58 of which fall in the PACS coverage), we successfully extracted 40, 44, 94, 77 and 55 sources at 100, 160, 250, 350 and 500\,\micron\ at above a $2\sigma$ significance, respectively.
The SPIRE detection rate in the ``deep''-mode clusters is higher than that in the ``snapshot''-mode clusters by $\sim10\%$.
113 (99) ALCS sources were detected at $>2\sigma$ ($>3\sigma$) in one \herschel\ band at least (including eight BCGs), and 91 sources were detected at $>2\sigma$ in two \herschel\ bands at least.

Only 28 secure ALCS sources (20\% of the total 141 sources) can not be extracted at S/N$>$2 in any \herschel\ band. 
The 16-50-84th percentiles of 1.15\,mm flux densities of these sources (0.45--0.92--1.41\,mJy) are smaller than those of \herschel-detected sources (0.66--1.22--3.00\,mJy).
Except for M1206-ID58 as a brightest cluster galaxy, BCG, at $z=0.440$, we refer to the remaining 27 sources as \herschel-faint galaxies in later analysis (Section~\ref{ss:04c_dropout}; also called as \herschel-dropout galaxies in \citealt{boone13}).



Among the 39 tentative ALCS sources at $\mathrm{S/N}_\mathrm{nat} = 4-5$ and $\mathrm{S/N}_\mathrm{tap} < 4.5$, 22 of them can be extracted above a $2\sigma$ significance in at least one \herschel\ band, including one BCG and one mismatched source.
The remaining 17 sources are undetected in any \herschel\ band including one BCG (R0032-ID162).
These sources are excluded for further analysis {because of a higher false ID rate}.

We further justify such a $2\sigma$ detection threshold by calculating the joint probability of spurious sources through a $\chi^2$ statistic of the detection significance in all available \herschel\ bands. 
For all ALCS sources extracted at S/N\,$>$\,2 in any \herschel\ band, only five sources ($4\pm2$\%) exhibit $p$-values of spurious detection at above 0.01, including three secure ALCS sources (ACT0102-ID118, P171-ID69, P171-ID177) and two tentative sources (AS295-ID269, M1115-ID33).
The largest $p$-value\,$=0.03$ is seen for source P171-ID69 with $\mathrm{S/N}_\mathrm{ALMA}=$\,19.7.
Therefore, we conclude that with the single-band $2\sigma$ detection threshold, the number of spurious \herschel\ sources will be on the order of unity.

\section{SED Fitting}
\label{sec:04_sed}

\subsection{Methodology}
\label{ss:04a_method}

\begin{figure*}
\centering
\includegraphics[width=0.424\linewidth]{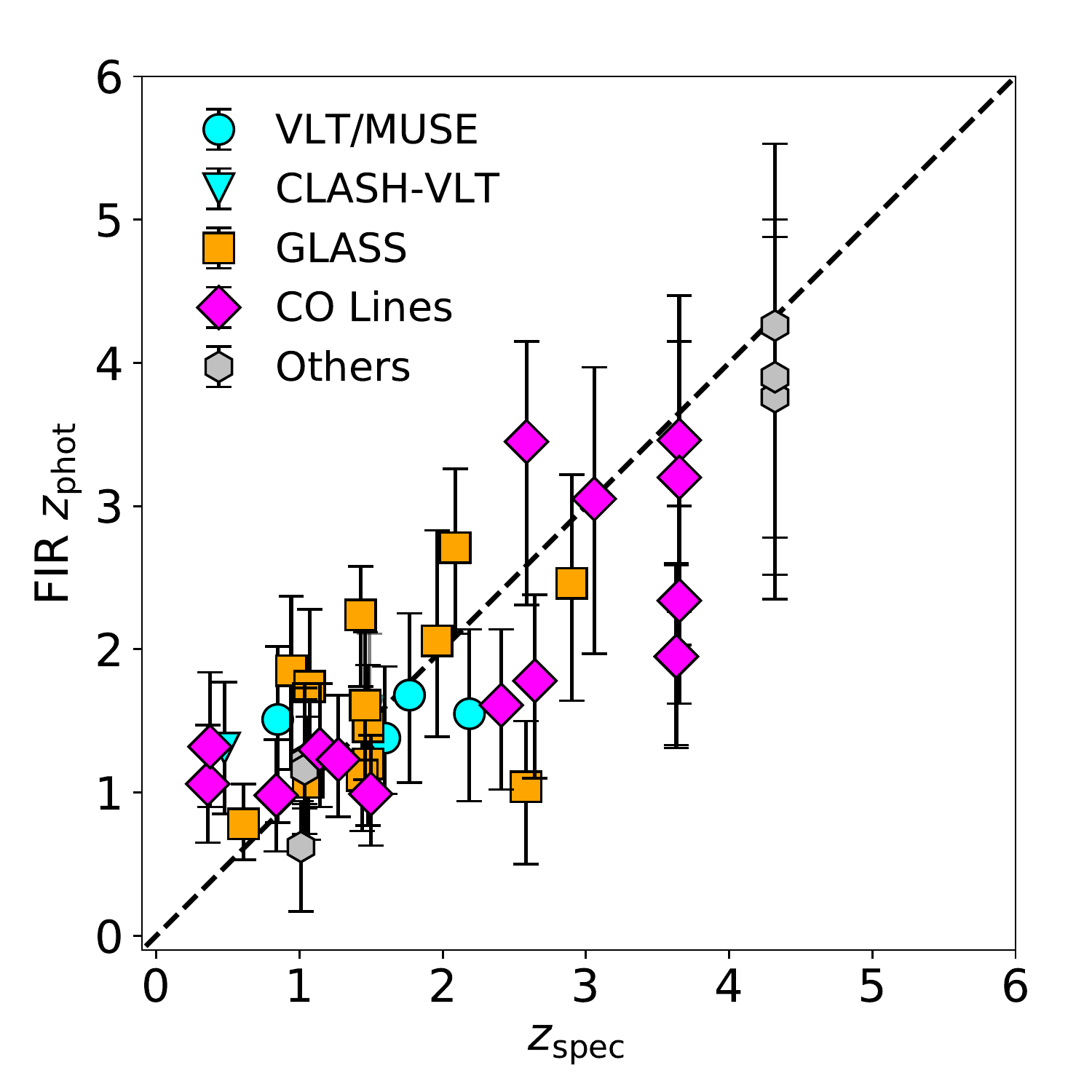}
\includegraphics[width=0.566\linewidth]{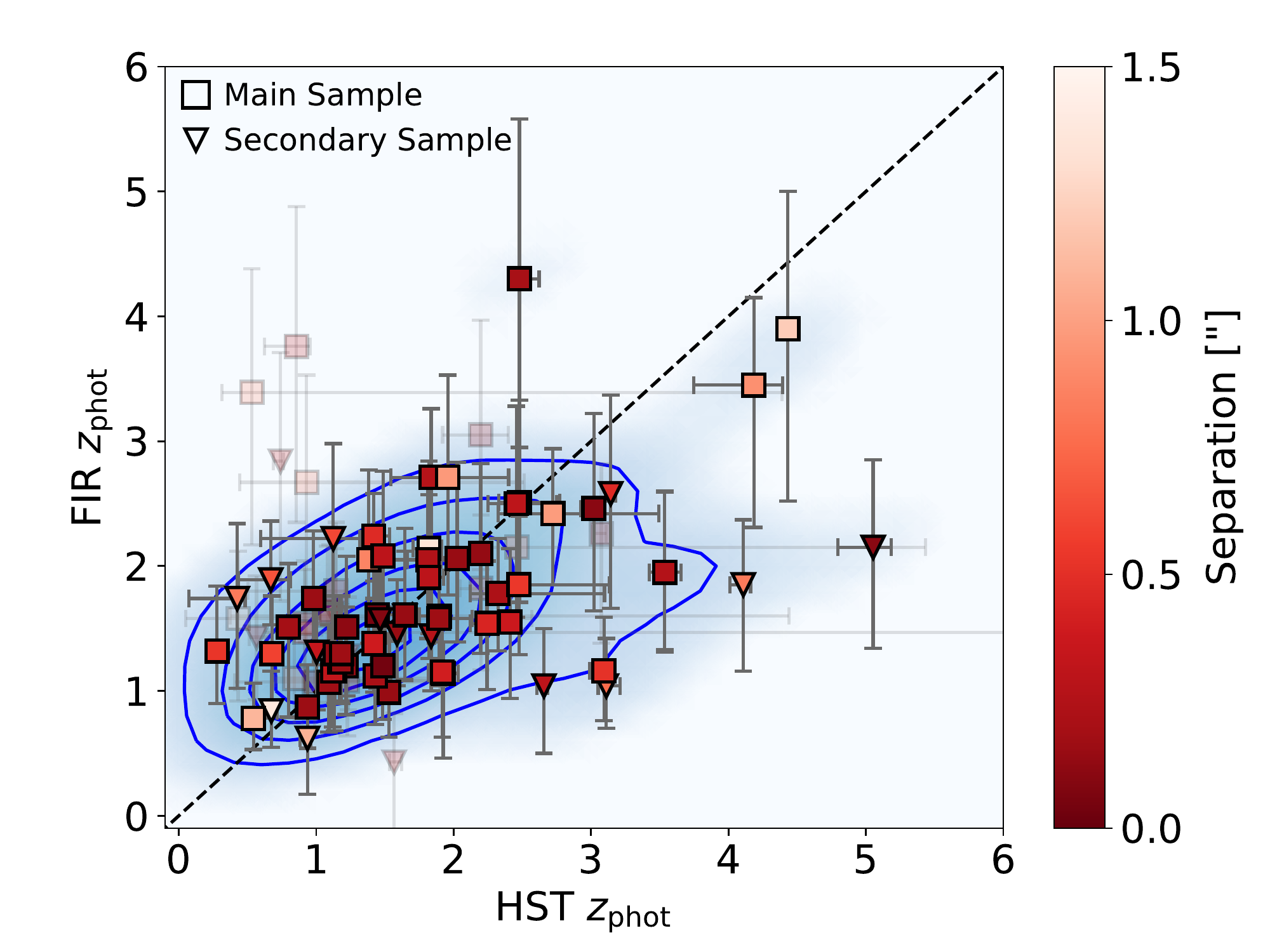}
\vspace{-5mm}
\caption{
\textit{Left}: Comparison of far-IR photometric redshifts and spectroscopic redshifts (see the references of \zsp's in Table~\ref{tab:04_sed}).
Dashed black line indicates the cases where far-IR \zph's are identical to \zsp's.
\textit{Right}: Comparison of photometric redshifts obtained through far-IR (\herschel\ and ALMA) and optical/near-IR (\hst) SED fitting.
Squares denote ALCS sources in the main sample, and downward triangles denote those in the secondary sample.
All the sources are color-coded with their observed offsets between the \hst\ and ALMA counterparts.
Blue density map and contours represent the overall distribution of the cross-matched sample, except for 22 sources that are either out of \hst/WFC3-IR coverage or detected at above $2\sigma$ in only one \herschel\ band (shown with transparent symbols).
Dashed black line indicates the cases where \zph's obtained through far-IR and \hst\ SED fitting are identical.
}
\label{fig:08_photz}
\end{figure*}

We perform far-IR SED modeling of our sample with the best available  redshifts ($z_\mathrm{best}$) using \textsc{magphys} \citep{dacunha08,dacunha15}.
Here the $z_\mathrm{best}$ is either spectroscopic redshift ($z_\mathrm{spec}$), published \hst-derived $z_\mathrm{phot}$ (Section~\ref{ss:02e_redshift}) or far-IR $z_\mathrm{phot}$ (priority from high to low). 
Redshift uncertainty is propagated into the uncertainties of derived physical properties through a Monte-Carlo sampling of $z_\mathrm{phot}$ likelihood when $z_\mathrm{spec}$ is not available.
In order to derive and validate far-IR $z_\mathrm{phot}$ for sources without $z_\mathrm{spec}$ or \hst\ \zph, we also perform simultaneous far-IR SED fitting and photometric redshift estimate of our sample using \textsc{magphys+photo-z} \citep{battisti19}, the photo-z extension of \textsc{magphys}.

\textsc{magphys} assumes a \citet{chabrier03} initial mass function (IMF), a continuous delayed exponential star-formation history (SFH) with random starburst, and energy balance between dust absorption in the UV and re-emission in the IR.
At far-IR wavelengths, the dust model assumed by \textsc{magphys} includes a warm (30--80\,K) and a cold (20--40\,K) component, and the prior distribution of luminosity-weighted dust temperature ($T_\mathrm{dust}$) peaks around 37\,K with a $1\sigma$ dispersion of $\sim$\,20\%.
Such a  dust temperature is comparable to those of widely adopted spectral templates at around median $L_\mathrm{IR}\sim10^{12}$\,\lsun\ including \citet{chary01} and \citet[which is based on \citealt{draine07} model]{magdis12}.
For a full description of the models and parameters assumed by \textsc{magphys}, see \citet{dacunha08,dacunha15} and \citet{battisti19}.
Here, we only include five or three bands of \herschel\ data and ALMA 1.15\,mm flux densities for our SED modeling.
Further optical/near-IR counterpart matching, uniform \hst\ and \spitzer\ photometry and panchromatic SED fitting will be presented by another paper of the ALCS collaboration, {and certain conclusions on $T_\mathrm{dust}$ and redshift distribution depending on the far-IR \zph's may be further revised}. 

Here, we highlight several caveats of our SED modeling obtained with \textsc{magphys}. 
First of all, 
the accuracy of far-IR $z_\mathrm{phot}$ is subject to a well-known degeneracy between redshift and dust temperature, typically showing an error around $\Delta z \sim 0.2 (1+z)$.
In addition, \textsc{magphys+photo-z} assumes a prior redshift distribution peaking at $z \sim 1.7$, and in practice we find that such a prior will lead to an artificial shift of $z_\mathrm{phot}$ estimate towards such a redshift.
To address this issue, we adopt a uniform redshift prior instead.
Furthermore, the non-thermal emission of BCGs seen at 1.15\,mm cannot be properly modeled, and thus their boosted ALMA flux densities (e.g., M1931-ID41, \citealt{forgarty19}) 
will lead to a wrong estimate of IR luminosity and SFR.
Therefore, we do not perform SED modeling for all the known BCGs.
We also note that only three lensed ALCS sources are detected in X-ray {among 31 clusters fields with publicly available \textit{Chandra} data (A370-ID110, M0416-ID117, M0329-ID11; these sources will be discussed by Uematsu et al.\ from the collaboration).
We also estimate the upper limit of AGN contribution to the derived IR luminosities.
We assume the SKIRTOR model \citep{stalevski12,stalevski16} for a type-2 AGN SED with an inclination angle of 70\arcdeg.
To estimate the upper limit of  X-ray luminosity, we used a simple absorbed power-law model with a photon index of 1.9 and an intrinsic absorption of $\log(N_\mathrm{H}/\si{cm^2})=23$.
The median X-ray luminosity is $L_X<2\times10^{43}$\,\si{erg.s^{-1}} for X-ray undetected sources, corresponding to an IR luminosity of $\lesssim 6\times 10^9$\,\lsun. Therefore, the AGN} contamination will not be a concern for the majority ($\gtrsim$\,95\%) of ALCS sources, {but we also note that in the case of a Compton thick AGN ($\log(N_\mathrm{H}/\si{cm^2})>24$), the upper limit on $L_X$ can be larger by more than an order of magnitude.}
Finally, \textsc{magphys} {can only provide weak} constraints on the physical properties of \herschel-faint galaxies individually, which are specifically discussed in Section~\ref{ss:04c_dropout}. 

Table~\ref{tab:04_sed} presents a summary of the best-fit galaxy properties of 125 ALCS sources detected at $\mathrm{S/N} \geq 4$, including 47 sources that are spectroscopically confirmed and additional 42 sources with catalogued \hst\ \zph.
This sample, further referred to as the ALCS-\herschel\ joint sample, includes 105 secure (i.e., the main sample) and 20 tentative ALCS sources (the secondary sample) detected above $2\sigma$ in one \herschel\ band at least, except for 11 BCGs and one special source (A2537-ID06) due to the poorness of SED fitting.
A2537-ID06 is only detected with ALMA at S/N=4.2 and offset from a passive cluster dwarf galaxy by 0\farcs9, and therefore it is likely a false detection with \herschel/SPIRE fluxes coming from an ALMA-faint \herschel\ source.
The definition of this sample is also visualized as the red filled regions in Figure~\ref{fig:catalog}.
The postage stamp images and best-fit far-IR SEDs of these 125 sources are shown in Appendix~\ref{apd:02_fs}.

We further model the dust temperature of sources in Table~\ref{tab:04_sed} with the \herschel\ and ALMA data. 
We fit the dust continuum emissions of all sources with modified blackbody (MBB) using the best available redshifts. 
The dust absorption coefficient is assumed to be $\kappa = 0.40 \times (\nu /250)^\beta$ in unit of \si{cm^2. g^{-1}}, where $\nu$ is the frequency in GHz in the rest frame. 
We assume a fixed dust emissivity of $\beta = 1.8$, which was widely adopted in previous studies (e.g., \citealt{ds17}; \citealt{dudze20}; \citealt{sun21a}) and supported by a recent 2\,mm study of SMGs at $z\simeq 1-3$ \citep{dacunha21}. 
Following previous work including \citet{greve12} and \citet{sun21a}, we only fit the SED over a rest-frame wavelength of 50\,\micron\ to avoid optically thick regime and eliminate any possible contribution of warm dust component at shorter wavelength.
We note that luminous SMGs ($L_\mathrm{IR} \simeq 10^{12.5} - 10^{13}$\,\lsun) are found to be optically thick at $\lambda_\mathrm{thick} \sim 100$\,\micron\ \citep[e.g.,][]{spilker16,simpson17,dudze20}, but we argue that less luminous ALCS sources have lower dust mass densities \citep[e.g.,][]{sun21a} and therefore smaller optical depths at the same wavelength in the rest frame. 
The best-fit dust temperatures and masses ($M_\mathrm{dust}$) are also presented in Table~\ref{tab:04_sed}.
The uncertainty of redshift is propagated into that of the dust temperature if the $z_\mathrm{spec}$ is unknown. 
The dust masses derived from this fitting procedure are also consistent with those from \textsc{magphys}. 

{We also note that many literature use the rest-frame wavelength of far-IR SED peak ($\lambda_\mathrm{peak}$) to quantify the luminosity-weighted dust temperature \citep[e.g.,][]{casey18a,reuter20,burnham21}. 
This is because $\lambda_\mathrm{peak}$ is less dependent on dust opacity assumption compared with $T_\mathrm{dust}$. 
Under the dust absorption coefficient assumption that we adopt, the conversion between $\lambda_\mathrm{peak}$ and $T_\mathrm{dust}$ is $\lambda_\mathrm{peak} = 3\times 10^3 T_\mathrm{dust}^{-1}$\,\si{\micro\meter.K}.
}

\subsection{Validity of Far-IR Photometric Redshifts}
\label{ss:04b_zphot}



To validate the far-IR photometric redshifts derived with \textsc{magphys+photo-z}, we first compare the far-IR \zph's of 47 spectroscopically confirmed sources with their \zsp's in the left panel of Figure~\ref{fig:08_photz}.
The median redshift discrepancy is $(z_\mathrm{phot}-z_\mathrm{spec})/(1+z_\mathrm{spec}) = -0.01 \pm 0.04$, indicating an excellent agreement.
We apply a Kolmogorov--Smirnov (K--S) test on the discrepancy between $z_\mathrm{phot}$ and $z_\mathrm{spec}$ divided by the uncertainty of far-IR $z_\mathrm{phot}$, in comparison with the standard normal distribution.
We conclude that the standard deviation of $(z_\mathrm{phot}-z_\mathrm{spec})/(1+z_\mathrm{spec})$ is well predicted by the uncertainty of far-IR $z_\mathrm{phot}$ ($p$-value$=$0.59).




The comparison of far-IR and \hst\ \zph's is shown in the right panel of Figure~\ref{fig:08_photz}.
We find a general agreement of \zph\ estimate between \hst\ and far-IR SED modeling albeit with a significant dispersion. 
We note that the best-fit linear slope of $z_\mathrm{phot,FIR}(z_\mathrm{phot,HST})$ is only $0.31\pm0.08$. 
However, this is mainly contributed by $\sim$\,4 tentative sources in the secondary sample with $z_\mathrm{phot,HST}\sim4$ but $z_\mathrm{phot,FIR}\sim2$, {which may be subject to a high false-ID rate ($\sim$\,18\%)}.
The discrepancy between the two \zph's, defined as $(z_\mathrm{phot,HST}-z_\mathrm{phot,FIR})/[1 + (z_\mathrm{phot,HST}+z_\mathrm{phot,FIR})/2]$, is observed to be $0.01_{-0.23}^{+0.27}$ (16--50--84th percentile) with a typical uncertainty of 0.21.
We perform a K--S test on the photometric redshift discrepancies divided by their uncertainties of the cross-matched sources.
The null hypothesis that the far-IR-\hst\ \zph\ discrepancies relative to their uncertainties are drawn from a standard normal distribution cannot be ruled out ($p$-value$=$0.33), reinforcing the agreement between these two photometric redshift estimates.

\subsection{\herschel-Faint Galaxies}
\label{ss:04c_dropout}

\begin{figure}[t]
\includegraphics[width=\linewidth]{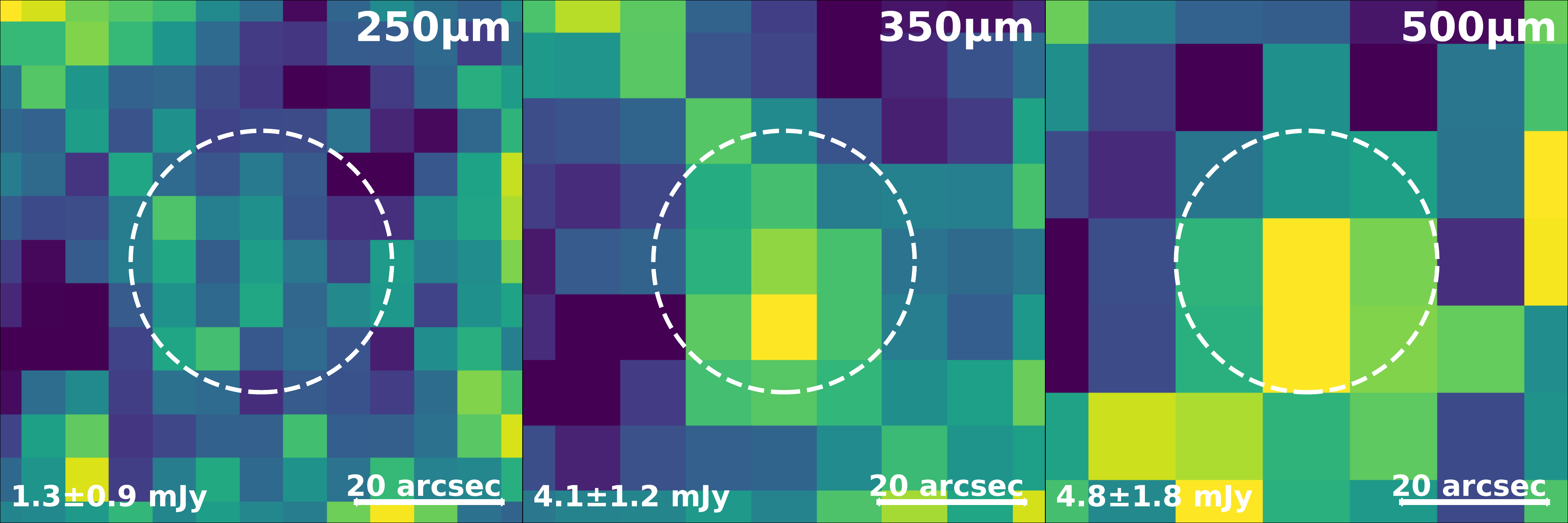}
\includegraphics[width=\linewidth]{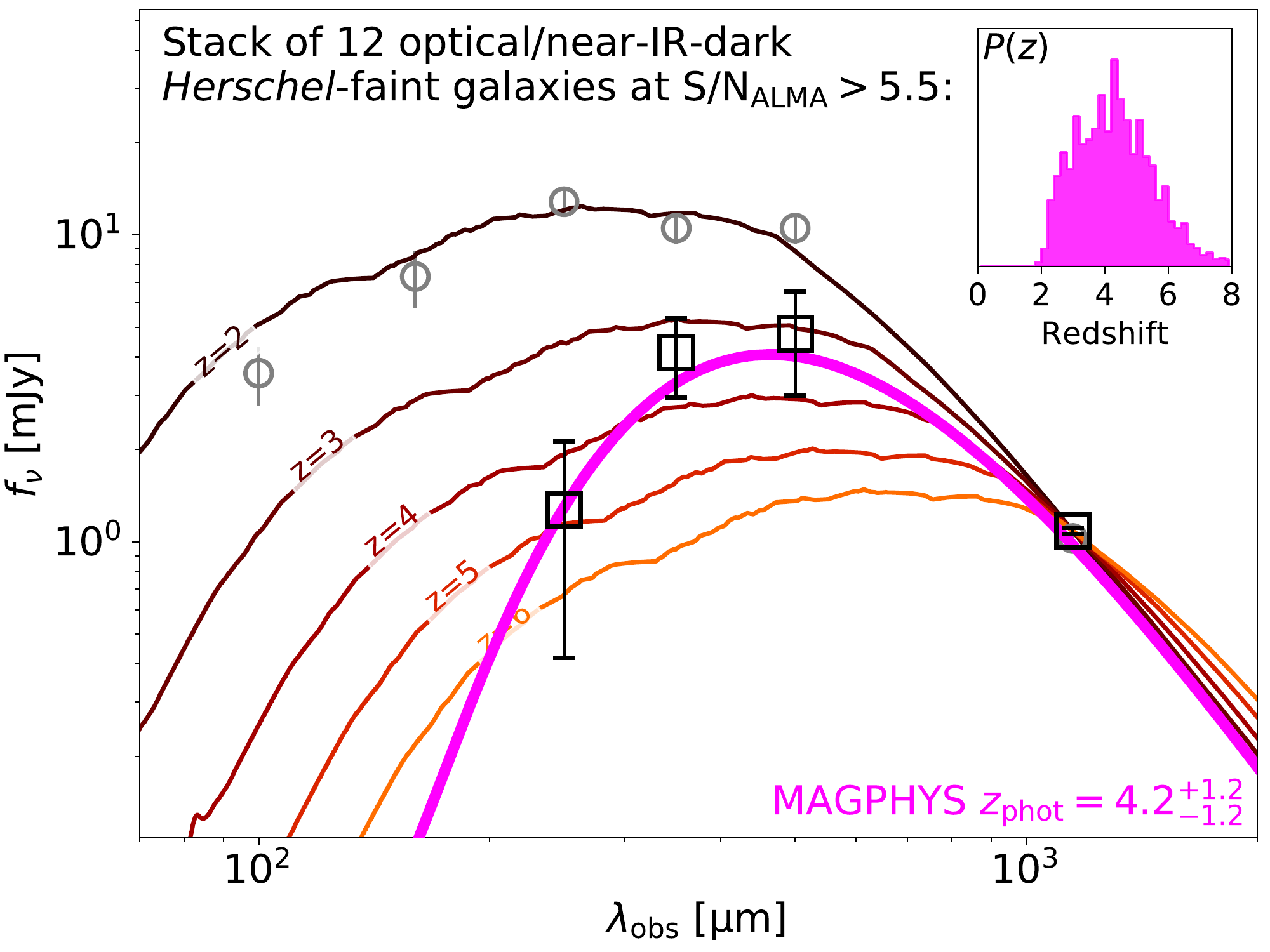}
\caption{\textit{Top}: Stacked \herschel/SPIRE images of 12 \herschel-faint galaxies ($\mathrm{S/N}_\mathrm{ALMA} > 5.5$) without $z_\mathrm{spec}$ or \hst\ $z_\mathrm{phot}$ at 250, 350 and 500\,\micron.
The white circles represent the apertures used for photometry. 
Flux density of the stacked source in each band is noted in the lower-left corner of each plot.
Scale bars of 20\arcsec\ are shown in the lower-right corners.
\textit{Bottom}: Far-IR SED of stacked sources (open black squares).
{The probability distribution of redshift is shown as the inset plot. }
The median far-IR SEDs of sources in the ALCS-\herschel\ joint sample are shown as open gray circles.
Composite SEDs of AS2UDS SMGs \citep[stack of $z>3$ sources;][]{dudze20} at $z=2-6$, normalized to the median 1.15\,mm flux density of \herschel-faint galaxies, are shown for comparison.
}
\label{fig:stack}
\end{figure}

Among the 27 \herschel-faint galaxies reported in Section~\ref{ss:03f_stats}, two of the sources have been spectroscopically confirmed. 
They are R0600-ID164 as a \cii-emitting lensed arc at $z=6.072$ (\citealt{fujimoto21,laporte21}) and R0032-ID32 as the faintest component of a multiply lensed arc at $z=3.631$ \citep{mdz17}.
In addition to this, eight sources exhibit \hst\ counterparts with tabulated photometric redshifts (median $z_\mathrm{phot} =2.0\pm 1.0$; \citealt{molino17,coe19}).
Postage stamp images of these ten sources are shown in Appendix~\ref{apd:02_fs} (Figure~\ref{fig:hdrop_z}).
Far-IR SEDs of these galaxies are also modeled with \textsc{magphys}, and we note that the adopted IR spectral templates are essentially MBB spectra at around $T_\mathrm{dust}=35\pm6$\,K without exceeding the \herschel\ non-detection limits except for SM0723-ID93 ($\mathrm{S/N}_\mathrm{ALMA}=4.6$ in 2\arcsec-tapered map), which is likely a random association between a cluster dwarf galaxy with spurious ALMA source.

The remaining 17 sources do not have cross-matched \hst\ $z_\mathrm{phot}$ because they are intrinsically faint shortward of 1.6\,\micron\ and/or out of \hst/WFC3-IR coverage (Figure~\ref{fig:hdrop_dark} in Appendix~\ref{apd:02_fs}).
Such an near-IR-dark (also often called as ``$H$-dropout/faint'') nature suggests that they are likely dust-obscured star-forming galaxies at $z \sim 4$ that have raised general interest in recent studies \citep[e.g.,][]{simpson14,Franco18, yamaguchi19,alcalde19,wang19,williams19,dudze20,smail21,gomez21,sun21b}.
For each individual sources, the non-detections in the \hst\ and \herschel\ bands prevent us from deriving useful constraints of their redshifts and physical properties.

To address this issue, we stack the \herschel\ residual images of \herschel-faint galaxies without spectroscopic or \hst\ photometric redshifts.
We note that four out of five sources at $\mathrm{S/N}_\mathrm{ALMA} < 5.5$ do not show any counterpart in \hst\ or \spitzer\ bands (ACT0102-ID11, M2129-ID24, ACT0102-ID251 and R1347-ID51).
Therefore, these sources could be spurious detections, or highly obscured galaxies at very high redshift (i.e., similar to R0600-ID67 and R0949-ID19, the brightest \herschel-faint galaxies in Figure~\ref{fig:hdrop_dark} that do not show any \hst\ or \spitzer\ counterpart; and also the  ALMA-only \cii-emitters at $z>6$ reported recently by \citealt{fudamoto21}).
Our stacking analysis suggests that including these sources will lead to a lower S/N in SPIRE 350 and 500\,\micron\ bands, and therefore we only present the stacked SEDs of 12 sources at $\mathrm{S/N}_\mathrm{ALMA} > 5.5$.

We first normalize the \herschel/SPIRE residual and uncertainty images of all sources by their ALMA flux densities.
Here the residual images are the scientific images with all the other \herschel\ sources subtracted assuming point-source models as described in Section~\ref{ss:03c_iter}.
PACS images are not stacked because of the unavailability for most sources.
We stack all the images in each SPIRE band using an inverse-variance-weighting method.
The stacked SPIRE images are presented in Figure~\ref{fig:stack}.
We measure the flux densities of stacked sources using an aperture of $r_\mathrm{aper}=18$\arcsec\ with appropriate aperture-correction factors.
The sky background is subtracted using the median of sigma-clipped local annulus, and
photometric uncertainty is estimated from the RMS of that.

The stacked source can be detected at $\sim 3\sigma$ in SPIRE 350 and 500\,\micron\ band while remains undetected at 250\,\micron\ ($<2\sigma$).
The stacked far-IR SED is shown below the stacked SPIRE images in Figure~\ref{fig:stack}.
We also show the median far-IR SEDs of ALCS sources that are detected with \herschel\ to visualize the clear difference in the continuum strength at below 500\,\micron.
Here all the photometric data are normalized to the median ALMA flux density (1.09\,mJy). 
With \textsc{magphys+photo-z}, we derive a median \zph\ of $4.2\pm1.2$ (uncertainties denote the 16-84th percentiles of the likelihood distribution), IR luminosity of $10^{12.1\pm0.3}$\,$\mu^{-1}$\lsun\ and SFR of $100_{-50}^{+100}$\,$\mu^{-1}$\smpy\ before lensing magnification correction.
The derived redshift is consistent with those of \hst\ $H$-faint galaxies in previous studies \citep[e.g.,][]{simpson14,wang19,alcalde19,dudze20,sun21b}.

Using the composite SED templates of AS2UDS SMGs at $z>3$ \citep{dudze20} and ALESS SMGs at $z>3.5$ \citep{dacunha15}, we derive best-fit $z_\mathrm{phot}$ of $3.8\pm0.4$ and $5.2\pm0.6$, respectively.
Given the large scattering of $T_\mathrm{dust}$ for sources at given luminosity and redshift \citep[e.g.,][]{schreiber18,dudze20}, the redshift uncertainty can be significantly underestimated with single template matching techniques.
In addition to this, the median IR luminosity of SMGs in the ALESS $z>3.5$ sample is $\sim10$ times higher than that of stacked \herschel-faint sources and therefore likely exhibits a higher $T_\mathrm{dust}$.
This likely leads to an overestimated $z_\mathrm{phot}$ through template matching for the ALESS $z>3.5$ sample.




\section{Lens Modeling}
\label{sec:05_lens}

\subsection{Models, Magnifications and Multiple Images}
\label{ss:05a_lens}
\begin{figure}[!t]
\centering
\includegraphics[width=\linewidth]{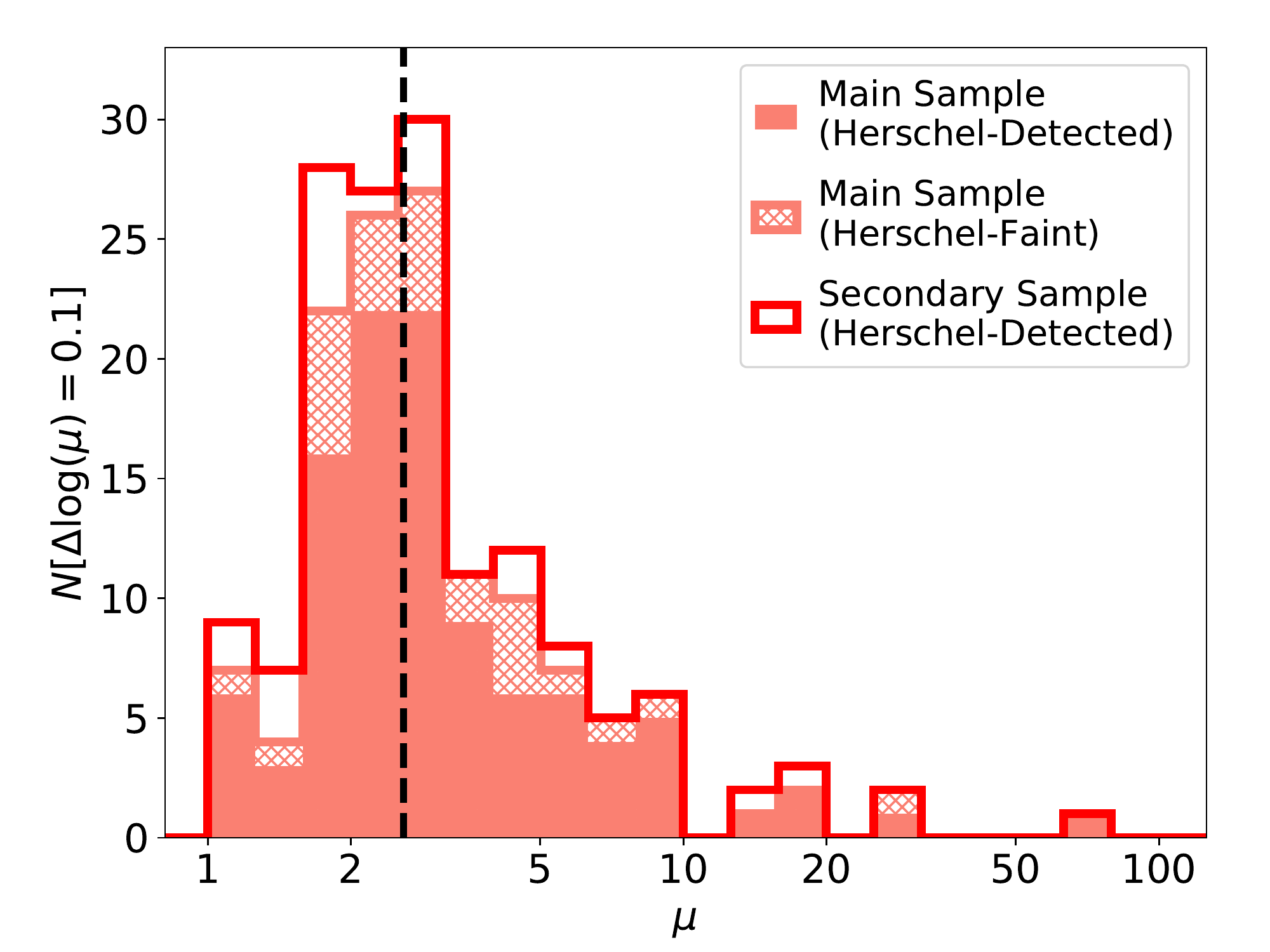}
\caption{Stacked histogram of lensing magnification factor ($\mu$) of 105 \herschel-detected ALCS sources in the main sample (filled light red bars), 27 \herschel-faint sources (hatched light red bars) and 20 \herschel-detected sources in the secondary sample (solid red steps).
The dashed black line denotes the median magnification factor ($\mu_\mathrm{med}=2.6$) of the whole sample.
No significant difference in $\mu_\mathrm{med}$ can be identified for all three subsamples. 
The bin size is $\Delta\log(\mu) = 0.1$.
}
\label{fig:mag}
\end{figure}

We calculate the lensing magnification factor ($\mu$) based on the best available redshifts.
We adopt two sets of parametric lens models, the so-called Zitrin-NFW lens models \citep{zitrin13, zitrin15} for all the HFF and CLASH clusters, and GLAFIC lens models \citep{oguri10,okabe20} for the RELICS clusters.  
The lensing magnification is derived using the maps\footnote{These maps are made available on Mikulski Archive for Space Telescopes (MAST) as high-level science products (HLSP; \href{https://archive.stsci.edu/hlsp/}{https://archive.stsci.edu/hlsp/}) and scaled to $D_{ls}/D_s=1$.} of projected cluster mass surface density ($\kappa$) and weak lensing shear ($\gamma$) at the centroid of ALCS source as $\mu = [(1-\kappa \cdot D_{ls} / D_{s})^2 - (\gamma \cdot D_{ls}/D_{s})^2]^{-1}$, where $D_{ls}$ is the angular diameter distance between the lens and the source, and $D_{s}$ is the angular diameter distance to the source.
We assume no magnification ($\mu=1$) for sources within or in front of the cluster fields ($z_s < z_{cl} + 0.1$), following \citetalias{rawle16}.
For 16 \herschel-faint galaxies without catalogued redshifts (Section~\ref{ss:04c_dropout}), we calculate their magnification at $z_s=4.2$ uniformly.

The distribution of magnification factors of the ALCS-\herschel\ joint sample and \herschel-faint sample is shown in Figure~\ref{fig:mag}.
The 16--50--84th percentiles of the distribution of $\mu$ are 1.8--2.6--5.2, and seven sources exhibit a strong magnification with $\mu > 10$.
The median lensing magnification suggests that the ALCS has reached a great depth that typical surveys in blank fields would require a $\sim 7 \times$ longer observing time to achieve. 

Among the 125 sources in our ALCS-\herschel\ joint sample, six groups of sources have been spectroscopically confirmed as multiply imaged systems.
These include (\romannumeral1) ACT0102-ID118/215/224 \citep{caputi21}, 
(\romannumeral2) M0417-ID46/58/121 \citep{kkprep}, 
(\romannumeral3) M0553-ID133/190/249 \citep{ebeling17}, 
(\romannumeral4) M1206-ID27/55/60/61 \citep{ebeling09}, 
(\romannumeral5) R0032-ID53/55/57/58 \citep{mdz17} and 
(\romannumeral6) R1347-ID145/148 \citep{richard21}. 
Additionally, M1931-ID47/55/61, R0032-ID208/281/304, and R0032-ID127/131/198 are also multiply lensed candidates yet to be spectroscopically confirmed, including \hst\ $H$-faint sources M1931-ID47 and R0032-ID208/281/304 that are part of the \herschel-faint sample. 
This reduces the number of independent sources in Table~\ref{tab:04_sed} to 109.
Multiply imaged sources of the same system are shown separately in diagrams for individual sources (e.g., Figure~\ref{fig:08_photz}), 
but only counted once in all statistics of physical properties (e.g., redshifts, SFRs and dust temperatures) in Section~\ref{sec:06_dis}.


\subsection{Uncertainties}
\label{ss:05b_unc}
\begin{figure*}[!htb]
\centering
\includegraphics[width=0.49\linewidth]{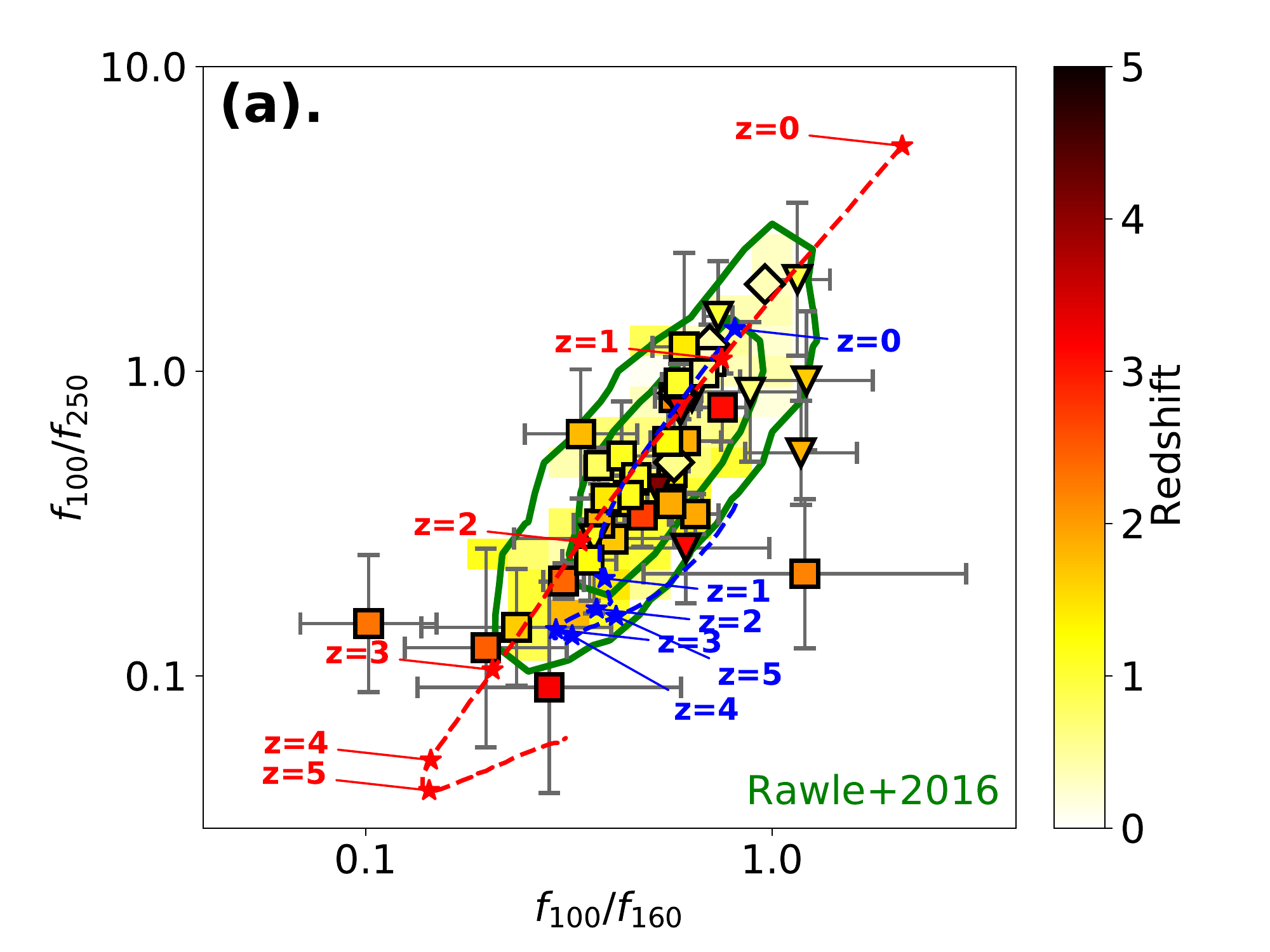}
\includegraphics[width=0.49\linewidth]{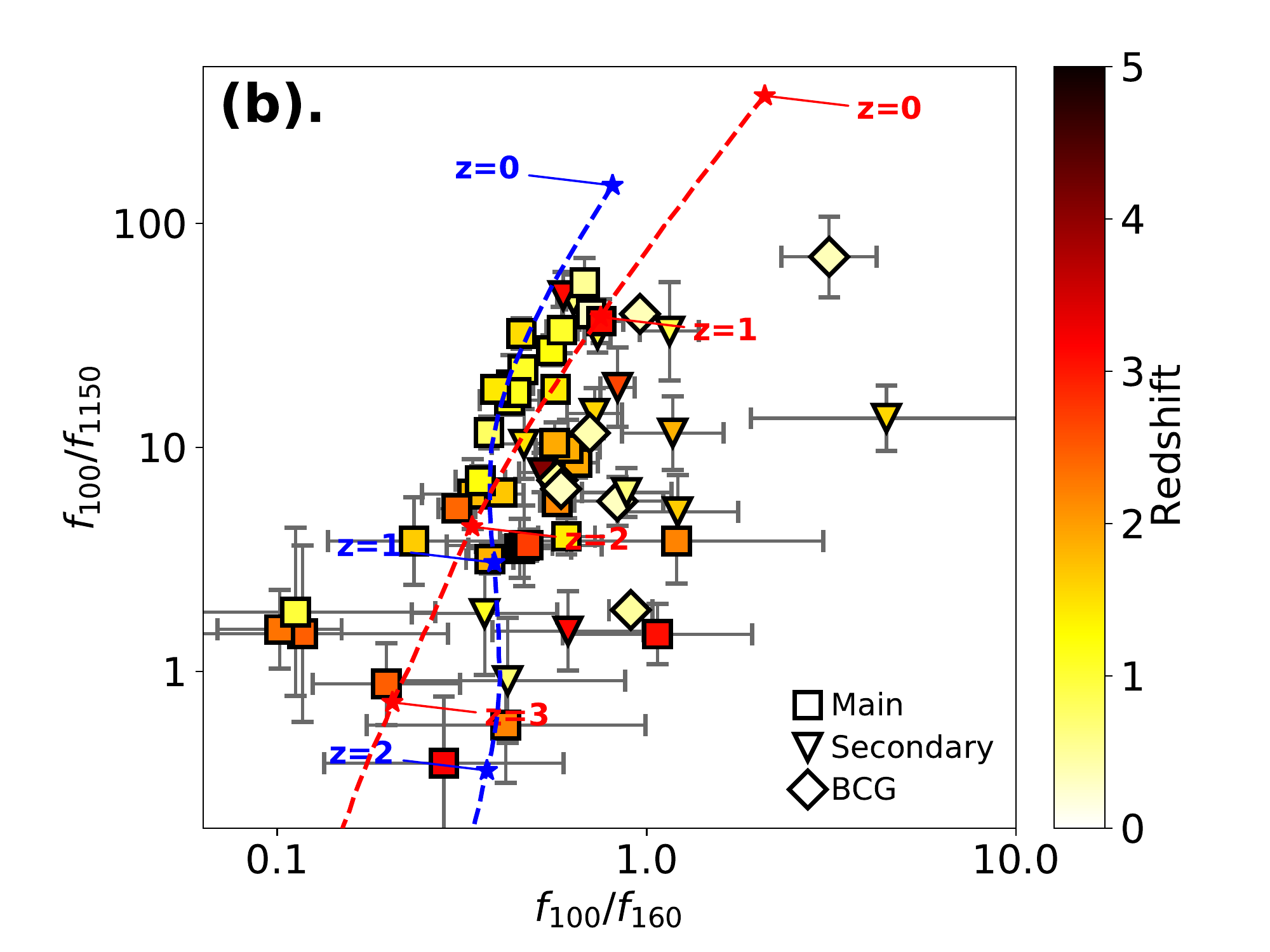}
\includegraphics[width=0.49\linewidth]{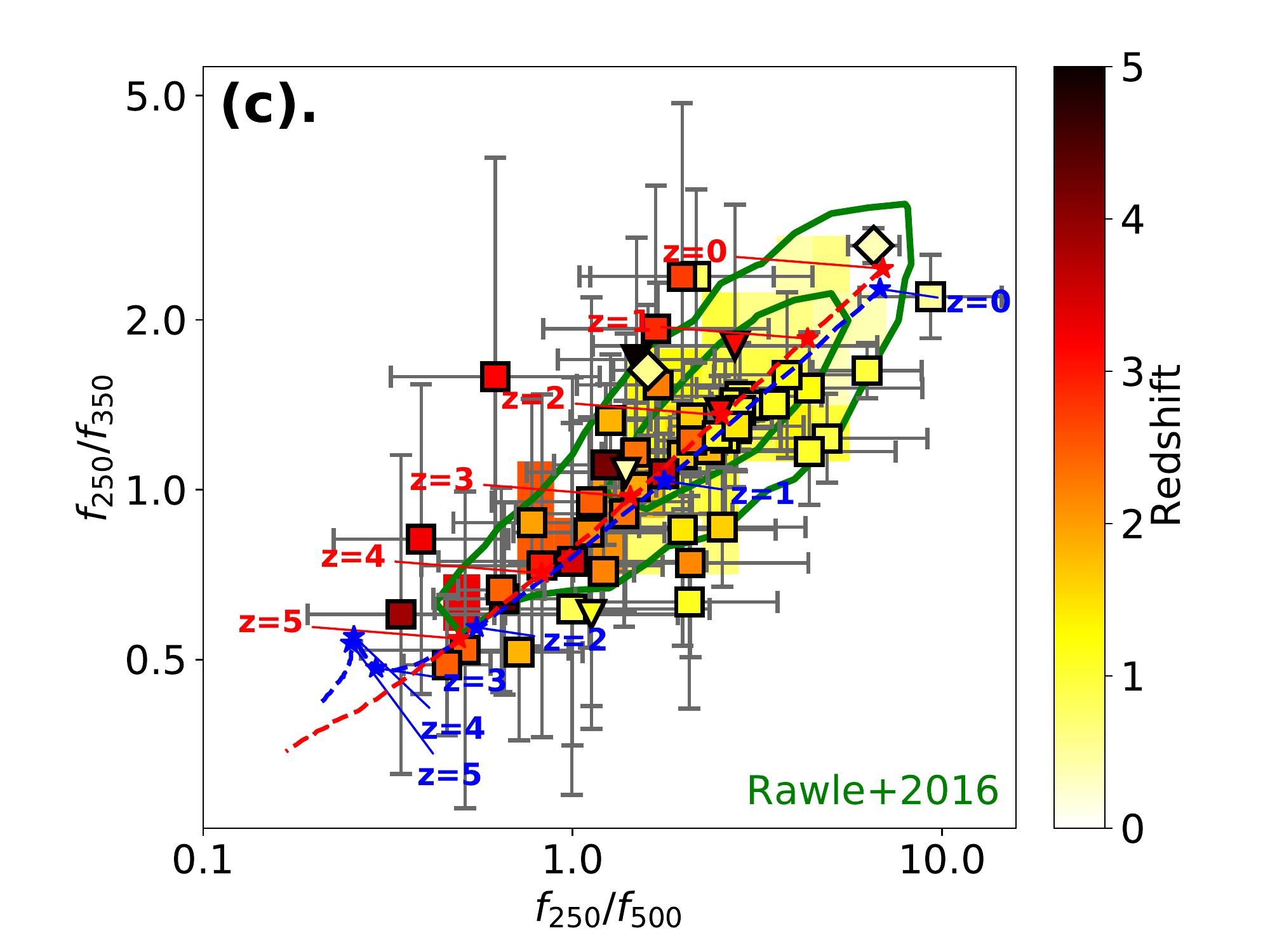}
\includegraphics[width=0.49\linewidth]{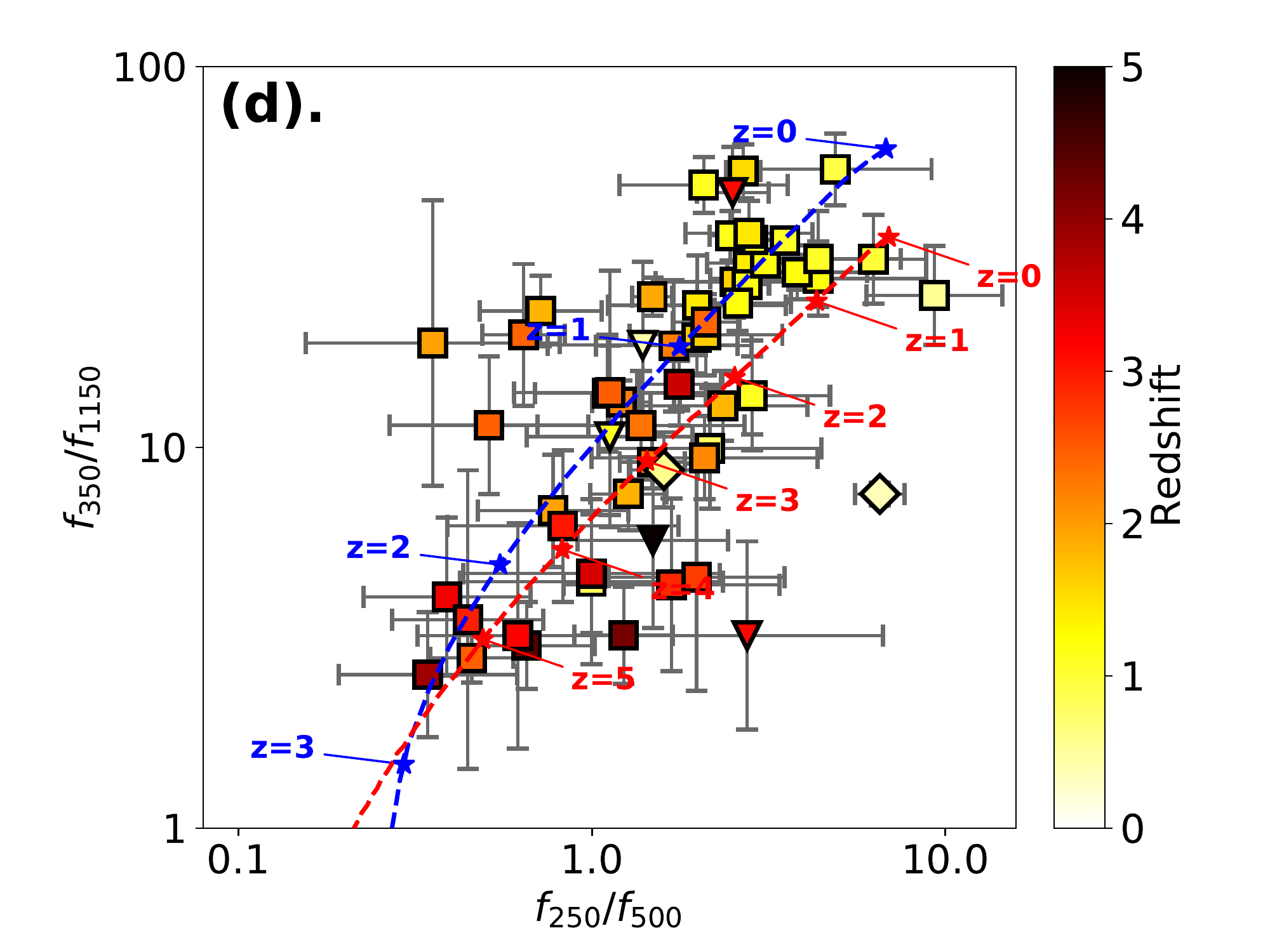}
\caption{Far-IR colors of \herschel-detected ALCS sources, color-coded with the redshifts.
(a)\ PACS--SPIRE color ($f_{100}/f_{250}$) versus PACS color ($f_{100}/f_{160}$).
(b)\ PACS--ALMA color ($f_{100}/f_\mathrm{1150}$) versus PACS color ($f_{100}/f_{160}$).
(c)\ SPIRE color ($f_{250}/f_{350}$) versus SPIRE color ($f_{250}/f_{500}$).
(d)\ SPIRE--ALMA color ($f_{350}/f_\mathrm{1150}$) versus SPIRE color ($f_{250}/f_{500}$).
Sources shown as squares are galaxies in the main sample ($\mathrm{S/N}\geq 5$), and downward triangles denote those in the secondary sample ($\mathrm{S/N} = 4 -5$).
BCGs are shown in diamonds.
We also compute and plot the redshift tracks of star-forming galaxies ($L_\mathrm{IR} = 10^{10.25}$\,\lsun, $T_\mathrm{dust}=20$\,K; blue tracks)
and ULIRGs ($L_\mathrm{IR}=10^{12}$\,\lsun, $T_\mathrm{dust}=40$\,K; red tracks) based on the spectral templates of \citet{rieke09}, and far-IR colors at redshifts of 0, 1, 2, 3, 4 and 5 are labeled out with pentagrams. 
In panel (a) and (c), the green contours enclose 68\% and 95\% of the far-IR color distribution of \herschel\ sources detected in \textit{Hubble} Frontier Fields \citepalias{rawle16}, and the background color maps denote the median redshift of these sources in each bin.
}
\label{fig:9_fir_color}
\end{figure*}

The detailed characterization of the lensing magnification uncertainty ($\sigma_\mu$) will be presented by a forthcoming source count paper of the ALCS collaboration \citep{sfprep}, and here we only present qualitative estimate with a few quick methods.
First of all, we estimate the $\sigma_\mu$ based on the uncertainty of photometric redshift for sources without spectroscopic confirmation. 
The typical lensing magnification error propagated from $z_\mathrm{phot}$ uncertainty is $\sigma_\mu / \mu \sim 5 \%$.
In addition to this, we also compare the derived magnification factors using ALMA and \hst\ source centroids, and the typical difference is found to be less than 1\%.
In order to quantify the $\sigma_\mu$ caused by extended source profile, we also measure the average magnification factor within a radius of 0\farcs6 from the ALMA source centroid.
Such an effect is negligible ($\sigma_\mu / \mu \lesssim 2\%$) in most cases except for M0159-ID24 and R0032-ID57 (Figure~\ref{fig:a1_sp}) because of the galaxy-galaxy lensing effect, which provides a stronger magnification gradient over a smaller angular scale.

As an alternative method, we also analyze the uncertainty map of magnification presented by \citet{zitrin13, zitrin15} in the CLASH cluster fields, obtained through a Monte Carlo Markov Chain (MCMC) fitting routines when the cluster mass models were constructed.
We find that the $\sigma_\mu / \mu$ is around $\sim 5\%$ at $\mu=3$ assuming a fixed sky position and redshift.
However, this uncertainty could be as large as $\sim10\%$ at $\mu=10$ and $\sim 50\%$ at $\mu = 100$, indicating that sources with larger magnifications are subject to a larger relative uncertainty and therefore their intrinsic physical properties (e.g., \lir\ and SFR) are more uncertain.

Finally, based on the standard deviation of magnifications predicted by different lens models of the same clusters produced by different methods and groups (i.e., GLAFIC, CATS, Zitrin-NFW and Zitrin Light-Traces-Mass (LTM) for Frontier Field clusters; \citealt{oguri10,richard14,zitrin15,kawamata16,kawamata18}), we find a typical magnification uncertainty of $\sigma_\mu / \mu \sim 20$\%.
This is comparable to the $\sigma_\mu$ reported in \citetalias{rawle16} ($\sigma_\mu = 0.5$, $\sim 20\%$ of the median magnification for sources in this work).

\section{Discussion}
\label{sec:06_dis}

\subsection{Far-IR Colors}
\label{ss:06a_color}


We further study the far-IR colors of \herschel-detected ALCS sources.
Figure~\ref{fig:9_fir_color} displays four color-color diagrams of our sample, covering the full wavelength range from 100\,\micron\ to 1.15\,mm.  
Based on the average IR spectral templates in \citet{rieke09}, we superimpose the redshift-evolution tracks of a typical star-forming galaxy (SFG; \lir\,$= 10^{10.25}$\,\lsun, corresponding to a $\mathrm{SFR}_\mathrm{IR}$ of $\sim 2$\,\smpy\ assuming the conversion factor in \citealt{ke12}) and a ULIRG (\lir\,$ = 10^{12}$\,\lsun, corresponding to a $\mathrm{SFR}_\mathrm{IR}$ of $\gtrsim 100$\,\smpy) on all of the four color-color diagrams.
The dust temperatures of the SFG and ULIRG templates are around 20\,K and 40\,K.
We also compare the distribution of our ALMA-selected sample with the \citetalias{rawle16} sample in two of the color-color diagrams where the 1.15\,mm flux density is not invoked.

\subsubsection{Selection bias}
\label{sss:06a_bias}

Through a comparison with the redshift-evolution tracks computed from the SED templates in \citet{rieke09}, the color-color distributions of our sample match those of SFGs at $0< z \lesssim 2.5$ and ULIRGs at $1\lesssim z \lesssim 5$.
At $z \gtrsim 2.5$, the depth of \herschel\ data is not deep enough shallow to select low-$T_\mathrm{dust}$ sources (i.e., $\sim$20\,K for typical SFGs) even with a lensing magnification factor of a few tens.

{The selection function of ALCS is nearly flat in terms of cold dust mass across $z \simeq 1 - 6$ ($M_\mathrm{dust} \gtrsim 10^{8}$\,\si{\mu^{-1}.M_\odot}).
However, with regard to a fixed $L_\mathrm{IR}$, the ALMA-\herschel\ joint selection} do bias against low-redshift ($z\lesssim 1$) galaxies with high dust temperature ($\sim$\,40\,K, ULIRG-like; see panel a and b {of Figure~\ref{fig:9_fir_color}}).
This could be interpreted as a combined effect of survey volume and 1.15\,mm selection-limit.
At $z<1$, the volume density of ULIRGs is $\sim$100 times lower than that of SFGs \citep[e.g.,][]{gruppioni13}, and therefore 
the expected number of ULIRG detectios with ALCS in this redshift range is only on the order of unity.
In addition, if we compare two galaxies with $T_\mathrm{dust}=20$\,K and 40\,K at the same redshift (e.g., $z=0.5$) and the same intrinsic IR luminosity, the warmer galaxy should be fainter at 1.15\,mm and thus less likely to be selected by ALCS.
These galaxies with high $T_\mathrm{dust}$ are likely to be selected as ALMA-faint \herschel\ sources, which are not included in our ALCS-\herschel\ joint sample.
This further suggests that the observed-frame 1.15\,mm selection has more bias in $L_\mathrm{IR}$ than $M_\mathrm{dust}$ \citep[e.g.,][]{scoville14,dudze21} because the $M_\mathrm{dust} - f_{1150}$ relation has less dependence on the dust temperature.

\subsubsection{Comparison with \citet{rawle16}}

We find that the majority of the \herschel-detected ALCS sources follow the similar distribution of far-IR colors as the \herschel-selected sources in \citetalias{rawle16} (Figure~\ref{fig:9_fir_color}, panel a and c).
However, only 4\% of the sources in \citetalias{rawle16} are at $z > 2$, in contrast to a large fraction of 43\% in our ALCS-\herschel\ joint sample.
Sources in \citetalias{rawle16} were selected based on \spitzer\ and \textit{WISE} mid-IR priors, and $>4\sigma$ detections in at least two \herschel\ bands were required. 
Such a selection is biased against high-redshift sources which are faint in bluer \herschel\ bands (e.g., PACS 100\,\micron) but likely detectable with SPIRE 500\,\micron\ and ALMA Band 6.
As shown in the panel (a) of Figure~\ref{fig:9_fir_color}, PACS-selected sources in \citetalias{rawle16} are mostly at $z\lesssim1$, which are populated by ALMA-faint \herschel\ sources that are not included in the ALCS sample.
Meanwhile, as discussed earlier in this subsection, the ALCS-\herschel\ joint sample is biased against low-redshift sources with moderately low IR luminosity ($L_\mathrm{IR} < 10^{11}$\,\lsun) but high dust temperature.
The combined selection effects lead to the significant difference in the redshift distributions of \herschel\ source samples of these two works.

\subsection{Redshift versus 1.15\,mm Flux Density}
\label{ss:06b_redshift}

\begin{figure*}
\centering
\includegraphics[width=0.49\linewidth]{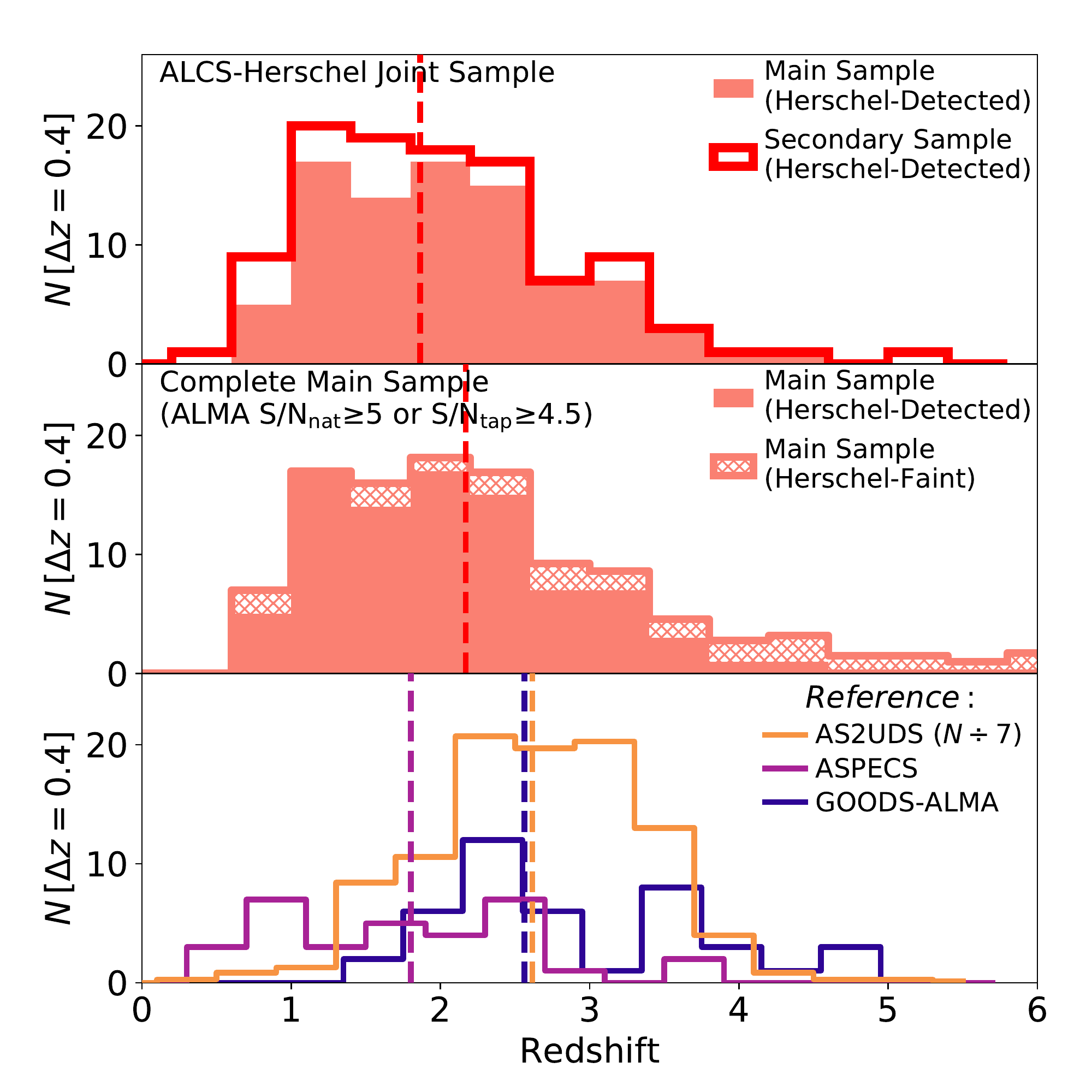}
\includegraphics[width=0.49\linewidth]{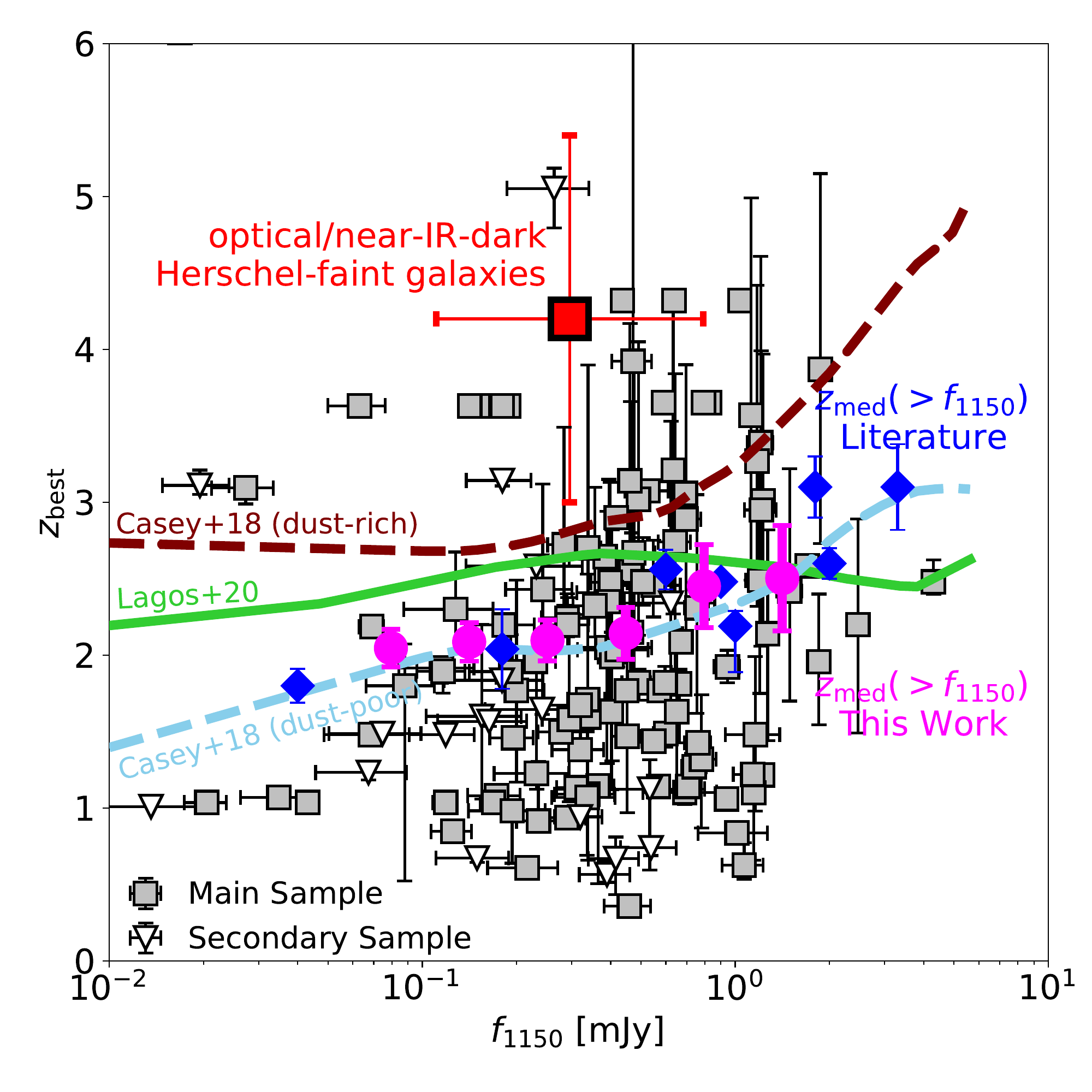}
\vspace{-2mm}
\caption{
\textit{Left}: Distribution of best available redshifts of ALCS sources. 
ALCS-\herschel\ sources in the main (filled light red bars) and secondary sample (solid red steps), catalogued in Table~\ref{tab:04_sed} (i.e., excluding \herschel-faint galaxies), are shown in the top panel. 
After assuming {the probability distribution for the redshifts of optical/near-IR-dark \herschel-faint galaxies
derived with {magphys+photo-z} ($z_\mathrm{phot}=4.2\pm1.2$, Section~\ref{ss:04c_dropout})}, the redshift distribution of sources in the main sample ($\mathrm{S/N}_\mathrm{nat}\geq5$ or $\mathrm{S/N}_\mathrm{tap}\geq4.5$) is shown in middle panel. 
Note that cluster member galaxies are not included.
The redshift distributions of reference samples, including AS2UDS (\citealt{dudze20}, orange steps; scaled by a factor of 1/7 for the purpose of displaying), ASPECS (\citealt{aravena20}, purple steps) and GOODS-ALMA (\citealt{gomez21}, blue steps) are shown in the lower panel.
Vertical dashed line indicates median redshift for each sample.
The bin sizes are $\Delta z = 0.4$.
\textit{Right}: Best available redshifts versus ALMA 1.15\,mm flux densities corrected for lensing magnification.
Sources in the main (secondary) sample are shown as filled gray squares (open downward triangles).
The median $z_\mathrm{best}$ and $f_{1150}$ of 17 optical/near-IR-dark \herschel-faint galaxies are visualized as the filled red square.
Magenta circles denote the median redshifts of secure ALCS sources at above given 1.15\,mm flux density thresholds.
Blue diamonds denote $z_\mathrm{med}$ at various 1.1--1.2\,mm flux density thresholds in literature \citep{michalowski12,yun12,miettinen17,brisbin17,dunlop17,aravena20,gomez21}.
The median redshifts as functions of 1.1\,mm flux density cut, modeled by \citet[both dust-rich and dust-poor scenarios for $z>4$ Universe]{casey18a} and \citet{lagos20}, are plotted for comparison.
}
\label{fig:hist_sed}
\end{figure*}

The top-left panel of Figure~\ref{fig:hist_sed} displays the redshift distribution of sources in our ALCS-\herschel\ joint sample.
The {16-50-84th percentiles of redshifts are 1.11--1.90--2.95 (1.05--1.81--2.96)} 
for the main (full) joint sample, and no obvious difference in redshift distribution can be found among sources in clusters observed with ``deep'' and ``snapshot'' mode.
{
The uncertainty of photometric redshifts are accounted for by the derivation of percentiles through a Monte-Carlo sampling of the probability distributions.
Among the joint sample, spectroscopically confirmed sources are generally at slightly lower redshifts. 
The 16-50-84th percentiles of $z_\mathrm{spec}$'s are 1.06--1.55--2.90 (1.03--1.49--2.70) for 26 (31) independent sources in the main (full) sample.}
However, we note that this ALCS-\herschel\ joint sample does not include 27 \herschel-faint galaxies that likely reside at higher redshifts (Section~\ref{ss:04c_dropout}). 

{To estimate the median redshift of the full ALCS sample, we must include the \herschel-faint sources. 
Based on the probability distribution of} 17 optical/near-IR-dark \herschel-faint galaxies (intrinsically 15 sources after removing multiply lensed images) {derived with \textsc{magphys+photo-z} ($z_\mathrm{phot}=4.2\pm1.2$)}, the 16--50--84th percentiles of redshifts are {1.15--2.08--3.59} for secure ALCS sources in the main sample (middle-left panel of Figure~\ref{fig:hist_sed}; note that cluster member galaxies are not included).

The median redshift of secure ALCS sources is higher than that of the main ASPECS sample ($z_\mathrm{med} = 1.80\pm0.15$) in the \textit{Hubble} Ultra Deep Field (HUDF; \citealt{aravena20}), but smaller than those of slightly shallower surveys including AS2UDS ($z_\mathrm{med} = 2.61\pm0.08$ down to $\sim$1\,mJy at 850\,\micron; \citealt{dudze20}), ASAGAO ($z_\mathrm{med} = 2.38\pm0.14$ down to $\sim$0.2\,mJy at 1.2\,mm, note that near-IR-dark ALMA sources at $z\sim4$ are not included; \citealt{yamaguchi19,yamaguchi20}) and GOODS-ALMA ($z_\mathrm{med} = 2.56\pm0.13$ down to $\sim$0.5\,mJy\ at 1.1\,mm in their main sample; \citealt{Franco18,gomez21}).
{The median redshift of ALCS sources is also smaller than those of millimeter-selected strongly lensed SMGs with wider but shallower surveys (e.g., $z_\mathrm{med}=2.9\pm0.1$ with the \textit{Planck}'s dusty GEMS sample, \citealt{canameras15}; $z_\mathrm{med}=3.9\pm0.2$ with the SPT sample, \citealt{reuter20}), as well as that of unlensed sources selected with the MORA survey at 2\,mm ($z_\mathrm{med}=3.6\pm0.3$ down to $\sim$\,0.3\,mJy, \citealt{casey21}).
}

{The median redshift of millimeter and submillimeter sources as a function of flux density limit is a key test for the evolution model of dust-obscured star-formation history of the Universe (see a review by \citealt{hodge20}).}
The right panel of Figure~\ref{fig:hist_sed} shows the best available redshifts versus intrinsic 1.15\,mm\ flux densities of all ALCS sources in the main sample (excluding cluster member galaxies; filled squares) and \herschel-detected ALCS sources in the secondary sample (open downward triangles).
Assuming {the probability distribution of redshifts for optical/near-IR-dark \herschel-faint galaxies derived with \textsc{magphys+photo-z}}, we are able to derive the median redshift of secure ALCS sources above given 1.15\,mm flux density cuts (i.e., $z_\mathrm{med}(>f_{1150})$; magenta circles).
{The uncertainty of photometric redshift is propagated into the uncertainty of  $z_\mathrm{med}(>f)$ through a Monte-Carlo sampling of $z_\mathrm{phot}$ likelihood.
}

The median redshifts of secure ALCS sources decrease from {$z_\mathrm{med} = 2.40\pm0.29$} at $f_{1150} > 1$\,mJy to {$z_\mathrm{med} = 2.04\pm0.12$} at $f_{1150} > 0.1$\,mJy, where our survey is $\sim50$\% complete given the depth and median lensing magnification.
{We also perform K-S test for the redshift distributions of ALCS and ASPECS sources above 0.1\,mJy, and no obvious difference can be found (p-value=0.64).}
{For spectroscopically confirmed sample, the decrease of median redshift is not conspicuous because of a smaller sample size ($z_\mathrm{med} = 1.71\pm0.87$ at $>$\,1\,mJy to $z_\mathrm{med} = 1.60\pm0.21$ at $>$\,0.1\,mJy).}
{We note that a higher redshift assumption (e.g., $z_\mathrm{med}\sim 6$) of optical/near-IR-dark \herschel-faint galaxies will not change our measurements of $z_\mathrm{med}(>f)$ despite a larger standard error.}
{We also compute the median redshifts of ALCS sources in logarithmic flux density bins from 0.1 to 2\,mJy (bin size is 0.1\,dex), and the null hypothesis that there is no monotonic relation between $z_\mathrm{med}$ and $f_{1150}$ can be ruled out ($p$-value\,$=$\,0.03, computed from Spearman's rank correlation coefficient $\rho=0.56$).}

Further compared with previous 1.1--1.2\,mm surveys of SMGs \citep{michalowski12,yun12,miettinen17,brisbin17,dunlop17,aravena20,gomez21}, the derived $z_\mathrm{med}(>f_{1150})$ function suggests {an increasing fraction} of $z\simeq 1 - 2$ galaxies at $f_{1150} < 1$\,mJy (also shown with ASPECS and semi-empirical model presented in \citealt{popping20}).
{A linear fitting to the $z_\mathrm{med}[>\log(f_{1150})]$ measurements suggests that the positive correlation is significant ($>5\sigma$).}
We note that $z_\mathrm{med}(>f_{1150})$ below 0.5\,mJy was poorly probed with previous surveys because of either limited volume (ASPECS-like) or relatively shallower depth (e.g, GOODS-ALMA and single-dish surveys).
Similar $z_\mathrm{med}(>f)$ trends were also reported at 870\,\micron\ {\citep{ivison07,stach19,simpson20,birkin21,chen22}}.

The overall $z_\mathrm{med}(>f_{1150})$ function obtained with the main ALCS sample is lower than that predicted by the \textsc{shark} semi-analytic model presented in \citet[the offset is $\Delta z \sim -0.4$]{lagos20}, {but very close to} that {empirically} modeled by \citet[$\Delta z \sim 0.1$]{casey18a} assuming a negligible contribution ($<10\%$) of SMGs to the cosmic SFR density at $z>4$, also known as the ``dust-poor'' scenario.
Nevertheless, the observed $z_\mathrm{med}(>f_{1150})$ of our sample is much lower than that of ``dust-rich'' scenario modeled by \citet[$\Delta z \sim -0.7$]{casey18a} in which SMGs dominate the star formation ($\sim90\%$) at $z>4$.
This {may} suggest that the majority of cosmic star formation at $z > 4$ will be hosted in unobscured environment, similar to the conclusions drawn in
\citet{dudze20}, \citet{bouwens20}, \citet{casey21} and \citet{zavala21}.
{However, we note that the decreasing obscured fraction of cosmic SFR density at $z>4$ cannot be considered as the unique cause of positive $z_\mathrm{med}(>f_{1150})$ relation.
Other scenarios, e.g., a steep faint-end slope of dust mass function at $z=1\sim2$, can lead to a similar $z_\mathrm{med}(>f_{1150})$ function as observed.
Detailed characterization of IR luminosity function and cosmic SFR density evolution will be presented in a future work.
}

\subsection{Statistics of Intrinsic (Source-Plane) Properties}
\label{ss:06c_intrinsic}
\begin{figure}[!t]
\centering
\includegraphics[width=\linewidth]{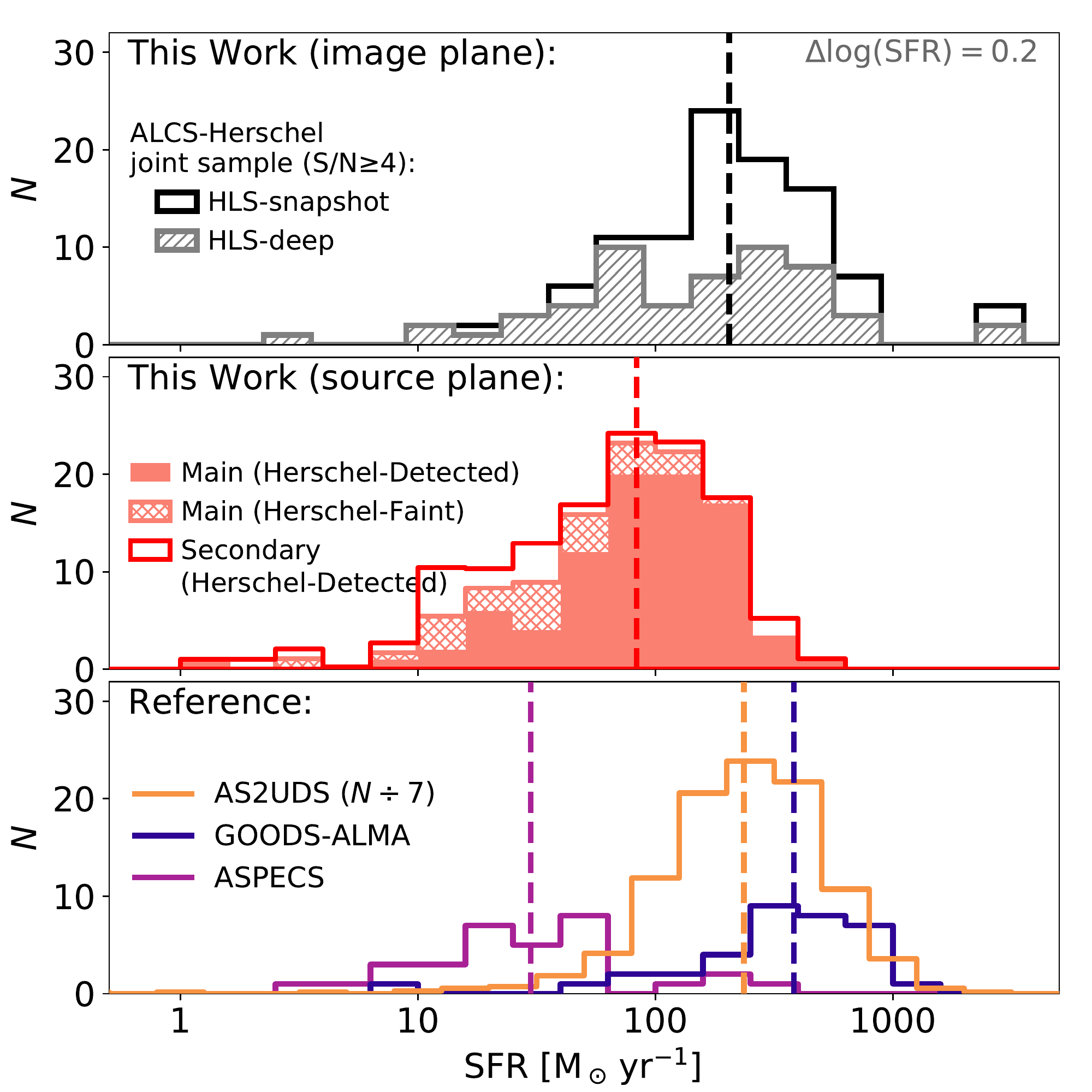}
\caption{Stacked histograms of observed (top panel) and intrinsic SFRs (middle panel) of ALCS sources.
In the top panel, we show the image-plane SFR distributions of sources in the ALCS-\herschel\ joint sample at $\mathrm{S/N}_\mathrm{ALMA}\geq 4$ detected in both the ``deep'' mode (hatched gray bars) and ``snapshot'' mode (solid black steps).
In the middle panel, we show the source-plane (intrinsic) SFR distributions of \herschel-detected sources in the main sample (shallow red filled bars), \herschel-faint sources (hatched bars) and \herschel-detected sources in the secondary sample (solid red steps).
Cluster member galaxies are not included.
We also show the SFR distributions of ALMA sources reported by AS2UDS (\citealt{dudze20}, orange steps; scaled by a factor of 1/7 for the purpose of displaying), GOODS-ALMA (\citealt{franco20}, blue steps) and ASPECS (\citealt{aravena20}, purple steps) in the bottom panel.
Vertical dashed line in each panel indicates the median SFR of each sample.
The bin size is $\Delta\log(\mathrm{SFR}) = 0.2$.
}
\label{fig:intrinsic}
\end{figure}

\begin{figure*}[!tbh]
\centering
\includegraphics[width=0.49\linewidth]{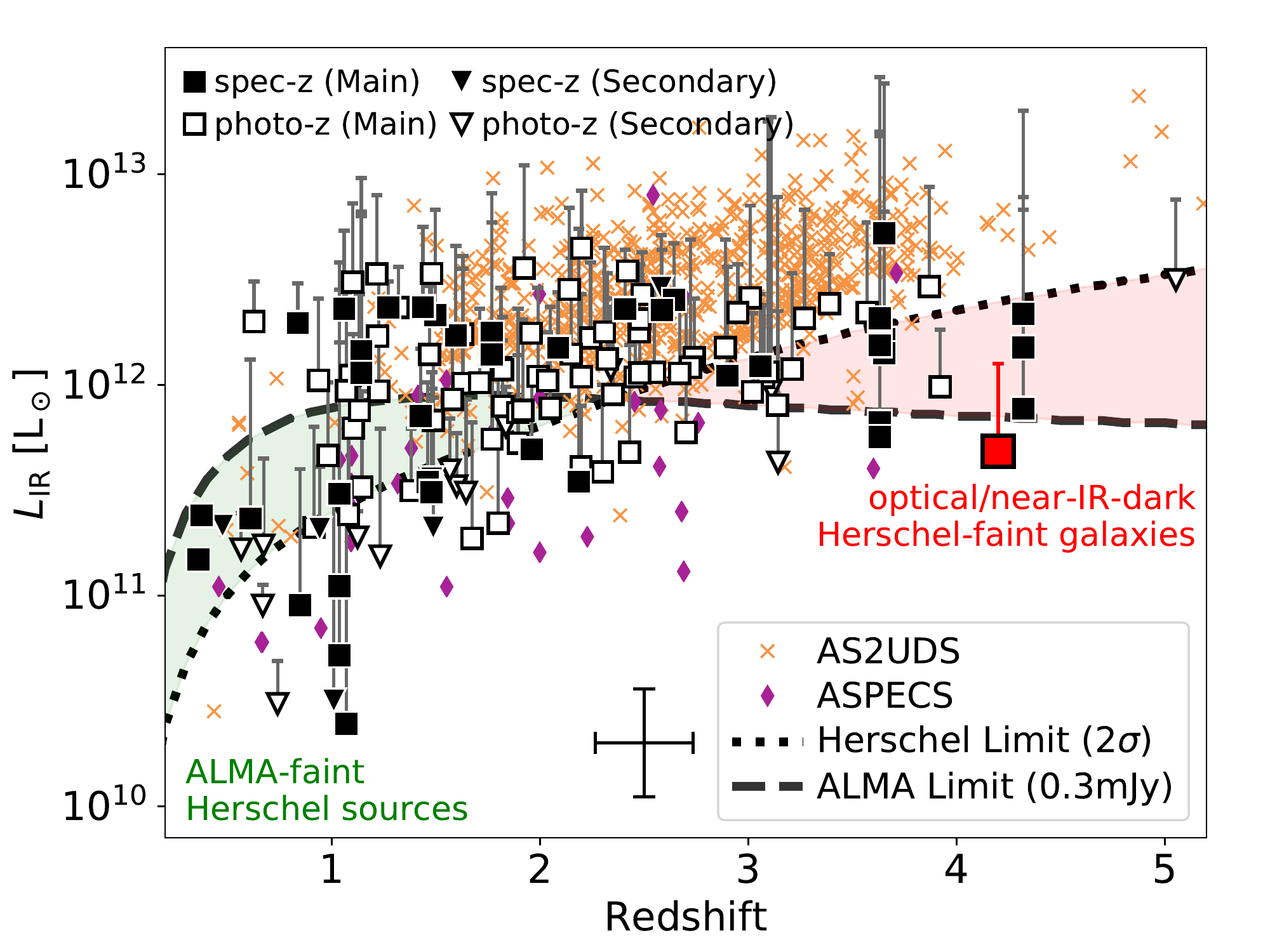}
\includegraphics[width=0.49\linewidth]{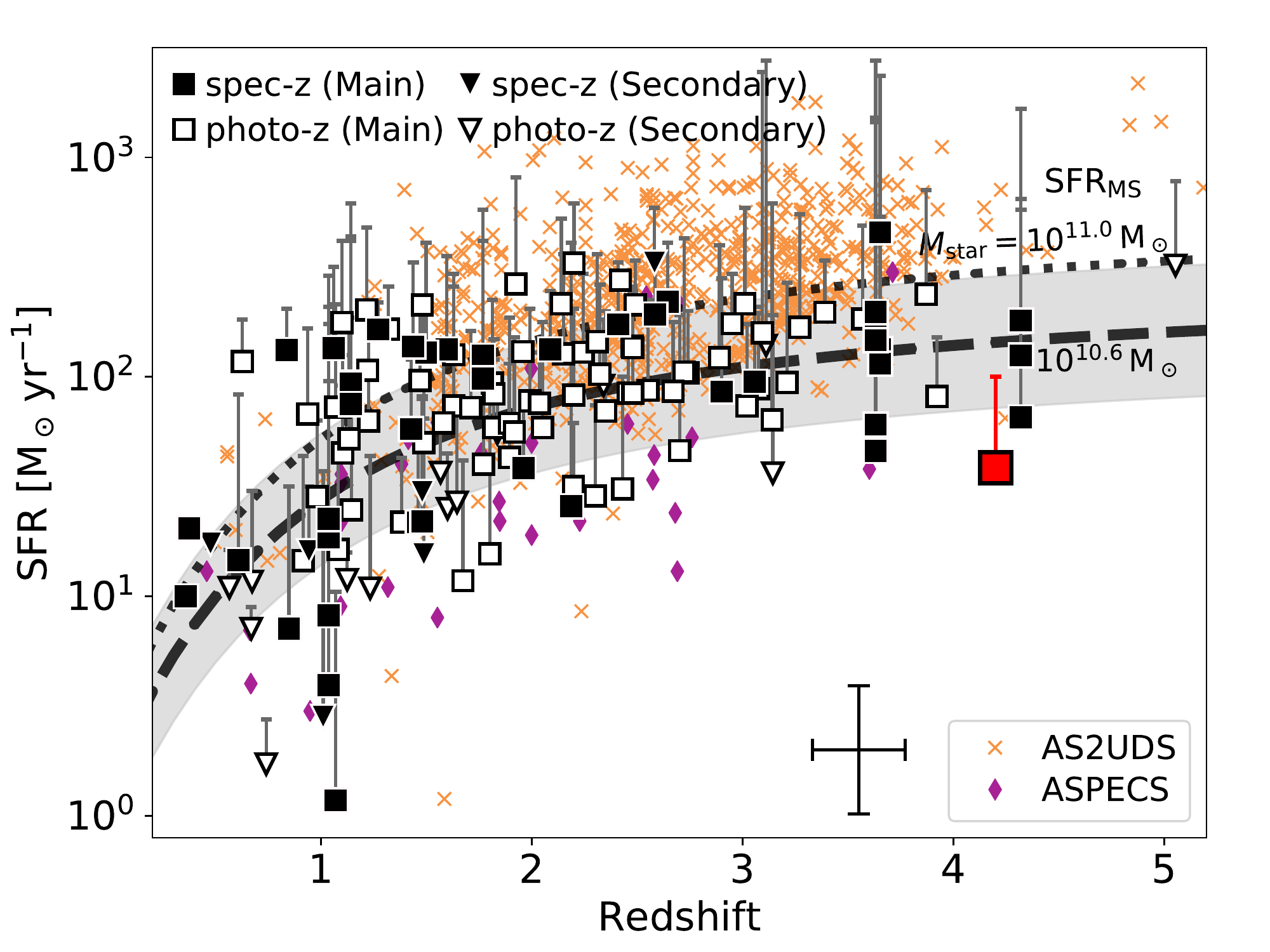}
\vspace{-3mm}
\caption{Intrinsic IR luminosity (left) and SFR (right) of ALCS sources versus redshift, including all sources in the main sample (\herschel-detected sources and \herschel-faint galaxies with \zsp\ or \hst-derived \zph; shown as squares) and \herschel-detected sources in the secondary sample (downward triangles).
Spectroscopically confirmed sources are shown with filled black symbols and the open symbols denote \zph-only ones.
The stacked optical/near-IR-dark \herschel-faint sources (Section~\ref{ss:04c_dropout}) are shown as the filled red squares.
Vertical bars indicate the logarithmic magnification ($\log \mu$) corrected for each source.
In the left panel, we plot the detection limit of ALMA ($f>0.3$\,mJy) and \herschel\ ($\mathrm{S/N} > 2$, $f\gtrsim5$\,mJy) in the ``deep'' mode without lensing magnification, assuming a ULIRG spectral template ($L_\mathrm{IR}=10^{12}\,$\lsun\ and $T_\mathrm{dust}\sim 40$\,K; \citealt{rieke09}).
ALMA-faint \herschel\ sources and \herschel-faint galaxies that are not included in Table~\ref{tab:04_sed} (ALCS-\herschel\ joint sample) are mostly detected in the shaded green and red regions, respectively.
In the right panel, we plot the SFRs of ``main sequence'' (MS) galaxies as a function of redshift {at fixed stellar masses of $10^{11}$ and $10^{10.6}$\,\msun\ \citep[dotted and dashed line;][]{speagle14}. 
The $1\sigma$ dispersion of $\mathrm{SFR}_\mathrm{MS}$ ($\sim$\,0.3\,dex) at $M_\mathrm{star}=10^{10.6}$\,\msun\ is shown as the filled grey region. }
ALMA sources reported by AS2UDS \citep[][orange crosses]{dudze20} and ASPECS \citep[][purple diamonds]{aravena20} are also plotted for comparison.
The typical uncertainties of redshifts, $L_\mathrm{IR}$ and SFR are shown to the left of the lower-right legends.
}
\label{fig:lir_vs_z}
\end{figure*}

We study the distribution of the intrinsic SFRs (total SFR derived by \textsc{magphys} and corrected for lensing magnification) for ALCS sources in Figure~\ref{fig:intrinsic}.
For all ALCS sources in the main sample (excluding cluster member galaxies and including \herschel-faint galaxies), the 16--50--84 percentiles of the distribution are 40, 94 and 178\,\smpy, slightly larger than those of the GOALS sample \citep[16--50--84 percentiles of SFR as 25, 45 and 175\,\smpy;][]{armus09,howell10}, which is mostly comprised of LIRGs in the local Universe.
\herschel-detected sources in the secondary sample exhibit LIRG-like SFRs with 16--50--84th percentiles of the distribution as 10, 25 and 97\,\smpy.

We also compare the intrinsic SFRs with those of ALMA sources reported by AS2UDS (707 SMGs at $z_\mathrm{med} = 2.6$; \citealt{stach19,dudze20}), GOODS-ALMA (35 sources at $z_\mathrm{med}=2.7$ studied in \citealt{franco20}; note that the SFRs are recomputed assuming \citealt{chabrier03} IMF) and ASPECS survey (32 sources in the main sample at $z_\mathrm{med}=1.8$; \citealt{glopez20,aravena20}).
Although the SFR distribution is similar in the image plane (lensing uncorrected), the median source-plane SFR (intrinsic) of ALCS sources is lower than that of conventional SMGs (i.e., $f>1$\,mJy at 850\,\micron) in the AS2UDS sample by a factor of $\sim3$ (median $\mathrm{SFR} = 236\pm8$\,\smpy; although 92 sources in AS2UDS sample are at $<100$\,\smpy\ with a median redshift of $2.0\pm0.1$).
Below 30\,\smpy, which is the median $\mathrm{SFR}$ of 32 sources in the main ASPECS sample down to a fidelity of 50\%, our sample contain more sources than ASPECS ($N=16$), including 11 \herschel-detected and $\sim$\,10 \herschel-faint sources at $\mathrm{S/N}\geq5$, as well as 12 tentative sources in the secondary sample.
Because of the same software (\textsc{magphys}) and SFR tracer (far-IR) being used, these comparisons are fair. 
{The median ratio between SFR and $L_\mathrm{IR}$ is found to be $10^{-10.1}$\,\si{M_\odot.yr^{-1}.L_\odot^{-1}}for both the main samples of ALCS and ASPECS, which is only offset from the conversion factor in \citet{ke12} by 0.1\,dex (assuming Chabrier IMF).}
We also note that the ASPECS sample presented by \citet{aravena20} did not enforce any \herschel\ detection, and no SPIRE flux density information was given because of blending issues, {making it difficult to directly and reliably constrain the dust temperature from the far-IR SED}.
This highlights the uniqueness of our ALCS-\herschel\ joint sample as the probe of (sub-)LIRG population at $z \lesssim 2$.

Figure~\ref{fig:lir_vs_z} shows the distribution of intrinsic IR luminosity and SFR as functions of redshift, highlighting the less vigorous star formation among the ALCS sources compared with unlensed SMGs \citep{dudze20}.
We calculate the nominal detection limits of infrared galaxies in the ALMA ($>$\,0.3\,mJy at 1.15\,mm) and \herschel\ ($>2\sigma$ in the ``deep'' mode; $\gtrsim$\,5\,mJy) bands assuming a ULIRG template at $L_\mathrm{IR} = 10^{12}$\,\lsun\ \citep{rieke09}.
We find that the intrinsic luminosities of 29\% (37\%) sources in the ALCS-\herschel\ joint sample are lower than the nominal detection limit of \herschel\ (ALMA).
This demonstrates the benefit of sky survey in lensing cluster fields especially for \herschel\ because the lensing magnification will allow us to extract and study sources below the nominal confusion-noise limit.

Sources whose observed flux densities are below either of the ALMA or \herschel\ detection thresholds are excluded from the ALCS-\herschel\ joint sample (Table~\ref{tab:04_sed}).
These sources are referred to as ALMA-faint \herschel\ sources (Section~\ref{ss:03c_iter}) or \herschel-faint galaxies (Section~\ref{ss:04c_dropout}; also included in Figure~\ref{fig:lir_vs_z}) in our study.
Based on the left panel of Figure~\ref{fig:lir_vs_z}, we find out that the ALMA-faint \herschel\ sources are likely distributed at $z\simeq 0 - 2$, consistent with the typical redshifts of \herschel\ sources detected in the Frontier Fields \citepalias{rawle16}.
ALCS sources without secure \herschel\ detection primarily reside at $z \gtrsim 3$, consistent with the redshifts of \herschel-faint galaxies presented in Section~\ref{ss:04c_dropout}.

We also find that $\sim$77\% of ALCS sources (excluding cluster member galaxies) host a lower SFR than the star-formation ``main sequence'' (MS) at a fixed stellar mass of $M_\mathrm{star} = 10^{11}$\,\msun\ \citep{speagle14}.
{Assuming that the majority of ALCS sources are on the MS \citep[e.g.,][]{aravena20, sun21a}, this suggests} a lower intrinsic stellar mass than unlensed SMG samples in literature (e.g., median $M_\mathrm{star} = 10^{11.1}$\,\msun\ in \citealt{dudze20}).
The median stellar mass is likely around $10^{10.6\pm0.2}$\,\msun\ based on a comparison with the MS across $z\simeq 1- 4$ \citep{speagle14}, {which will be further analyzed and constrained by a future work from the collaboration}.
The $1\sigma$ scattering of measured SFRs around the $\mathrm{SFR}_\mathrm{MS}(M_\mathrm{star} = 10^{10.6}\,\msun)$ is 0.4\,dex. 

\subsection{Dust Temperature versus IR Luminosity}
\label{ss:06d_temp}

\begin{figure*}[t]
\centering
\includegraphics[width=0.49\linewidth]{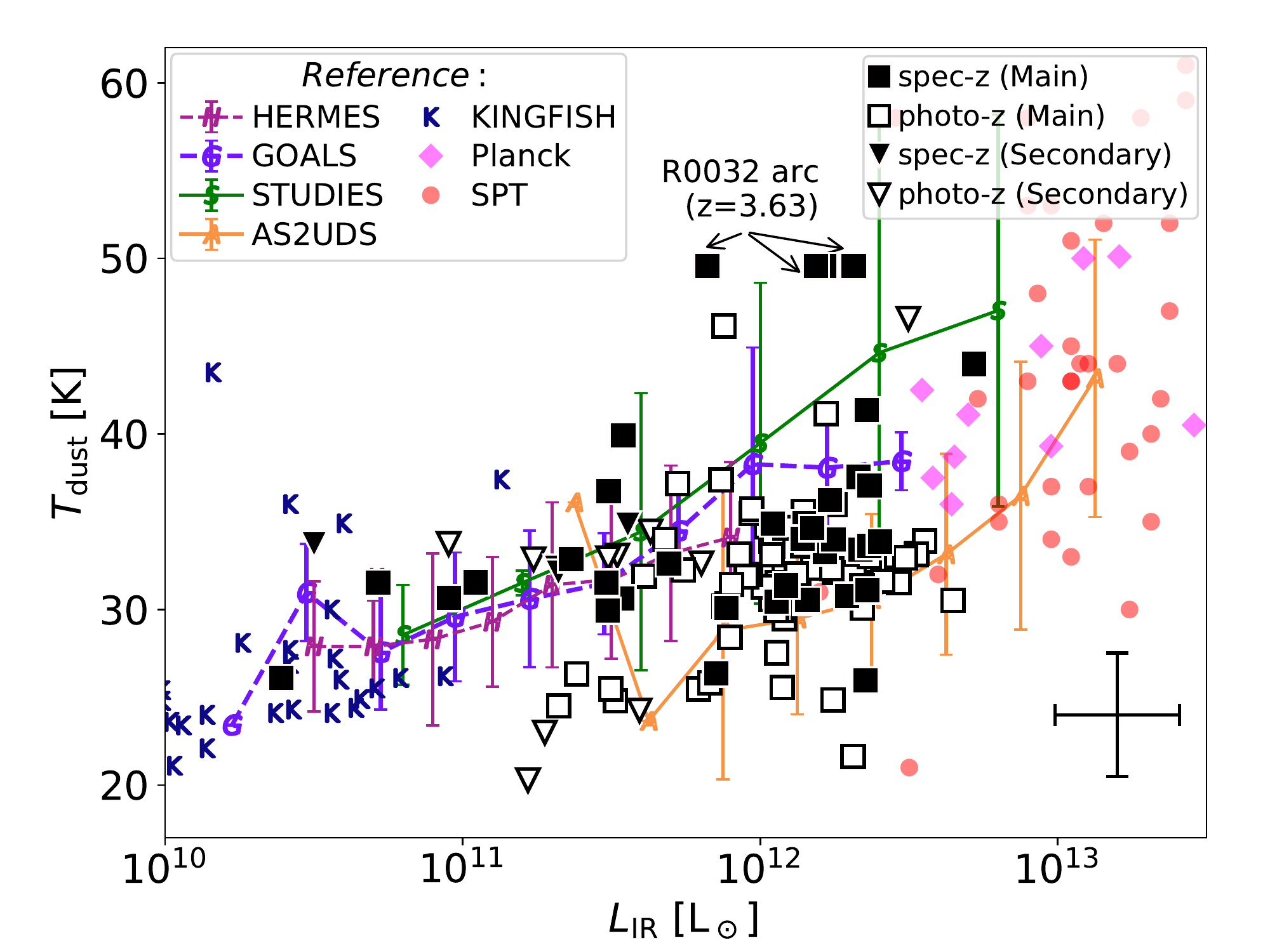}
\includegraphics[width=0.49\linewidth]{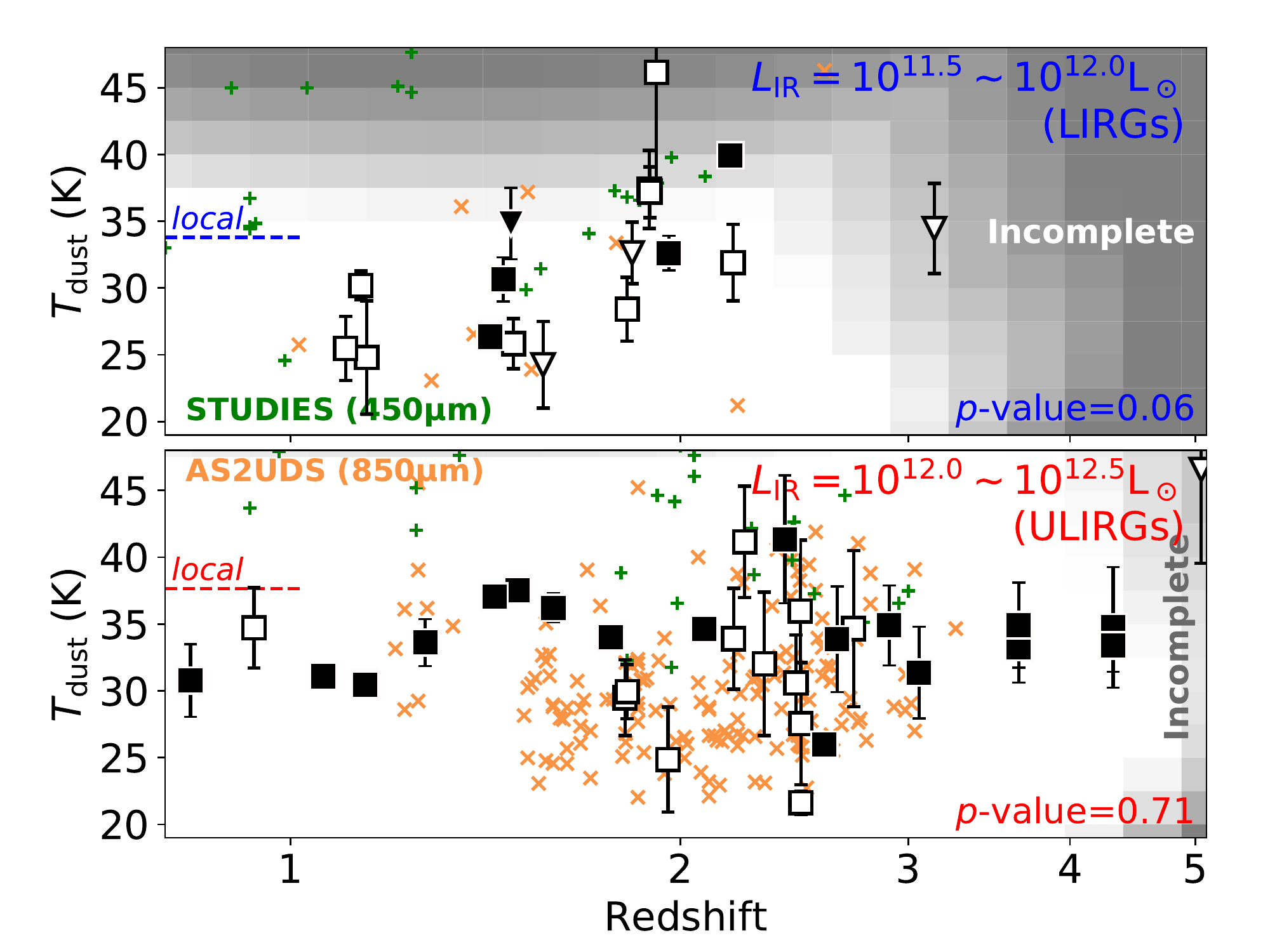}
\vspace{0mm}
\caption{\textit{Left}: Dust temperature versus intrinsic IR luminosity of sources in the ALCS-\herschel\ joint sample. 
Symbols of ALCS sources are the same as those in Figure~\ref{fig:lir_vs_z}, and the typical uncertainty is shown in the lower-right corner.
We also compare our sample with 850-\micron-selected SMGs (AS2UDS sample at $z_\mathrm{med}=2.6$; \citealt{dudze20}), 450-\micron-selected SMGs (STUDIES sample at $z_\mathrm{med}=1.8$; \citealt{lim20}), lensed SMGs (SPT sample at {$z_\mathrm{med}=3.9$; \citealt{spilker16,strandet16,reuter20}; \textit{Planck}'s dusty GEMS sample at $z_\mathrm{med}=2.9$; \citealt{canameras15,canameras18})}, low-redshift ($z \lesssim 1$) LIRGs \citep[HerMES/PEP sample;][]{symeonidis13}, local (U)LIRGs \citep[GOALS sample;][]{ds17} and nearby galaxies \citep[KINGFISH sample;][]{skibba11,hunt15}.
Symbols of reference sample are labeled in the legend in the upper-left corner.
\textit{Right}: Dust temperature versus redshift of ALCS-\herschel\ sources in the intrinsic IR luminosity bins of $10^{11.5} - 10^{12}$\,\lsun\ (i.e., LIRGs; top) and $10^{12} - 10^{12.5}$\,\lsun\ (i.e., ULIRGs; bottom). 
Note that we only consider sources either with \zsp\ or accurate \hst\ \zph, and therefore the uncertainty of $T_\mathrm{dust}$ for each source is less than 20\%.
The $p$-value of Spearman's rank correlation is shown in the lower-right corner of each panel, suggesting no conspicuous evolution of $T_\mathrm{dust}$ for the ULIRG bin but a tentative redshift dependence for the LIRG bin.
Shaded gray regions indicate the $T_\mathrm{dust}-z$ space that the samples are incomplete assuming 16th percentile lensing magnification ($\mu=1.8$), where the sources are not included for $p$-value calculation.
The gradation in gray scale indicates the incompleteness from the highest to lowest $L_\mathrm{IR}$ in each luminosity bin.
450-\micron-selected SMGs (STUDIES sample, \citealt{lim20}; green pluses) and 850-\micron-selected SMGs (AS2UDS sample, \citealt{dudze20}; orange crosses) with relatively accurate $T_\mathrm{dust}$ measurements (uncertainty $<$15\%) are also shown for comparison.
The median $T_\mathrm{dust}$ of local galaxies in the GOALS sample \citep{ds17} is shown as the horizontal short dashed line in each bin.
}
\label{fig:temp}
\end{figure*}

We study the relation between the dust temperature and intrinsic IR luminosity of sources in the ALCS-\herschel\ joint sample as shown in the left panel of Figure~\ref{fig:temp}.
{Sources with redistributed Herschel fluxes (Section~\ref{ss:03d_und}) are considered only once among those in each blended group.}
We also compare our sample with a variety of galaxies, including nearby galaxies \citep[KINGFISH sample;][]{skibba11,hunt15}, local (U)LIRGs \citep[GOALS sample;][]{ds17}, low-redshift LIRGs \citep[HerMES/PEP sample at $z\lesssim 1$;][]{symeonidis13}, 850-\micron-selected SMGs at $z_\mathrm{med}=2.6$ \citep[AS2UDS sample;][]{dudze20}, 450-\micron-selected SMGs at $z_\mathrm{med}=1.8$ \citep[STUDIES sample;][]{lim20} and galaxy-lensed SMGs at {$z_\mathrm{med} = 3.9$ \citep[SPT sample;][]{spilker16,strandet16,reuter20}}.

The median dust temperature is $32.0\pm0.5$\,K for the ALCS-\herschel\ joint sample (with or without far-IR-\zph\ galaxies included).
{Spectroscopically confirmed sources have a slightly higher median dust temperature ($33.6\pm0.9$\,K) than that of sources without $z_\mathrm{spec}$'s ($31.9\pm0.5$\,K).
}
We identify a weak positive $L_\mathrm{IR}-T_\mathrm{dust}$ relation as $T_\mathrm{dust} = (2.2\pm1.7)\log(L_\mathrm{IR}/10^{12}) + (32.6\pm0.7)$ where the units of $L_\mathrm{IR}$ and $T_\mathrm{dust}$ are \lsun\ and K, respectively.
Such a weak $L_\mathrm{IR}-T_\mathrm{dust}$ relation {(Spearman's $\rho=0.22$, $p$-value$=$0.03; uncertainties of $T_\mathrm{dust}$ are considered)} is not consistent with the {strong relations} drawn from most of the high-redshift reference samples {\citep[see also][]{burnham21}}.
It is, however, consistent with some cosmological simulation results including \citet{liang19}.

We argue that a weak observed $L_\mathrm{IR}-T_\mathrm{dust}$ relation is a consequence of inhomogeneous $L_\mathrm{IR}(T_\mathrm{dust})$ selection threshold caused by lensing effect.
The $L_\mathrm{IR} - T_\mathrm{dust}$ relation is a joint effect of both physics {(the Stefan–Boltzmann law)} and selection.
If lensing magnification is not applied, the detection threshold of $L_\mathrm{IR}$ at each given redshift and millimeter flux density (e.g., $z=2$ and $f_{1150}=0.3$\,mJy) will be exactly a monotonic function of $T_\mathrm{dust}$ \citep[e.g.,][]{lim20}.
Combined with the selection bias towards high-$L_\mathrm{IR}$ and high-$T_\mathrm{dust}$ source at higher redshifts in the \herschel\ bands, a positive $L_\mathrm{IR} - T_\mathrm{dust}$ relation could be identified.
In the image plane, we do find a positive $\mu L_\mathrm{IR}-T_\mathrm{dust}$ correlation at $4\sigma$ significance.
However, with the lensing magnification, we are able to detect sources with lower $L_\mathrm{IR}$ at given $T_\mathrm{dust}$ inhomogeneously, leading to a weaker $L_\mathrm{IR}-T_\mathrm{dust}$ relation in the source plane.
{This is also seen with the strongly lensed SPT sources \citep{spilker16,reuter20} where the significance of $L_\mathrm{IR}-T_\mathrm{dust}$ relation is also around $2\sigma$, despite that SPT sources are far more luminous.
We also note that a strong $L_\mathrm{IR}-T_\mathrm{dust}$ relation can be identified for the joint sample of ALCS and SPT. 
This is because the sources of two surveys are selected in distinct ranges of IR luminosities, and the increase of $T_\mathrm{dust}$ over a wider range of $L_\mathrm{IR}$ becomes significant enough ($T\propto L^{0.25} R^{-0.5}$ according to the Stefan-Boltzmann law).
}

We find that at an intrinsic IR luminosity between $10^{11} - 10^{12}\,$\lsun\ (i.e., LIRGs; $z=1.5_{-0.4}^{+0.5}$ for 16-50-84th percentiles of the redshift distribution), the dust temperatures of ALCS sources resemble those of local analogs \citep{ds17} and low-redshift LIRGs selected by \herschel/SPIRE in cosmological deep fields \citep{symeonidis13}. 
This indicates no or weak evolution of the dust temperature of LIRG-like galaxies from $z \sim 2$ to the local Universe, consistent with the conclusion made based on lensed HLS sources at $z_\mathrm{med}=1.9$ on $\Sigma_\mathrm{IR} - T_\mathrm{dust}$ plane \citep{sun21a}.

At $L_\mathrm{IR}\gtrsim 10^{12}$\,\lsun, ALCS sources ($z=2.3_{-0.9}^{+0.8}$ for 16-50-84th percentiles of the redshift distribution) show lower dust temperatures than those of local ULIRGs, resembling 850-\micron-selected SMGs at $z_\mathrm{med}=2.6$ \citep[AS2UDS sample;][]{dudze20} except for one warm outlying system (R0032 lensed arc at $z=3.63$, $T_\mathrm{dust}\sim50$\,K; Figure~\ref{fig:a1_sp} and \citealt{mdz17}).
As a result, the difference in $T_\mathrm{dust}$ between LIRGs and ULIRGs in the ALCS-\herschel\ joint sample at $z\gtrsim 1$ is not significant, which is reflected by the weak slope of $L_\mathrm{IR}-T_\mathrm{dust}$ relation.
Previous works have reported a lower dust temperature or larger optical depth in $z\simeq 1-3$ SMGs compared to their local analogs {\citep[e.g.,][]{symeonidis09, symeonidis13, hwang10,swinbank14}}.
As also pointed out by a few of these studies, the high IR luminosity with relatively low dust temperature is likely caused by a larger size of star-forming region in SMGs ($R_\mathrm{e}=1\sim 2$\,kpc, e.g., \citealt{simpson15,ikarashi15,hodge16,rujopakarn16,fujimoto17,elbaz18,lang19,gullberg19,sun21a,gomez21}) in contrast to the compact size often seen in local ULIRGs \citep[e.g., $\sim 0.1$\,kpc in Arp\,220;][]{soifer00,bm17,sakamoto17}.

We further study the redshift evolution of dust temperature in the IR luminosity bins of $10^{11.5} - 10^{12}$\,\lsun\ (LIRGs) and $10^{12} - 10^{12.5}$\,\lsun\ (ULIRGs).
Because 84\% of ALCS sources have lensing magnification factors greater than 1.8, our joint sample is $\gtrsim$\,80\% complete for galaxies in the LIRG bin at $z\lesssim2.5$ and $T_\mathrm{dust} \lesssim 35$\,K, and galaxies in the ULIRG bin at $z\lesssim4$ and $T_\mathrm{dust} \lesssim 45$\,K.
The right panel of Figure~\ref{fig:temp} shows the dust temperatures versus redshifts of galaxies in the two $L_\mathrm{IR}$ bins, with the incomplete region on $T_\mathrm{dust} - z$ plane visualized with gray shade.
Note that we only include galaxies with uncertainties of $\Delta T / T$ less than 20\%, which only consist of sources with \zsp\ or accurate \hst\ \zph.

To test the existence of any monotonic redshift evolution of $T_\mathrm{dust}$, we compute the Spearman's rank correlation coefficient and $p$-value for sources within the redshift ranges of high completeness.
A weak positive $T_\mathrm{dust}(z)$ relation can be tentatively drawn for sources in the $L_\mathrm{IR}=10^{11.5} - 10^{12}$\,\lsun\ bin ($p$-value\,$=$\,0.06), with sources at $z\lesssim2$ showing lower $T_\mathrm{dust}$ than that of local LIRGs \citep{ds17}.
We argue that the selection bias against low-redshift sources with warm dust temperatures ($T_\mathrm{dust}\gtrsim35$\,K; discussed in Section~\ref{ss:06a_color}) is likely the cause of this weak $T_\mathrm{dust}(z)$ relation.

We note that similar trend is also seen with \citet{symeonidis13} for galaxies with $L_\mathrm{IR}=10^{11.6}-10^{11.8}$\,\lsun\ at $z<1$, and positive $T_\mathrm{dust}(z)$ trend has been suggested by certain stacking analyses \citep[e.g.,][]{magnelli14,schreiber18,simpson19} and simulations (e.g., \citealt{liang19}; also \citealt{lagos20} but a weak $T_\mathrm{dust}(z)$ evolution).
However, {the samples used by most of the stacking analyses are stellar-mass-selected, which are different in total SFR from $L_\mathrm{IR}$- or $M_\mathrm{dust}$-selected samples obtained with submillimeter/millimeter surveys.
This means that the evolution of $T_\mathrm{dust}(z)$ at fixed stellar mass can be a combined effect of the weak evolution of $L_\mathrm{IR}-T_\mathrm{dust}$ relation and strong evolution of star-forming main sequence (i.e., increasing $\mathrm{SFR} / M_\mathrm{star}$ and thus $L_\mathrm{IR} / M_\mathrm{star}$ towards higher redshifts; see also a recent study by \citealt{drew22}).
The positive $T_\mathrm{dust}(z)$ trend} is not seen in 450/850-\micron-selected SMGs in the AS2UDS and STUDIES sample {\citep{dudze20,dudze21,lim20}}. 

For ULIRGs in the bin of $L_\mathrm{IR}=10^{12}-10^{12.5}$\,\lsun, no redshift evolution of $T_\mathrm{dust}$ can be identified ($p$-value\,$=$\,0.71).
Similar conclusion can also be drawn with the STUDIES and AS2UDS SMGs with relatively accurate $T_\mathrm{dust}$ measurements (uncertainty $<$15\%; typically requires $\geq$1-band SPIRE detection) in this $L_\mathrm{IR}$ bin.


\subsection{Gas Depletion Time Scale}
\label{ss:06f_tdep}
\begin{figure}
\centering
\includegraphics[width = \linewidth]{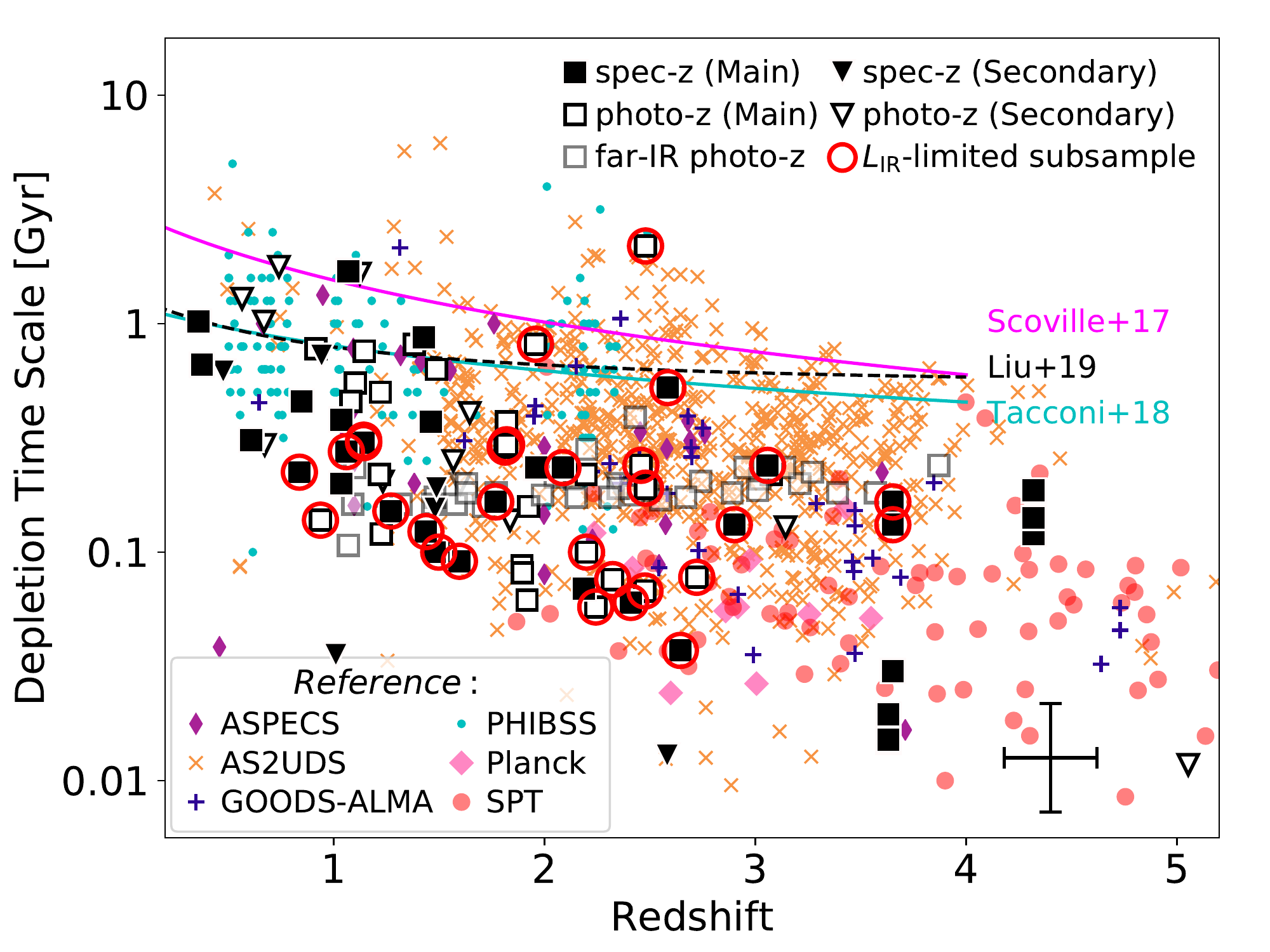}
\caption{Molecular gas depletion time scale ($t_\mathrm{dep} = \delta_\mathrm{GDR} M_\mathrm{dust} / \mathrm{SFR}$) of sources in the ALCS-\herschel\ joint sample as a function of redshift.
The typical uncertainty is shown in the lower-right corner.
Symbols are the same as those in Figure~\ref{fig:lir_vs_z}, but we de-emphasize far-IR-\zph\ sources with semi-transparent symbols to address the degeneracy between $T_\mathrm{dust}$ priors and $t_\mathrm{dep}$.
Sources in the $L_\mathrm{IR}$-limited ULIRG subsample at $z\simeq1-4$ where ALMA and SPIRE detections are $\gtrsim$\,80\% complete (see Section~\ref{ss:06d_temp} and the lower-right panel of Figure~\ref{fig:temp}) are highlighted with red open circles, and no redshift dependence of $t_\mathrm{dep}$ can be identified.
We also plot sources in ASPECS main sample \citep[][purple diamonds]{aravena20}, AS2UDS \citep[][orange crosses]{dudze20}, GOODS-ALMA \citep[][blue pluses]{franco20}, PHIBSS \citep[][cyan dots]{tacconi18}, 
{\textit{Planck}'s dusty GEMs \citep[][pink diamonds]{canameras18} and SPT sample \citep[][shallow red circles]{reuter20}} for comparisons.
$t_\mathrm{dep}(z)$ relations based on the prescriptions of \citet{scoville17}, \citet{tacconi18} and \citet{liu19} are shown in magenta, cyan and dashed black lines, respectively, assuming $M_\mathrm{star}=10^{10.6}\,$\msun\ and main-sequence SFR.
}
\label{fig:16_tdep}
\end{figure}

Assuming a canonical gas-to-dust mass ratio (GDR) of $\delta_\mathrm{GDR}=100$, we can estimate a median gas depletion time scale for sources in the ALCS-\herschel\ joint sample as $t_\mathrm{dep} = \delta_\mathrm{GDR} M_\mathrm{dust} / \mathrm{SFR} = 190_{-95}^{+266}$\,Myr (error bar denotes $1\sigma$ dispersion).
Sources with higher $T_\mathrm{dust}$ generally show shorter $t_\mathrm{dep}$ (equivalently higher star-forming efficiency, $\mathrm{SFE}=\mathrm{SFR} / M_\mathrm{gas}$).
For sources only with far-IR \zph, their dust temperatures are close to the peak of prior $T_\mathrm{dust}$ distribution as assumed by \textsc{magphys}, and therefore the $t_\mathrm{dep}$ is nearly constant.
Despite such a degeneracy, the derived median $t_\mathrm{dep}$ does not change {significantly} if we exclude sources without \zsp\ {($166\pm25$\,Myr)} or \hst\ \zph\ {($219\pm30$\,Myr)} from the statistics.
We further note that $t_\mathrm{dep}$ is a conserved quantity independent of lensing magnification effect.

The derived $t_\mathrm{dep}$ of our sample is broadly consistent with those derived with AS2UDS sample ($295^{+307}_{-185}$\,Myr; \citealt{dudze20}), ASPECS main sample ($299^{+431}_{-167}$\,Myr; \citealt{aravena20}) and HLS bright lensed source sample ($226^{+196}_{-73}$\,Myr; \citealt{sun21a}) using the same method (i.e., $\delta_\mathrm{GDR}=100$ with $M_\mathrm{dust}$ derived from \textsc{magphys} SED modeling).
Although the requirement of \herschel\ detections will result in a selection bias towards sources with higher $T_\mathrm{dust}$ and shorter $t_\mathrm{dep}$, the use of \herschel\ data provides characterization of obscured SFR and $t_\mathrm{dep}$ in higher precision.

Figure~\ref{fig:16_tdep} shows the gas depletion time scale as a function of redshift for sources in the ALCS-\herschel\ joint sample.
We also compare ALCS sources with ASPECS main sample \citep{aravena20}, AS2UDS \citep{dudze20}, GOODS-ALMA \citep{franco20} and PHIBSS sample \citep{tacconi18} across comparable redshift ranges.
The best-fit models of $t_\mathrm{dep}(z)$ at $M_\mathrm{star}=10^{10.6}$\,\msun\ and main-sequence SFR \citep{speagle14}, based on the prescriptions of \citet{scoville17}, \citet{tacconi18} and \citet{liu19}, are also plotted for comparison.
Sources in our sample exhibit a shorter depletion time scale than the model predictions.
This is caused by the IR-selection nature (i.e., favoring sources above the star-forming MS with higher $L_\mathrm{IR}$ and $T_\mathrm{dust}$) and potentially low $T_\mathrm{dust}$ assumptions for single-band millimeter continuum sources in previous studies (e.g., 25\,K in \citealt{scoville17}).
Additionally, $\delta_\mathrm{GDR}$ is observed to be a function of metallicity (e.g., \citealt{leroy11}; $\delta_\mathrm{GDR} \sim 200$ at half solar metallicity), which could introduce further uncertainty to $M_\mathrm{gas}$ and $t_\mathrm{dep}$.

Although a declining trend of $t_\mathrm{dep}$ towards higher redshift can be identified, we argue that this is a selection effect of sources with higher $L_\mathrm{IR}$ and $T_\mathrm{dust}$ towards higher redshift.
If we restrict the sample to an intrinsic IR luminosity of $10^{12}-10^{12.5}$\,\lsun\ at $z\simeq1-4$ where the ALMA and \herschel/SPIRE detections are $\gtrsim$\,80\% complete (Section~\ref{ss:06d_temp} and the lower-right panel of Figure~\ref{fig:temp}), we find no significant redshift dependence of $t_\mathrm{dep}$ (the $p$-value of Spearman's rank correlation is 0.21).
This is consistent with the conclusion of \citet{dudze20} drawn upon $L_\mathrm{IR}$-limited sample with both ALMA and SPIRE detections.

\subsection{Dust Temperature of a \cii-Emitter at $z=6.07$}
\label{ss:06e_z6}

\begin{figure}
\centering
\includegraphics[width=\linewidth]{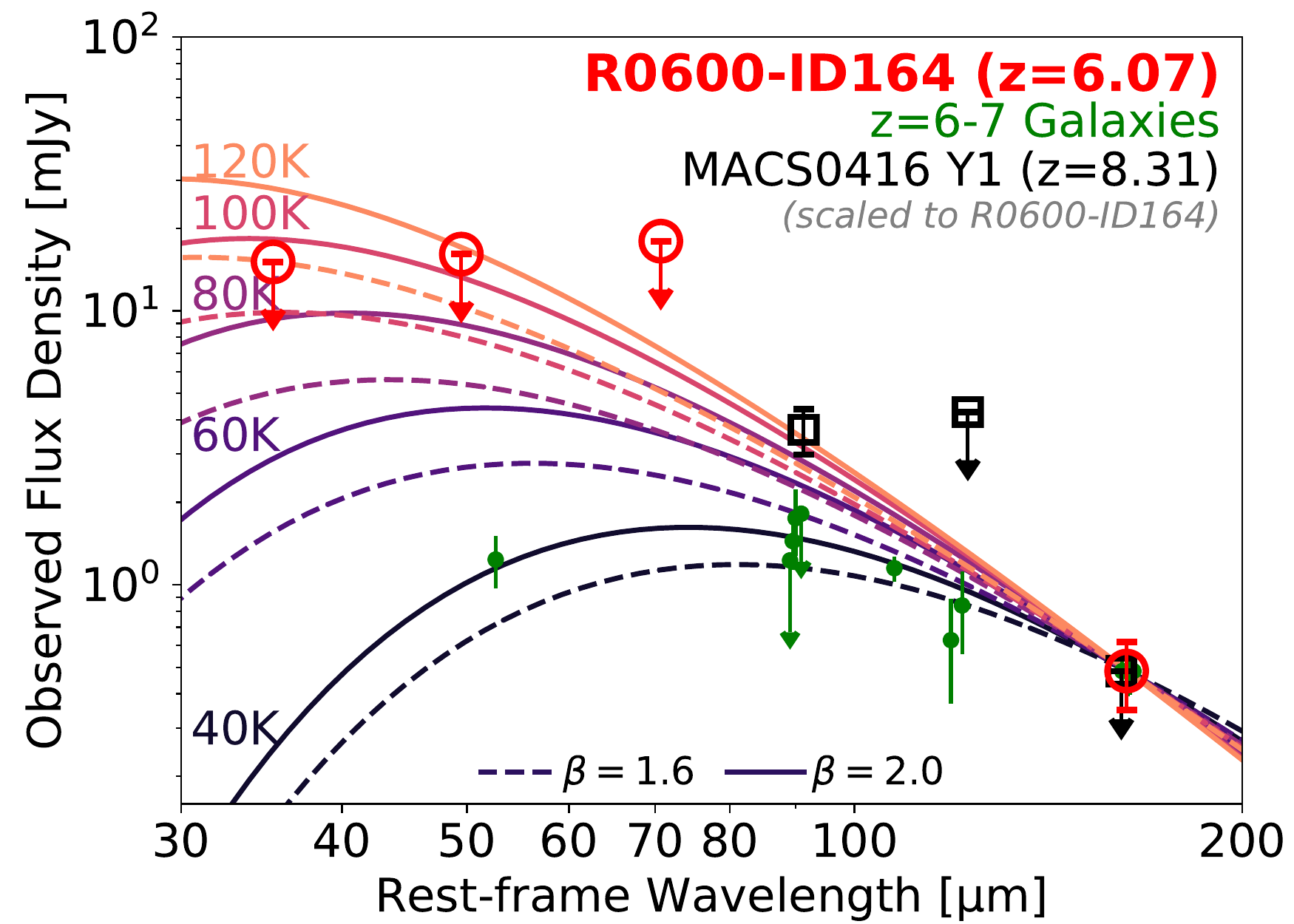}
\caption{Rest-frame far-IR SED of R0600-ID164 (red open circles), a \cii-emitting lensed arc at $z=6.07$ blindly discovered with the ALCS \citep{fujimoto21,laporte21}.
Modified blackbody spectra at {$T_\mathrm{dust}=40  - 120$\,K} (with dust emissivity $\beta = 1.6$ and 2.0 as dashed and solid lines) are plotted for comparison.
Compared with the MACS0416\,Y1 (open black squares; \citealt{bakx2020}) with unusually warm dust temperature ($T_\mathrm{dust}>80$\,K) and/or steep dust emissivity index ($\beta > 2$), we are able to rule out such a scenario with \herschel/SPIRE non-detections.
SEDs of galaxies at $z\simeq6-7$ with normal dust temperature (40--50\,K; \citealt{hashimoto19,harikane20,bakx21}) are shown as green dots. 
Note: all SEDs of galaxies in literature are scaled to the same flux density of R0600-ID184 at 160\,\micron\ in the rest frame. 
Upper limits are at $3\sigma$.
}
\label{fig:z6_T}
\end{figure}

R0600-ID164 is a \cii-emitting lensed arc at $z=6.072$ discovered blindly by the ALCS \citep[named as RXCJ0600--2007 z6.1/z6.2 in][]{fujimoto21}. 
\citet{laporte21} reported the ALMA dust continuum detection of this source ($\mathrm{S/N} = 4.84$ on the 2\arcsec-tapered map).
Although this source is not detected in any \herschel\ band, the upper limits of flux densities at 250--500\,\micron\ can be used to constrain its dust temperature.

Figure~\ref{fig:z6_T} displays the observed far-IR SED of R0600-ID164 before lensing magnification correction (red open circles; upper limits are at $3\sigma$).
We compare the SED with modified blackbody spectral templates at intrinsic $T_\mathrm{dust} = 50 - 125$\,K with the CMB effects considered following the prescription of \citet{dacunha13}.
Based on the dust mass, dust continuum size reported in \citet{laporte21} and dust absorption coefficient adopted in Section~\ref{ss:04a_method}, we find a low dust mass surface density of $(6\pm3)\times10^{6}$\,\si{M_\odot.kpc^{-2}}. Therefore, the optical depth of dust continuum is on the order of unity at $\lambda_\mathrm{thick} \lesssim 30$\,\micron\ in the rest frame.
The dust emissivities are assumed as $\beta =1.6$ and 2.0.
We also compare R0600-ID164 with galaxies at $z\simeq 6- 7$ with normal dust temperatures (40--50\,K; shown as green dots) confirmed with ALMA continuum detections in two bands at least, including J1211--0118, J0217--0208 \citep[$z=6.03$ and 6.20, respectively;][]{harikane20},  A1689-zD1 \citep[$z=7.13$;][]{watson15,knudsen17,inoue20,bakx21} and B14--65666 \citep[$z=7.15$;][]{hashimoto19,sugahara21}.

\citet{bakx2020} showed that MACS0416\,Y1, a lensed galaxy at $z=8.31$ (black open squares in Figure~\ref{fig:z6_T}; also \citealt{tamura19}), exhibits abnormally warm dust temperature ($T_\mathrm{dust} > 80$\,K, 90\% confidence) and/or steep dust emissivity index ($\beta > 2$).
With the three-band SPIRE flux density upper limits, we calculate the $\chi^2$ of non-detections assuming a $T_\mathrm{dust} = 80$\,K, $\beta = 2.0$ and $\lambda_\mathrm{thick} = 30$\,\micron\ dust continuum model.
The derived $\chi^2$ is 6.6 {(reduced $\chi^2_\nu=2.2$)}, indicating that such a model can be ruled out at $>90\%$ confidence level.
Although assuming lower dust emissivity and longer $\lambda_\mathrm{thick}$ will lead to a smaller $\chi^2$, we can draw the conclusion that the $T_\mathrm{dust}$ and $\beta$ of R0600-ID164 are not as extreme as those observed for MACS0416\,Y1. 
However, we also note that the differential lensing effect seen around the caustic line ($\mu=30-160$; \citealt{fujimoto21}) may introduce further uncertainty on the dust temperature constraint.


\section{Summary}
\label{sec:07_sum}

We present an ALMA-\herschel\ joint study of 1.15\,mm continuum sources detected by the ALMA Lensing Cluster Survey (ALCS), an ALMA Cycle-6 large program (PI: Kohno) dedicated for surveying intrinsically faint millimeter sources with the assistance of gravitational lensing.
All of the 33 ALCS cluster fields have been observed by \herschel/SPIRE at 250, 350 and 500\,\micron; 18 of them were observed down to confusion-limited noise levels, and have high-quality \herschel/PACS 100 and 160\,\micron\ coverages (i.e., in the ``deep'' mode). 
We conducted PSF flux extraction for all 141 secure ALCS sources ($\mathrm{S/N}_\mathrm{nat} \geq 5$ in the native-resolution maps or $\mathrm{S/N}_\mathrm{tap} \geq 4.5$ in the 2\arcsec-tapered maps; the main sample) and 39 tentative ALCS sources with near/mid-IR counterparts ($\mathrm{S/N}_\mathrm{nat}=4-5$ and $\mathrm{S/N}_\mathrm{tap}<4.5$; the secondary sample).
We then performed far-infrared SED modeling for 125 of them, which are detected in the \herschel\ bands at S/N\,$\geq$\,2 (the ALCS-\herschel\ joint sample). 
The main results are the following:

\begin{enumerate}

\setlength\itemsep{-0.25em}

\item 113 out of the 141 secure ALCS sources were detected at $>2\sigma$ in at least one \herschel\ band, and 22 out of 39 tentative ALCS sources were detected above the same threshold (Table~\ref{tab:03_phot}).
The single-band \herschel\ detection rate remains nearly constant as a function of $\mathrm{S/N}_\mathrm{ALMA}$ at 100 and 250\,\micron, but clearly correlated with $\mathrm{S/N}_\mathrm{ALMA}$ at longer wavelength (350 and 500\,\micron).

\item We conducted far-IR SED modeling and derived photometric redshifts for 125 \herschel-detected ALCS sources (109 independently) excluding BCGs. 
Among them, 47 sources are spectroscopically confirmed, and additional 42 sources have catalogued \hst\ photometric redshifts.
Physical properties ($L_\mathrm{IR}$, SFR, $M_\mathrm{dust}$, $T_\mathrm{dust}$) of sources in this ALCS-\herschel\ joint sample are presented in Table~\ref{tab:04_sed}, derived based on the best available redshifts ($z_\mathrm{best}$).

\item 27 lensed ALCS sources in the main sample are not detected in any \herschel\ band ($<2\sigma$).
Among these \herschel-faint galaxies, ten sources have catalogued \zsp\ ($z=$\,3.631, 6.072; two sources) or \hst\ \zph\ ($z_\mathrm{med}=2.0\pm1.0$; eight sources).
The remaining 17 optical/near-IR-dark sources likely reside at $z_\mathrm{phot} = 4.2 \pm 1.2$, hosting a typical IR luminosity of $10^{11.7\pm0.3}(2.6/\mu)$\,\lsun\ and obscured SFR of $40_{-20}^{+40}(2.6/\mu)$\,\smpy\ (median lensing magnification factor $\mu_\mathrm{med}=2.6$).

\item 
{ALCS sources are cold-dust-mass-selected ($M_\mathrm{dust}\gtrsim 10^{8}$\,\si{\mu^{-1}.M_\odot}) across $z\simeq 1 - 6$. 
However, at a fixed $L_\mathrm{IR}$, the} ALMA-\herschel\ joint selection is biased against galaxies with local-ULIRG-like dust temperature ($T_\mathrm{dust}\sim 40$\,K) at $z \lesssim 1$. 
This is because (\romannumeral1) the effective survey volume of ALCS at $z<1$ is limited for the selection of ULIRGs ($L_\mathrm{IR}>10^{12}$\,\lsun), and (\romannumeral2) at given redshift and intrinsic $L_\mathrm{IR}$, galaxies with higher dust temperatures (i.e., ULIRG-like) will appear fainter in ALMA Band 6, and thus it is more challenging to detect them with the ALCS.

\item The 16--50--84th percentiles of the redshift distribution of the ALCS sources in the main sample (excluding cluster member galaxies) are {1.15--2.08--3.59}.
The median redshift of secure ALCS sources is higher than that of sources selected with the deep ASPECS survey \citep{aravena20}, but lower than those of sources in shallower ALMA Band-6/7 surveys \citep[e.g.,][]{dudze20,yamaguchi20,gomez21}.
Together with the median redshift as a function of 1.15\,mm flux density cut ($z_\mathrm{med}(>f_{1150})$), this suggests an increasing fraction of $z\simeq 1 - 2$ galaxies among fainter millimeter sources ($\sim$\,0.1\,mJy) and {potentially} decreasing obscured fraction of cosmic star formation at $z>4$.

\item With a median lensing magnification of $\mu=2.6_{-0.8}^{+2.6}$, we derive an intrinsic SFR distribution of $94_{-54}^{+84}$\,\smpy\ for sources at $\mathrm{S/N}\geq5$ (errorbar denotes $1\sigma$ dispersion).
The intrinsic SFRs and IR luminosities of our sample are slightly higher than those of local LIRGs \citep[e.g., GOALS sample;][]{howell10} but lower than those of conventional unlensed SMGs selected at $z\sim2-3$ by a factor of $\sim3$ \citep{dudze20}.

\item We compare the dust temperatures (modeled by a modified blackbody spectrum with $\beta = 1.8$) versus IR luminosities with various galaxy samples from the literature.
The median $T_\mathrm{dust}$ is $32.0\pm0.5$\,K for the ALCS-\herschel\ joint sample.
Our result suggests no or weak evolution of the $T_\mathrm{dust}$ of LIRGs ($L_\mathrm{IR}<10^{12}$\,\lsun) from $z\sim 2$ to the local Universe at a given \lir.
At $L_\mathrm{IR}\gtrsim 10^{12}$\,\lsun, ALCS sources exhibit cooler dust temperature compared with local ULIRGs, and no evolution of $T_\mathrm{dust}$ can be found at $z\simeq 1- 4$, as has been reported by previous SMG surveys \citep{dudze20}.

\item Assuming a canonical gas-to-dust ratio of 100, the gas depletion time scales for sources in the ALCS-\herschel\ sample are found to be $190_{-95}^{+265}$\,Myr.
For sources in the IR luminosity range of $10^{12}-10^{12.5}$\,\lsun, no redshift evolution of $t_\mathrm{dep}$ can be identified across $z\simeq1-4$.

\item The $z=6.072$ \cii-emitting lensed arc, R0600-ID164, is the highest-redshift source in our ALCS continuum source sample ($\mathrm{S/N}\geq 4$) confirmed so far \citep{fujimoto21,laporte21}.
With the \herschel/SPIRE non-detections, we can rule out a MACS0416\,Y1-like warm dust temperature ($T_\mathrm{dust} > 80$\,K; \citealt{bakx2020}) at $>90$\% confidence level.


\end{enumerate}

With the lensing magnification provided by massive galaxy clusters, our joint analysis based on ALMA and \herschel\ observations reveal the population of galaxies at $z \simeq 1 - 3$ with moderate star formation rate at a few tens of solar masses per year.
ALMA surveys in blank fields would require $\sim 7$ times longer observation times to reach the same depth.
In the high-redshift regime, we discover 17 optical/near-IR-dark \herschel-faint sources that are likely dust-obscured star-forming galaxies at $z\sim 4$.
These sources are excellent targets for multi-wavelength follow-up.
Future ALMA spectral line-scan and \textit{JWST}/NIRSpec observations will provide key insights into the nature of these distant dusty star-forming galaxies, revealing their contribution to the obscured cosmic SFR density at $z > 4$.



\acknowledgments

\vspace{-3.0cm}


{We thank the anonymous referee for helpful comments.}
FS acknowledges support from the NRAO Student Observing Surport (SOS) award SOSPA7-022.
FS and EE acknowledge funding from JWST/NIRCam contract to the University of Arizona, NAS5-02105.
{KK acknowledges support from JSPS KAKENHI Grant Number JP17H06130
and the NAOJ ALMA Scientific Research Grant Number 2017-06B.}
IRS acknowledges support from STFC (ST/T000244/1).
P.G.P.-G. acknowledges support from Spanish Government grant PGC2018-093499-B-I00.
{MO acknowledges support from JSPS KAKENHI Grant Numbers JP18K03693, JP20H00181, JP20H05856, JP22H01260.
}
AZ acknowledges support from the Ministry of Science and Technology, Israel.
We thank Ugn{\.{e}} Dudzevi{\v{c}}i{\={u}}t{\.{e}} for sharing the composite SEDs of AS2UDS SMGs. 
We thank Claudia Lagos for helpful discussion.

This paper makes use of the following ALMA data: ADS/JAO.ALMA\#2018.1.00035.L, 2013.1.00999.S and  2015.1.01425S. 
ALMA is a partnership of ESO (representing its member states), NSF (USA) and NINS (Japan), together with NRC (Canada), MOST and ASIAA (Taiwan), and KASI (Republic of Korea), in cooperation with the Republic of Chile. The Joint ALMA Observatory is operated by ESO, AUI/NRAO and NAOJ.
The National Radio Astronomy Observatory is a facility of the National Science Foundation operated under cooperative agreement by Associated Universities, Inc.
This work is based on observations made with \textit{Herschel}.
\textit{Herschel} is an ESA space observatory with science instruments provided by European-led Principal Investigator consortia and with important participation from NASA.
This work is based (in part) on observations made with the \textit{Spitzer Space Telescope}, which was operated by the Jet Propulsion Laboratory, California Institute of Technology under a contract with NASA.
This research is based on observations made with the NASA/ESA \textit{Hubble Space Telescope} obtained from the Space Telescope Science Institute, which is operated by the Association of Universities for Research in Astronomy, Inc., under NASA contract NAS 5–26555.
{Some of the data presented in this paper were obtained from the Mikulski Archive for Space Telescopes (MAST) at the Space Telescope Science Institute. 
The specific observations analyzed can be accessed via \dataset[Frontier Fields]{https://doi.org/10.17909/T9KK5N}, \dataset[CLASH]{https://doi.org/10.17909/t90w2b} and \dataset[RELICS]{https://doi.org/10.17909/T9SP45}.}


\facilities{HST (ACS and WFC3); Spitzer (IRAC); Herschel (PACS and SPIRE); ALMA.}


\software{astropy \citep{astropy},
          Photutils \citep{photutils},
          MAGPHYS \citep{dacunha08,dacunha15,battisti19}.
          }

\vspace{-5mm}



\startlongtable
\begin{deluxetable*}{@{\extracolsep{2pt}}lrrrrrrrr} 
\tablecaption{Summary of \herschel\ observations for 33 ALCS galaxy cluster fields
\label{tab:01_obs}}
\tablewidth{0pt}
\tabletypesize{\scriptsize}
\tablehead{
\colhead{Cluster Name} & 
\colhead{Group\tablenotemark{a}} & 
\multicolumn2c{Coordinates} &
\colhead{Short Name} & 
\multicolumn2c{\herschel/PACS 100/160\,\micron}
& \multicolumn2c{\herschel/SPIRE 250/350/500\,\micron}
\\\cline{3-4}\cline{6-7}\cline{8-9}
\colhead{}  & \colhead{} 
& \colhead{RA} & \colhead{Dec} & \colhead{} & 
\colhead{Observation ID} & \colhead{$t_\mathrm{obs}$ (h)\tablenotemark{b}} 
& \colhead{Observation ID} & \colhead{$t_\mathrm{obs}$ (s)\tablenotemark{b}} 
}
\startdata
\multicolumn2c{18 HLS-``deep'' clusters:} \\\hline
            Abell209 &  CLASH & 01:31:52.5 & --13:36:38 & A209 & 134218841[8,9] & 4.28 & 1342188581 & 5803 \\
            Abell383 &  CLASH & 02:48:03.3 & --03:31:44 & A383 & 134218915[1,2] & 4.28 & 1342189503 & 5803 \\
    MACS0329.7--0211 &  CLASH & 03:29:41.6 & --02:11:47 & M0329 & 134224928[0,1] & 4.28 & 13422[14564,39844] & 1580 \\
    MACS0429.6--0253 &  CLASH & 04:29:36.1 & --02:53:08 & M0429 & 1342250[641,836] & 4.28 & 13422[39932,41124] & 1580 \\
     MACS1115.9+0129 &  CLASH & 11:15:52.0 &  01:29:56 & M1115 & 13422476[72,91] & 4.28 & 13422[23226,56866] & 1580 \\
    MACS1206.2--0847 &  CLASH & 12:06:12.2 & --08:48:02 & M1206 & 134225745[5,6] & 4.28 & 13422[34856,47273] & 1580 \\
    MACS1311.0--0310 &  CLASH & 13:11:01.6 & --03:10:39 & M1311 & 13422486[26,56] & 4.28 & 13422[34800,59416] & 1580 \\
       RXJ1347--1145 &  CLASH & 13:47:30.5 & --11:45:10 & R1347 & 134221383[6,7] & 5.23 & 13422[01256--63,47859--61] & 12728 \\
     MACS1423.8+2404 &  CLASH & 14:23:47.7 &  24:04:40 & M1423 & 134218821[5,6] & 5.47 & 1342188159 & 6636 \\
    MACS1931.8--2635 &  CLASH & 19:31:49.6 & --26:34:34 & M1931 & 13422416[19,81] & 4.28 & 13422[15993,54639] & 1580 \\
    MACS2129.4--0741 &  CLASH & 21:29:26.2 & --07:41:26 & M2129 & 134218780[1,2] & 5.47 & 1342195710 & 5786 \\
      RXJ2129.7+0005 &  CLASH & 21:29:39.9 &  00:05:18 & R2129 & 134218725[6,7] & 5.32 & 1342188167 & 6636 \\
           Abell2744 &    HFF & 00:14:21.2 & --30:23:50 & A2744 & 134218825[1,2] & 5.47 & 1342188584 & 5803 \\
            Abell370 &    HFF & 02:39:52.9 & --01:34:36 & A370 & 134222333[2,3] & 5.23 & 13422[01311--18,48002--04] & 12728 \\
   MACSJ0416.1--2403 &    HFF & 04:16:08.9 & --24:04:28 & M0416 & 134225029[1,2] & 4.28 & 13422[39858,41122] & 1580 \\
    MACSJ1149.5+2223 &    HFF & 11:49:36.3 &  22:23:58 & M1149 & 134221179[7,8] & 4.28 & 1342222841 & 5786 \\
          AbellS1063 &    HFF & 22:48:44.4 & --44:31:48 & AS1063 & 134218822[2,3] & 5.47 & 1342188165 & 6636 \\
         Abell2537 & RELICS & 23:08:22.2 & --02:11:32 & A2537 & 1342187[799,800] & 5.47 & 1342188179 & 6636 \\
\hline
\multicolumn2c{15 HLS-``snapshot'' clusters:} \\\hline
     RXCJ0032.1+1808 & RELICS & 00:32:11.0 &  18:07:49 & R0032 & \nodata & \nodata    & 1342234685 & 169 \\
   MACSJ0035.4--2015 & RELICS & 00:35:26.9 & --20:15:40 & M0035 & \nodata & \nodata    & 1342234697 & 169 \\
   ACTCLJ0102--49151 & RELICS & 01:03:00.0 & --49:16:22 & ACT0102 & \nodata & \nodata    & 1342258408 & 169 \\
   MACSJ0159.8--0849 & RELICS & 01:59:49.4 & --08:50:00 & M0159 & \nodata & \nodata    & 1342237535 & 169 \\
           AbellS295 & RELICS & 02:45:31.3 & --53:02:24 & AS295 & \nodata & \nodata    & 1342236215 & 169 \\
   MACSJ0257.1--2325 & RELICS & 02:57:10.2 & --23:26:11 & M0257 & \nodata & \nodata    & 1342214559 & 169 \\
    PLCKG171.9--40.7 & RELICS & 03:12:56.9 &  08:22:19 & P171 & \nodata & \nodata    & 1342239833 & 169 \\
           Abell3192 & RELICS & 03:58:53.0 & --29:55:44 & A3192 & \nodata & \nodata    & 1342239861 & 169 \\
   MACSJ0417.5--1154 & RELICS & 04:17:33.7 & --11:54:22 & M0417 & \nodata & \nodata    & 1342239855 & 169 \\
   MACSJ0553.4--3342 & RELICS & 05:53:23.0 & --33:42:29 & M0553 & \nodata & \nodata    & 1342227700 & 169 \\
    RXCJ0600.1--2007 & RELICS & 06:00:09.7 & --20:08:08 & R0600 & \nodata & \nodata    & 1342230801 & 169 \\
  SMACSJ0723.3--7327 & RELICS & 07:23:19.4 & --73:27:15 & SM0723 & \nodata & \nodata    & 1342229668 & 169 \\
     RXCJ0949.8+1707 & RELICS & 09:49:50.8 &  17:07:15 & R0949 & \nodata & \nodata    & 1342246604 & 169 \\
           Abell2163 & RELICS & 16:15:48.3 & --06:07:36 & A2163 & \nodata & \nodata    & 1342229566 & 169 \\
    RXCJ2211.7--0350 & RELICS & 22:11:45.9 & --03:49:44 & R2211 & \nodata & \nodata    & 1342211362 & 169 \\
\enddata
\tablecomments{Clusters with \herschel/PACS data are considered as observed in the ``deep'' mode, and the remaining ones are considered as observed in the ``snapshot'' mode (see Section~\ref{ss:02c_spire}). 
\herschel\ observation IDs in brackets indicate the difference in the last a few digits, for example, the SPIRE data of Abell370 was taken with observation IDs 1342201311--1342201318 and 1342248002--1342248004. 
}
\vspace{-2mm}
\tablenotetext{a}{Group name of \hst\ Program.}
\vspace{-2mm}
\tablenotetext{b}{Total scan time of all observations. }
\end{deluxetable*}

\startlongtable
\begin{deluxetable*}{lrrrrrr} 
\tablecaption{Summary of median $1\sigma$ depth of prior-based \herschel\ catalogues within the ALCS footprints
\label{tab:02_unc}}
\tablewidth{0pt}
\tabletypesize{\scriptsize}
\tablehead{
\colhead{Cluster Name} & $N(\mathrm{main})$\tablenotemark{a}
 & \colhead{PACS 100\,\micron}  & \colhead{PACS 160\,\micron}  & \colhead{SPIRE 250\,\micron}  & \colhead{SPIRE 350\,\micron}  & \colhead{SPIRE 500\,\micron} 
\\
\colhead{} & \colhead{}  & \colhead{(mJy)} 
 & \colhead{(mJy)}  & \colhead{(mJy)}  & \colhead{(mJy)}  & \colhead{(mJy)} 
}
\startdata
Abell209 &    1  & \nodata (1.5) & \nodata (3.0) & \nodata (5.0) & \nodata (5.1) & \nodata (5.7) \\
Abell383 &    1  & 0.6 (1.5) & 1.8 (2.9) & 2.8 (5.5) & 2.7 (5.1) & \nodata (5.9) \\
MACS0329.7--0211 &    1  & 0.5 (1.4) & \nodata (2.8) & \nodata (7.1) & \nodata (6.8) & \nodata (7.5) \\
MACS0429.6--0253 &    3  & 0.6 (1.4) & 1.0 (2.9) & 3.3 (7.5) & \nodata (6.9) & \nodata (8.0) \\
MACS1115.9+0129 &    4  & 0.8 (1.4) & 0.8 (2.9) & 3.2 (7.8) & 4.3 (6.5) & 3.9 (7.1) \\
MACS1206.2--0847 &    6  & 0.5 (1.4) & 1.7 (3.1) & 15.5 (8.6) & 2.2 (7.5) & 4.4 (8.2) \\
MACS1311.0--0310 &    2  & 0.5 (1.4) & 1.3 (2.8) & 3.2 (7.4) & 2.7 (6.5) & 2.6 (7.2) \\
RXJ1347--1145 &    6  & 0.5 (1.3) & 0.7 (2.4) & 2.4 (4.3) & 6.7 (4.3) & 2.0 (5.2) \\
MACS1423.8+2404 &    2  & 0.6 (1.5) & 1.3 (2.9) & 3.4 (5.0) & 3.0 (5.0) & 2.3 (5.6) \\
MACS1931.8--2635 &    4  & 1.0 (1.4) & 1.2 (2.8) & 2.8 (7.0) & 3.0 (6.5) & 3.0 (7.2) \\
MACS2129.4--0741 &    2  & 0.5 (1.5) & 0.8 (3.4) & 2.1 (6.8) & 2.9 (6.9) & \nodata (6.9) \\
RXJ2129.7+0005 &    2  & \nodata (1.3) & 1.1 (3.1) & 3.2 (5.5) & \nodata (5.7) & 4.1 (6.1) \\
Abell2744 &    6  & 0.9 (1.5) & 1.5 (3.0) & 2.4 (5.1) & 3.3 (5.4) & 2.8 (5.8) \\
Abell370 &    5  & 0.5 (1.3) & 1.3 (2.8) & 2.8 (5.0) & 3.2 (5.3) & 2.2 (6.2) \\
MACSJ0416.1--2403 &    4  & 0.5 (1.4) & 0.8 (2.7) & 2.7 (7.0) & 1.9 (6.0) & 1.7 (7.3) \\
MACSJ1149.5+2223 &    1  & 0.9 (1.5) & 3.2 (3.0) & 2.1 (5.4) & 2.2 (4.9) & 3.5 (5.6) \\
AbellS1063 &    4  & 0.9 (1.4) & 0.8 (2.3) & 3.3 (5.5) & 5.0 (5.6) & 3.2 (6.4) \\
Abell2537 &    2  & 1.5 (1.5) & 2.2 (3.1) & 3.2 (5.9) & 4.0 (6.1) & 2.6 (6.0) \\
RXCJ0032.1+1808 &   17  &  \nodata &  \nodata & 4.5 (14.4) & 6.4 (16.6) & 4.8 (16.9) \\
MACSJ0035.4--2015 &    2  &  \nodata &  \nodata & 3.6 (10.7) & \nodata (11.6) & \nodata (12.9) \\
ACTCLJ0102--49151 &   12  &  \nodata &  \nodata & 4.7 (11.8) & 4.8 (11.7) & 5.0 (15.7) \\
MACSJ0159.8--0849 &    4  &  \nodata &  \nodata & 4.5 (11.1) & 4.1 (12.0) & 5.2 (13.7) \\
AbellS295 &    0  &  \nodata &  \nodata & \nodata (11.7) & \nodata (12.4) & \nodata (15.2) \\
MACSJ0257.1--2325 &    1  &  \nodata &  \nodata & 4.4 (11.8) & 4.5 (11.8) & 4.7 (15.3) \\
PLCKG171.9--40.7 &    3  &  \nodata &  \nodata & 5.1 (12.3) & \nodata (12.8) & \nodata (13.7) \\
Abell3192 &    4  &  \nodata &  \nodata & 4.4 (11.1) & 4.7 (12.1) & 4.6 (14.1) \\
MACSJ0417.5--1154 &    7  &  \nodata &  \nodata & 4.7 (10.9) & 5.7 (13.0) & 5.2 (15.2) \\
MACSJ0553.4--3342 &   13  &  \nodata &  \nodata & 4.5 (10.6) & 4.6 (11.7) & 6.1 (14.5) \\
RXCJ0600.1--2007 &    4  &  \nodata &  \nodata & 4.4 (11.5) & 5.6 (12.4) & \nodata (13.8) \\
SMACSJ0723.3--7327 &    2  &  \nodata &  \nodata & \nodata (12.2) & \nodata (13.2) & \nodata (15.5) \\
RXCJ0949.8+1707 &    4  &  \nodata &  \nodata & 5.0 (12.0) & 5.3 (12.1) & 5.7 (14.6) \\
Abell2163 &    0  &  \nodata &  \nodata & \nodata (11.1) & \nodata (10.9) & \nodata (13.8) \\
RXCJ2211.7--0350 &    3  &  \nodata &  \nodata & 5.0 (12.5) & 6.3 (12.6) & 6.9 (14.4) \\
\enddata
\tablecomments{Values in parenthesis are RMS noises directly measured from 2D \herschel\ uncertainty maps within the ALCS footprints (i.e., without any dedicated positional prior; Section~\ref{ss:03d_und}).}
\vspace{-2mm}
\tablenotetext{a}{Number of ALMA sources in the main sample ($\mathrm{S/N}_\mathrm{nat}\geq5$ in the native-resolution maps, or $\mathrm{S/N}_\mathrm{nat}\geq4.5$ in the 2\arcsec-tapered maps).}
\end{deluxetable*}



\appendix

\vspace{-5mm}



\section{Quality of the Photometric Results}
\label{apd:01_xid}

\renewcommand{\thefigure}{A\arabic{figure}}
\renewcommand{\theHfigure}{A\arabic{figure}}
\setcounter{figure}{0}

\begin{figure}[!htb]
\centering
\includegraphics[width=0.49\linewidth]{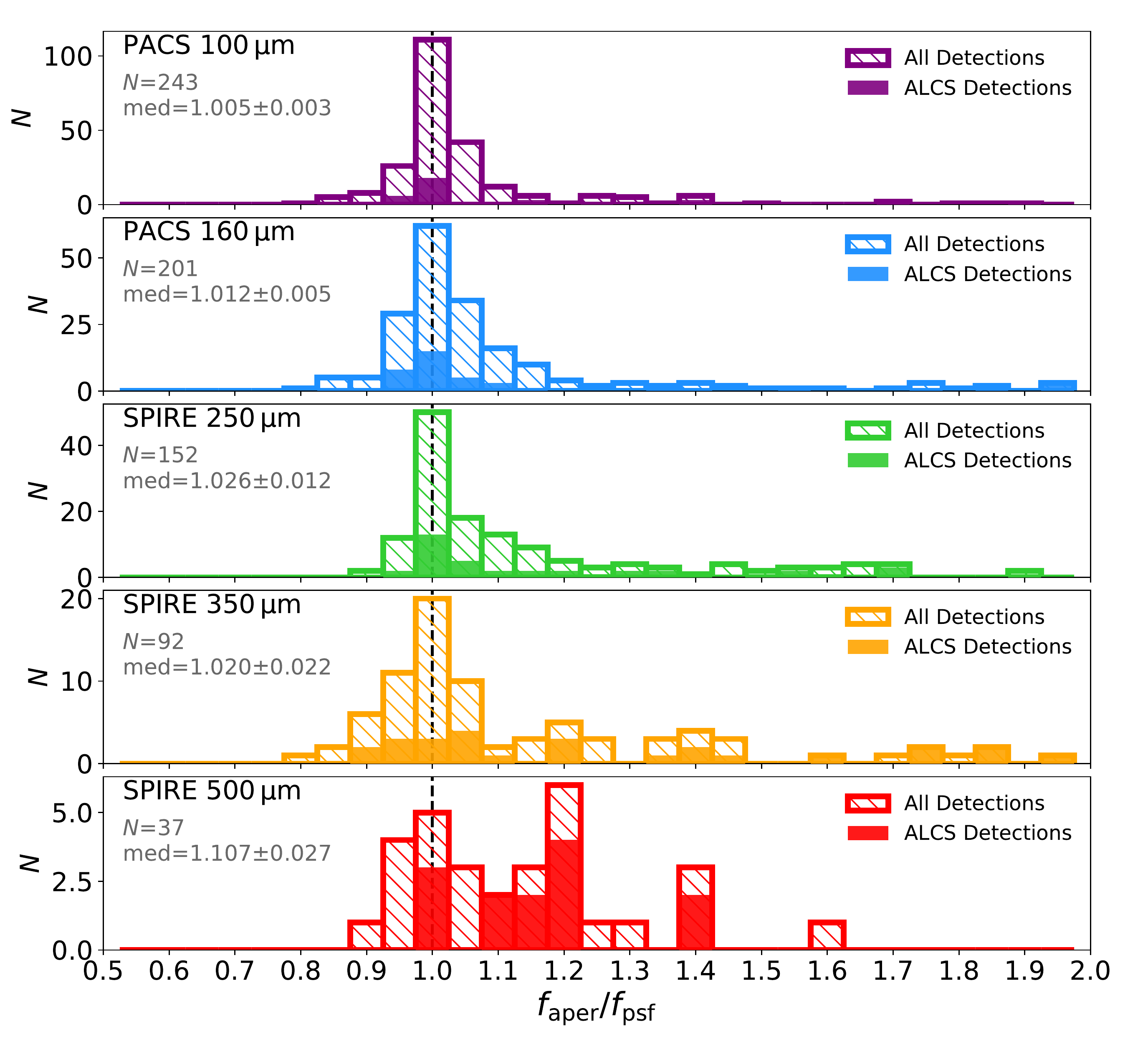}
\includegraphics[width=0.49\linewidth]{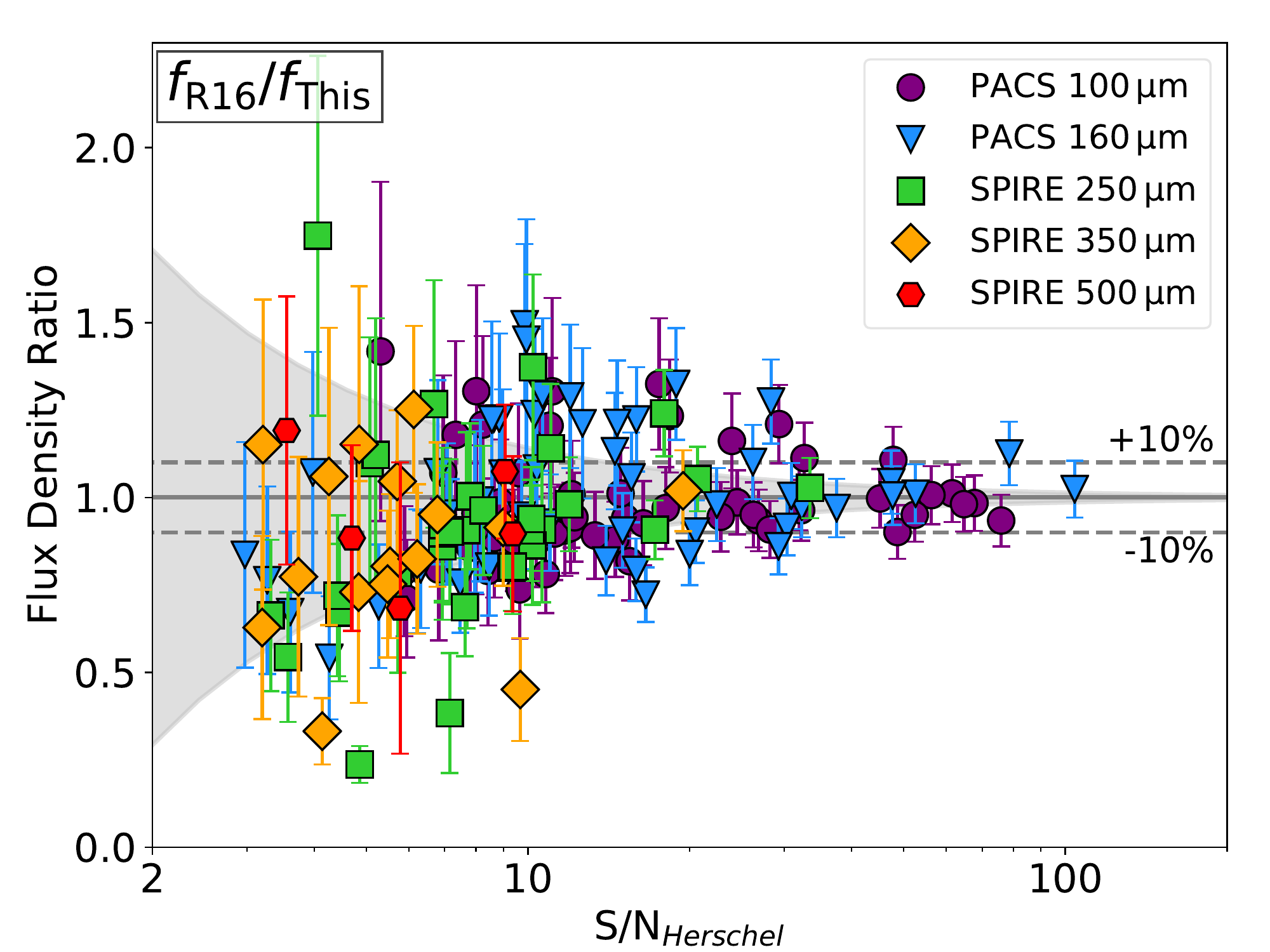}
\caption{\textit{Left}: Ratios of flux densities measured with aperture ($f_\mathrm{aper}$) and PSF ($f_\mathrm{psf}$) photometry in the five \herschel\ bands (100--500\,\micron\ from the top to the bottom).
All the sources detected in the \herschel\ images at S/N$>$5 are shown as the empty hatched histograms, and the ALCS sources are shown as the solid filled ones.
Dashed black lines indicate the cases where the flux densities obtained through PSF and aperture photometry are identical. 
The number of sources and the median $f_\mathrm{aper}/f_\mathrm{psf}$ ratio (with $1\sigma$ standard error) are also shown in each panel for each \herschel\ band.
\textit{Right}:
\herschel\ flux density ratios of submillimeter sources in \textit{Hubble} Frontier Fields measured by \citet{rawle16} and this work (i.e., both detected and undetected by the ALCS), plotted as a function of S/N in the corresponding \herschel\ band ($\mathrm{S/N}_\mathit{Herschel}$).
Symbols are the same as those in Figure~\ref{fig:05_hers_unc} and also labeled in the upper-right corner.
The solid gray line indicates the case in which the two measurements are identical, 
and the dashed lines indicate the cases of $\pm10$\% deviation.
The shaded region denotes the $1\sigma$ dispersion range of flux ratios for sources detected at any given $\mathrm{S/N}_\mathit{Herschel}$.
}
\label{fig:04_flux_comp}
\end{figure}


In order to examine the quality of our PSF photometric results, we first compared our \herschel\ flux densities ($f_\mathrm{psf}$) with those obtained with aperture photometry ($f_\mathrm{aper}$) in the left panel of Figure~\ref{fig:04_flux_comp}.
This comparison includes both the ALCS sources and ALMA-undetected \herschel\ sources extracted at S/N\,$>$\,5 in each \herschel\ band.
We find that the majority of \herschel\ sources exhibit comparable flux densities obtained through both methods, i.e., $f_\mathrm{aper}/f_\mathrm{psf}\sim 1$, up to a wavelength of 350\,\micron\ (see statistics for each band in the left panel of Figure~\ref{fig:04_flux_comp}).
Because of source blending, a number excess of sources can be found at $f_\mathrm{aper} / f_\mathrm{psf} > 1$ especially at longer wavelength (i.e., 500\,\micron).
Such an effect is inevitable for the majority of ALCS sources given the large beam size of 35\arcsec\ at 500\,\micron.
We then conclude that our flux densities obtained through PSF photometry are not subject to any obvious systematic offset from those by aperture photometry.

We also compare our flux density measurements of sources in Frontier Fields with \citetalias{rawle16}.
\citetalias{rawle16} reported 263 secure \herschel\ detections within the \hst/ACS footprints of the six HFF clusters (both central regions and parallel footprints).
We cross-match our photometric measurements ($f_\mathrm{This}$; including both ALCS sources and ALMA-undetected \herschel\ sources) with those of \citetalias{rawle16} ($f_\mathrm{R16}$) allowing a maximum angular separation of 1\farcs5. 
The flux density ratios are plotted as a function of \herschel-band S/N's in the right panel of Figure~\ref{fig:04_flux_comp}.
We find a general consistency between the flux densities of these two works (median flux ratio is $f_\mathrm{R16}/f_\mathrm{This}=0.98\pm0.02$) without any conspicuous systematic offset.

To evaluate the accuracy of SPIRE photometry without PACS information, we also experimented extraction from 250\,\micron\ (instead of 100\,\micron) for the ``deep''-mode clusters. 
The resultant SPIRE flux densities are generally consistent with those extracted using PACS information at the $1\sigma$ confidence level.



\begin{figure}[t]
\centering
\includegraphics[width=\linewidth]{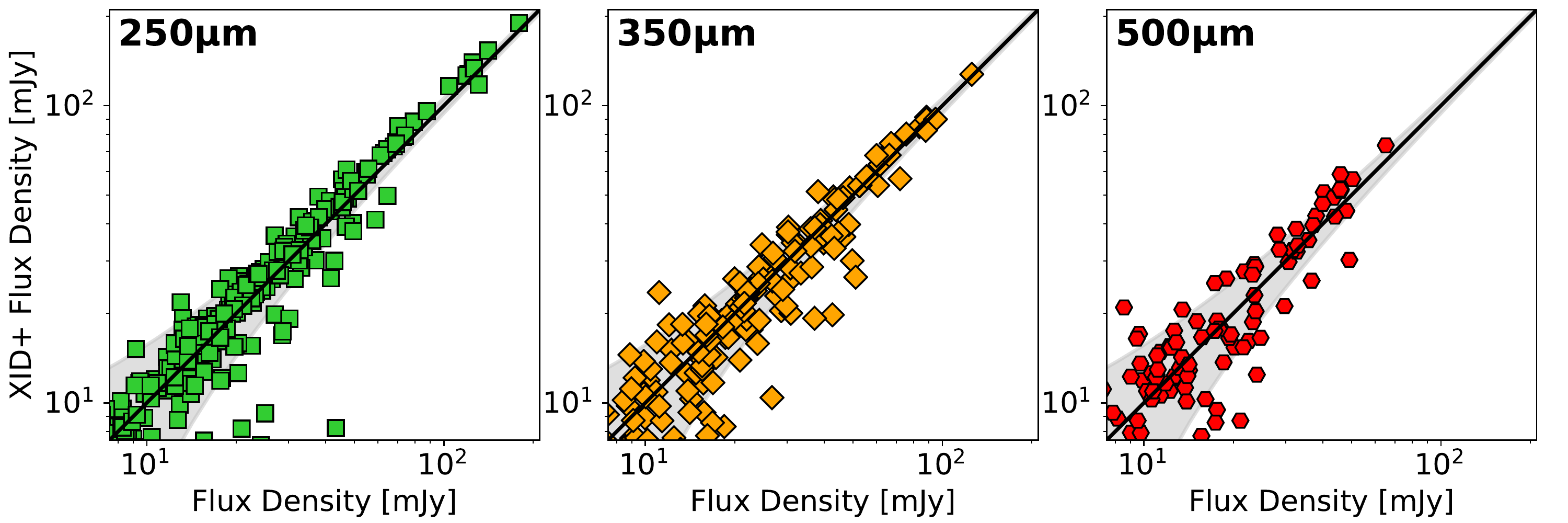}
\caption{Comparisons of \herschel/SPIRE flux densities, including both submillimeter sources detected and undetected by the ALCS, measured with \textsc{xid+} \citep{hurley17} and our iterative extraction procedure (Section~\ref{sec:03_res}). 
Comparison in the SPIRE 250, 350 and 500\,\micron\ band are shown in the left, central and right panel, respectively.
In each panel, the diagonal solid black line indicates identical flux densities measured with the two procedures, and the shaded gray region indicates the expected $1\sigma$ dispersion range of flux densities given the depth of data taken in the ``snapshot'' mode. 
}
\label{fig:xid}
\end{figure}

We also extracted the fluxes of ALCS sources using \textsc{xid+} \citep{hurley17}.
\textsc{xid+} is a prior-based \herschel/SPIRE flux extraction software built upon a probabilistic Bayesian framework.
The positional prior that we used was the same as described in Section~\ref{sec:03_res}, i.e., including both ALCS-detected sources at $\mathrm{S/N}_\mathrm{ALMA} \geq 4$ and ALMA-undetected \herschel\ sources.
Compared with our iterative extraction procedure, \textsc{xid+} does not distinguish the priority of input sources.

Figure~\ref{fig:xid} shows the comparisons of SPIRE flux densities measured by both de-blending routines.
We observe a general consistency between the flux density estimate using both methods.
For ALCS sources in the main sample, the median difference of flux densities estimated with two routines is around 1\,mJy in all the three bands, and the standard deviation of flux density differences are well predicted by the joint uncertainty of both measurements.
We further analyze ALCS sources whose \textsc{xid+}-derived flux densities are different from those extracted iteratively by a factor of $>2$.
The median separation of these sources to their nearest neighbors is found to be 7\arcsec$\pm$6\arcsec, significantly smaller than the typical separation of 14\arcsec$\pm$10\arcsec\ for the full sample.
Therefore, we conclude that the large photometric discrepancy among these sources is mainly caused by strong blending effect. 
With prior ALMA flux density information, iterative extraction procedure has the potential to enhance the photometric accuracy in crowded lensing cluster fields.


\section{Postage Stamp Images and Far-IR SEDs of ALCS Sources}
\label{apd:02_fs}

\renewcommand{\thefigure}{B\arabic{figure}}
\renewcommand{\theHfigure}{B\arabic{figure}}
\setcounter{figure}{0}

In the online journal, we show the figure sets of \hst-\spitzer-ALMA postage stamp images and best-fit far-IR SED plots of all 125 sources in the ALCS-\herschel\ joint sample (Table~\ref{tab:04_sed}). 
These include:

\figsetstart
\figsetnum{B1}
\figsettitle{Postage stamp images and far-IR SEDs of 105 sources in the main ALCS-\herschel\ sample.}
\figsetgrpstart
\figsetgrpnum{B1.1}
\figsetgrptitle{A2163-ID11}
\figsetplot{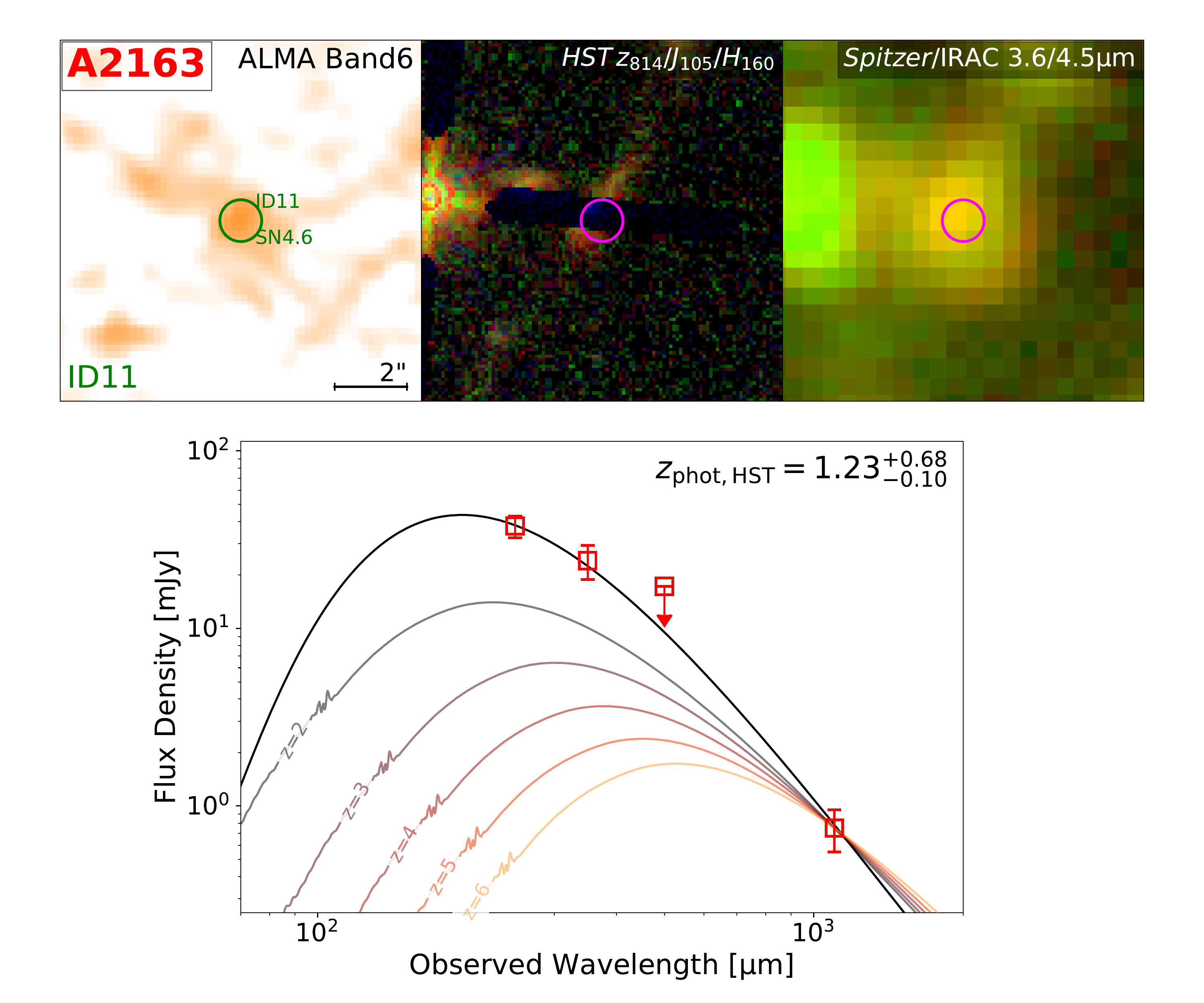}
\figsetgrpnote{Postage stamp images (top) and far-IR SED (bottom) of A2163-ID11.}
\figsetgrpend

\figsetgrpstart
\figsetgrpnum{B1.2}
\figsetgrptitle{A2537-ID42}
\figsetplot{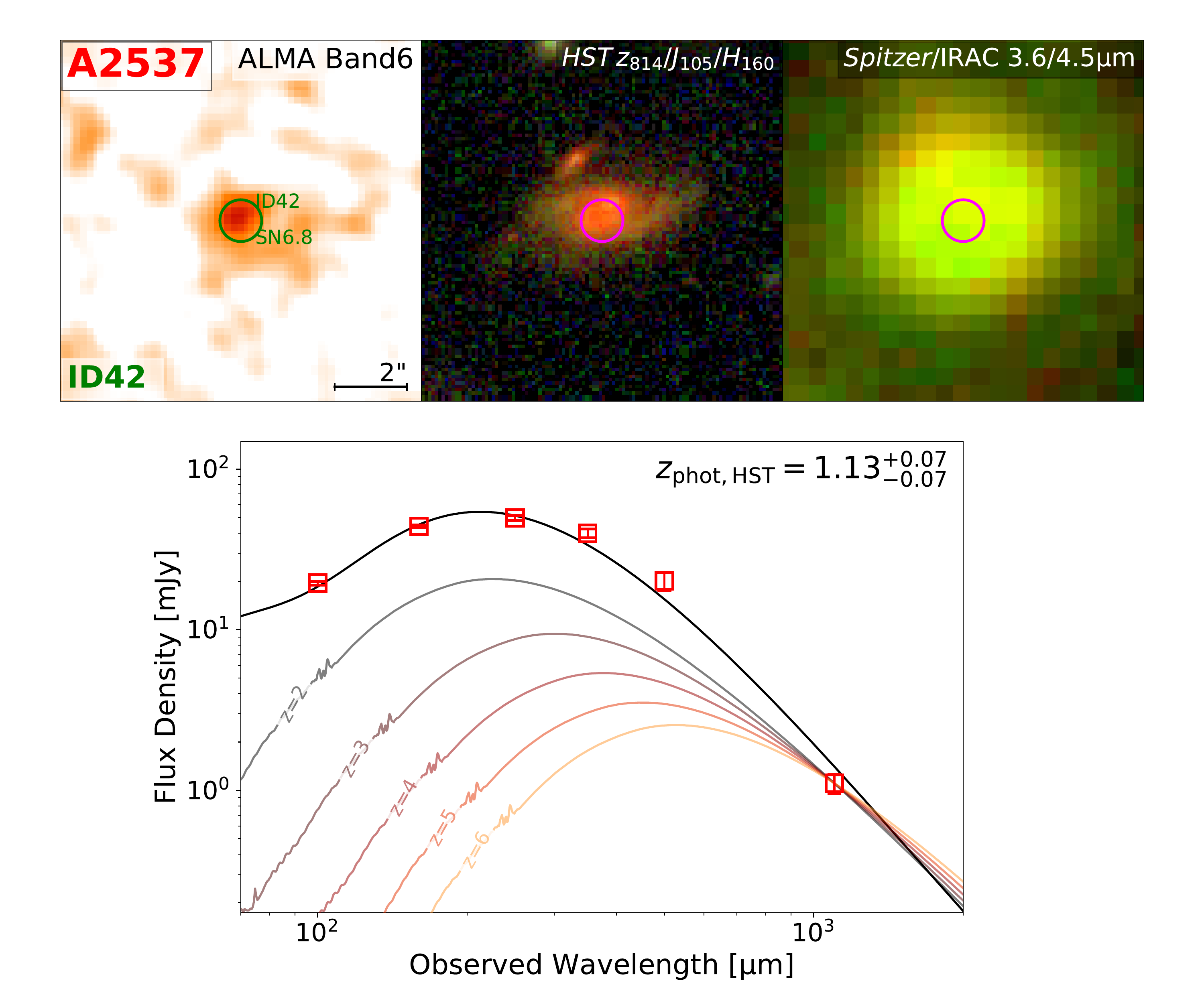}
\figsetgrpnote{Postage stamp images (top) and far-IR SED (bottom) of A2537-ID42.}
\figsetgrpend

\figsetgrpstart
\figsetgrpnum{B1.3}
\figsetgrptitle{A2537-ID49}
\figsetplot{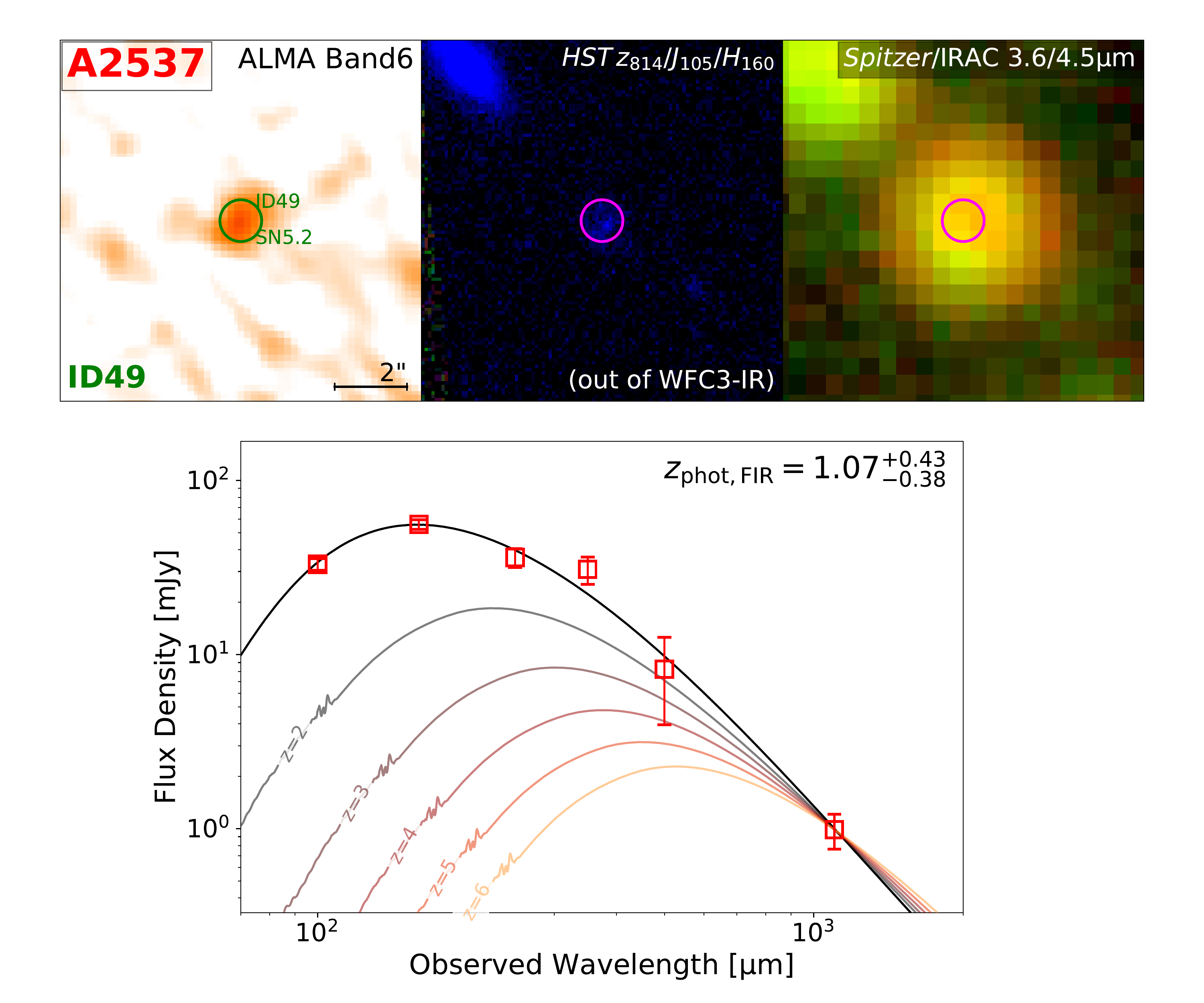}
\figsetgrpnote{Postage stamp images (top) and far-IR SED (bottom) of A2537-ID49.}
\figsetgrpend

\figsetgrpstart
\figsetgrpnum{B1.4}
\figsetgrptitle{A2537-ID66}
\figsetplot{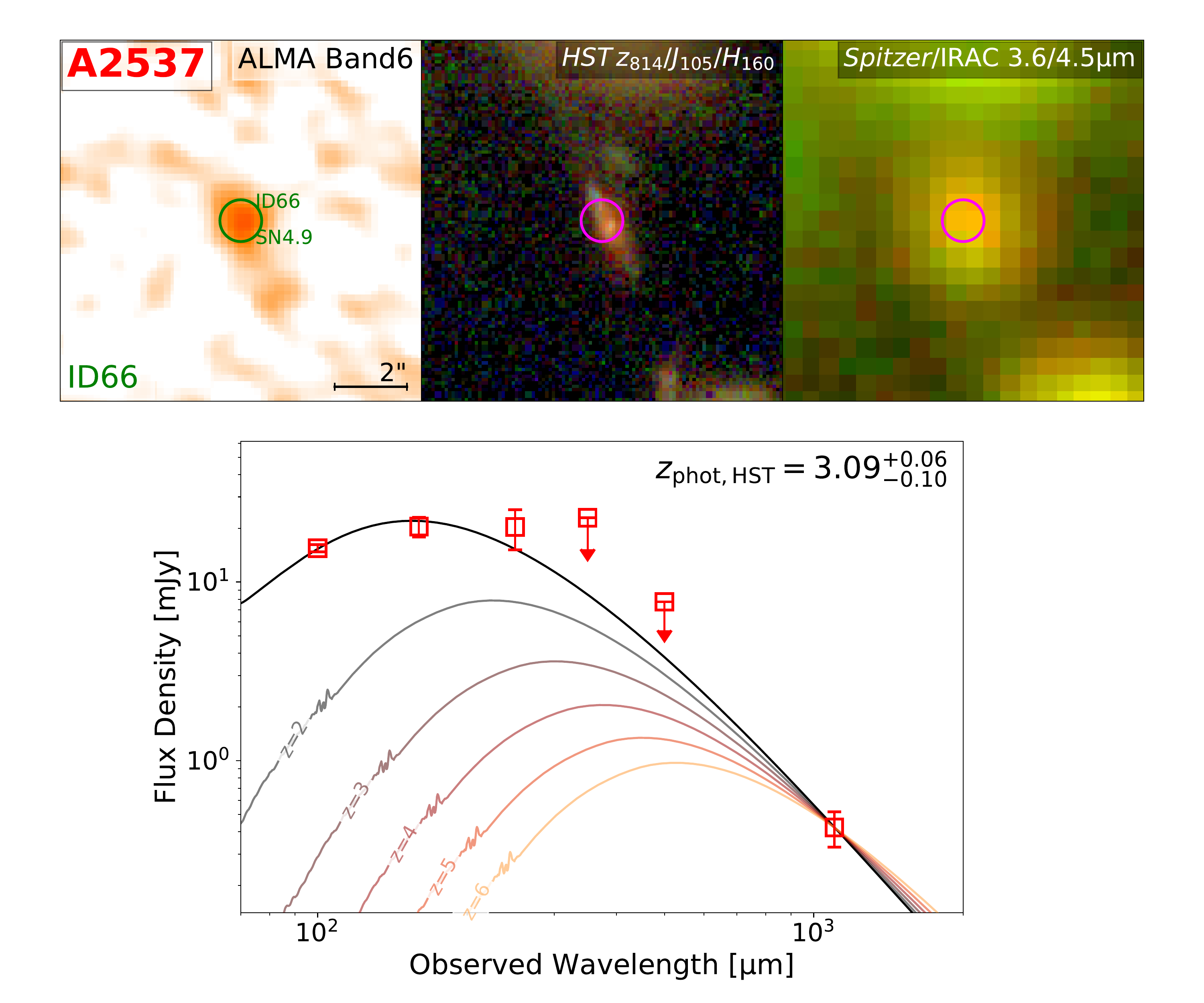}
\figsetgrpnote{Postage stamp images (top) and far-IR SED (bottom) of A2537-ID66.}
\figsetgrpend

\figsetgrpstart
\figsetgrpnum{B1.5}
\figsetgrptitle{A2744-ID07}
\figsetplot{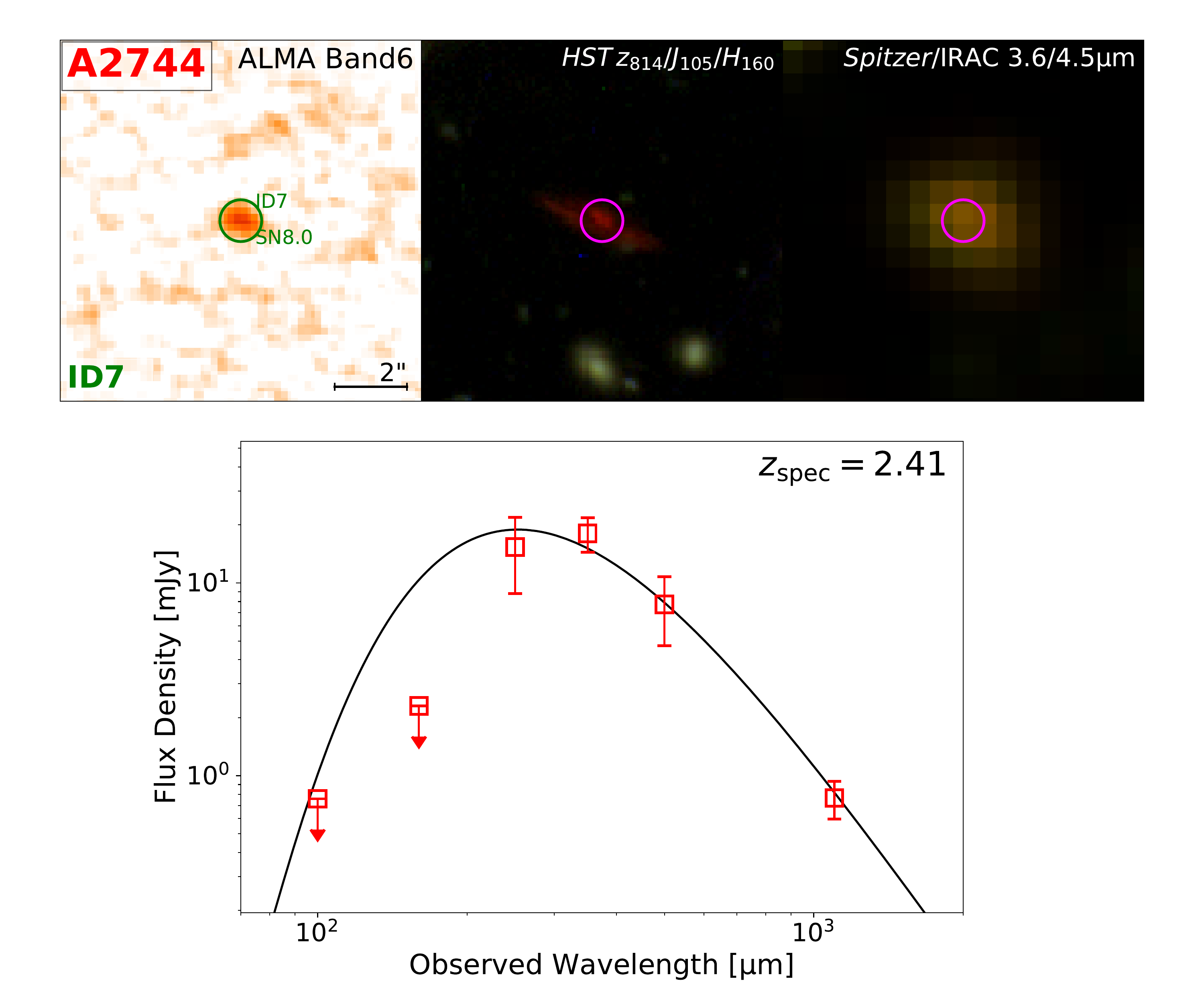}
\figsetgrpnote{Postage stamp images (top) and far-IR SED (bottom) of A2744-ID07.}
\figsetgrpend

\figsetgrpstart
\figsetgrpnum{B1.6}
\figsetgrptitle{A2744-ID21}
\figsetplot{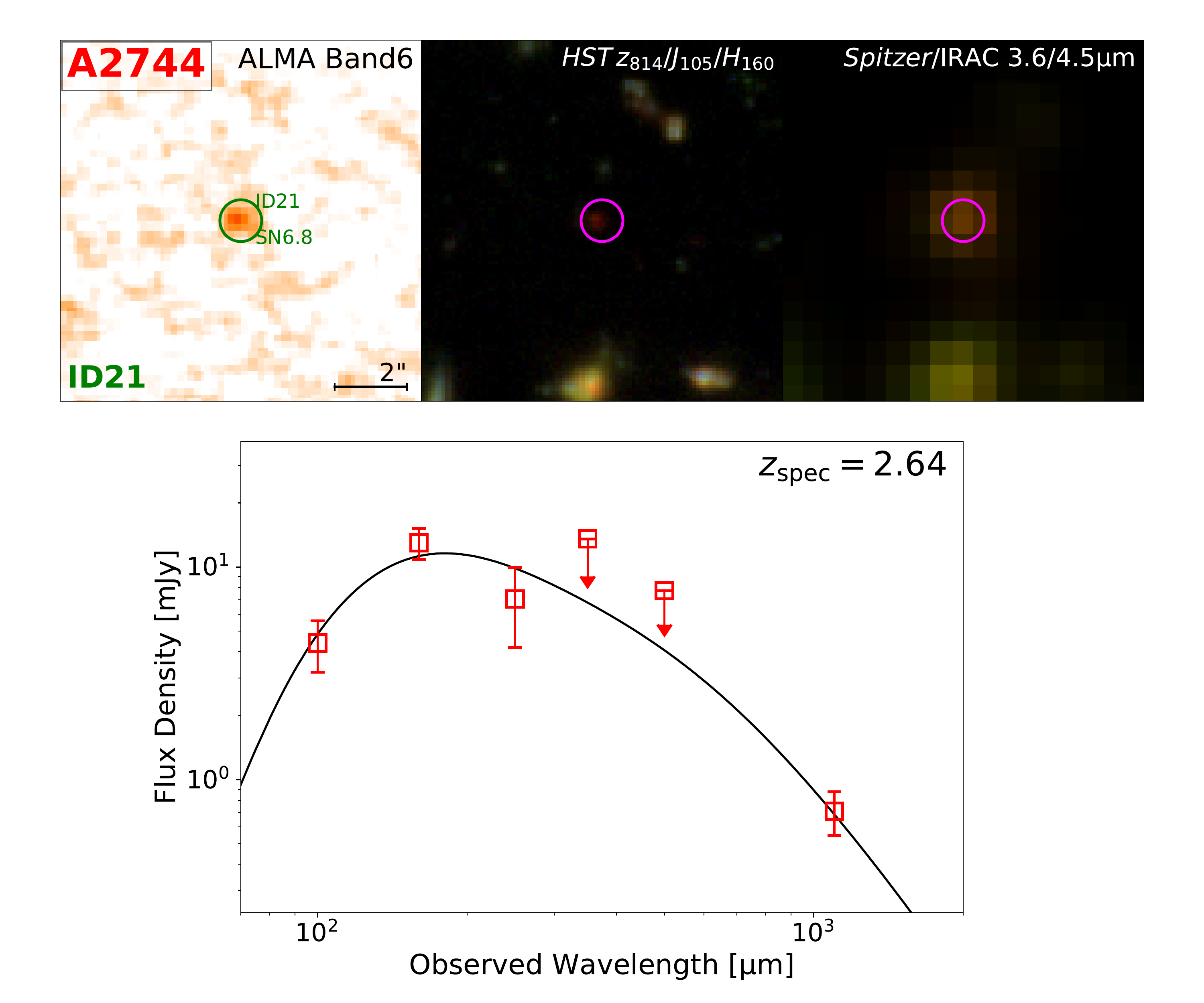}
\figsetgrpnote{Postage stamp images (top) and far-IR SED (bottom) of A2744-ID21.}
\figsetgrpend

\figsetgrpstart
\figsetgrpnum{B1.7}
\figsetgrptitle{A2744-ID33}
\figsetplot{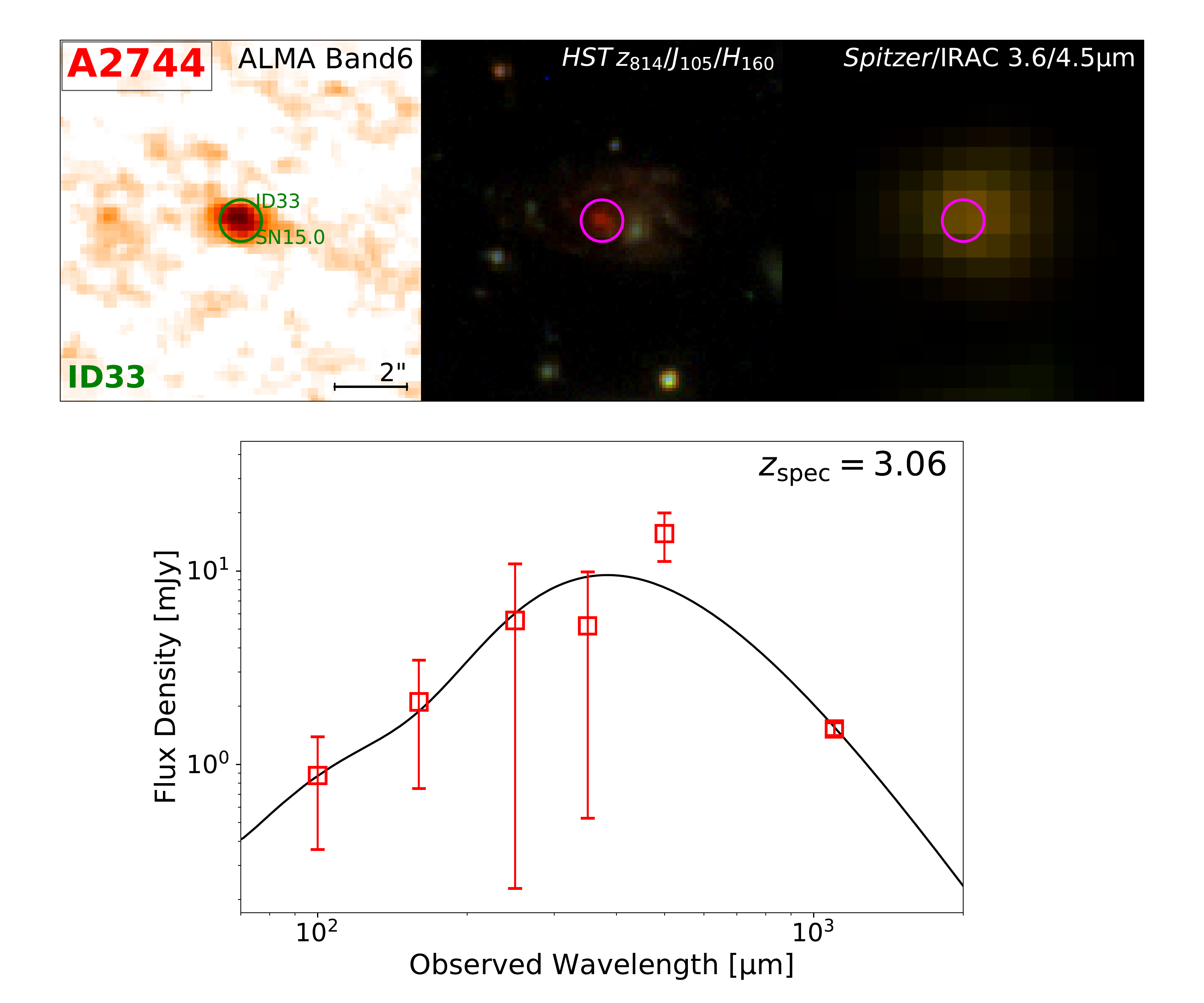}
\figsetgrpnote{Postage stamp images (top) and far-IR SED (bottom) of A2744-ID33.}
\figsetgrpend

\figsetgrpstart
\figsetgrpnum{B1.8}
\figsetgrptitle{A2744-ID56}
\figsetplot{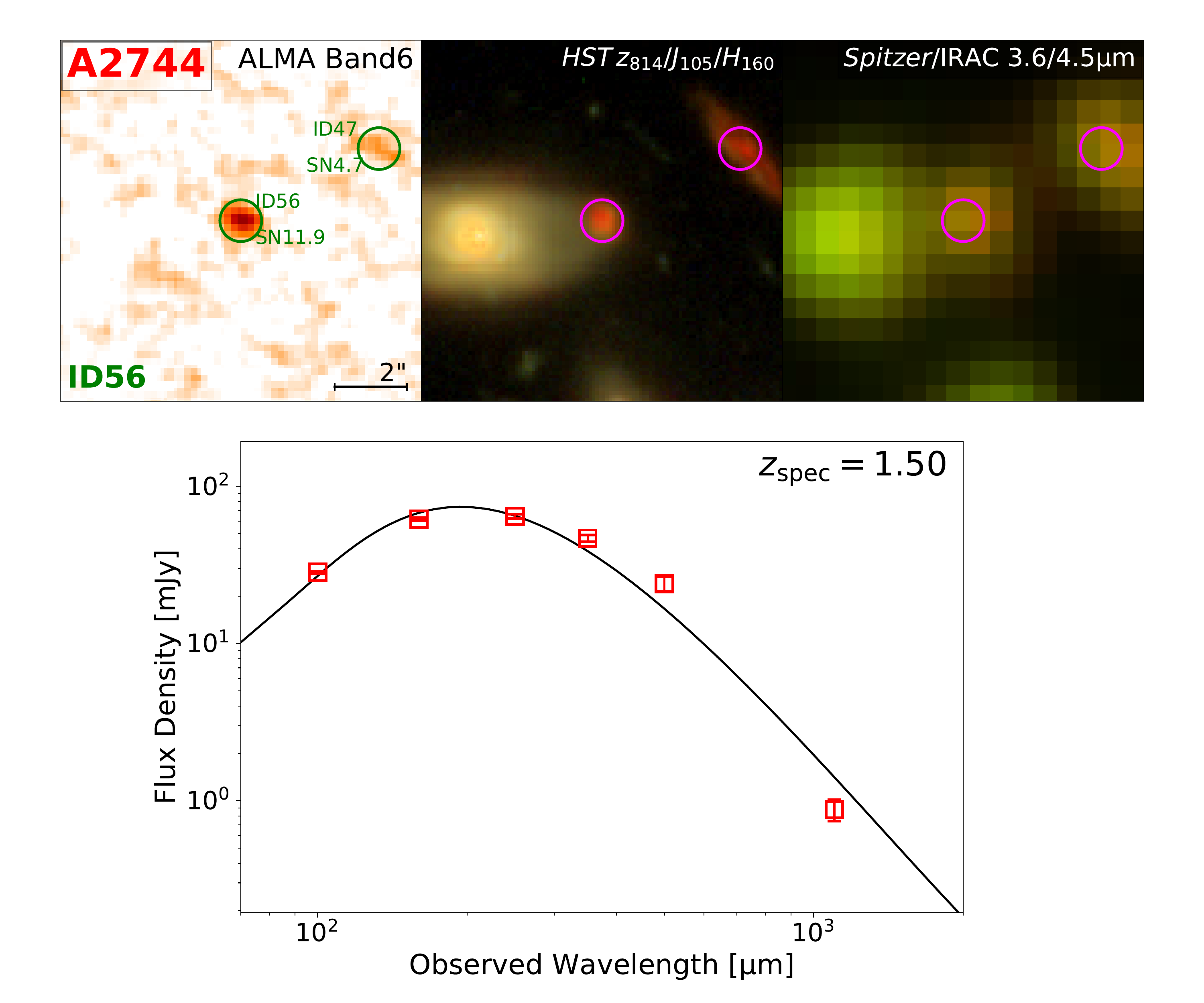}
\figsetgrpnote{Postage stamp images (top) and far-IR SED (bottom) of A2744-ID56.}
\figsetgrpend

\figsetgrpstart
\figsetgrpnum{B1.9}
\figsetgrptitle{A2744-ID81}
\figsetplot{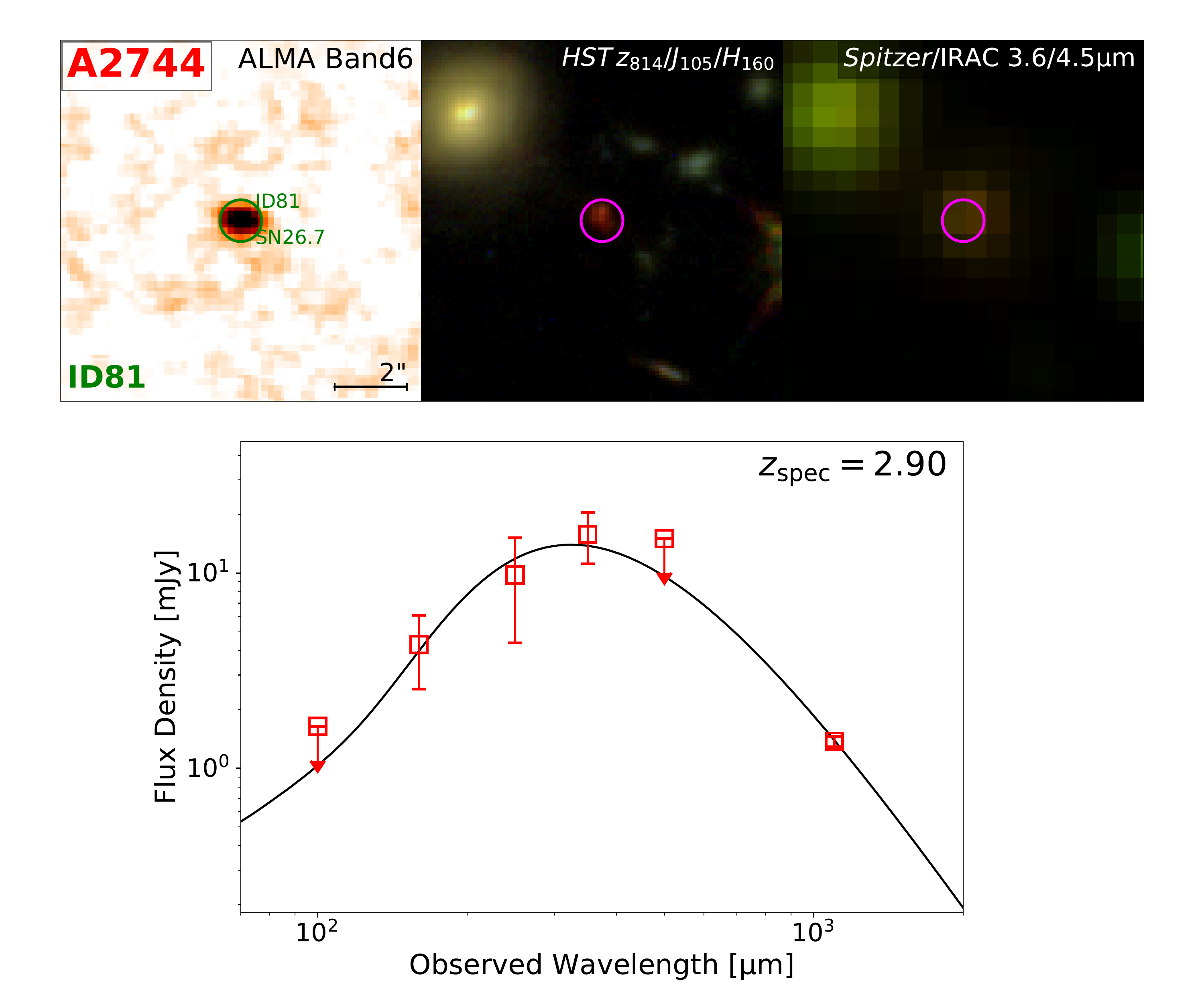}
\figsetgrpnote{Postage stamp images (top) and far-IR SED (bottom) of A2744-ID81.}
\figsetgrpend

\figsetgrpstart
\figsetgrpnum{B1.10}
\figsetgrptitle{A2744-ID319}
\figsetplot{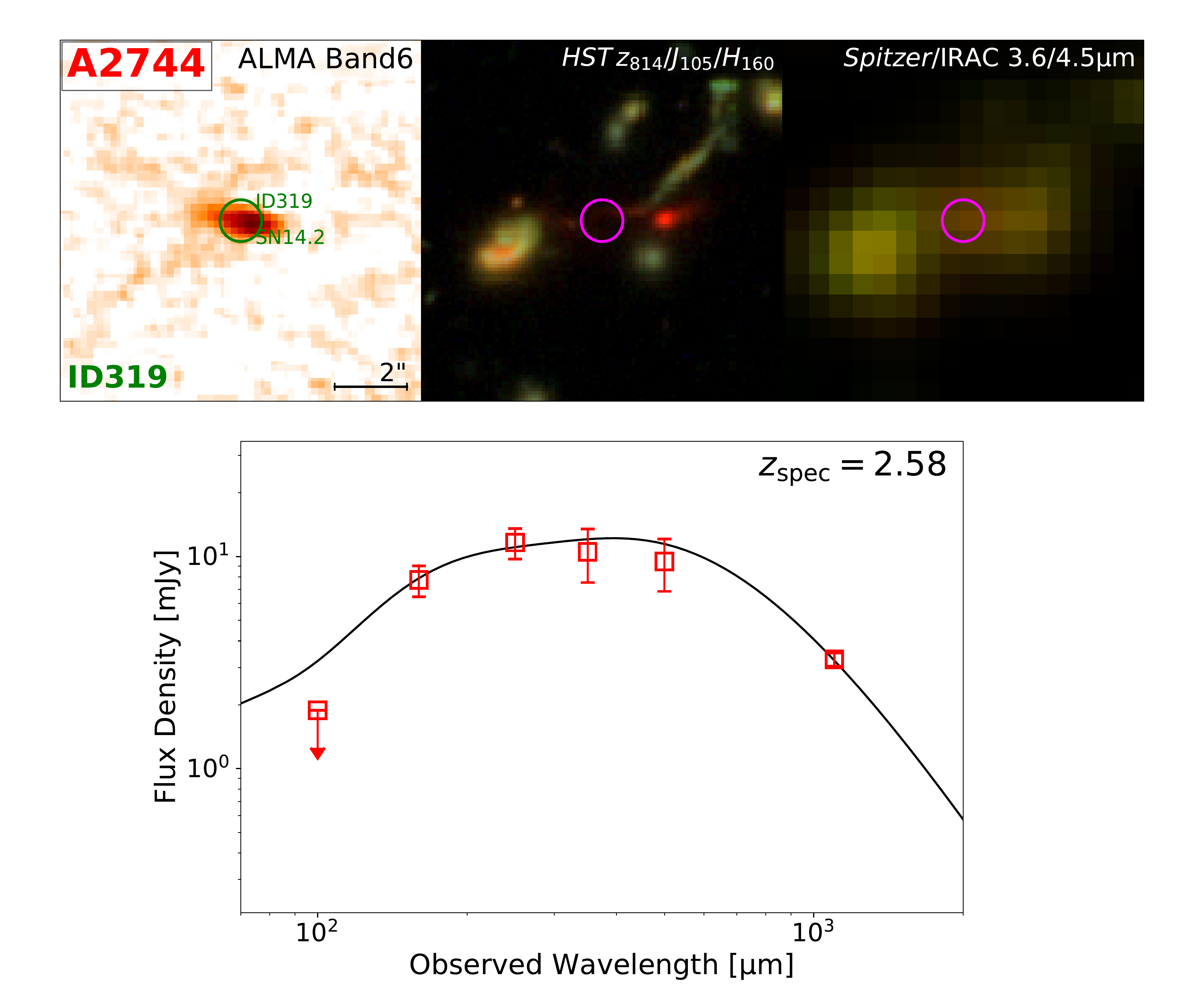}
\figsetgrpnote{Postage stamp images (top) and far-IR SED (bottom) of A2744-ID319.}
\figsetgrpend

\figsetgrpstart
\figsetgrpnum{B1.11}
\figsetgrptitle{A3192-ID31}
\figsetplot{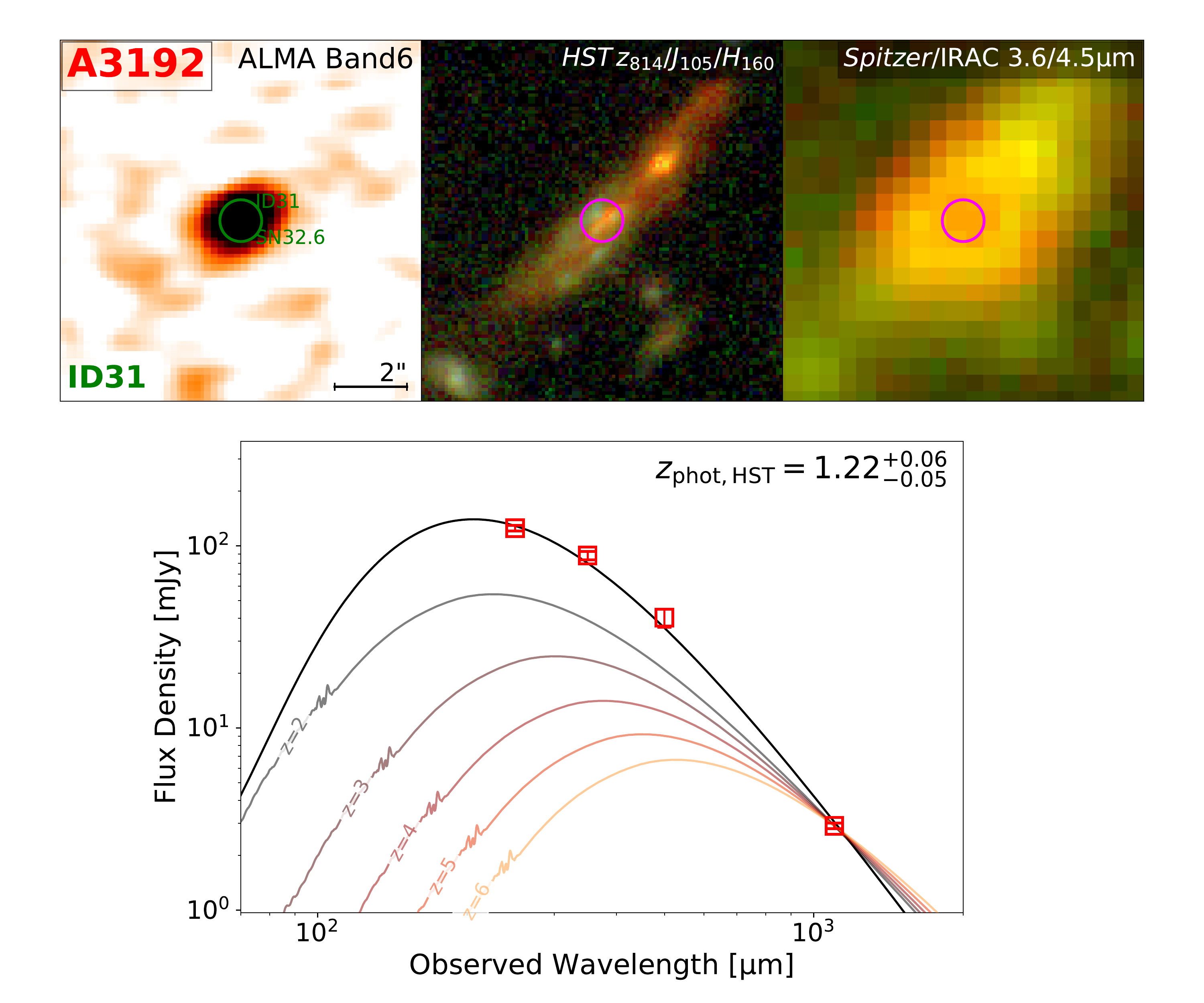}
\figsetgrpnote{Postage stamp images (top) and far-IR SED (bottom) of A3192-ID31.}
\figsetgrpend

\figsetgrpstart
\figsetgrpnum{B1.12}
\figsetgrptitle{A3192-ID40}
\figsetplot{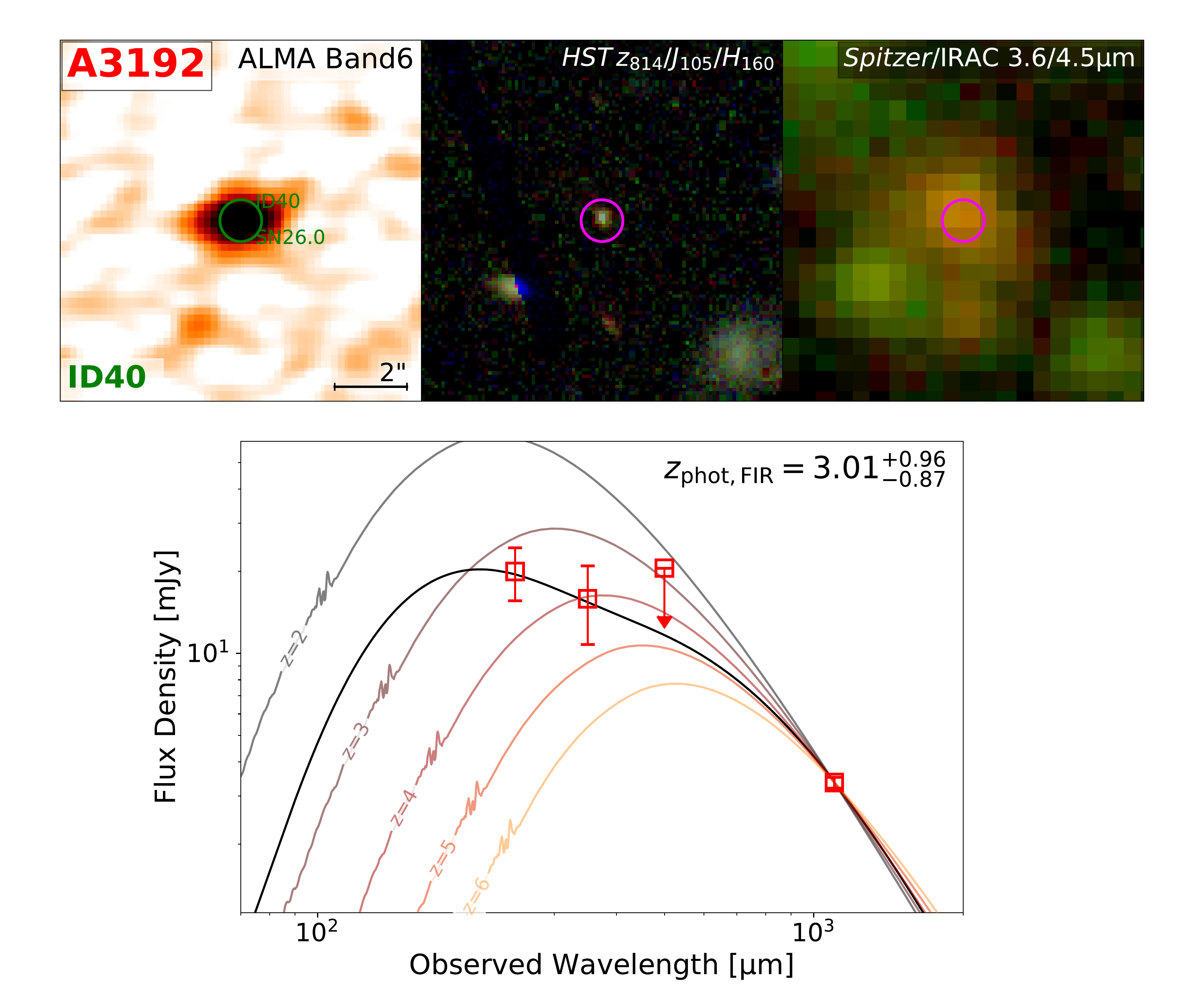}
\figsetgrpnote{Postage stamp images (top) and far-IR SED (bottom) of A3192-ID40.}
\figsetgrpend

\figsetgrpstart
\figsetgrpnum{B1.13}
\figsetgrptitle{A3192-ID131}
\figsetplot{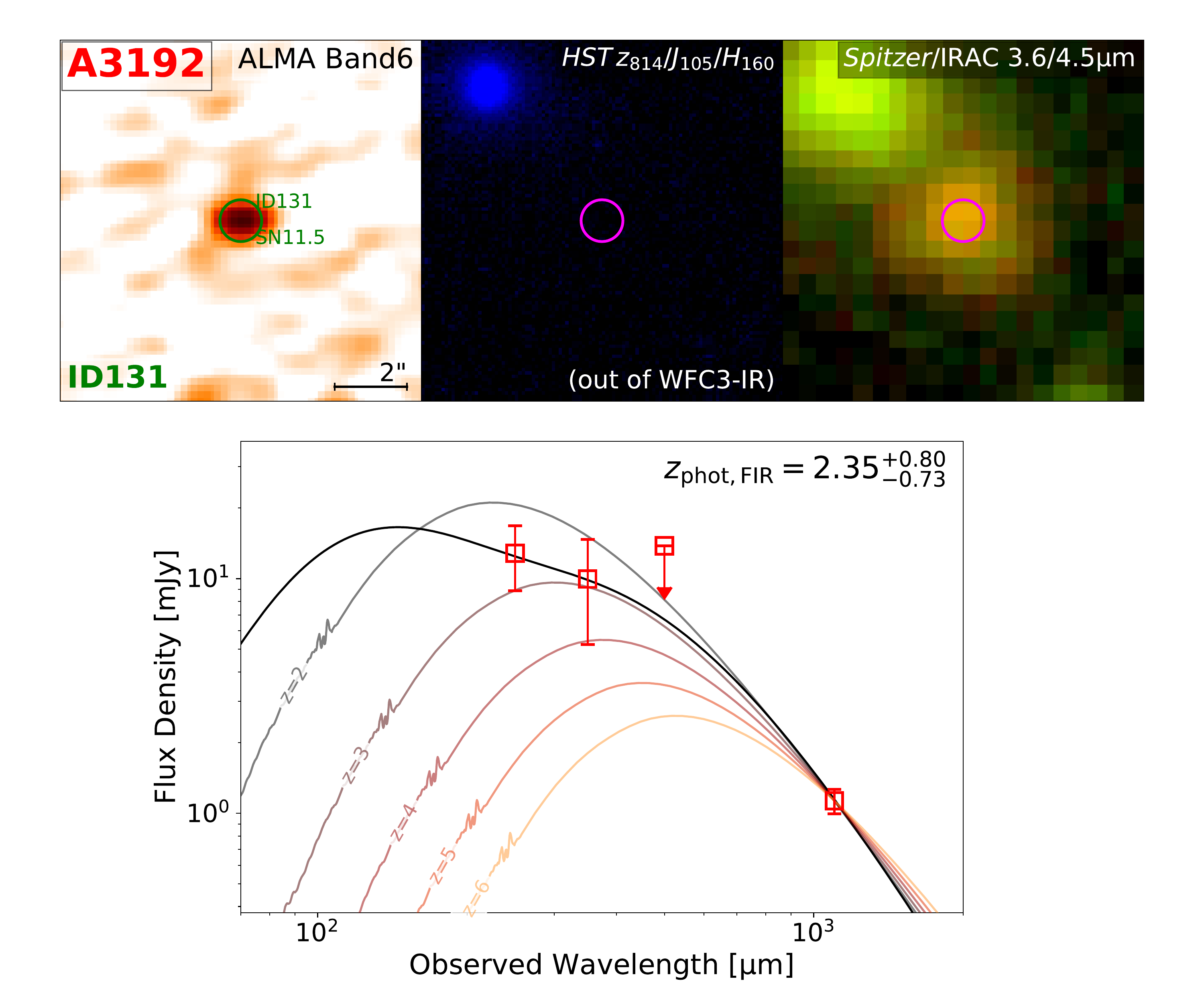}
\figsetgrpnote{Postage stamp images (top) and far-IR SED (bottom) of A3192-ID131.}
\figsetgrpend

\figsetgrpstart
\figsetgrpnum{B1.14}
\figsetgrptitle{A3192-ID138}
\figsetplot{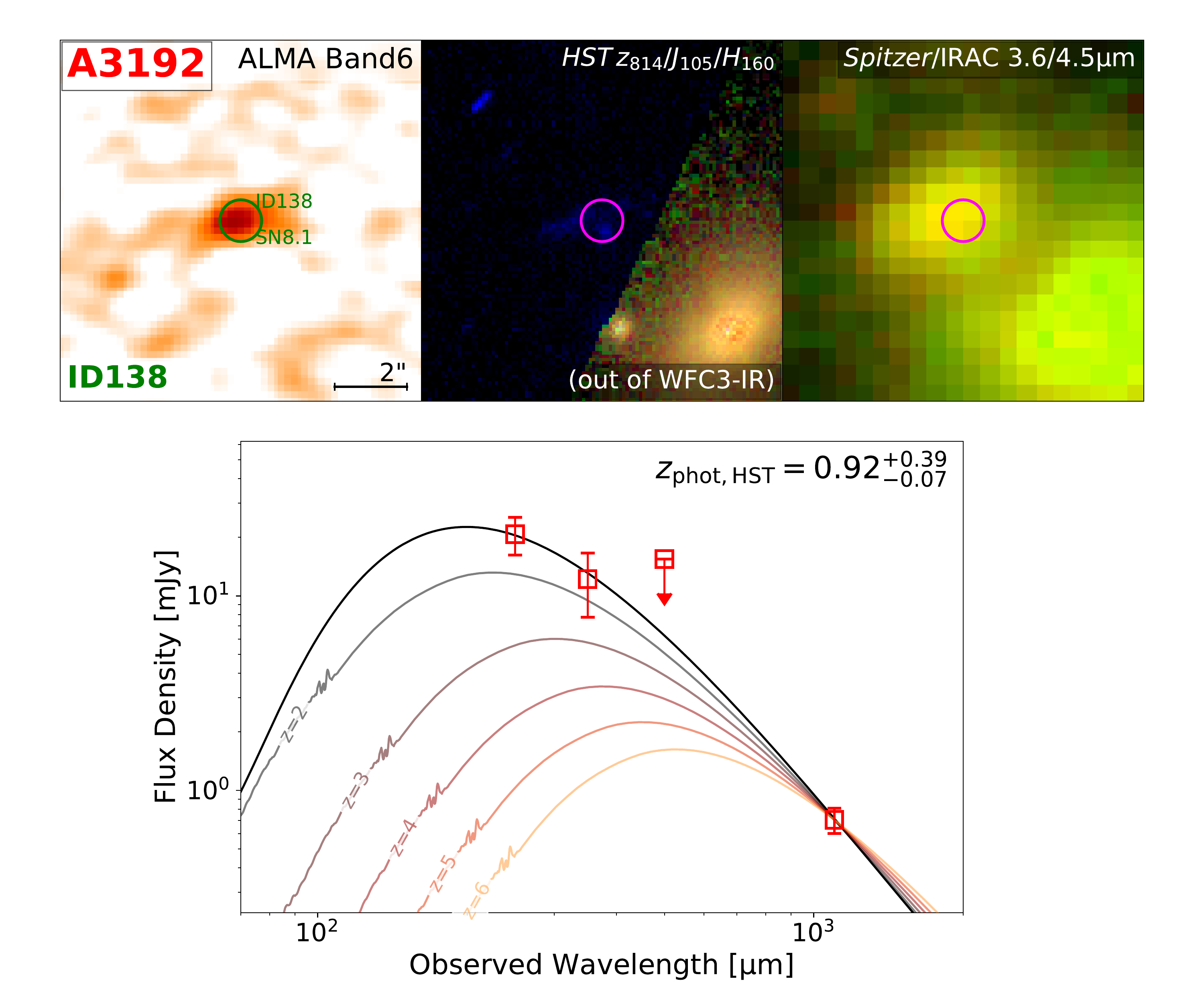}
\figsetgrpnote{Postage stamp images (top) and far-IR SED (bottom) of A3192-ID138.}
\figsetgrpend

\figsetgrpstart
\figsetgrpnum{B1.15}
\figsetgrptitle{A370-ID18}
\figsetplot{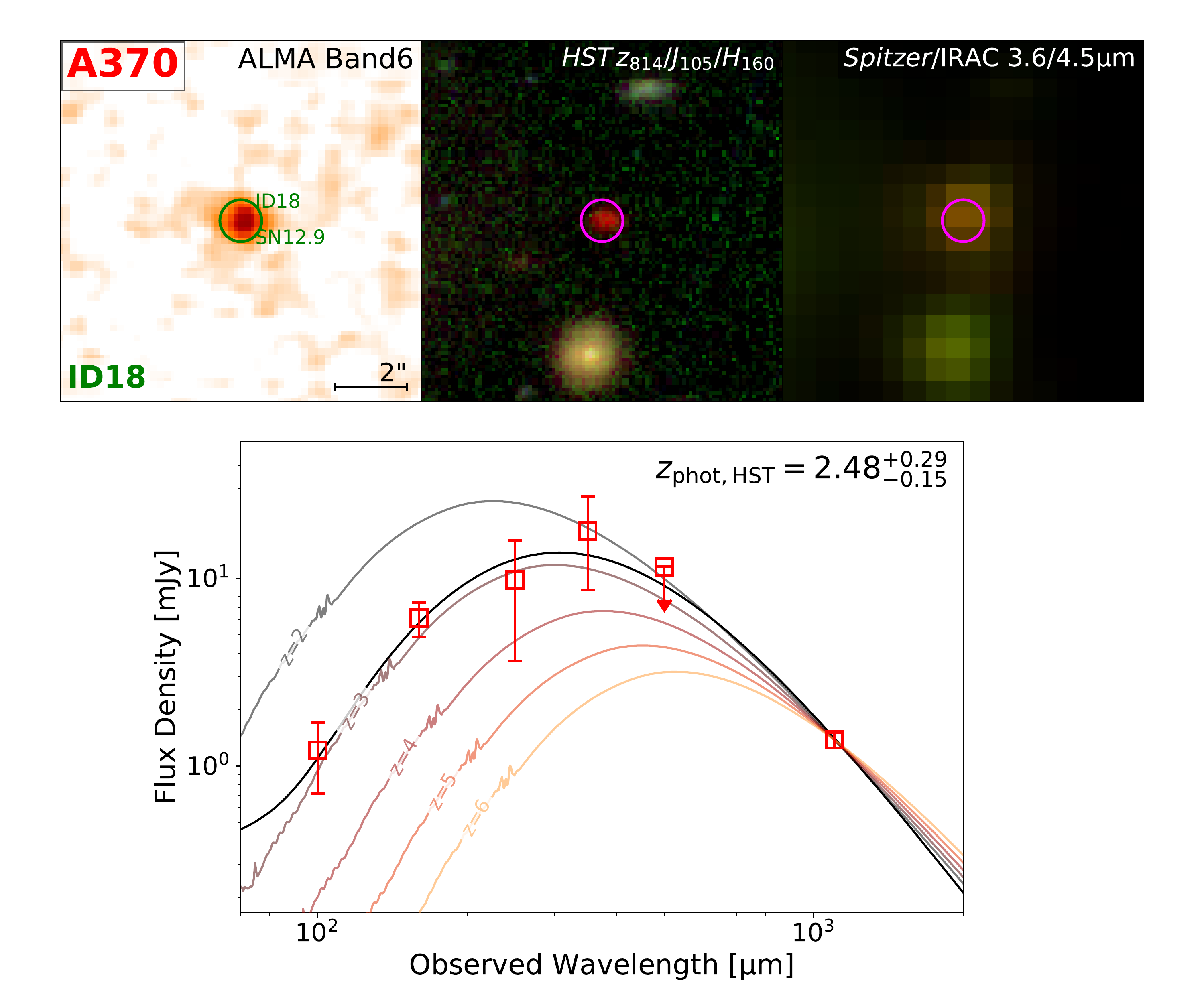}
\figsetgrpnote{Postage stamp images (top) and far-IR SED (bottom) of A370-ID18.}
\figsetgrpend

\figsetgrpstart
\figsetgrpnum{B1.16}
\figsetgrptitle{A370-ID31}
\figsetplot{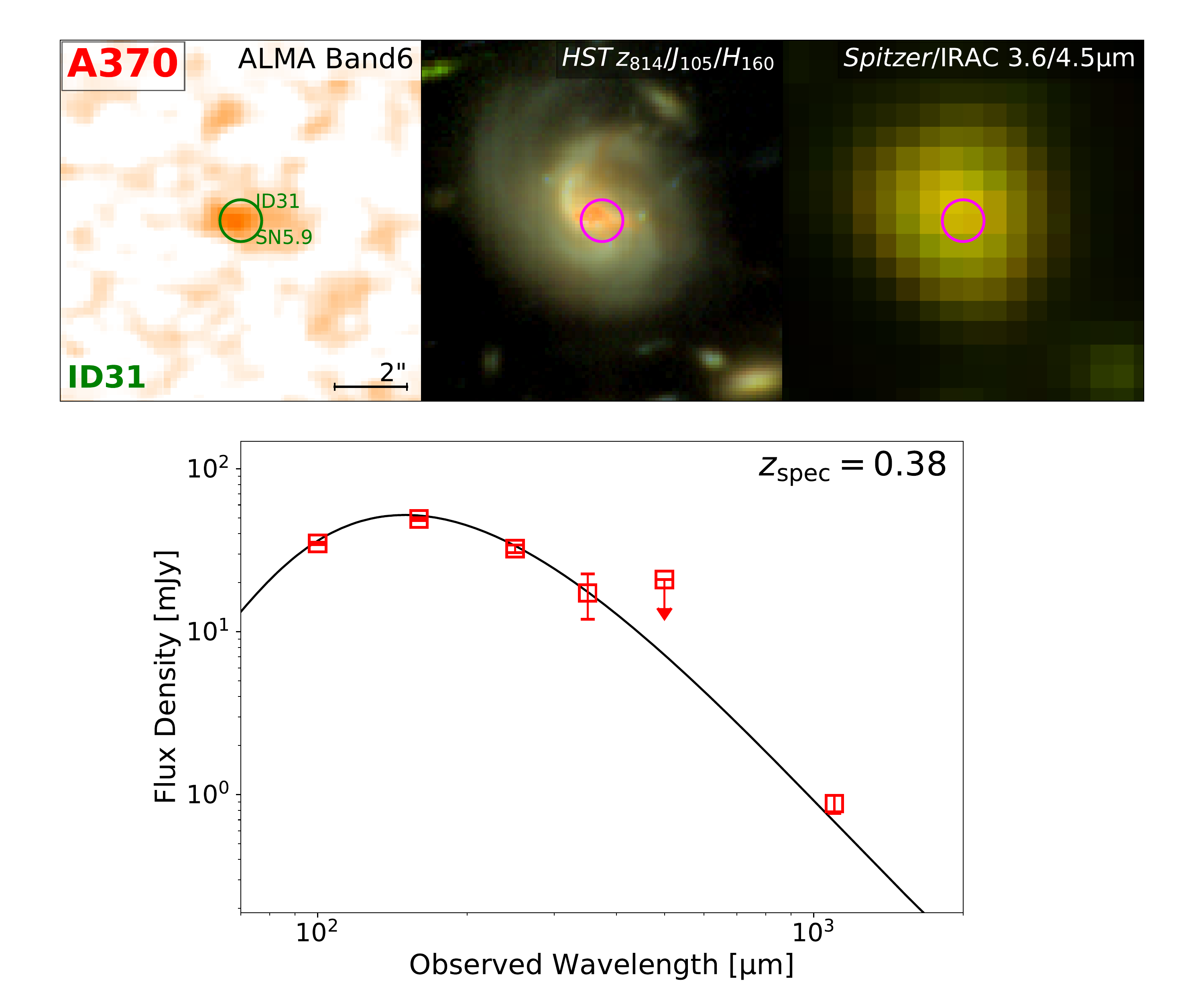}
\figsetgrpnote{Postage stamp images (top) and far-IR SED (bottom) of A370-ID31.}
\figsetgrpend

\figsetgrpstart
\figsetgrpnum{B1.17}
\figsetgrptitle{A370-ID103}
\figsetplot{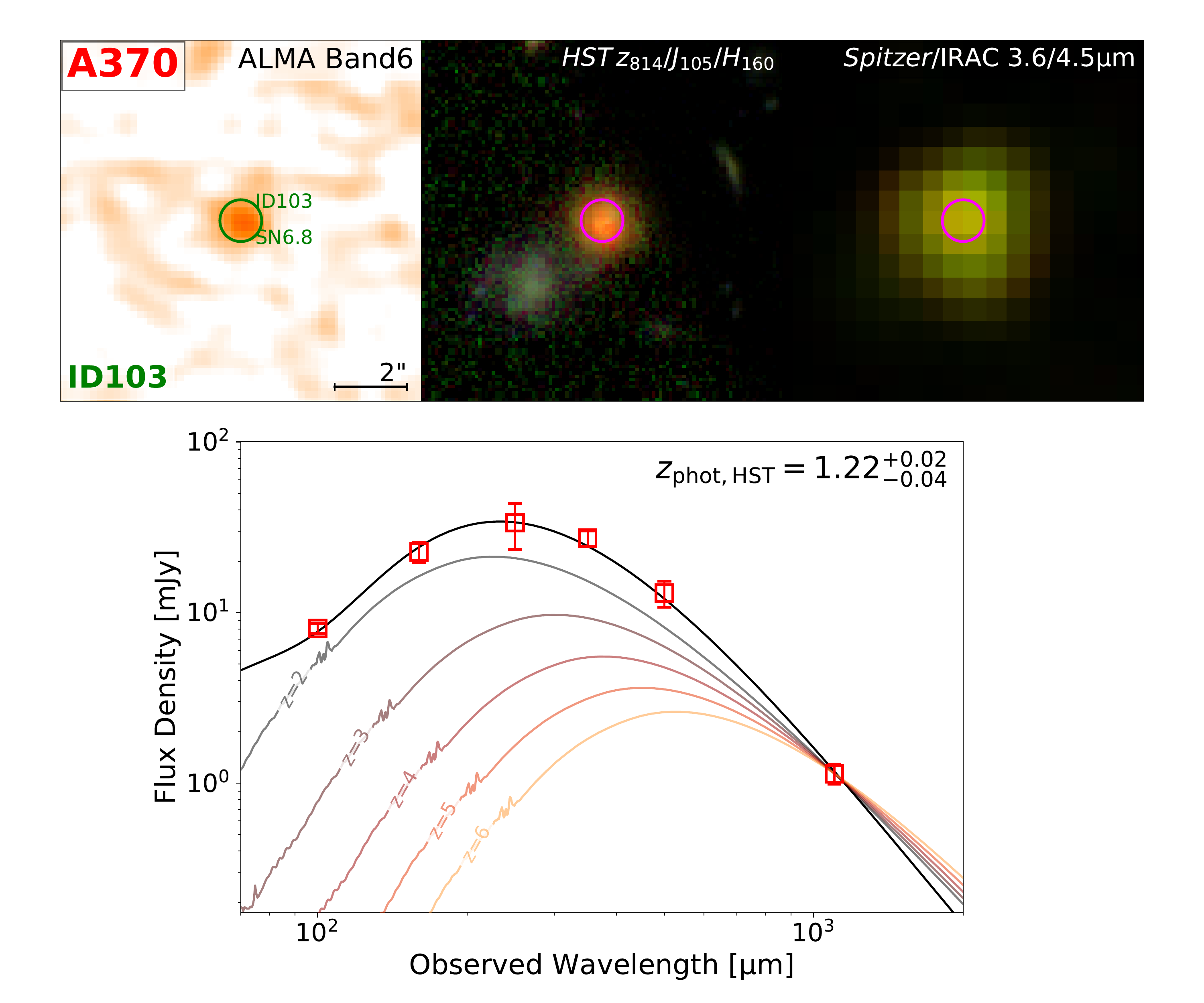}
\figsetgrpnote{Postage stamp images (top) and far-IR SED (bottom) of A370-ID103.}
\figsetgrpend

\figsetgrpstart
\figsetgrpnum{B1.18}
\figsetgrptitle{A370-ID110}
\figsetplot{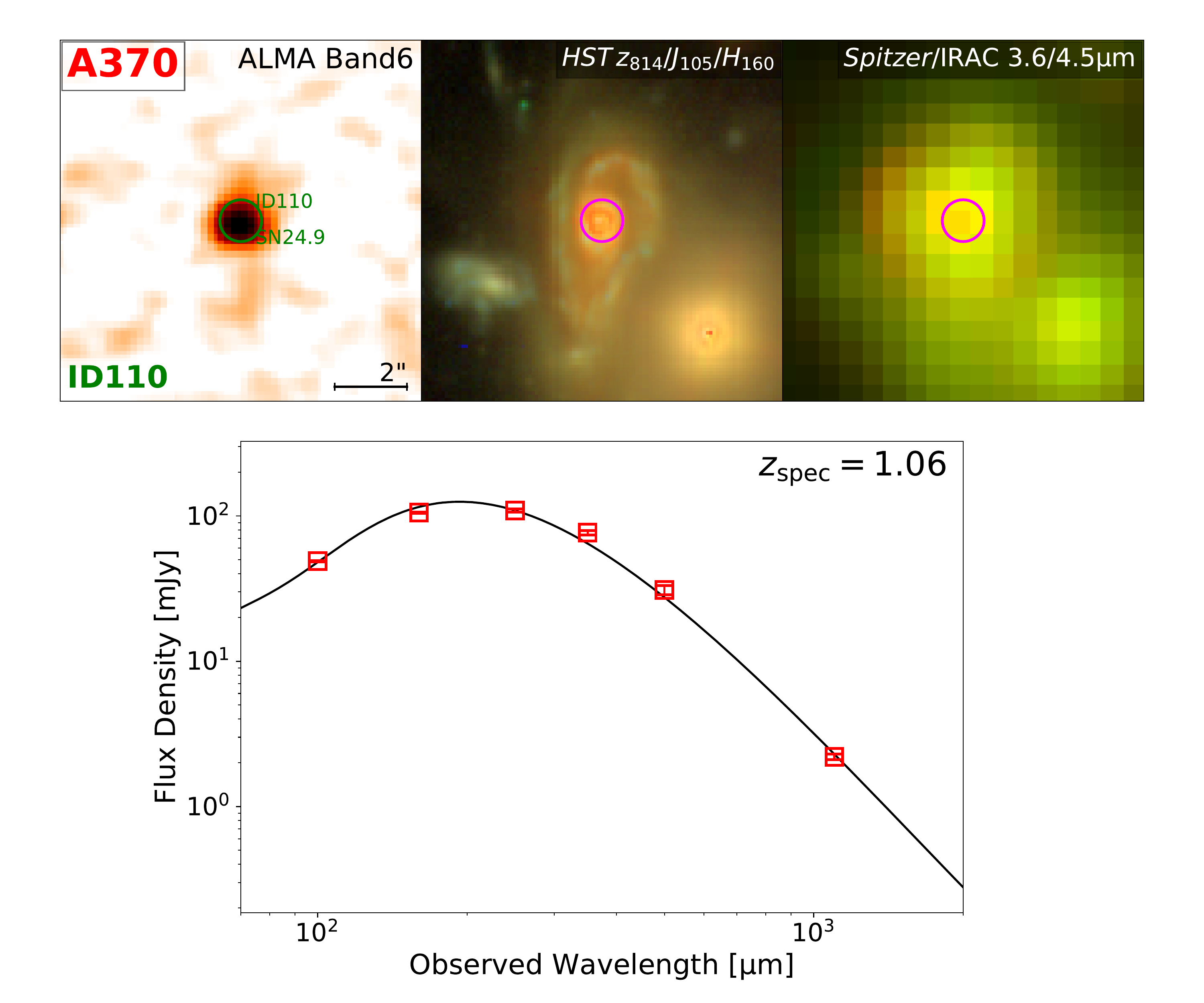}
\figsetgrpnote{Postage stamp images (top) and far-IR SED (bottom) of A370-ID110.}
\figsetgrpend

\figsetgrpstart
\figsetgrpnum{B1.19}
\figsetgrptitle{A370-ID146}
\figsetplot{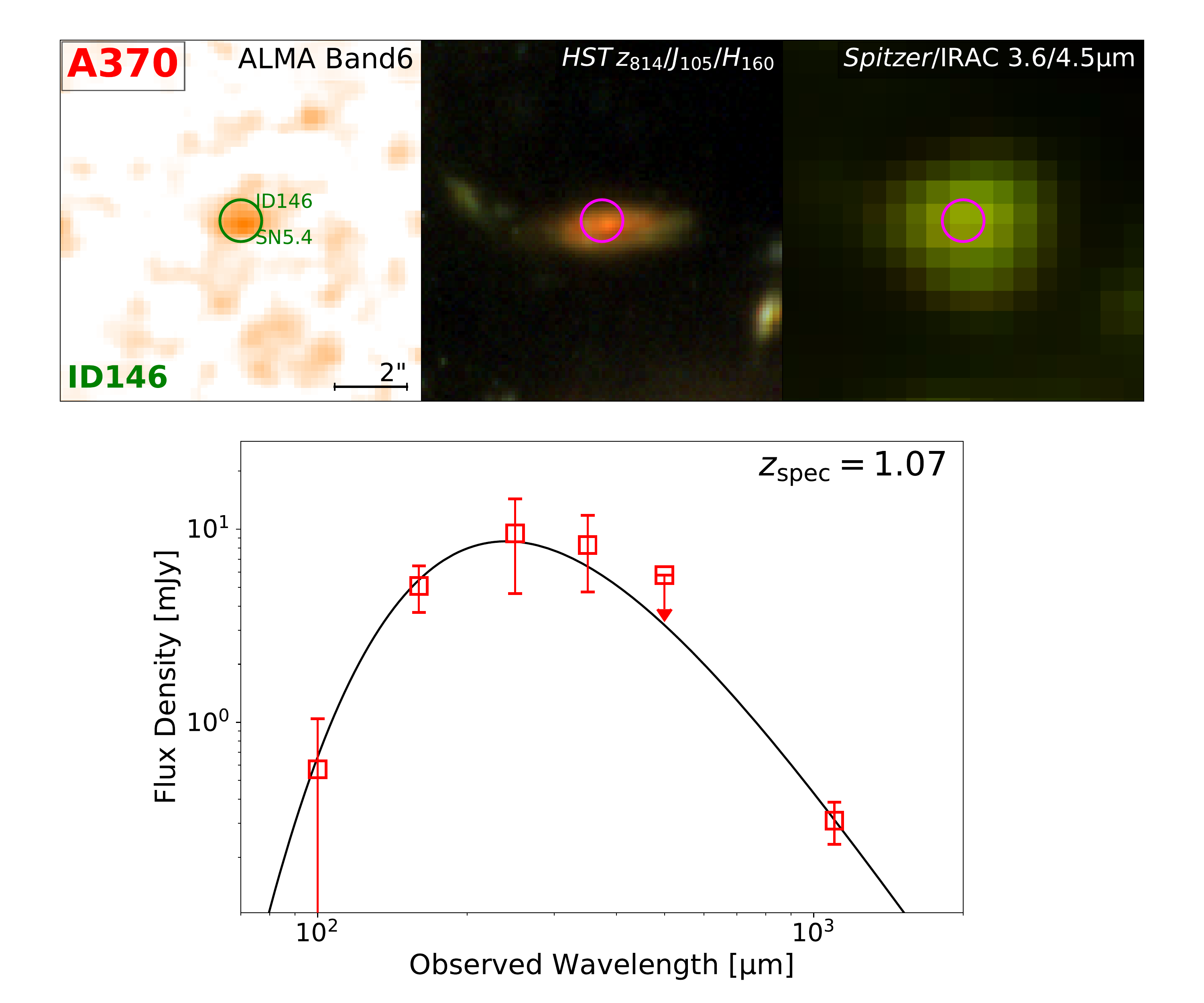}
\figsetgrpnote{Postage stamp images (top) and far-IR SED (bottom) of A370-ID146.}
\figsetgrpend

\figsetgrpstart
\figsetgrpnum{B1.20}
\figsetgrptitle{ACT0102-ID22}
\figsetplot{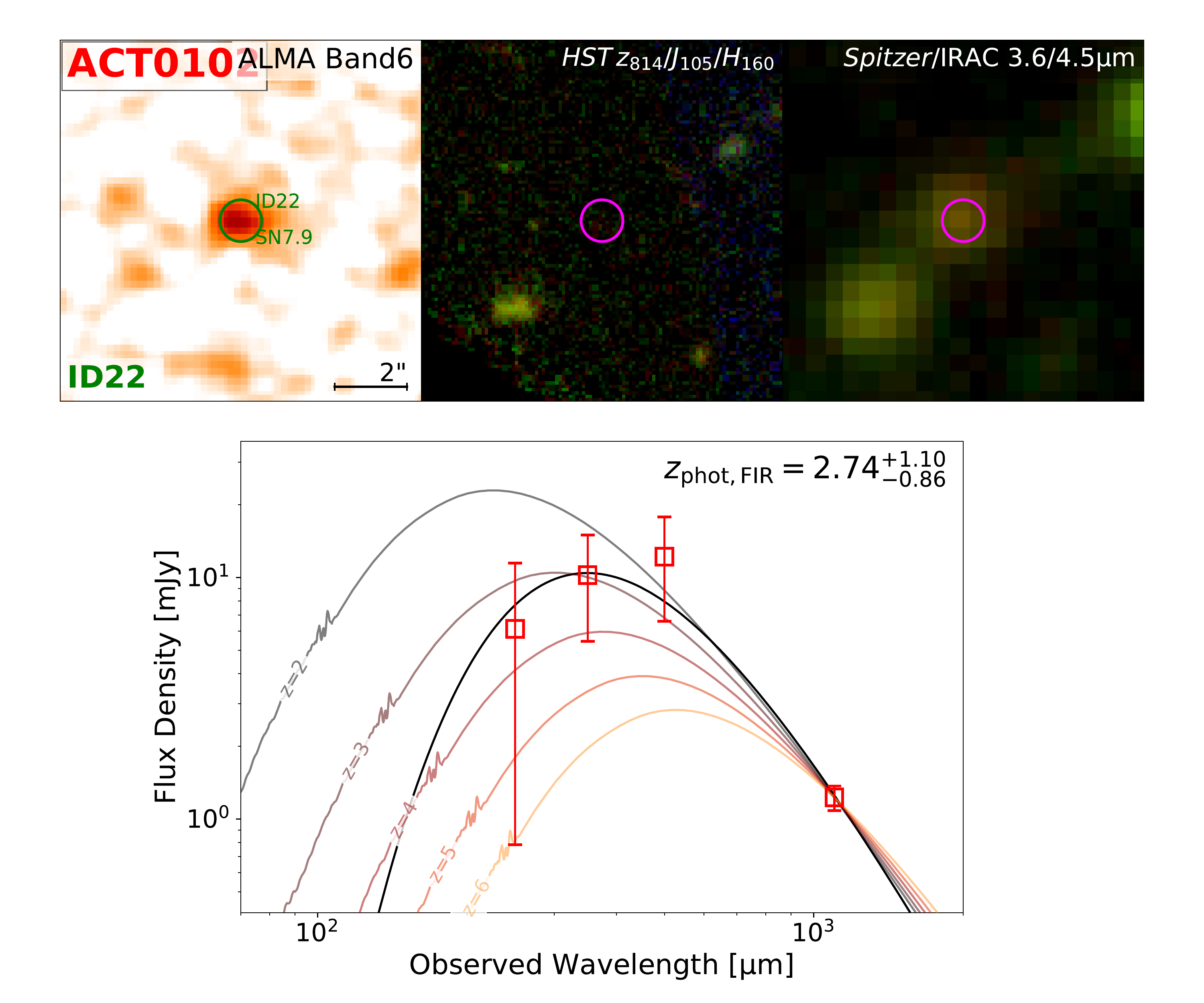}
\figsetgrpnote{Postage stamp images (top) and far-IR SED (bottom) of ACT0102-ID22.}
\figsetgrpend

\figsetgrpstart
\figsetgrpnum{B1.21}
\figsetgrptitle{ACT0102-ID50}
\figsetplot{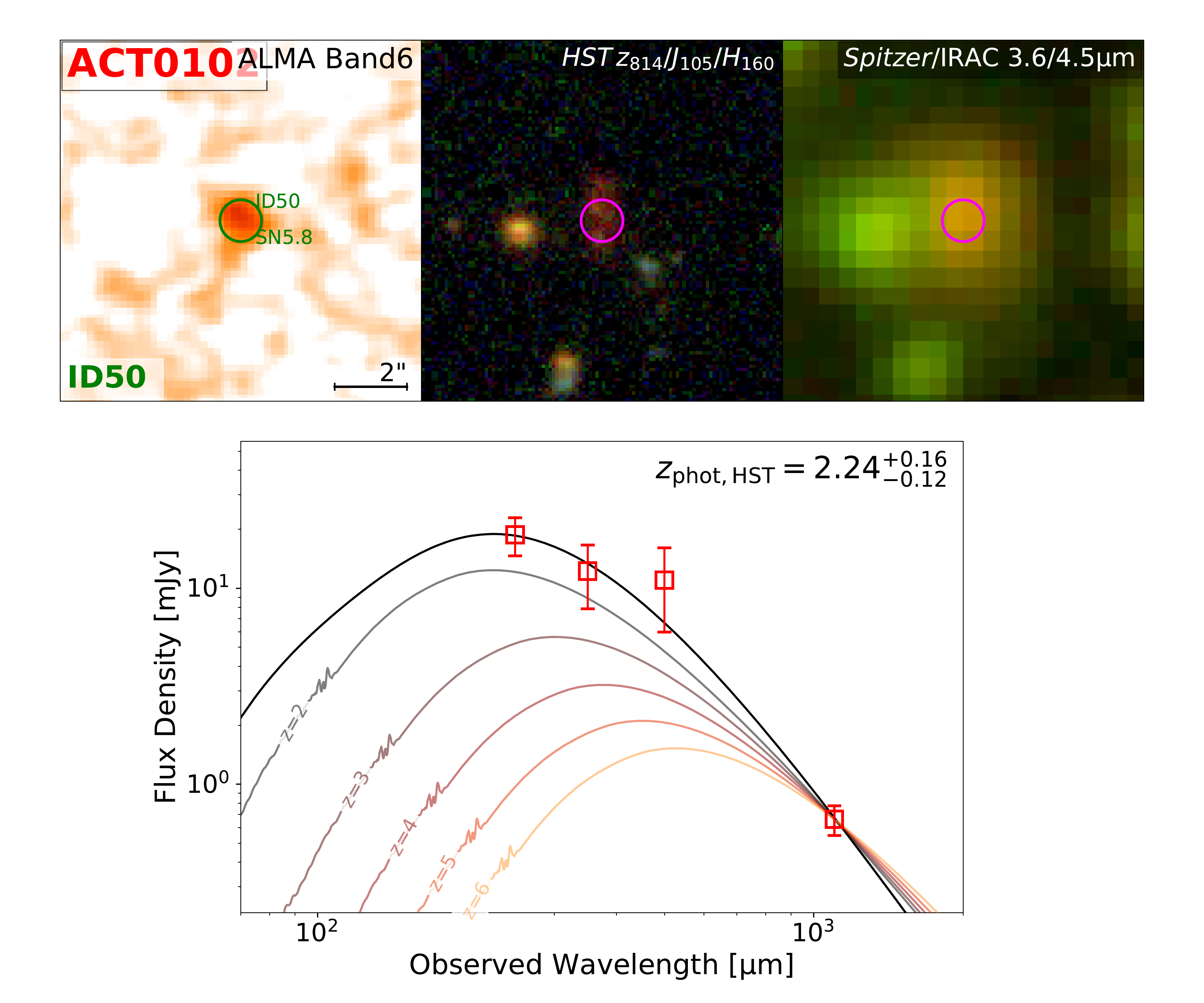}
\figsetgrpnote{Postage stamp images (top) and far-IR SED (bottom) of ACT0102-ID50.}
\figsetgrpend

\figsetgrpstart
\figsetgrpnum{B1.22}
\figsetgrptitle{ACT0102-ID118}
\figsetplot{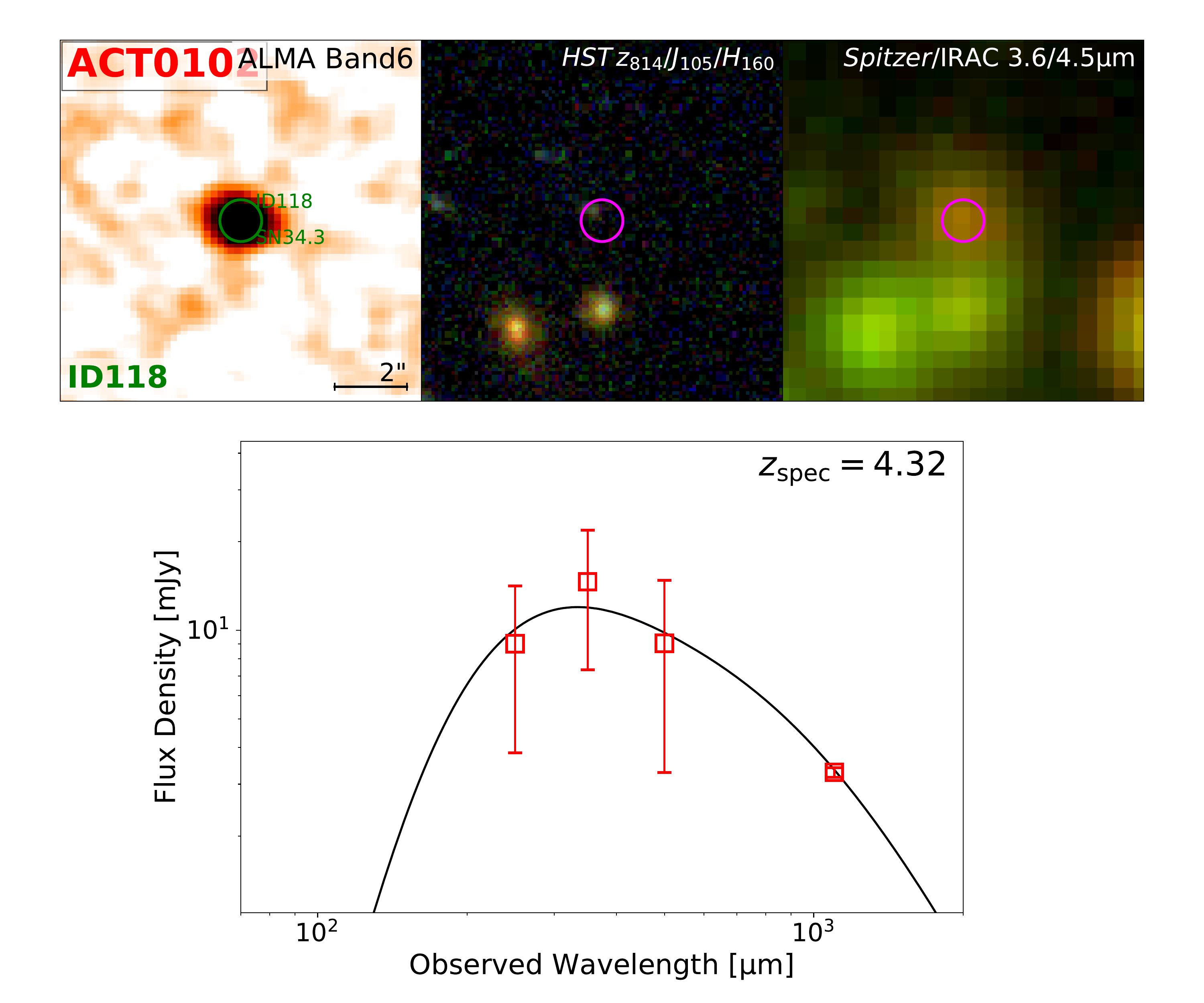}
\figsetgrpnote{Postage stamp images (top) and far-IR SED (bottom) of ACT0102-ID118.}
\figsetgrpend

\figsetgrpstart
\figsetgrpnum{B1.23}
\figsetgrptitle{ACT0102-ID160}
\figsetplot{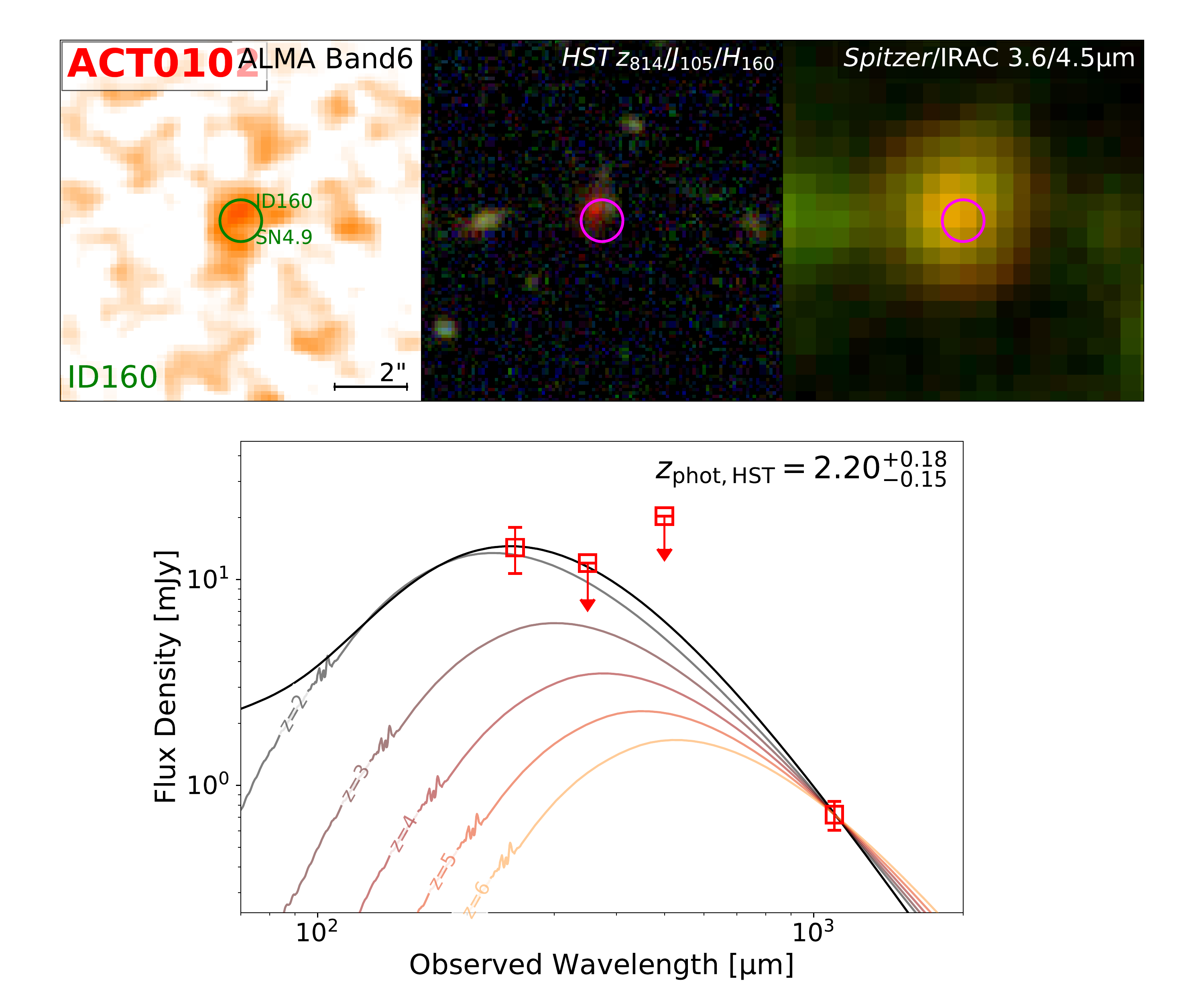}
\figsetgrpnote{Postage stamp images (top) and far-IR SED (bottom) of ACT0102-ID160.}
\figsetgrpend

\figsetgrpstart
\figsetgrpnum{B1.24}
\figsetgrptitle{ACT0102-ID215}
\figsetplot{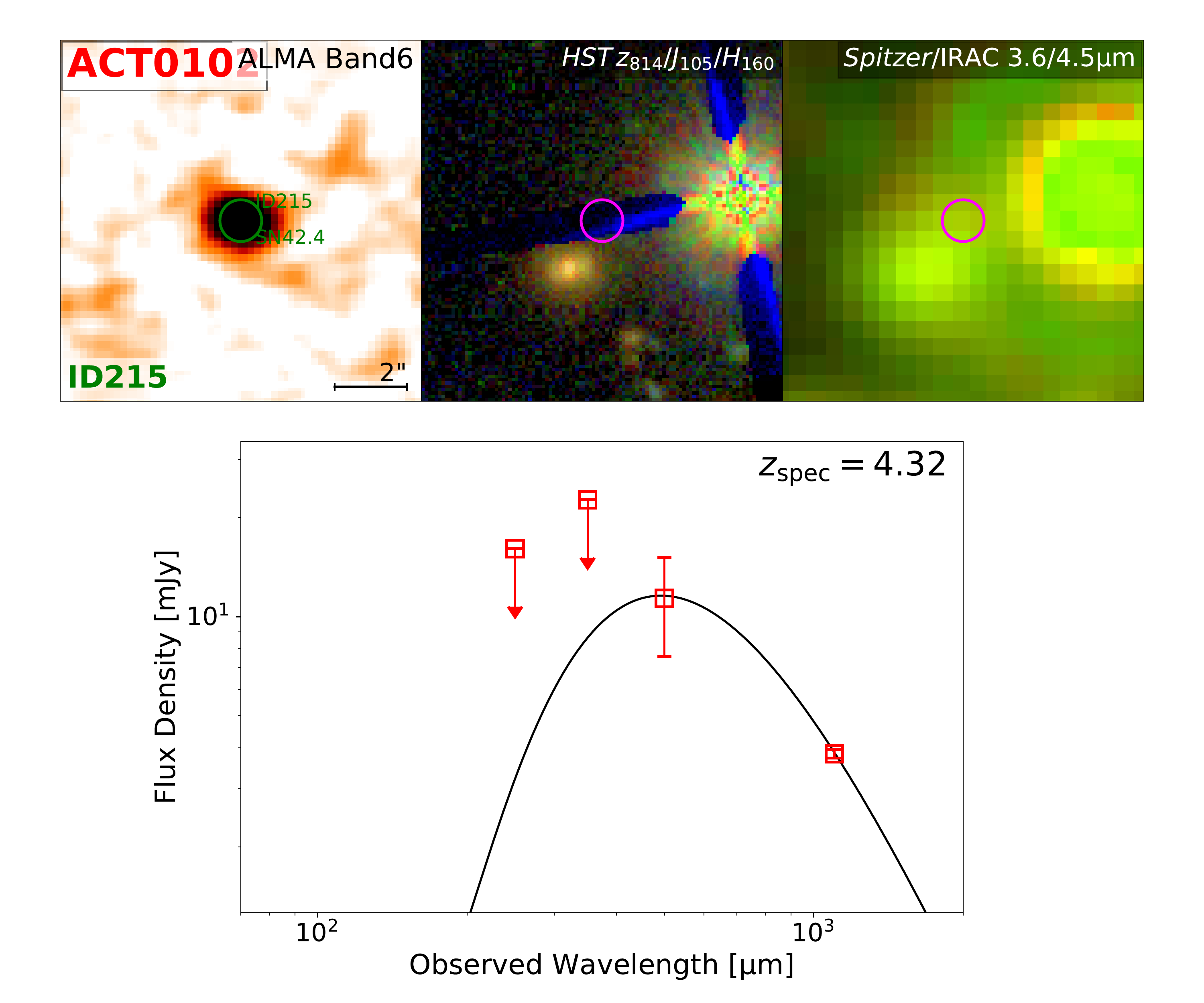}
\figsetgrpnote{Postage stamp images (top) and far-IR SED (bottom) of ACT0102-ID215.}
\figsetgrpend

\figsetgrpstart
\figsetgrpnum{B1.25}
\figsetgrptitle{ACT0102-ID223}
\figsetplot{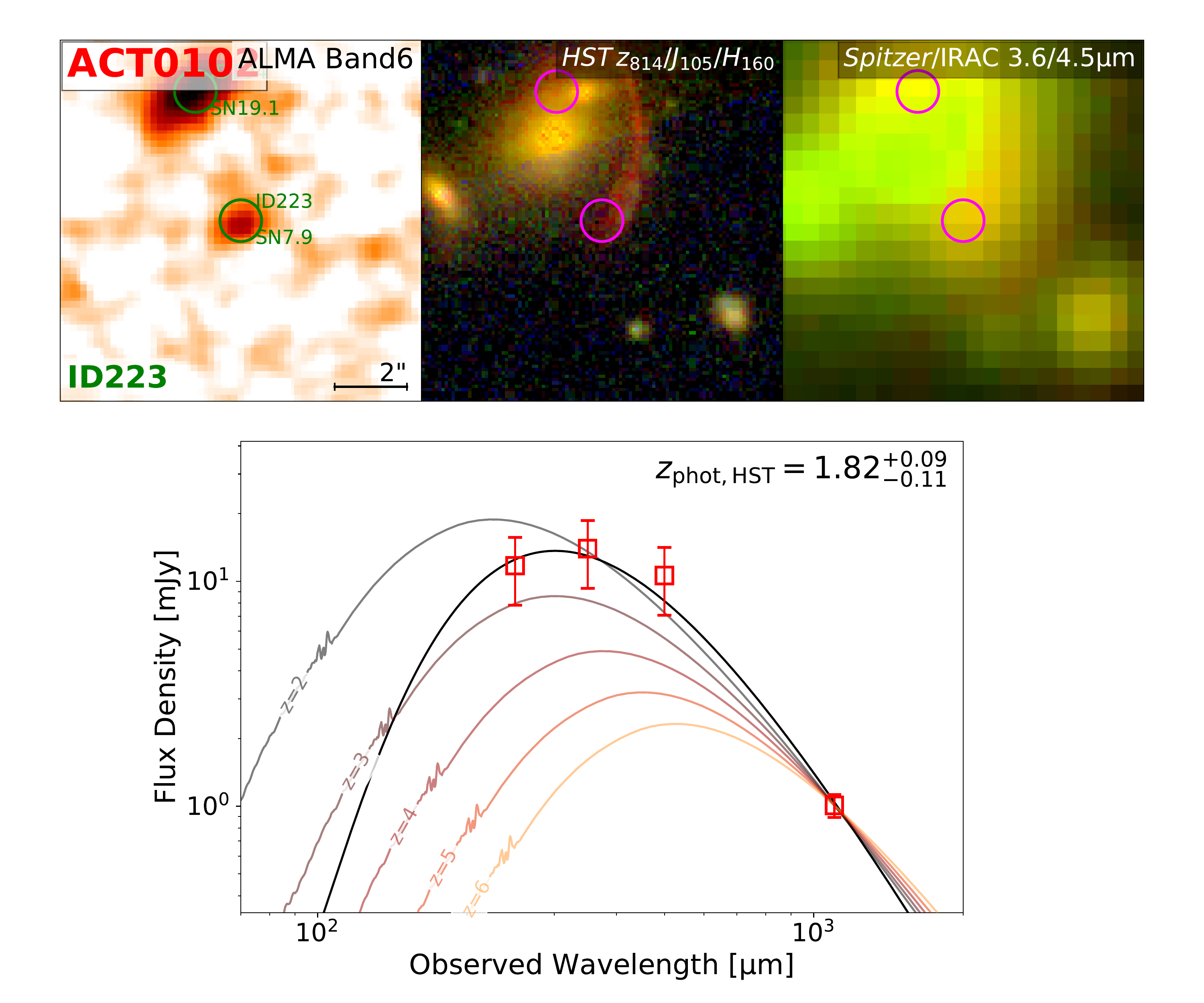}
\figsetgrpnote{Postage stamp images (top) and far-IR SED (bottom) of ACT0102-ID223.}
\figsetgrpend

\figsetgrpstart
\figsetgrpnum{B1.26}
\figsetgrptitle{ACT0102-ID224}
\figsetplot{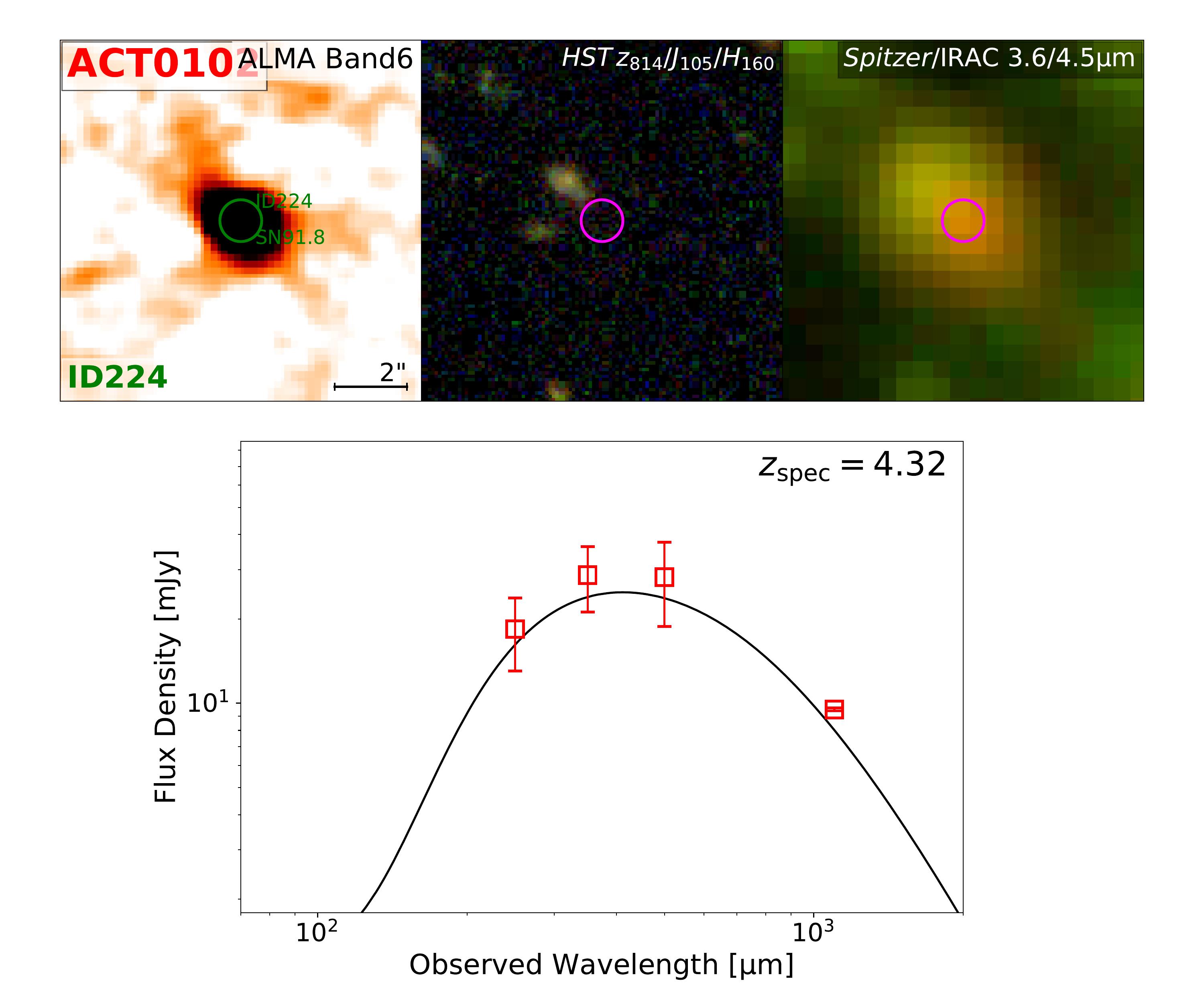}
\figsetgrpnote{Postage stamp images (top) and far-IR SED (bottom) of ACT0102-ID224.}
\figsetgrpend

\figsetgrpstart
\figsetgrpnum{B1.27}
\figsetgrptitle{ACT0102-ID241}
\figsetplot{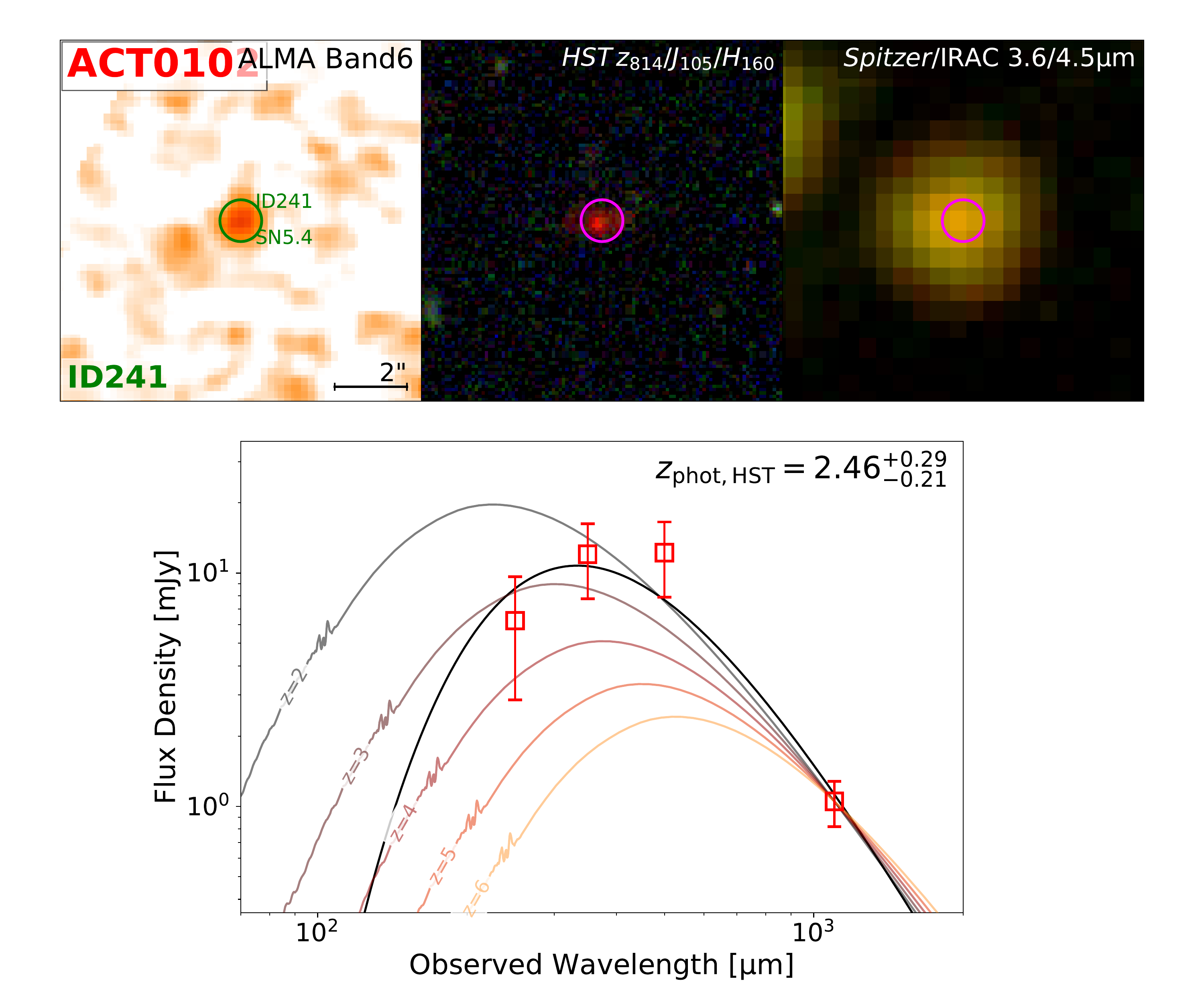}
\figsetgrpnote{Postage stamp images (top) and far-IR SED (bottom) of ACT0102-ID241.}
\figsetgrpend

\figsetgrpstart
\figsetgrpnum{B1.28}
\figsetgrptitle{ACT0102-ID294}
\figsetplot{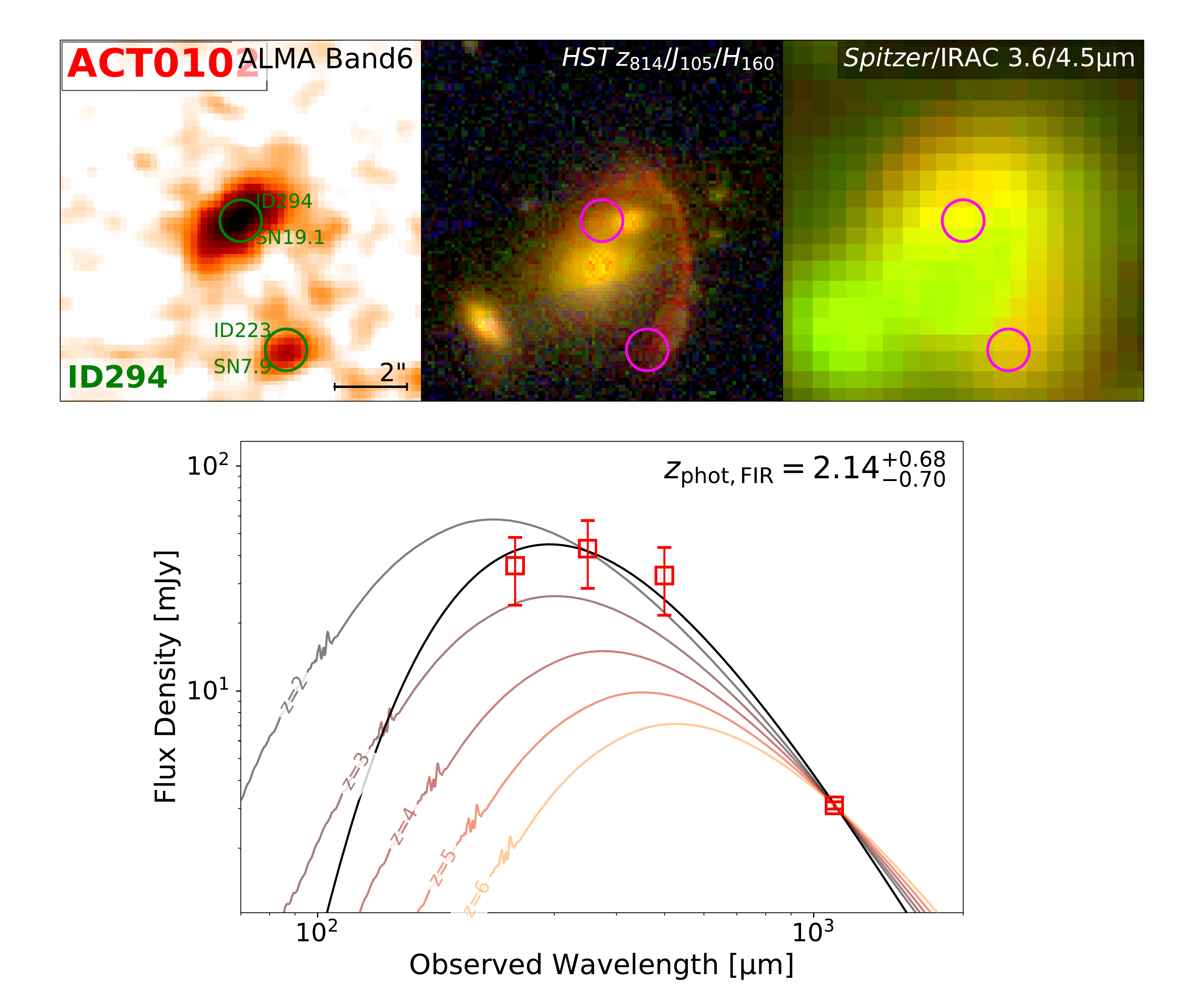}
\figsetgrpnote{Postage stamp images (top) and far-IR SED (bottom) of ACT0102-ID294.}
\figsetgrpend

\figsetgrpstart
\figsetgrpnum{B1.29}
\figsetgrptitle{AS1063-ID15}
\figsetplot{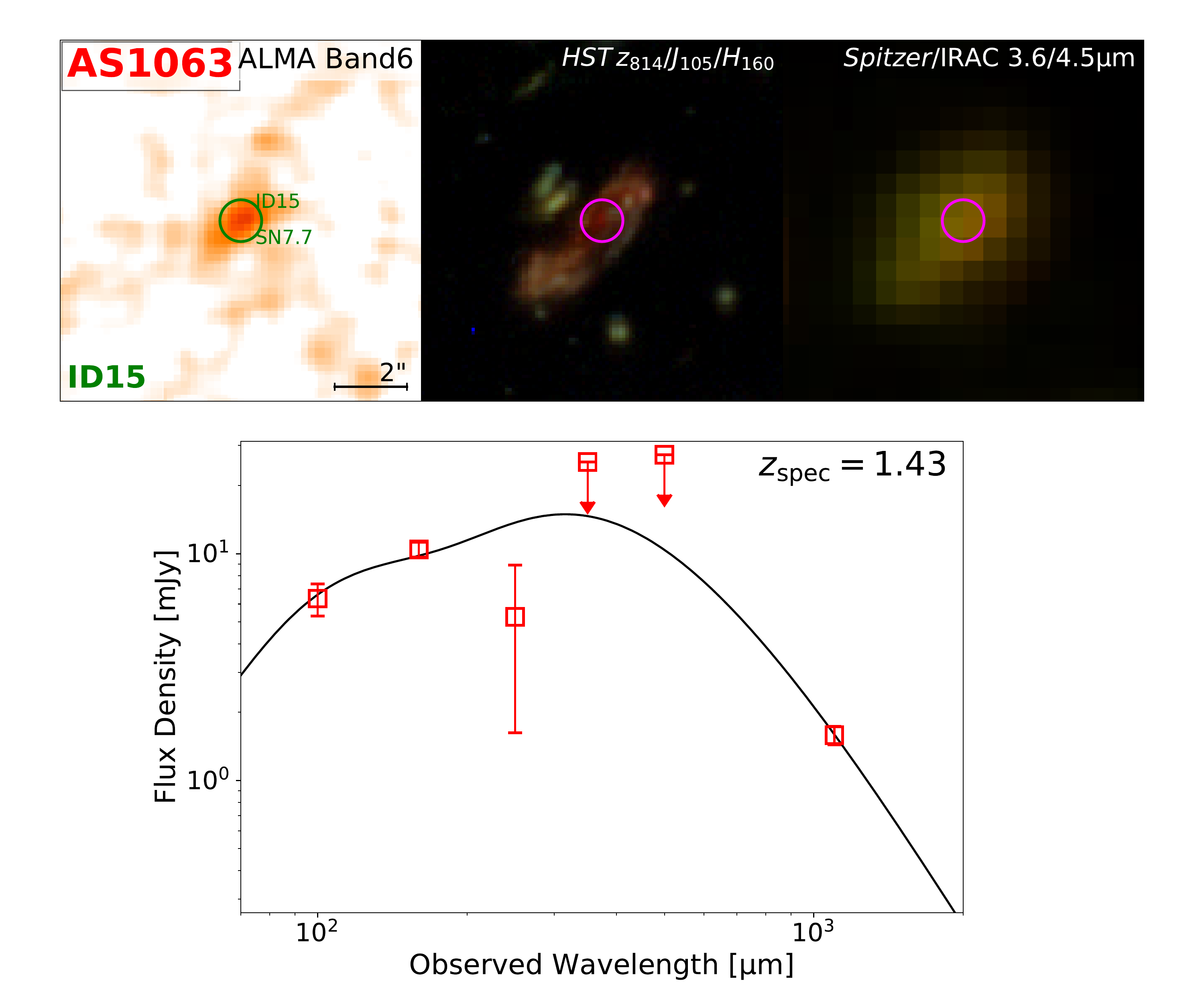}
\figsetgrpnote{Postage stamp images (top) and far-IR SED (bottom) of AS1063-ID15.}
\figsetgrpend

\figsetgrpstart
\figsetgrpnum{B1.30}
\figsetgrptitle{AS1063-ID17}
\figsetplot{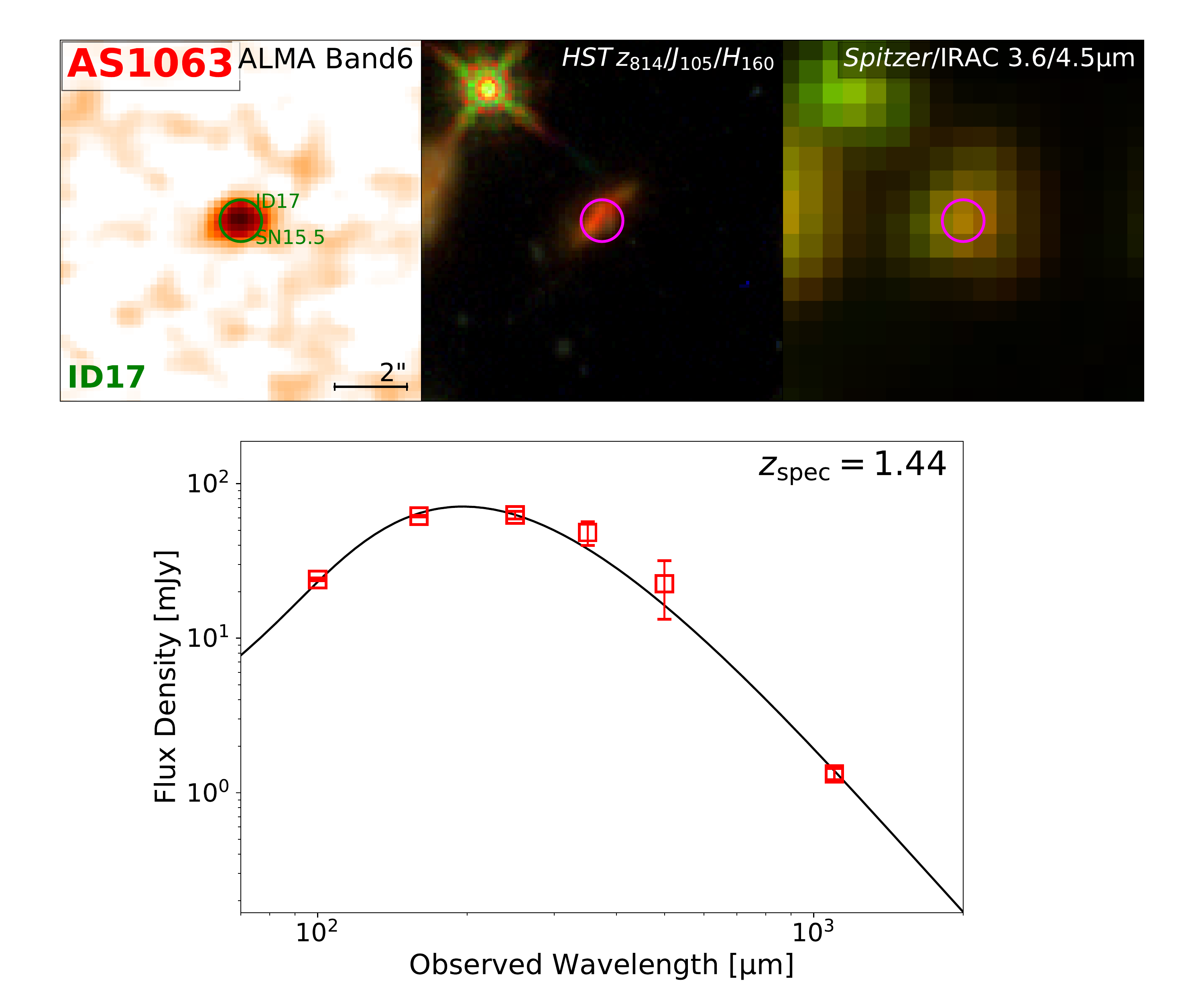}
\figsetgrpnote{Postage stamp images (top) and far-IR SED (bottom) of AS1063-ID17.}
\figsetgrpend

\figsetgrpstart
\figsetgrpnum{B1.31}
\figsetgrptitle{AS1063-ID147}
\figsetplot{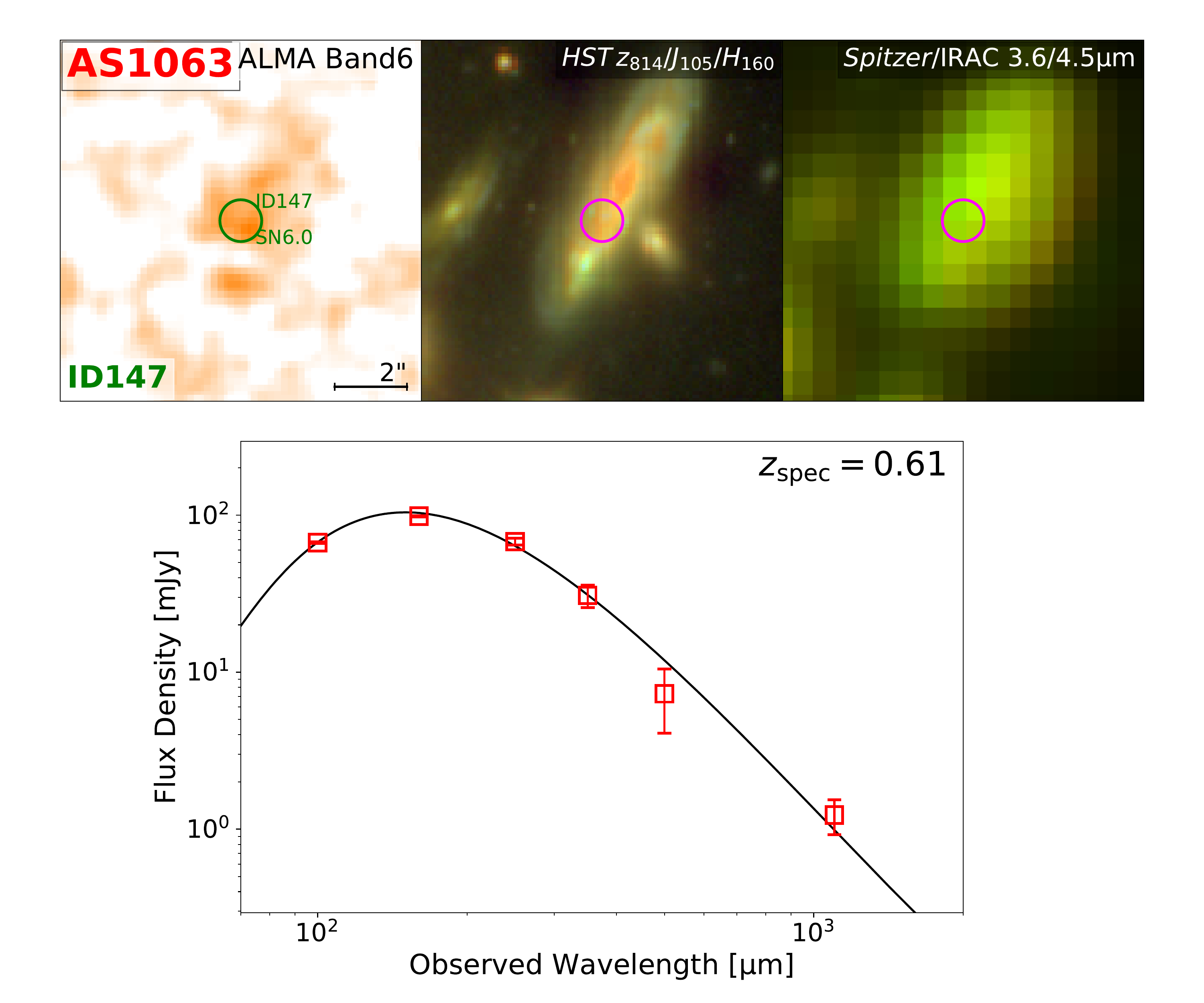}
\figsetgrpnote{Postage stamp images (top) and far-IR SED (bottom) of AS1063-ID147.}
\figsetgrpend

\figsetgrpstart
\figsetgrpnum{B1.32}
\figsetgrptitle{AS1063-ID222}
\figsetplot{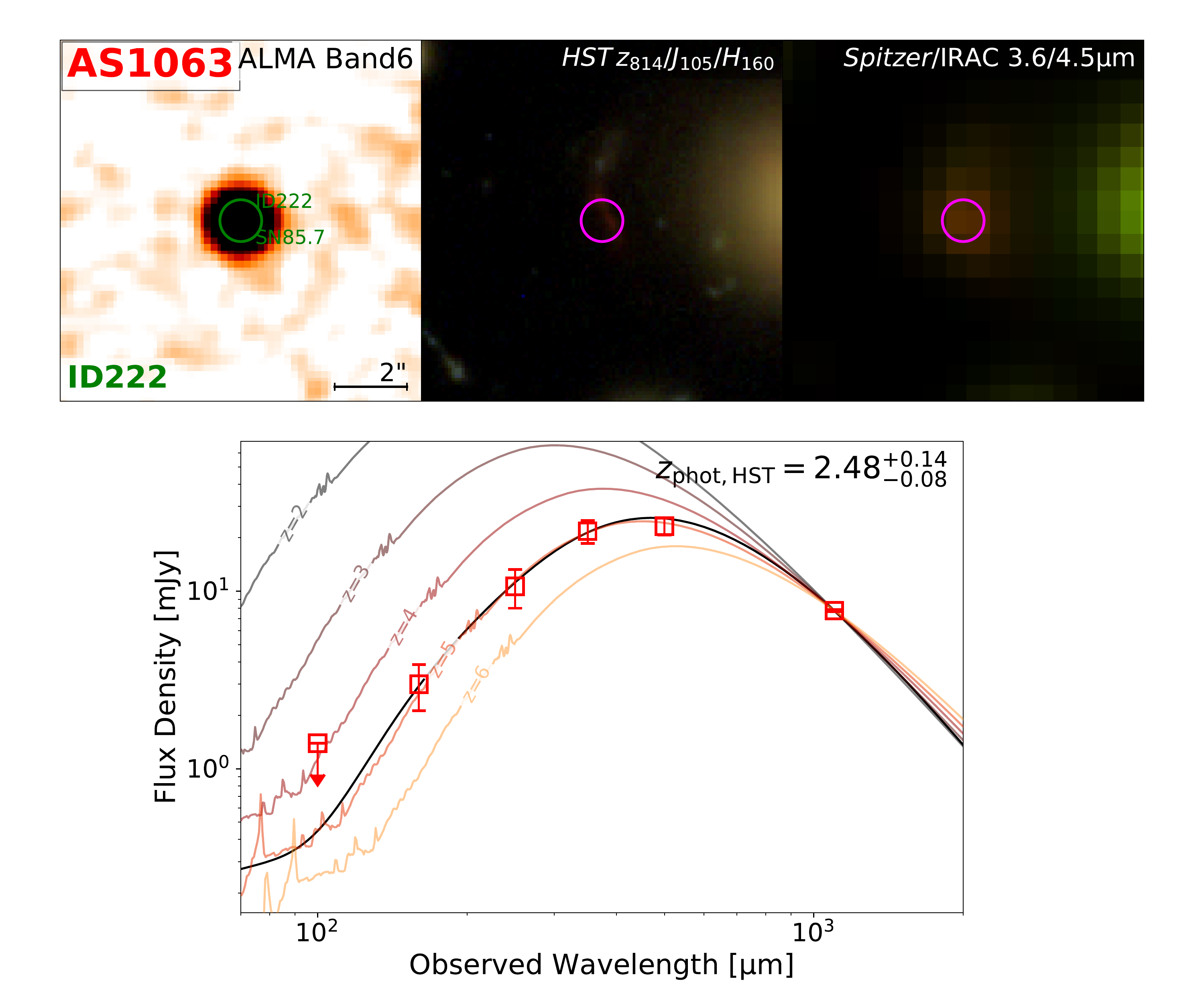}
\figsetgrpnote{Postage stamp images (top) and far-IR SED (bottom) of AS1063-ID222.}
\figsetgrpend

\figsetgrpstart
\figsetgrpnum{B1.33}
\figsetgrptitle{M0035-ID41}
\figsetplot{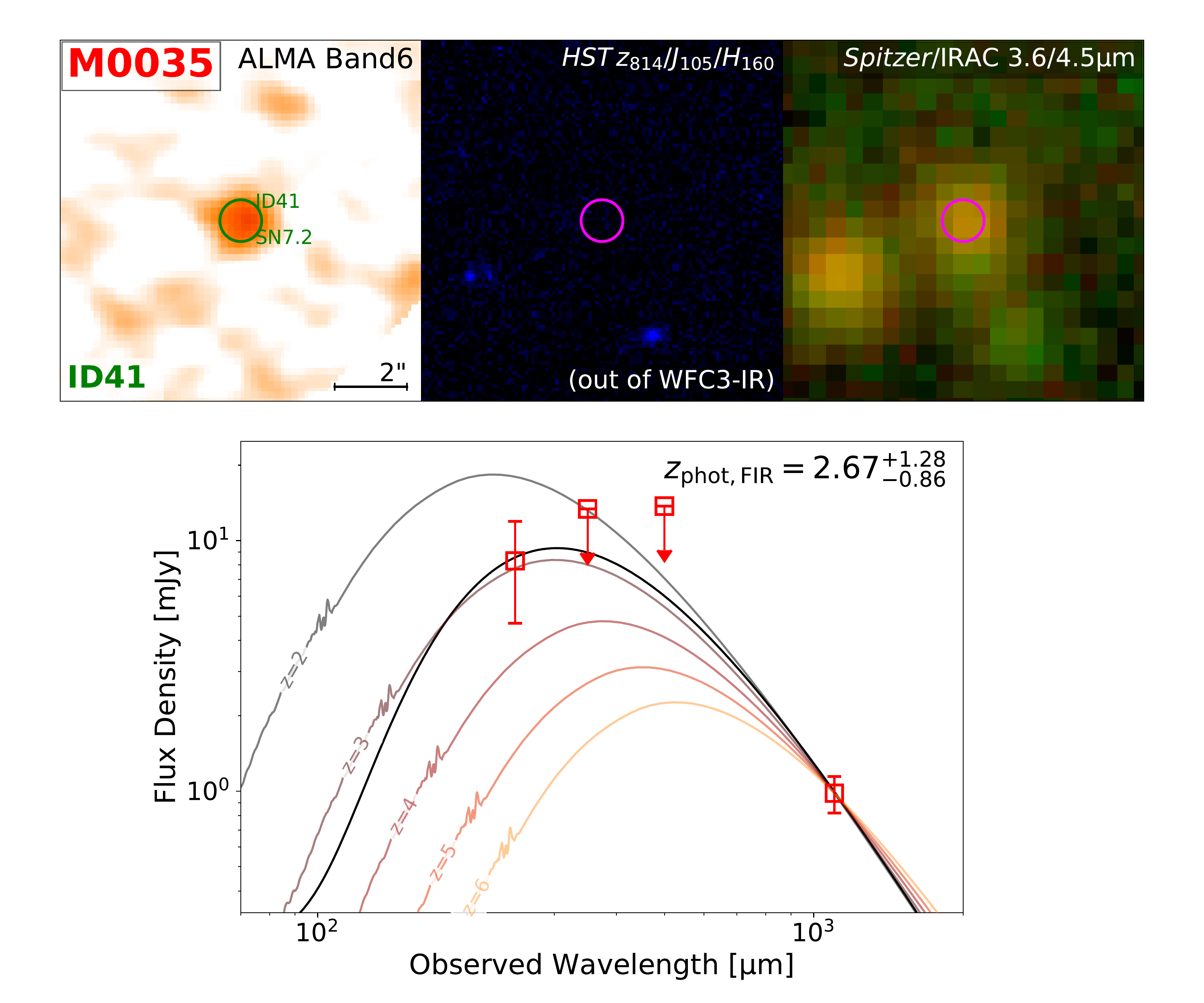}
\figsetgrpnote{Postage stamp images (top) and far-IR SED (bottom) of M0035-ID41.}
\figsetgrpend

\figsetgrpstart
\figsetgrpnum{B1.34}
\figsetgrptitle{M0035-ID94}
\figsetplot{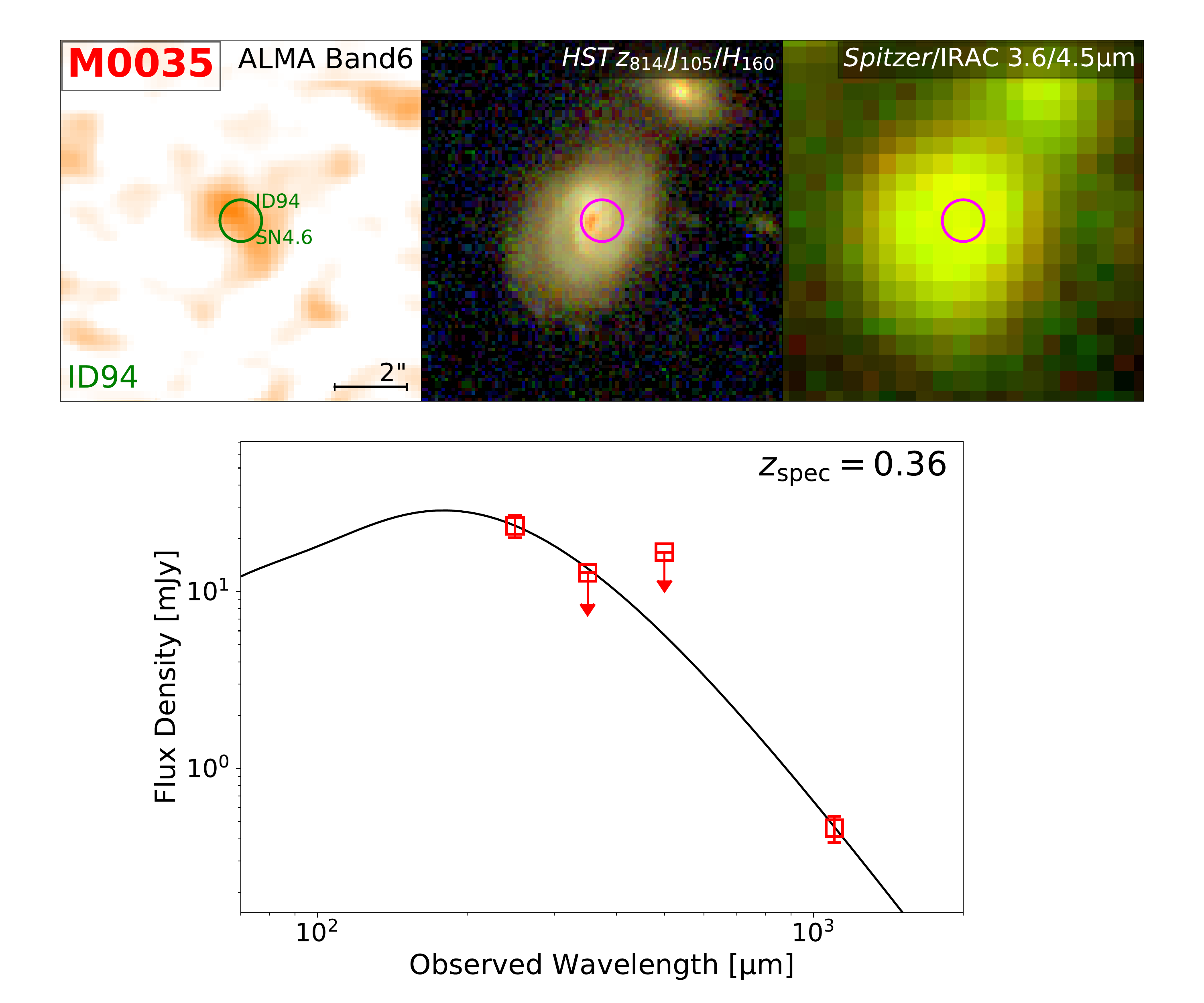}
\figsetgrpnote{Postage stamp images (top) and far-IR SED (bottom) of M0035-ID94.}
\figsetgrpend

\figsetgrpstart
\figsetgrpnum{B1.35}
\figsetgrptitle{M0159-ID05}
\figsetplot{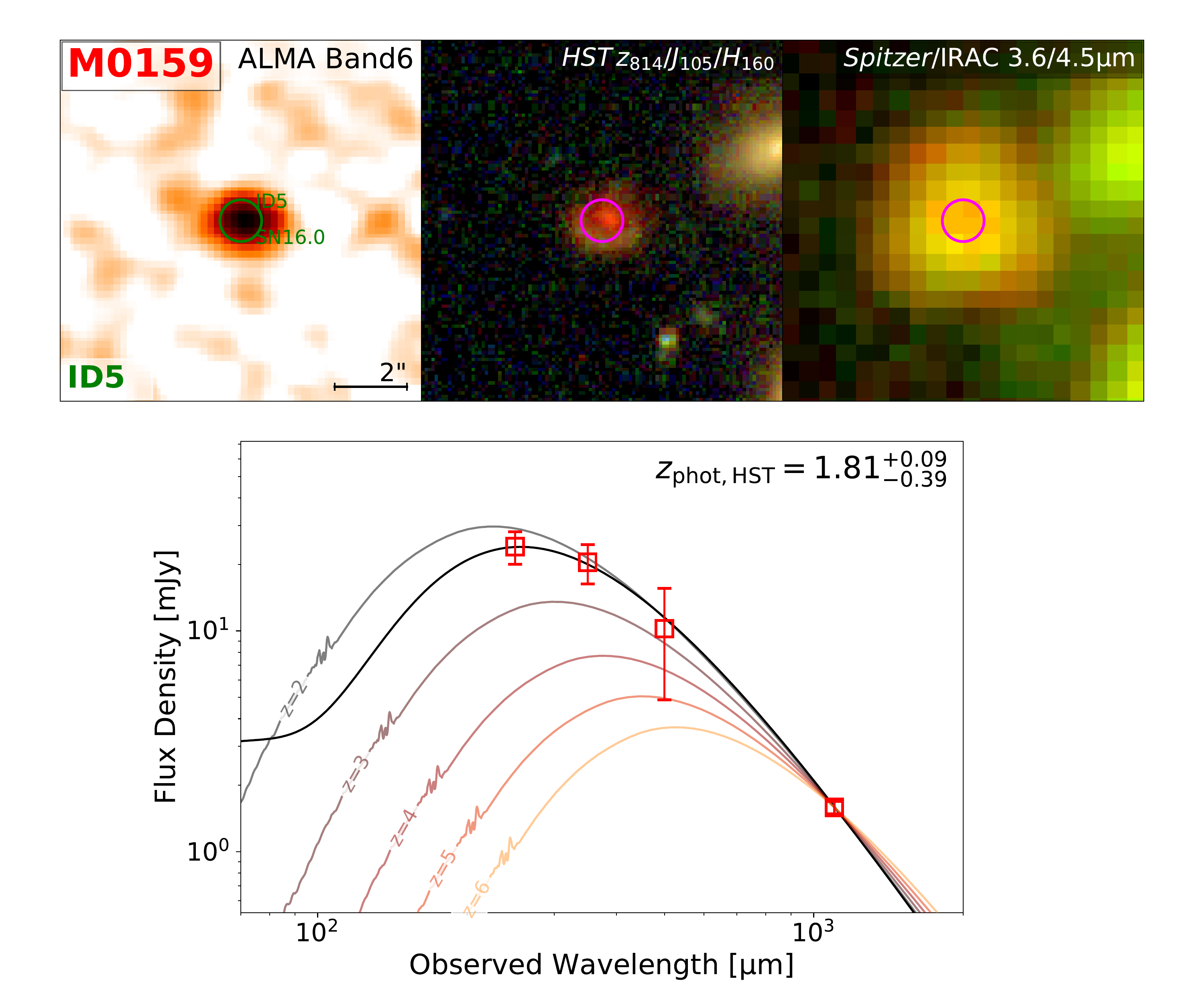}
\figsetgrpnote{Postage stamp images (top) and far-IR SED (bottom) of M0159-ID05.}
\figsetgrpend

\figsetgrpstart
\figsetgrpnum{B1.36}
\figsetgrptitle{M0159-ID24}
\figsetplot{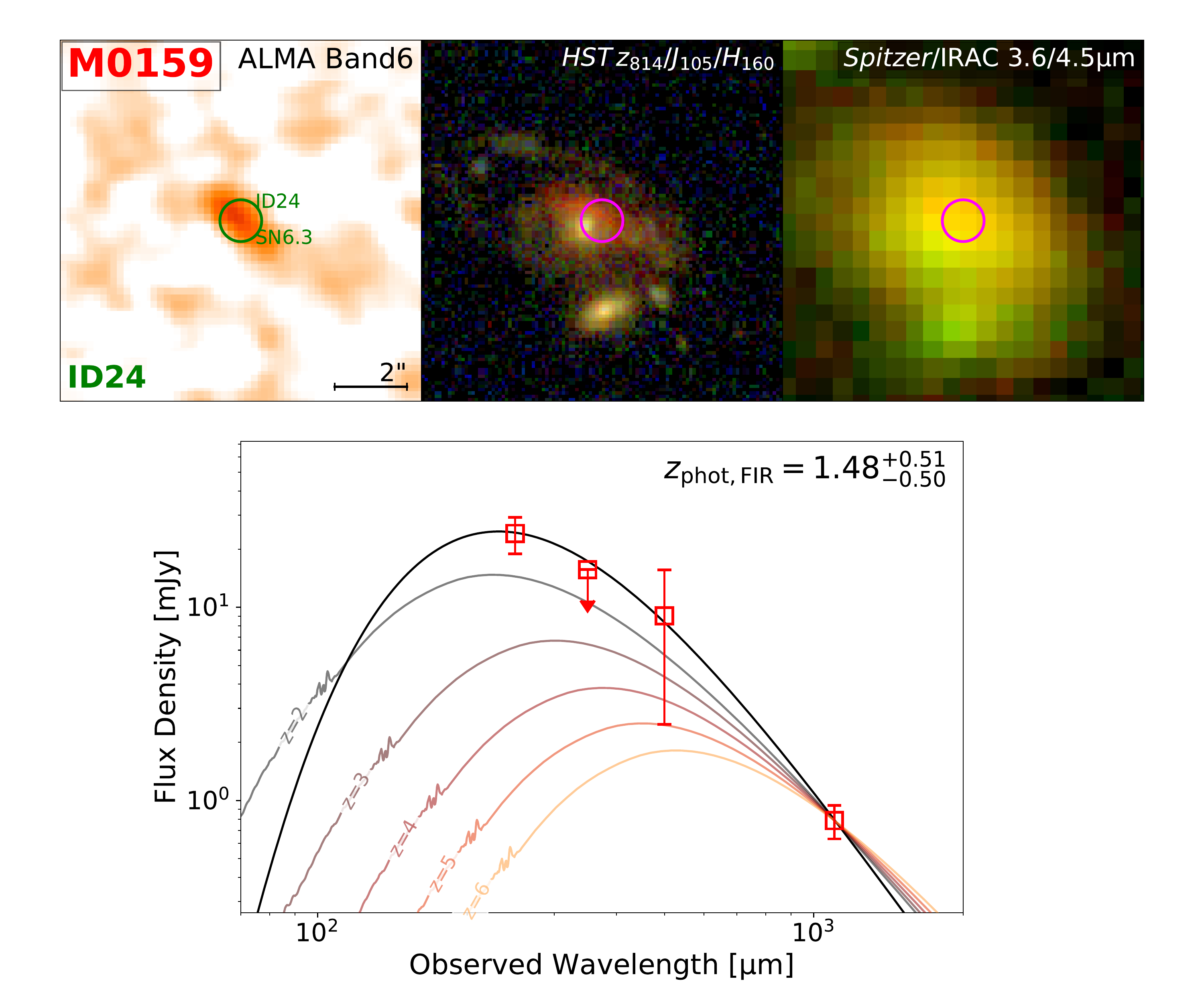}
\figsetgrpnote{Postage stamp images (top) and far-IR SED (bottom) of M0159-ID24.}
\figsetgrpend

\figsetgrpstart
\figsetgrpnum{B1.37}
\figsetgrptitle{M0159-ID61}
\figsetplot{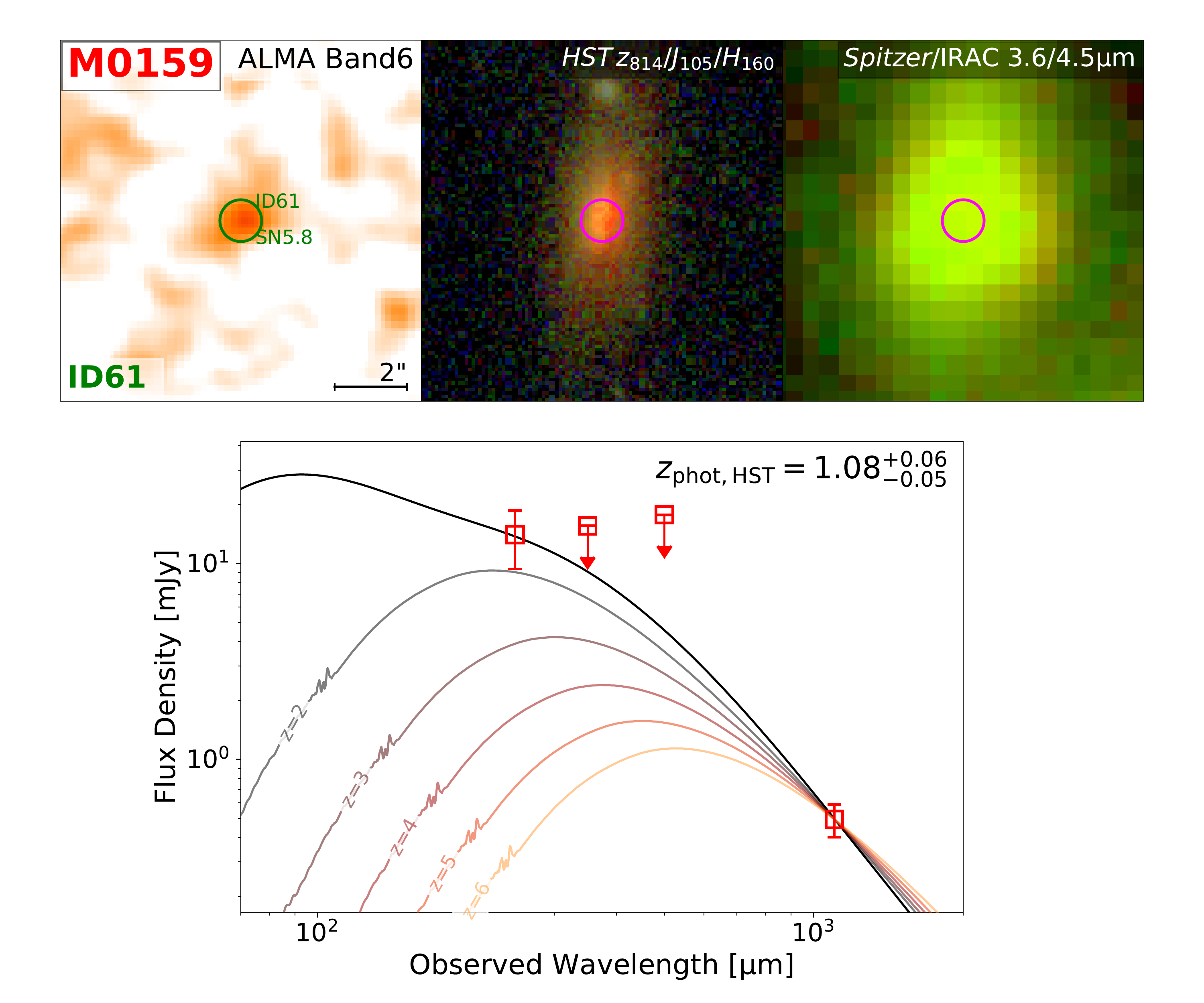}
\figsetgrpnote{Postage stamp images (top) and far-IR SED (bottom) of M0159-ID61.}
\figsetgrpend

\figsetgrpstart
\figsetgrpnum{B1.38}
\figsetgrptitle{M0257-ID13}
\figsetplot{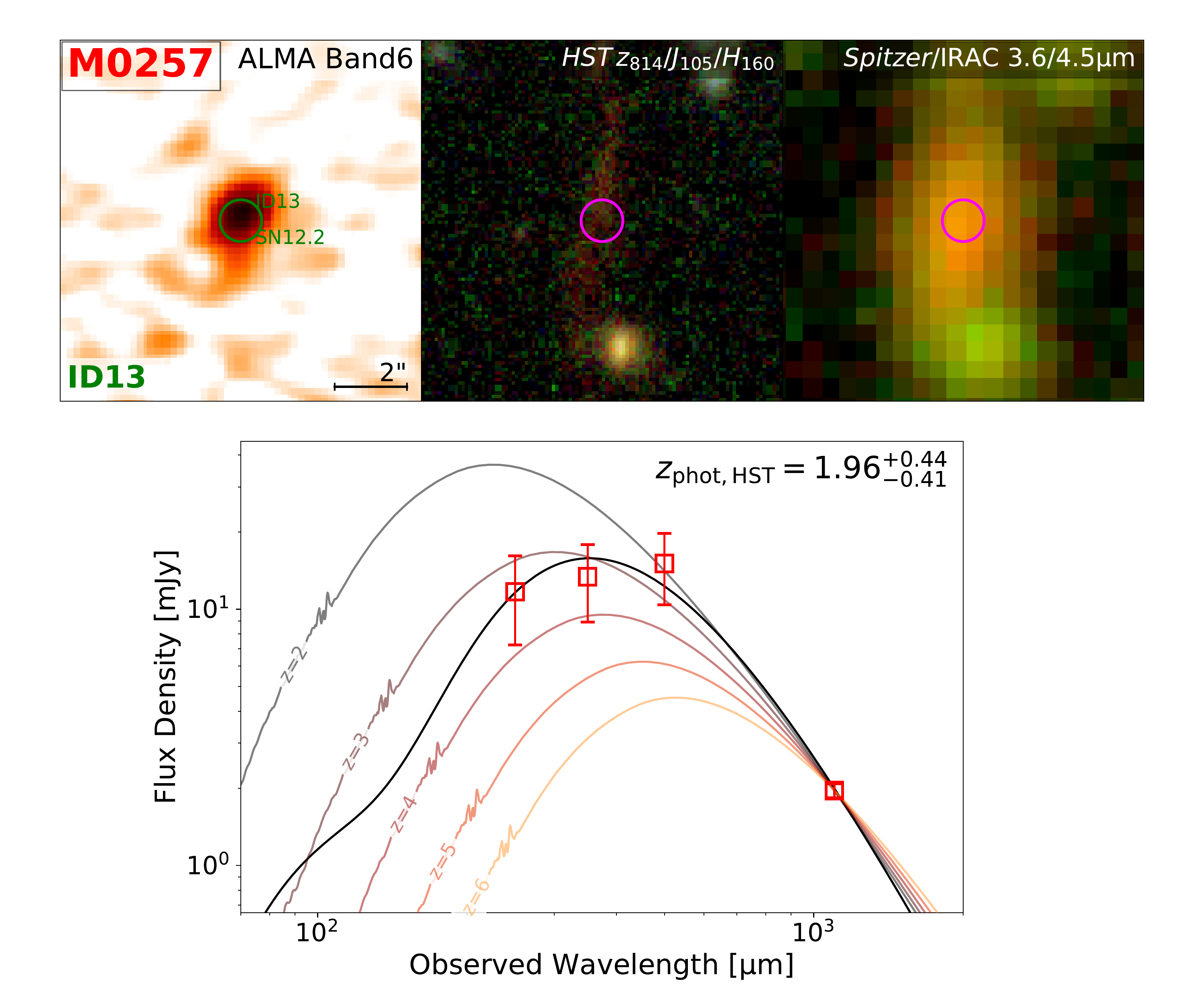}
\figsetgrpnote{Postage stamp images (top) and far-IR SED (bottom) of M0257-ID13.}
\figsetgrpend

\figsetgrpstart
\figsetgrpnum{B1.39}
\figsetgrptitle{M0329-ID11}
\figsetplot{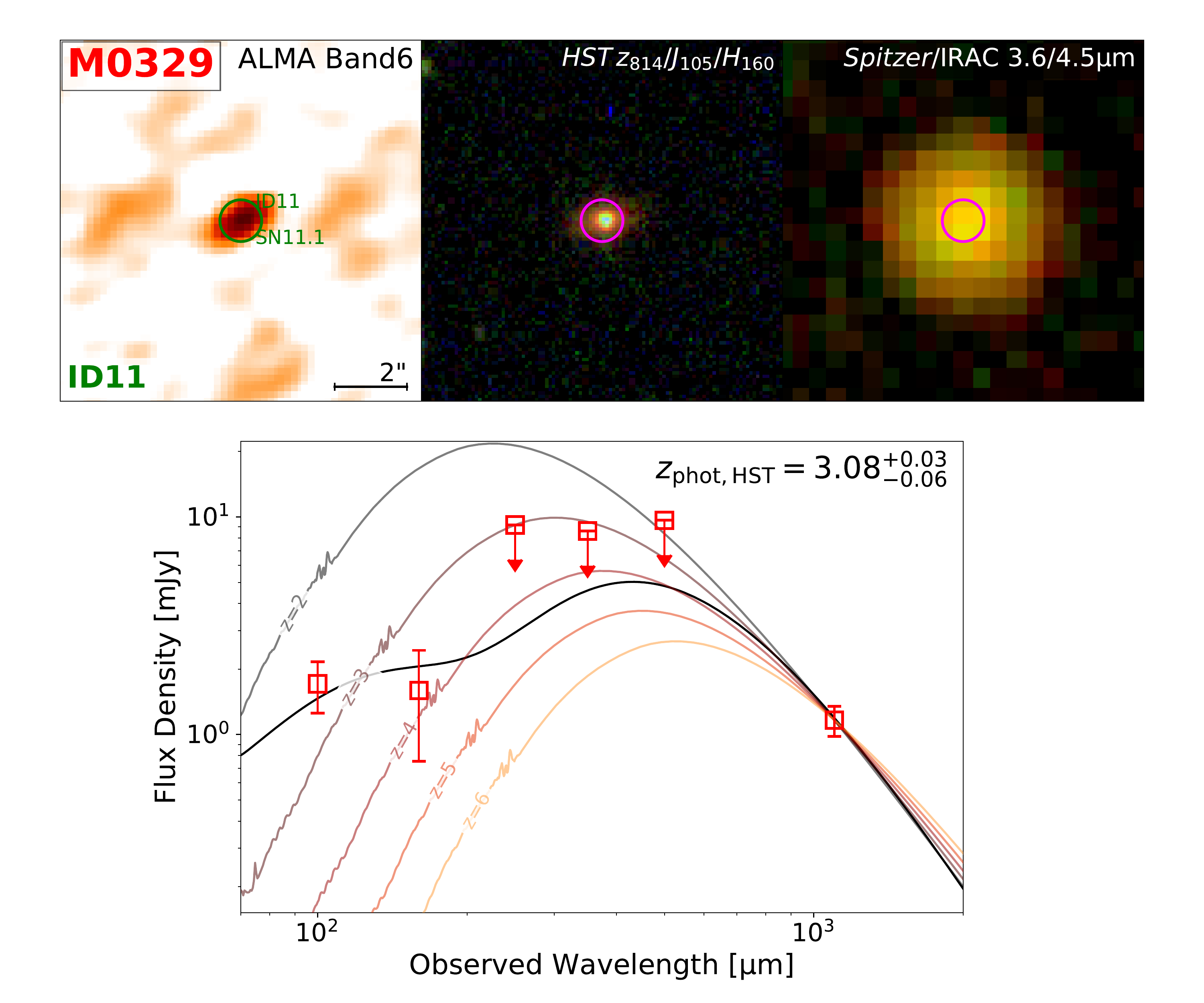}
\figsetgrpnote{Postage stamp images (top) and far-IR SED (bottom) of M0329-ID11.}
\figsetgrpend

\figsetgrpstart
\figsetgrpnum{B1.40}
\figsetgrptitle{M0416-ID51}
\figsetplot{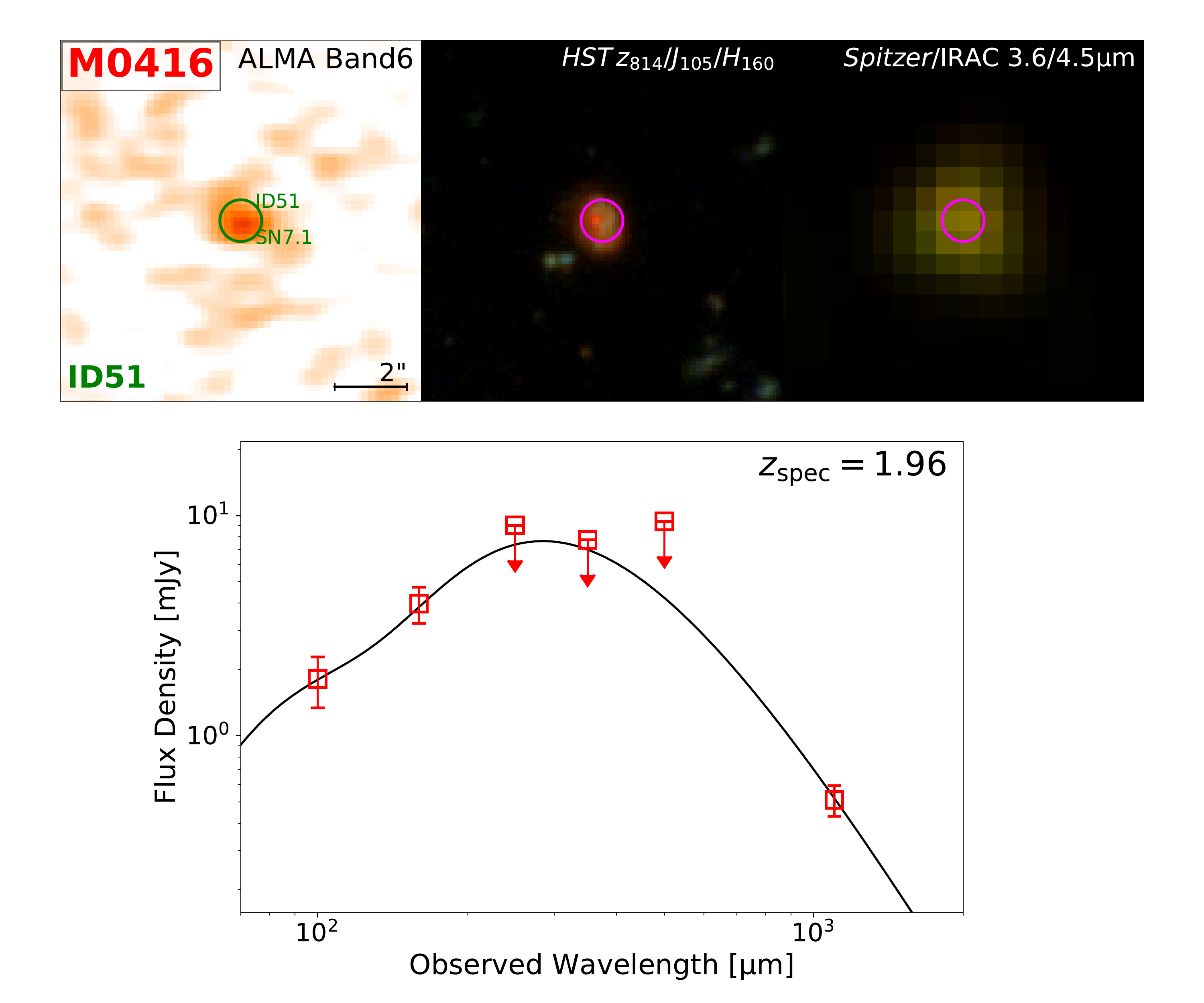}
\figsetgrpnote{Postage stamp images (top) and far-IR SED (bottom) of M0416-ID51.}
\figsetgrpend

\figsetgrpstart
\figsetgrpnum{B1.41}
\figsetgrptitle{M0416-ID79}
\figsetplot{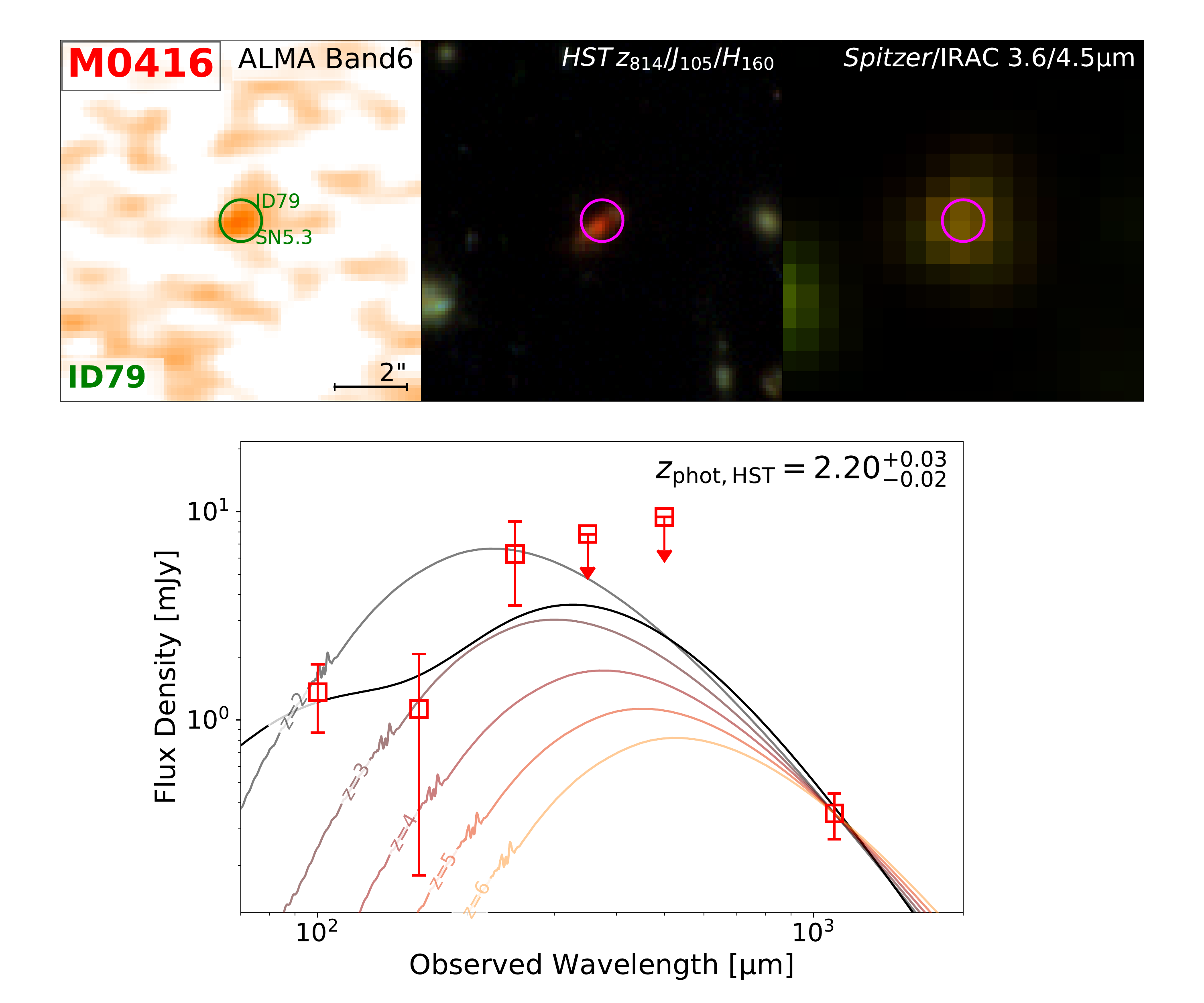}
\figsetgrpnote{Postage stamp images (top) and far-IR SED (bottom) of M0416-ID79.}
\figsetgrpend

\figsetgrpstart
\figsetgrpnum{B1.42}
\figsetgrptitle{M0416-ID117}
\figsetplot{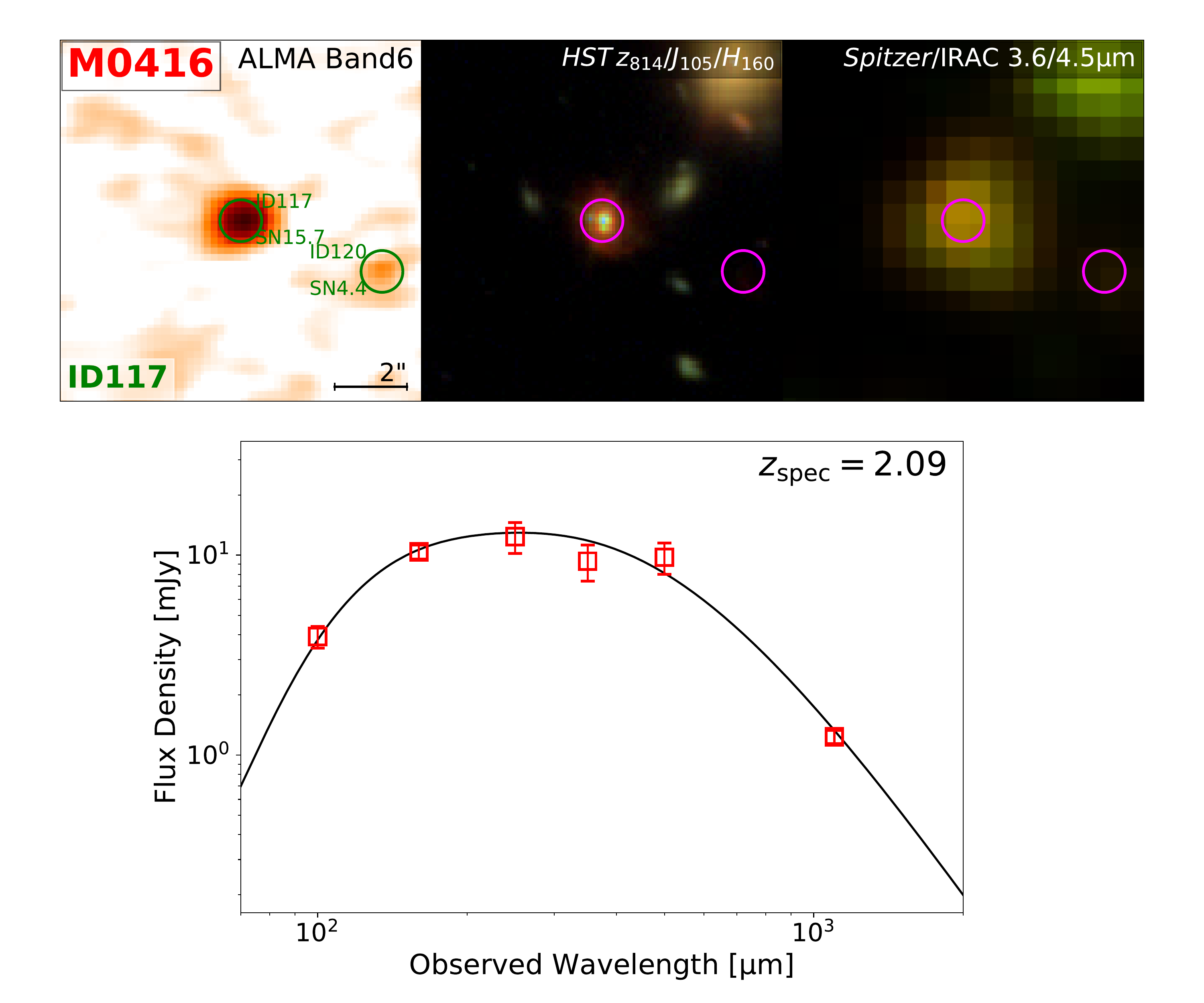}
\figsetgrpnote{Postage stamp images (top) and far-IR SED (bottom) of M0416-ID117.}
\figsetgrpend

\figsetgrpstart
\figsetgrpnum{B1.43}
\figsetgrptitle{M0416-ID160}
\figsetplot{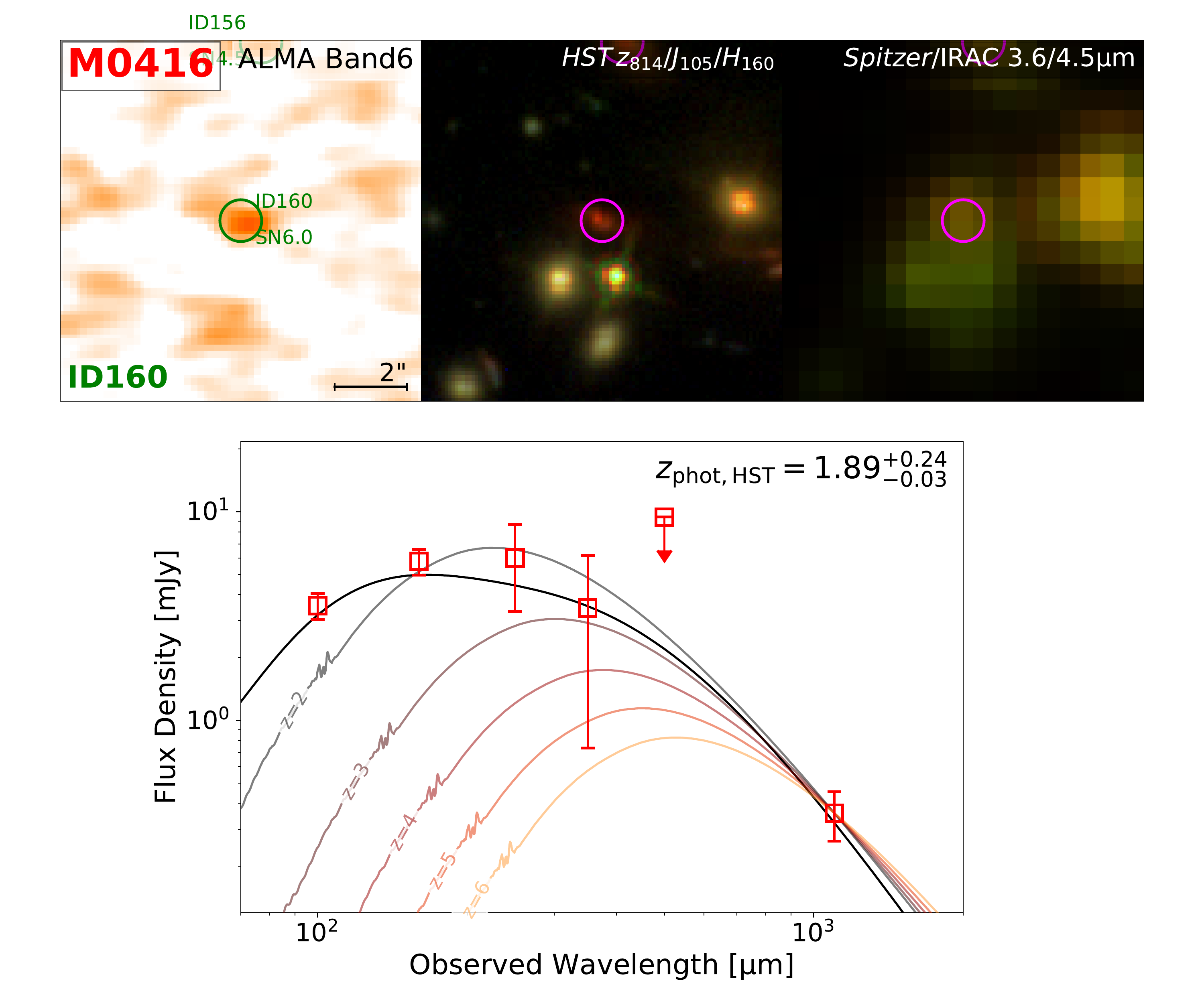}
\figsetgrpnote{Postage stamp images (top) and far-IR SED (bottom) of M0416-ID160.}
\figsetgrpend

\figsetgrpstart
\figsetgrpnum{B1.44}
\figsetgrptitle{M0417-ID46}
\figsetplot{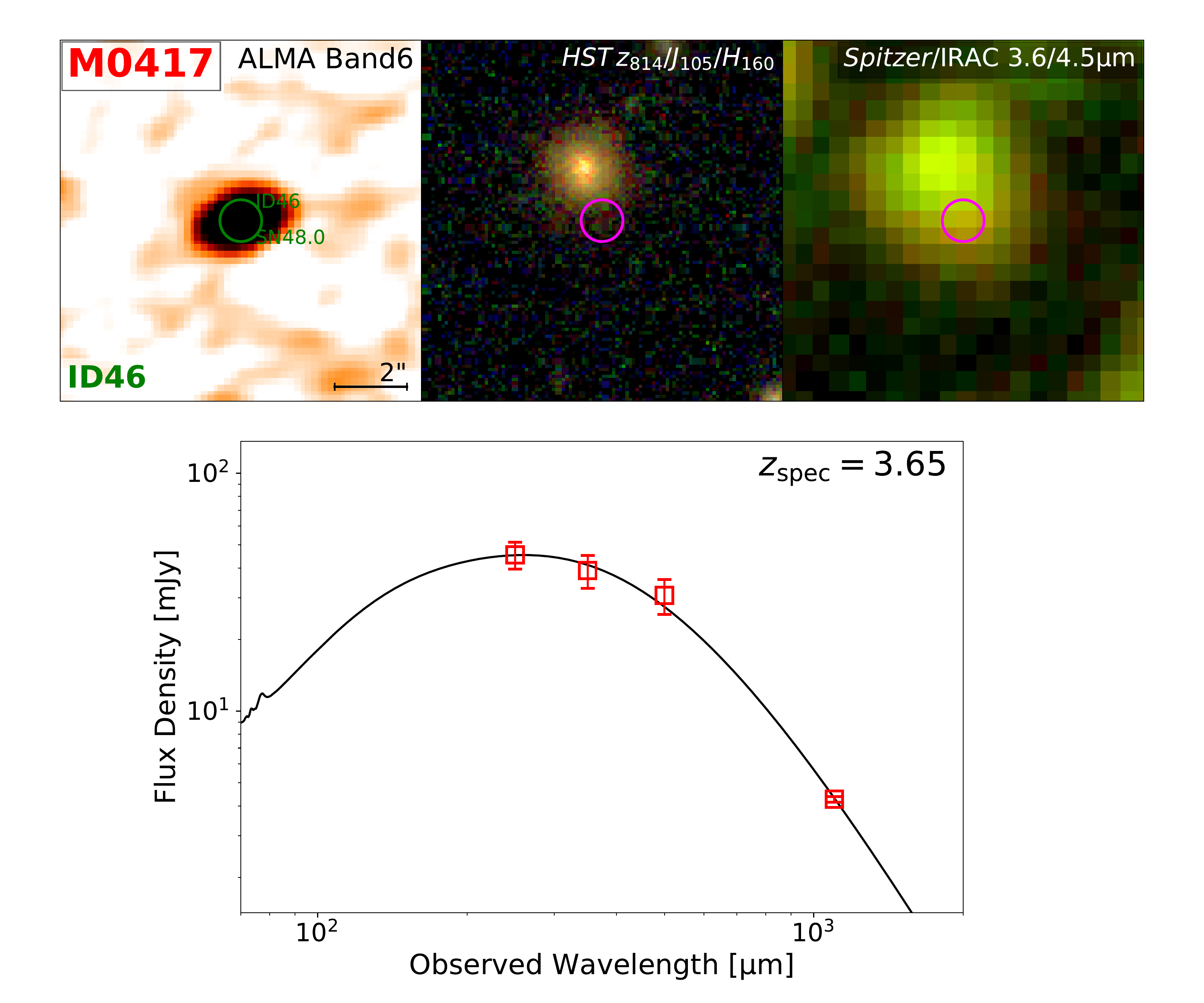}
\figsetgrpnote{Postage stamp images (top) and far-IR SED (bottom) of M0417-ID46.}
\figsetgrpend

\figsetgrpstart
\figsetgrpnum{B1.45}
\figsetgrptitle{M0417-ID49}
\figsetplot{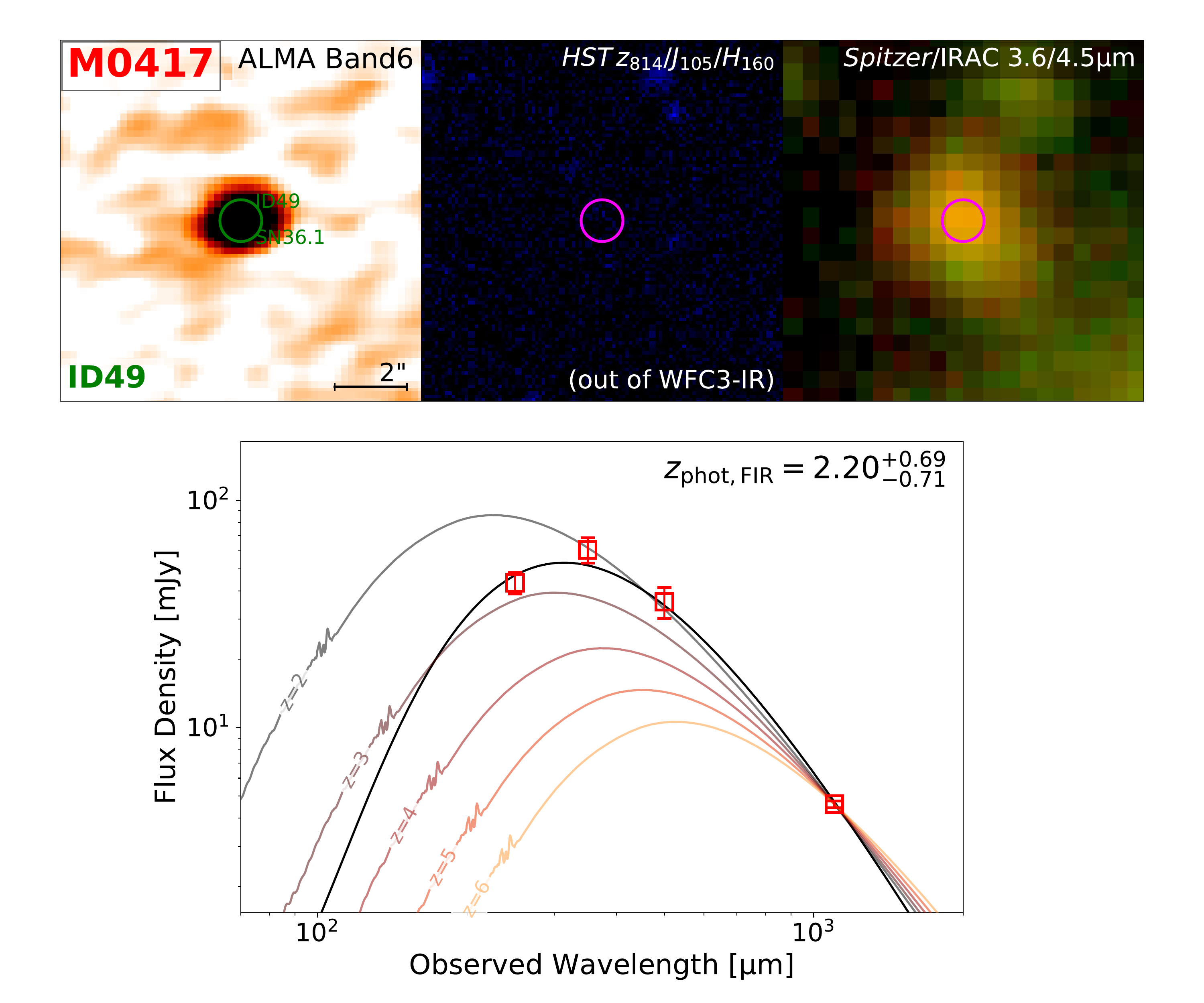}
\figsetgrpnote{Postage stamp images (top) and far-IR SED (bottom) of M0417-ID49.}
\figsetgrpend

\figsetgrpstart
\figsetgrpnum{B1.46}
\figsetgrptitle{M0417-ID58}
\figsetplot{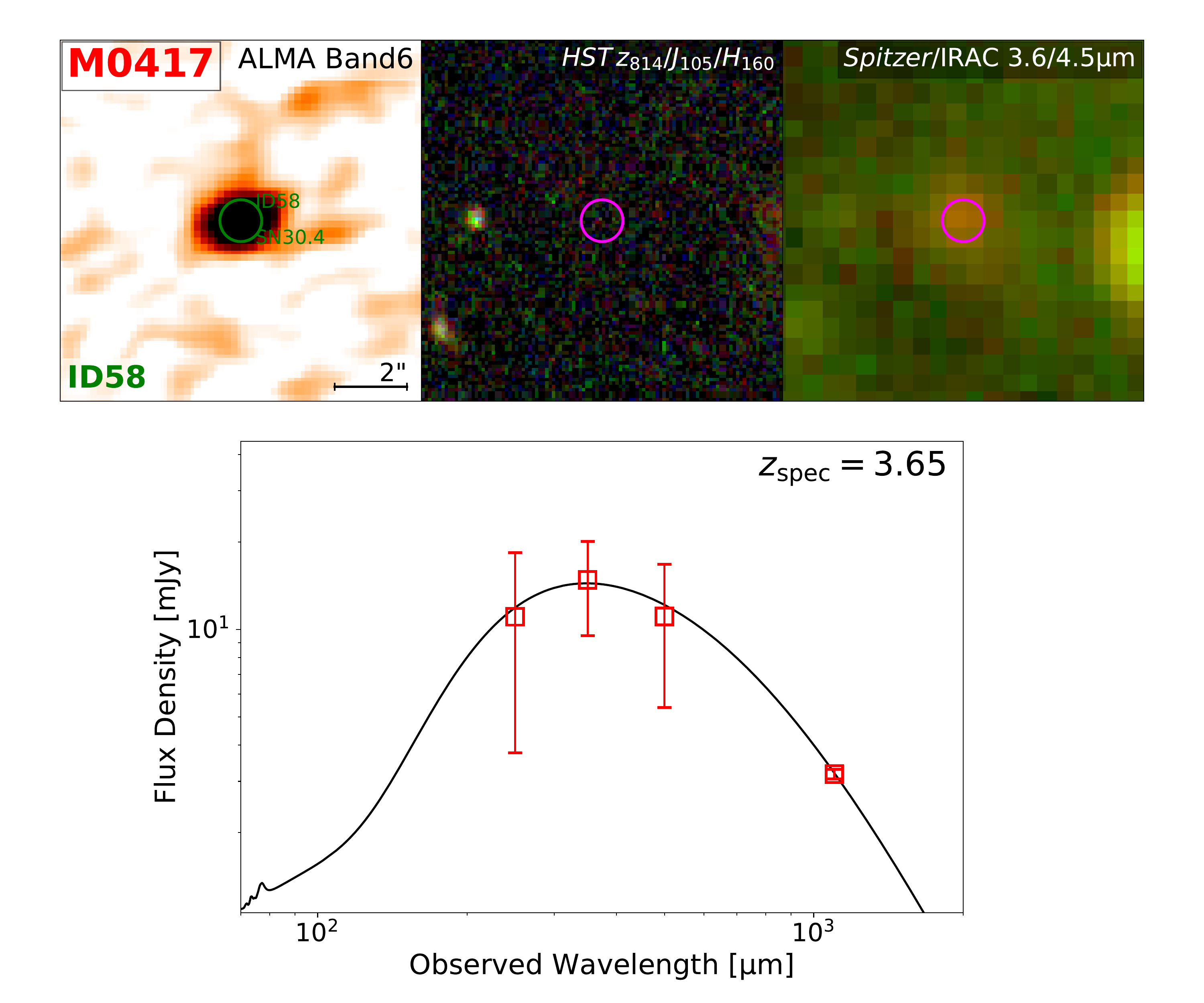}
\figsetgrpnote{Postage stamp images (top) and far-IR SED (bottom) of M0417-ID58.}
\figsetgrpend

\figsetgrpstart
\figsetgrpnum{B1.47}
\figsetgrptitle{M0417-ID121}
\figsetplot{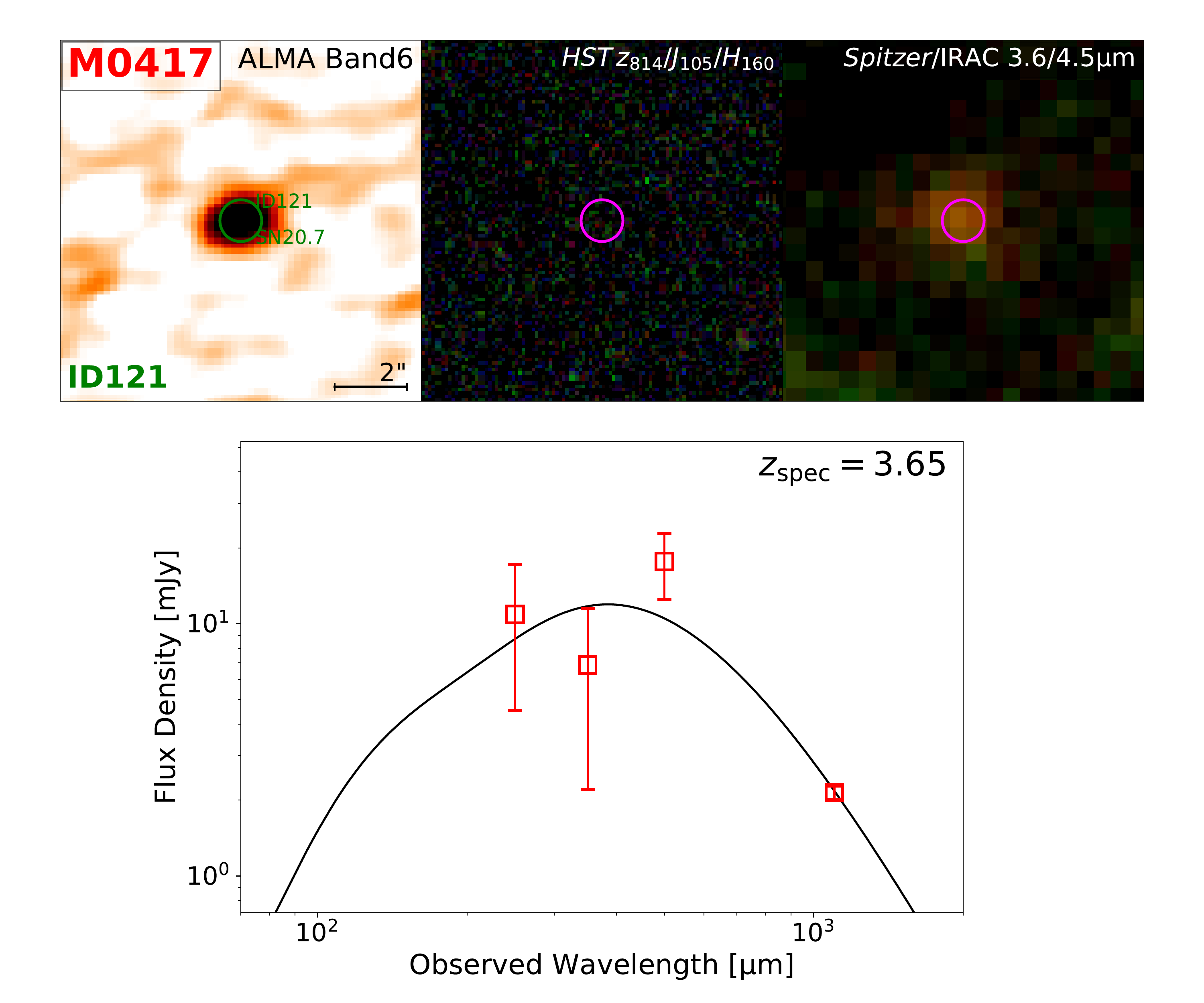}
\figsetgrpnote{Postage stamp images (top) and far-IR SED (bottom) of M0417-ID121.}
\figsetgrpend

\figsetgrpstart
\figsetgrpnum{B1.48}
\figsetgrptitle{M0417-ID204}
\figsetplot{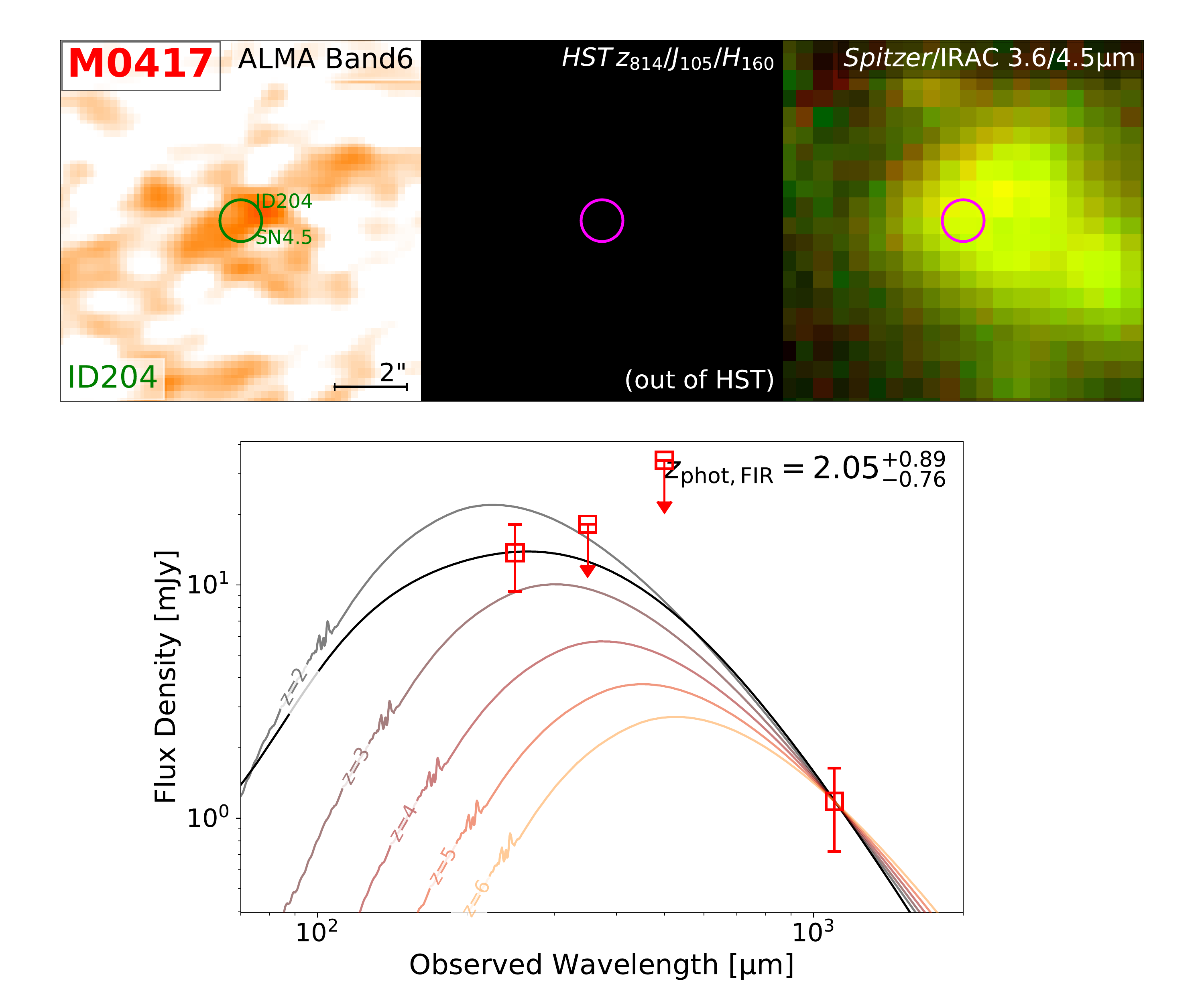}
\figsetgrpnote{Postage stamp images (top) and far-IR SED (bottom) of M0417-ID204.}
\figsetgrpend

\figsetgrpstart
\figsetgrpnum{B1.49}
\figsetgrptitle{M0417-ID218}
\figsetplot{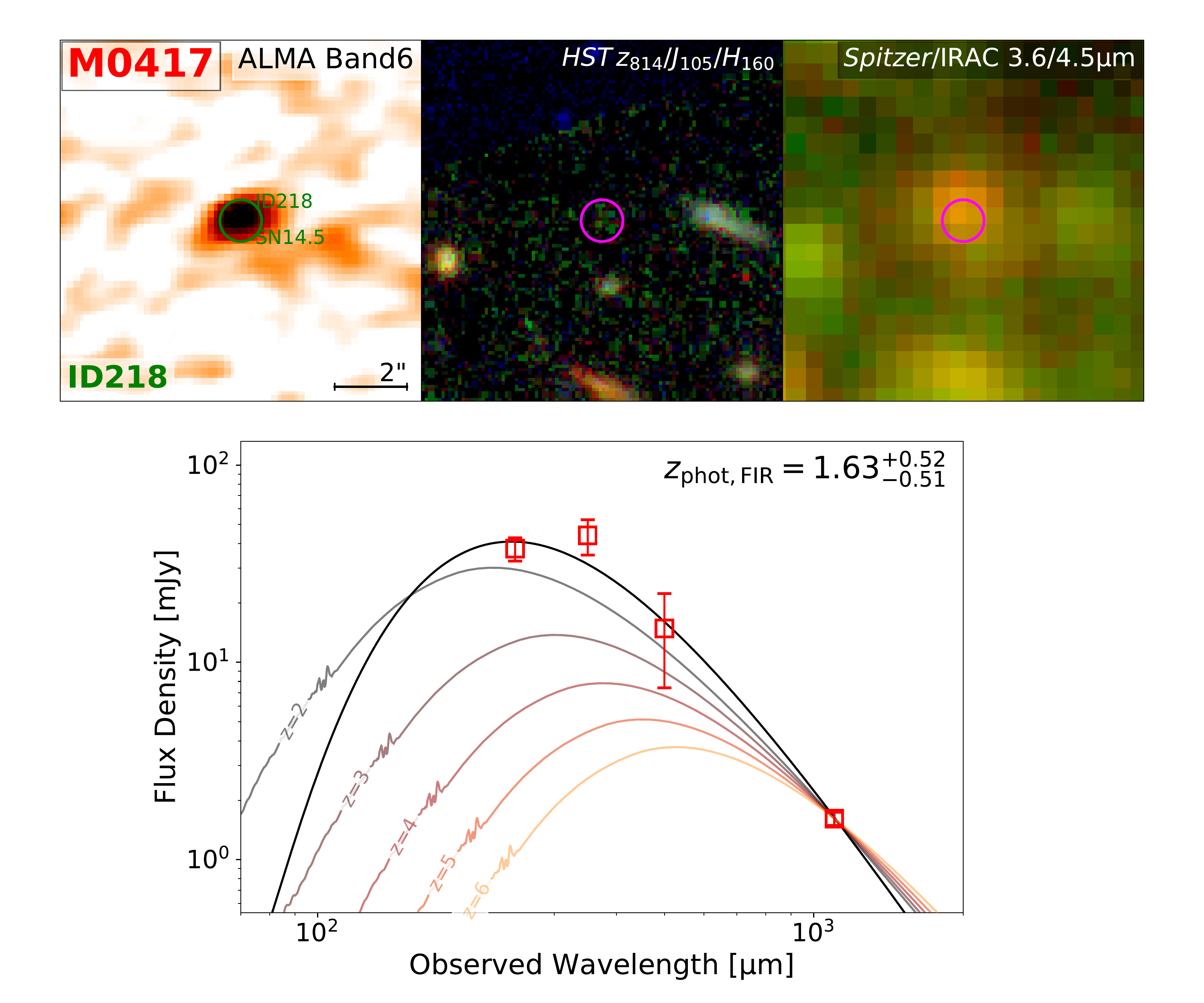}
\figsetgrpnote{Postage stamp images (top) and far-IR SED (bottom) of M0417-ID218.}
\figsetgrpend

\figsetgrpstart
\figsetgrpnum{B1.50}
\figsetgrptitle{M0417-ID221}
\figsetplot{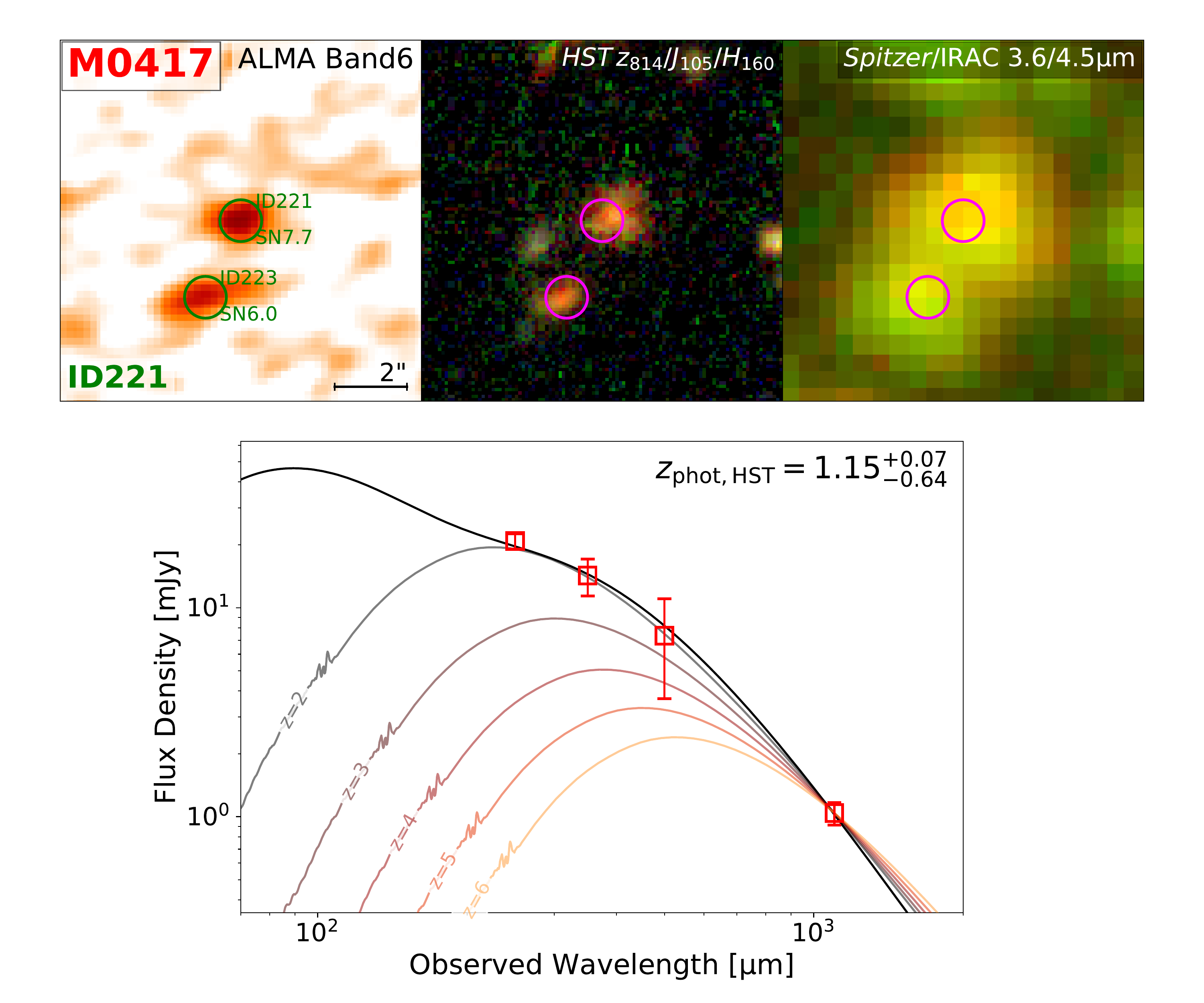}
\figsetgrpnote{Postage stamp images (top) and far-IR SED (bottom) of M0417-ID221.}
\figsetgrpend

\figsetgrpstart
\figsetgrpnum{B1.51}
\figsetgrptitle{M0417-ID223}
\figsetplot{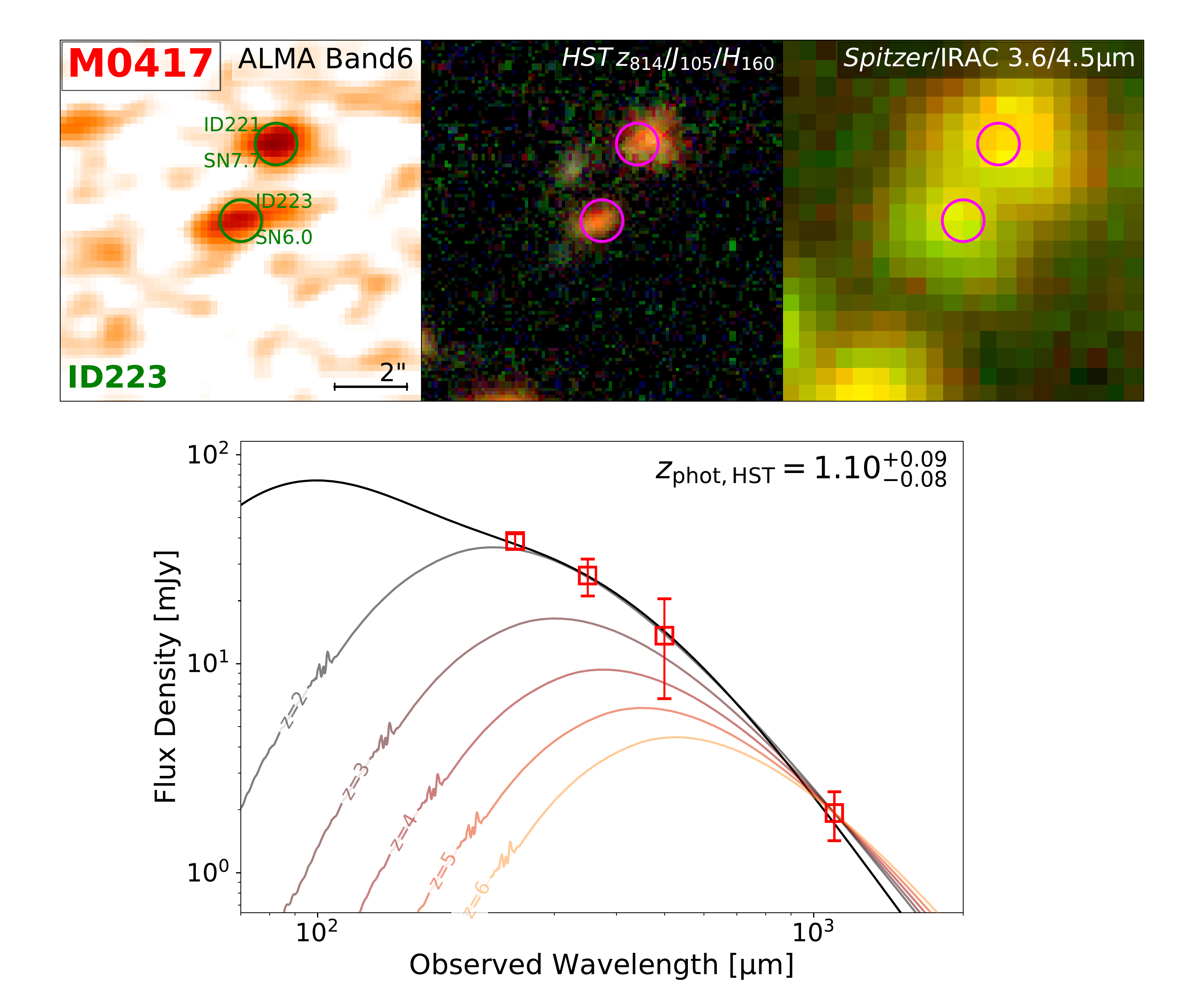}
\figsetgrpnote{Postage stamp images (top) and far-IR SED (bottom) of M0417-ID223.}
\figsetgrpend

\figsetgrpstart
\figsetgrpnum{B1.52}
\figsetgrptitle{M0429-ID27}
\figsetplot{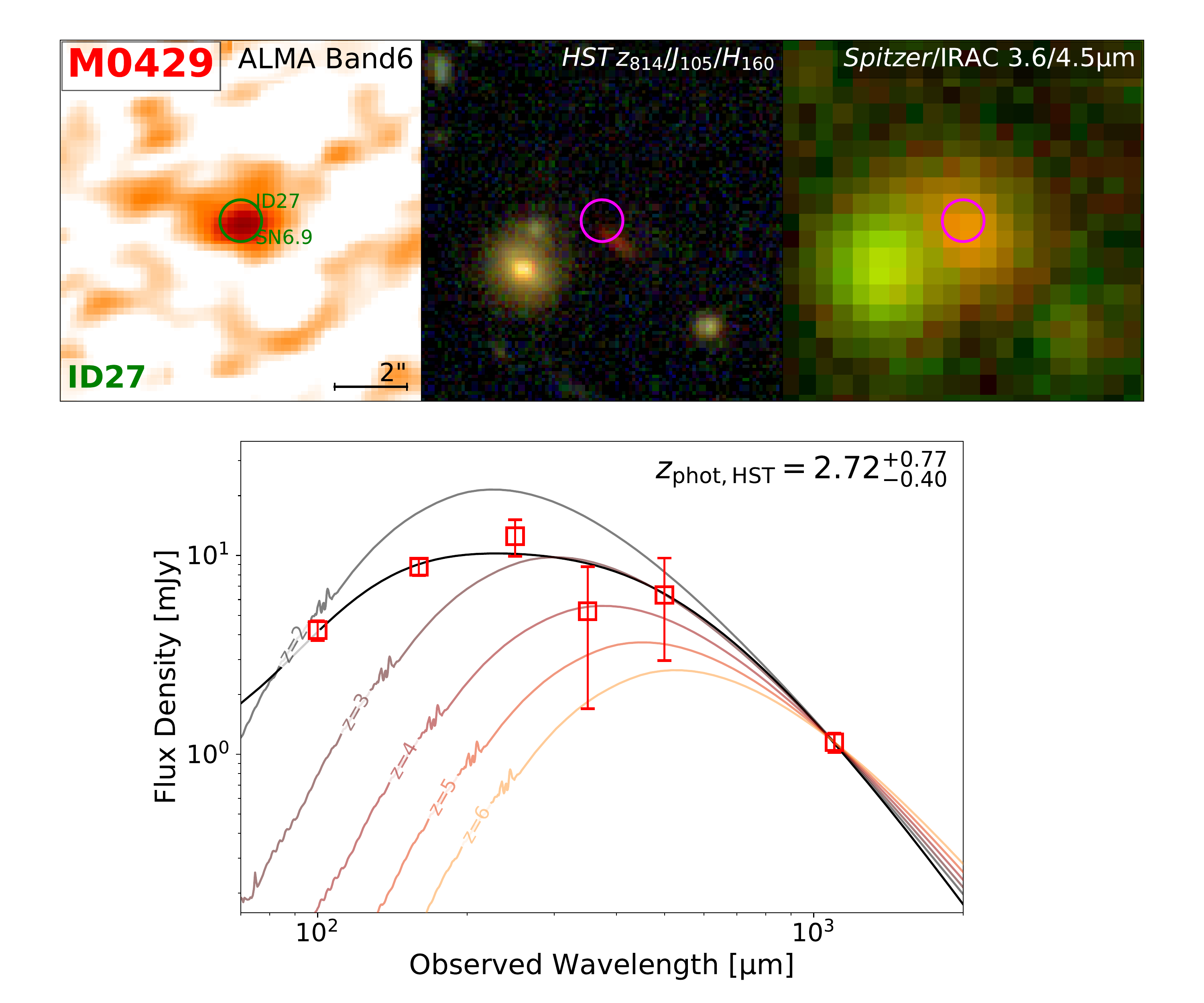}
\figsetgrpnote{Postage stamp images (top) and far-IR SED (bottom) of M0429-ID27.}
\figsetgrpend

\figsetgrpstart
\figsetgrpnum{B1.53}
\figsetgrptitle{M0553-ID17}
\figsetplot{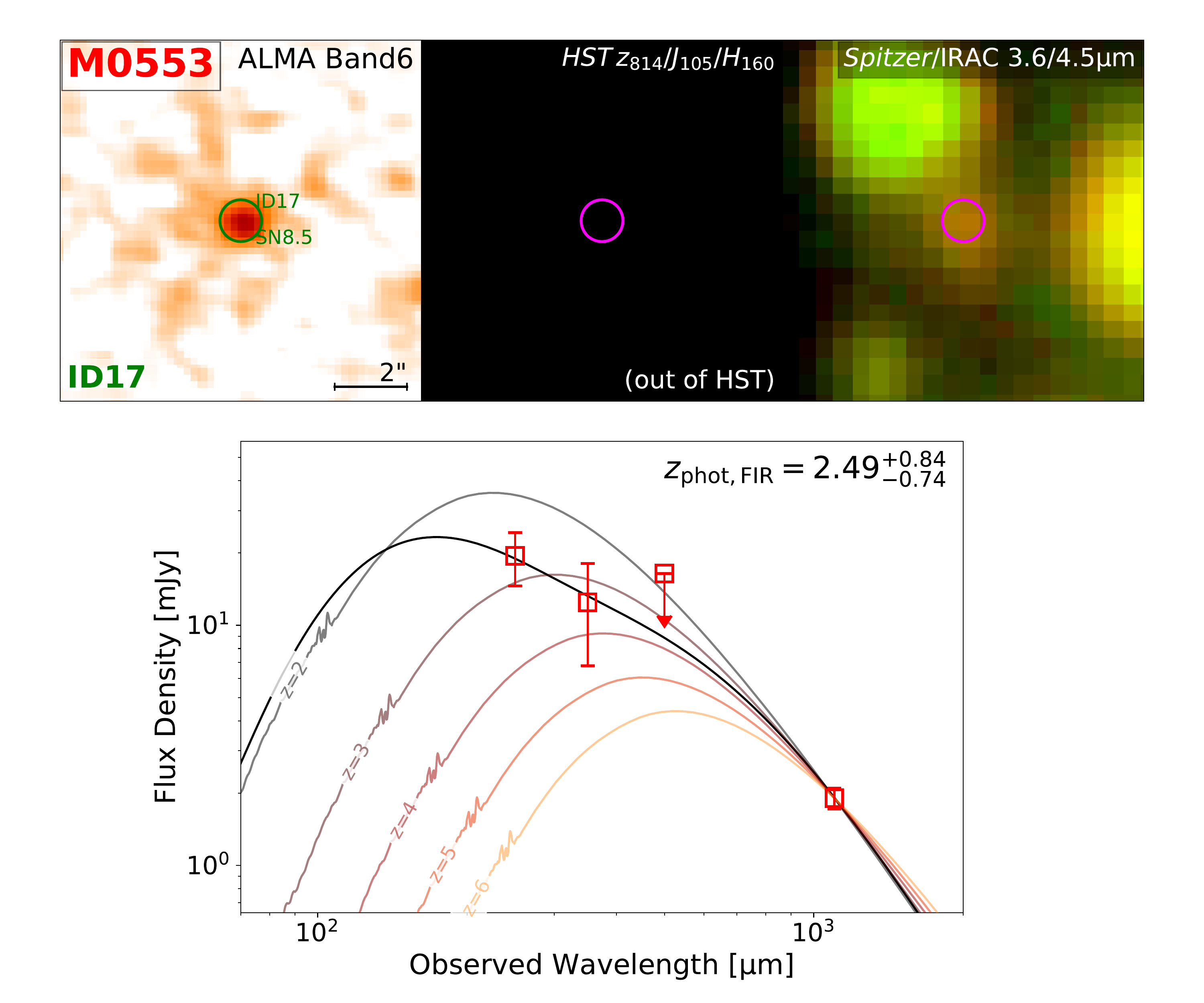}
\figsetgrpnote{Postage stamp images (top) and far-IR SED (bottom) of M0553-ID17.}
\figsetgrpend

\figsetgrpstart
\figsetgrpnum{B1.54}
\figsetgrptitle{M0553-ID61}
\figsetplot{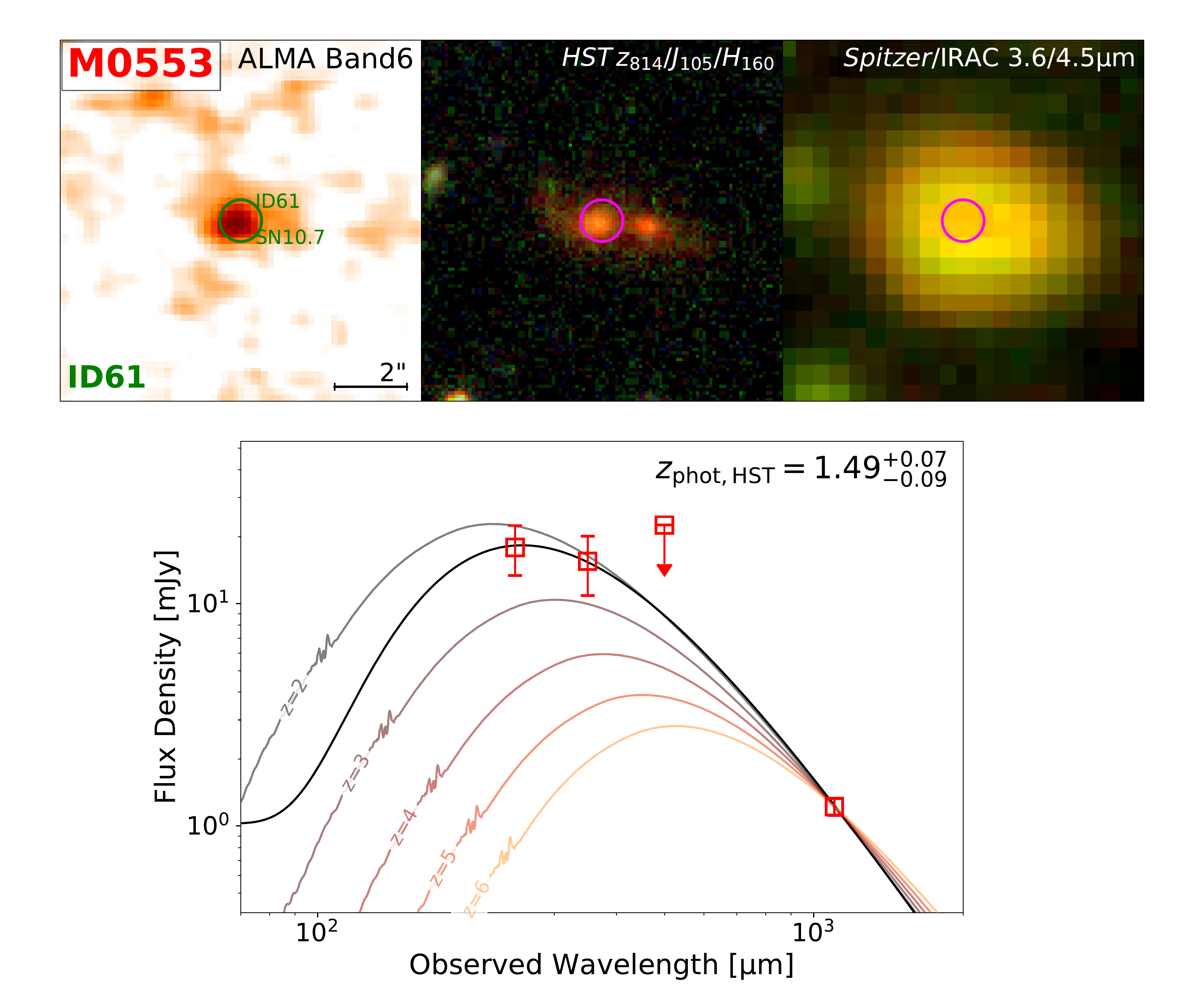}
\figsetgrpnote{Postage stamp images (top) and far-IR SED (bottom) of M0553-ID61.}
\figsetgrpend

\figsetgrpstart
\figsetgrpnum{B1.55}
\figsetgrptitle{M0553-ID133}
\figsetplot{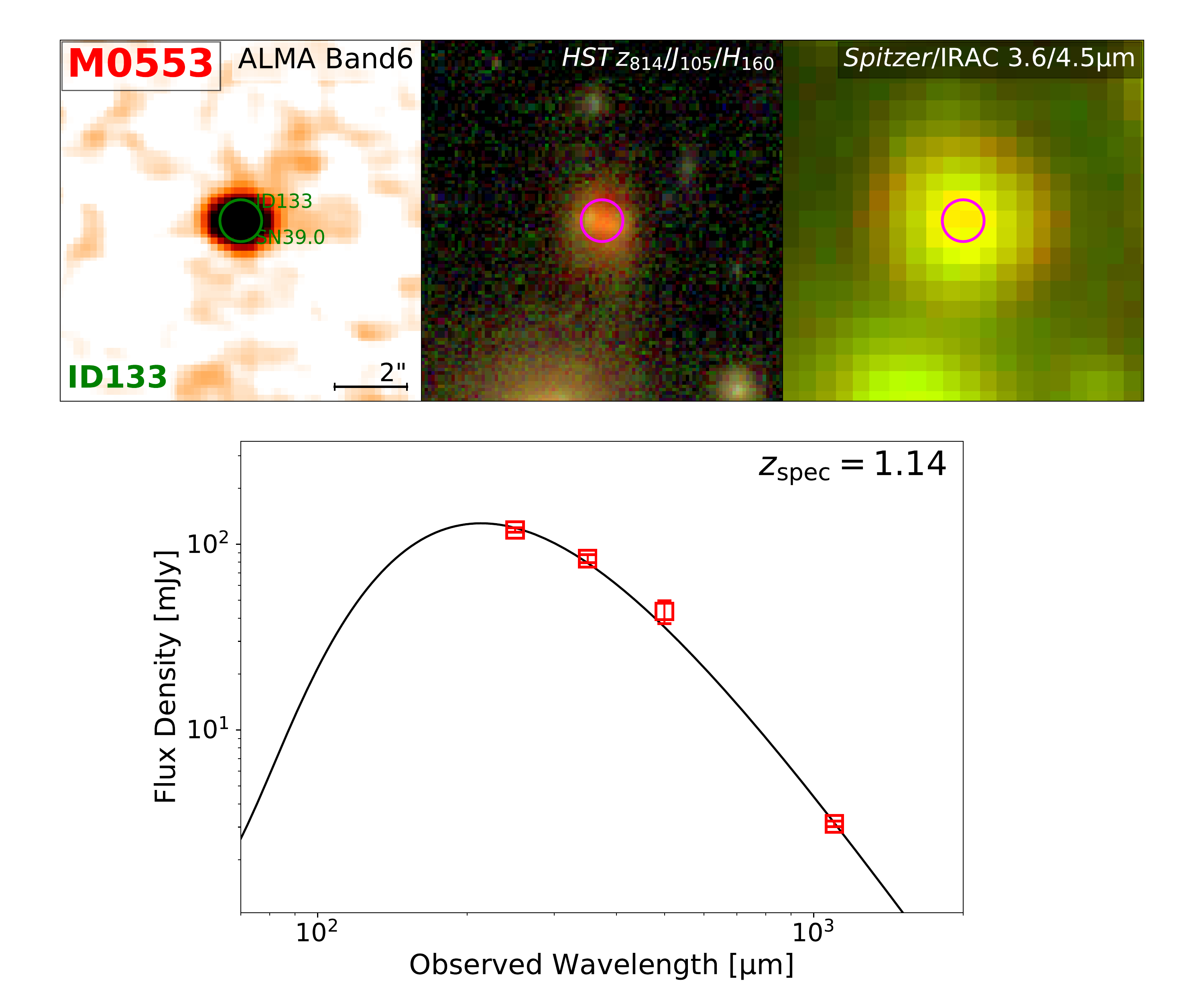}
\figsetgrpnote{Postage stamp images (top) and far-IR SED (bottom) of M0553-ID133.}
\figsetgrpend

\figsetgrpstart
\figsetgrpnum{B1.56}
\figsetgrptitle{M0553-ID190}
\figsetplot{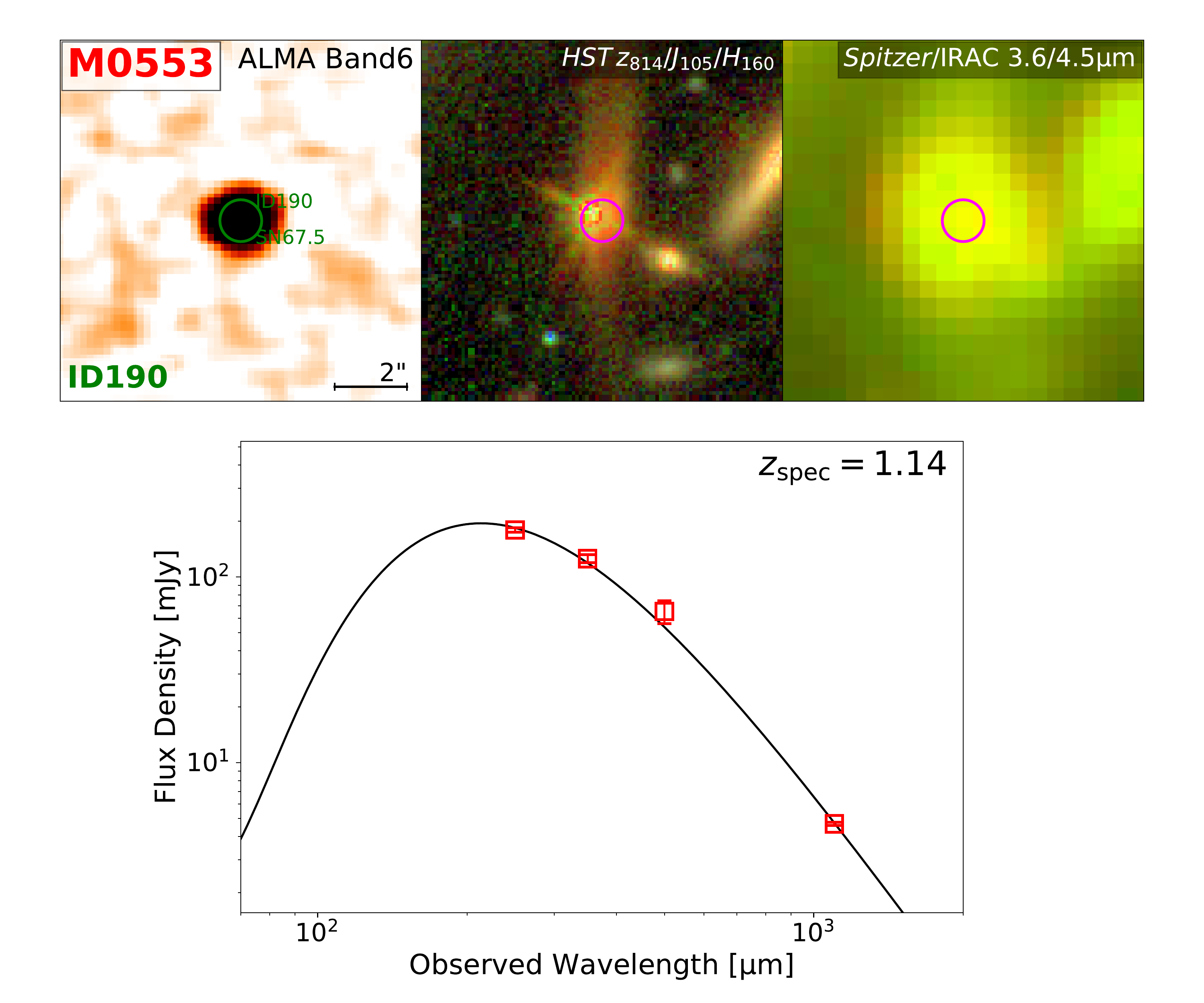}
\figsetgrpnote{Postage stamp images (top) and far-IR SED (bottom) of M0553-ID190.}
\figsetgrpend

\figsetgrpstart
\figsetgrpnum{B1.57}
\figsetgrptitle{M0553-ID200}
\figsetplot{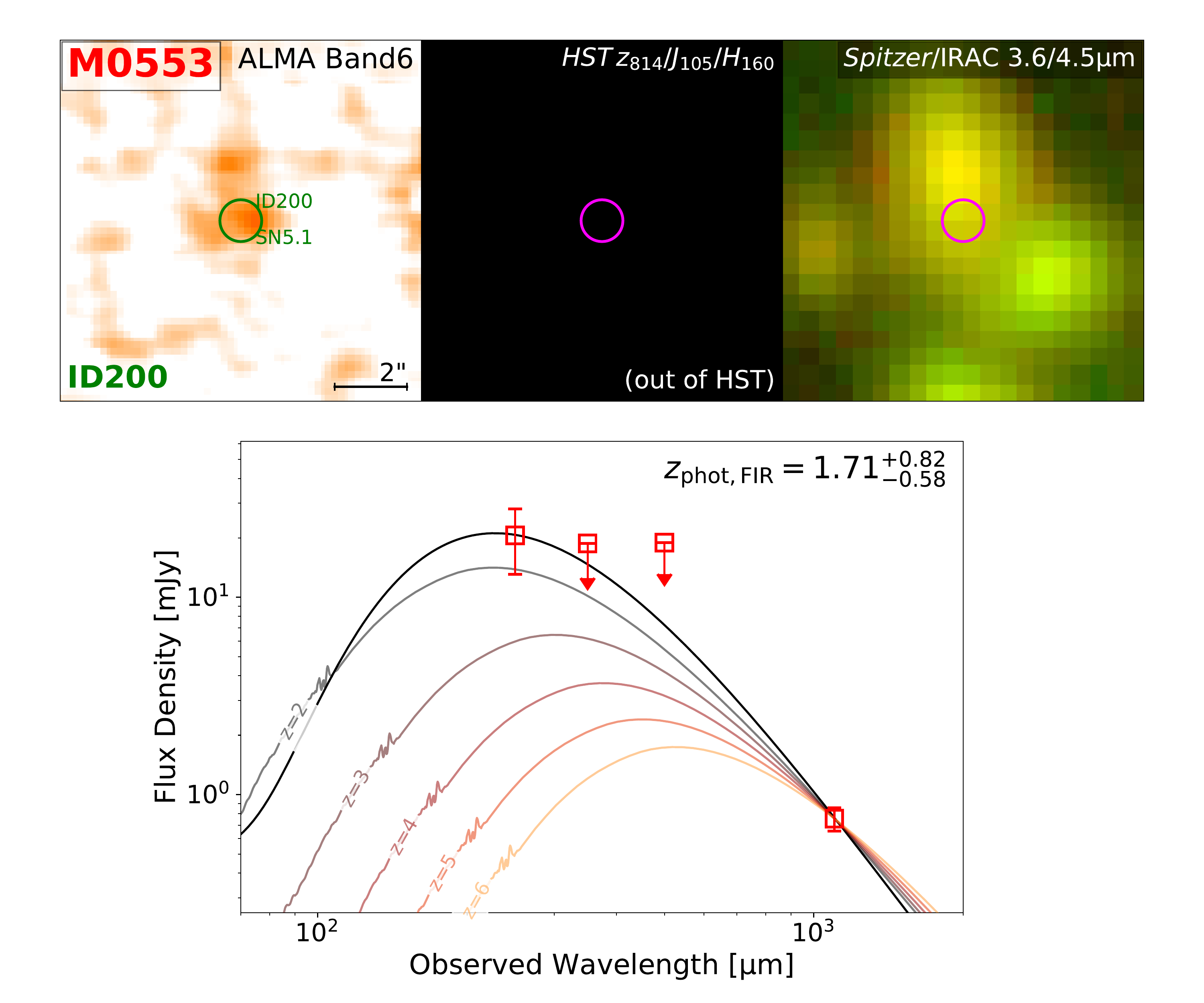}
\figsetgrpnote{Postage stamp images (top) and far-IR SED (bottom) of M0553-ID200.}
\figsetgrpend

\figsetgrpstart
\figsetgrpnum{B1.58}
\figsetgrptitle{M0553-ID249}
\figsetplot{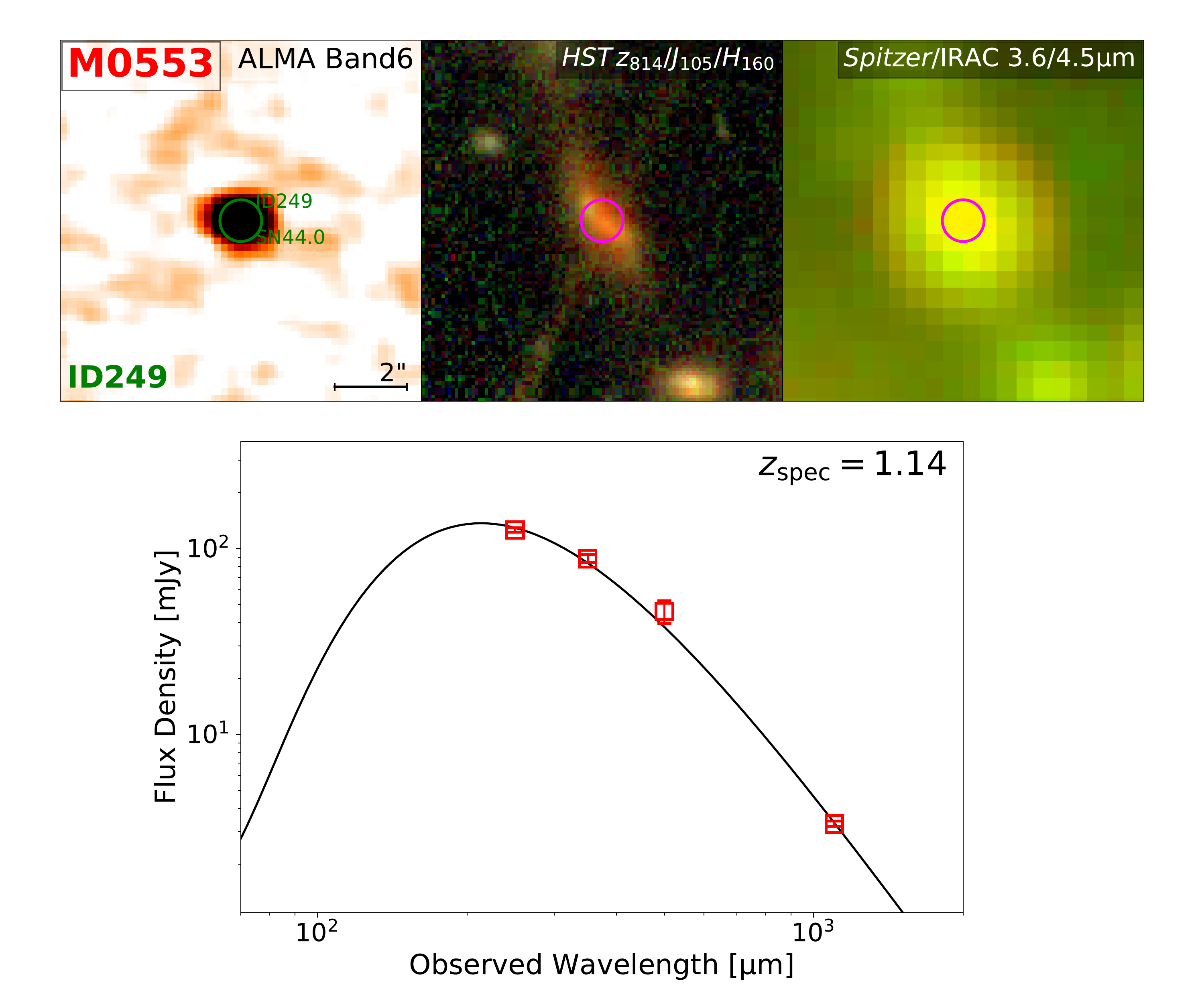}
\figsetgrpnote{Postage stamp images (top) and far-IR SED (bottom) of M0553-ID249.}
\figsetgrpend

\figsetgrpstart
\figsetgrpnum{B1.59}
\figsetgrptitle{M0553-ID275}
\figsetplot{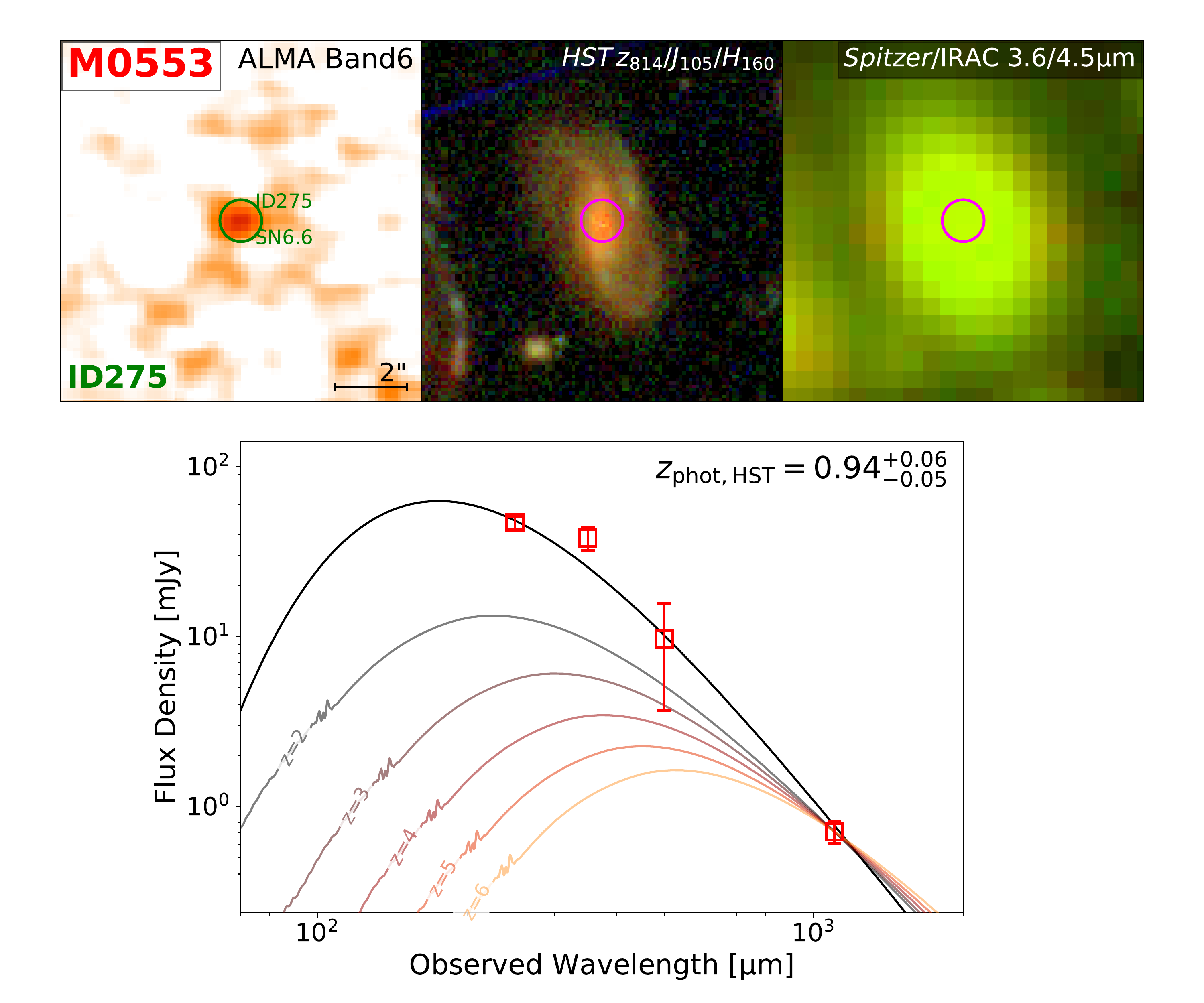}
\figsetgrpnote{Postage stamp images (top) and far-IR SED (bottom) of M0553-ID275.}
\figsetgrpend

\figsetgrpstart
\figsetgrpnum{B1.60}
\figsetgrptitle{M0553-ID303}
\figsetplot{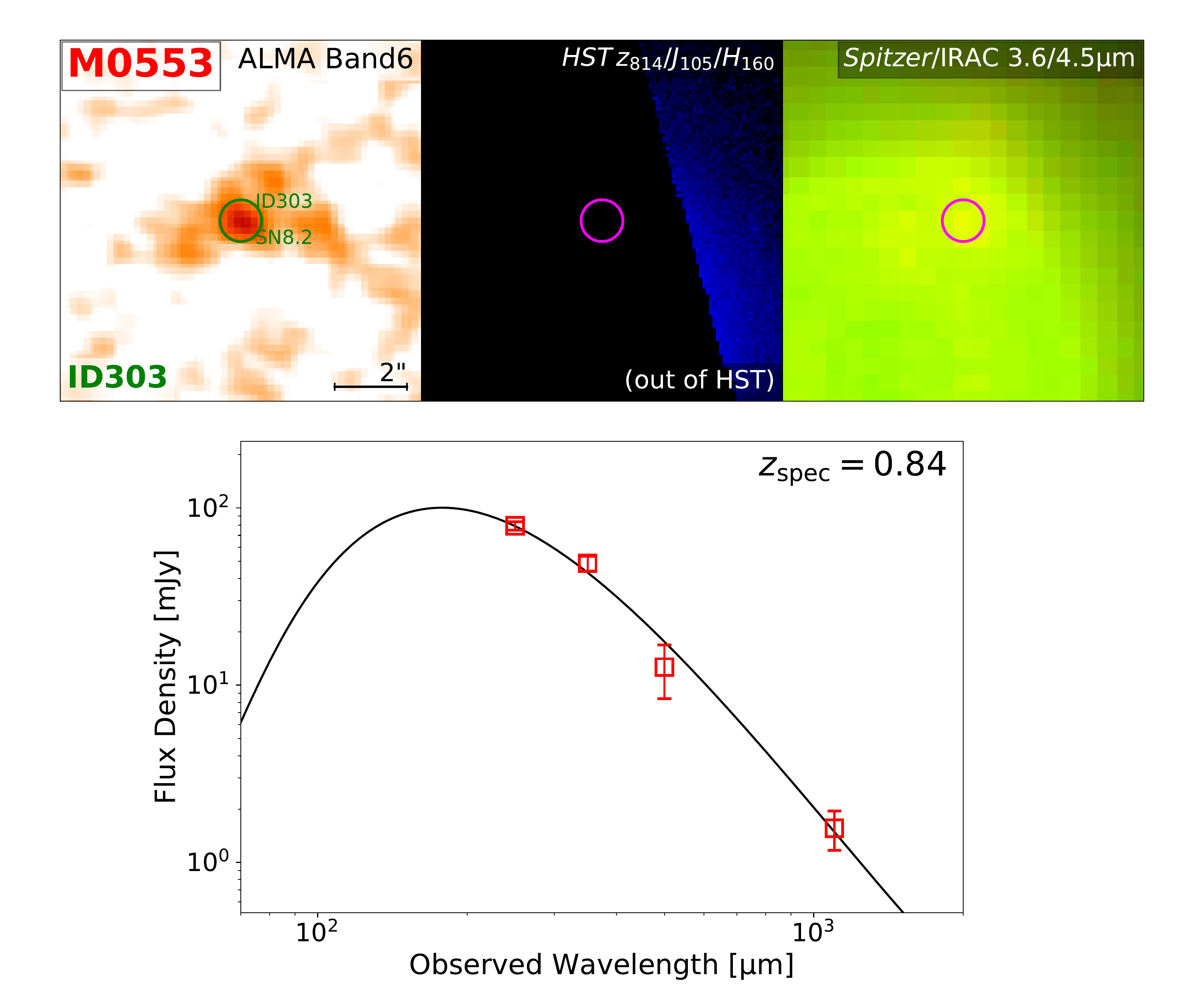}
\figsetgrpnote{Postage stamp images (top) and far-IR SED (bottom) of M0553-ID303.}
\figsetgrpend

\figsetgrpstart
\figsetgrpnum{B1.61}
\figsetgrptitle{M0553-ID355}
\figsetplot{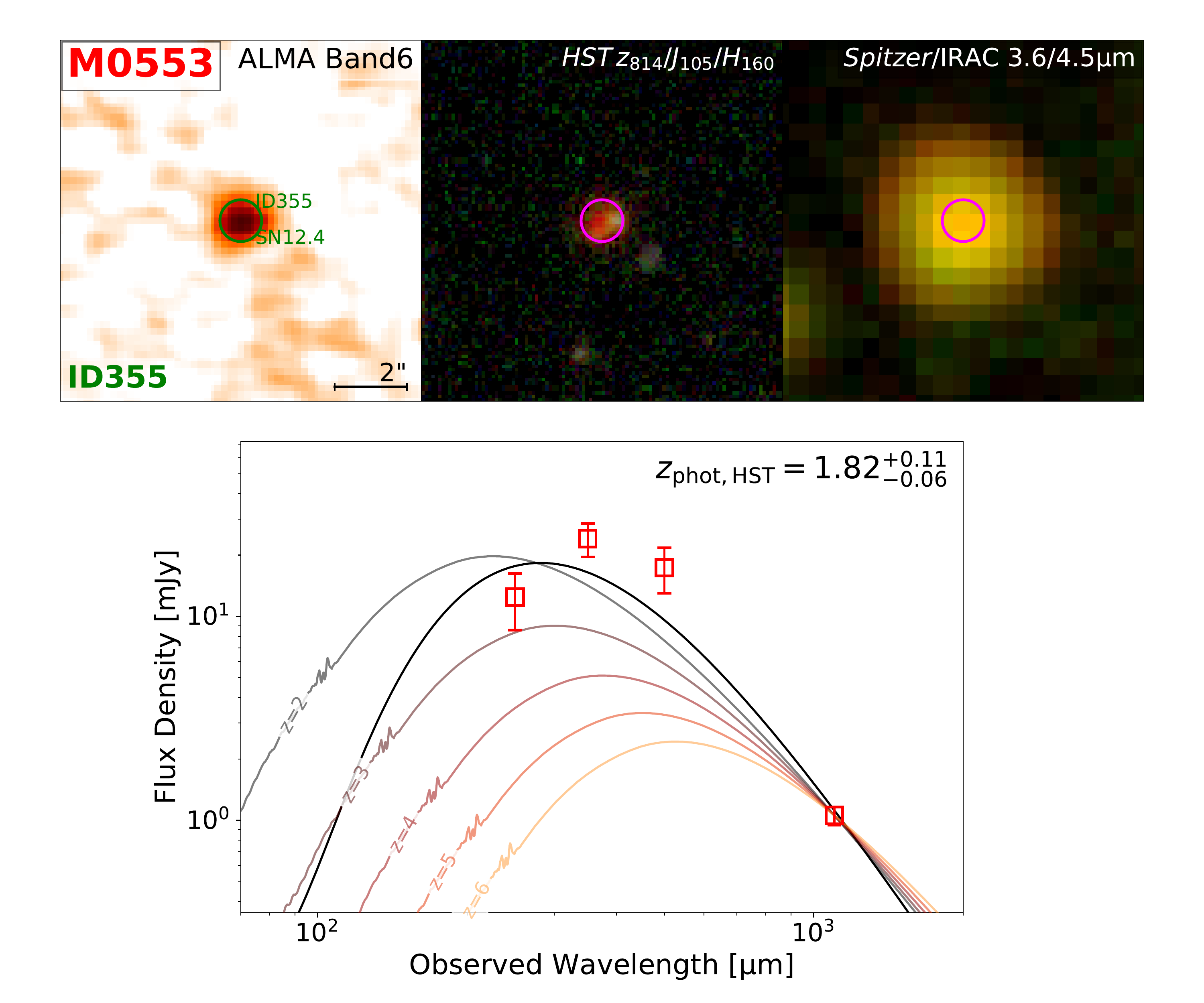}
\figsetgrpnote{Postage stamp images (top) and far-IR SED (bottom) of M0553-ID355.}
\figsetgrpend

\figsetgrpstart
\figsetgrpnum{B1.62}
\figsetgrptitle{M0553-ID375}
\figsetplot{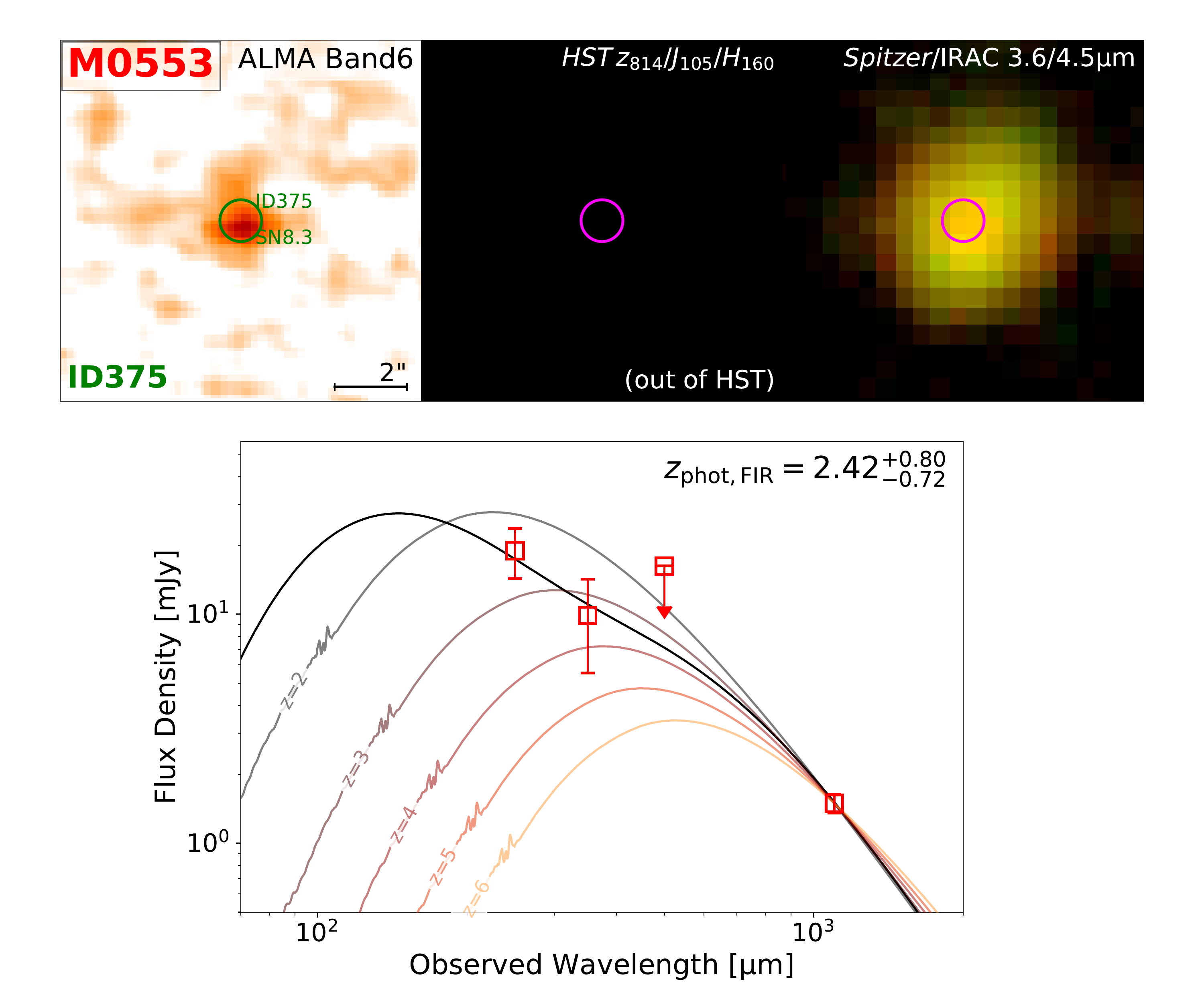}
\figsetgrpnote{Postage stamp images (top) and far-IR SED (bottom) of M0553-ID375.}
\figsetgrpend

\figsetgrpstart
\figsetgrpnum{B1.63}
\figsetgrptitle{M0553-ID398}
\figsetplot{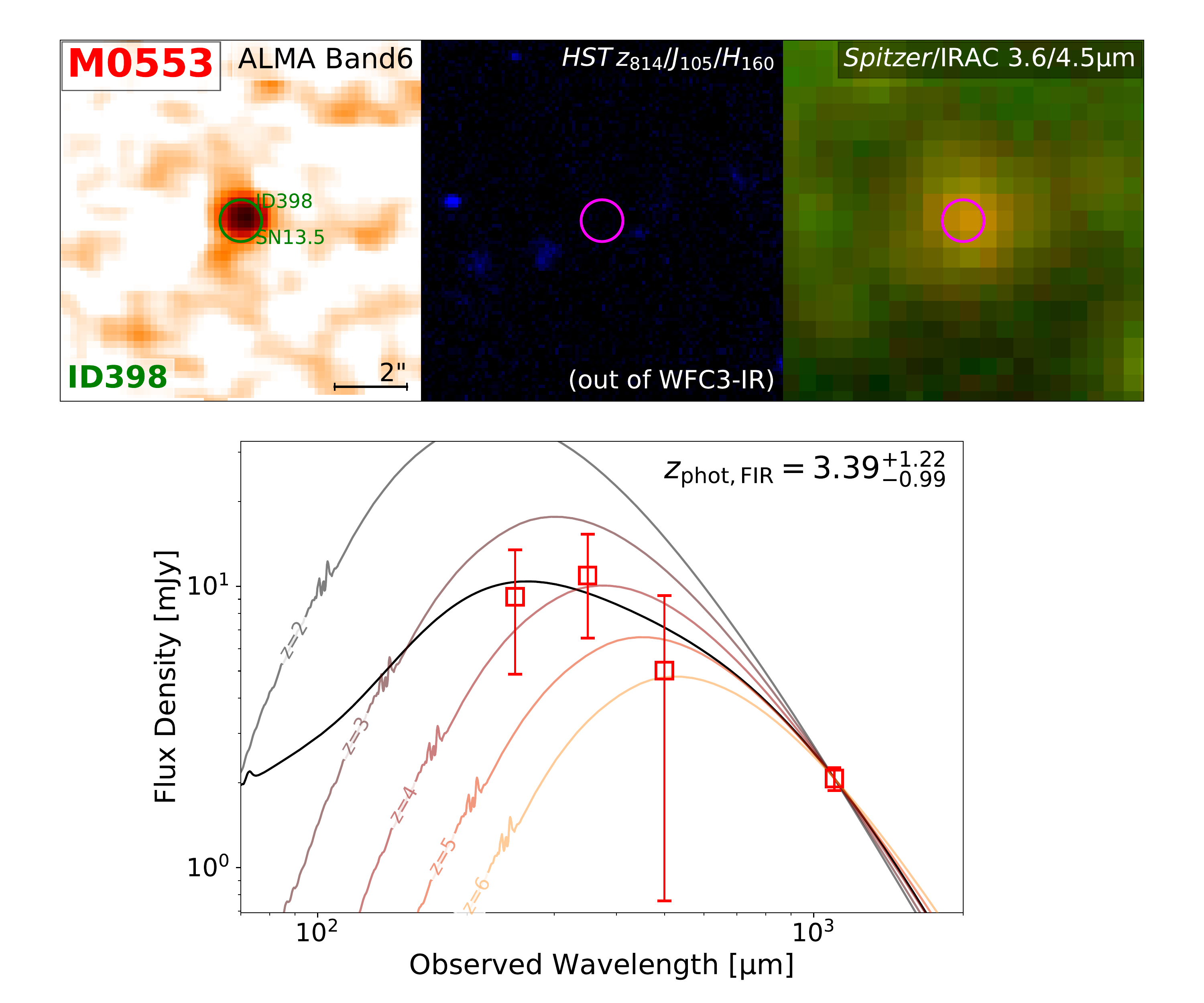}
\figsetgrpnote{Postage stamp images (top) and far-IR SED (bottom) of M0553-ID398.}
\figsetgrpend

\figsetgrpstart
\figsetgrpnum{B1.64}
\figsetgrptitle{M1115-ID02}
\figsetplot{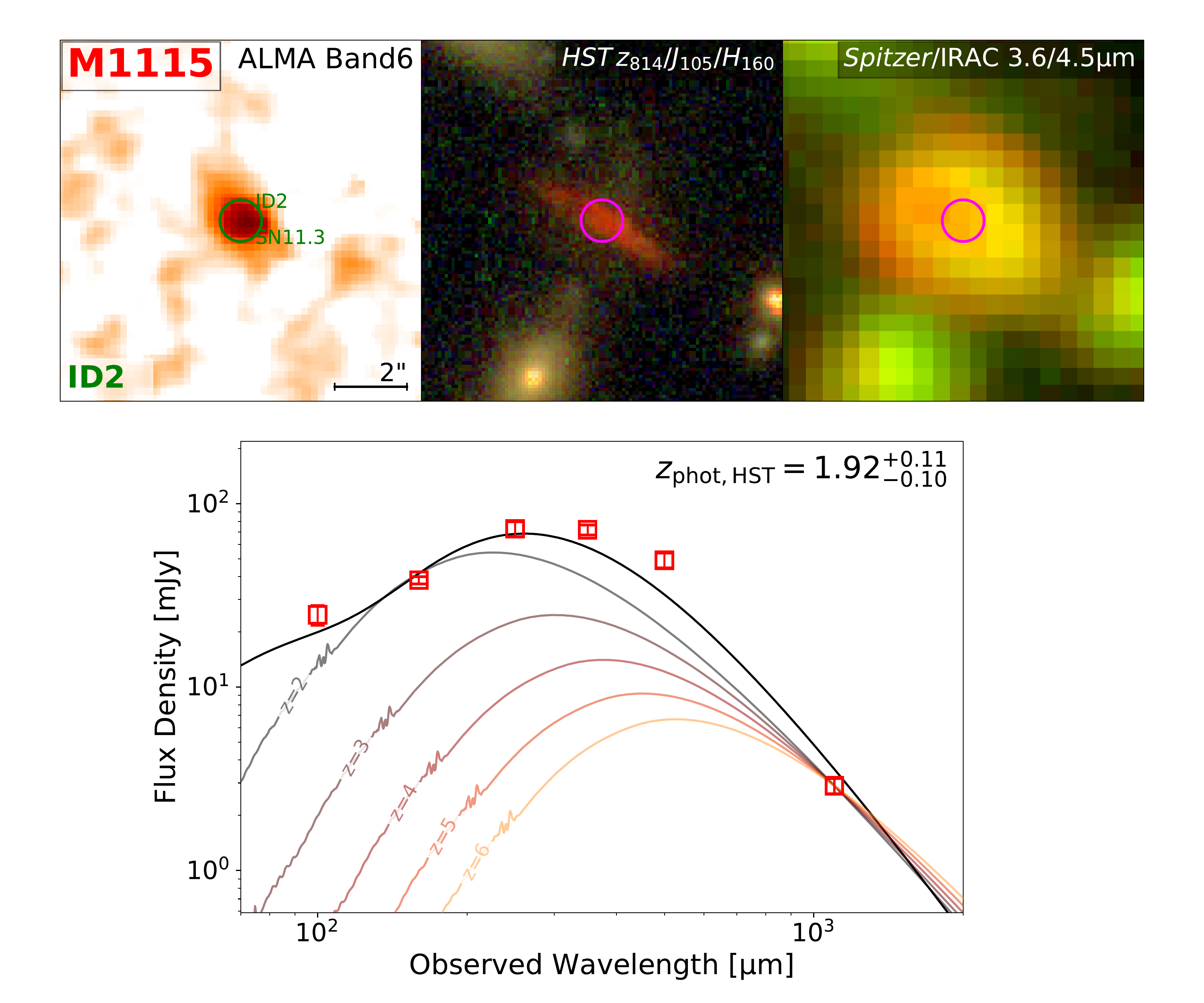}
\figsetgrpnote{Postage stamp images (top) and far-IR SED (bottom) of M1115-ID02.}
\figsetgrpend

\figsetgrpstart
\figsetgrpnum{B1.65}
\figsetgrptitle{M1115-ID04}
\figsetplot{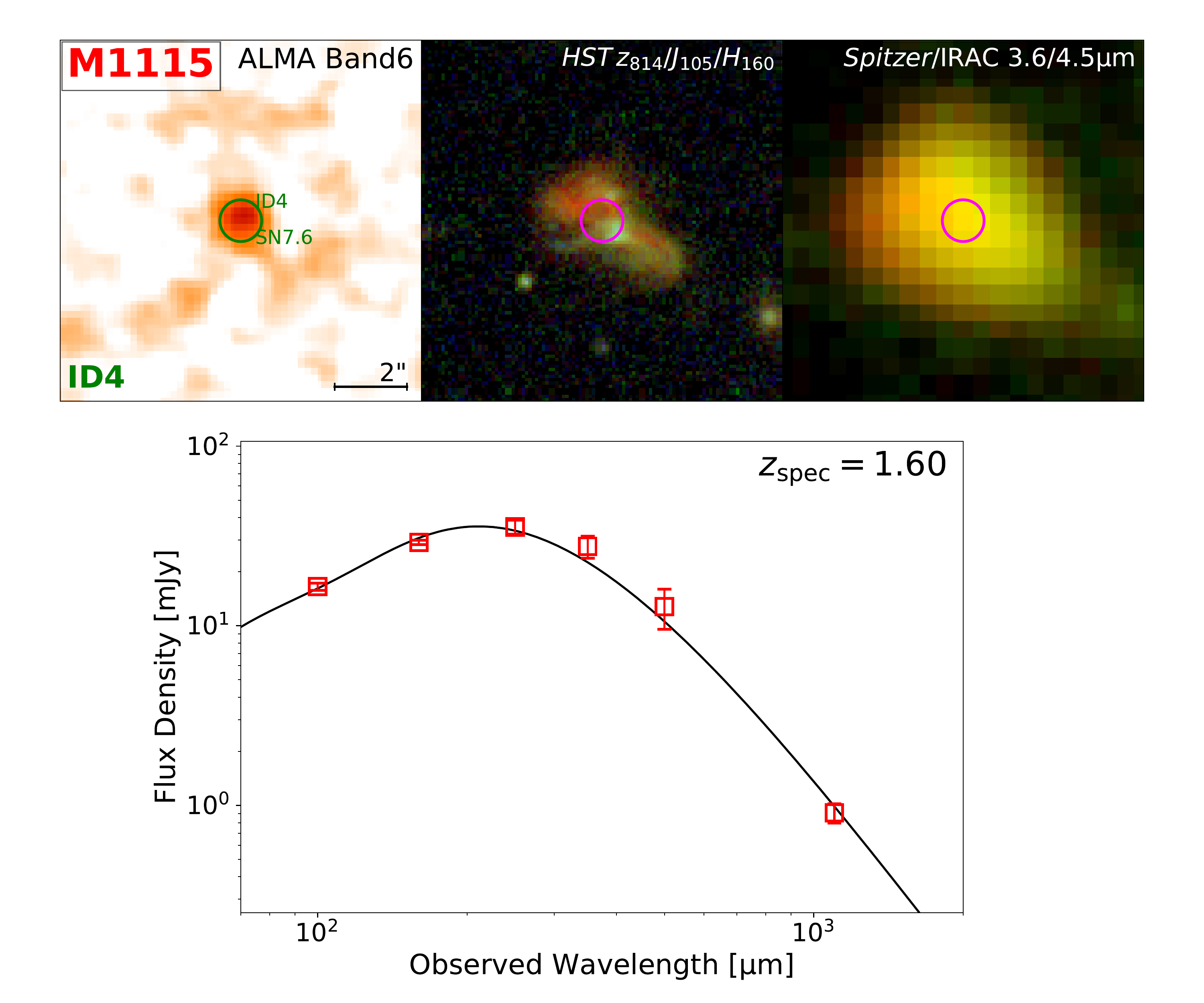}
\figsetgrpnote{Postage stamp images (top) and far-IR SED (bottom) of M1115-ID04.}
\figsetgrpend

\figsetgrpstart
\figsetgrpnum{B1.66}
\figsetgrptitle{M1115-ID34}
\figsetplot{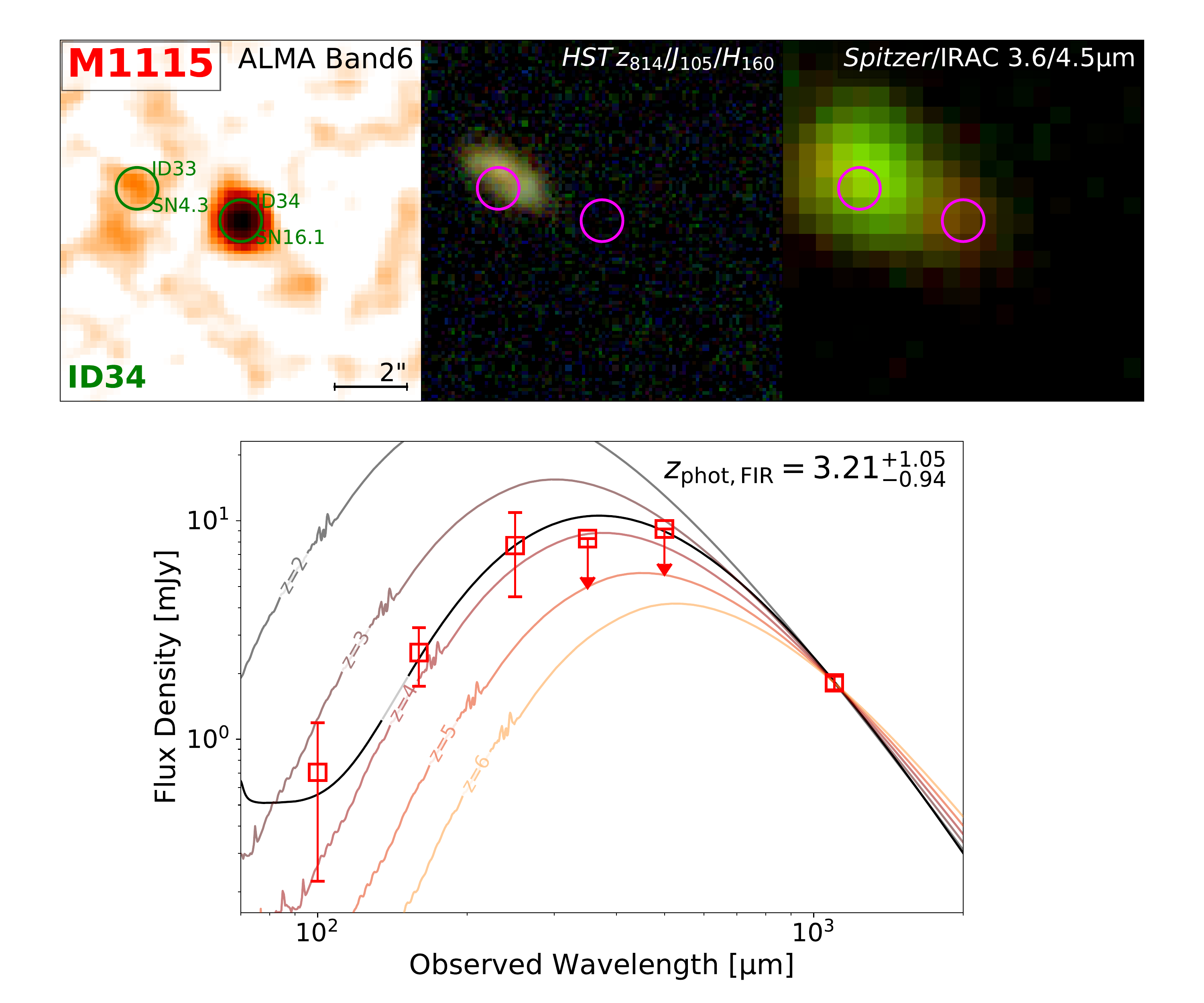}
\figsetgrpnote{Postage stamp images (top) and far-IR SED (bottom) of M1115-ID34.}
\figsetgrpend

\figsetgrpstart
\figsetgrpnum{B1.67}
\figsetgrptitle{M1149-ID77}
\figsetplot{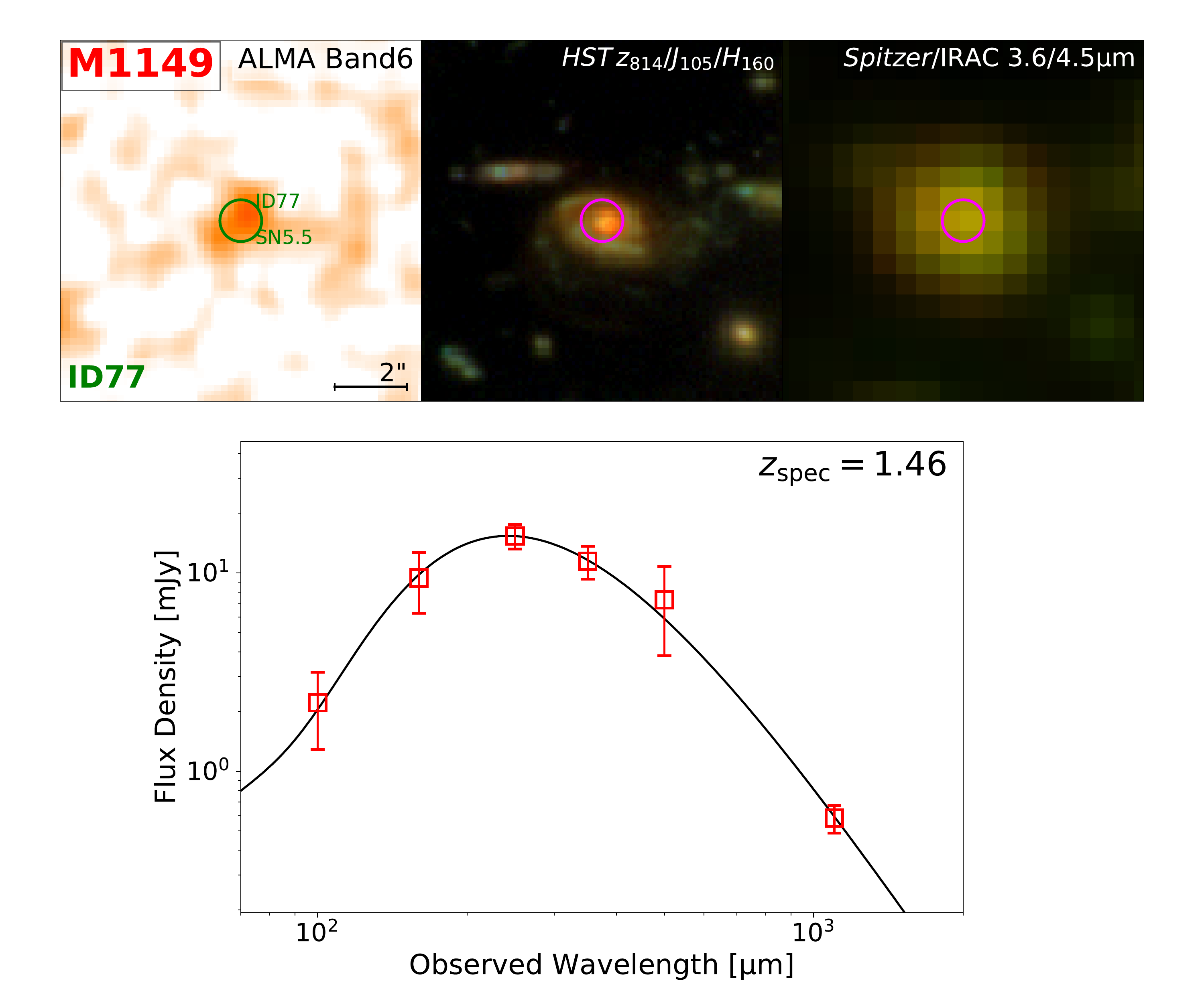}
\figsetgrpnote{Postage stamp images (top) and far-IR SED (bottom) of M1149-ID77.}
\figsetgrpend

\figsetgrpstart
\figsetgrpnum{B1.68}
\figsetgrptitle{M1149-ID229}
\figsetplot{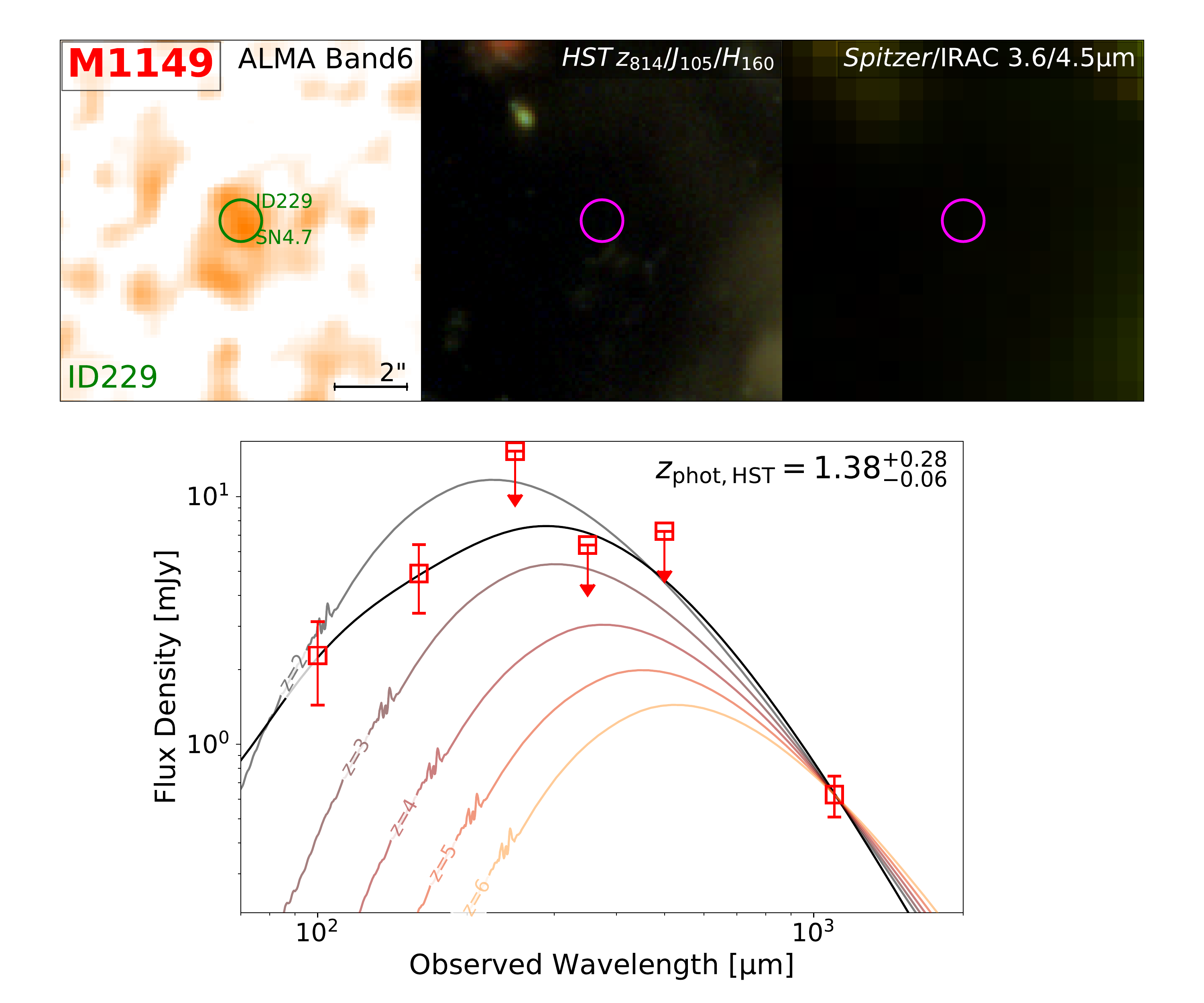}
\figsetgrpnote{Postage stamp images (top) and far-IR SED (bottom) of M1149-ID229.}
\figsetgrpend

\figsetgrpstart
\figsetgrpnum{B1.69}
\figsetgrptitle{M1206-ID27}
\figsetplot{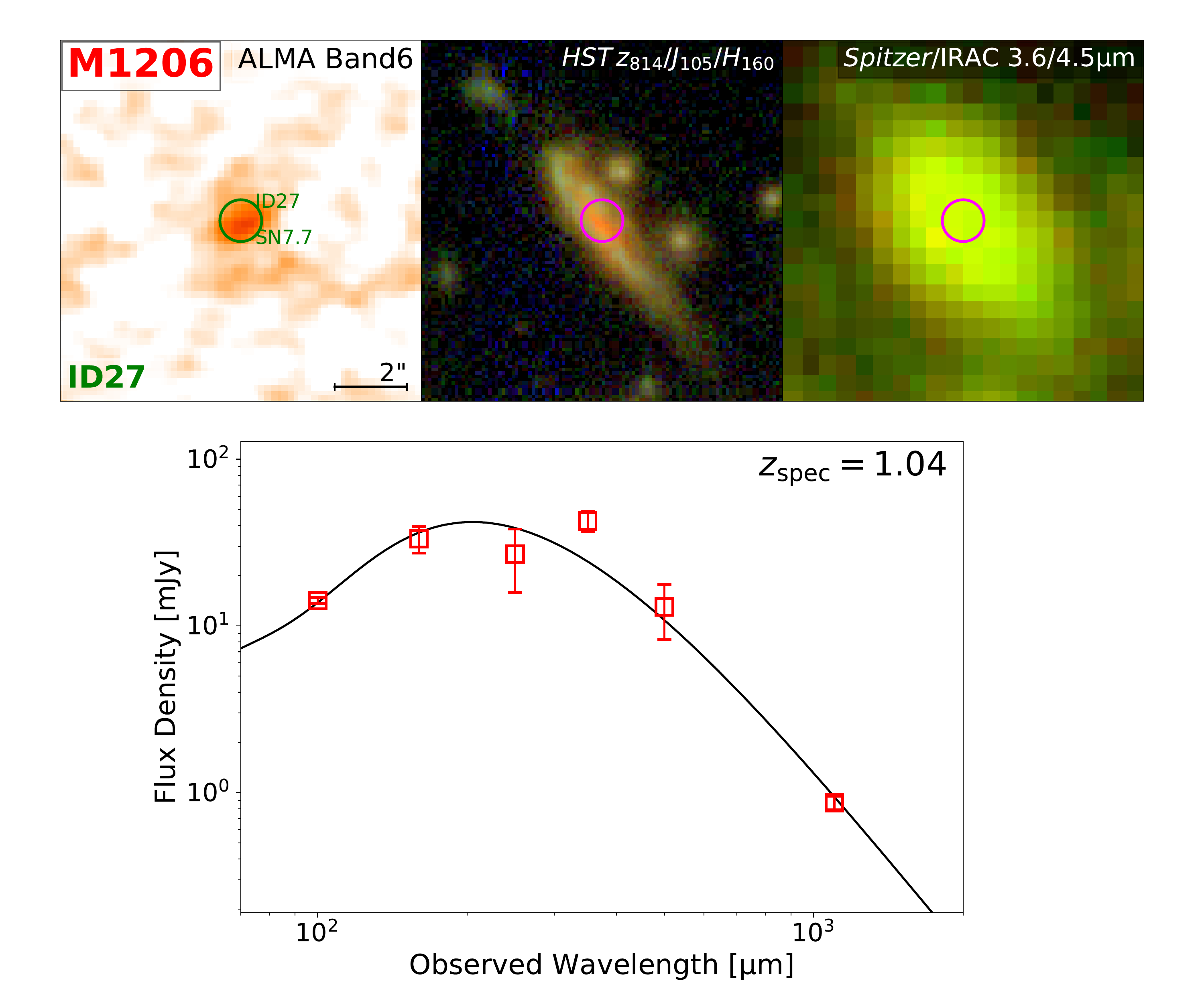}
\figsetgrpnote{Postage stamp images (top) and far-IR SED (bottom) of M1206-ID27.}
\figsetgrpend

\figsetgrpstart
\figsetgrpnum{B1.70}
\figsetgrptitle{M1206-ID55}
\figsetplot{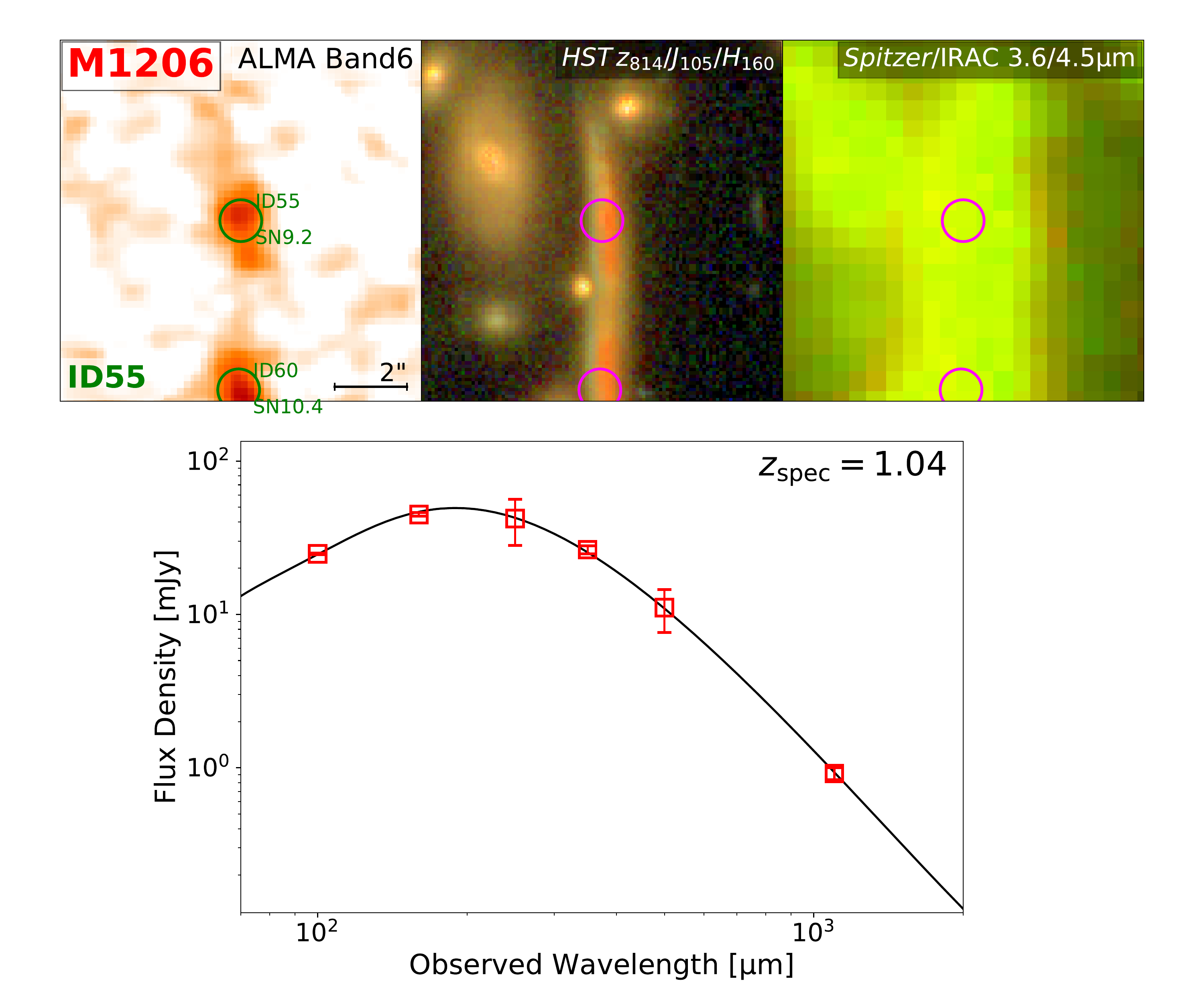}
\figsetgrpnote{Postage stamp images (top) and far-IR SED (bottom) of M1206-ID55.}
\figsetgrpend

\figsetgrpstart
\figsetgrpnum{B1.71}
\figsetgrptitle{M1206-ID60}
\figsetplot{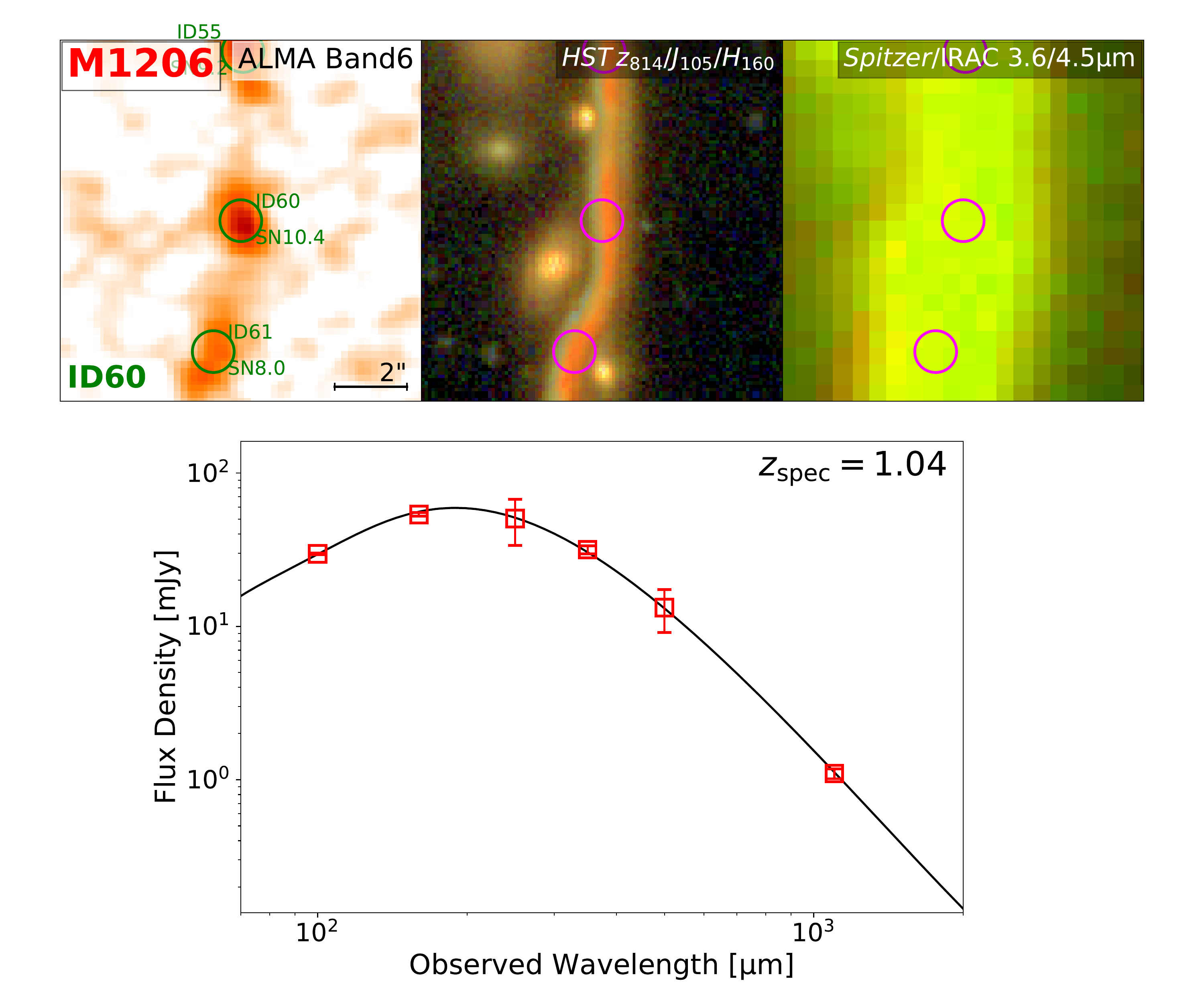}
\figsetgrpnote{Postage stamp images (top) and far-IR SED (bottom) of M1206-ID60.}
\figsetgrpend

\figsetgrpstart
\figsetgrpnum{B1.72}
\figsetgrptitle{M1206-ID61}
\figsetplot{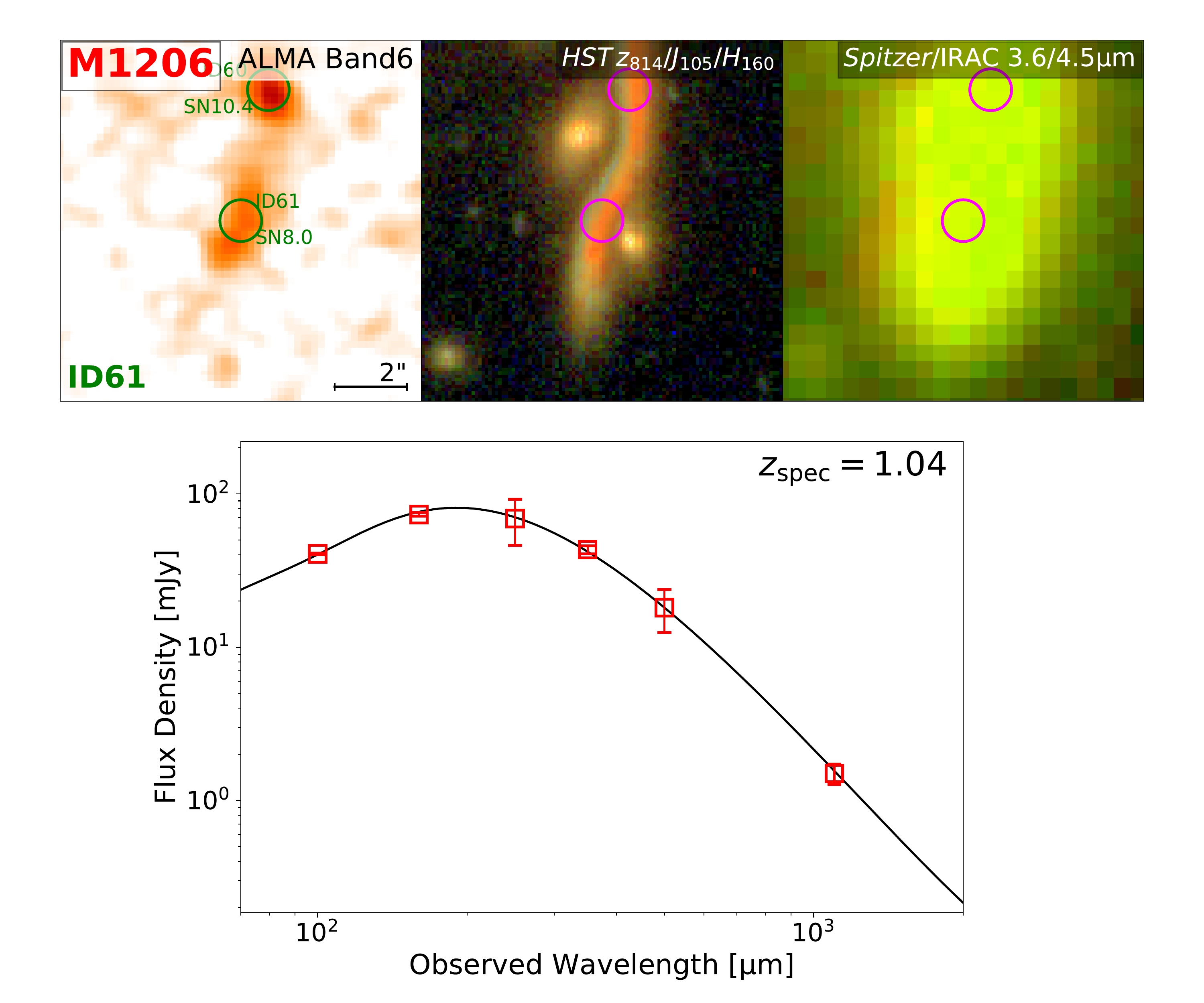}
\figsetgrpnote{Postage stamp images (top) and far-IR SED (bottom) of M1206-ID61.}
\figsetgrpend

\figsetgrpstart
\figsetgrpnum{B1.73}
\figsetgrptitle{M1311-ID27}
\figsetplot{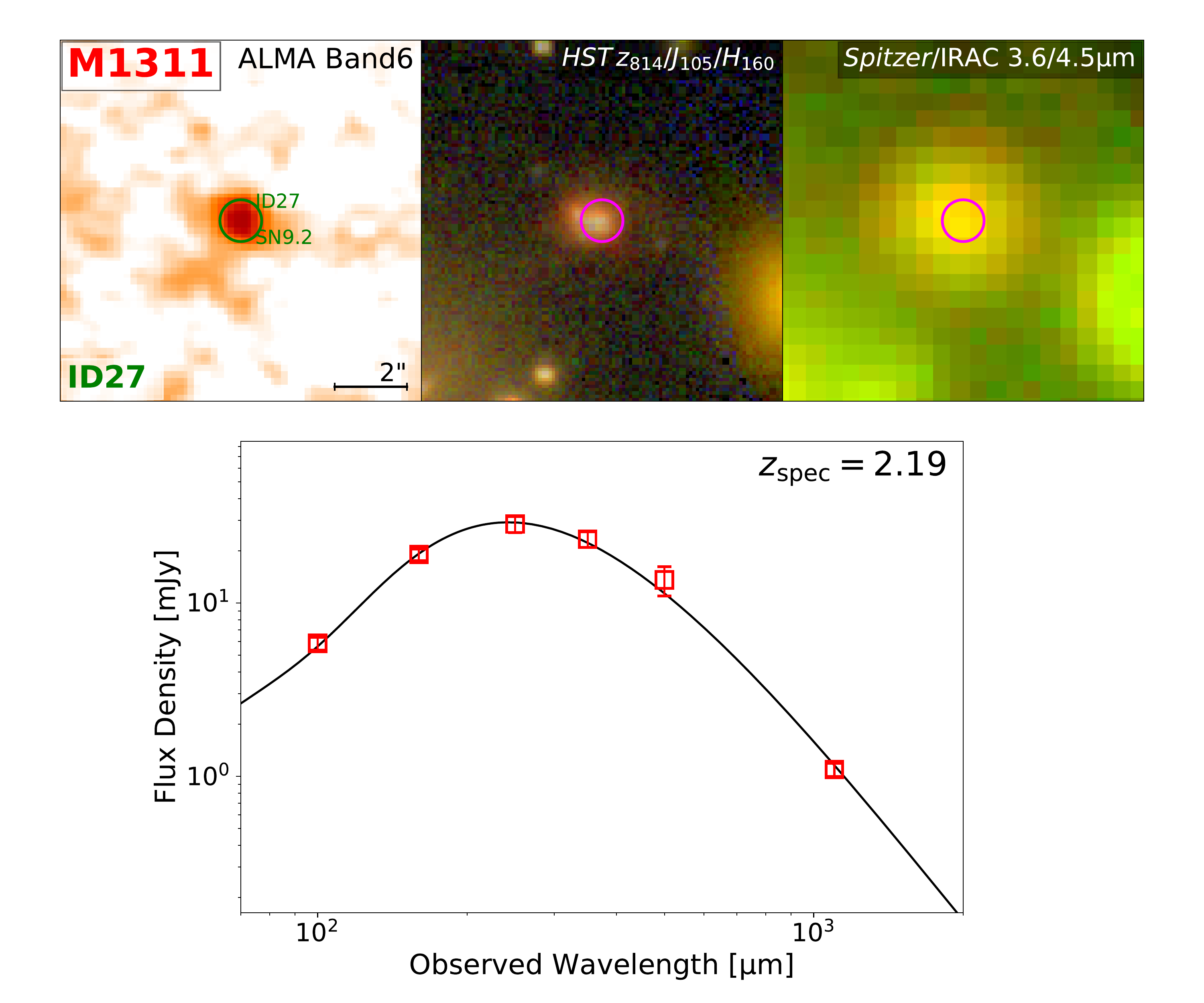}
\figsetgrpnote{Postage stamp images (top) and far-IR SED (bottom) of M1311-ID27.}
\figsetgrpend

\figsetgrpstart
\figsetgrpnum{B1.74}
\figsetgrptitle{M1311-ID33}
\figsetplot{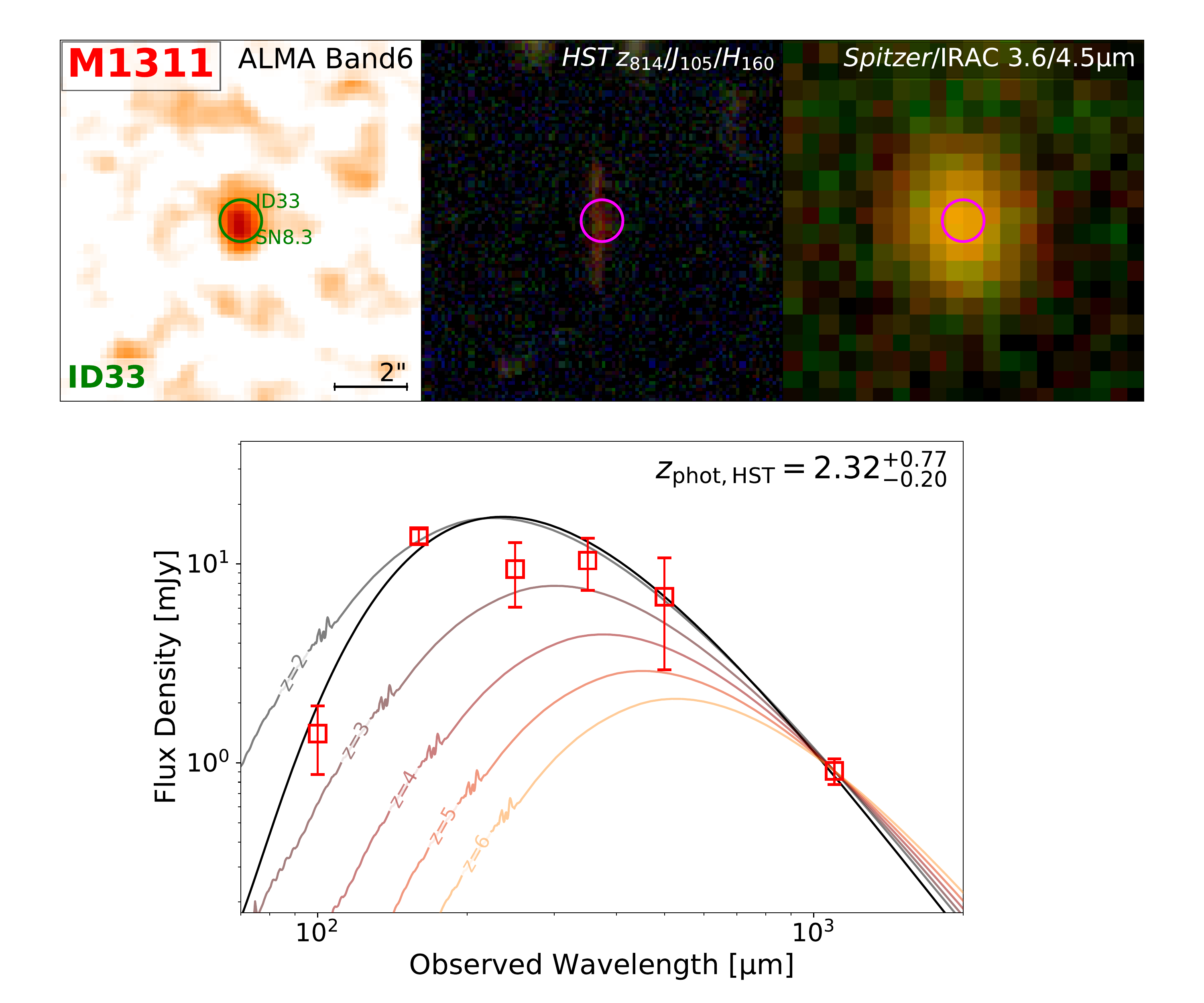}
\figsetgrpnote{Postage stamp images (top) and far-IR SED (bottom) of M1311-ID33.}
\figsetgrpend

\figsetgrpstart
\figsetgrpnum{B1.75}
\figsetgrptitle{M1423-ID38}
\figsetplot{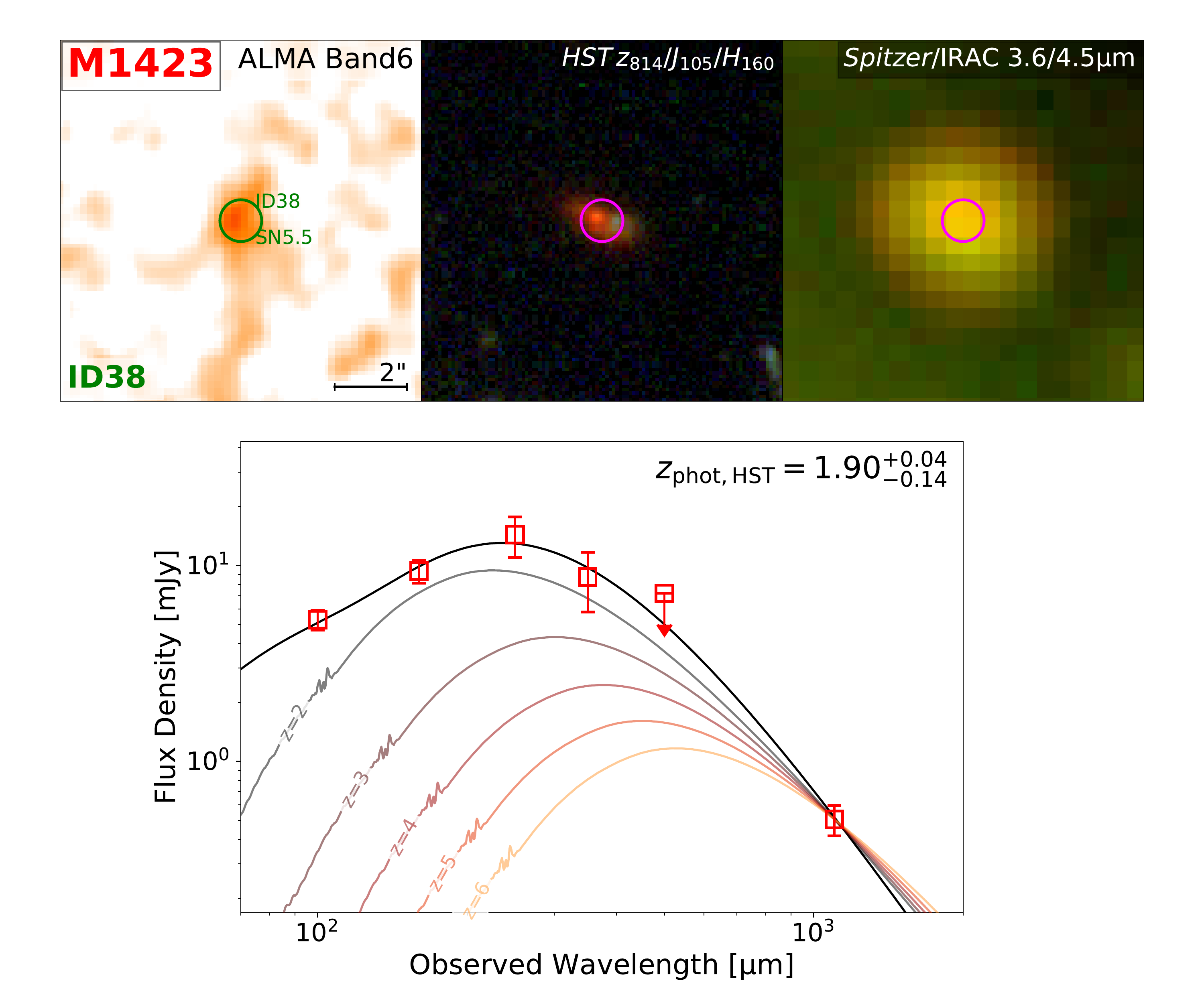}
\figsetgrpnote{Postage stamp images (top) and far-IR SED (bottom) of M1423-ID38.}
\figsetgrpend

\figsetgrpstart
\figsetgrpnum{B1.76}
\figsetgrptitle{M1931-ID55}
\figsetplot{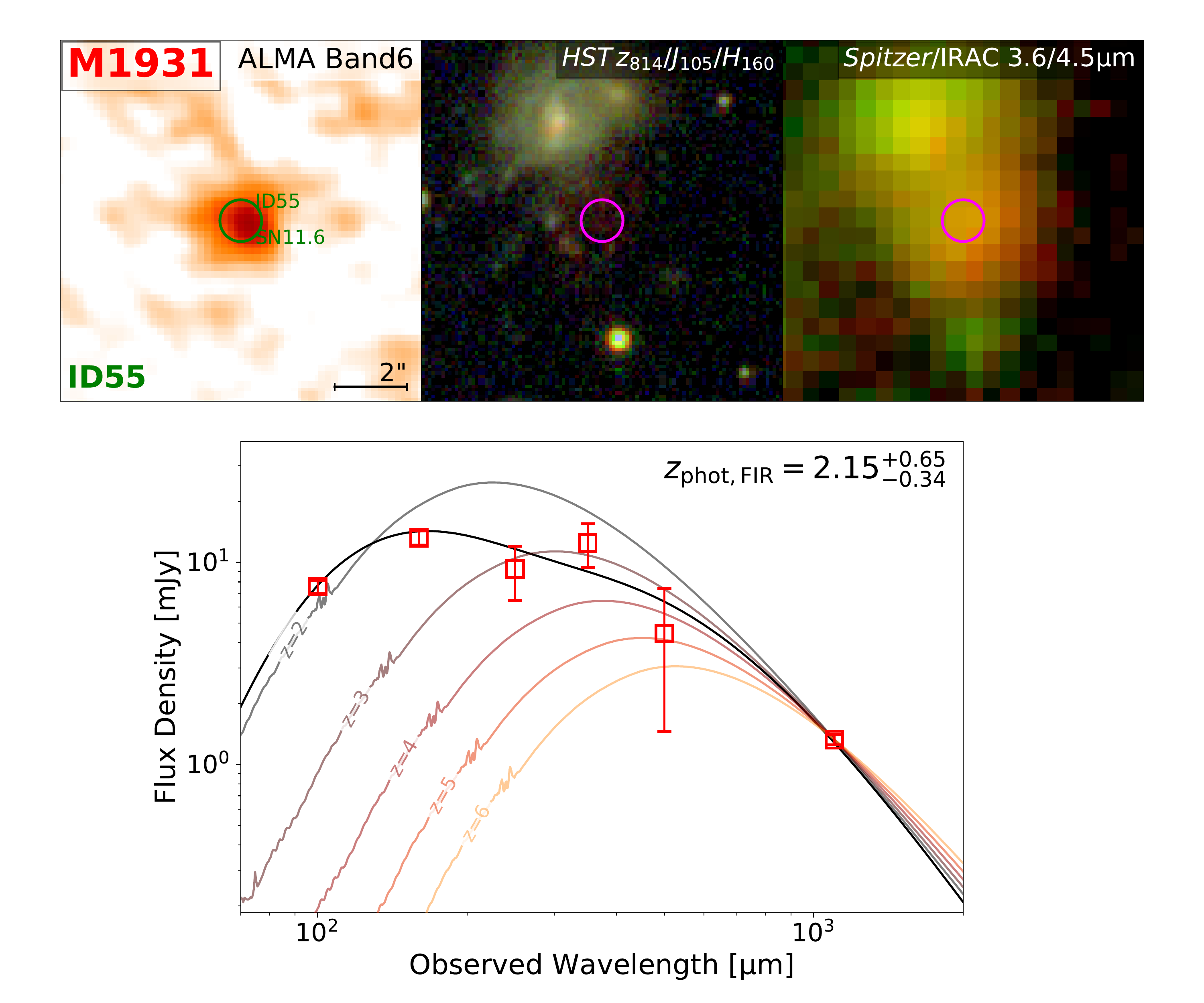}
\figsetgrpnote{Postage stamp images (top) and far-IR SED (bottom) of M1931-ID55.}
\figsetgrpend

\figsetgrpstart
\figsetgrpnum{B1.77}
\figsetgrptitle{M1931-ID61}
\figsetplot{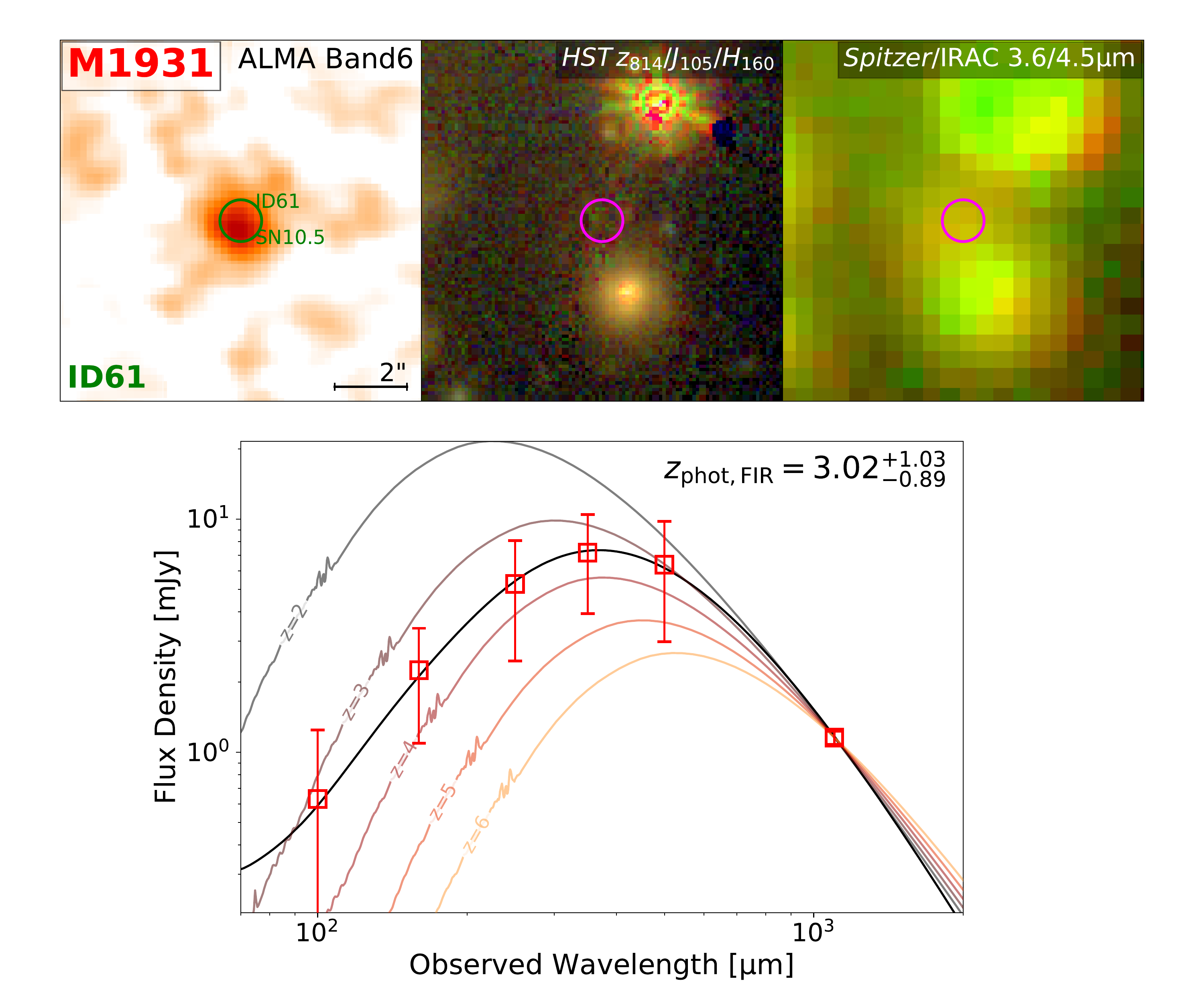}
\figsetgrpnote{Postage stamp images (top) and far-IR SED (bottom) of M1931-ID61.}
\figsetgrpend

\figsetgrpstart
\figsetgrpnum{B1.78}
\figsetgrptitle{M2129-ID46}
\figsetplot{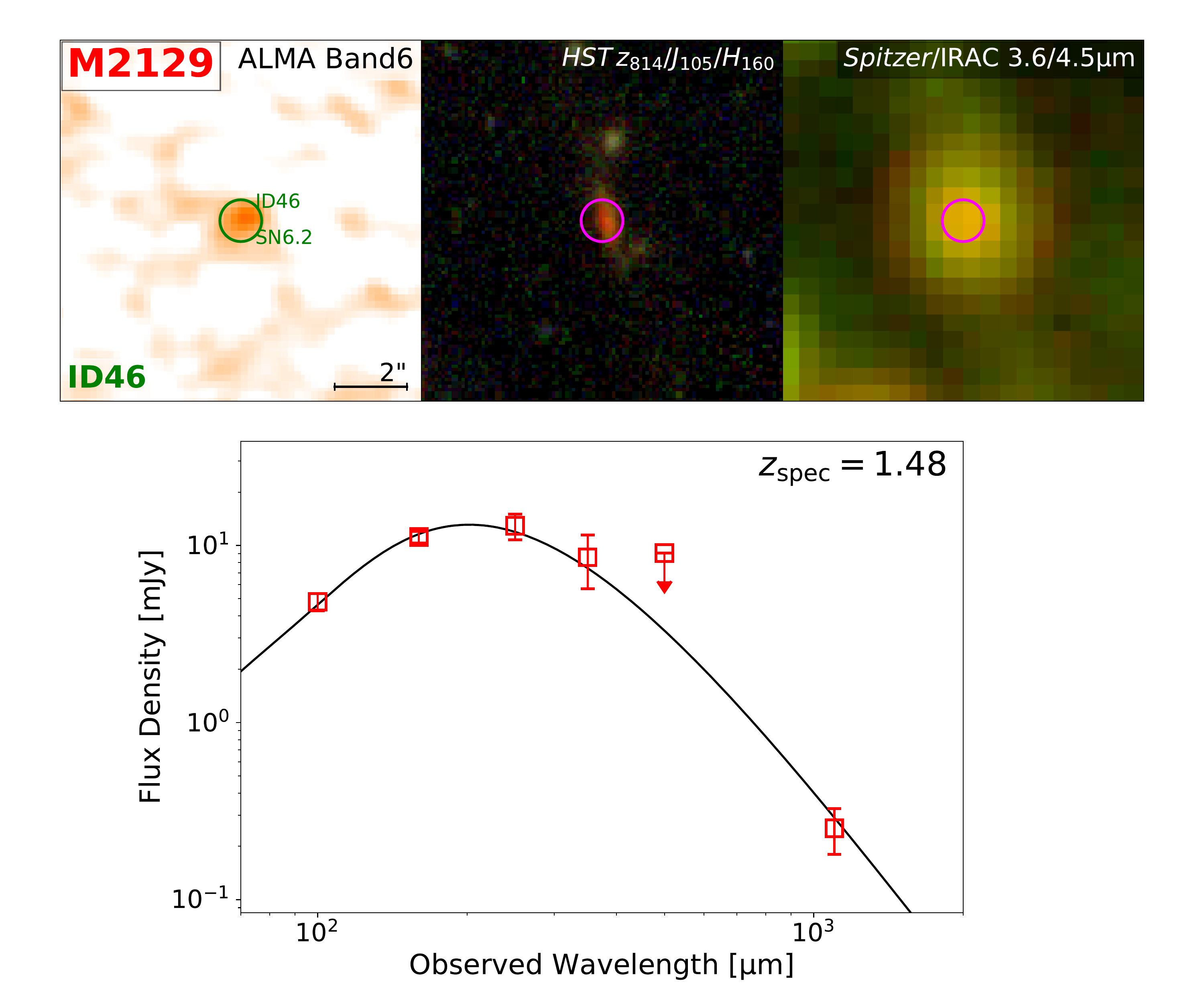}
\figsetgrpnote{Postage stamp images (top) and far-IR SED (bottom) of M2129-ID46.}
\figsetgrpend

\figsetgrpstart
\figsetgrpnum{B1.79}
\figsetgrptitle{P171-ID69}
\figsetplot{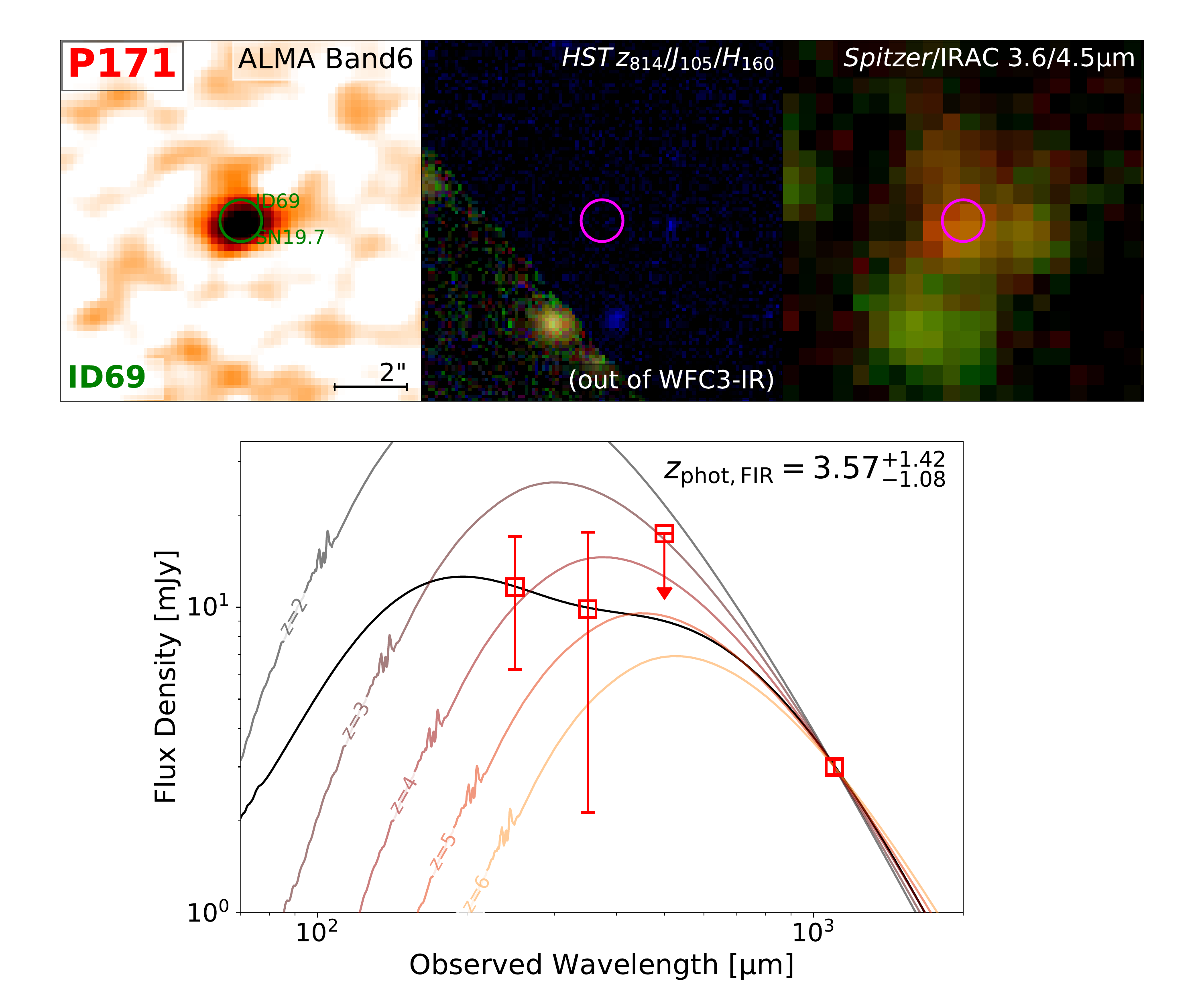}
\figsetgrpnote{Postage stamp images (top) and far-IR SED (bottom) of P171-ID69.}
\figsetgrpend

\figsetgrpstart
\figsetgrpnum{B1.80}
\figsetgrptitle{P171-ID177}
\figsetplot{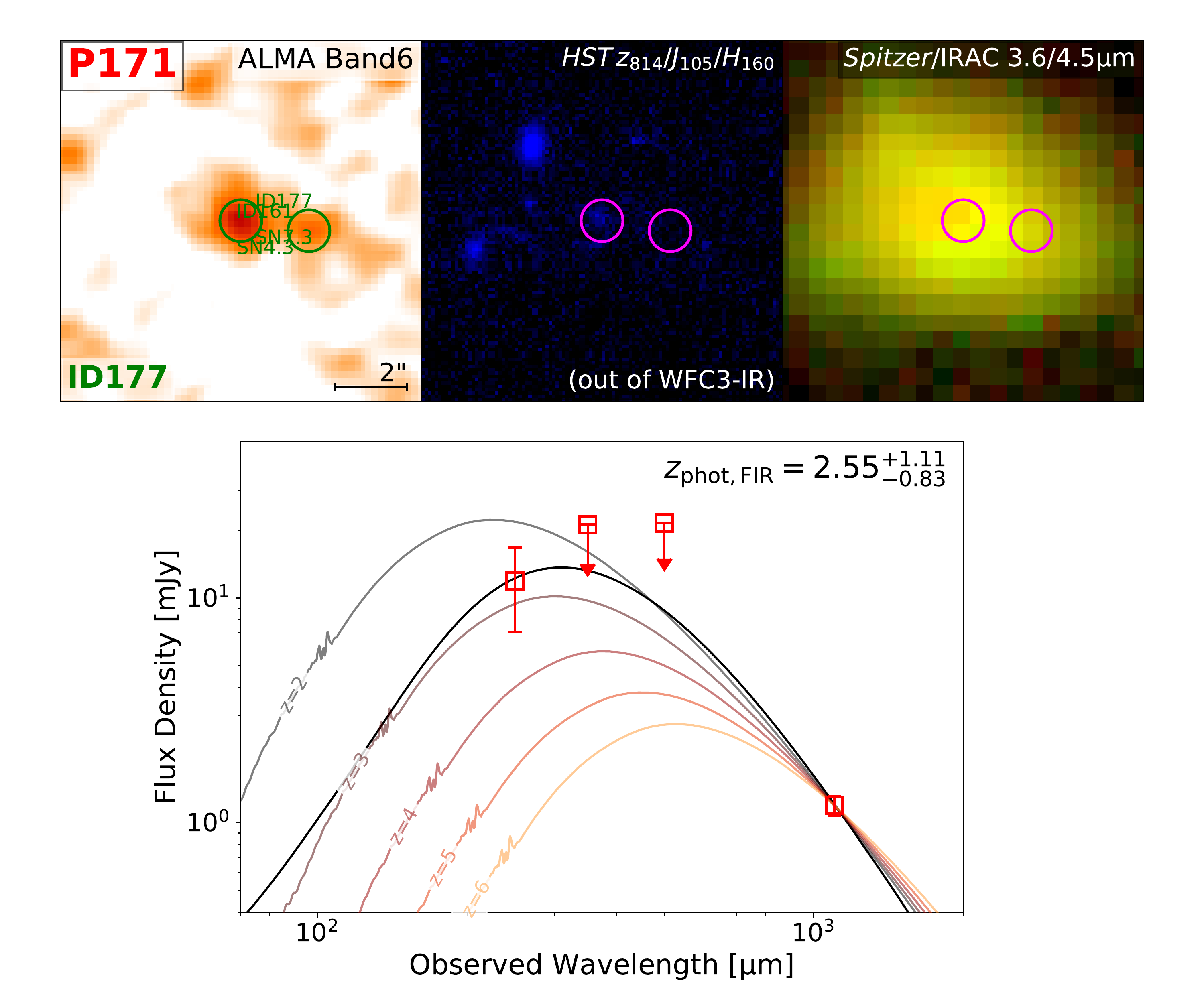}
\figsetgrpnote{Postage stamp images (top) and far-IR SED (bottom) of P171-ID177.}
\figsetgrpend

\figsetgrpstart
\figsetgrpnum{B1.81}
\figsetgrptitle{R0032-ID53}
\figsetplot{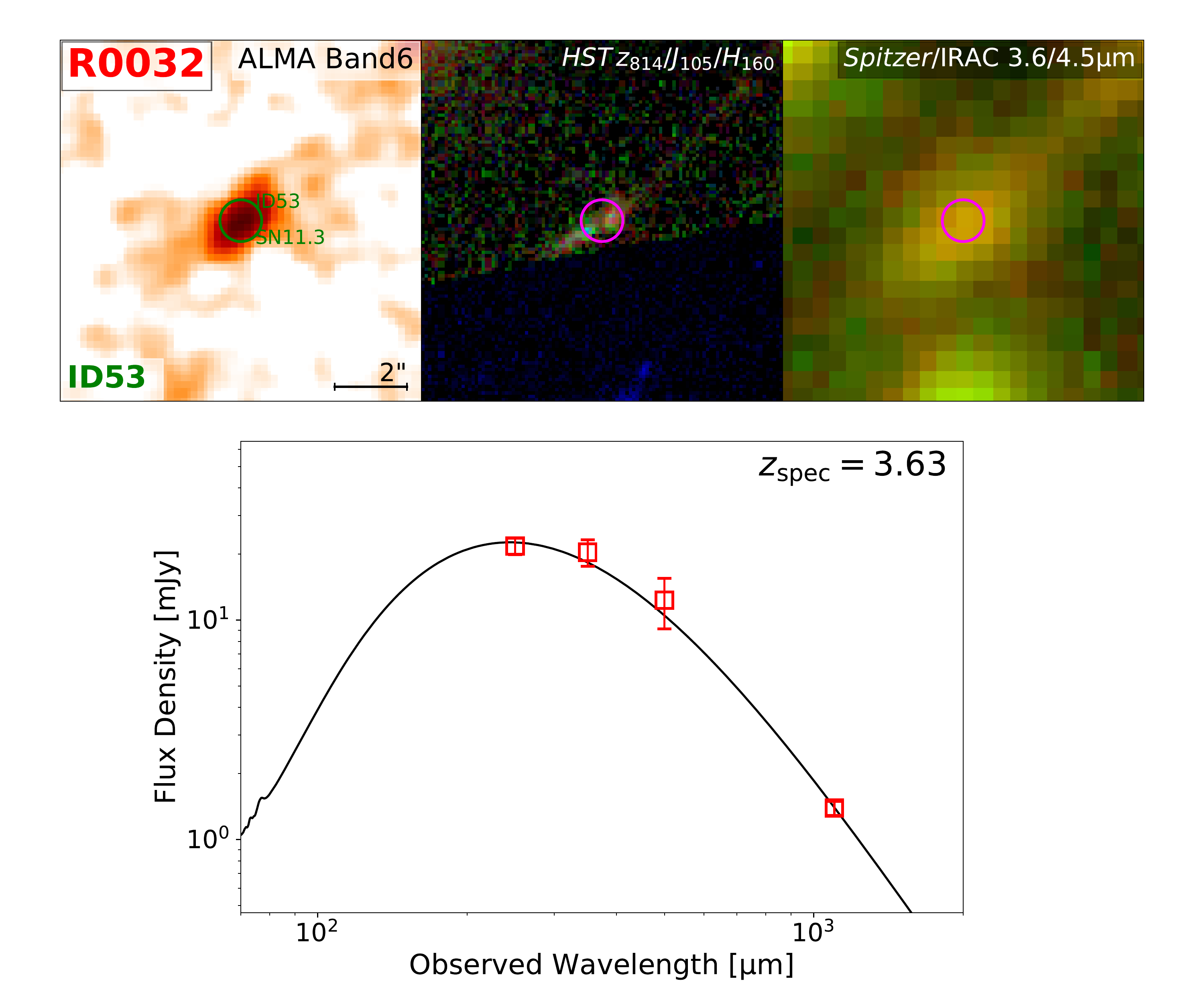}
\figsetgrpnote{Postage stamp images (top) and far-IR SED (bottom) of R0032-ID53.}
\figsetgrpend

\figsetgrpstart
\figsetgrpnum{B1.82}
\figsetgrptitle{R0032-ID55}
\figsetplot{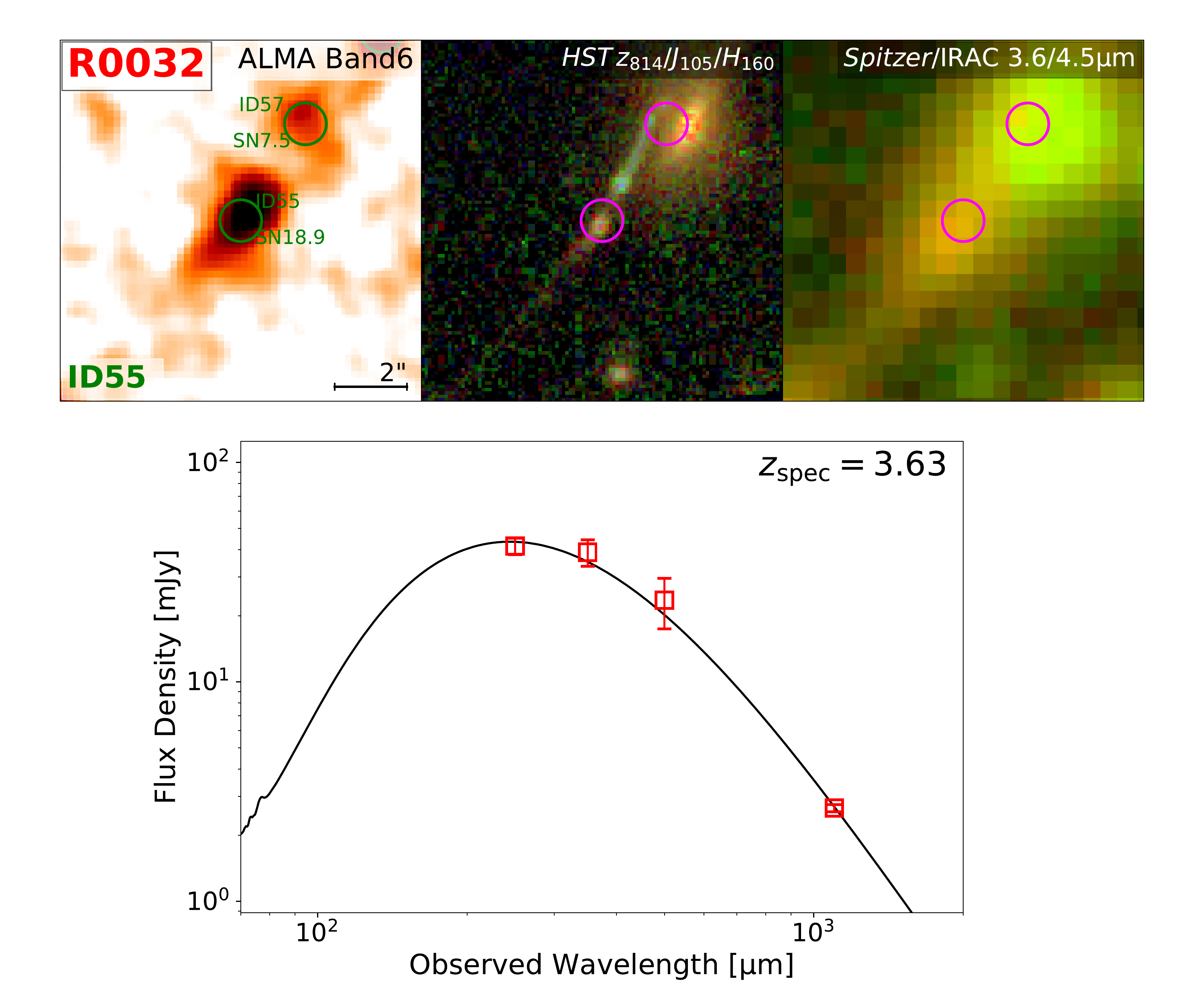}
\figsetgrpnote{Postage stamp images (top) and far-IR SED (bottom) of R0032-ID55.}
\figsetgrpend

\figsetgrpstart
\figsetgrpnum{B1.83}
\figsetgrptitle{R0032-ID57}
\figsetplot{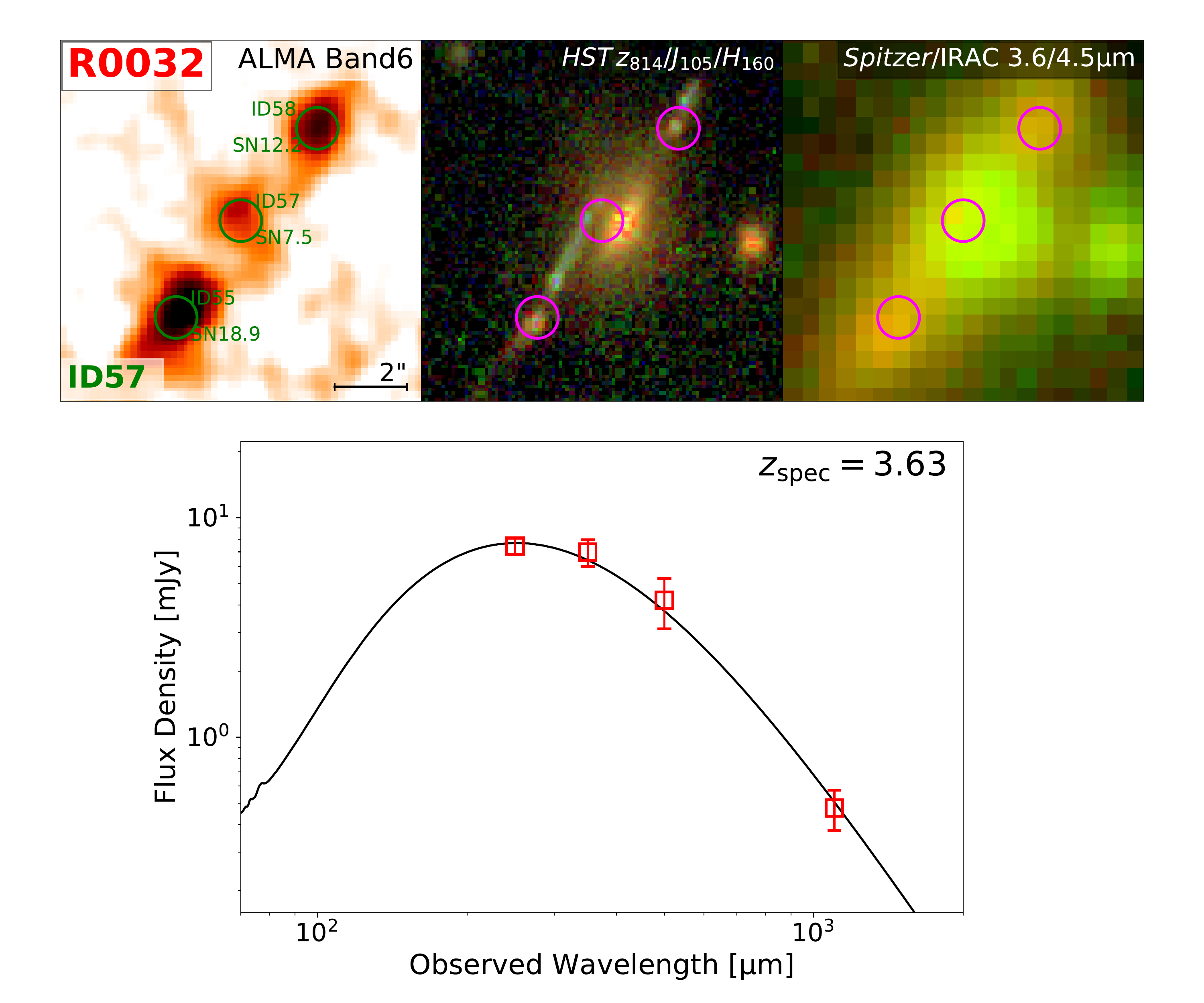}
\figsetgrpnote{Postage stamp images (top) and far-IR SED (bottom) of R0032-ID57.}
\figsetgrpend

\figsetgrpstart
\figsetgrpnum{B1.84}
\figsetgrptitle{R0032-ID58}
\figsetplot{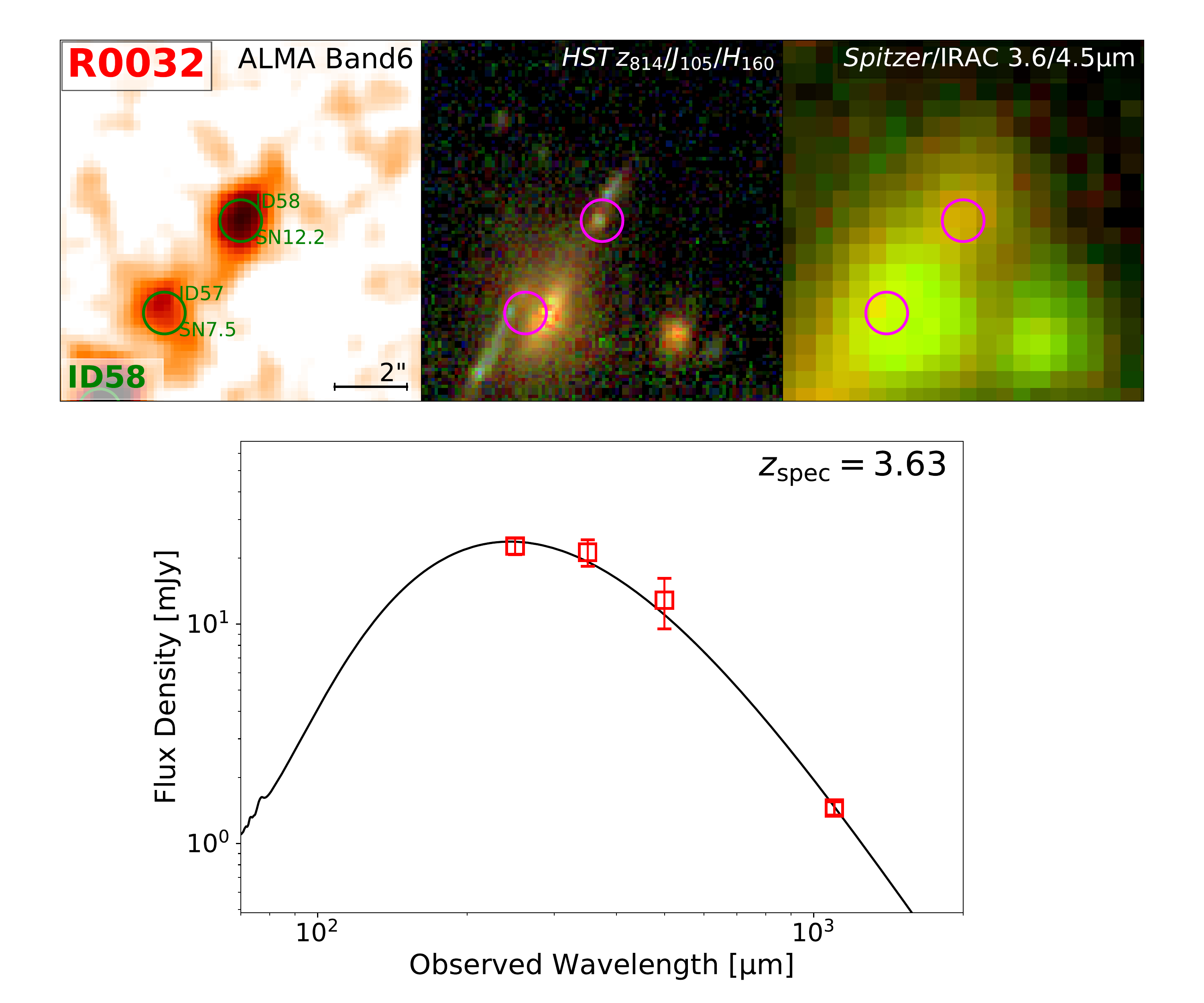}
\figsetgrpnote{Postage stamp images (top) and far-IR SED (bottom) of R0032-ID58.}
\figsetgrpend

\figsetgrpstart
\figsetgrpnum{B1.85}
\figsetgrptitle{R0032-ID127}
\figsetplot{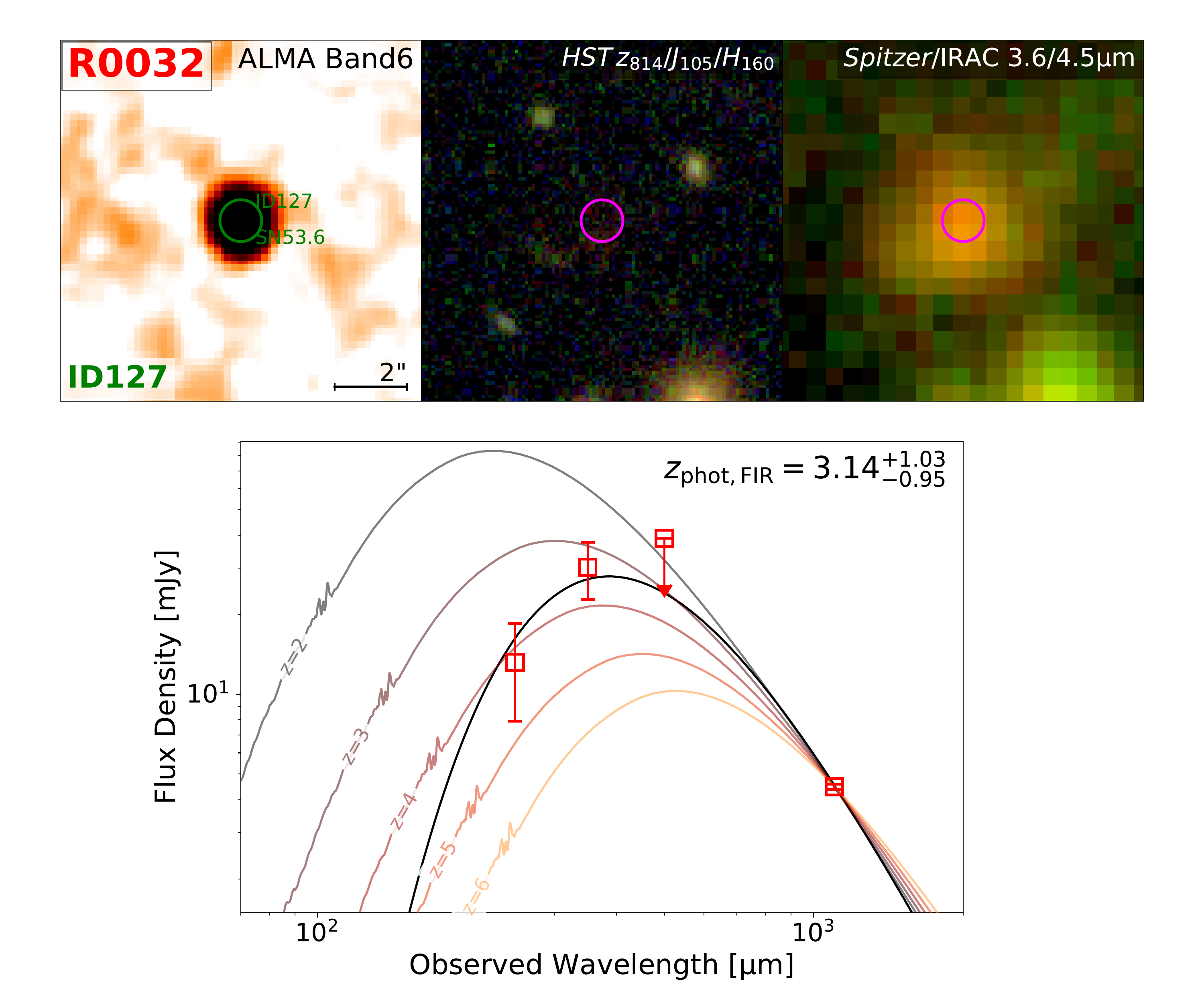}
\figsetgrpnote{Postage stamp images (top) and far-IR SED (bottom) of R0032-ID127.}
\figsetgrpend

\figsetgrpstart
\figsetgrpnum{B1.86}
\figsetgrptitle{R0032-ID131}
\figsetplot{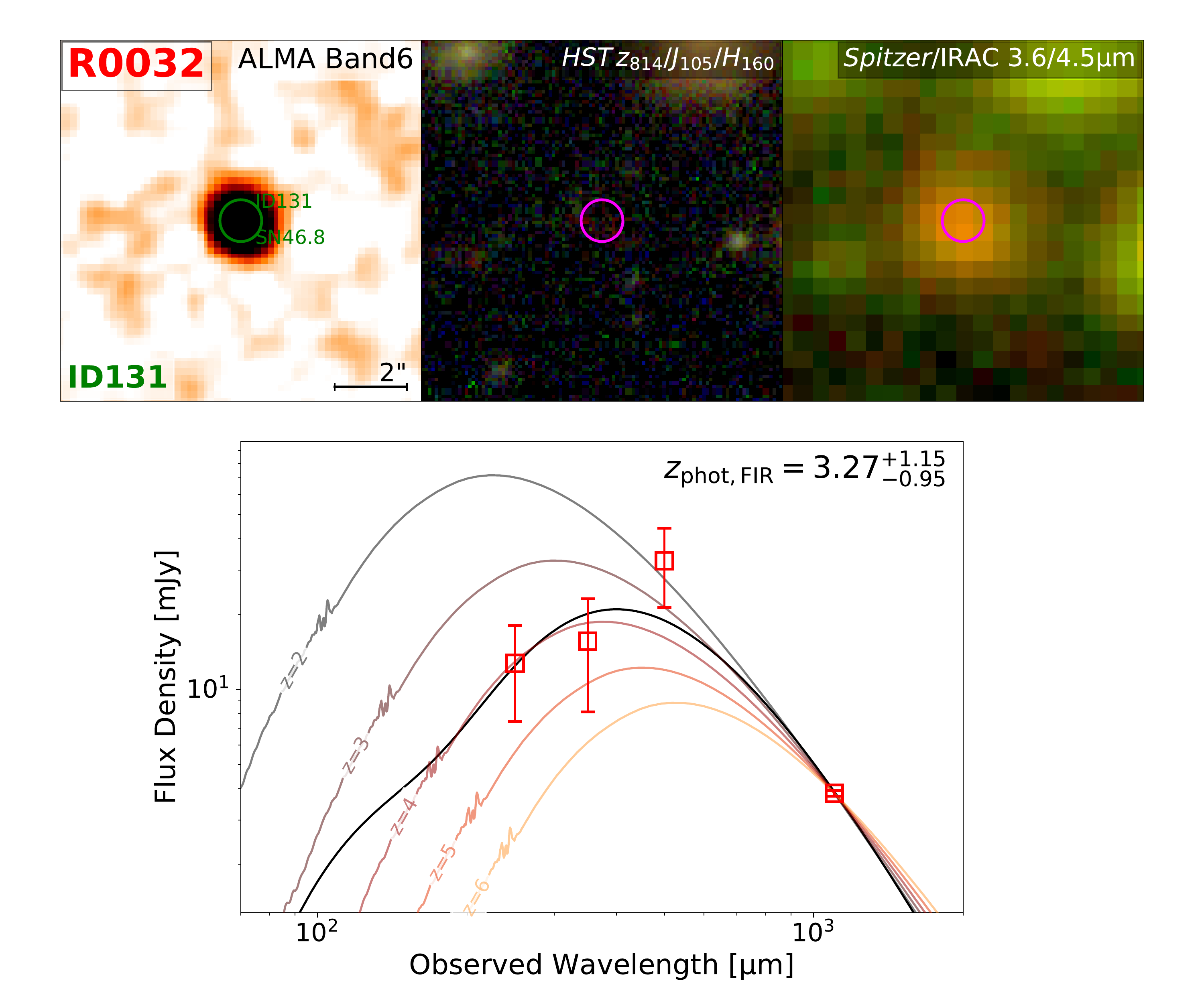}
\figsetgrpnote{Postage stamp images (top) and far-IR SED (bottom) of R0032-ID131.}
\figsetgrpend

\figsetgrpstart
\figsetgrpnum{B1.87}
\figsetgrptitle{R0032-ID198}
\figsetplot{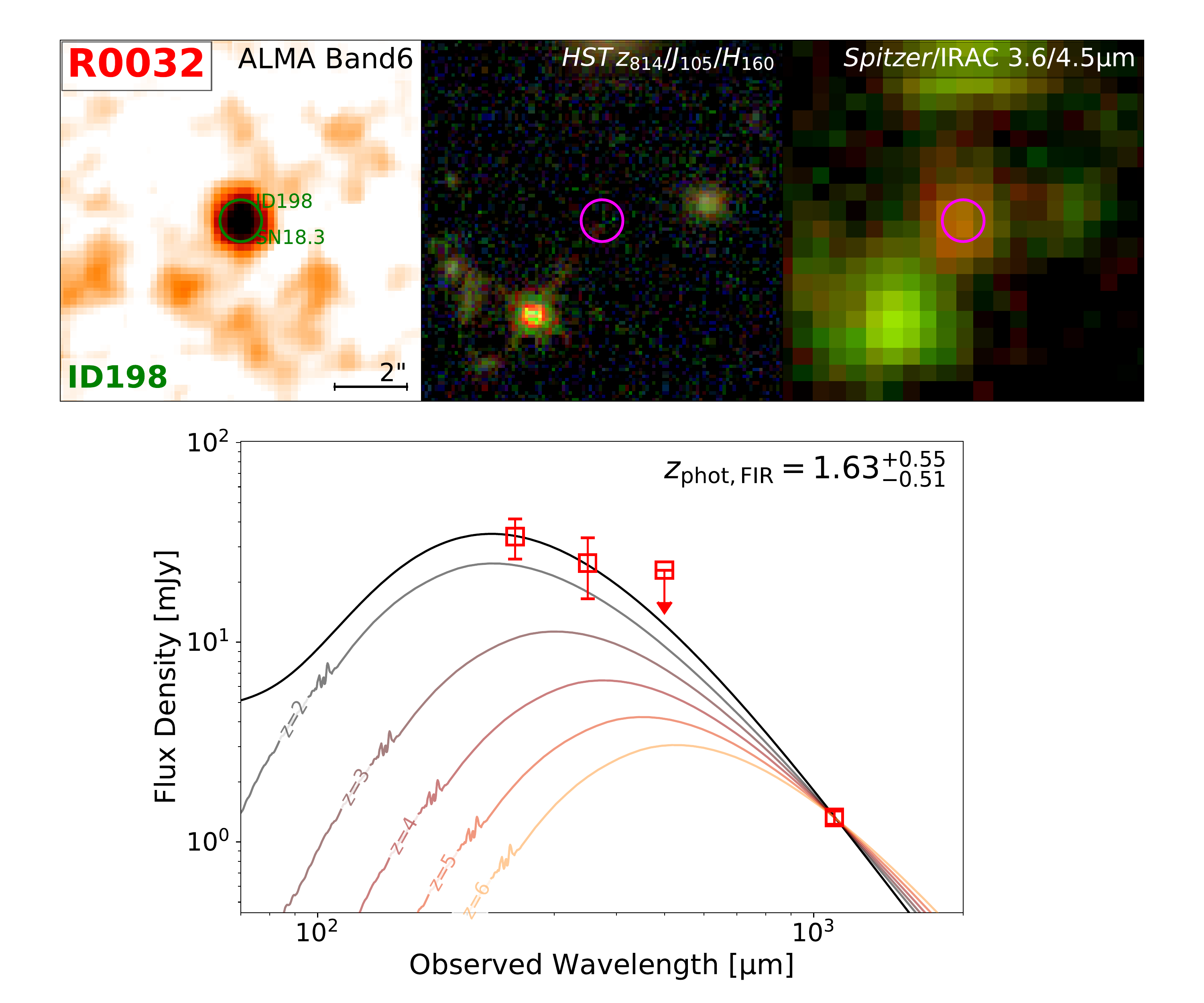}
\figsetgrpnote{Postage stamp images (top) and far-IR SED (bottom) of R0032-ID198.}
\figsetgrpend

\figsetgrpstart
\figsetgrpnum{B1.88}
\figsetgrptitle{R0032-ID220}
\figsetplot{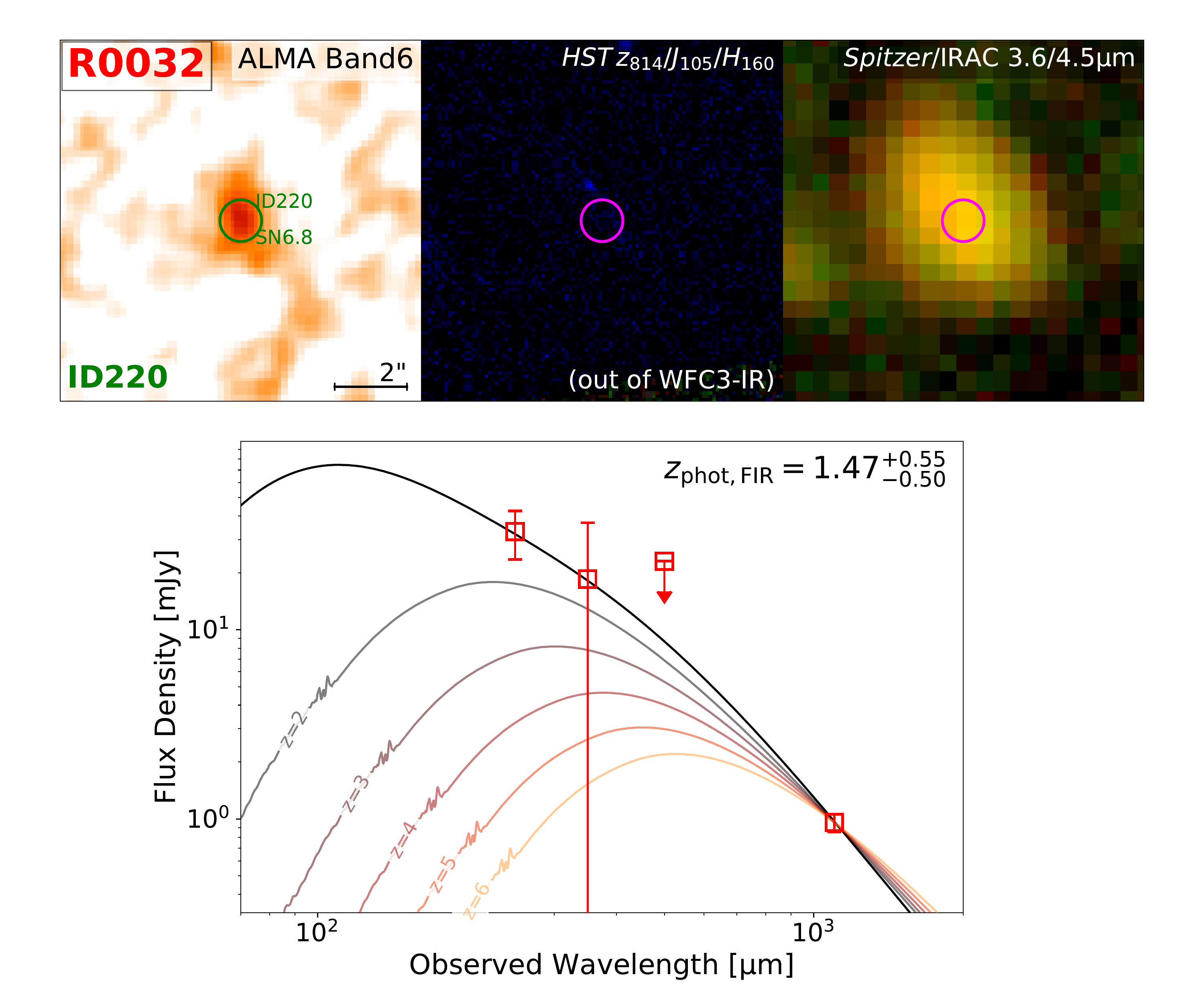}
\figsetgrpnote{Postage stamp images (top) and far-IR SED (bottom) of R0032-ID220.}
\figsetgrpend

\figsetgrpstart
\figsetgrpnum{B1.89}
\figsetgrptitle{R0032-ID238}
\figsetplot{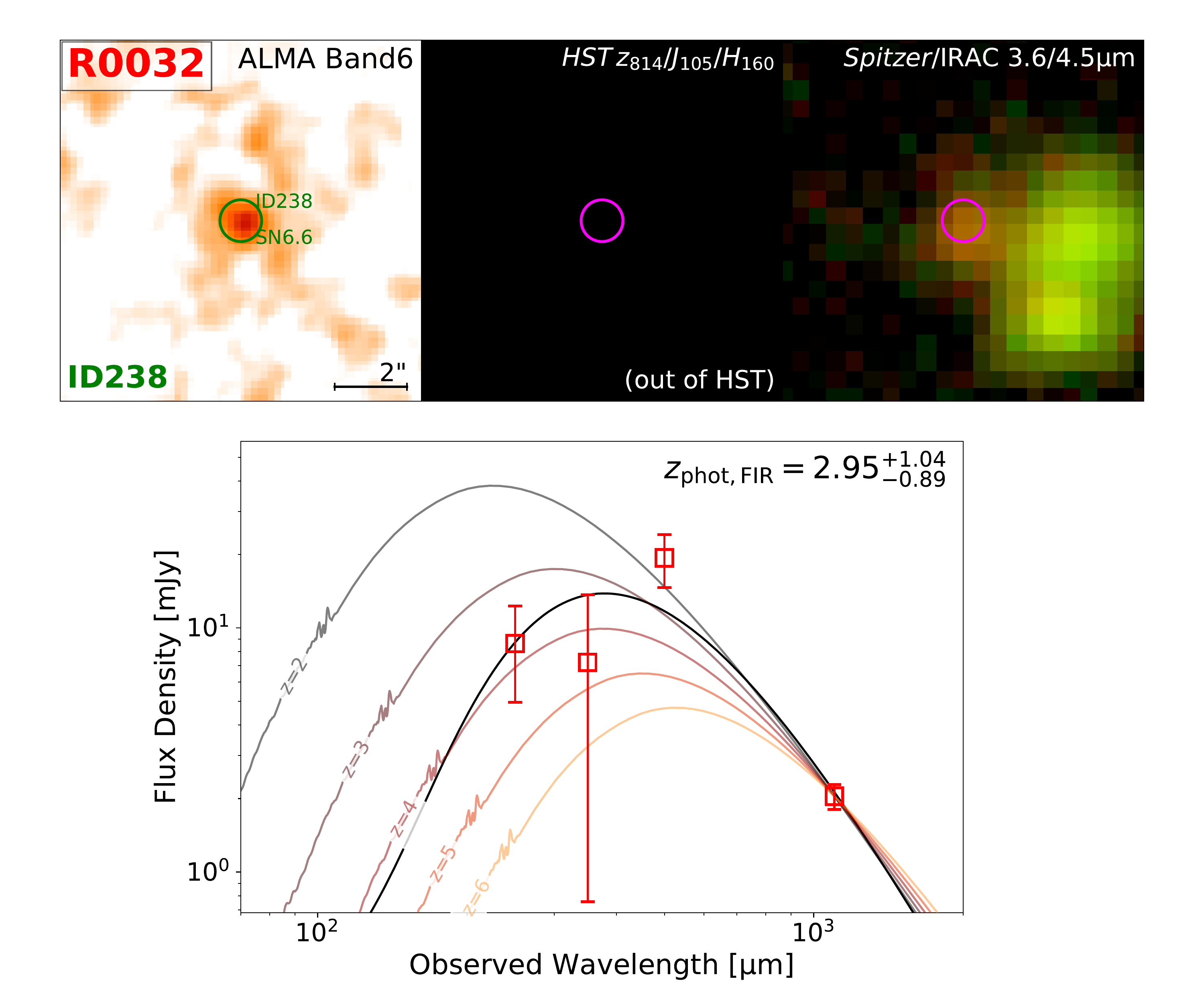}
\figsetgrpnote{Postage stamp images (top) and far-IR SED (bottom) of R0032-ID238.}
\figsetgrpend

\figsetgrpstart
\figsetgrpnum{B1.90}
\figsetgrptitle{R0032-ID276}
\figsetplot{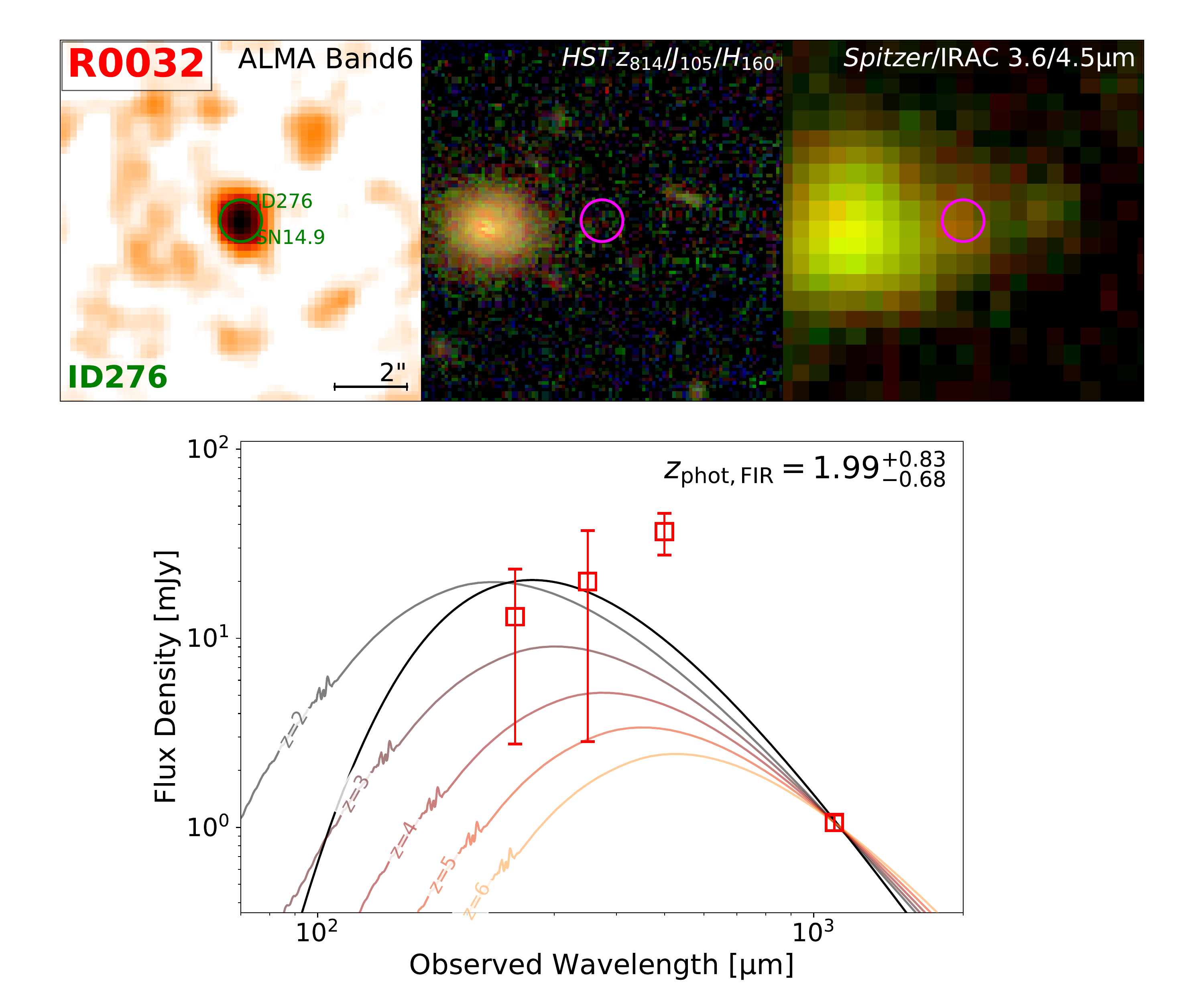}
\figsetgrpnote{Postage stamp images (top) and far-IR SED (bottom) of R0032-ID276.}
\figsetgrpend

\figsetgrpstart
\figsetgrpnum{B1.91}
\figsetgrptitle{R0032-ID287}
\figsetplot{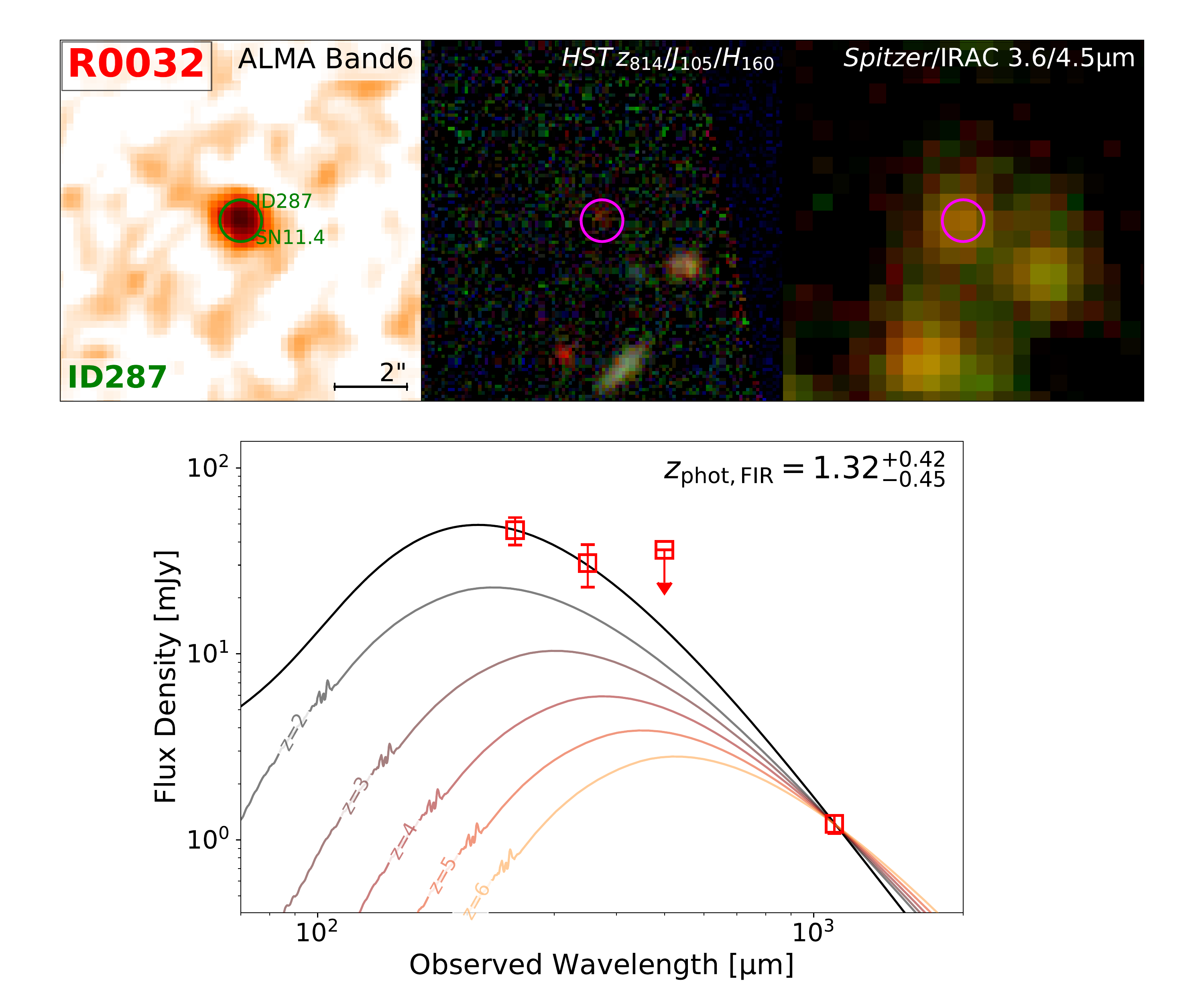}
\figsetgrpnote{Postage stamp images (top) and far-IR SED (bottom) of R0032-ID287.}
\figsetgrpend

\figsetgrpstart
\figsetgrpnum{B1.92}
\figsetgrptitle{R0600-ID12}
\figsetplot{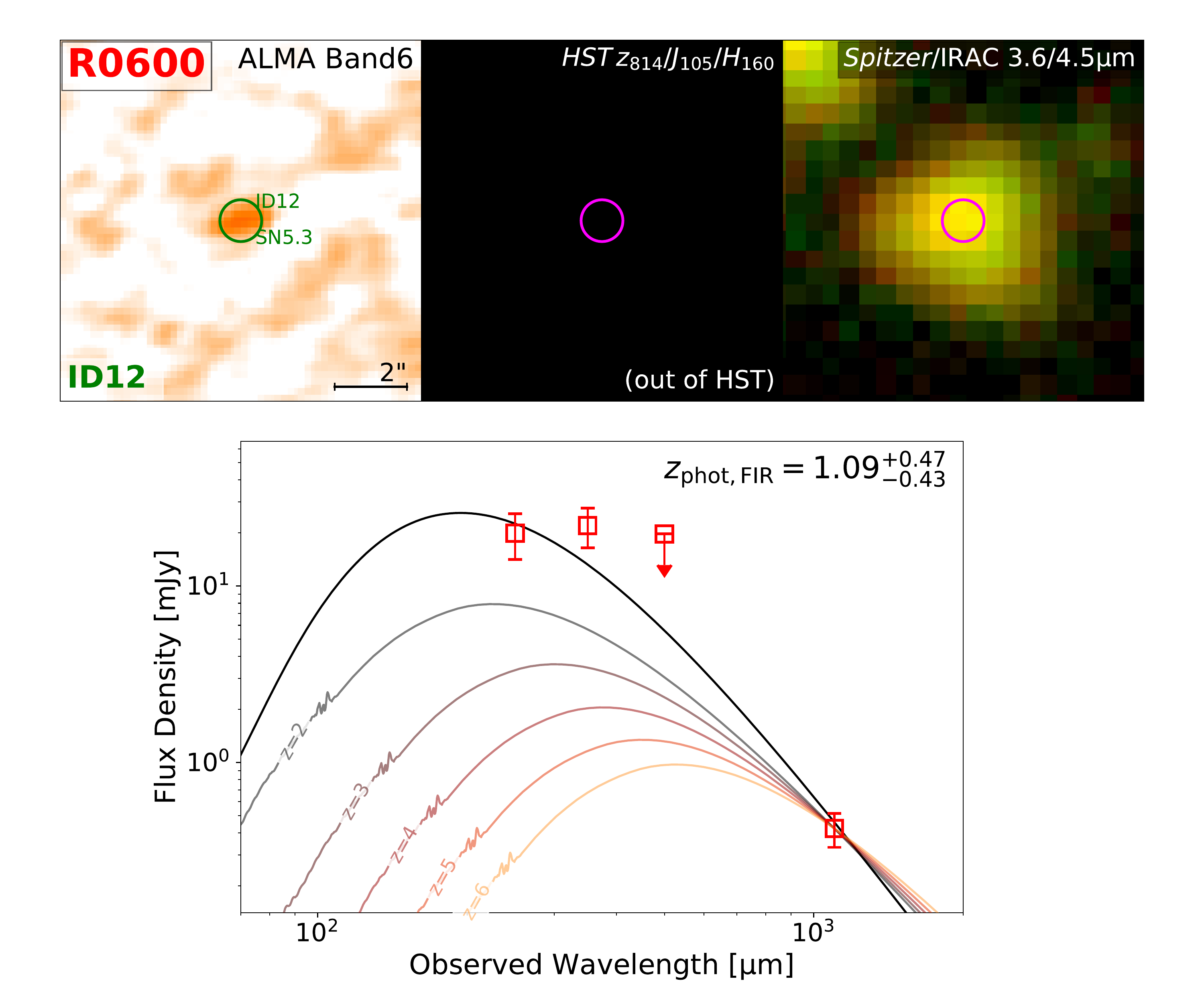}
\figsetgrpnote{Postage stamp images (top) and far-IR SED (bottom) of R0600-ID12.}
\figsetgrpend

\figsetgrpstart
\figsetgrpnum{B1.93}
\figsetgrptitle{R0600-ID13}
\figsetplot{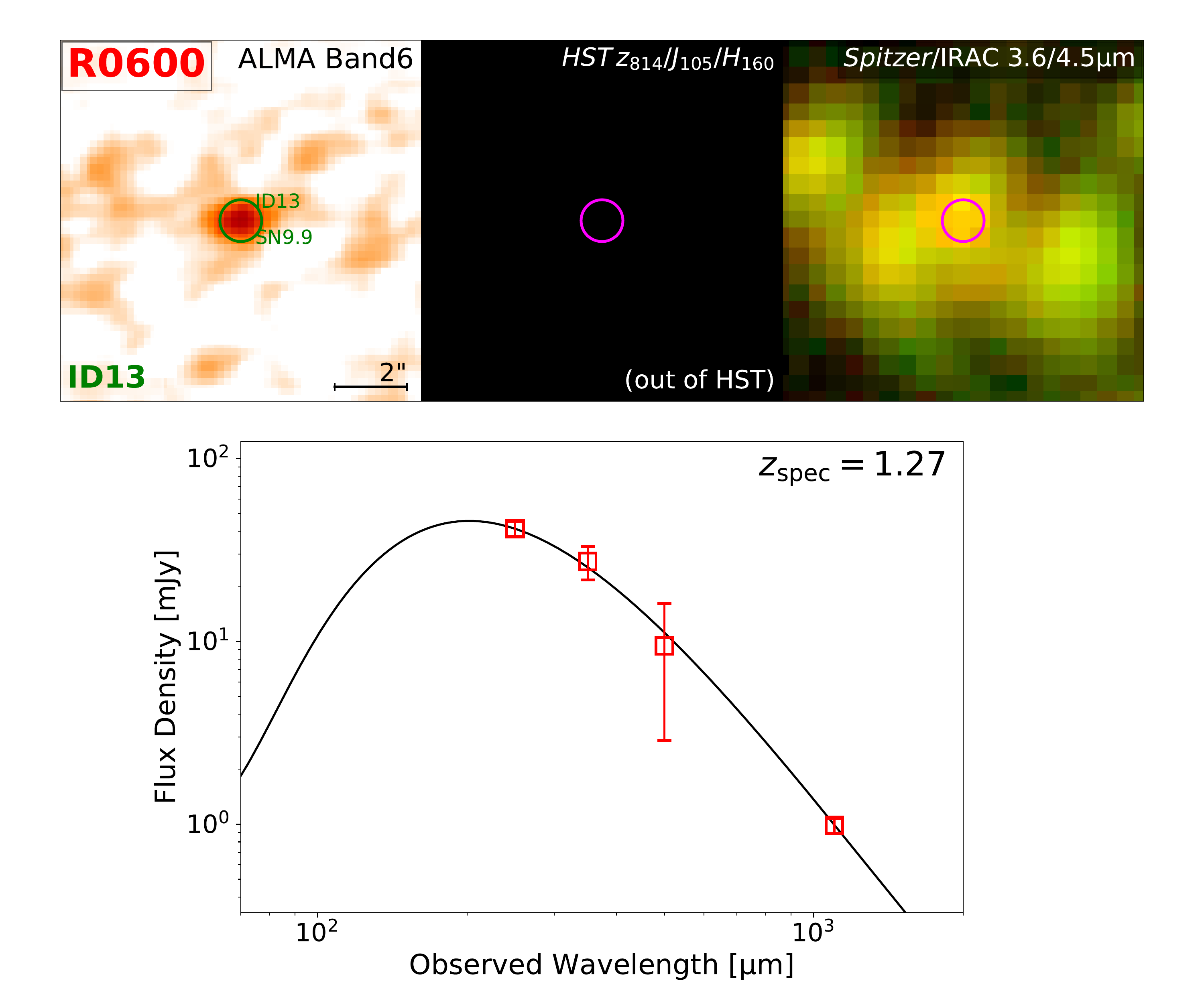}
\figsetgrpnote{Postage stamp images (top) and far-IR SED (bottom) of R0600-ID13.}
\figsetgrpend

\figsetgrpstart
\figsetgrpnum{B1.94}
\figsetgrptitle{R0600-ID111}
\figsetplot{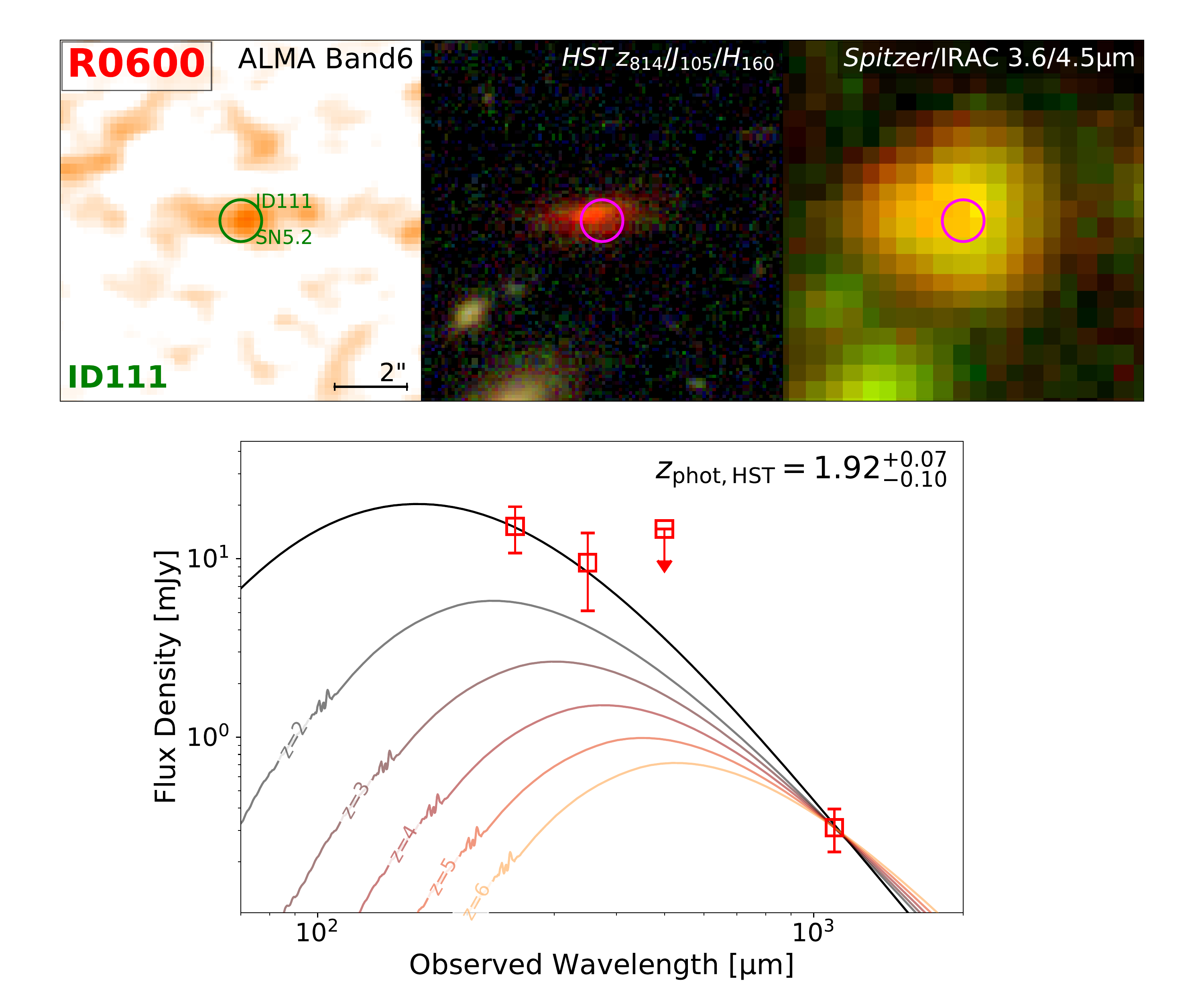}
\figsetgrpnote{Postage stamp images (top) and far-IR SED (bottom) of R0600-ID111.}
\figsetgrpend

\figsetgrpstart
\figsetgrpnum{B1.95}
\figsetgrptitle{R0949-ID10}
\figsetplot{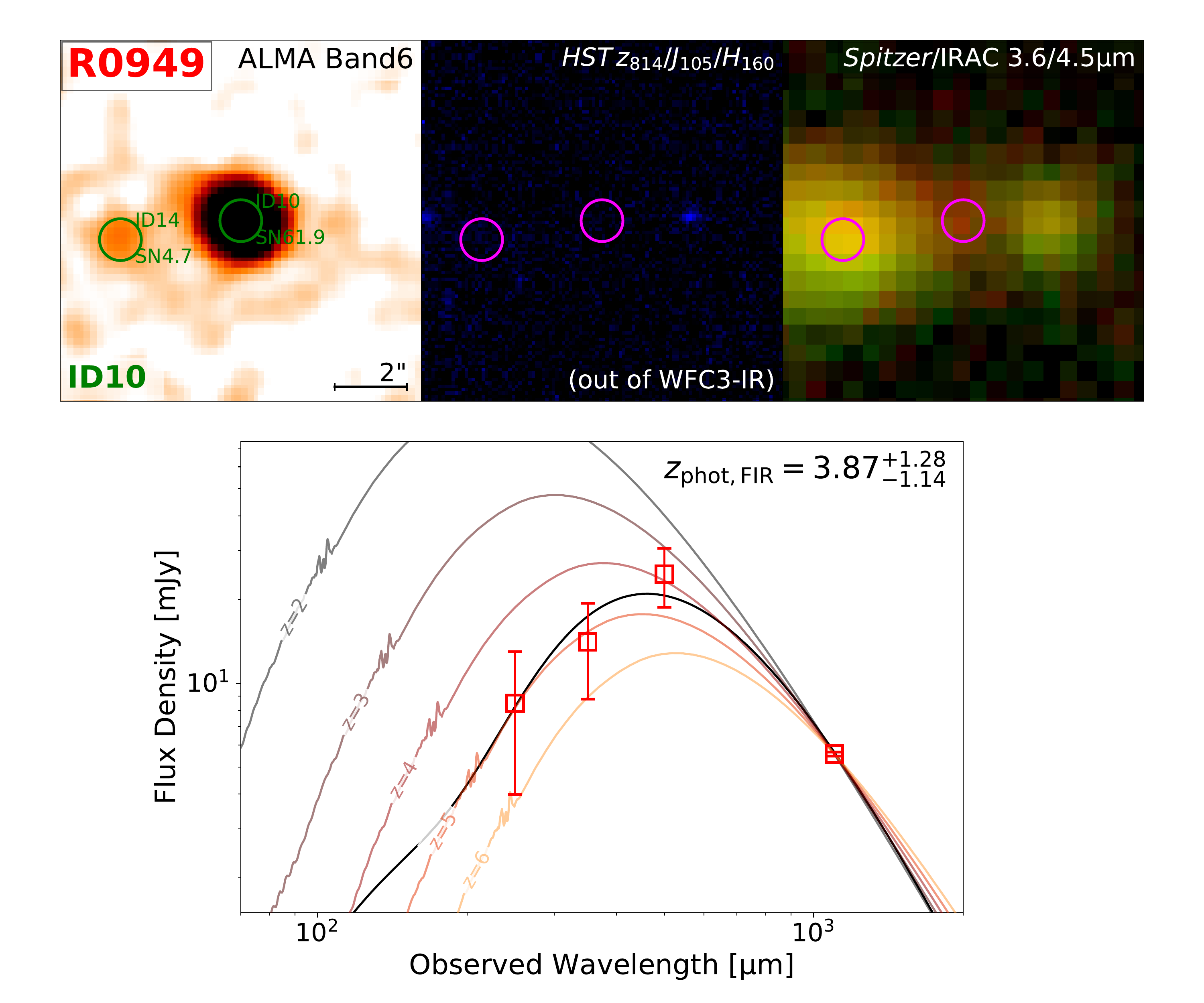}
\figsetgrpnote{Postage stamp images (top) and far-IR SED (bottom) of R0949-ID10.}
\figsetgrpend

\figsetgrpstart
\figsetgrpnum{B1.96}
\figsetgrptitle{R0949-ID122}
\figsetplot{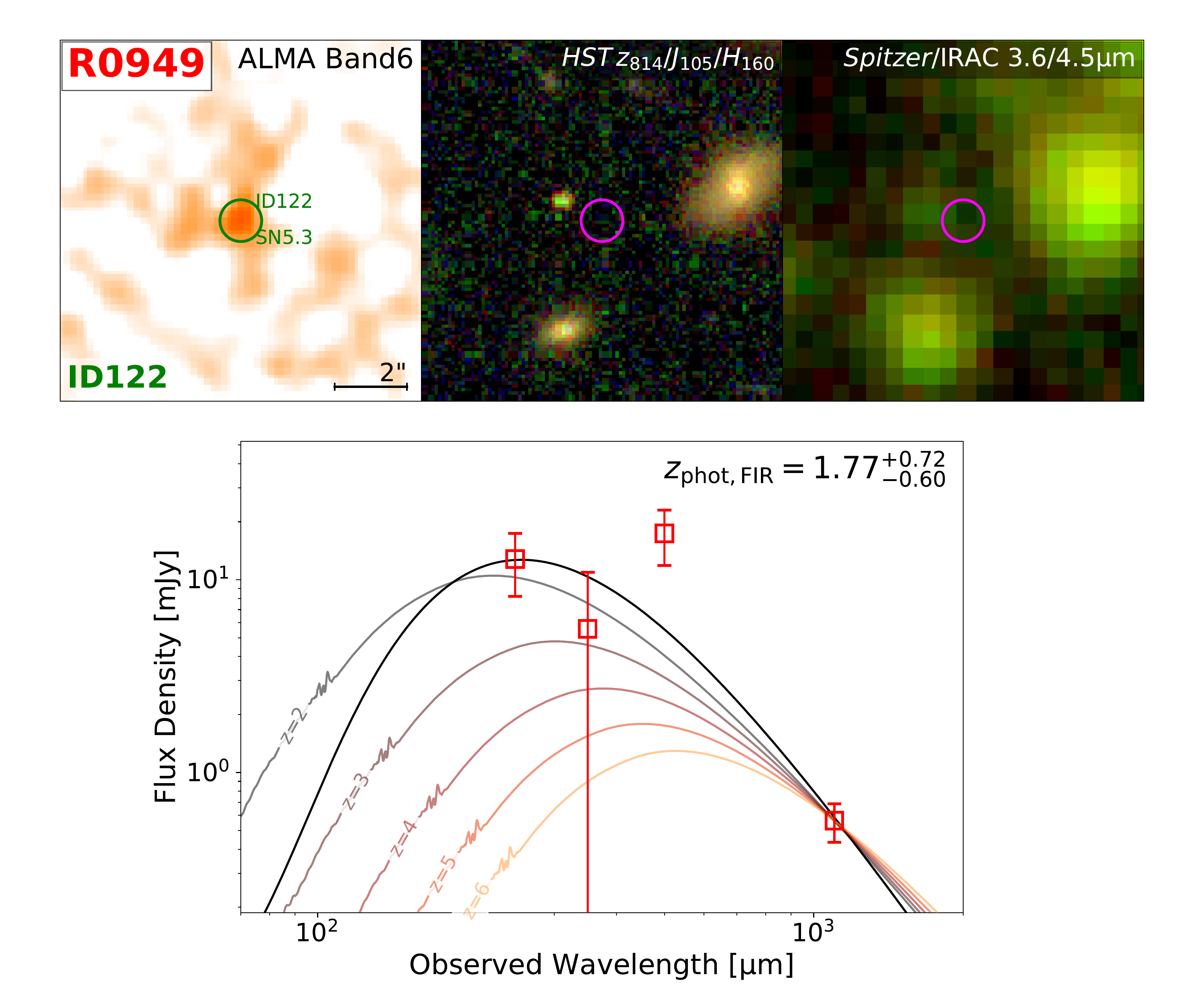}
\figsetgrpnote{Postage stamp images (top) and far-IR SED (bottom) of R0949-ID122.}
\figsetgrpend

\figsetgrpstart
\figsetgrpnum{B1.97}
\figsetgrptitle{R0949-ID124}
\figsetplot{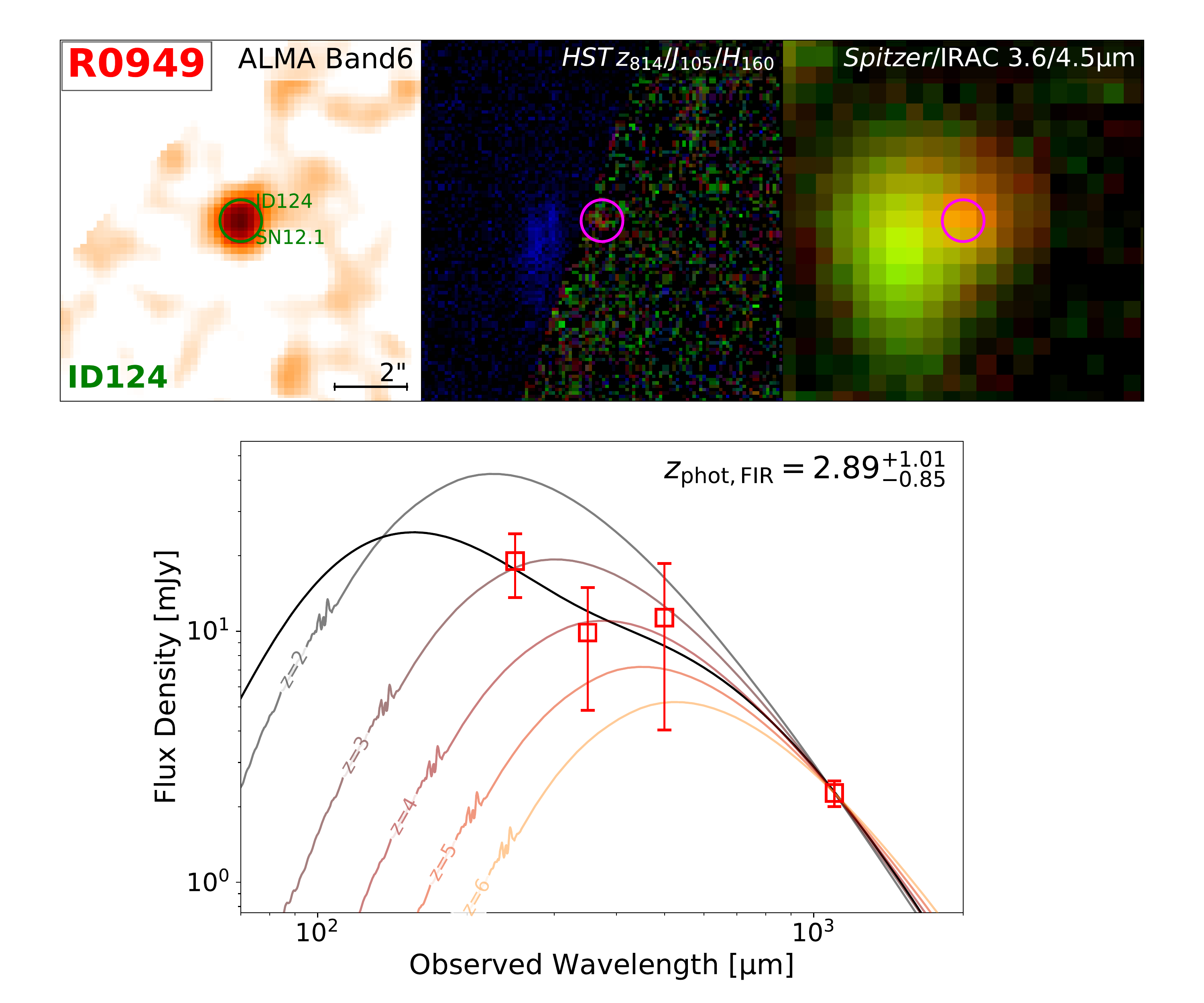}
\figsetgrpnote{Postage stamp images (top) and far-IR SED (bottom) of R0949-ID124.}
\figsetgrpend

\figsetgrpstart
\figsetgrpnum{B1.98}
\figsetgrptitle{R1347-ID41}
\figsetplot{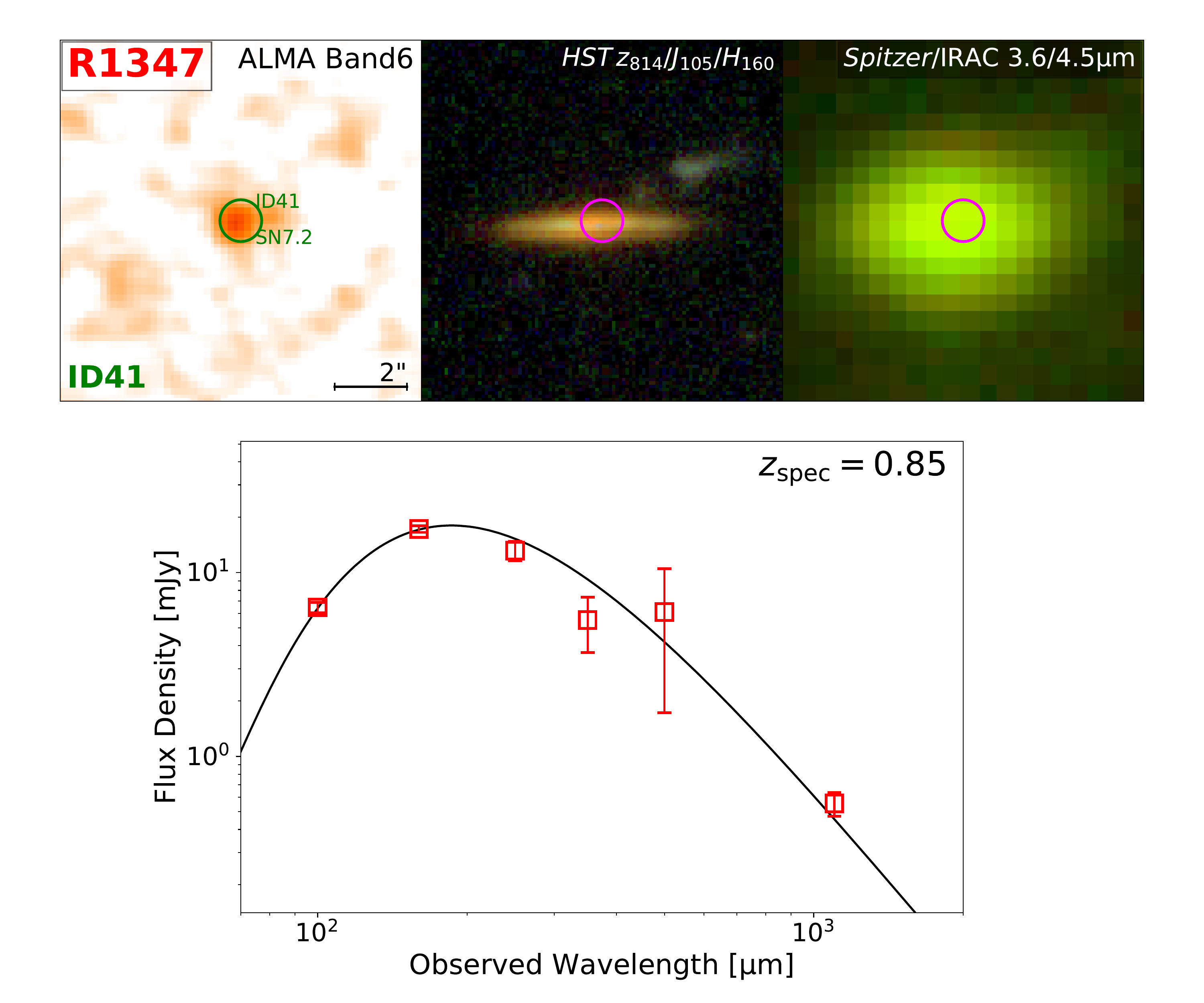}
\figsetgrpnote{Postage stamp images (top) and far-IR SED (bottom) of R1347-ID41.}
\figsetgrpend

\figsetgrpstart
\figsetgrpnum{B1.99}
\figsetgrptitle{R1347-ID145}
\figsetplot{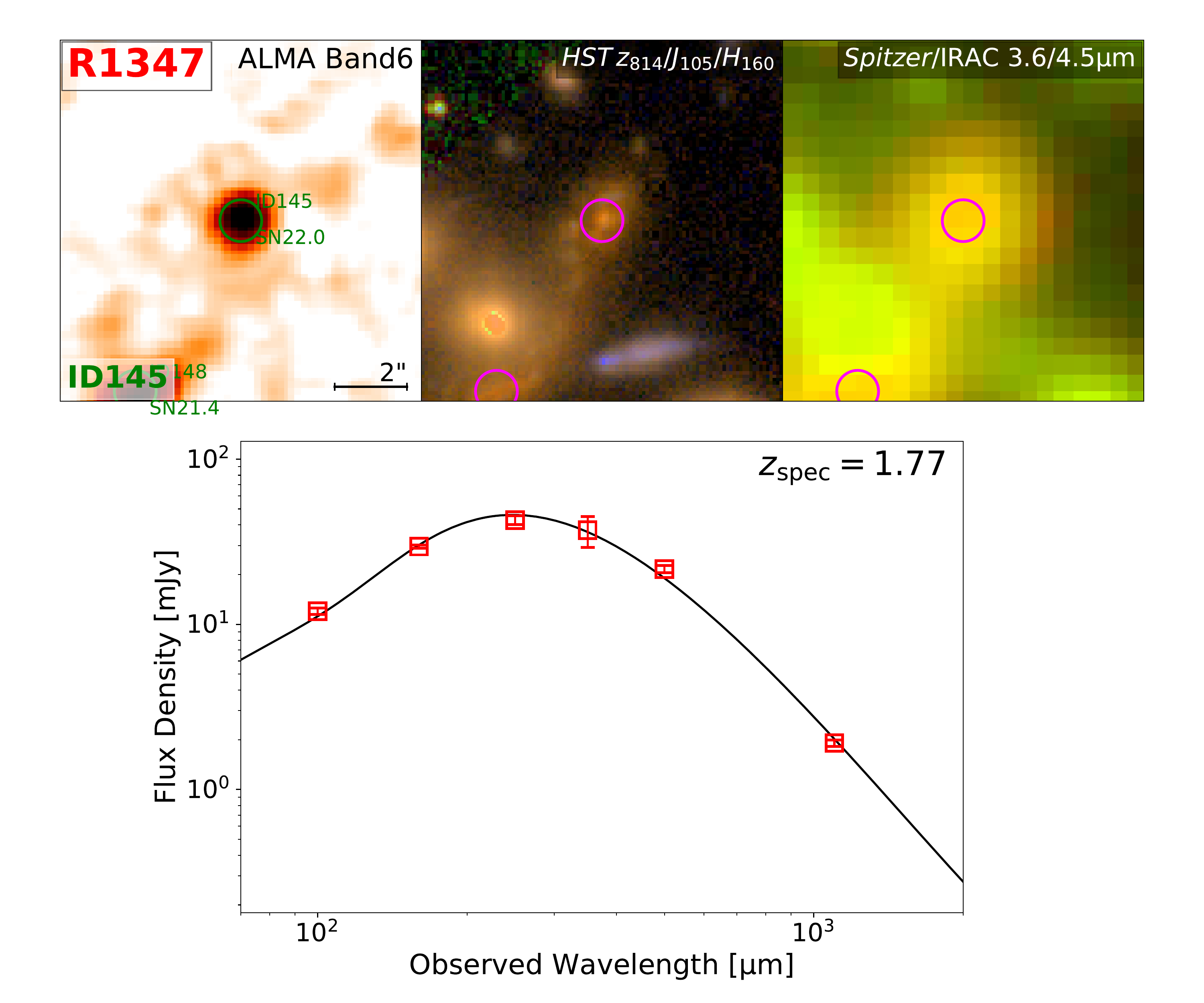}
\figsetgrpnote{Postage stamp images (top) and far-IR SED (bottom) of R1347-ID145.}
\figsetgrpend

\figsetgrpstart
\figsetgrpnum{B1.100}
\figsetgrptitle{R1347-ID148}
\figsetplot{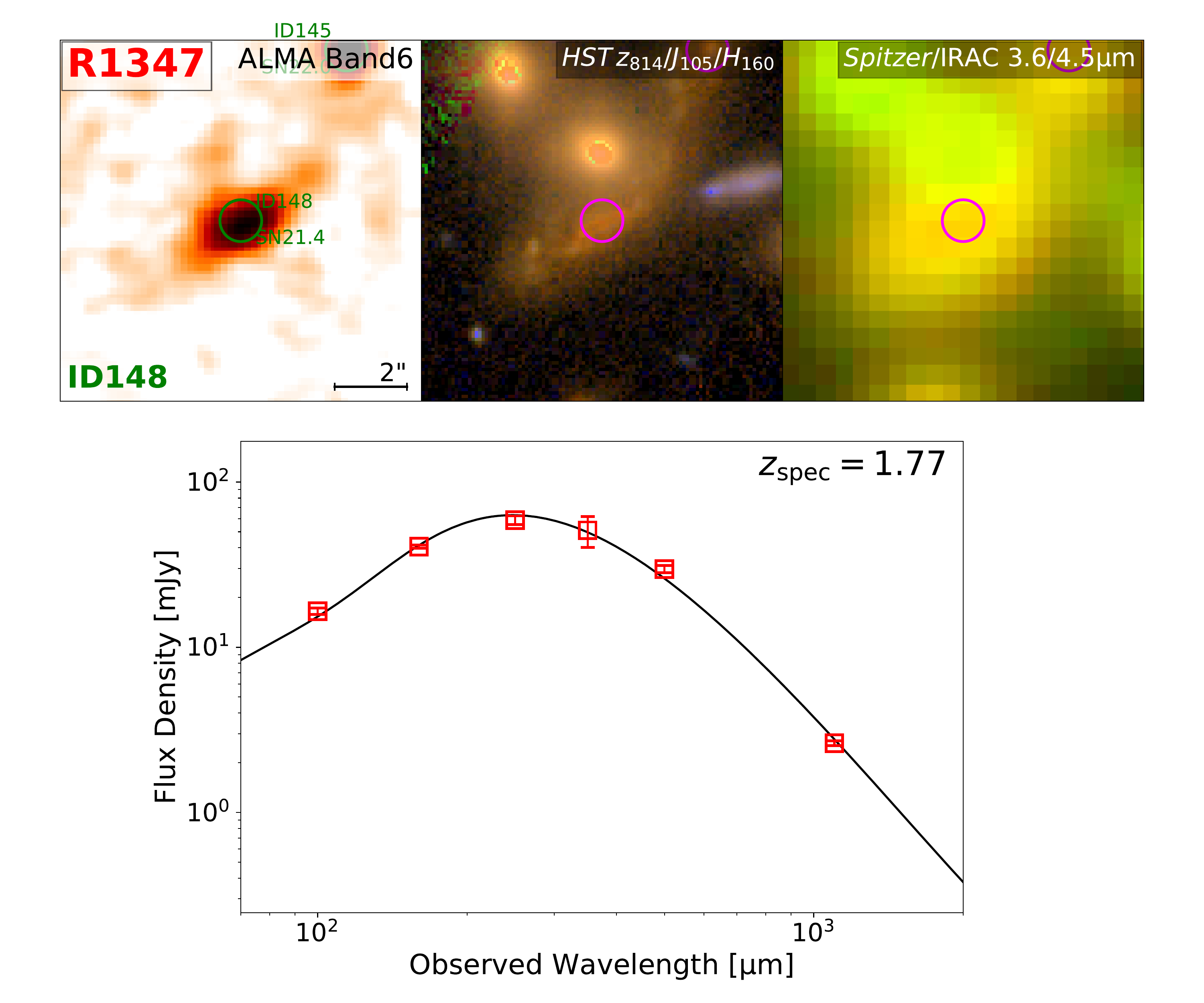}
\figsetgrpnote{Postage stamp images (top) and far-IR SED (bottom) of R1347-ID148.}
\figsetgrpend

\figsetgrpstart
\figsetgrpnum{B1.101}
\figsetgrptitle{R1347-ID166}
\figsetplot{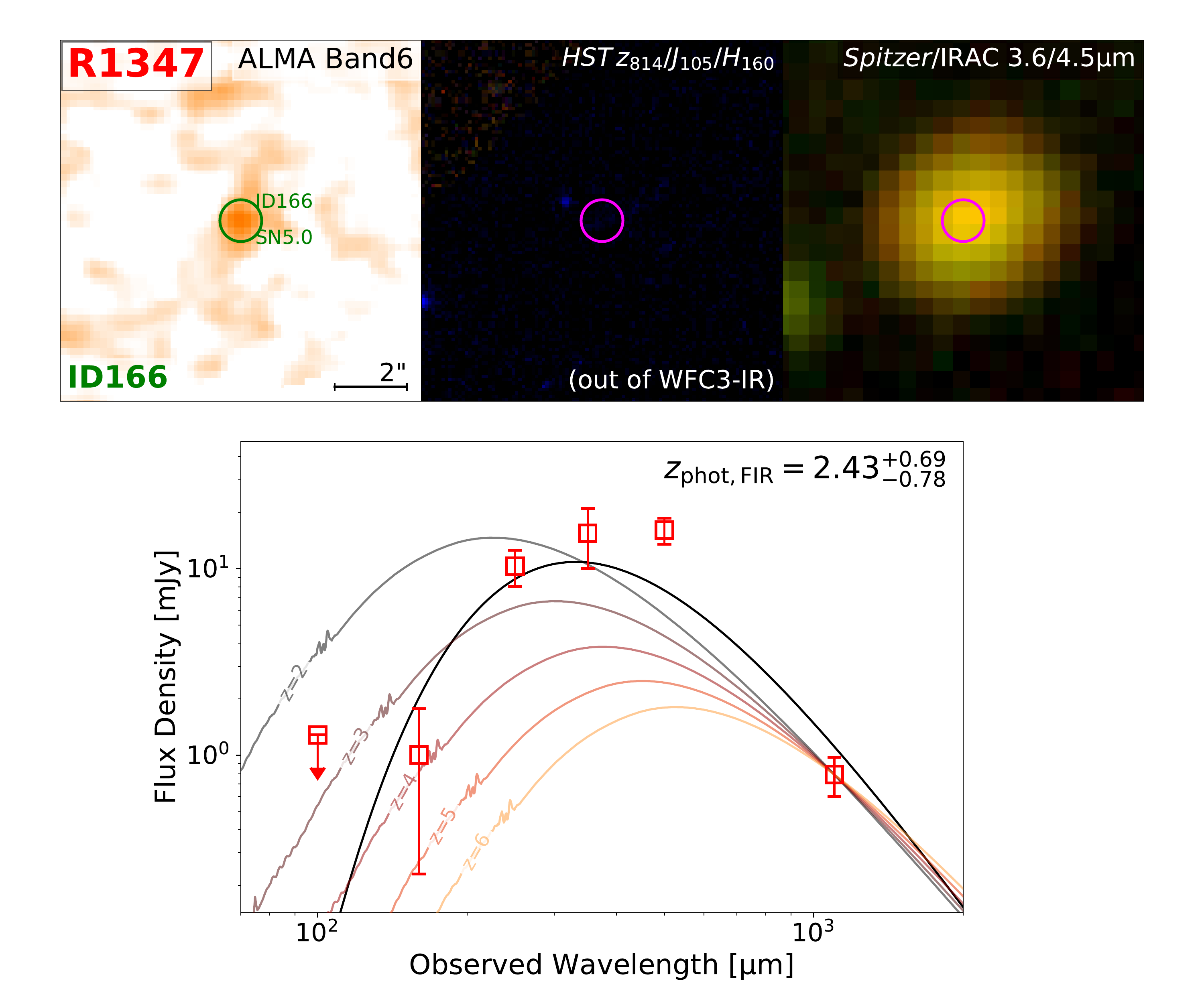}
\figsetgrpnote{Postage stamp images (top) and far-IR SED (bottom) of R1347-ID166.}
\figsetgrpend

\figsetgrpstart
\figsetgrpnum{B1.102}
\figsetgrptitle{R2129-ID37}
\figsetplot{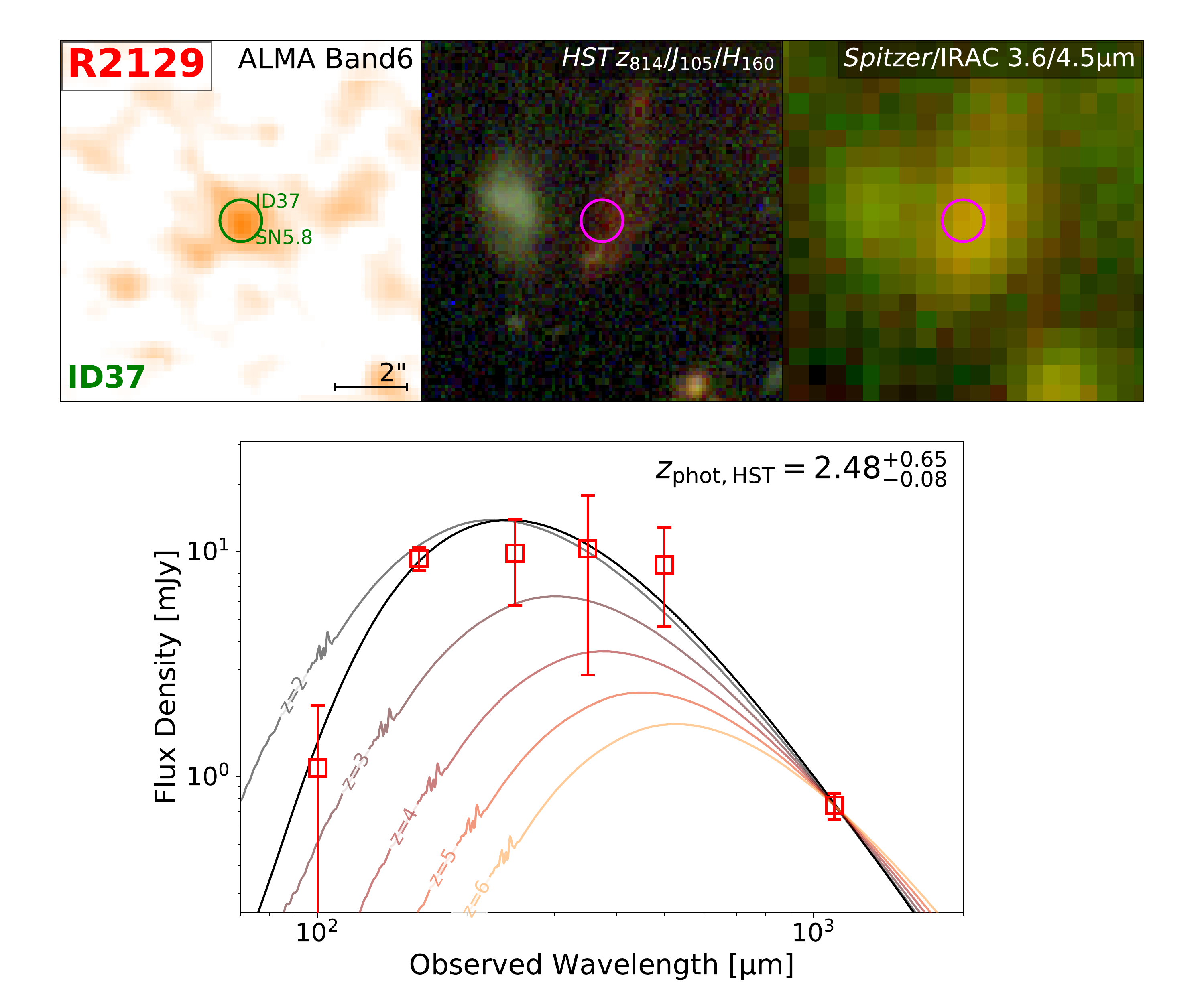}
\figsetgrpnote{Postage stamp images (top) and far-IR SED (bottom) of R2129-ID37.}
\figsetgrpend

\figsetgrpstart
\figsetgrpnum{B1.103}
\figsetgrptitle{R2211-ID19}
\figsetplot{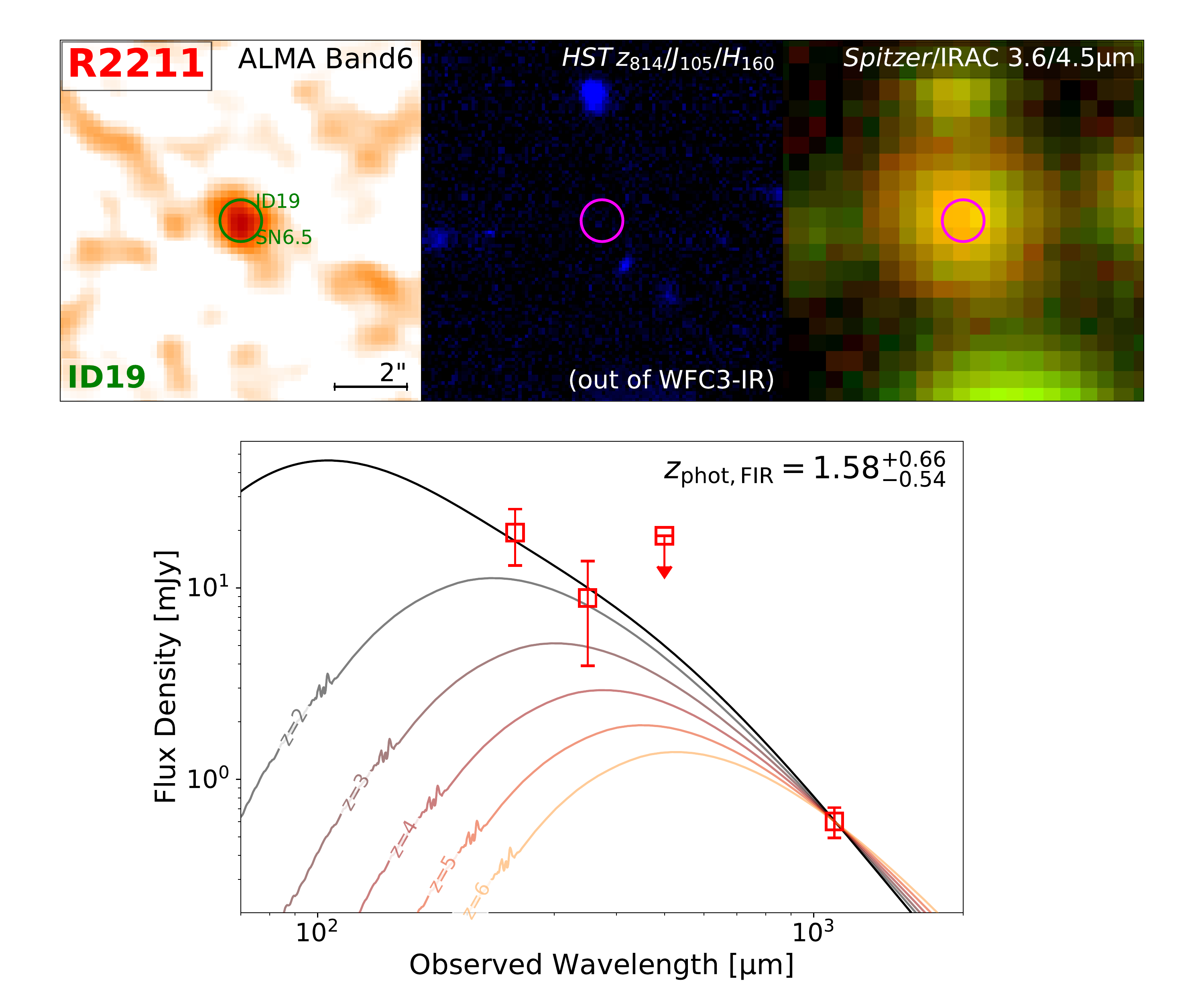}
\figsetgrpnote{Postage stamp images (top) and far-IR SED (bottom) of R2211-ID19.}
\figsetgrpend

\figsetgrpstart
\figsetgrpnum{B1.104}
\figsetgrptitle{R2211-ID35}
\figsetplot{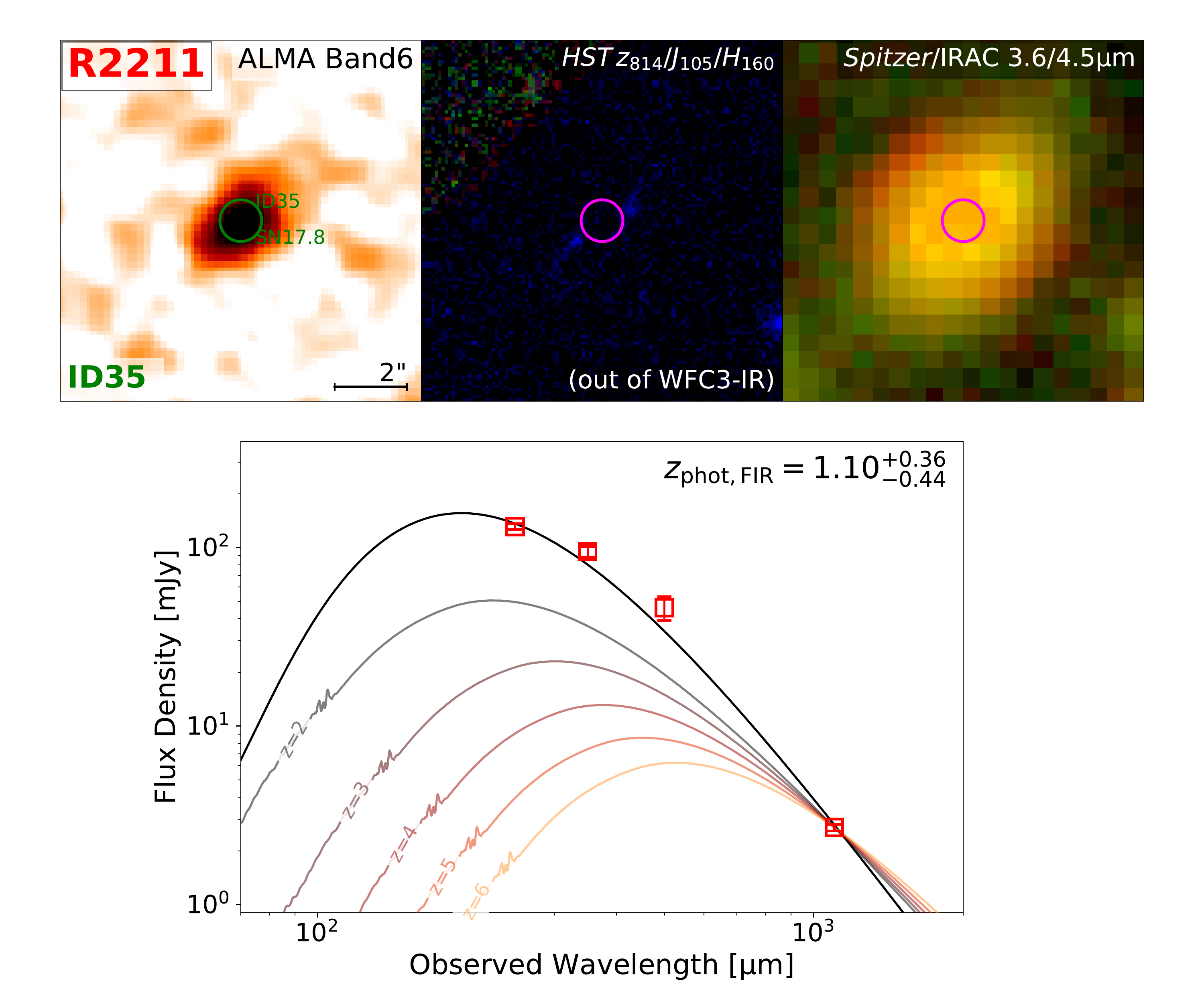}
\figsetgrpnote{Postage stamp images (top) and far-IR SED (bottom) of R2211-ID35.}
\figsetgrpend

\figsetgrpstart
\figsetgrpnum{B1.105}
\figsetgrptitle{R2211-ID171}
\figsetplot{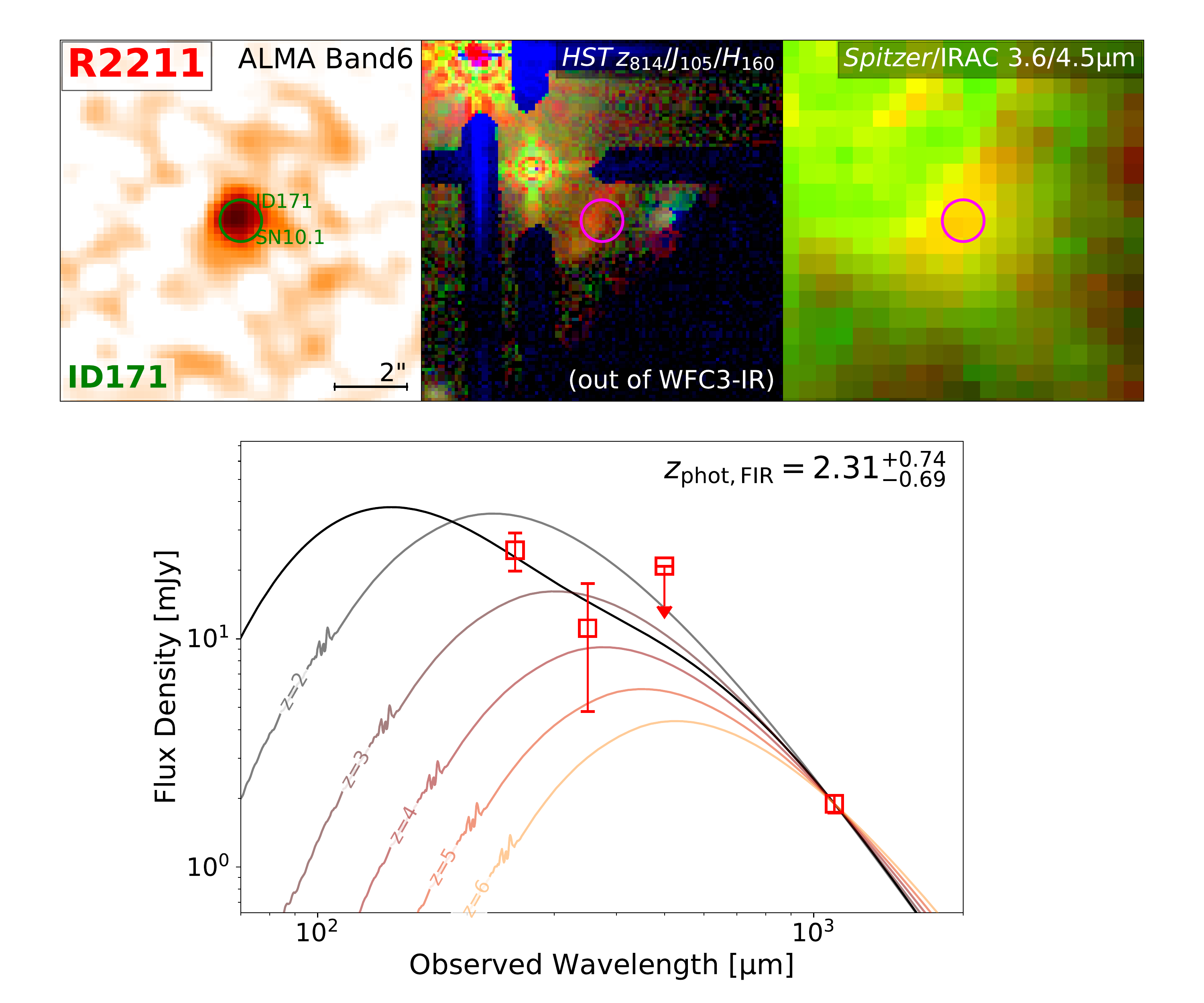}
\figsetgrpnote{Postage stamp images (top) and far-IR SED (bottom) of R2211-ID171.}
\figsetgrpend

\figsetend
\begin{figure}
\figurenum{B1}
\plotone{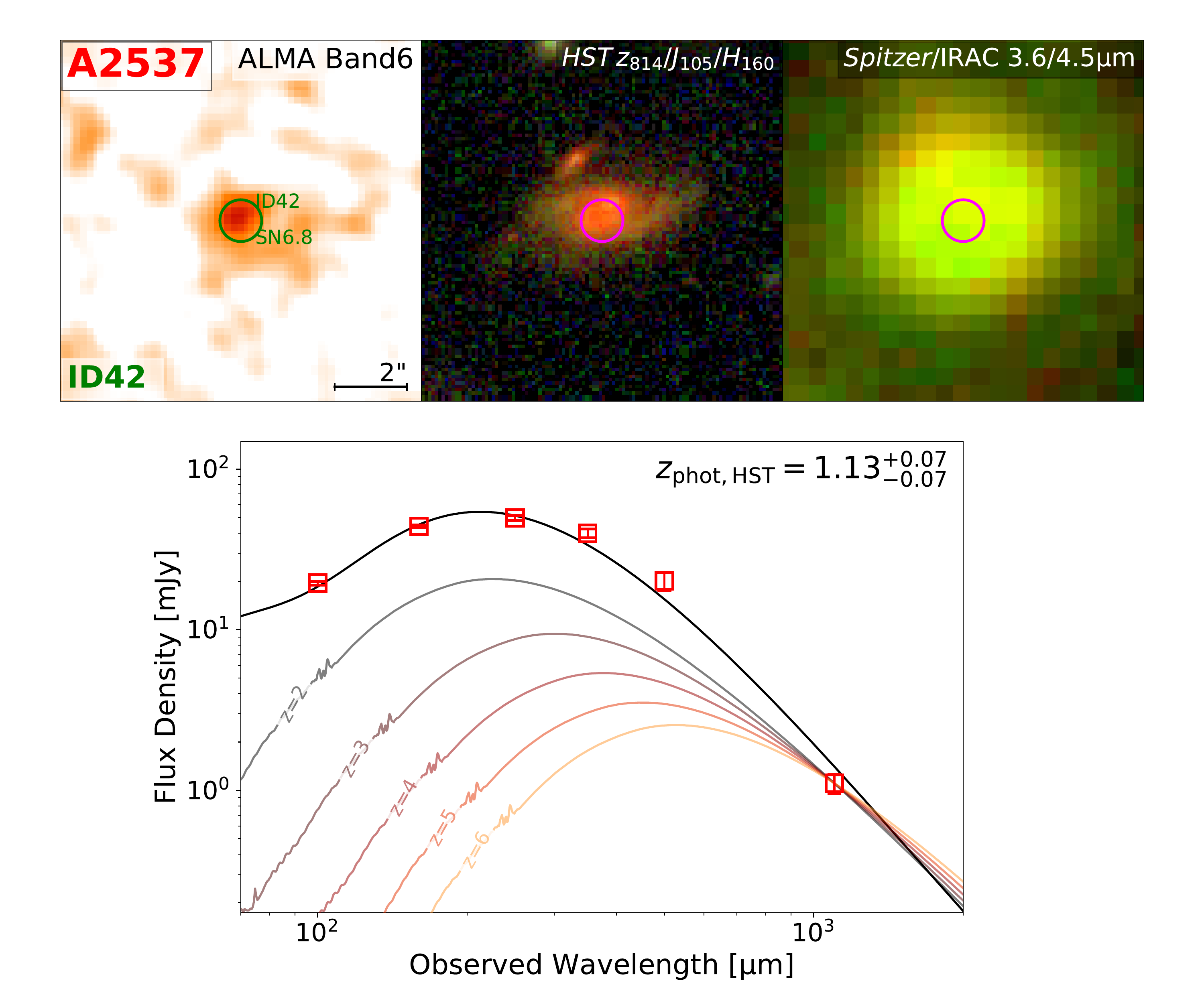}
\caption{\textit{Top}: Postage stamp images of A2537-ID42 ($\mathrm{S/N}_\mathrm{ALMA}=6.8$), one of the 105 sources in the main ALCS-\herschel\ joint sample.
From the left to the right we show the ALMA Band-6 continuum image, \hst\ F814W/F105W/F160W true color image and \spitzer/IRAC 3.6/4.5\,\micron\ true color image.
The galaxy is in the center of each image labeled by green or magenta circles.
Other ALCS sources in the field, if exist, are also labeled out with their S/N and ID noted.
A two-arcsec scale bar is plotted in the lower right corner of ALMA continuum image.
\textit{Bottom}: Far-IR SED of A2537-ID42.
\herschel\ and ALMA photometric measurements are shown as open red squares (upper limits are at $3\sigma$).
Best-fit SED model derived with \textsc{magphys} at $z_\mathrm{best}$ is shown as solid black line.
For sources without spectroscopic confirmation (\zsp), we also plot the spectral templates of (U)LIRG ($L_\mathrm{IR}=10^{12}$\,\lsun; \citealt{rieke09}) at $z=2-6$ (normalized to the 1.15\,mm flux density) for comparison.
The complete figure set (105 images) is available in the online journal.
}
\label{fs:01_main}
\end{figure}

\figsetstart
\figsetnum{B2}
\figsettitle{Postage stamp images and far-IR SEDs of 20 sources in the secondary ALCS-\herschel\ sample.}
\figsetgrpstart
\figsetgrpnum{B2.1}
\figsetgrptitle{A2537-ID24}
\figsetplot{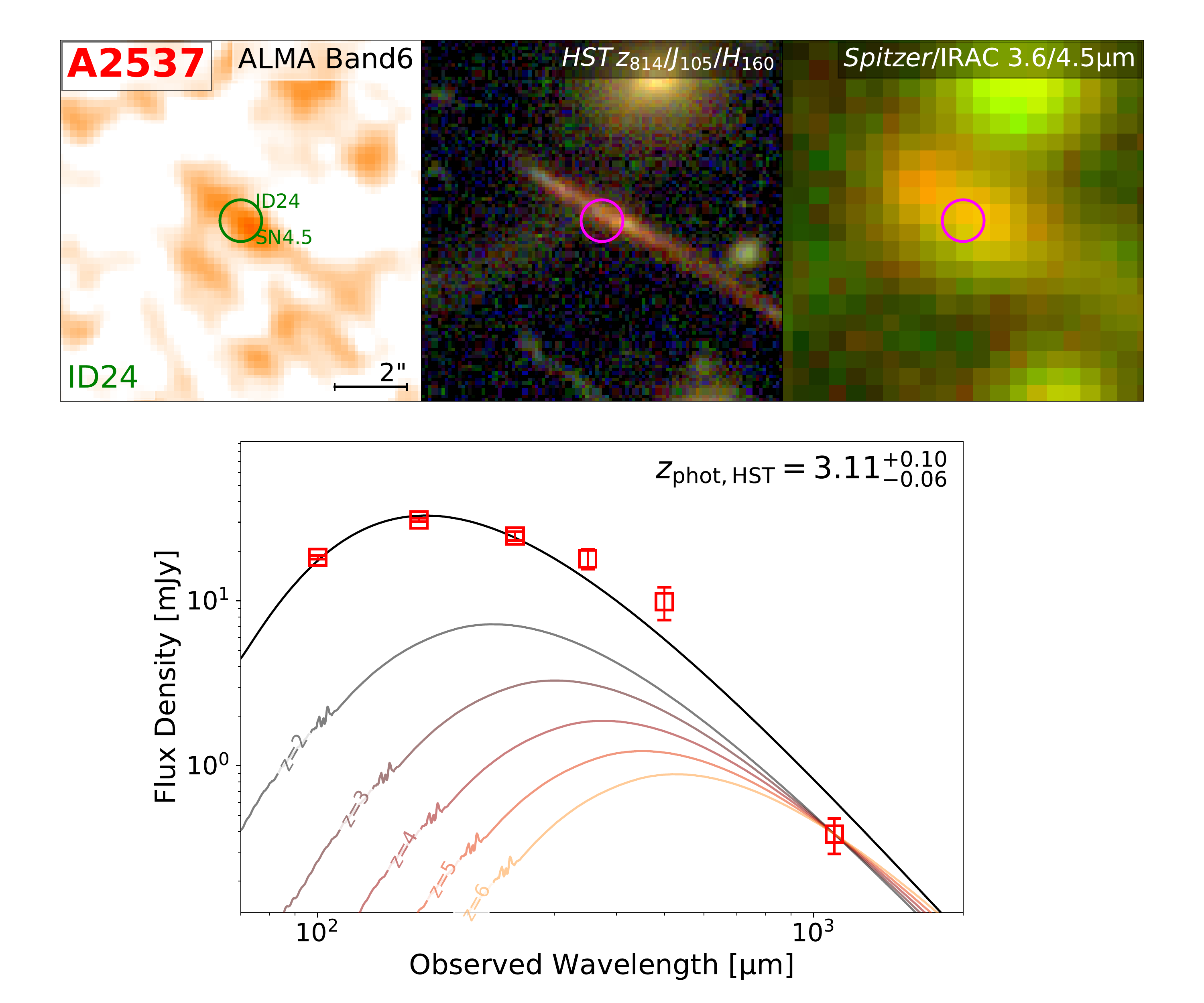}
\figsetgrpnote{Postage stamp images (top) and far-IR SED (bottom) of A2537-ID24.}
\figsetgrpend

\figsetgrpstart
\figsetgrpnum{B2.2}
\figsetgrptitle{A2744-ID47}
\figsetplot{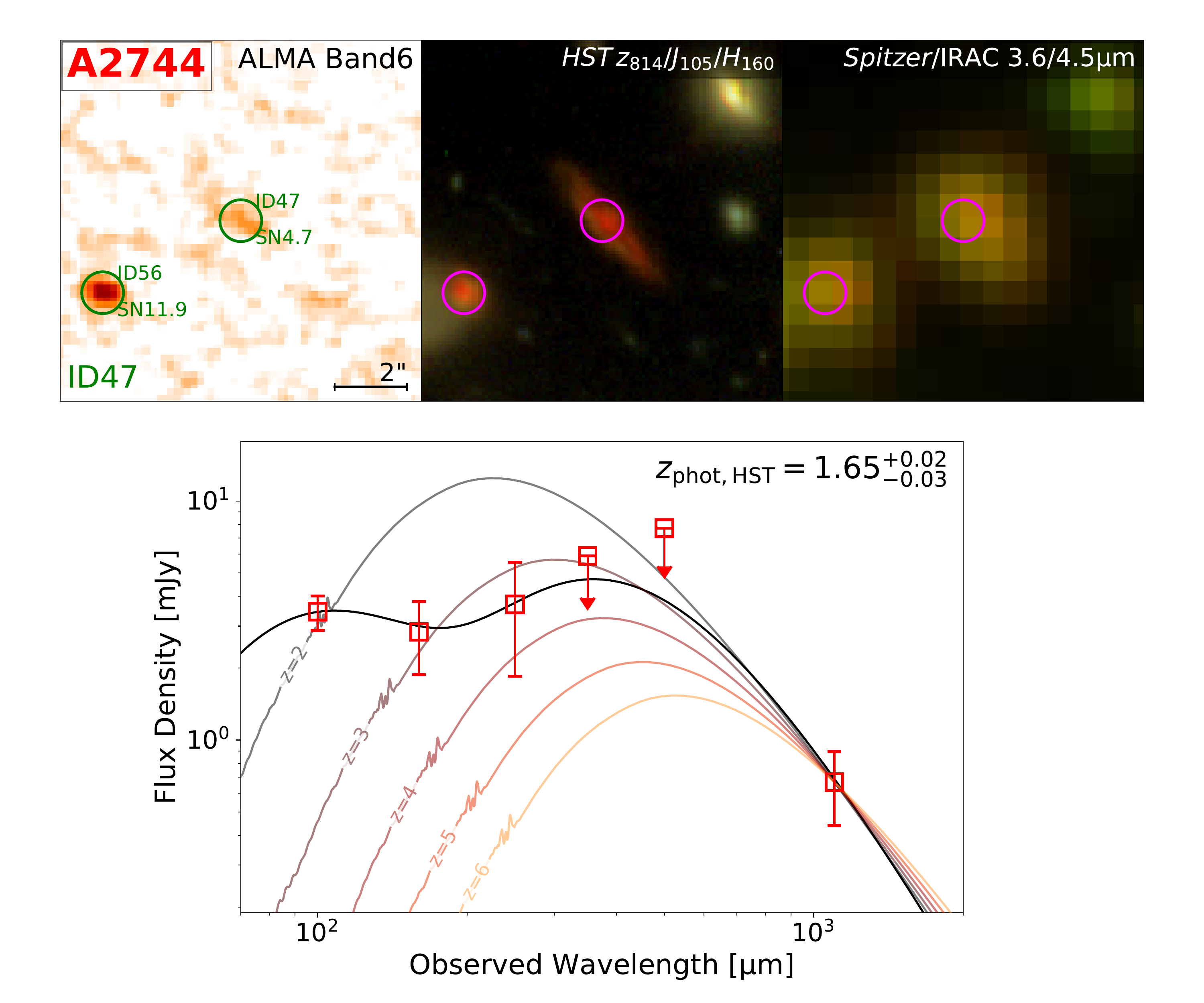}
\figsetgrpnote{Postage stamp images (top) and far-IR SED (bottom) of A2744-ID47.}
\figsetgrpend

\figsetgrpstart
\figsetgrpnum{B2.3}
\figsetgrptitle{A2744-ID178}
\figsetplot{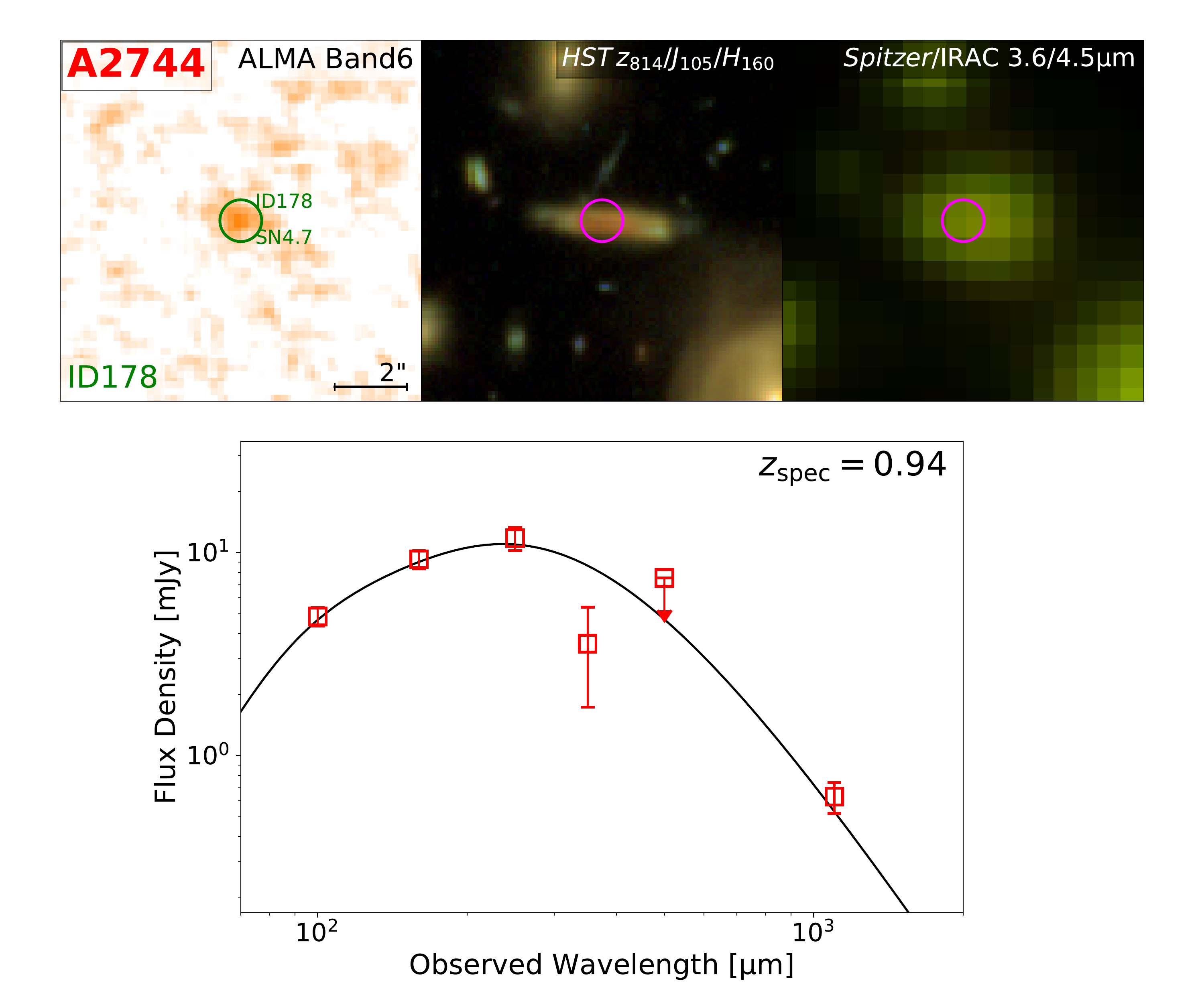}
\figsetgrpnote{Postage stamp images (top) and far-IR SED (bottom) of A2744-ID178.}
\figsetgrpend

\figsetgrpstart
\figsetgrpnum{B2.4}
\figsetgrptitle{A2744-ID227}
\figsetplot{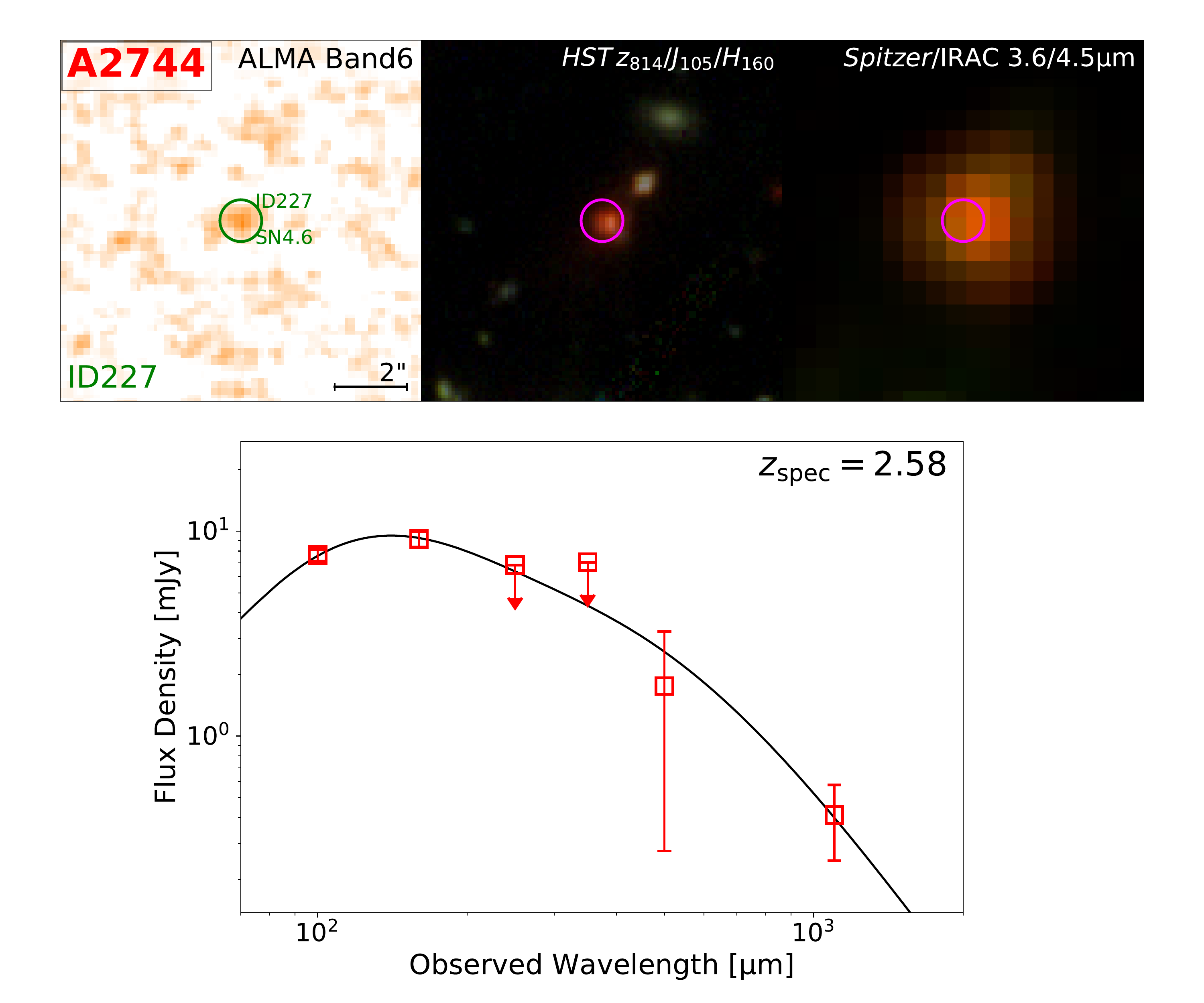}
\figsetgrpnote{Postage stamp images (top) and far-IR SED (bottom) of A2744-ID227.}
\figsetgrpend

\figsetgrpstart
\figsetgrpnum{B2.5}
\figsetgrptitle{A370-ID27}
\figsetplot{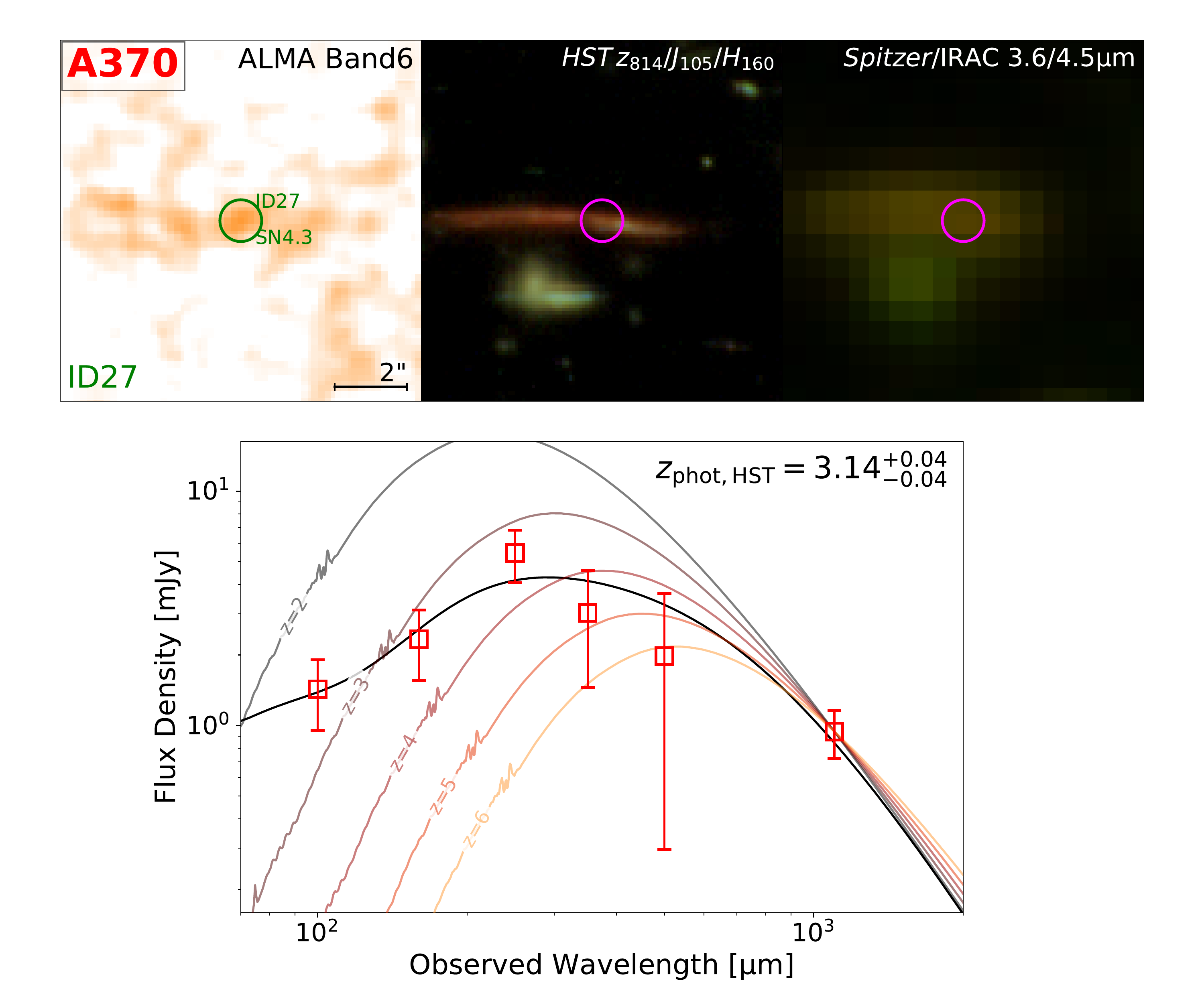}
\figsetgrpnote{Postage stamp images (top) and far-IR SED (bottom) of A370-ID27.}
\figsetgrpend

\figsetgrpstart
\figsetgrpnum{B2.6}
\figsetgrptitle{A383-ID24}
\figsetplot{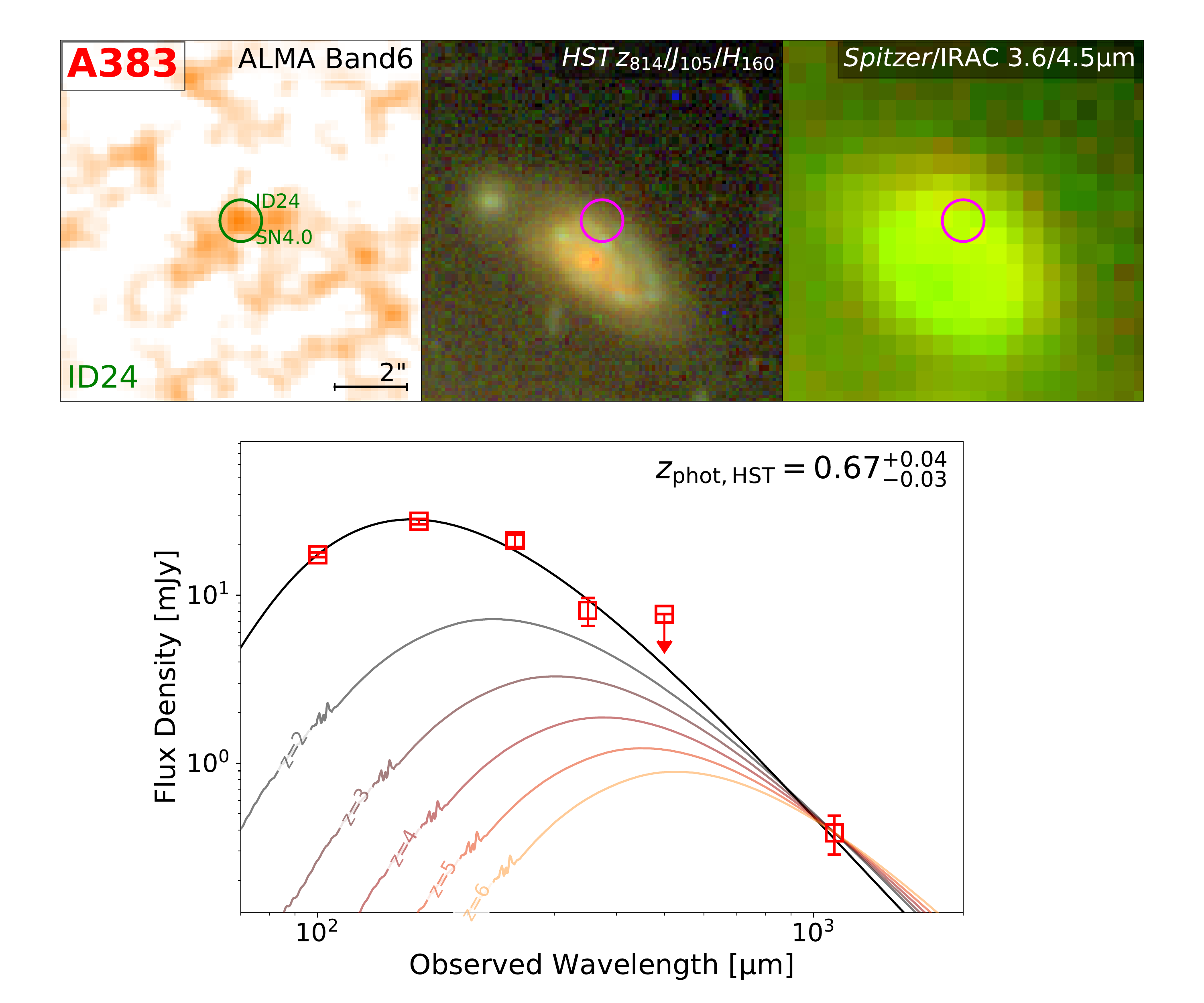}
\figsetgrpnote{Postage stamp images (top) and far-IR SED (bottom) of A383-ID24.}
\figsetgrpend

\figsetgrpstart
\figsetgrpnum{B2.7}
\figsetgrptitle{A383-ID50}
\figsetplot{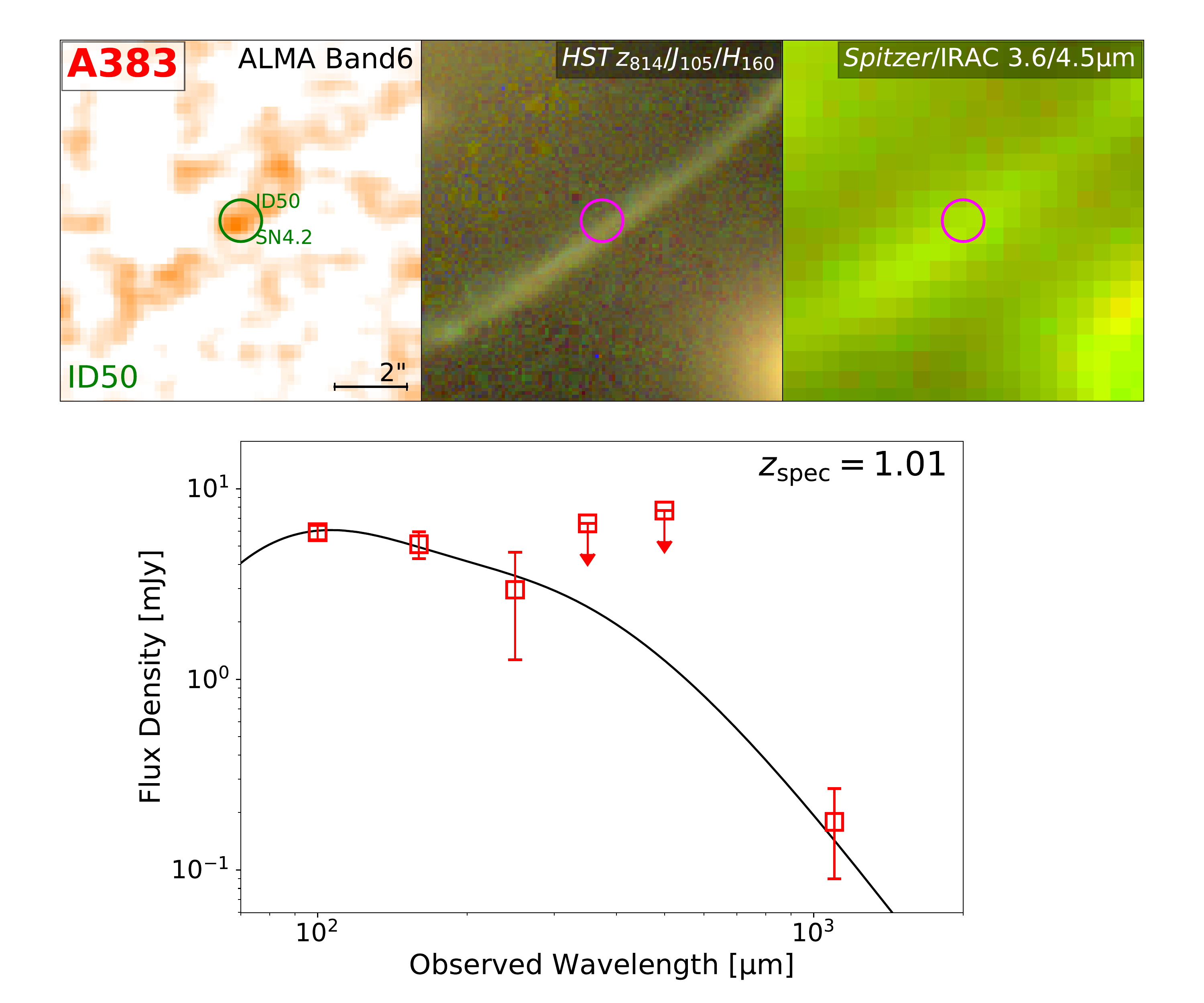}
\figsetgrpnote{Postage stamp images (top) and far-IR SED (bottom) of A383-ID50.}
\figsetgrpend

\figsetgrpstart
\figsetgrpnum{B2.8}
\figsetgrptitle{AS295-ID269}
\figsetplot{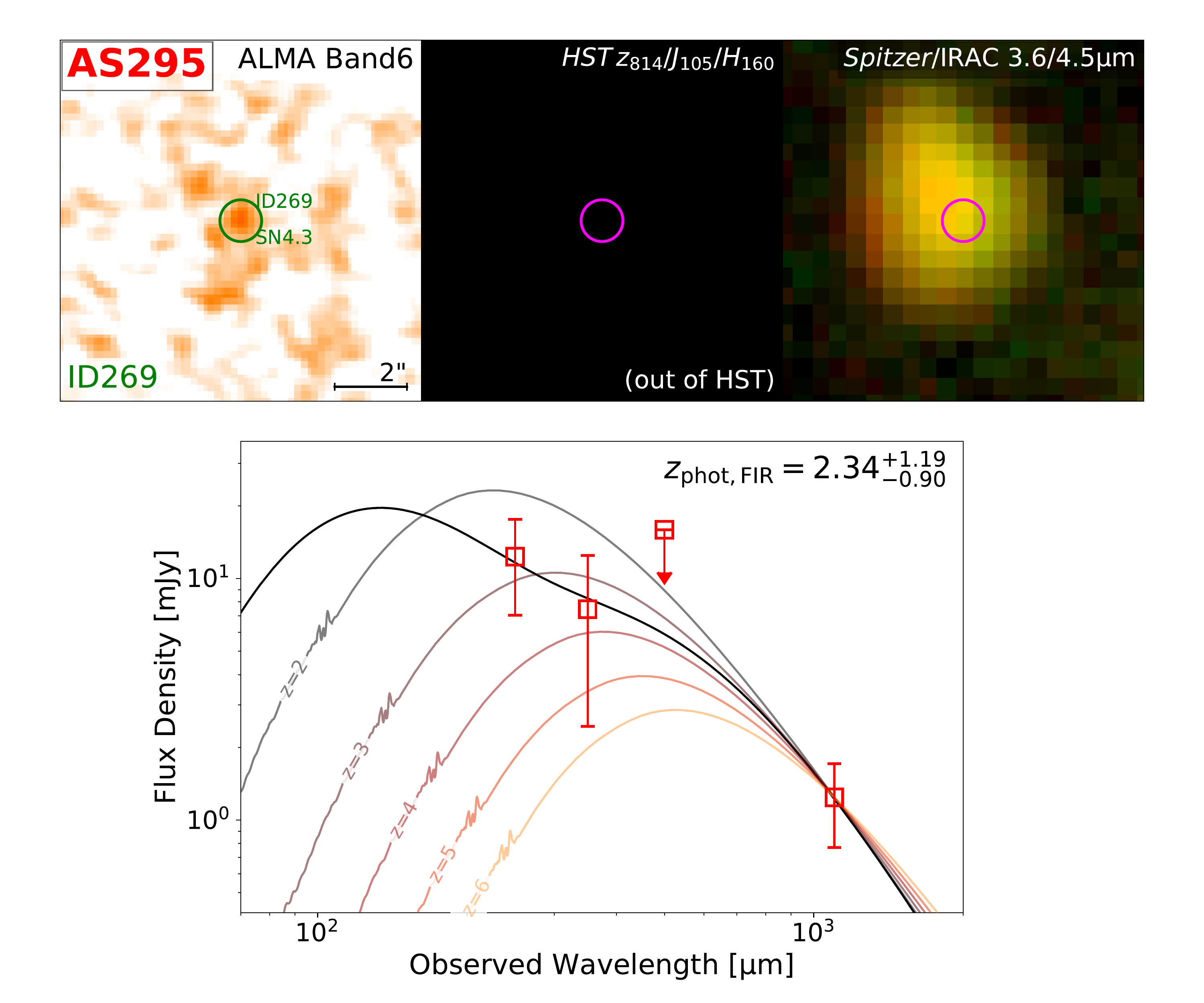}
\figsetgrpnote{Postage stamp images (top) and far-IR SED (bottom) of AS295-ID269.}
\figsetgrpend

\figsetgrpstart
\figsetgrpnum{B2.9}
\figsetgrptitle{M0416-ID138}
\figsetplot{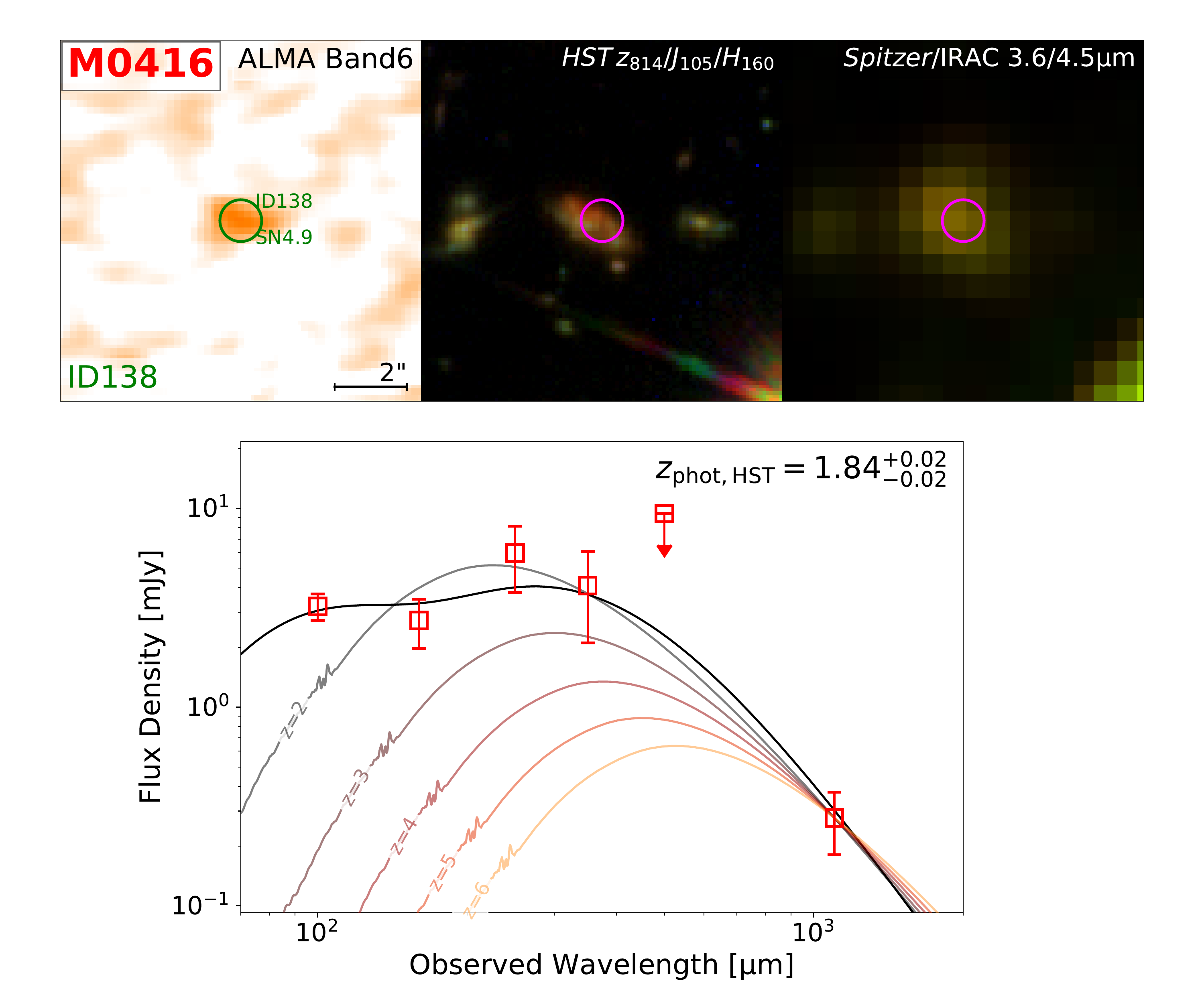}
\figsetgrpnote{Postage stamp images (top) and far-IR SED (bottom) of M0416-ID138.}
\figsetgrpend

\figsetgrpstart
\figsetgrpnum{B2.10}
\figsetgrptitle{M0416-ID156}
\figsetplot{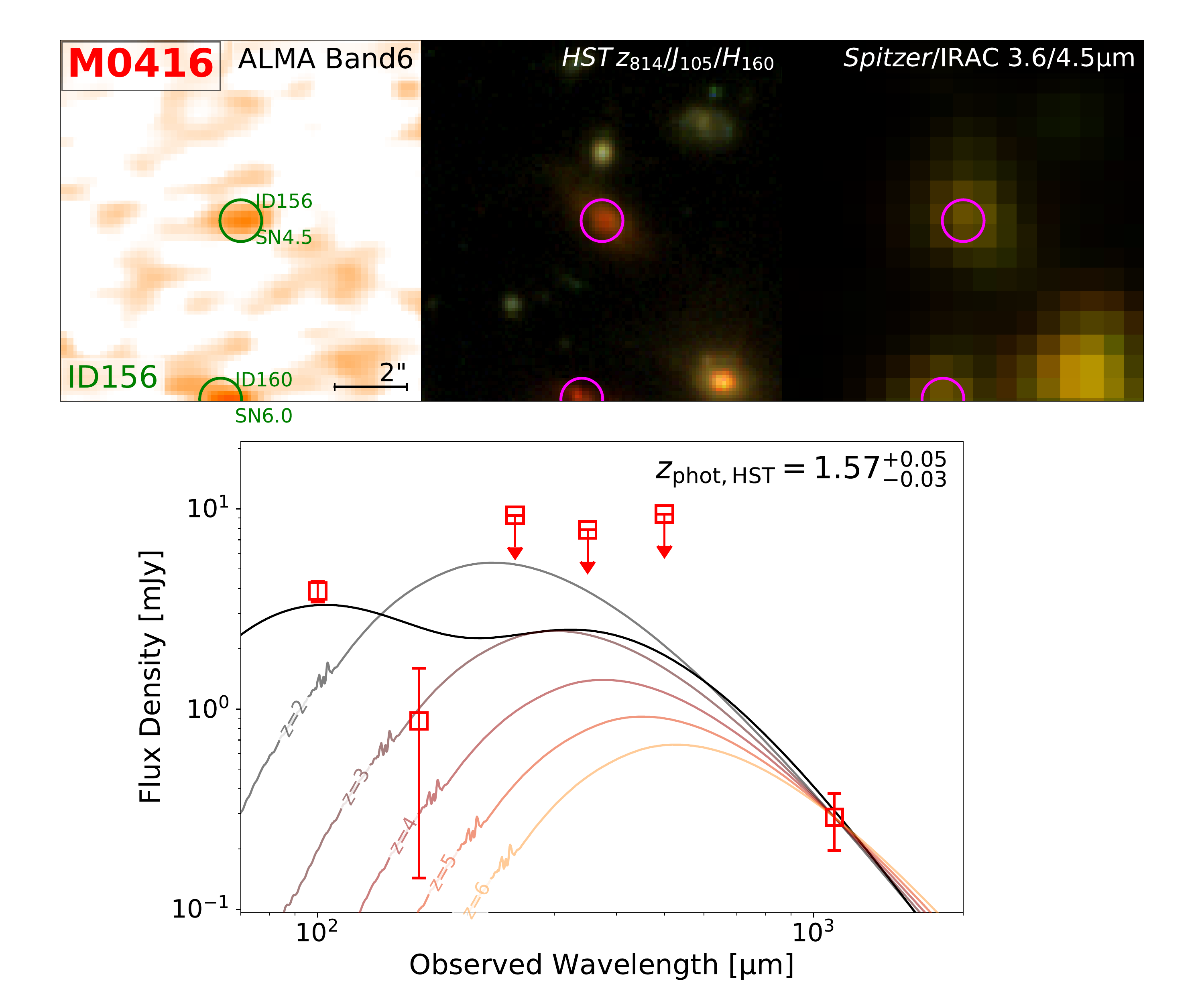}
\figsetgrpnote{Postage stamp images (top) and far-IR SED (bottom) of M0416-ID156.}
\figsetgrpend

\figsetgrpstart
\figsetgrpnum{B2.11}
\figsetgrptitle{M1115-ID33}
\figsetplot{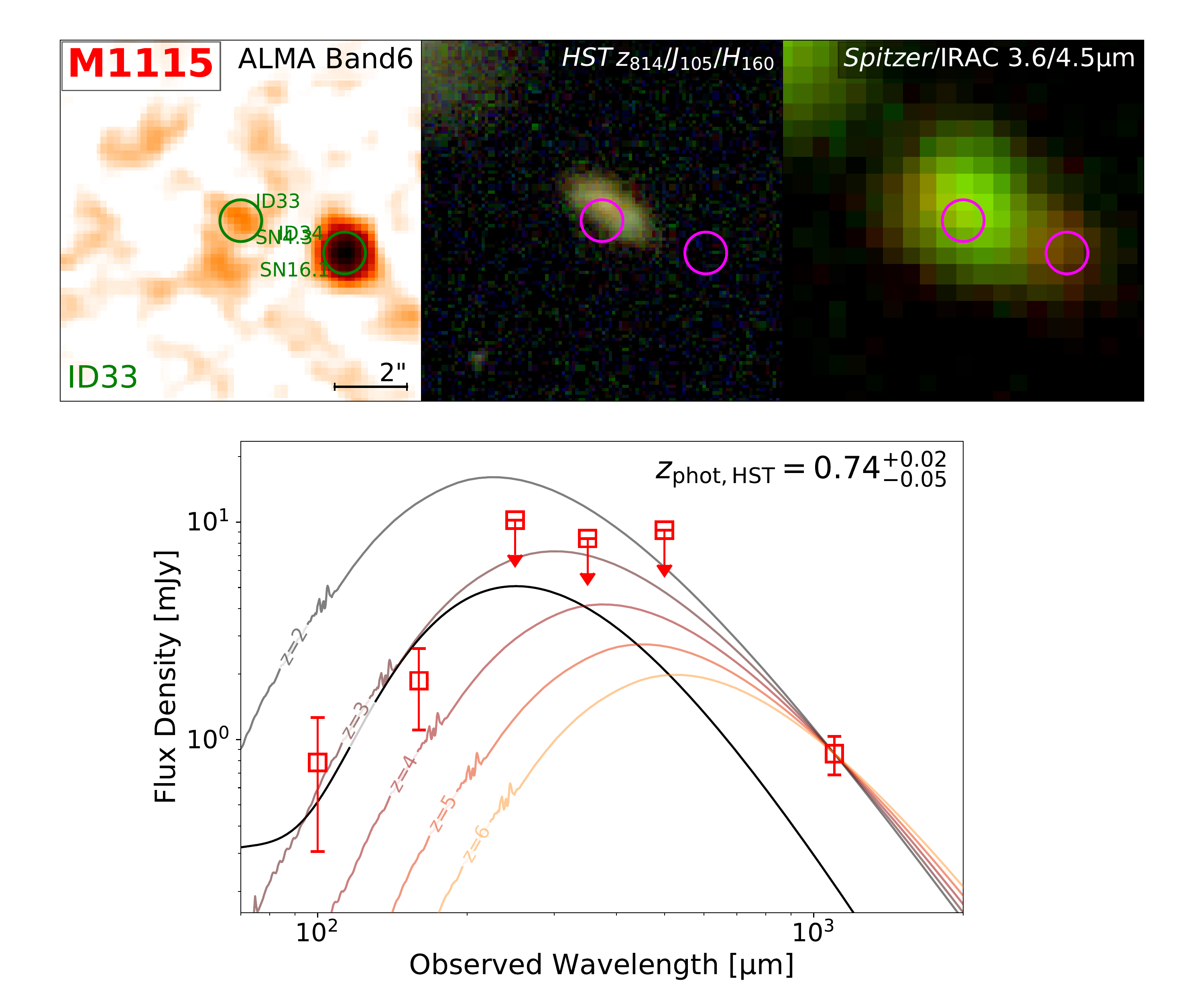}
\figsetgrpnote{Postage stamp images (top) and far-IR SED (bottom) of M1115-ID33.}
\figsetgrpend

\figsetgrpstart
\figsetgrpnum{B2.12}
\figsetgrptitle{M1149-ID27}
\figsetplot{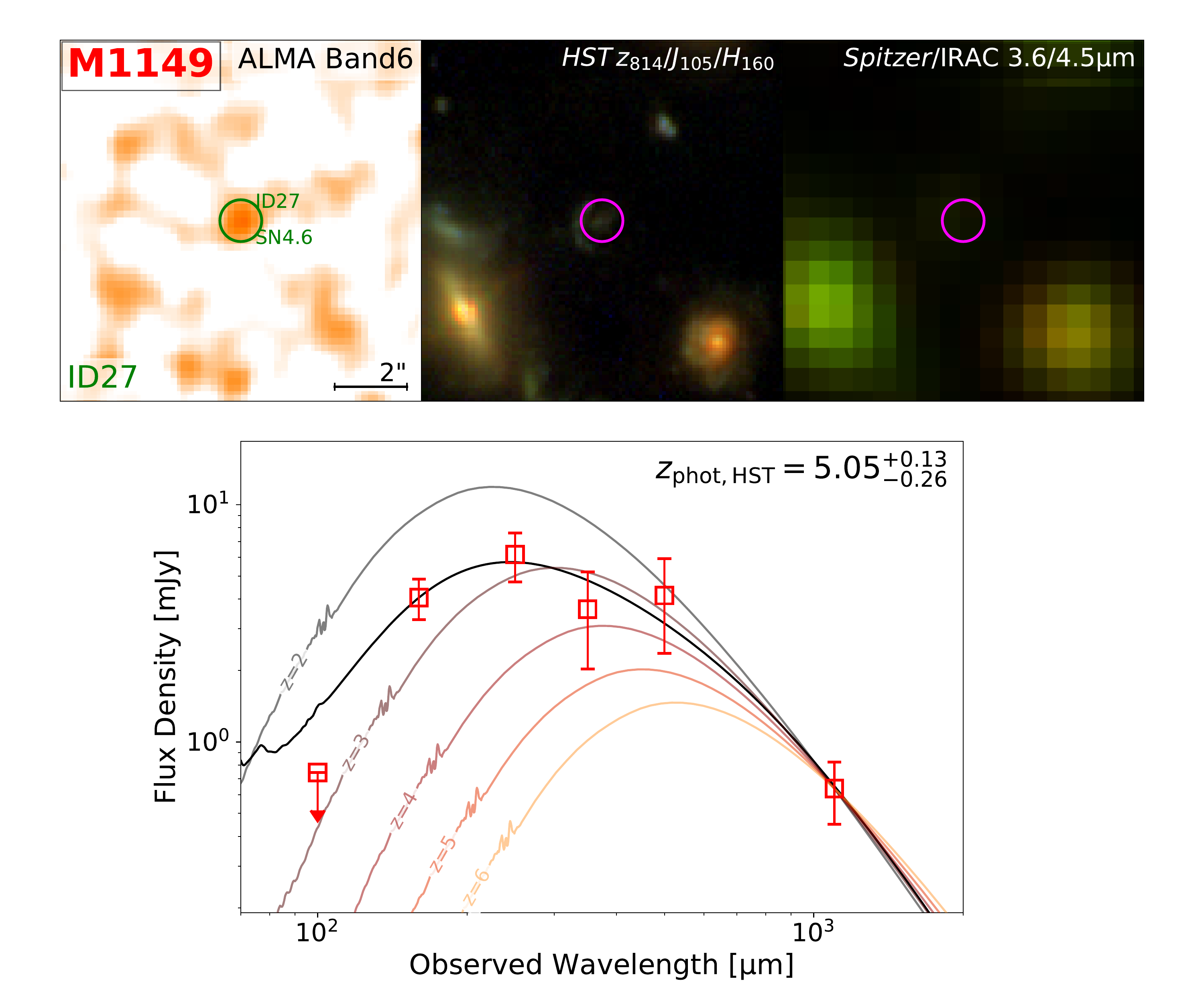}
\figsetgrpnote{Postage stamp images (top) and far-IR SED (bottom) of M1149-ID27.}
\figsetgrpend

\figsetgrpstart
\figsetgrpnum{B2.13}
\figsetgrptitle{M1149-ID95}
\figsetplot{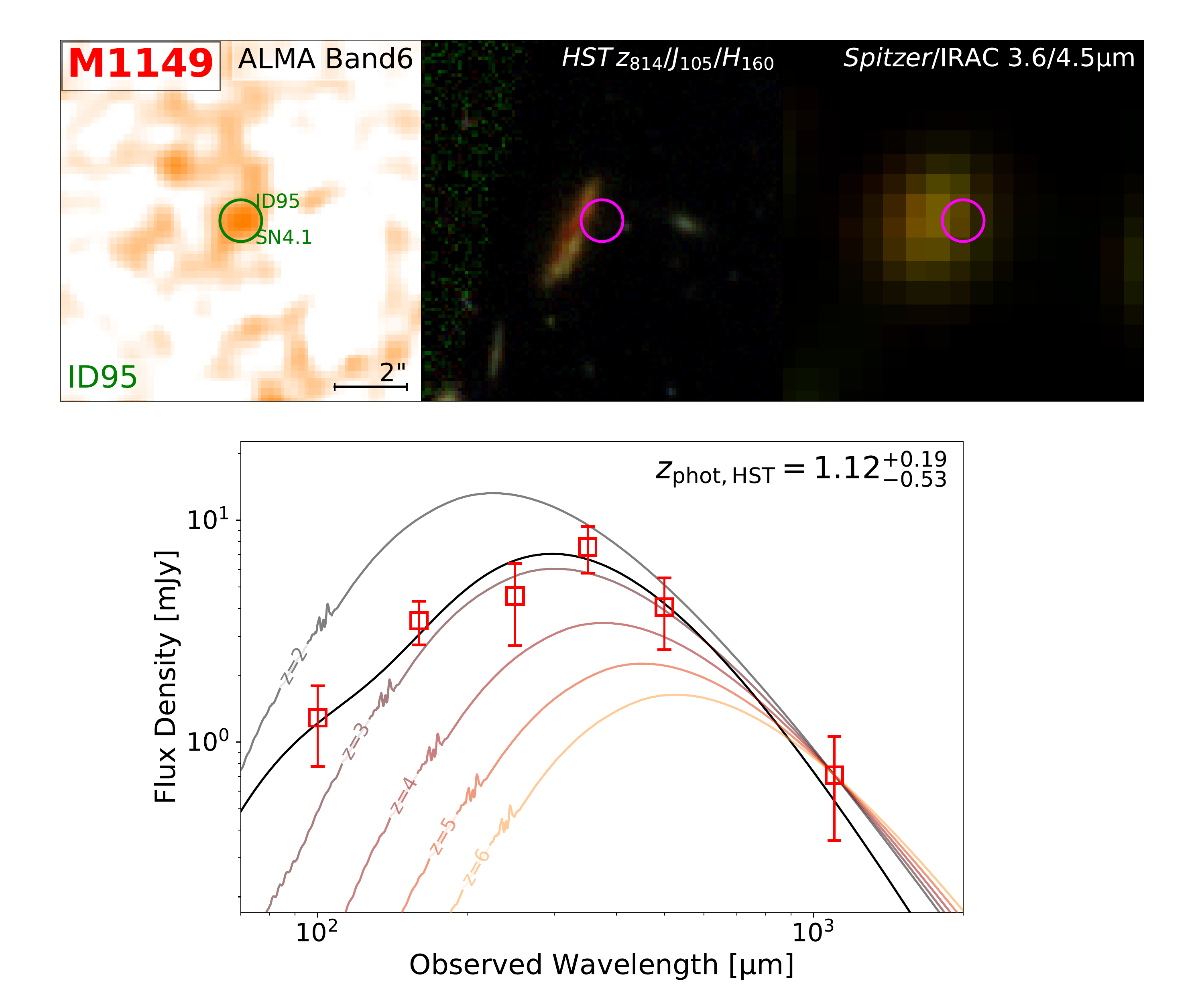}
\figsetgrpnote{Postage stamp images (top) and far-IR SED (bottom) of M1149-ID95.}
\figsetgrpend

\figsetgrpstart
\figsetgrpnum{B2.14}
\figsetgrptitle{M1206-ID38}
\figsetplot{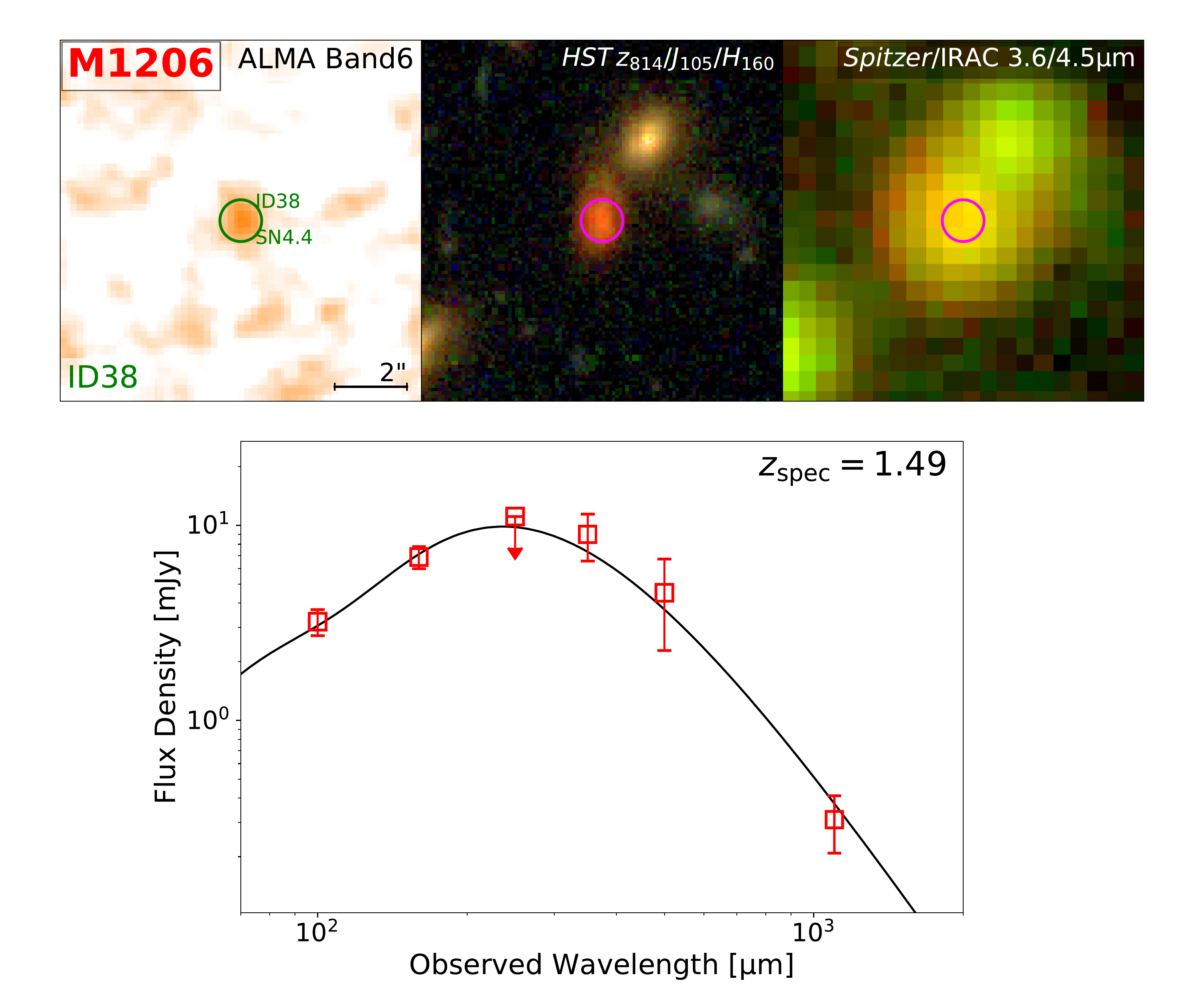}
\figsetgrpnote{Postage stamp images (top) and far-IR SED (bottom) of M1206-ID38.}
\figsetgrpend

\figsetgrpstart
\figsetgrpnum{B2.15}
\figsetgrptitle{M1206-ID84}
\figsetplot{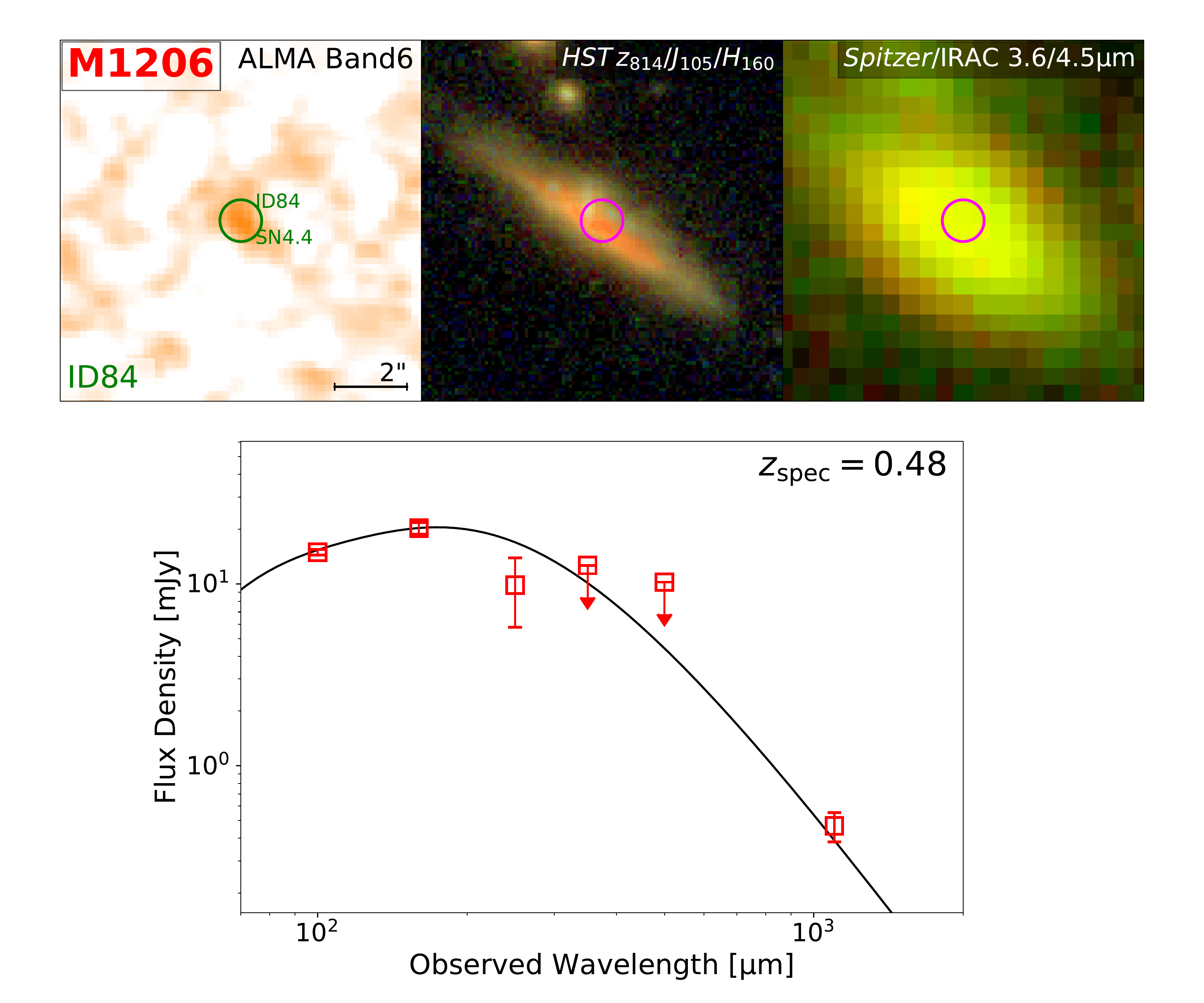}
\figsetgrpnote{Postage stamp images (top) and far-IR SED (bottom) of M1206-ID84.}
\figsetgrpend

\figsetgrpstart
\figsetgrpnum{B2.16}
\figsetgrptitle{M1423-ID52}
\figsetplot{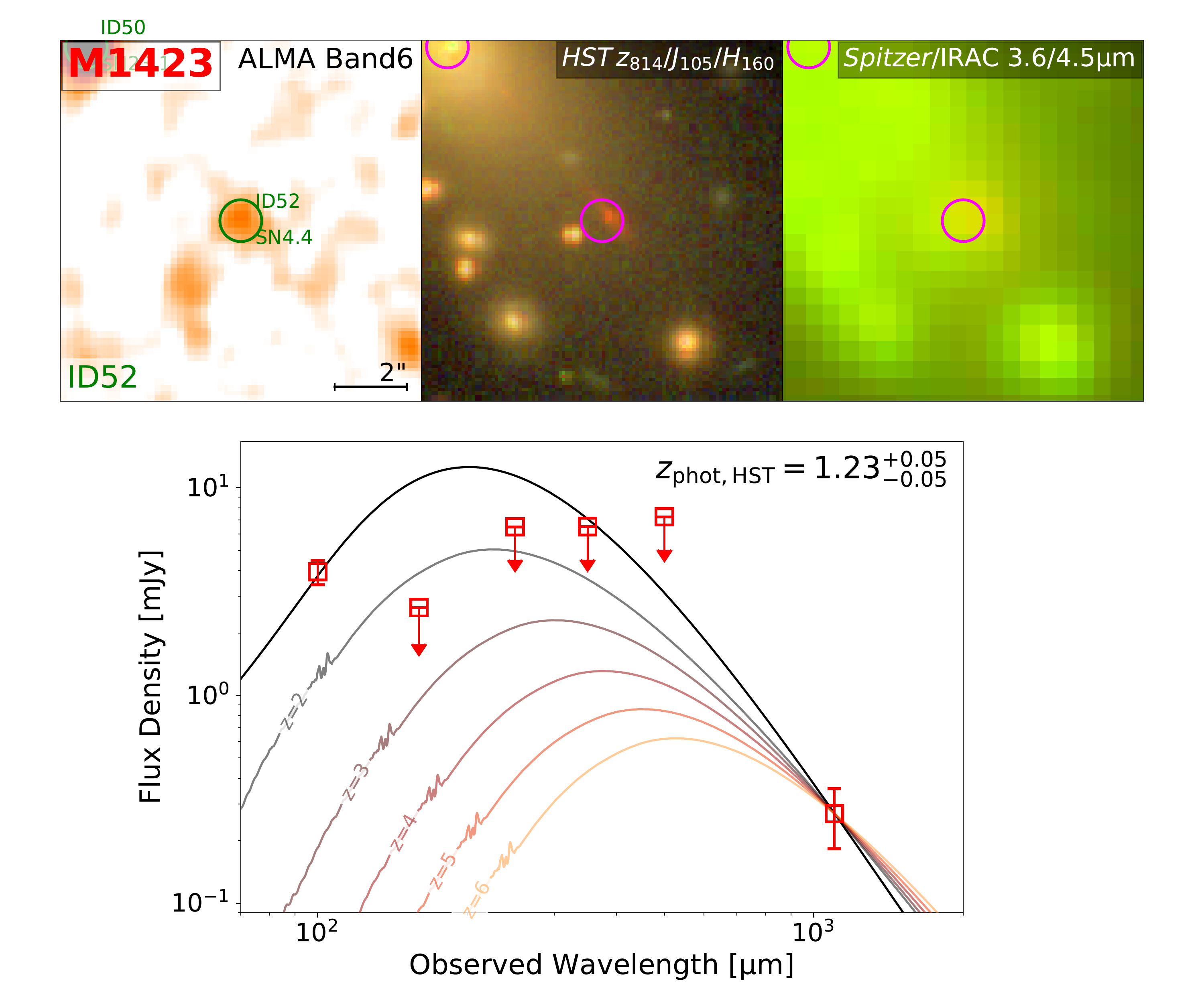}
\figsetgrpnote{Postage stamp images (top) and far-IR SED (bottom) of M1423-ID52.}
\figsetgrpend

\figsetgrpstart
\figsetgrpnum{B2.17}
\figsetgrptitle{M1423-ID76}
\figsetplot{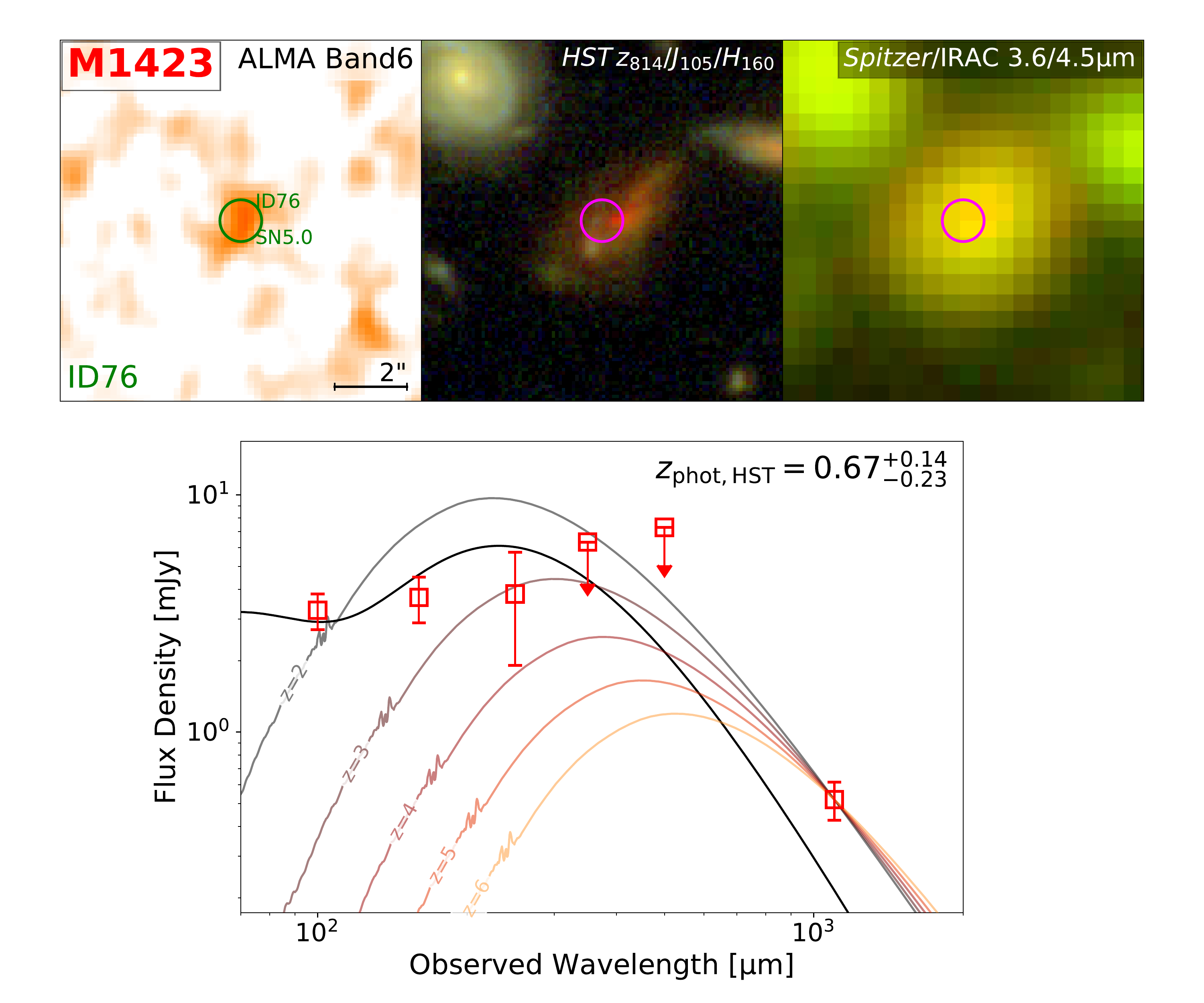}
\figsetgrpnote{Postage stamp images (top) and far-IR SED (bottom) of M1423-ID76.}
\figsetgrpend

\figsetgrpstart
\figsetgrpnum{B2.18}
\figsetgrptitle{M1931-ID69}
\figsetplot{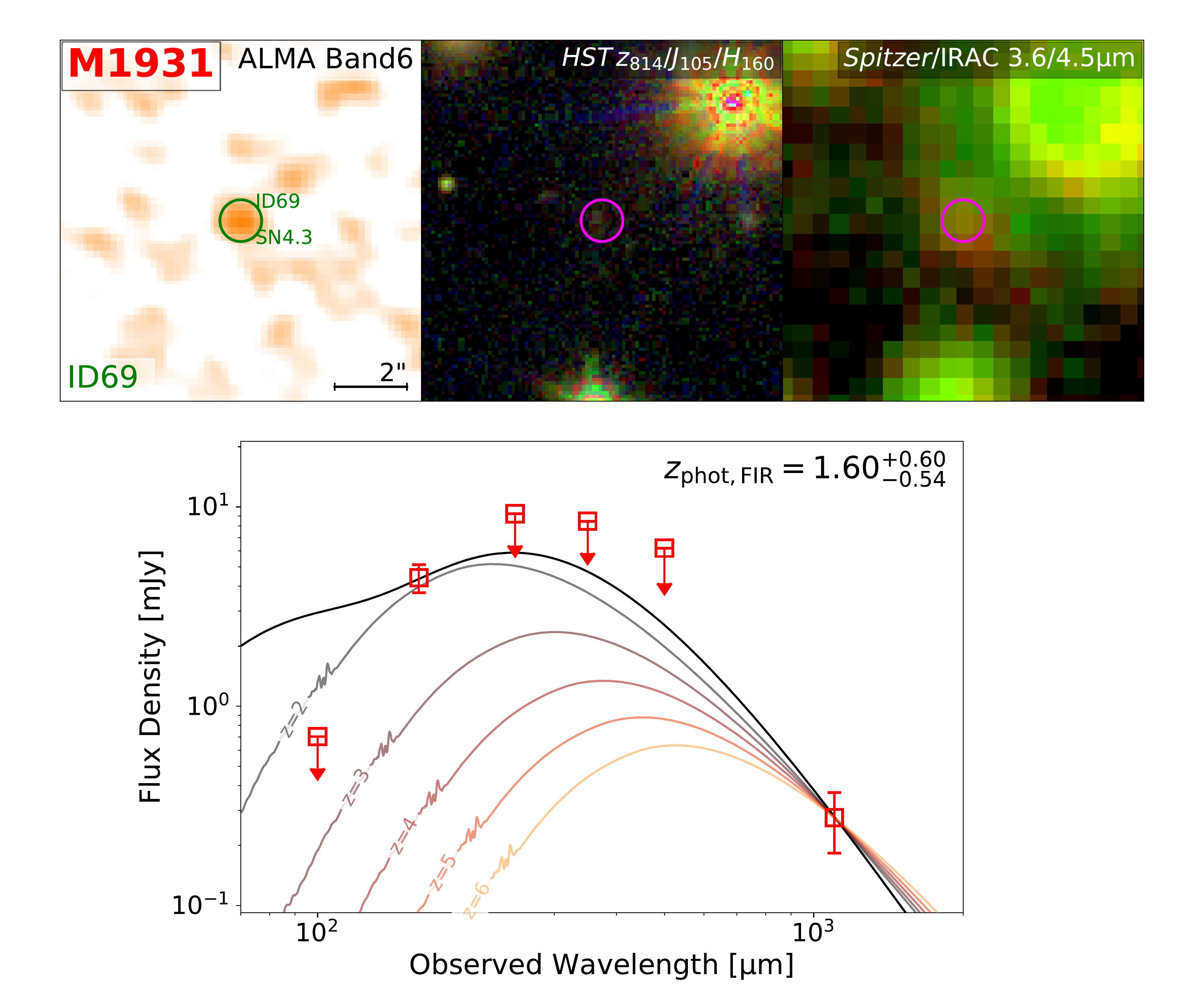}
\figsetgrpnote{Postage stamp images (top) and far-IR SED (bottom) of M1931-ID69.}
\figsetgrpend

\figsetgrpstart
\figsetgrpnum{B2.19}
\figsetgrptitle{M2129-ID62}
\figsetplot{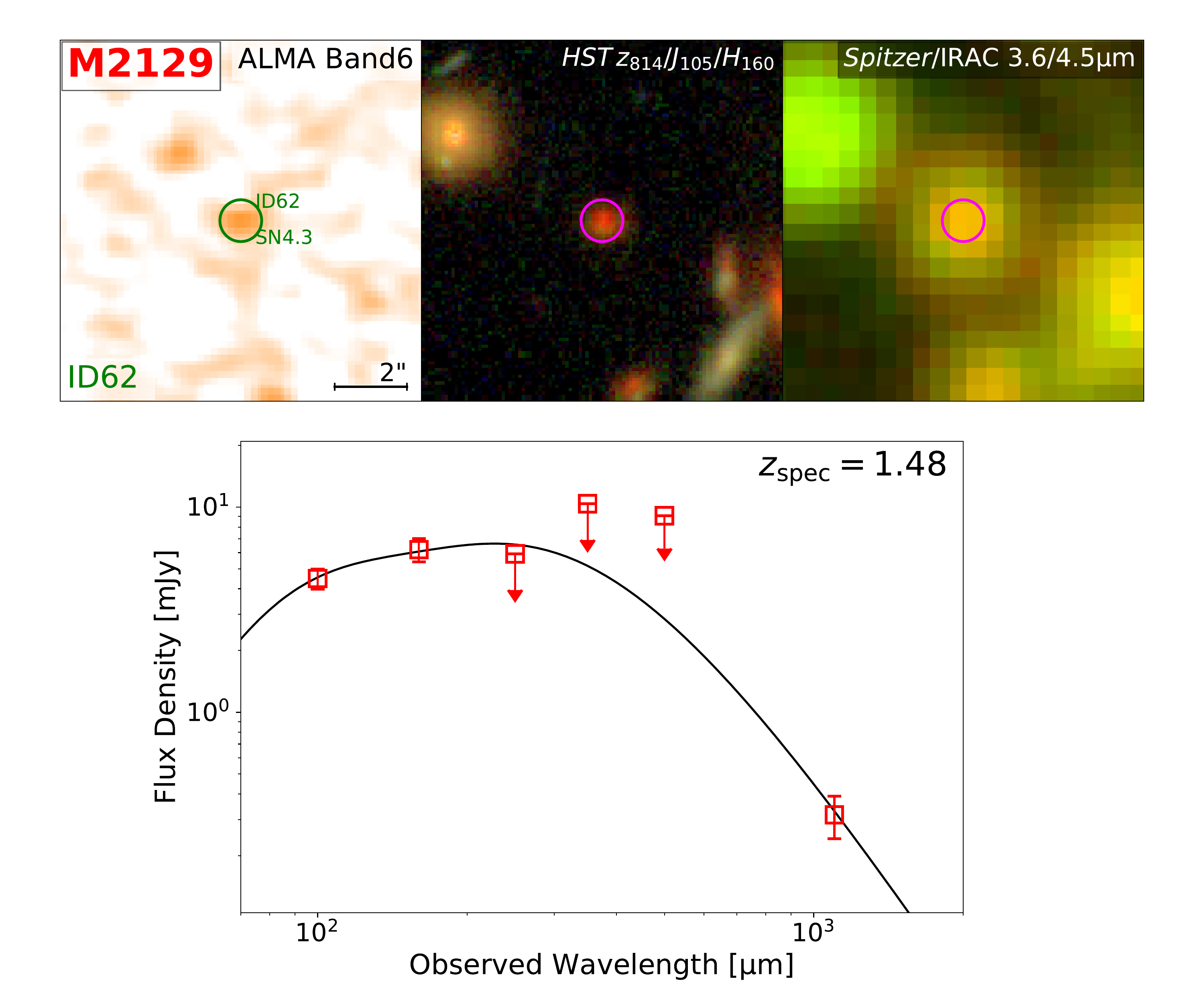}
\figsetgrpnote{Postage stamp images (top) and far-IR SED (bottom) of M2129-ID62.}
\figsetgrpend

\figsetgrpstart
\figsetgrpnum{B2.20}
\figsetgrptitle{R0949-ID119}
\figsetplot{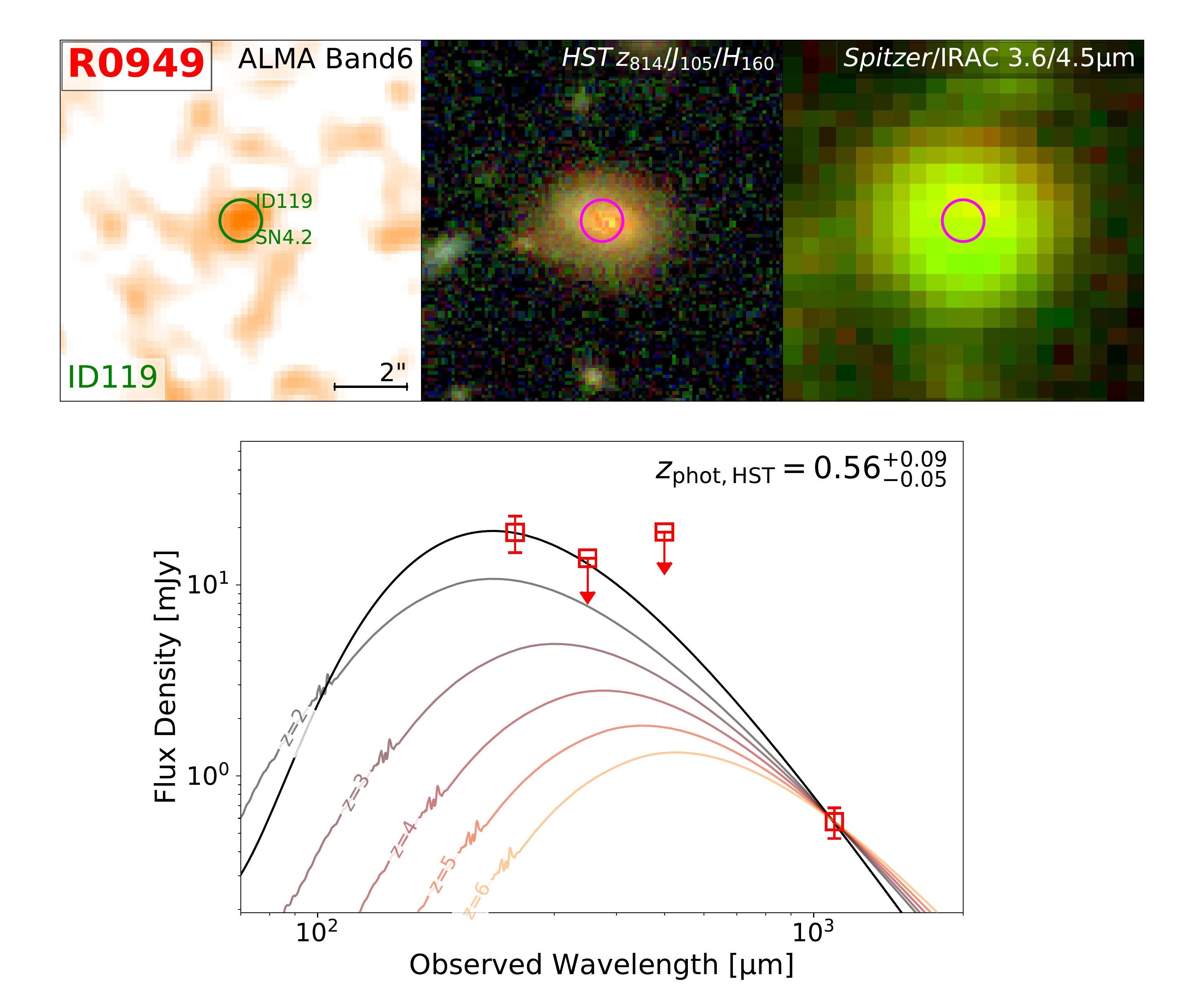}
\figsetgrpnote{Postage stamp images (top) and far-IR SED (bottom) of R0949-ID119.}
\figsetgrpend

\figsetend
\begin{figure}
\figurenum{B2}
\plotone{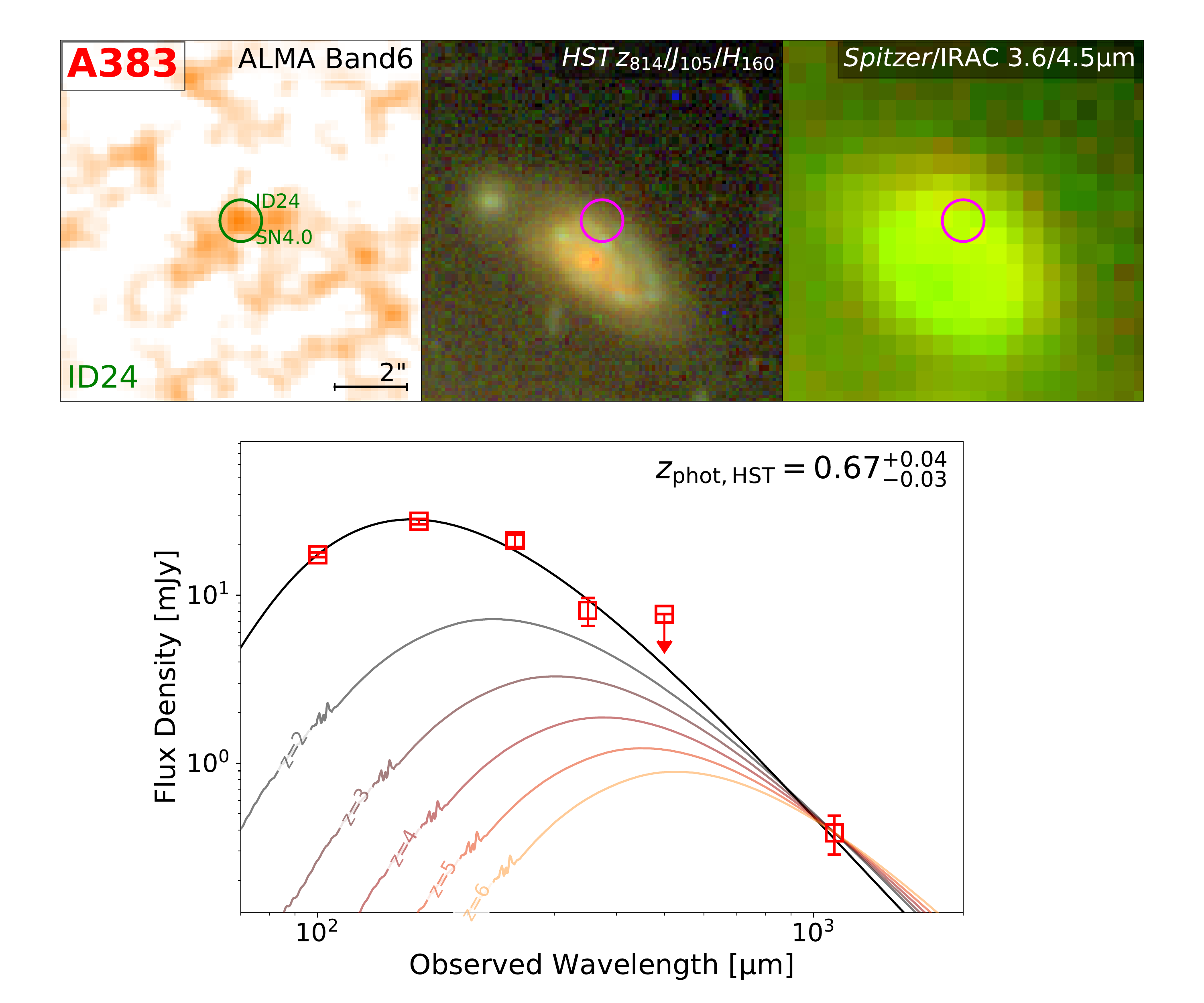}
\caption{Same as Figure~\ref{fs:01_main} but for A383-ID24 ($\mathrm{S/N}_\mathrm{ALMA}=4.0$), one of the 20 sources in the secondary ALCS-\herschel\ joint sample.
The complete figure set (20 images) is available in the online journal.
}
\label{fs:02_supp}
\end{figure}

In addition to these, we also show the ALMA-\hst-\spitzer\ postage stamps images of ten \herschel-faint galaxies that have \zsp\ or catalogued \hst\ \zph\ (Figure~\ref{fig:hdrop_z}), and the remaining 17 \herschel-faint galaxies without \zsp\ or catalogued \hst\ \zph\ (i.e., optical/near-IR-dark; Figure~\ref{fig:hdrop_dark}).

\renewcommand{\thefigure}{B3}
\begin{figure*}
\centering
\includegraphics[width=0.9\linewidth]{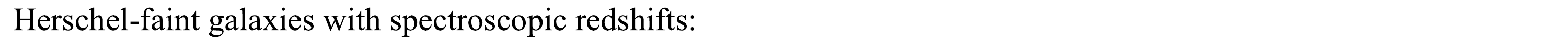}
\includegraphics[width=0.45\linewidth]{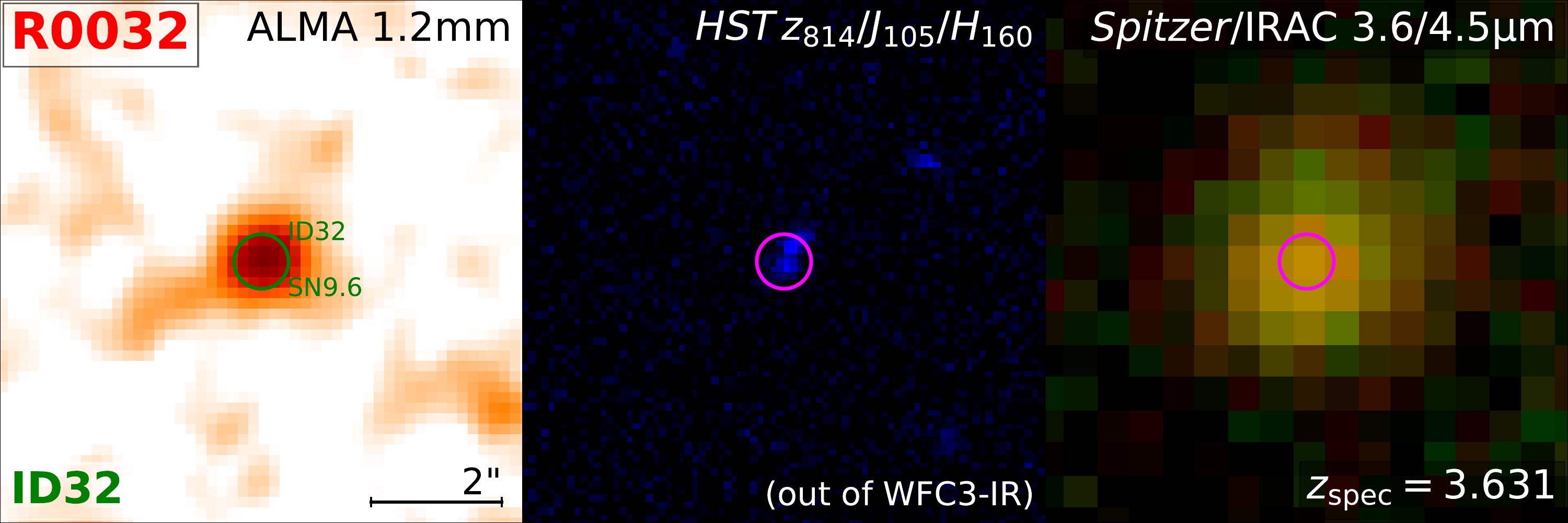}
\includegraphics[width=0.45\linewidth]{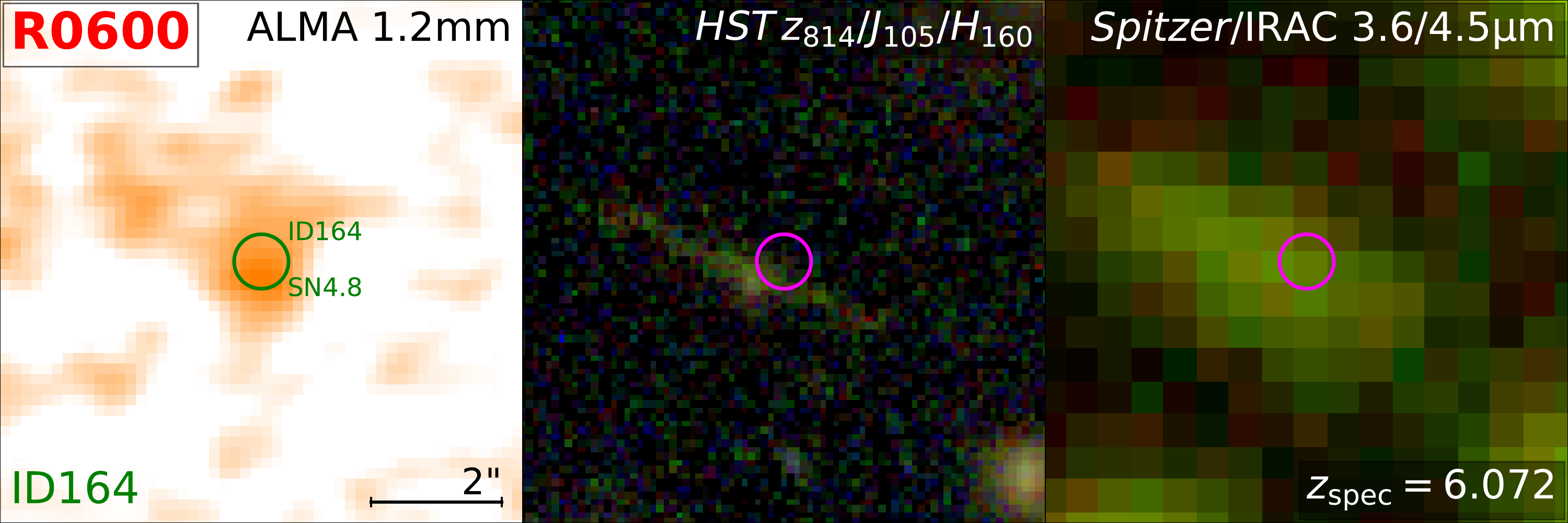}
\includegraphics[width=0.9\linewidth]{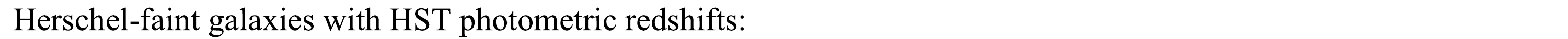}
\includegraphics[width=0.45\linewidth]{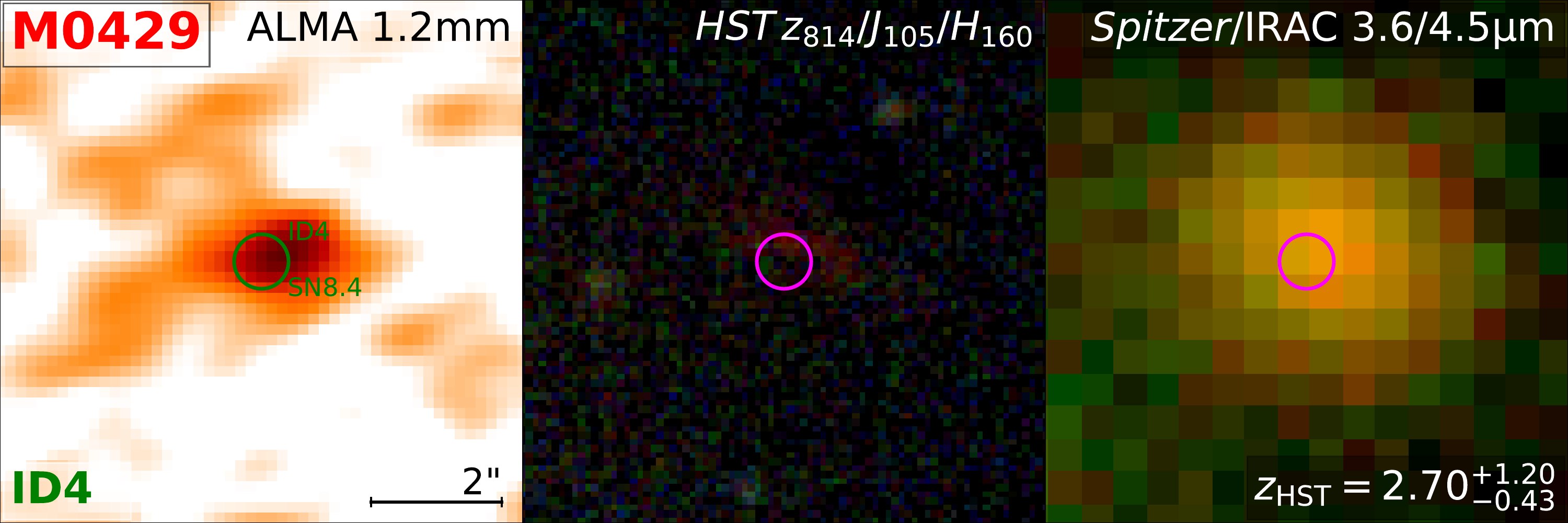}
\includegraphics[width=0.45\linewidth]{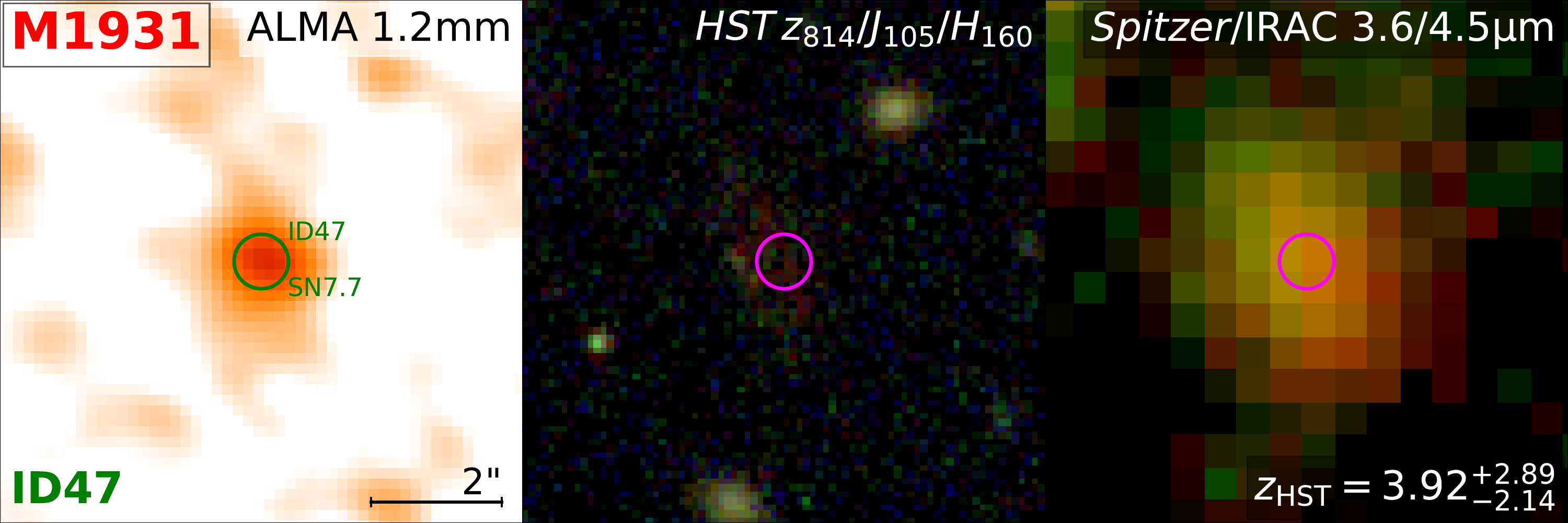}
\includegraphics[width=0.45\linewidth]{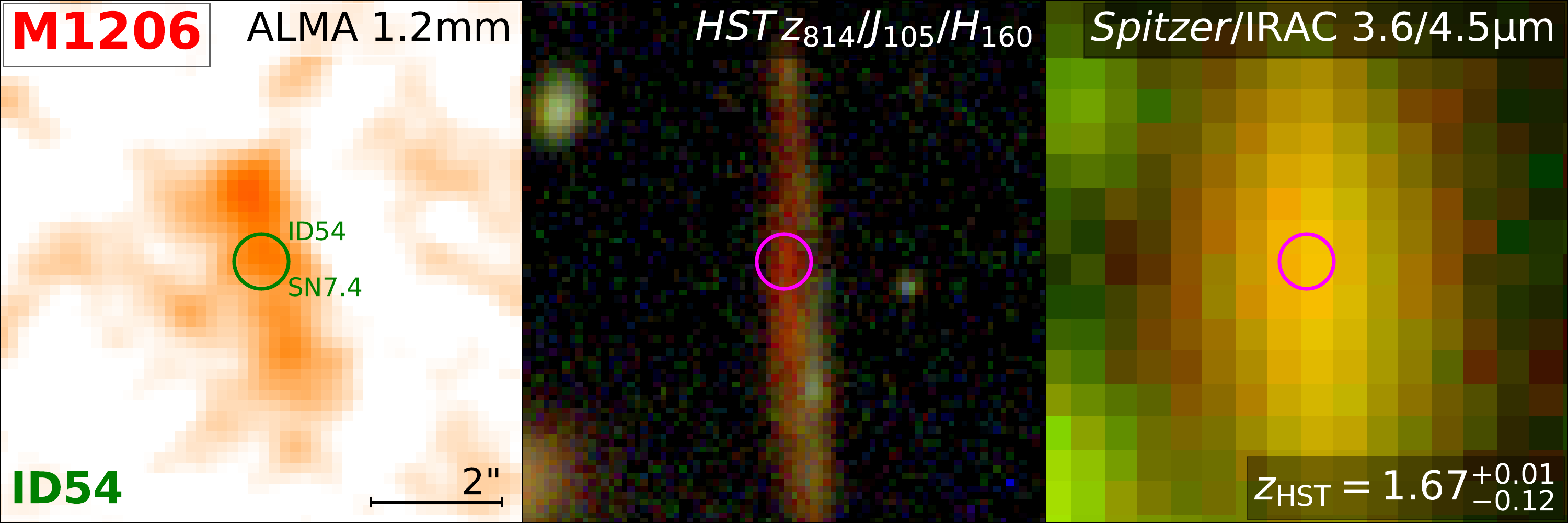}
\includegraphics[width=0.45\linewidth]{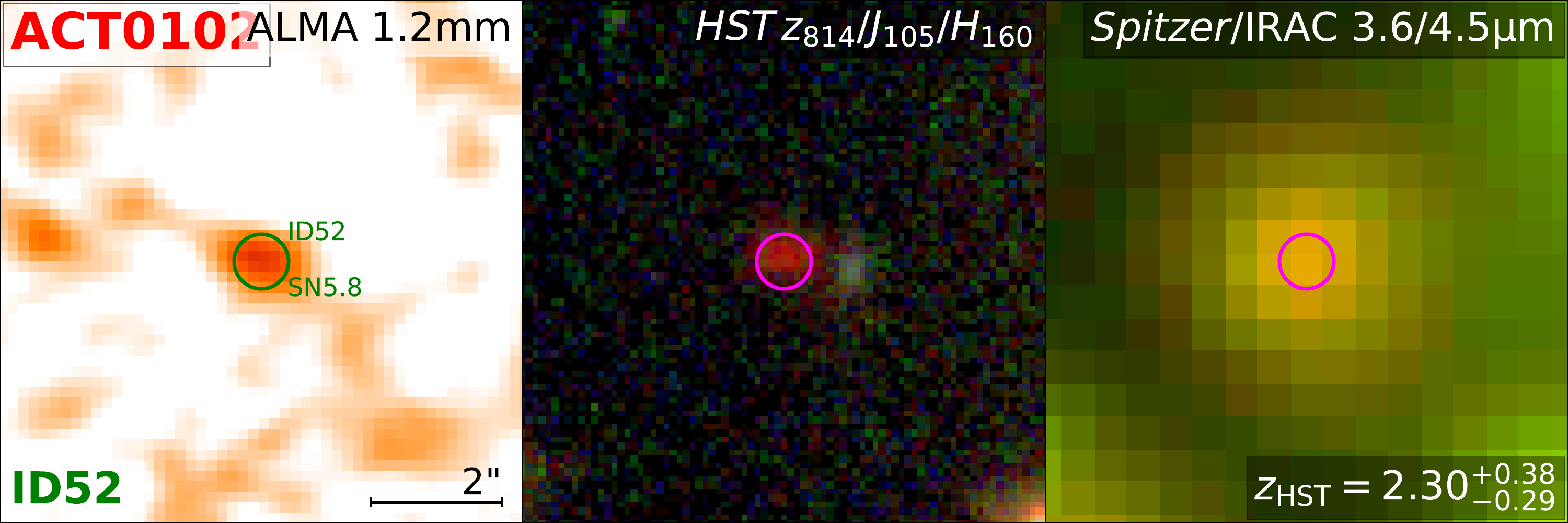}
\includegraphics[width=0.45\linewidth]{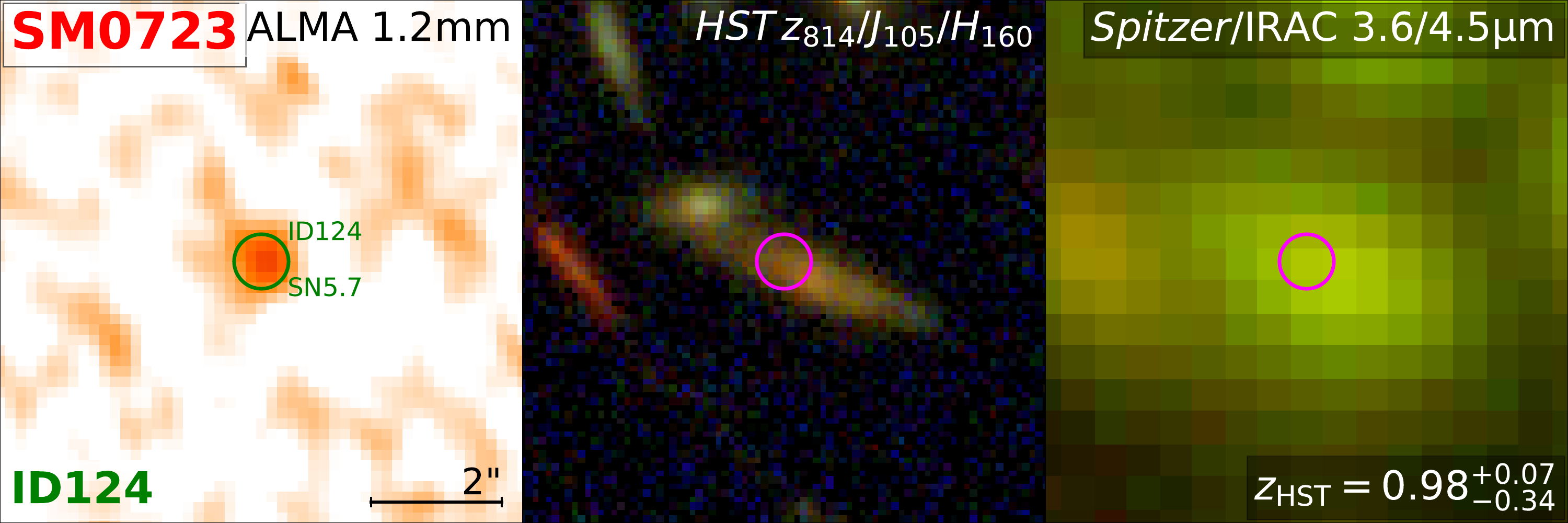}
\includegraphics[width=0.45\linewidth]{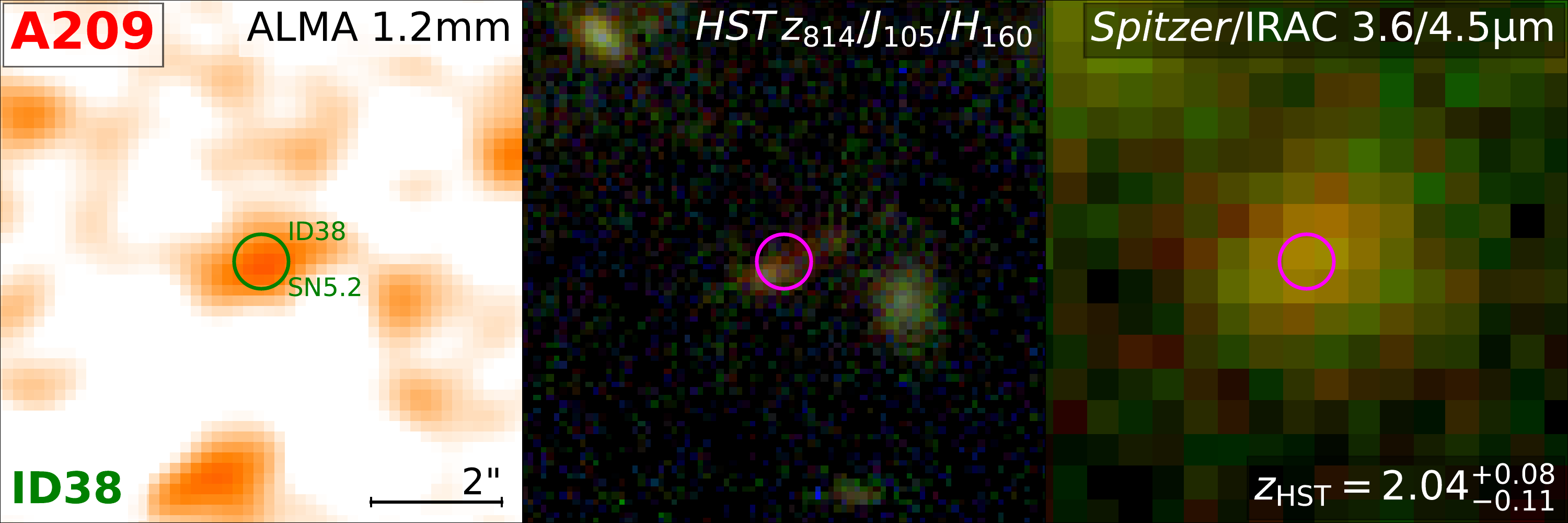}
\includegraphics[width=0.45\linewidth]{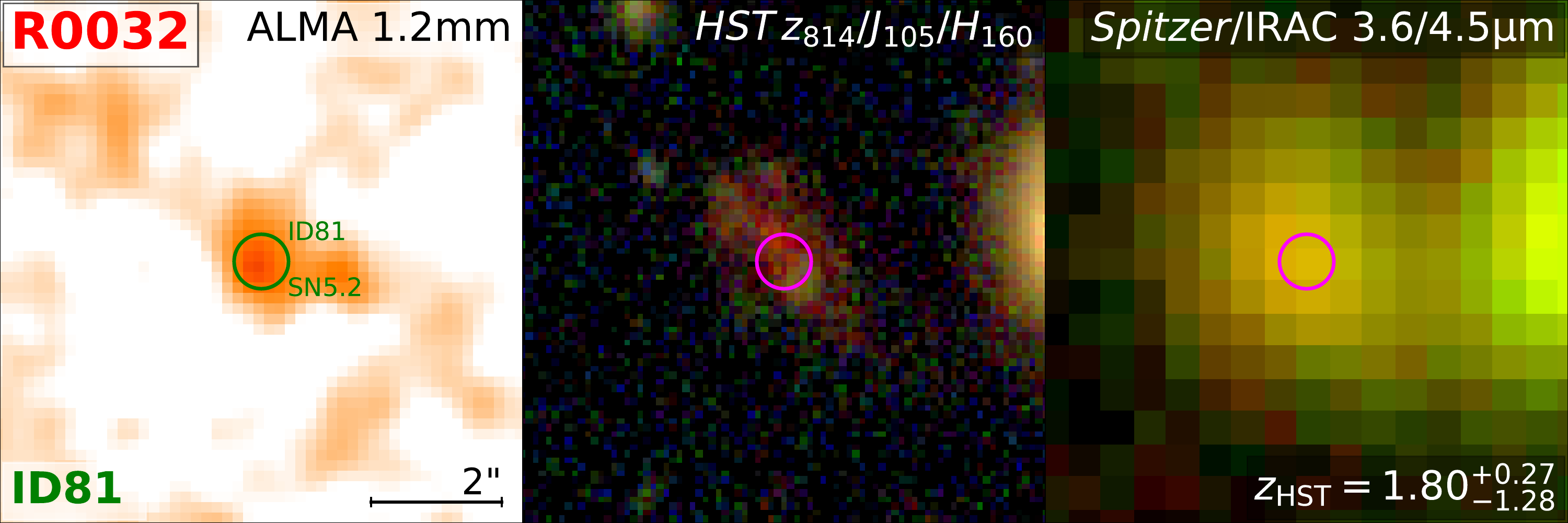}
\includegraphics[width=0.45\linewidth]{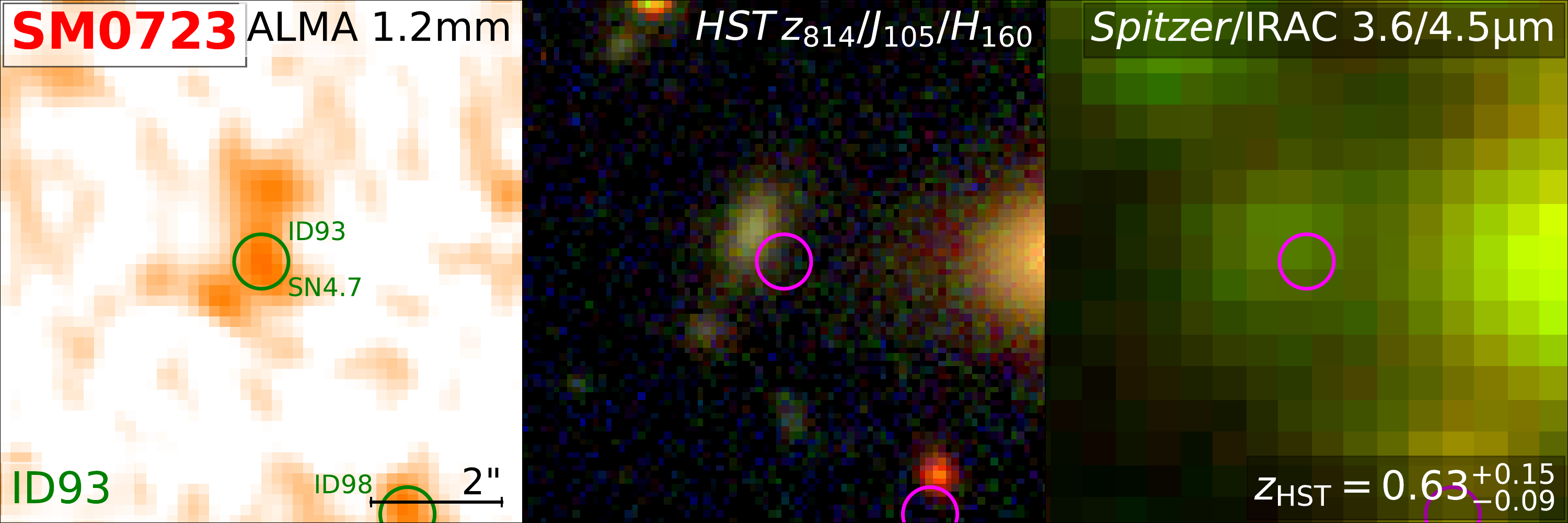}
\caption{Postage stamp images of ten \herschel-faint galaxies with either spectroscopic redshifts (R0032-ID32 and R0600-ID164) or \hst\ photometric redshifts (the remaining eight sources).
In each category, sources are shown in descending order of ALMA S/N.
The layout of these postage stamp images is the same as that in Figure~\ref{fs:01_main}.
The best available redshift is noted in the lower right corner of \spitzer/IRAC image.
}
\label{fig:hdrop_z}
\end{figure*}

\renewcommand{\thefigure}{B4}
\begin{figure}[phtb]
\centering
{
\includegraphics[width=1.0\linewidth]{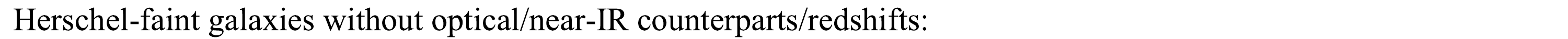}
\includegraphics[width=0.325\linewidth]{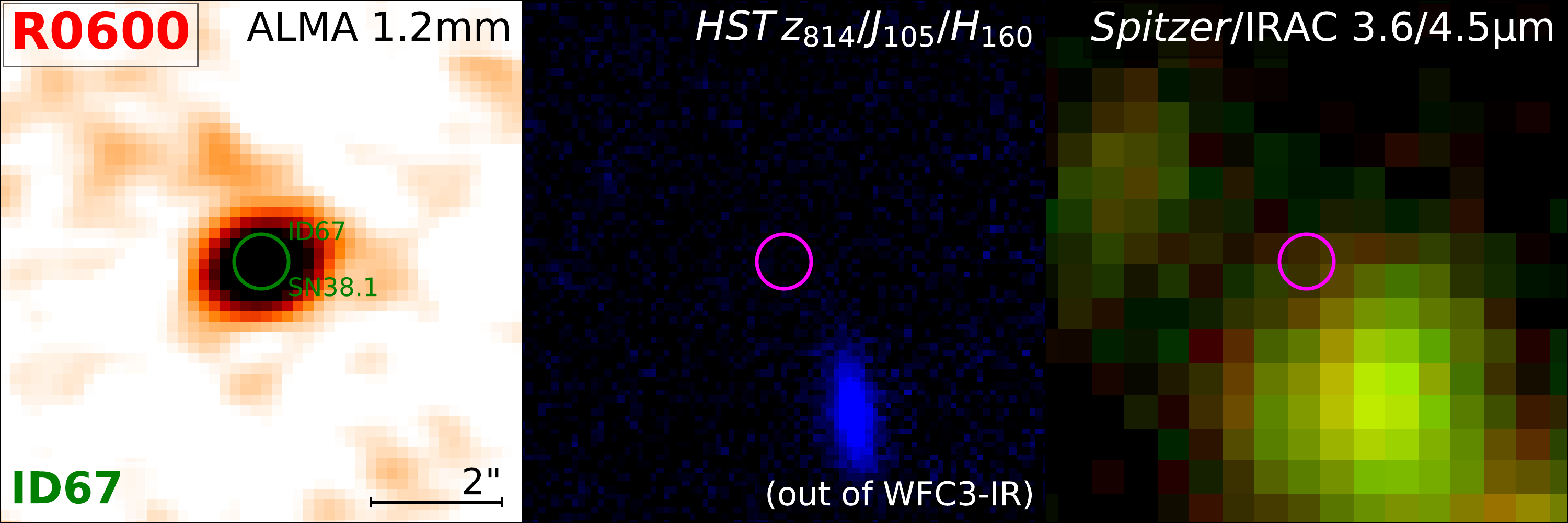}
\includegraphics[width=0.325\linewidth]{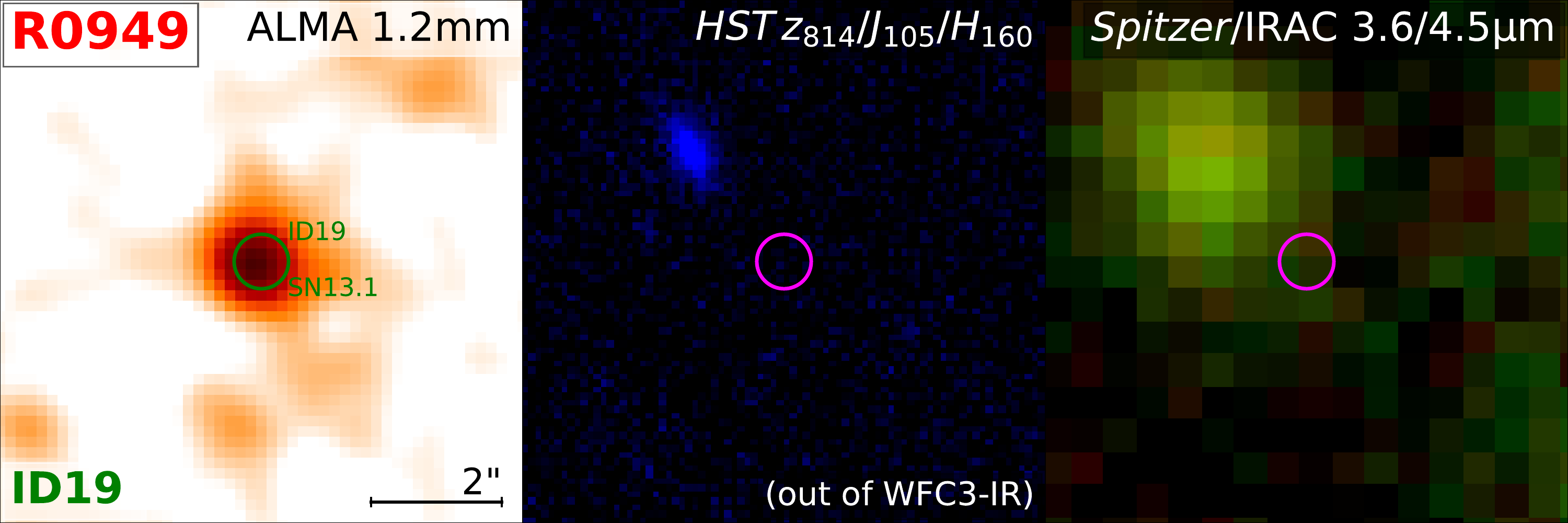}
\includegraphics[width=0.325\linewidth]{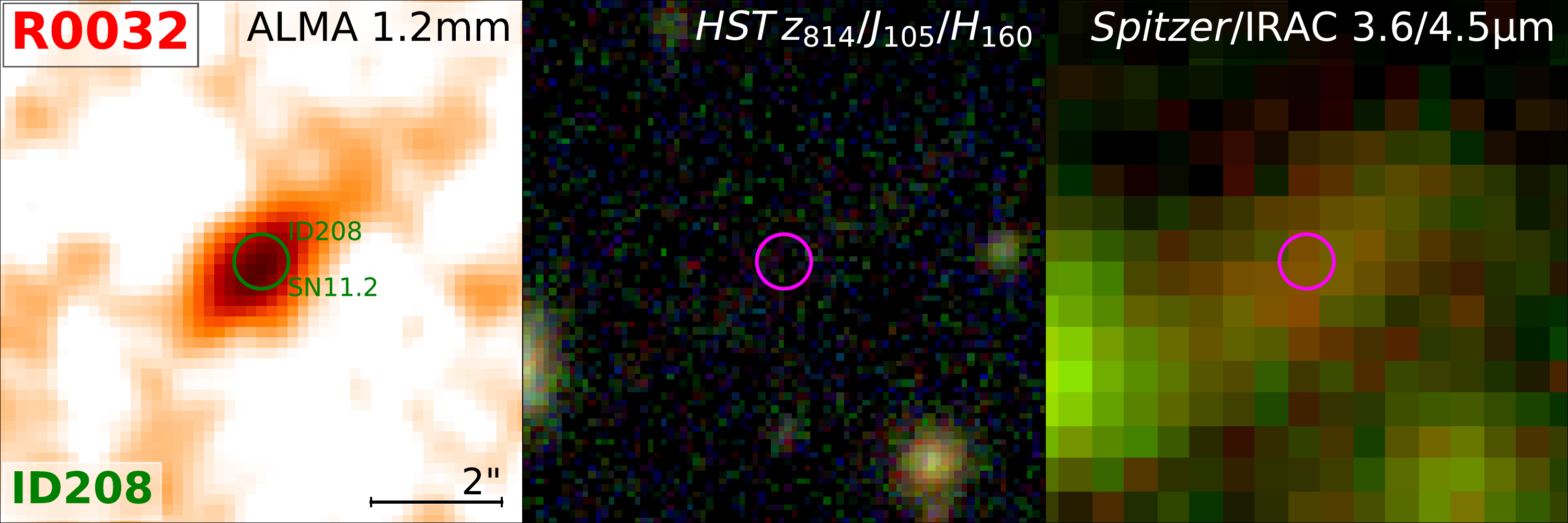}
\includegraphics[width=0.325\linewidth]{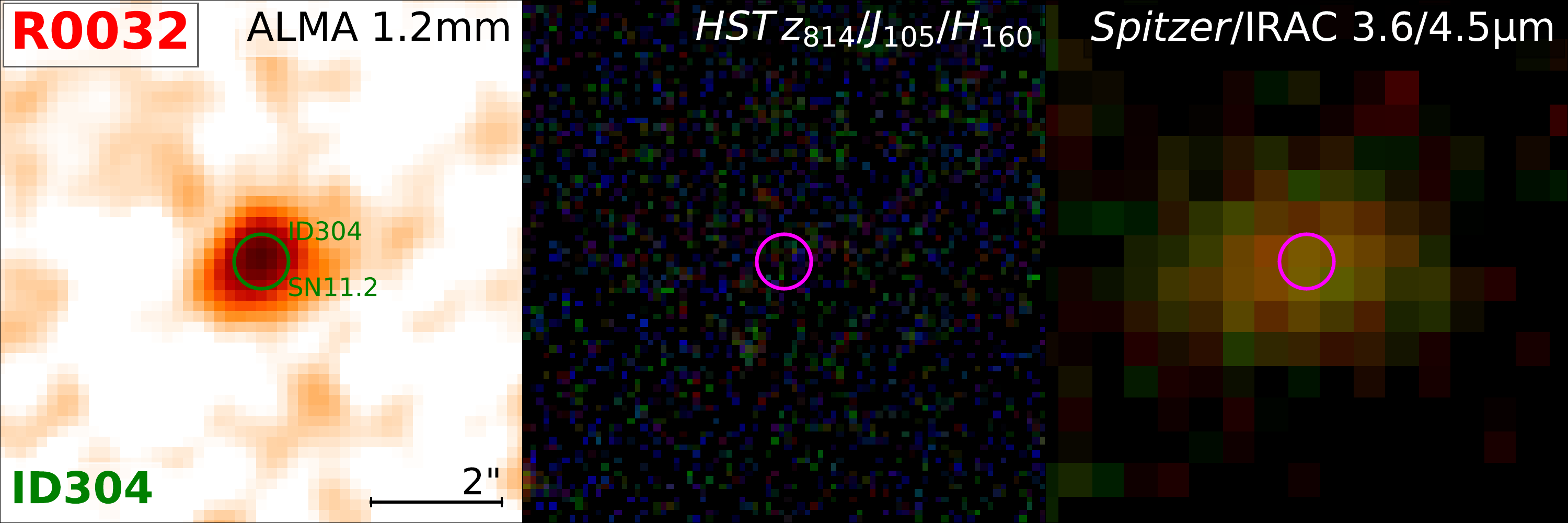}
\includegraphics[width=0.325\linewidth]{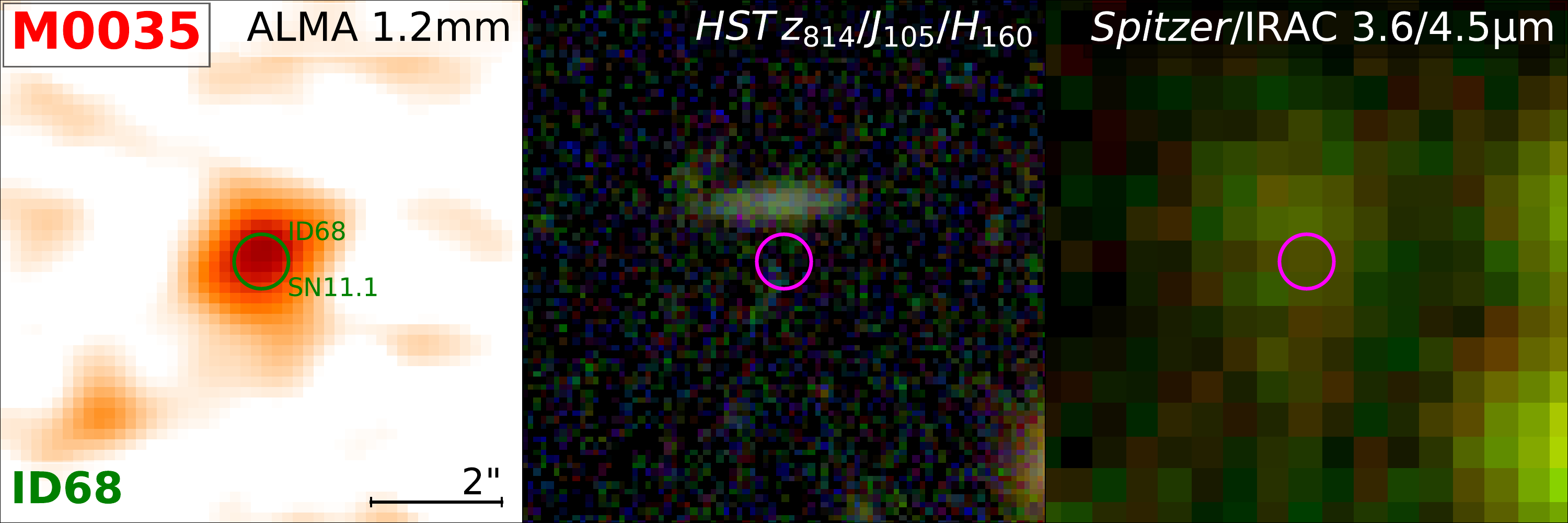}
\includegraphics[width=0.325\linewidth]{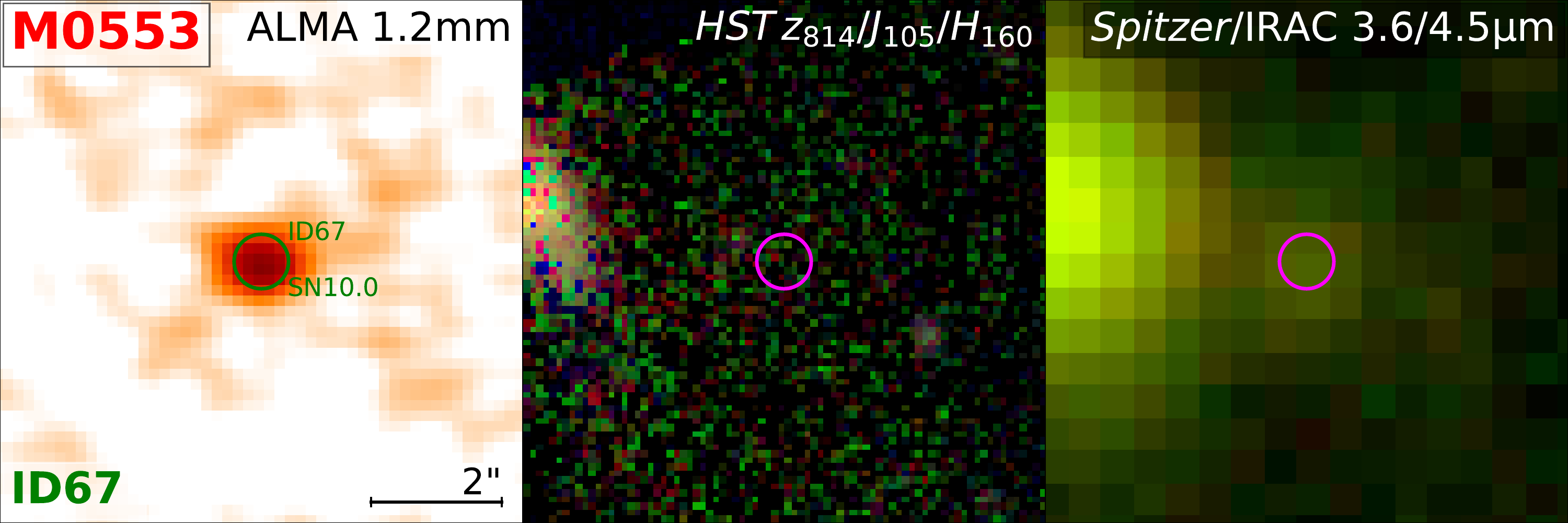}
\includegraphics[width=0.325\linewidth]{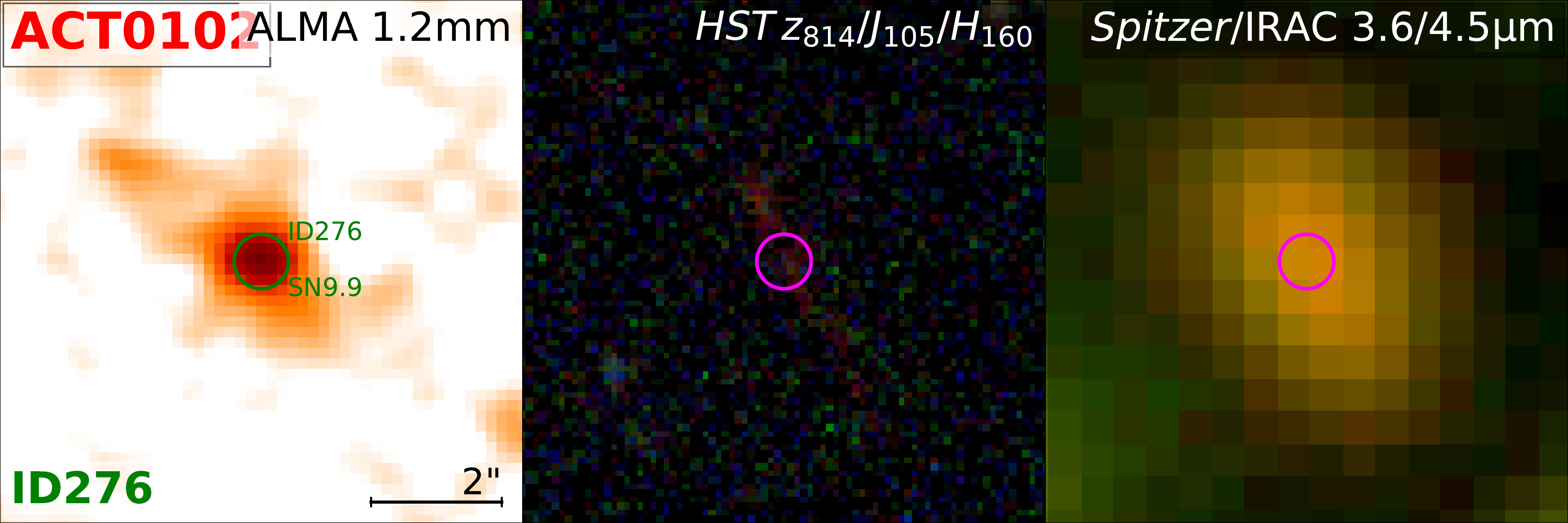}
\includegraphics[width=0.325\linewidth]{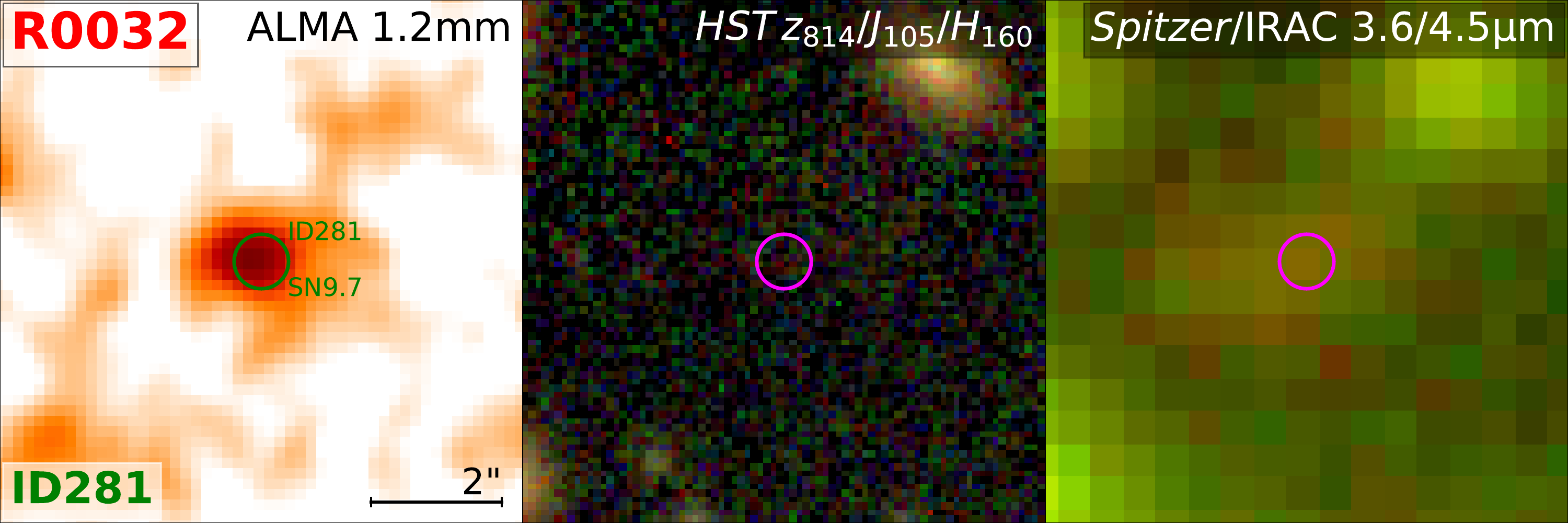}
\includegraphics[width=0.325\linewidth]{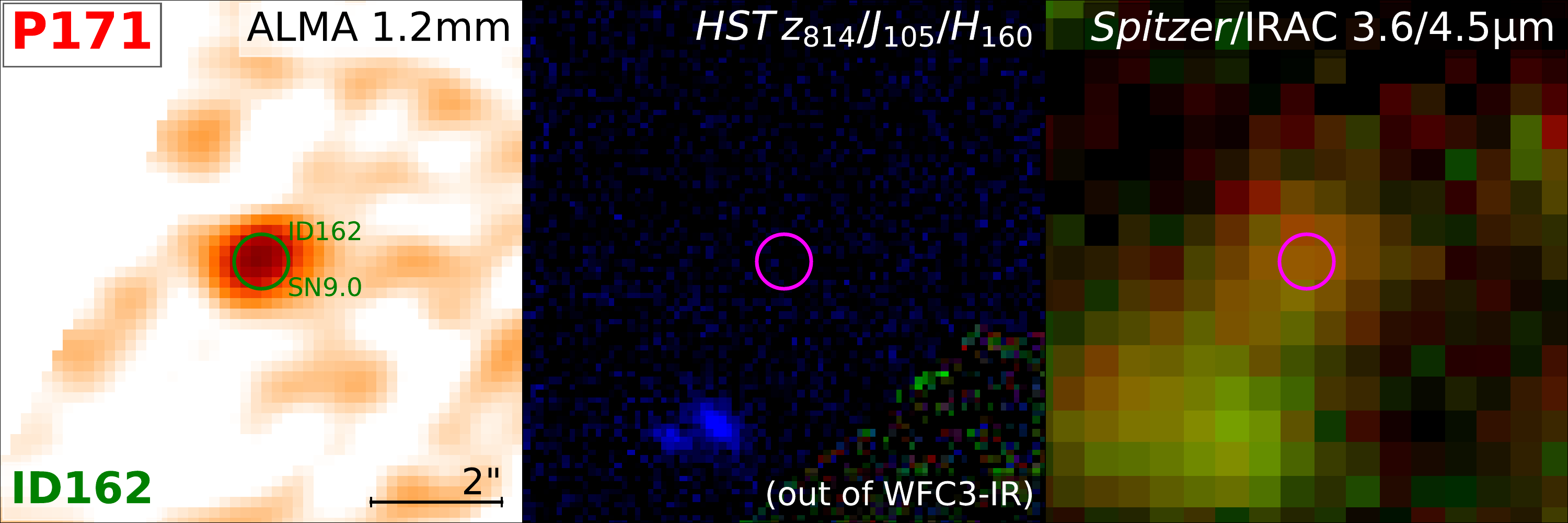}
\includegraphics[width=0.325\linewidth]{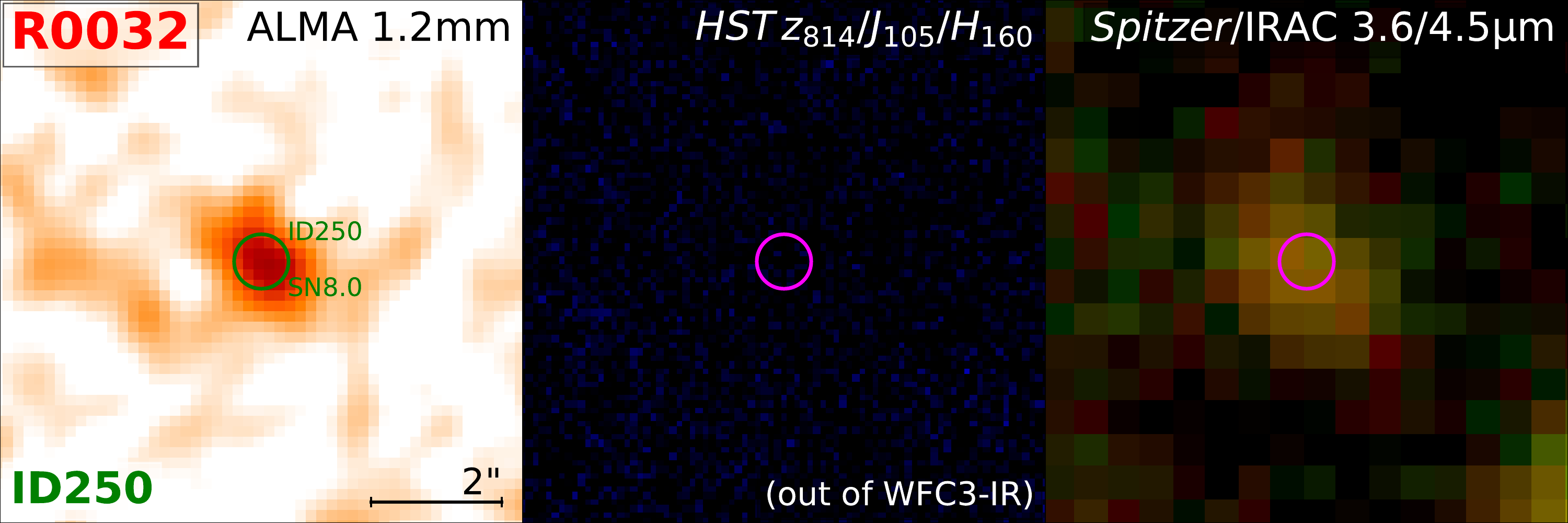}
\includegraphics[width=0.325\linewidth]{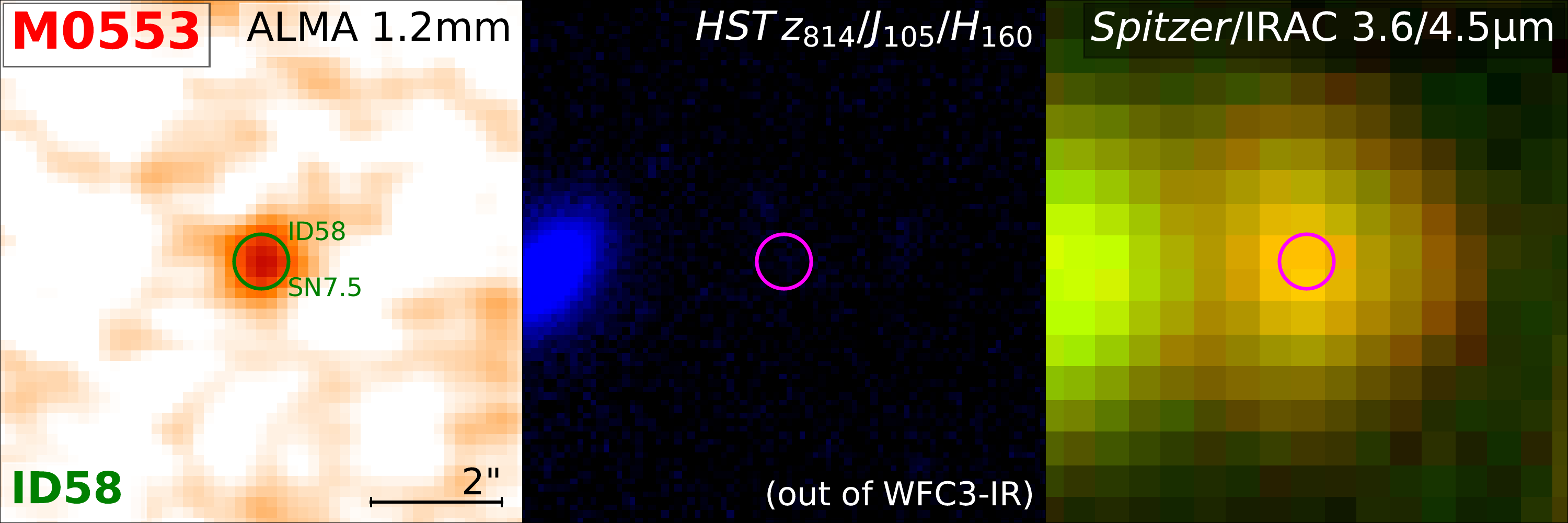}
\includegraphics[width=0.325\linewidth]{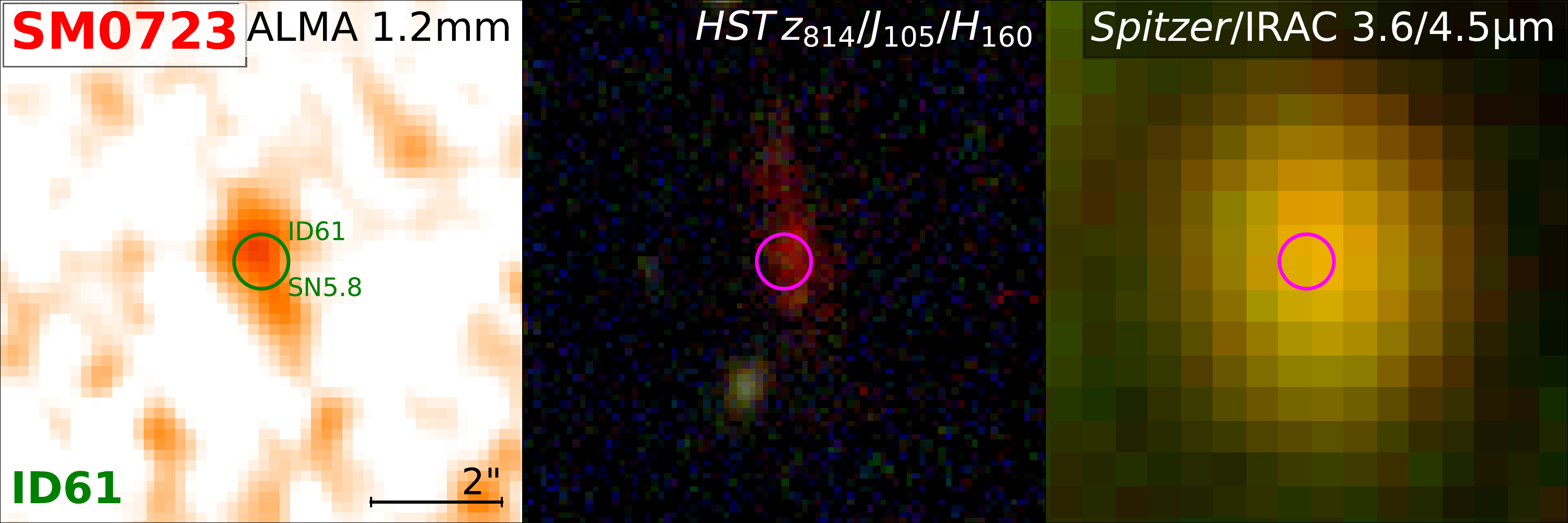}
\includegraphics[width=0.325\linewidth]{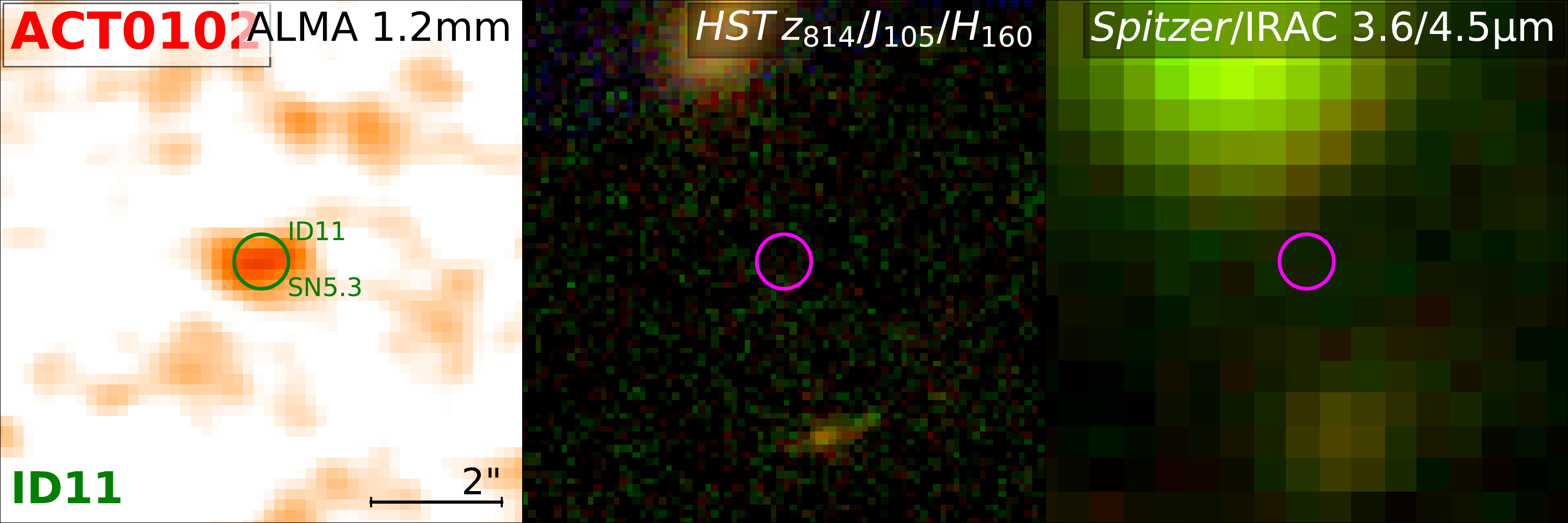}
\includegraphics[width=0.325\linewidth]{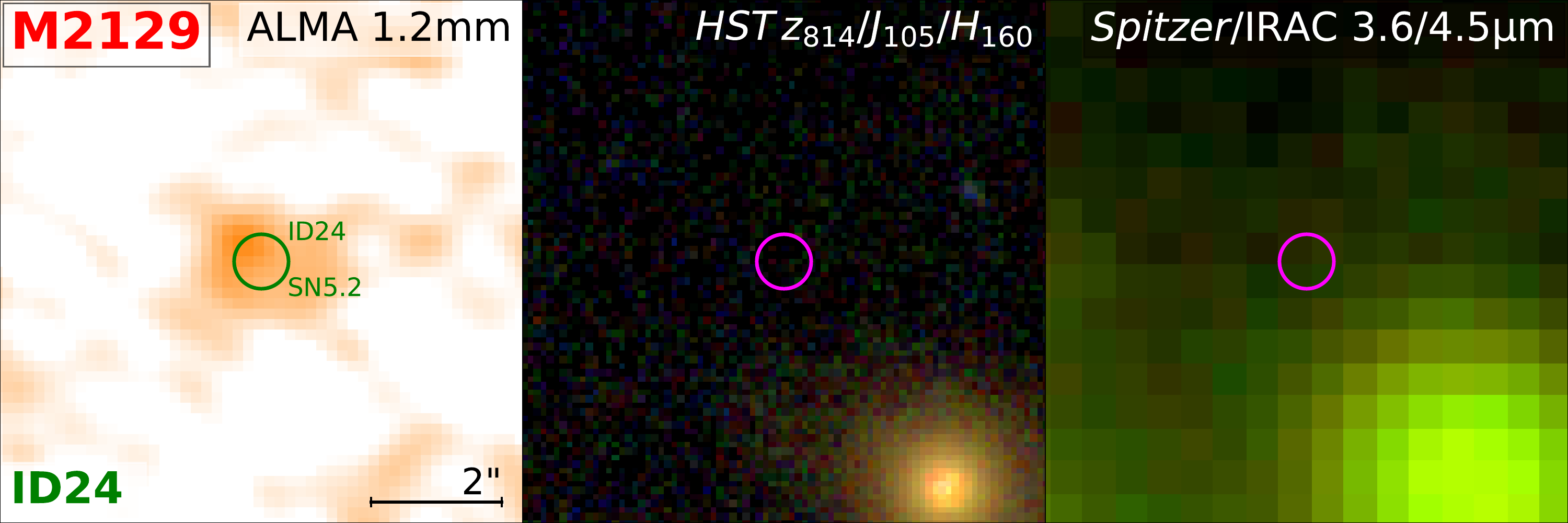}
\includegraphics[width=0.325\linewidth]{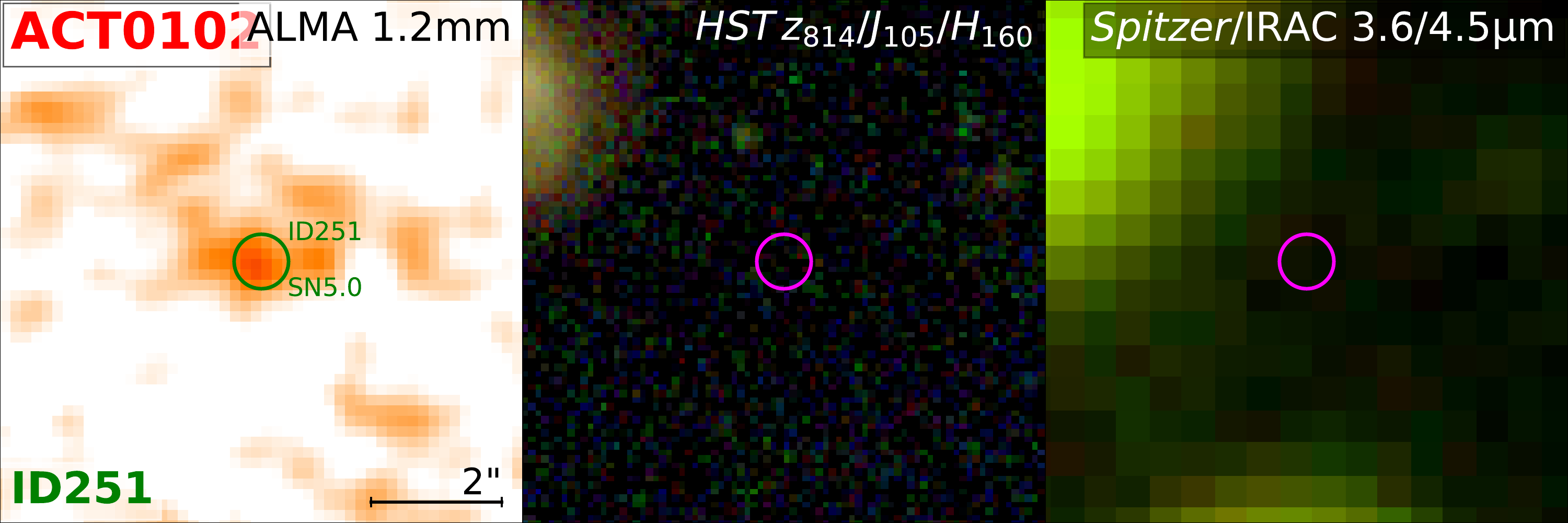}
\includegraphics[width=0.325\linewidth]{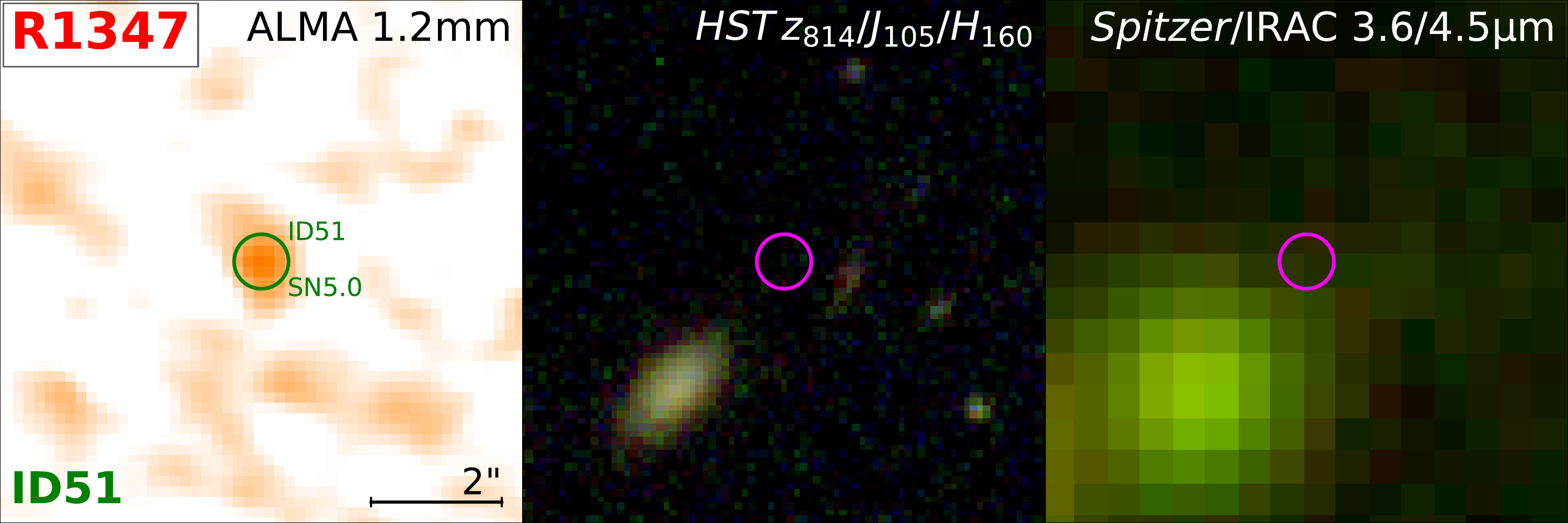}
\includegraphics[width=0.325\linewidth]{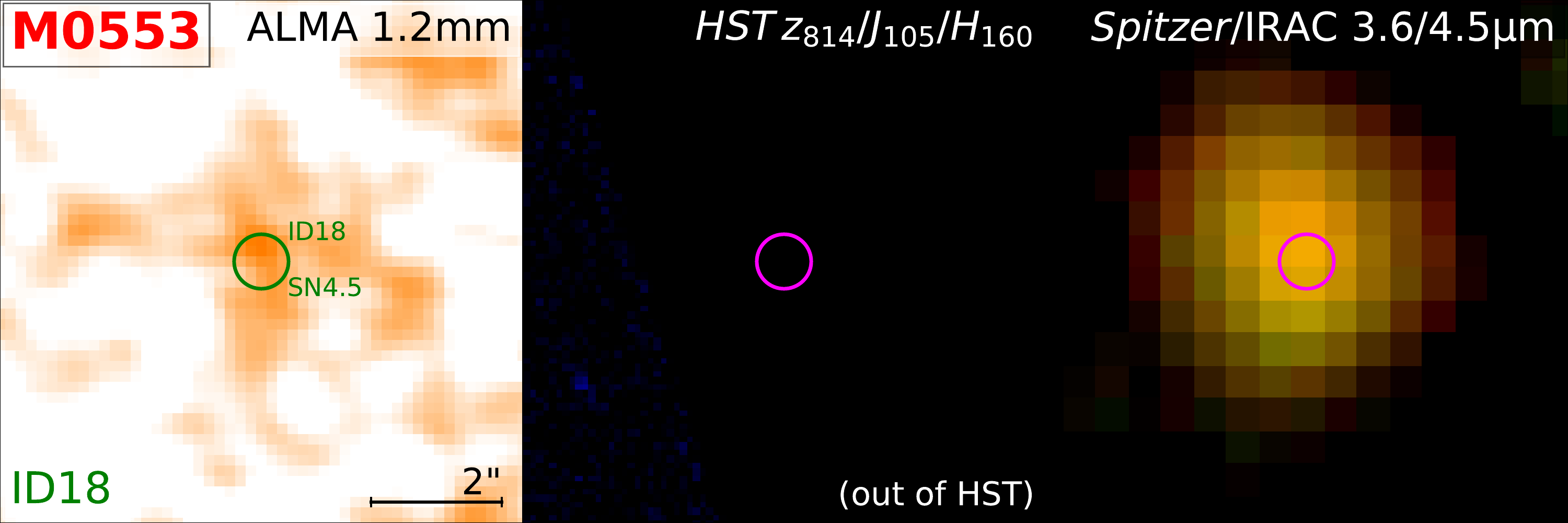}
\includegraphics[width=0.325\linewidth]{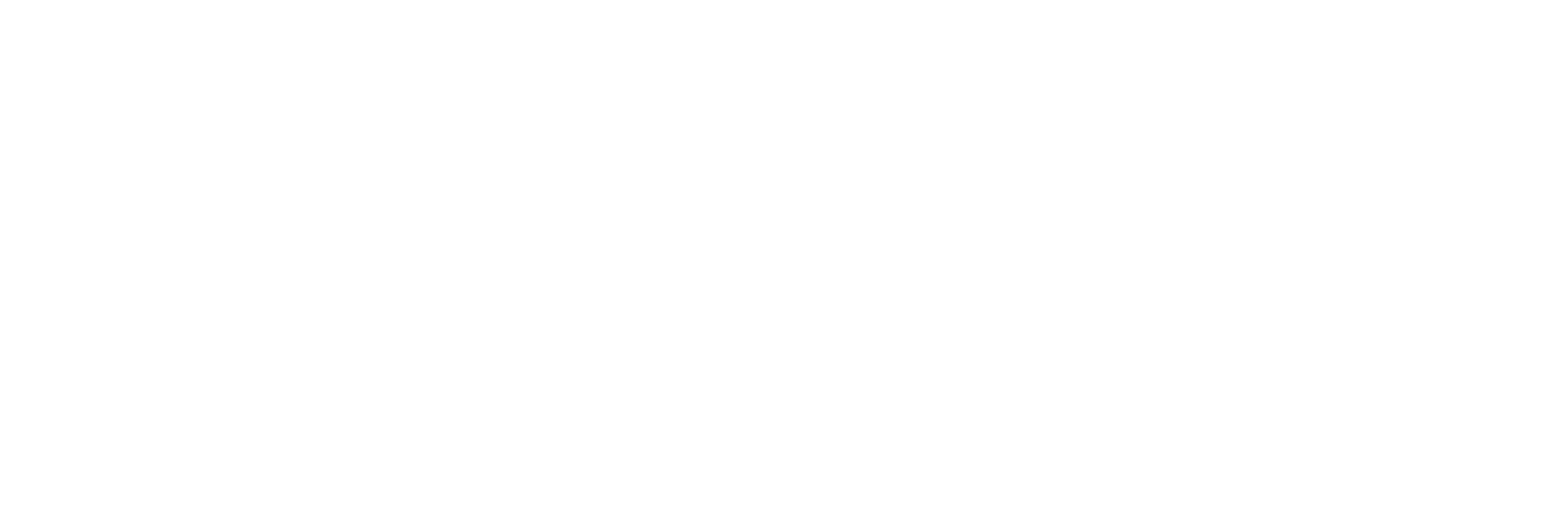}
}
\caption{Same as Figure~\ref{fig:hdrop_z} but for 17 \herschel-faint sources without optical/near-IR counterparts or catalogued redshifts.}
\label{fig:hdrop_dark}
\end{figure}

\section{Catalogs of Herschel Photometry and Derived Physical Properties}
\label{apd:03_tab}

\setcounter{table}{0}
\renewcommand{\thetable}{C\arabic{table}}
\renewcommand{\theHtable}{C\arabic{table}}

Table~\ref{tab:03_phot} presents the \herschel\ flux density measurements of (\romannumeral1) 141 secure ALCS sources at $\mathrm{S/N}_\mathrm{nat} \geq 5$ or $\mathrm{S/N}_\mathrm{tap} \geq 4.5$ (the main sample), and (\romannumeral2) 39 tentative ALCS sources at $\mathrm{S/N}_\mathrm{nat} = 4-5$ and $\mathrm{S/N}_\mathrm{tap}<4.5$ with cross-matched \hst\ or \spitzer/IRAC counterparts within 1\arcsec\ separation (the secondary sample).
Table~\ref{tab:04_sed} presents the galaxy properties of 125 ALCS sources that are detected at $>2\sigma$ in $\geq1$ \herschel\ band (i.e., ALCS-\herschel\ joint sample).
Note that BCGs and \herschel-faint galaxies are not included in this table.
The definition of the samples in these two tables are also visualized in Figure~\ref{fig:catalog}.



\startlongtable
\begin{deluxetable*}{@{\extracolsep{2pt}}rlrrrrrrrr}
\tablecaption{Summary of \herschel\ photometry
\label{tab:03_phot}}
\tablewidth{0pt}
\tabletypesize{\scriptsize}
\tablehead{
\colhead{\#} & 
\colhead{Name} & \multicolumn2c{Coordinates} &
\colhead{S/N$_\mathrm{ALMA}$} & \multicolumn2c{\herschel/PACS}  & \multicolumn3c{\herschel/SPIRE} \\
\cline{3-4}\cline{6-7}\cline{8-10}
\colhead{}  & \colhead{} & \colhead{RA} & \colhead{Dec} & \colhead{} 
 & \colhead{100\,\micron}  & \colhead{160\,\micron}  & \colhead{250\,\micron}  & \colhead{350\,\micron} & \colhead{500\,\micron} \\
 \colhead{}  & \colhead{} & \colhead{} & \colhead{} & \colhead{} & \colhead{(mJy)} 
 & \colhead{(mJy)}  & \colhead{(mJy)}  & \colhead{(mJy)}  & \colhead{(mJy)}
}
\startdata
 & \multicolumn4l{\textit{Main Sample} ($\mathrm{S/N}_\mathrm{nat}\geq5$ or $\mathrm{S/N}_\mathrm{tap}\geq4.5$)}\\\hline
  1 & A209-ID38$^\downarrow$ & 01:31:53.38 & --13:37:17.4 & 5.2 & $<$1.5 & $<$2.3 & $<$6.4 & $<$6.6 & $<$7.4 \\
  2 & A2163-ID11 & 16:15:49.06 & --06:08:13.9 & 4.6$^\dagger$ & \nodata & \nodata & 37.7$\pm$5.3 & 24.1$\pm$5.2 & $<$17.2 \\
  3 & A2537-ID42 & 23:08:24.43 & --02:11:05.5 & 6.8 & 19.5$\pm$0.5 & 44.0$\pm$0.9 & 49.7$\pm$1.9 & 39.8$\pm$2.5 & 20.2$\pm$2.6 \\
  4 & A2537-ID49 & 23:08:17.92 & --02:11:12.7 & 5.2 & 33.0$\pm$2.5 & 55.9$\pm$3.4 & 36.1$\pm$4.5 & 30.9$\pm$5.5 & 8.2$\pm$4.3 \\
  5 & A2537-ID66 & 23:08:19.68 & --02:11:25.3 & 4.9 & 15.5$\pm$0.7 & 20.5$\pm$2.6 & 20.3$\pm$5.2 & $<$22.9 & $<$7.8 \\
  6 & A2744-ID07 & 00:14:19.12 & --30:22:42.2 & 8.0 & $<$0.8 & $<$2.3 & 15.4$\pm$6.6 & 18.1$\pm$3.7 & 7.7$\pm$3.0 \\
  7 & A2744-ID21 & 00:14:22.09 & --30:22:49.7 & 6.8 & 4.4$\pm$1.2 & 13.0$\pm$2.1 & 7.1$\pm$2.9 & $<$13.6 & $<$7.7 \\
  8 & A2744-ID33 & 00:14:20.40 & --30:22:54.5 & 15.0 & 0.9$\pm$0.5 & 2.1$\pm$1.4 & 5.5$\pm$5.3 & 5.2$\pm$4.7 & 15.6$\pm$4.4 \\
  9 & A2744-ID56 & 00:14:17.58 & --30:23:00.6 & 11.9 & 28.3$\pm$0.6 & 61.7$\pm$1.2 & 64.3$\pm$1.9 & 46.6$\pm$2.4 & 24.0$\pm$2.6 \\
 10 & A2744-ID81 & 00:14:19.80 & --30:23:07.7 & 26.7 & $<$1.6 & 4.3$\pm$1.8 & 9.8$\pm$5.4 & 15.8$\pm$4.6 & $<$15.0 \\
 11 & A2744-ID319 & 00:14:18.26 & --30:24:47.5 & 14.2 & $<$1.9 & 7.8$\pm$1.3 & 11.7$\pm$1.9 & 10.5$\pm$3.0 & 9.5$\pm$2.6 \\
 12 & A3192-ID31 & 03:58:50.02 & --29:56:30.8 & 32.6 & \nodata & \nodata & 124.9$\pm$4.4 & 88.4$\pm$4.7 & 40.3$\pm$4.6 \\
 13 & A3192-ID40 & 03:58:51.13 & --29:55:10.9 & 26.0 & \nodata & \nodata & 20.0$\pm$4.4 & 15.8$\pm$5.1 & $<$20.5 \\
 14 & A3192-ID131 & 03:58:47.63 & --29:56:33.5 & 11.5 & \nodata & \nodata & 12.8$\pm$4.0 & 10.0$\pm$4.7 & $<$13.8 \\
 15 & A3192-ID138 & 03:58:57.13 & --29:54:53.5 & 8.1 & \nodata & \nodata & 20.8$\pm$4.5 & 12.2$\pm$4.4 & $<$15.4 \\
 16 & A370-ID18 & 02:39:50.20 & --01:35:42.1 & 12.9 & 1.2$\pm$0.5 & 6.1$\pm$1.3 & 9.8$\pm$6.2 & 17.9$\pm$9.2 & $<$11.5 \\
 17 & A370-ID31 & 02:39:54.43 & --01:33:36.5 & 5.9 & 34.8$\pm$0.6 & 49.1$\pm$1.0 & 32.3$\pm$1.8 & 17.3$\pm$5.4 & $<$20.9 \\
 18 & A370-ID103 & 02:39:58.15 & --01:34:24.5 & 6.8 & 8.1$\pm$0.5 & 22.7$\pm$3.2 & 33.6$\pm$10.1 & 27.3$\pm$2.8 & 13.0$\pm$2.2 \\
 19 & A370-ID110 & 02:39:56.57 & --01:34:26.3 & 24.9 & 48.8$\pm$0.6 & 105.5$\pm$0.9 & 108.8$\pm$2.8 & 76.7$\pm$2.8 & 30.9$\pm$2.2 \\
 20 & A370-ID146 & 02:39:51.44 & --01:34:41.5 & 5.4 & 0.6$\pm$0.5 & 5.1$\pm$1.4 & 9.5$\pm$4.9 & 8.3$\pm$3.5 & $<$5.8 \\
 21 & A383-ID40$^\bullet$ & 02:48:03.39 & --03:31:44.8 & 34.2 & 14.1$\pm$0.6 & 24.0$\pm$1.8 & 16.6$\pm$2.8 & 7.3$\pm$2.7 & $<$7.7 \\
 22 & ACT0102-ID11$^\downarrow$ & 01:02:59.47 & --49:17:25.4 & 5.3 & \nodata & \nodata & $<$13.6 & $<$15.0 & $<$20.5 \\
 23 & ACT0102-ID22 & 01:03:03.04 & --49:17:11.2 & 7.9 & \nodata & \nodata & 6.1$\pm$5.3 & 10.2$\pm$4.8 & 12.2$\pm$5.6 \\
 24 & ACT0102-ID50 & 01:02:59.40 & --49:16:55.6 & 5.8 & \nodata & \nodata & 18.8$\pm$4.1 & 12.2$\pm$4.4 & 11.0$\pm$5.0 \\
 25 & ACT0102-ID52$^\downarrow$ & 01:03:01.47 & --49:16:56.6 & 5.8 & \nodata & \nodata & 6.3$\pm$4.1 & $<$15.3 & $<$20.5 \\
 26 & ACT0102-ID118 & 01:02:51.09 & --49:15:38.9 & 34.3 & \nodata & \nodata & 9.0$\pm$5.2 & 14.6$\pm$7.3 & 9.0$\pm$5.7 \\
 27 & ACT0102-ID160 & 01:03:00.20 & --49:16:03.7 & 4.9$^\dagger$ & \nodata & \nodata & 14.3$\pm$3.6 & $<$12.0 & $<$20.4 \\
 28 & ACT0102-ID215 & 01:02:54.92 & --49:15:14.7 & 42.4 & \nodata & \nodata & $<$16.1 & $<$22.6 & 11.4$\pm$3.8 \\
 29 & ACT0102-ID223 & 01:02:49.24 & --49:15:08.9 & 7.9 & \nodata & \nodata & 11.7$\pm$3.9 & 14.0$\pm$4.7 & 10.6$\pm$3.5 \\
 30 & ACT0102-ID224 & 01:02:55.68 & --49:15:09.1 & 91.8 & \nodata & \nodata & 18.4$\pm$5.4 & 28.7$\pm$7.5 & 28.2$\pm$9.4 \\
 31 & ACT0102-ID241 & 01:02:58.13 & --49:14:56.2 & 5.4 & \nodata & \nodata & 6.2$\pm$3.4 & 12.0$\pm$4.3 & 12.2$\pm$4.3 \\
 32 & ACT0102-ID251$^\downarrow$ & 01:02:55.07 & --49:14:19.2 & 5.0 & \nodata & \nodata & $<$14.8 & $<$15.5 & $<$20.1 \\
 33 & ACT0102-ID276$^\downarrow$ & 01:02:49.29 & --49:14:38.2 & 9.9 & \nodata & \nodata & $<$14.6 & 6.3$\pm$4.0 & $<$20.1 \\
 34 & ACT0102-ID294 & 01:02:49.37 & --49:15:05.3 & 19.1$^\dagger$ & \nodata & \nodata & 36.0$\pm$12.0 & 42.9$\pm$14.3 & 32.6$\pm$10.9 \\
 35 & AS1063-ID15 & 22:48:46.58 & --44:30:46.9 & 7.7$^\dagger$ & 6.3$\pm$1.0 & 10.4$\pm$0.8 & 5.3$\pm$3.6 & $<$25.4 & $<$27.3 \\
 36 & AS1063-ID17 & 22:48:47.29 & --44:30:48.2 & 15.5 & 24.0$\pm$0.5 & 61.6$\pm$0.8 & 62.7$\pm$3.6 & 48.4$\pm$8.5 & 22.5$\pm$9.2 \\
 37 & AS1063-ID147 & 22:48:41.82 & --44:31:57.5 & 6.0$^\dagger$ & 66.9$\pm$0.9 & 98.4$\pm$0.9 & 67.8$\pm$3.3 & 30.8$\pm$5.0 & 7.3$\pm$3.2 \\
 38 & AS1063-ID222 & 22:48:49.06 & --44:32:25.4 & 85.7 & $<$1.4 & 3.0$\pm$0.9 & 10.6$\pm$2.6 & 21.8$\pm$3.2 & 23.3$\pm$2.6 \\
 39 & M0035-ID41 & 00:35:20.46 & --20:15:35.2 & 7.2 & \nodata & \nodata & 8.3$\pm$3.6 & $<$13.4 & $<$13.7 \\
 40 & M0035-ID68$^\downarrow$ & 00:35:30.76 & --20:15:54.9 & 11.1 & \nodata & \nodata & $<$14.3 & $<$15.1 & $<$14.5 \\
 41 & M0035-ID94 & 00:35:27.24 & --20:16:17.4 & 4.6$^\dagger$ & \nodata & \nodata & 23.6$\pm$3.4 & $<$12.8 & $<$16.7 \\
 42 & M0159-ID05 & 01:59:51.69 & --08:49:10.1 & 16.0 & \nodata & \nodata & 24.0$\pm$4.0 & 20.4$\pm$4.1 & 10.2$\pm$5.4 \\
 43 & M0159-ID24 & 01:59:47.33 & --08:49:33.2 & 6.3 & \nodata & \nodata & 24.1$\pm$5.2 & $<$15.7 & 9.0$\pm$6.6 \\
 44 & M0159-ID46$^\bullet$ & 01:59:49.35 & --08:49:59.0 & 32.8 & \nodata & \nodata & 11.7$\pm$4.4 & $<$15.6 & 14.2$\pm$5.2 \\
 45 & M0159-ID61 & 01:59:51.56 & --08:50:13.3 & 5.8 & \nodata & \nodata & 14.1$\pm$4.7 & $<$15.6 & $<$17.8 \\
 46 & M0257-ID13 & 02:57:11.25 & --23:25:43.6 & 12.2$^\dagger$ & \nodata & \nodata & 11.7$\pm$4.4 & 13.4$\pm$4.5 & 15.1$\pm$4.7 \\
 47 & M0329-ID11 & 03:29:41.76 & --02:10:56.8 & 11.1 & 1.7$\pm$0.5 & 1.6$\pm$0.8 & $<$9.2 & $<$8.6 & $<$9.7 \\
 48 & M0416-ID51 & 04:16:06.97 & --24:04:00.0 & 7.1 & 1.8$\pm$0.5 & 4.0$\pm$0.7 & $<$9.0 & $<$7.8 & $<$9.4 \\
 49 & M0416-ID79 & 04:16:11.67 & --24:04:19.5 & 5.3 & 1.4$\pm$0.5 & 1.1$\pm$0.9 & 6.3$\pm$2.7 & $<$7.8 & $<$9.4 \\
 50 & M0416-ID117 & 04:16:10.80 & --24:04:47.6 & 15.7 & 3.9$\pm$0.5 & 10.4$\pm$0.8 & 12.4$\pm$2.2 & 9.3$\pm$1.9 & 9.8$\pm$1.7 \\
 51 & M0416-ID160 & 04:16:08.83 & --24:05:22.5 & 6.0 & 3.5$\pm$0.5 & 5.8$\pm$0.8 & 6.0$\pm$2.7 & 3.5$\pm$2.7 & $<$9.4 \\
 52 & M0417-ID46 & 04:17:33.32 & --11:55:03.3 & 48.0 & \nodata & \nodata & 45.4$\pm$5.8 & 39.0$\pm$6.2 & 30.7$\pm$5.1 \\
 53 & M0417-ID49 & 04:17:40.15 & --11:55:00.8 & 36.1 & \nodata & \nodata & 43.4$\pm$4.7 & 60.6$\pm$7.7 & 35.8$\pm$5.6 \\
 54 & M0417-ID58 & 04:17:35.65 & --11:54:53.3 & 30.4 & \nodata & \nodata & 11.1$\pm$7.3 & 14.8$\pm$5.3 & 11.1$\pm$5.7 \\
 55 & M0417-ID121 & 04:17:37.02 & --11:54:19.8 & 20.7 & \nodata & \nodata & 10.9$\pm$6.3 & 6.9$\pm$4.7 & 17.6$\pm$5.2 \\
 56 & M0417-ID204 & 04:17:32.23 & --11:52:39.1 & 4.5$^\dagger$ & \nodata & \nodata & 13.8$\pm$4.4 & $<$18.2 & $<$34.2 \\
 57 & M0417-ID218 & 04:17:32.01 & --11:52:57.4 & 14.5 & \nodata & \nodata & 37.7$\pm$5.1 & 44.0$\pm$8.9 & 14.9$\pm$7.4 \\
 58 & M0417-ID221 & 04:17:32.48 & --11:53:02.8 & 7.7 & \nodata & \nodata & 20.9$\pm$1.8 & 14.3$\pm$2.9 & 7.3$\pm$3.7 \\
 59 & M0417-ID223 & 04:17:32.55 & --11:53:05.0 & 6.0 & \nodata & \nodata & 38.6$\pm$3.2 & 26.4$\pm$5.3 & 13.6$\pm$6.8 \\
 60 & M0429-ID04$^\downarrow$ & 04:29:35.01 & --02:52:39.3 & 8.4 & $<$0.7 & $<$2.3 & $<$9.7 & $<$9.0 & 5.7$\pm$4.1 \\
 61 & M0429-ID19$^\bullet$ & 04:29:36.03 & --02:53:06.8 & 27.6 & 31.2$\pm$0.6 & 44.4$\pm$1.1 & 25.5$\pm$3.9 & 7.3$\pm$5.1 & $<$10.4 \\
 62 & M0429-ID27 & 04:29:37.10 & --02:53:35.2 & 6.9 & 4.2$\pm$0.5 & 8.8$\pm$0.9 & 12.5$\pm$2.6 & 5.2$\pm$3.5 & 6.3$\pm$3.4 \\
 63 & M0553-ID17 & 05:53:15.40 & --33:41:35.4 & 8.5 & \nodata & \nodata & 19.5$\pm$4.9 & 12.4$\pm$5.7 & $<$16.4 \\
 64 & M0553-ID18$^\downarrow$ & 05:53:16.75 & --33:41:35.6 & 4.5$^\dagger$ & \nodata & \nodata & $<$14.6 & 5.7$\pm$5.7 & $<$18.7 \\
 65 & M0553-ID58$^\downarrow$ & 05:53:30.18 & --33:41:52.1 & 7.5 & \nodata & \nodata & $<$13.3 & $<$15.3 & $<$18.9 \\
 66 & M0553-ID61 & 05:53:21.49 & --33:41:54.7 & 10.7 & \nodata & \nodata & 17.9$\pm$4.5 & 15.5$\pm$4.6 & $<$22.6 \\
 67 & M0553-ID67$^\downarrow$ & 05:53:29.30 & --33:41:56.3 & 10.0 & \nodata & \nodata & $<$59.2 & 27.3$\pm$22.9 & $<$18.9 \\
 68 & M0553-ID133 & 05:53:27.79 & --33:42:16.0 & 39.0 & \nodata & \nodata & 119.3$\pm$3.1 & 83.7$\pm$4.2 & 43.5$\pm$6.1 \\
 69 & M0553-ID190 & 05:53:27.85 & --33:42:30.6 & 67.5 & \nodata & \nodata & 179.1$\pm$4.7 & 125.6$\pm$6.3 & 65.2$\pm$9.2 \\
 70 & M0553-ID200 & 05:53:33.40 & --33:42:31.7 & 5.1$^\dagger$ & \nodata & \nodata & 20.6$\pm$7.5 & $<$18.8 & $<$18.9 \\
 71 & M0553-ID249 & 05:53:27.63 & --33:42:43.9 & 44.0 & \nodata & \nodata & 126.1$\pm$3.3 & 88.4$\pm$4.4 & 45.9$\pm$6.5 \\
 72 & M0553-ID275 & 05:53:26.05 & --33:42:51.6 & 6.6 & \nodata & \nodata & 47.0$\pm$4.2 & 38.1$\pm$6.0 & 9.6$\pm$6.0 \\
 73 & M0553-ID303 & 05:53:31.98 & --33:43:00.5 & 8.2$^\dagger$ & \nodata & \nodata & 79.2$\pm$4.5 & 48.7$\pm$4.7 & 12.6$\pm$4.3 \\
 74 & M0553-ID355 & 05:53:23.28 & --33:43:16.0 & 12.4 & \nodata & \nodata & 12.4$\pm$3.8 & 24.1$\pm$4.5 & 17.3$\pm$4.3 \\
 75 & M0553-ID375 & 05:53:13.29 & --33:43:21.3 & 8.3 & \nodata & \nodata & 18.9$\pm$4.7 & 9.9$\pm$4.3 & $<$16.2 \\
 76 & M0553-ID398 & 05:53:29.87 & --33:43:26.0 & 13.5 & \nodata & \nodata & 9.2$\pm$4.3 & 10.9$\pm$4.4 & 5.0$\pm$4.3 \\
 77 & M1115-ID02 & 11:15:50.69 & 01:30:35.5 & 11.3 & 24.8$\pm$3.0 & 38.2$\pm$1.8 & 72.8$\pm$6.6 & 72.0$\pm$4.6 & 49.2$\pm$4.6 \\
 78 & M1115-ID04 & 11:15:52.03 & 01:30:28.2 & 7.6 & 16.5$\pm$0.8 & 29.1$\pm$0.9 & 35.5$\pm$3.2 & 27.7$\pm$3.9 & 12.8$\pm$3.2 \\
 79 & M1115-ID34 & 11:15:54.14 & 01:29:56.3 & 16.1 & 0.7$\pm$0.5 & 2.5$\pm$0.8 & 7.7$\pm$3.2 & $<$8.3 & $<$9.2 \\
 80 & M1115-ID36$^\bullet$ & 11:15:51.90 & 01:29:54.9 & 7.5 & 2.8$\pm$0.5 & 3.4$\pm$0.8 & $<$10.2 & $<$8.4 & $<$7.6 \\
 81 & M1149-ID77 & 11:49:36.10 & 22:24:24.5 & 5.5$^\dagger$ & 2.2$\pm$0.9 & 9.4$\pm$3.2 & 15.4$\pm$2.1 & 11.5$\pm$2.2 & 7.3$\pm$3.5 \\
 82 & M1149-ID229 & 11:49:34.04 & 22:23:16.7 & 4.7$^\dagger$ & 2.3$\pm$0.8 & 4.9$\pm$1.5 & $<$15.3 & $<$6.4 & $<$7.3 \\
 83 & M1206-ID27 & 12:06:11.26 & --08:47:43.6 & 7.7$^\dagger$ & 14.2$\pm$0.6 & 33.3$\pm$6.0 & 26.9$\pm$11.1 & 42.6$\pm$6.0 & 13.0$\pm$4.7 \\
 84 & M1206-ID54$^\downarrow$ & 12:06:15.94 & --08:47:59.4 & 7.4$^\dagger$ & 0.7$\pm$0.5 & $<$2.4 & $<$11.4 & $<$9.4 & 3.6$\pm$2.6 \\
 85 & M1206-ID55 & 12:06:10.74 & --08:48:00.6 & 9.2$^\dagger$ & 24.8$\pm$0.3 & 44.9$\pm$1.2 & 42.2$\pm$14.1 & 26.4$\pm$1.6 & 11.1$\pm$3.5 \\
 86 & M1206-ID58$^\bullet$ & 12:06:12.15 & --08:48:03.5 & 6.5$^\dagger$ & $<$0.7 & $<$2.4 & $<$11.1 & $<$9.8 & $<$10.7 \\
 87 & M1206-ID60 & 12:06:10.75 & --08:48:05.3 & 10.4$^\dagger$ & 29.7$\pm$0.4 & 53.7$\pm$1.5 & 50.5$\pm$16.8 & 31.6$\pm$1.9 & 13.3$\pm$4.1 \\
 88 & M1206-ID61 & 12:06:10.80 & --08:48:08.9 & 8.0$^\dagger$ & 40.6$\pm$0.6 & 73.4$\pm$2.0 & 69.1$\pm$23.0 & 43.3$\pm$2.6 & 18.1$\pm$5.6 \\
 89 & M1311-ID27 & 13:11:01.32 & --03:10:33.6 & 9.2$^\dagger$ & 5.8$\pm$0.5 & 19.0$\pm$1.5 & 28.5$\pm$3.0 & 23.3$\pm$2.4 & 13.6$\pm$2.6 \\
 90 & M1311-ID33 & 13:11:00.14 & --03:10:42.4 & 8.3 & 1.4$\pm$0.5 & 13.8$\pm$1.2 & 9.4$\pm$3.4 & 10.4$\pm$3.0 & 6.8$\pm$3.9 \\
 91 & M1423-ID38 & 14:23:47.01 & 24:04:54.2 & 5.5 & 5.3$\pm$0.6 & 9.4$\pm$1.2 & 14.4$\pm$3.4 & 8.7$\pm$3.0 & $<$7.2 \\
 92 & M1423-ID50$^\bullet$ & 14:23:47.88 & 24:04:42.4 & 23.1 & 11.3$\pm$0.6 & 19.7$\pm$1.3 & 22.5$\pm$3.4 & 13.8$\pm$3.1 & 14.1$\pm$2.3 \\
 93 & M1931-ID41$^\bullet$ & 19:31:49.63 & --26:34:33.2 & 83.3 & 233.6$\pm$1.4 & 243.3$\pm$1.3 & 121.0$\pm$2.8 & 44.7$\pm$3.0 & 18.5$\pm$3.0 \\
 94 & M1931-ID47$^\downarrow$ & 19:31:52.08 & --26:34:36.4 & 7.7 & $<$0.7 & $<$2.9 & $<$9.2 & $<$8.5 & $<$6.9 \\
 95 & M1931-ID55 & 19:31:48.10 & --26:34:43.4 & 11.6$^\dagger$ & 7.6$\pm$0.6 & 13.2$\pm$1.0 & 9.3$\pm$2.8 & 12.5$\pm$3.0 & 4.4$\pm$3.0 \\
 96 & M1931-ID61 & 19:31:49.28 & --26:34:50.9 & 10.5$^\dagger$ & 0.6$\pm$0.6 & 2.2$\pm$1.2 & 5.3$\pm$2.8 & 7.2$\pm$3.3 & 6.4$\pm$3.4 \\
 97 & M2129-ID24$^\downarrow$ & 21:29:25.28 & --07:41:07.7 & 5.2$^\dagger$ & $<$0.7 & $<$2.7 & $<$9.7 & $<$9.2 & $<$8.9 \\
 98 & M2129-ID46 & 21:29:29.45 & --07:41:31.2 & 6.2 & 4.8$\pm$0.5 & 11.1$\pm$0.8 & 12.9$\pm$2.1 & 8.6$\pm$2.9 & $<$9.1 \\
 99 & P171-ID69 & 03:12:52.76 & 08:22:45.0 & 19.7 & \nodata & \nodata & 11.6$\pm$5.4 & 9.8$\pm$7.7 & $<$17.4 \\
100 & P171-ID162$^\downarrow$ & 03:12:58.70 & 08:23:29.2 & 9.0 & \nodata & \nodata & $<$16.6 & $<$17.4 & $<$23.6 \\
101 & P171-ID177 & 03:12:55.48 & 08:23:44.0 & 7.3$^\dagger$ & \nodata & \nodata & 11.9$\pm$4.8 & $<$21.2 & $<$21.6 \\
102 & R0032-ID32$^\downarrow$ & 00:32:09.77 & 18:06:24.5 & 9.6 & \nodata & \nodata & 9.0$\pm$7.5 & 14.3$\pm$8.7 & $<$23.3 \\
103 & R0032-ID53 & 00:32:08.22 & 18:06:40.3 & 11.3 & \nodata & \nodata & 21.8$\pm$1.9 & 20.4$\pm$2.8 & 12.3$\pm$3.2 \\
104 & R0032-ID55 & 00:32:07.84 & 18:06:46.1 & 18.9$^\dagger$ & \nodata & \nodata & 41.6$\pm$3.6 & 38.9$\pm$5.4 & 23.5$\pm$6.1 \\
105 & R0032-ID57 & 00:32:07.71 & 18:06:48.8 & 7.5 & \nodata & \nodata & 7.4$\pm$0.6 & 7.0$\pm$1.0 & 4.2$\pm$1.1 \\
106 & R0032-ID58 & 00:32:07.56 & 18:06:51.4 & 12.2 & \nodata & \nodata & 22.7$\pm$2.0 & 21.2$\pm$2.9 & 12.9$\pm$3.3 \\
107 & R0032-ID81$^\downarrow$ & 00:32:12.86 & 18:07:02.0 & 5.2 & \nodata & \nodata & $<$17.6 & $<$19.6 & $<$21.3 \\
108 & R0032-ID127 & 00:32:13.16 & 18:08:14.3 & 53.6 & \nodata & \nodata & 13.2$\pm$5.3 & 30.2$\pm$7.4 & $<$38.9 \\
109 & R0032-ID131 & 00:32:12.19 & 18:08:13.3 & 46.8 & \nodata & \nodata & 12.7$\pm$5.3 & 15.6$\pm$7.4 & 32.7$\pm$11.4 \\
110 & R0032-ID198 & 00:32:08.54 & 18:08:26.7 & 18.3 & \nodata & \nodata & 33.8$\pm$7.7 & 24.9$\pm$8.4 & $<$23.0 \\
111 & R0032-ID208$^\downarrow$ & 00:32:13.97 & 18:08:33.7 & 11.2 & \nodata & \nodata & $<$15.6 & $<$21.7 & 10.6$\pm$8.8 \\
112 & R0032-ID220 & 00:32:08.61 & 18:09:08.9 & 6.8$^\dagger$ & \nodata & \nodata & 33.1$\pm$9.5 & 18.5$\pm$18.3 & $<$23.1 \\
113 & R0032-ID238 & 00:32:07.83 & 18:09:33.8 & 6.6 & \nodata & \nodata & 8.6$\pm$3.7 & 7.2$\pm$6.5 & 19.4$\pm$4.8 \\
114 & R0032-ID250$^\downarrow$ & 00:32:08.76 & 18:09:20.8 & 8.0 & \nodata & \nodata & $<$17.1 & 11.4$\pm$11.3 & $<$23.1 \\
115 & R0032-ID276 & 00:32:08.61 & 18:08:59.2 & 14.9 & \nodata & \nodata & 13.0$\pm$10.3 & 20.0$\pm$17.1 & 36.7$\pm$9.1 \\
116 & R0032-ID281$^\downarrow$ & 00:32:12.25 & 18:08:49.0 & 9.7 & \nodata & \nodata & $<$15.7 & $<$21.8 & $<$21.8 \\
117 & R0032-ID287 & 00:32:07.70 & 18:08:52.9 & 11.4 & \nodata & \nodata & 46.3$\pm$7.8 & 30.8$\pm$7.9 & $<$36.2 \\
118 & R0032-ID304$^\downarrow$ & 00:32:08.95 & 18:08:41.8 & 11.2 & \nodata & \nodata & 13.2$\pm$7.7 & $<$27.0 & $<$22.7 \\
119 & R0600-ID12 & 06:00:06.32 & --20:10:04.6 & 5.3 & \nodata & \nodata & 19.9$\pm$5.8 & 22.0$\pm$5.6 & $<$19.7 \\
120 & R0600-ID13 & 06:00:05.05 & --20:09:51.1 & 9.9 & \nodata & \nodata & 41.4$\pm$3.9 & 27.3$\pm$5.7 & 9.5$\pm$6.6 \\
121 & R0600-ID67$^\downarrow$ & 06:00:05.01 & --20:09:24.3 & 38.1 & \nodata & \nodata & $<$15.5 & $<$16.3 & $<$18.0 \\
122 & R0600-ID111 & 06:00:08.91 & --20:08:55.9 & 5.2 & \nodata & \nodata & 15.2$\pm$4.4 & 9.5$\pm$4.4 & $<$14.7 \\
123 & R0600-ID164$^\downarrow$ & 06:00:09.10 & --20:08:26.5 & 4.8$^\dagger$ & \nodata & \nodata & $<$15.1 & $<$16.1 & $<$18.0 \\
124 & R0949-ID10 & 09:49:52.13 & 17:05:47.2 & 61.9 & \nodata & \nodata & 8.5$\pm$4.5 & 14.1$\pm$5.3 & 24.7$\pm$5.9 \\
125 & R0949-ID19$^\downarrow$ & 09:49:53.85 & 17:05:57.7 & 13.1 & \nodata & \nodata & $<$16.2 & $<$15.9 & $<$16.5 \\
126 & R0949-ID122 & 09:49:51.16 & 17:08:12.8 & 5.3 & \nodata & \nodata & 12.8$\pm$4.6 & 5.6$\pm$5.4 & 17.4$\pm$5.6 \\
127 & R0949-ID124 & 09:49:54.62 & 17:08:10.5 & 12.1 & \nodata & \nodata & 19.0$\pm$5.4 & 9.9$\pm$5.0 & 11.3$\pm$7.3 \\
128 & R1347-ID41 & 13:47:31.11 & --11:44:38.8 & 7.2 & 6.5$\pm$0.4 & 17.3$\pm$0.7 & 13.2$\pm$1.6 & 5.5$\pm$1.8 & 6.1$\pm$4.4 \\
129 & R1347-ID51$^\downarrow$ & 13:47:33.74 & --11:44:48.9 & 5.0 & $<$0.6 & $<$1.9 & $<$5.6 & $<$5.5 & $<$13.2 \\
130 & R1347-ID75$^\bullet$ & 13:47:30.63 & --11:45:09.5 & 57.0 & 5.8$\pm$0.4 & 6.4$\pm$0.7 & $<$5.6 & $<$5.5 & 9.4$\pm$4.3 \\
131 & R1347-ID145 & 13:47:27.65 & --11:45:51.1 & 22.0 & 12.0$\pm$0.5 & 29.6$\pm$0.7 & 42.8$\pm$2.6 & 37.1$\pm$7.9 & 21.6$\pm$1.1 \\
132 & R1347-ID148 & 13:47:27.85 & --11:45:55.8 & 21.4$^\dagger$ & 16.5$\pm$0.7 & 40.7$\pm$0.9 & 58.8$\pm$3.5 & 51.0$\pm$10.8 & 29.7$\pm$1.4 \\
133 & R1347-ID166 & 13:47:28.00 & --11:46:12.2 & 5.0 & $<$1.3 & 1.0$\pm$0.8 & 10.3$\pm$2.3 & 15.5$\pm$5.5 & 16.1$\pm$2.6 \\
134 & R2129-ID20$^\bullet$ & 21:29:39.96 & 00:05:21.0 & 17.8 & 1.1$\pm$0.6 & $<$2.4 & 6.2$\pm$2.4 & 5.9$\pm$3.8 & $<$12.3 \\
135 & R2129-ID37 & 21:29:38.24 & 00:04:52.6 & 5.8 & 1.1$\pm$1.0 & 9.3$\pm$1.1 & 9.8$\pm$4.0 & 10.3$\pm$7.5 & 8.7$\pm$4.1 \\
136 & R2211-ID19 & 22:11:47.70 & --03:51:03.6 & 6.5 & \nodata & \nodata & 19.5$\pm$6.4 & 8.9$\pm$5.0 & $<$18.7 \\
137 & R2211-ID35 & 22:11:42.91 & --03:50:41.7 & 17.8$^\dagger$ & \nodata & \nodata & 131.1$\pm$5.0 & 94.8$\pm$6.3 & 46.0$\pm$6.9 \\
138 & R2211-ID171 & 22:11:42.20 & --03:50:23.8 & 10.1 & \nodata & \nodata & 24.4$\pm$4.7 & 11.1$\pm$6.3 & $<$20.8 \\
139 & SM0723-ID61$^\downarrow$ & 07:23:03.86 & --73:27:06.2 & 5.8 & \nodata & \nodata & $<$16.2 & $<$17.4 & $<$20.2 \\
140 & SM0723-ID93$^\downarrow$ & 07:23:03.15 & --73:27:21.0 & 4.7$^\dagger$ & \nodata & \nodata & $<$16.2 & $<$17.4 & $<$20.2 \\
141 & SM0723-ID124$^\downarrow$ & 07:23:25.08 & --73:27:38.9 & 5.7 & \nodata & \nodata & $<$15.7 & $<$17.1 & $<$20.2 \\
\hline & \multicolumn4l{\textit{Secondary Sample} ($\mathrm{S/N}_\mathrm{nat}=4-5$ and $\mathrm{S/N}_\mathrm{tap}<4.5$)}\\\hline
142 & A2537-ID06 & 23:08:20.81 & --02:10:34.6 & 4.2 & $<$0.8 & $<$2.4 & 6.8$\pm$2.8 & 6.3$\pm$2.5 & 4.9$\pm$2.2 \\
143 & A2537-ID24 & 23:08:21.17 & --02:10:54.5 & 4.5 & 18.4$\pm$0.5 & 30.9$\pm$0.8 & 24.7$\pm$1.5 & 18.0$\pm$2.5 & 9.9$\pm$2.2 \\
144 & A2744-ID17 & 00:14:19.51 & --30:22:48.6 & 4.4 & $<$0.8 & $<$2.3 & $<$6.6 & $<$6.9 & $<$7.7 \\
145 & A2744-ID47 & 00:14:17.29 & --30:22:58.6 & 4.7 & 3.4$\pm$0.6 & 2.8$\pm$1.0 & 3.7$\pm$1.8 & $<$5.9 & $<$7.7 \\
146 & A2744-ID176 & 00:14:17.36 & --30:23:45.4 & 4.2$^\dagger$ & 0.7$\pm$0.5 & $<$2.3 & $<$6.8 & $<$7.0 & 2.3$\pm$1.5 \\
147 & A2744-ID178 & 00:14:24.12 & --30:23:46.0 & 4.7 & 4.9$\pm$0.5 & 9.3$\pm$0.9 & 11.8$\pm$1.5 & 3.6$\pm$1.8 & $<$7.5 \\
148 & A2744-ID227 & 00:14:16.56 & --30:24:10.0 & 4.6 & 7.6$\pm$0.5 & 9.2$\pm$0.8 & $<$6.8 & $<$7.1 & 1.8$\pm$1.5 \\
149 & A3192-ID83 & 03:58:53.92 & --29:55:55.4 & 4.2 & \nodata & \nodata & $<$14.4 & $<$13.2 & $<$18.4 \\
150 & A3192-ID154 & 03:58:56.69 & --29:54:20.1 & 4.2 & \nodata & \nodata & $<$11.7 & $<$12.8 & $<$13.1 \\
151 & A370-ID27 & 02:39:53.54 & --01:33:35.3 & 4.3 & 1.4$\pm$0.5 & 2.3$\pm$0.8 & 5.4$\pm$1.4 & 3.0$\pm$1.6 & 2.0$\pm$1.7 \\
152 & A370-ID172 & 02:39:47.38 & --01:34:53.1 & 4.3 & $<$0.7 & $<$2.3 & $<$6.7 & $<$6.9 & $<$8.1 \\
153 & A383-ID24 & 02:48:02.83 & --03:31:26.5 & 4.0 & 17.5$\pm$0.6 & 27.6$\pm$1.0 & 21.2$\pm$1.8 & 8.1$\pm$1.5 & $<$7.7 \\
154 & A383-ID50 & 02:48:02.82 & --03:31:57.5 & 4.2 & 5.9$\pm$0.5 & 5.1$\pm$0.8 & 3.0$\pm$1.7 & $<$6.6 & $<$7.7 \\
155 & ACT0102-ID128 & 01:02:50.47 & --49:15:41.7 & 4.5 & \nodata & \nodata & $<$14.6 & $<$14.8 & $<$20.2 \\
156 & AS295-ID09 & 02:45:36.63 & --53:03:27.4 & 4.3 & \nodata & \nodata & $<$15.8 & $<$13.6 & 8.2$\pm$6.1 \\
157 & AS295-ID269 & 02:45:22.33 & --53:00:48.9 & 4.3 & \nodata & \nodata & 12.3$\pm$5.3 & 7.5$\pm$5.0 & $<$15.9 \\
158 & M0035-ID33 & 00:35:25.50 & --20:15:29.5 & 4.1 & \nodata & \nodata & $<$12.7 & $<$15.1 & $<$16.7 \\
159 & M0416-ID120 & 04:16:10.51 & --24:04:49.0 & 4.4 & $<$1.5 & $<$2.3 & $<$6.5 & $<$7.8 & $<$9.4 \\
160 & M0416-ID138 & 04:16:10.52 & --24:05:04.8 & 4.9 & 3.2$\pm$0.5 & 2.7$\pm$0.8 & 6.0$\pm$2.2 & 4.1$\pm$2.0 & $<$9.4 \\
161 & M0416-ID156 & 04:16:08.79 & --24:05:17.6 & 4.5 & 3.9$\pm$0.5 & 0.9$\pm$0.7 & $<$9.3 & $<$7.9 & $<$9.4 \\
162 & M1115-ID33 & 11:15:54.34 & 01:29:57.1 & 4.3 & 0.8$\pm$0.5 & 1.9$\pm$0.8 & $<$10.2 & $<$8.4 & $<$9.2 \\
163 & M1149-ID27 & 11:49:40.00 & 22:24:57.1 & 4.6 & $<$0.7 & 4.1$\pm$0.8 & 6.2$\pm$1.4 & 3.6$\pm$1.6 & 4.1$\pm$1.8 \\
164 & M1149-ID95 & 11:49:41.05 & 22:24:16.9 & 4.1 & 1.3$\pm$0.5 & 3.5$\pm$0.8 & 4.6$\pm$1.8 & 7.6$\pm$1.8 & 4.1$\pm$1.4 \\
165 & M1206-ID38 & 12:06:08.25 & --08:47:51.1 & 4.4 & 3.2$\pm$0.5 & 6.9$\pm$0.9 & $<$11.1 & 9.0$\pm$2.4 & 4.5$\pm$2.2 \\
166 & M1206-ID84 & 12:06:13.13 & --08:48:27.2 & 4.4 & 14.9$\pm$0.5 & 20.2$\pm$1.5 & 9.8$\pm$4.1 & $<$12.6 & $<$10.2 \\
167 & M1423-ID52 & 14:23:47.57 & 24:04:37.6 & 4.4 & 3.9$\pm$0.5 & $<$2.7 & $<$6.5 & $<$6.5 & $<$7.2 \\
168 & M1423-ID76 & 14:23:48.03 & 24:04:12.2 & 5.0 & 3.3$\pm$0.6 & 3.7$\pm$0.8 & 3.8$\pm$1.9 & $<$6.3 & $<$7.3 \\
169 & M1931-ID42$^\bullet$ & 19:31:49.48 & --26:34:31.3 & 4.2 & 13.5$\pm$1.3 & 4.3$\pm$1.2 & $<$9.1 & $<$8.5 & $<$9.3 \\
170 & M1931-ID69 & 19:31:47.63 & --26:35:02.2 & 4.3 & $<$0.7 & 4.4$\pm$0.7 & $<$9.2 & $<$8.5 & $<$6.2 \\
171 & M2129-ID62 & 21:29:29.63 & --07:41:37.7 & 4.3 & 4.5$\pm$0.5 & 6.2$\pm$0.8 & $<$5.9 & $<$10.4 & $<$9.1 \\
172 & P171-ID161 & 03:12:55.35 & 08:23:43.7 & 4.3 & \nodata & \nodata & $<$16.6 & $<$21.2 & $<$21.5 \\
173 & R0032-ID63 & 00:32:11.39 & 18:06:52.2 & 4.5 & \nodata & \nodata & $<$18.1 & $<$19.6 & $<$22.1 \\
174 & R0032-ID162$^\bullet$ & 00:32:11.53 & 18:07:52.3 & 4.8 & \nodata & \nodata & 8.2$\pm$4.5 & $<$21.9 & $<$21.4 \\
175 & R0032-ID245 & 00:32:09.46 & 18:09:28.9 & 4.6 & \nodata & \nodata & 8.2$\pm$4.3 & $<$20.9 & $<$12.9 \\
176 & R0949-ID14 & 09:49:52.36 & 17:05:46.6 & 4.7 & \nodata & \nodata & $<$16.7 & $<$15.9 & $<$14.0 \\
177 & R0949-ID113 & 09:49:53.87 & 17:08:17.0 & 4.0 & \nodata & \nodata & $<$15.4 & $<$15.7 & $<$14.9 \\
178 & R0949-ID119 & 09:49:52.12 & 17:08:09.7 & 4.2 & \nodata & \nodata & 18.8$\pm$4.1 & $<$13.8 & $<$18.9 \\
179 & R2211-ID164 & 22:11:44.72 & --03:50:13.7 & 4.6 & \nodata & \nodata & $<$16.3 & $<$16.3 & $<$18.7 \\
180 & SM0723-ID98 & 07:23:02.63 & --73:27:24.9 & 4.1 & \nodata & \nodata & $<$16.2 & $<$17.4 & $<$20.2

\enddata
\tablecomments{Here we only present \herschel\ photometry for all secure ALMA sources (141 sources in total; main sample) and tentative ALMA sources with cross-matched \hst\ or \spitzer\ counterparts (39 sources in total; secondary sample).
}
\vspace{-2mm}
\tablenotetext{\downarrow}{\herschel-faint sources with low significance of detection (S/N$<$2) in all \herschel\ bands but secure detection (S/N\,$\geq$\,5) at 1.15\,mm (see Section~\ref{ss:04c_dropout}). 
}
\vspace{-2mm}
\tablenotetext{\bullet}{BCGs in corresponding clusters. M1931-ID42 was identified as the extended tail of the BCG M1931-ID41 \citep{forgarty19}.
}
\vspace{-2mm}
\tablenotetext{\dagger}{Measured from 2\arcsec-tapered ALMA maps instead of native-resolution maps due to a higher significance of detection.}
\end{deluxetable*}




\startlongtable

\begin{deluxetable*}{rlrrcrrrrrr}
\tablecaption{Summary of \textsc{MAGPHYS} SED fitting results}
\tablewidth{0pt}
\tabletypesize{\scriptsize}
\tablehead{
\colhead{\#} & 
\colhead{Name} & \colhead{$z_\mathrm{best}$\tablenotemark{a}} &
\colhead{$z_\mathrm{FIR}$\tablenotemark{b}} &
\colhead{Ref.\tablenotemark{a}} & \colhead{$\mu$\tablenotemark{c}} & \colhead{Model\tablenotemark{d}}  & \colhead{$\log(L_\mathrm{IR})$\tablenotemark{e}} & \colhead{$\log(\mathrm{SFR})$\tablenotemark{e}} & \colhead{$\log(M_\mathrm{dust})$\tablenotemark{e}} & \colhead{$T_\mathrm{dust}$\tablenotemark{f}} \\
\colhead{}  & \colhead{} & \colhead{} & \colhead{} & \colhead{} & \colhead{} & \colhead{} 
 & \colhead{($\mu^{-1}$\lsun)}  & \colhead{($\mu^{-1}$\smpy)}  & \colhead{($\mu^{-1}$\msun)} & \colhead{(K)}
}
\startdata
 & \multicolumn3c{Main Sample ($\mathrm{S/N}_\mathrm{ALMA}\geq5$)}\\\hline
  1 &    A2163-ID11 & 1.23$^{+0.68}_{-0.10}$ & 1.08$^{+0.44}_{-0.43}$ &  HST & 3.2 & (1) & 12.48$\pm$0.33 & 2.31$\pm$0.44 & 8.39$\pm$0.11 & 35.5$\pm$7.7 \\
  2 &    A2537-ID42 & 1.13$^{+0.07}_{-0.07}$ & 1.16$^{+0.48}_{-0.54}$ &  HST & 3.6 & (1) & 12.43$\pm$0.06 & 2.27$\pm$0.15 & 8.77$\pm$0.06 & 30.2$\pm$1.1 \\
  3 &    A2537-ID49 & 1.07$^{+0.43}_{-0.38}$ & \nodata &  FIR & 2.9 & (1) & 12.44$\pm$0.43 & 2.33$\pm$0.49 & 8.36$\pm$0.10 & 35.7$\pm$7.2 \\
  4 &    A2537-ID66 & 3.09$^{+0.06}_{-0.10}$ & 1.16$^{+0.40}_{-0.43}$ &  HST & 15.5 & (1) & 13.25$\pm$0.04 & 3.39$\pm$0.21 & 7.72$\pm$0.16 & 63.7$\pm$9.8 \\
  5 &    A2744-ID07 & 2.41 & 1.61$^{+0.59}_{-0.53}$ &  (1) & 1.9 & (2) & 12.64$\pm$0.16 & 2.52$\pm$0.23 & 8.30$\pm$0.09 & 41.4$\pm$4.8 \\
  6 &    A2744-ID21 & 2.64 & 1.78$^{+0.68}_{-0.60}$ &  (1) & 1.9 & (2) & 12.67$\pm$0.04 & 2.61$\pm$0.11 & 8.18$\pm$0.15 & 33.9$\pm$4.0 \\
  7 &    A2744-ID33 & 3.06 & 3.05$^{+1.08}_{-0.92}$ &  (1) & 2.2 & (2) & 12.43$\pm$0.11 & 2.32$\pm$0.16 & 8.70$\pm$0.10 & 31.4$\pm$3.4 \\
  8 &    A2744-ID56 & 1.50 & 0.99$^{+0.36}_{-0.41}$ &  (1) & 3.2 & (2) & 12.83$\pm$0.02 & 2.61$\pm$0.18 & 8.61$\pm$0.04 & 37.5$\pm$0.5 \\
  9 &    A2744-ID81 & 2.90 & 2.46$^{+0.82}_{-0.76}$ &  (2) & 3.3 & (2) & 12.56$\pm$0.10 & 2.45$\pm$0.15 & 8.57$\pm$0.07 & 34.9$\pm$3.0 \\
 10 &   A2744-ID319 & 2.58 & 3.45$^{+1.14}_{-0.70}$ &  (1) & 1.9 & (2) & 12.64$\pm$0.06 & 2.57$\pm$0.10 & 9.29$\pm$0.07 & 26.0$\pm$0.9 \\
 11 &    A3192-ID31 & 1.22$^{+0.06}_{-0.05}$ & 1.20$^{+0.39}_{-0.47}$ &  HST & 2.4 & (1) & 12.90$\pm$0.12 & 2.68$\pm$0.31 & 9.02$\pm$0.06 & 33.1$\pm$1.1 \\
 12 &    A3192-ID40 & 3.01$^{+0.96}_{-0.87}$ & \nodata &  FIR & 2.7 & (1) & 12.85$\pm$0.25 & 2.77$\pm$0.28 & 9.04$\pm$0.20 & 31.6$\pm$7.4 \\
 13 &   A3192-ID131 & 2.35$^{+0.80}_{-0.73}$ & \nodata &  FIR & 2.9 & (1) & 12.41$\pm$0.33 & 2.30$\pm$0.40 & 8.58$\pm$0.22 & 31.9$\pm$7.7 \\
 14 &   A3192-ID138 & 0.92$^{+0.39}_{-0.07}$ & 1.54$^{+0.53}_{-0.50}$ &  HST & 3.0 & (1) & 11.80$\pm$0.34 & 1.64$\pm$0.43 & 8.53$\pm$0.15 & 24.5$\pm$3.7 \\
 15 &     A370-ID18 & 2.48$^{+0.29}_{-0.15}$ & 2.51$^{+0.80}_{-0.82}$ &  HST & 2.7 & (2) & 12.49$\pm$0.13 & 2.36$\pm$0.18 & 8.64$\pm$0.11 & 27.5$\pm$4.6 \\
 16 &     A370-ID31 & 0.38 & 1.32$^{+0.42}_{-0.52}$ &  (3) & 1.0 & (-1) & 11.38$\pm$0.09 & 1.31$\pm$0.16 & 8.13$\pm$0.09 & 28.2$\pm$0.2 \\
 17 &    A370-ID103 & 1.22$^{+0.02}_{-0.04}$ & 1.51$^{+0.55}_{-0.57}$ &  HST & 1.0 & (-1) & 12.23$\pm$0.07 & 2.03$\pm$0.18 & 8.73$\pm$0.08 & 28.5$\pm$1.0 \\
 18 &    A370-ID110 & 1.06 & 1.07$^{+0.40}_{-0.46}$ &  (4) & 2.3 & (2) & 12.73$\pm$0.04 & 2.50$\pm$0.18 & 8.94$\pm$0.04 & 31.1$\pm$0.2 \\
 19 &    A370-ID146 & 1.07 & 1.74$^{+0.53}_{-0.54}$ &  (5) & 8.9 & (2) & 11.34$\pm$0.05 & 1.02$\pm$0.25 & 8.25$\pm$0.12 & 26.1$\pm$1.7 \\
 20 &  ACT0102-ID22 & 2.74$^{+1.10}_{-0.86}$ & \nodata &  FIR & 1.9 & (1) & 12.41$\pm$0.28 & 2.30$\pm$0.36 & 8.61$\pm$0.17 & 32.0$\pm$9.1 \\
 21 &  ACT0102-ID50 & 2.24$^{+0.16}_{-0.12}$ & 1.54$^{+0.53}_{-0.50}$ &  HST & 2.3 & (1) & 12.58$\pm$0.16 & 2.47$\pm$0.23 & 8.23$\pm$0.09 & 41.1$\pm$4.2 \\
 22 & ACT0102-ID118 & 4.32 & 3.76$^{+1.41}_{-1.12}$ &  (6) & 5.2 & (1) & 12.89$\pm$0.17 & 2.81$\pm$0.20 & 8.89$\pm$0.13 & 34.8$\pm$4.5 \\
 23 & ACT0102-ID160 & 2.20$^{+0.18}_{-0.15}$ & 1.82$^{+0.65}_{-0.59}$ &  HST & 2.5 & (1) & 12.43$\pm$0.17 & 2.31$\pm$0.25 & 8.31$\pm$0.11 & 33.9$\pm$3.8 \\
 24 & ACT0102-ID215 & 4.32 & 4.26$^{+1.48}_{-1.27}$ &  (6) & 8.8 & (1) & 12.83$\pm$0.19 & 2.76$\pm$0.23 & 9.03$\pm$0.15 & 30.1$\pm$3.6 \\
 25 & ACT0102-ID223 & 1.82$^{+0.09}_{-0.11}$ & 2.14$^{+0.68}_{-0.70}$ &  HST & 2.0 & (1) & 12.21$\pm$0.17 & 2.08$\pm$0.27 & 8.65$\pm$0.11 & 28.4$\pm$2.4 \\
 26 & ACT0102-ID224 & 4.32 & 3.90$^{+1.38}_{-1.10}$ &  (6) & 9.2 & (1) & 13.30$\pm$0.09 & 3.22$\pm$0.12 & 9.37$\pm$0.08 & 33.4$\pm$2.0 \\
 27 & ACT0102-ID241 & 2.46$^{+0.29}_{-0.21}$ & 2.50$^{+0.91}_{-0.78}$ &  HST & 1.9 & (1) & 12.32$\pm$0.17 & 2.21$\pm$0.24 & 8.59$\pm$0.15 & 30.6$\pm$3.6 \\
 28 & ACT0102-ID294 & 2.14$^{+0.68}_{-0.70}$ & \nodata &  FIR & 2.4 & (1) & 12.84$\pm$0.33 & 2.72$\pm$0.41 & 9.04$\pm$0.17 & 31.8$\pm$7.3 \\
 29 &   AS1063-ID15 & 1.43 & 2.24$^{+0.50}_{-0.34}$ &  (5) & 2.1 & (2) & 12.17$\pm$0.06 & 2.08$\pm$0.10 & 9.02$\pm$0.05 & 26.4$\pm$0.6 \\
 30 &   AS1063-ID17 & 1.44 & 1.12$^{+0.39}_{-0.44}$ &  (5) & 2.4 & (2) & 12.75$\pm$0.03 & 2.52$\pm$0.19 & 8.61$\pm$0.04 & 37.1$\pm$0.6 \\
 31 &  AS1063-ID147 & 0.61 & 0.78$^{+0.25}_{-0.28}$ &  (7) & 5.7 & (2) & 12.12$\pm$0.05 & 1.92$\pm$0.23 & 8.41$\pm$0.07 & 32.9$\pm$0.2 \\
 32 &  AS1063-ID222 & 2.48$^{+0.14}_{-0.08}$ & 4.30$^{+1.35}_{-1.28}$ &  HST & 1.8 & (2) & 12.57$\pm$0.07 & 2.39$\pm$0.16 & 9.73$\pm$0.06 & 21.6$\pm$0.9 \\
 33 &    M0035-ID41 & 2.67$^{+1.28}_{-0.86}$ & \nodata &  FIR & 2.1 & (1) & 12.37$\pm$0.33 & 2.25$\pm$0.40 & 8.49$\pm$0.24 & 29.9$\pm$9.6 \\
 34 &    M0035-ID94 & 0.36 & 1.06$^{+0.41}_{-0.41}$ &  (8) & 1.0 & (-1) & 11.17$\pm$0.22 & 1.00$\pm$0.39 & 8.01$\pm$0.16 & 21.9$\pm$2.4 \\
 35 &    M0159-ID05 & 1.81$^{+0.09}_{-0.39}$ & 2.05$^{+0.63}_{-0.67}$ &  HST & 2.4 & (1) & 12.46$\pm$0.18 & 2.35$\pm$0.26 & 8.81$\pm$0.10 & 29.5$\pm$2.8 \\
 36 &    M0159-ID24 & 1.48$^{+0.51}_{-0.50}$ & \nodata &  FIR & 0.7 & (1) & 12.36$\pm$0.29 & 2.16$\pm$0.40 & 8.40$\pm$0.14 & 33.2$\pm$8.7 \\
 37 &    M0159-ID61 & 1.08$^{+0.06}_{-0.05}$ & 1.61$^{+0.70}_{-0.55}$ &  HST & 2.9 & (1) & 11.84$\pm$0.22 & 1.67$\pm$0.34 & 8.33$\pm$0.16 & 26.3$\pm$3.5 \\
 38 &    M0257-ID13 & 1.96$^{+0.44}_{-0.41}$ & 2.71$^{+0.94}_{-0.82}$ &  HST & 1.1 & (1) & 12.27$\pm$0.25 & 2.14$\pm$0.36 & 9.05$\pm$0.13 & 24.9$\pm$3.9 \\
 39 &    M0329-ID11 & 3.08$^{+0.03}_{-0.06}$ & 2.26$^{+0.88}_{-0.56}$ &  HST & 2.2 & (2) & 12.38$\pm$0.12 & 2.29$\pm$0.14 & 8.63$\pm$0.16 &  \nodata \\
 40 &    M0416-ID51 & 1.96 & 2.06$^{+0.67}_{-0.77}$ &  (5) & 2.2 & (2) & 12.04$\pm$0.05 & 1.93$\pm$0.10 & 8.30$\pm$0.11 & 32.6$\pm$1.3 \\
 41 &    M0416-ID79 & 2.20$^{+0.03}_{-0.02}$ & 2.10$^{+0.80}_{-0.72}$ &  HST & 1.9 & (2) & 11.90$\pm$0.09 & 1.79$\pm$0.15 & 8.13$\pm$0.17 & 31.9$\pm$2.9 \\
 42 &   M0416-ID117 & 2.09 & 2.71$^{+0.60}_{-0.55}$ &  (5) & 1.8 & (2) & 12.44$\pm$0.04 & 2.39$\pm$0.09 & 8.76$\pm$0.09 & 34.6$\pm$0.7 \\
 43 &   M0416-ID160 & 1.89$^{+0.24}_{-0.03}$ & 1.58$^{+0.54}_{-0.47}$ &  HST & 1.9 & (2) & 12.14$\pm$0.10 & 2.06$\pm$0.14 & 8.00$\pm$0.23 & 37.4$\pm$2.9 \\
 44 &    M0417-ID46 & 3.65 & 2.34$^{+0.72}_{-0.66}$ &  (9) & 5.2 & (1) & 13.43$\pm$0.09 & 3.37$\pm$0.13 & 8.85$\pm$0.04 & 44.0$\pm$1.2 \\
 45 &    M0417-ID49 & 2.20$^{+0.69}_{-0.71}$ & \nodata &  FIR & 1.9 & (1) & 12.92$\pm$0.38 & 2.79$\pm$0.48 & 9.24$\pm$0.27 & 30.5$\pm$6.7 \\
 46 &    M0417-ID58 & 3.65 & 3.46$^{+1.20}_{-1.01}$ &  (9) & 4.0 & (1) & 12.82$\pm$0.15 & 2.73$\pm$0.19 & 8.95$\pm$0.11 & 33.3$\pm$2.6 \\
 47 &   M0417-ID121 & 3.65 & 3.20$^{+1.17}_{-0.95}$ &  (9) & 3.6 & (1) & 12.71$\pm$0.17 & 2.62$\pm$0.20 & 8.74$\pm$0.10 & 34.9$\pm$3.2 \\
 48 &   M0417-ID204 & 2.05$^{+0.89}_{-0.76}$ & \nodata &  FIR & 3.0 & (1) & 12.37$\pm$0.27 & 2.25$\pm$0.34 & 8.58$\pm$0.27 & 30.4$\pm$9.4 \\
 49 &   M0417-ID218 & 1.63$^{+0.52}_{-0.51}$ & \nodata &  FIR & 4.0 & (1) & 12.61$\pm$0.36 & 2.47$\pm$0.46 & 8.77$\pm$0.17 & 32.0$\pm$6.4 \\
 50 &   M0417-ID221 & 1.15$^{+0.07}_{-0.64}$ & 1.80$^{+0.57}_{-0.55}$ &  HST & 2.9 & (1) & 11.97$\pm$0.20 & 1.85$\pm$0.36 & 8.73$\pm$0.12 & 24.8$\pm$4.2 \\
 51 &   M0417-ID223 & 1.10$^{+0.09}_{-0.08}$ & 1.64$^{+0.58}_{-0.53}$ &  HST & 2.8 & (1) & 12.24$\pm$0.19 & 2.10$\pm$0.31 & 8.84$\pm$0.21 & 25.5$\pm$2.4 \\
 52 &    M0429-ID27 & 2.72$^{+0.77}_{-0.40}$ & 2.42$^{+0.57}_{-0.52}$ &  HST & 4.0 & (2) & 12.69$\pm$0.23 & 2.63$\pm$0.27 & 8.52$\pm$0.19 & 34.7$\pm$5.8 \\
 53 &    M0553-ID17 & 2.49$^{+0.84}_{-0.74}$ & \nodata &  FIR & 1.6 & (1) & 12.63$\pm$0.30 & 2.53$\pm$0.36 & 8.80$\pm$0.22 & 31.7$\pm$7.5 \\
 54 &    M0553-ID61 & 1.49$^{+0.07}_{-0.09}$ & 2.08$^{+0.65}_{-0.68}$ &  HST & 2.0 & (1) & 12.14$\pm$0.18 & 2.01$\pm$0.30 & 8.81$\pm$0.08 & 25.8$\pm$1.9 \\
 55 &   M0553-ID133 & 1.14 & 1.30$^{+0.40}_{-0.46}$ &  (10) & 4.4 & (1) & 12.80$\pm$0.12 & 2.62$\pm$0.30 & 9.10$\pm$0.04 & 30.4$\pm$0.5 \\
 56 &   M0553-ID190 & 1.14 & 1.30$^{+0.40}_{-0.46}$ &  (10) & 6.6 & (1) & 12.98$\pm$0.12 & 2.79$\pm$0.30 & 9.28$\pm$0.04 & 30.6$\pm$0.4 \\
 57 &   M0553-ID200 & 1.71$^{+0.82}_{-0.58}$ & \nodata &  FIR & 2.2 & (1) & 12.36$\pm$0.31 & 2.21$\pm$0.39 & 8.41$\pm$0.15 & 31.3$\pm$9.1 \\
 58 &   M0553-ID249 & 1.14 & 1.30$^{+0.40}_{-0.46}$ &  (10) & 5.8 & (1) & 12.82$\pm$0.12 & 2.64$\pm$0.30 & 9.12$\pm$0.04 & 30.5$\pm$0.5 \\
 59 &   M0553-ID275 & 0.94$^{+0.06}_{-0.05}$ & 0.87$^{+0.33}_{-0.34}$ &  HST & 2.5 & (1) & 12.41$\pm$0.17 & 2.22$\pm$0.34 & 8.36$\pm$0.11 & 34.7$\pm$3.0 \\
 60 &   M0553-ID303 & 0.84 & 0.98$^{+0.39}_{-0.39}$ &  (8) & 1.5 & (1) & 12.48$\pm$0.16 & 2.31$\pm$0.31 & 8.66$\pm$0.12 & 30.8$\pm$2.7 \\
 61 &   M0553-ID355 & 1.82$^{+0.11}_{-0.06}$ & 1.91$^{+0.65}_{-0.66}$ &  HST & 1.8 & (1) & 12.32$\pm$0.14 & 2.17$\pm$0.25 & 8.64$\pm$0.09 & 29.9$\pm$2.0 \\
 62 &   M0553-ID375 & 2.42$^{+0.80}_{-0.72}$ & \nodata &  FIR & 1.0 & (-1) & 12.54$\pm$0.29 & 2.44$\pm$0.34 & 8.69$\pm$0.20 & 31.9$\pm$7.4 \\
 63 &   M0553-ID398 & 3.39$^{+1.22}_{-0.99}$ & \nodata &  FIR & 1.7 & (1) & 12.62$\pm$0.27 & 2.53$\pm$0.32 & 8.79$\pm$0.22 & 32.8$\pm$8.7 \\
 64 &    M1115-ID02 & 1.92$^{+0.11}_{-0.10}$ & 1.14$^{+0.68}_{-0.25}$ &  HST & 3.1 & (2) & 13.04$\pm$0.09 & 2.91$\pm$0.14 & 9.11$\pm$0.07 & 33.9$\pm$1.3 \\
 65 &    M1115-ID04 & 1.60 & 1.38$^{+0.39}_{-0.50}$ &  (11) & 2.7 & (2) & 12.66$\pm$0.03 & 2.55$\pm$0.09 & 8.51$\pm$0.06 & 36.2$\pm$1.1 \\
 66 &    M1115-ID34 & 3.21$^{+1.05}_{-0.94}$ & \nodata &  FIR & 2.9 & (2) & 12.53$\pm$0.28 & 2.43$\pm$0.31 & 8.73$\pm$0.23 & 25.6$\pm$6.9 \\
 67 &    M1149-ID77 & 1.46 & 1.61$^{+0.52}_{-0.51}$ &  (2) & 3.0 & (2) & 12.01$\pm$0.05 & 1.81$\pm$0.22 & 8.38$\pm$0.09 & 30.6$\pm$1.6 \\
 68 &   M1149-ID229 & 1.38$^{+0.28}_{-0.06}$ & 2.05$^{+0.67}_{-0.72}$ &  HST & 2.0 & (2) & 11.79$\pm$0.13 & 1.63$\pm$0.20 & 8.54$\pm$0.14 & 25.5$\pm$2.5 \\
 69 &    M1206-ID27 & 1.04 & 1.16$^{+0.45}_{-0.49}$ &  (12) & 5.2 & (2) & 12.20$\pm$0.04 & 1.98$\pm$0.21 & 8.56$\pm$0.06 & 29.9$\pm$1.2 \\
 70 &    M1206-ID55 & 1.04 & 1.23$^{+0.31}_{-0.53}$ &  (12) & 7.7 & (2) & 12.37$\pm$0.05 & 2.24$\pm$0.11 & 8.54$\pm$0.05 & 31.5$\pm$0.5 \\
 71 &    M1206-ID60 & 1.04 & 1.23$^{+0.31}_{-0.53}$ &  (12) & 25.5 & (2) & 12.45$\pm$0.05 & 2.32$\pm$0.11 & 8.62$\pm$0.05 & 31.6$\pm$0.5 \\
 72 &    M1206-ID61 & 1.04 & 1.22$^{+0.33}_{-0.51}$ &  (12) & 73.1 & (2) & 12.58$\pm$0.05 & 2.46$\pm$0.11 & 8.76$\pm$0.07 & 31.5$\pm$0.6 \\
 73 &    M1311-ID27 & 2.19 & 1.55$^{+0.61}_{-0.59}$ &  (11) & 15.8 & (2) & 12.74$\pm$0.03 & 2.61$\pm$0.10 & 8.45$\pm$0.05 & 39.9$\pm$0.8 \\
 74 &    M1311-ID33 & 2.32$^{+0.77}_{-0.20}$ & 1.78$^{+0.46}_{-0.44}$ &  HST & 2.6 & (2) & 12.53$\pm$0.21 & 2.42$\pm$0.33 & 8.30$\pm$0.12 & 32.0$\pm$5.4 \\
 75 &    M1423-ID38 & 1.90$^{+0.04}_{-0.14}$ & 1.60$^{+0.49}_{-0.47}$ &  HST & 4.3 & (2) & 12.36$\pm$0.08 & 2.27$\pm$0.12 & 8.18$\pm$0.13 & 37.2$\pm$1.9 \\
 76 &    M1931-ID55 & 2.15$^{+0.65}_{-0.34}$ & \nodata &  FIR & 2.9 & (2) & 12.60$\pm$0.25 & 2.56$\pm$0.29 & 8.80$\pm$0.15 & 35.6$\pm$5.6 \\
 77 &    M1931-ID61 & 3.02$^{+1.03}_{-0.89}$ & \nodata &  FIR & 2.4 & (2) & 12.34$\pm$0.34 & 2.24$\pm$0.42 & 8.56$\pm$0.25 & 32.0$\pm$8.1 \\
 78 &    M2129-ID46 & 1.48 & 1.20$^{+0.43}_{-0.47}$ &  (5) & 3.7 & (2) & 12.06$\pm$0.04 & 1.91$\pm$0.16 & 7.91$\pm$0.12 & 36.7$\pm$1.8 \\
 79 &     P171-ID69 & 3.57$^{+1.42}_{-1.08}$ & \nodata &  FIR & 2.7 & (1) & 12.77$\pm$0.27 & 2.69$\pm$0.31 & 8.95$\pm$0.25 & 30.1$\pm$9.0 \\
 80 &    P171-ID177 & 2.55$^{+1.11}_{-0.83}$ & \nodata &  FIR & 2.6 & (1) & 12.47$\pm$0.27 & 2.35$\pm$0.33 & 8.58$\pm$0.18 & 31.5$\pm$9.3 \\
 81 &    R0032-ID53 & 3.63 & 1.95$^{+0.62}_{-0.64}$ &  (13) & 9.2 & (1) & 13.18$\pm$0.08 & 3.16$\pm$0.15 & 8.34$\pm$0.04 & 49.6$\pm$1.5 \\
 82 &    R0032-ID55 & 3.63 & 1.95$^{+0.62}_{-0.64}$ &  (13) & 18.8 & (1) & 13.46$\pm$0.08 & 3.44$\pm$0.16 & 8.61$\pm$0.04 & 49.6$\pm$1.2 \\
 83 &    R0032-ID57 & 3.63 & 1.95$^{+0.64}_{-0.65}$ &  (13) & 7.5 & (1) & 12.70$\pm$0.08 & 2.66$\pm$0.14 & 7.95$\pm$0.07 & 49.6$\pm$2.7 \\
 84 &    R0032-ID58 & 3.63 & 1.95$^{+0.62}_{-0.64}$ &  (13) & 7.7 & (1) & 13.20$\pm$0.07 & 3.18$\pm$0.15 & 8.36$\pm$0.04 & 49.6$\pm$1.5 \\
 85 &   R0032-ID127 & 3.14$^{+1.03}_{-0.95}$ & \nodata &  FIR & 9.7 & (1) & 12.89$\pm$0.23 & 2.79$\pm$0.29 & 9.16$\pm$0.18 & 31.4$\pm$7.7 \\
 86 &   R0032-ID131 & 3.27$^{+1.15}_{-0.95}$ & \nodata &  FIR & 3.3 & (1) & 12.83$\pm$0.32 & 2.74$\pm$0.36 & 9.09$\pm$0.28 & 31.9$\pm$8.1 \\
 87 &   R0032-ID198 & 1.63$^{+0.55}_{-0.51}$ & \nodata &  FIR & 2.0 & (1) & 12.55$\pm$0.54 & 2.41$\pm$0.63 & 8.67$\pm$0.16 & 32.3$\pm$6.9 \\
 88 &   R0032-ID220 & 1.47$^{+0.55}_{-0.50}$ & \nodata &  FIR & 2.1 & (1) & 12.47$\pm$0.34 & 2.31$\pm$0.42 & 8.52$\pm$0.14 & 33.6$\pm$8.1 \\
 89 &   R0032-ID238 & 2.95$^{+1.04}_{-0.89}$ & \nodata &  FIR & 1.7 & (1) & 12.57$\pm$0.29 & 2.47$\pm$0.35 & 8.84$\pm$0.25 & 31.3$\pm$8.0 \\
 90 &   R0032-ID276 & 1.99$^{+0.83}_{-0.68}$ & \nodata &  FIR & 2.6 & (1) & 12.46$\pm$0.37 & 2.31$\pm$0.45 & 8.56$\pm$0.20 & 33.1$\pm$9.6 \\
 91 &   R0032-ID287 & 1.32$^{+0.42}_{-0.45}$ & \nodata &  FIR & 1.6 & (1) & 12.56$\pm$0.29 & 2.41$\pm$0.38 & 8.62$\pm$0.11 & 33.0$\pm$6.7 \\
 92 &    R0600-ID12 & 1.09$^{+0.47}_{-0.43}$ & \nodata &  FIR & 1.2 & (1) & 12.14$\pm$0.36 & 1.96$\pm$0.46 & 8.17$\pm$0.12 & 33.6$\pm$8.7 \\
 93 &    R0600-ID13 & 1.27 & 1.23$^{+0.40}_{-0.45}$ &  (8) & 1.3 & (1) & 12.49$\pm$0.13 & 2.34$\pm$0.26 & 8.52$\pm$0.07 & 33.6$\pm$1.8 \\
 94 &   R0600-ID111 & 1.92$^{+0.07}_{-0.10}$ & 1.15$^{+0.52}_{-0.47}$ &  HST & 2.7 & (1) & 12.31$\pm$0.19 & 2.18$\pm$0.29 & 7.97$\pm$0.09 & 46.1$\pm$8.1 \\
 95 &    R0949-ID10 & 3.87$^{+1.28}_{-1.14}$ & \nodata &  FIR & 3.0 & (1) & 12.94$\pm$0.35 & 2.85$\pm$0.41 & 9.23$\pm$0.31 & 31.5$\pm$8.0 \\
 96 &   R0949-ID122 & 1.77$^{+0.72}_{-0.60}$ & \nodata &  FIR & 2.8 & (1) & 12.19$\pm$0.42 & 2.05$\pm$0.53 & 8.31$\pm$0.22 & 32.2$\pm$8.7 \\
 97 &   R0949-ID124 & 2.89$^{+1.01}_{-0.85}$ & \nodata &  FIR & 3.3 & (1) & 12.69$\pm$0.31 & 2.60$\pm$0.35 & 8.86$\pm$0.24 & 32.4$\pm$8.1 \\
 98 &    R1347-ID41 & 0.85 & 1.51$^{+0.72}_{-0.51}$ &  (11) & 4.4 & (2) & 11.60$\pm$0.04 & 1.50$\pm$0.15 & 8.16$\pm$0.08 & 30.7$\pm$0.8 \\
 99 &   R1347-ID145 & 1.77 & 1.68$^{+0.61}_{-0.57}$ &  (14) & 3.3 & (2) & 12.77$\pm$0.03 & 2.62$\pm$0.10 & 8.84$\pm$0.05 & 34.0$\pm$0.3 \\
100 &   R1347-ID148 & 1.77 & 1.68$^{+0.61}_{-0.57}$ &  (14) & 5.8 & (2) & 12.91$\pm$0.04 & 2.76$\pm$0.10 & 8.98$\pm$0.04 & 34.0$\pm$0.2 \\
101 &   R1347-ID166 & 2.43$^{+0.69}_{-0.78}$ & \nodata &  FIR & 3.2 & (2) & 12.19$\pm$0.37 & 2.00$\pm$0.54 & 8.59$\pm$0.20 & 34.0$\pm$7.6 \\
102 &    R2129-ID37 & 2.48$^{+0.65}_{-0.08}$ & 1.85$^{+0.56}_{-0.55}$ &  HST & 1.9 & (2) & 12.52$\pm$0.13 & 2.41$\pm$0.19 & 8.24$\pm$0.08 & 36.0$\pm$5.3 \\
103 &    R2211-ID19 & 1.58$^{+0.66}_{-0.54}$ & \nodata &  FIR & 2.0 & (1) & 12.24$\pm$0.31 & 2.10$\pm$0.39 & 8.31$\pm$0.16 & 33.1$\pm$8.9 \\
104 &    R2211-ID35 & 1.10$^{+0.36}_{-0.44}$ & \nodata &  FIR & 2.4 & (1) & 12.86$\pm$0.48 & 2.62$\pm$0.62 & 8.99$\pm$0.12 & 32.9$\pm$6.3 \\
105 &   R2211-ID171 & 2.31$^{+0.74}_{-0.69}$ & \nodata &  FIR & 2.5 & (1) & 12.65$\pm$0.43 & 2.56$\pm$0.50 & 8.80$\pm$0.23 & 32.1$\pm$7.2 \\
\hline & \multicolumn3c{Secondary Sample ($\mathrm{S/N}_\mathrm{ALMA}=4-5$)}\\\hline
106 &    A2537-ID24 & 3.11$^{+0.10}_{-0.06}$ & 1.04$^{+0.34}_{-0.38}$ &  HST & 19.8 & (1) & 13.27$\pm$0.08 & 3.44$\pm$0.22 & 7.83$\pm$0.06 & 65.9$\pm$5.3 \\
107 &    A2744-ID47 & 1.65$^{+0.02}_{-0.03}$ & 1.60$^{+0.52}_{-0.70}$ &  HST & 2.8 & (2) & 11.93$\pm$0.07 & 1.87$\pm$0.11 & 8.48$\pm$0.33 & 32.9$\pm$5.1 \\
108 &   A2744-ID178 & 0.94 & 1.85$^{+0.69}_{-0.52}$ &  (15) & 2.0 & (2) & 11.61$\pm$0.06 & 1.50$\pm$0.12 & 8.36$\pm$0.16 & 32.4$\pm$1.1 \\
109 &   A2744-ID227 & 2.58 & 1.04$^{+0.54}_{-0.46}$ &  (5) & 1.8 & (2) & 12.71$\pm$0.05 & 2.77$\pm$0.19 & 7.88$\pm$0.27 &  \nodata \\
110 &     A370-ID27 & 3.14$^{+0.04}_{-0.04}$ & 2.58$^{+0.92}_{-0.79}$ &  HST & 5.2 & (2) & 12.35$\pm$0.06 & 2.28$\pm$0.10 & 8.39$\pm$0.26 & 34.5$\pm$3.4 \\
111 &     A383-ID24 & 0.67$^{+0.04}_{-0.03}$ & 0.84$^{+0.29}_{-0.32}$ &  HST & 2.6 & (2) & 11.65$\pm$0.08 & 1.48$\pm$0.23 & 7.95$\pm$0.10 & 32.9$\pm$0.8 \\
112 &     A383-ID50 & 1.01 & 0.62$^{+0.45}_{-0.32}$ &  (16) & 13.2 & (2) & 11.62$\pm$0.07 & 1.57$\pm$0.16 & 7.12$\pm$0.34 & 33.7$\pm$6.4 \\
113 &   AS295-ID269 & 2.34$^{+1.19}_{-0.90}$ & \nodata &  FIR & 2.0 & (1) & 12.37$\pm$0.31 & 2.26$\pm$0.38 & 8.56$\pm$0.28 &  \nodata \\
114 &   M0416-ID138 & 1.84$^{+0.02}_{-0.02}$ & 1.44$^{+0.44}_{-0.41}$ &  HST & 1.5 & (2) & 11.99$\pm$0.07 & 1.93$\pm$0.11 & 8.07$\pm$0.22 & 32.6$\pm$2.3 \\
115 &   M0416-ID156 & 1.57$^{+0.05}_{-0.03}$ & 0.43$^{+0.83}_{-0.39}$ &  HST & 1.8 & (2) & 11.84$\pm$0.06 & 1.81$\pm$0.15 & 8.21$\pm$0.23 & 24.2$\pm$3.2 \\
116 &    M1115-ID33 & 0.74$^{+0.02}_{-0.05}$ & 2.84$^{+1.12}_{-0.87}$ &  HST & 1.6 & (2) & 10.69$\pm$0.14 & 0.44$\pm$0.27 & 7.69$\pm$0.59 & 15.8$\pm$1.1 \\
117 &    M1149-ID27 & 5.05$^{+0.13}_{-0.26}$ & 2.15$^{+0.81}_{-0.70}$ &  HST & 2.4 & (2) & 12.88$\pm$0.08 & 2.89$\pm$0.19 & 7.96$\pm$0.17 & 46.6$\pm$7.0 \\
118 &    M1149-ID95 & 1.12$^{+0.19}_{-0.53}$ & 2.22$^{+0.87}_{-0.76}$ &  HST & 1.3 & (2) & 11.40$\pm$0.43 & 1.20$\pm$0.56 & 8.42$\pm$0.36 & 23.0$\pm$4.2 \\
119 &    M1206-ID38 & 1.49 & 1.57$^{+0.52}_{-0.54}$ &  (17) & 4.2 & (2) & 11.94$\pm$0.05 & 1.81$\pm$0.12 & 8.09$\pm$0.17 & 32.1$\pm$1.6 \\
120 &    M1206-ID84 & 0.48 & 1.31$^{+0.46}_{-0.46}$ &  (17) & 1.0 & (-1) & 11.33$\pm$0.11 & 1.24$\pm$0.16 & 8.03$\pm$0.20 & 30.6$\pm$0.9 \\
121 &    M1423-ID52 & 1.23$^{+0.05}_{-0.05}$ & 1.28$^{+0.43}_{-0.47}$ &  HST & 4.0 & (2) & 11.79$\pm$0.08 & 1.64$\pm$0.18 & 7.95$\pm$0.21 &  \nodata \\
122 &    M1423-ID76 & 0.67$^{+0.14}_{-0.23}$ & 1.89$^{+0.36}_{-0.47}$ &  HST & 1.2 & (2) & 11.05$\pm$0.41 & 0.95$\pm$0.45 & 7.96$\pm$0.54 & 33.7$\pm$4.9 \\
123 &    M1931-ID69 & 1.60$^{+0.60}_{-0.54}$ & \nodata &  FIR & 1.8 & (2) & 11.77$\pm$0.41 & 1.65$\pm$0.49 & 7.95$\pm$0.27 & 33.1$\pm$7.6 \\
124 &    M2129-ID62 & 1.48 & 1.46$^{+0.42}_{-0.43}$ &  (5) & 2.7 & (2) & 11.98$\pm$0.05 & 1.90$\pm$0.10 & 8.09$\pm$0.17 & 34.8$\pm$2.7 \\
125 &   R0949-ID119 & 0.56$^{+0.09}_{-0.05}$ & 1.43$^{+0.50}_{-0.49}$ &  HST & 1.5 & (1) & 11.39$\pm$0.22 & 1.21$\pm$0.38 & 8.32$\pm$0.15 & 20.3$\pm$2.4

\enddata
\tablenotetext{a}{Best redshifts. \hst\ or far-IR photometric redshifts are presented as the median of their likelihood distributions with 1$\sigma$ confidence range. 
References of spectroscopic redshifts:
(1) CO spectroscopy (private communication from F.\ Bauer),
(2) GLASS and \citet{laporte17},
(3) \citet{wold12}, 
(4) \citet{soucail99}, 
(5) GLASS \citep{treu15,schmidt14},
(6) \citet{caputi21},
(7) \citet{gomez12}, GLASS \citep{treu15,schmidt14} and \citet{walth19}.
(8) CO spectroscopy (ALCS; \citealt{sfprep}),
(9) CO spectroscopy \citep{kkprep}, 
(10) \citet{ebeling17}, 
(11) \citet{caminha19}, 
(12) ``Cosmic Snake" \citep[e.g.,][]{ebeling09, biviano13, cava18}, 
(13) \citet{mdz17}, 
(14) \citet{richard21},
(15) GLASS \citep{wang15},
(16) \citet{sand04},
(17) \citet{biviano13}. 
\hst\ photometric redshifts are taken from \citet{molino17}, \citet{shipley18} and \citet{coe19} for sources detected in CLASH, HFF and RELICS cluster fields, respectively.
}
\vspace{-2mm}
\tablenotetext{b}{Far-IR photometric redshifts modeled with \textsc{magphys+photo-z} (\citealt{battisti19}; see Section~\ref{ss:04a_method}).
This column is left blank if $z_\mathrm{best} = z_\mathrm{FIR}$.
}
\vspace{-2mm}
\tablenotetext{c}{Lensing magnification factors are calculated based on the cluster mass models and the best redshifts of sources (see Section~\ref{ss:05a_lens}, also for the evaluation of uncertainty).}
\vspace{-2mm}
\tablenotetext{d}{Adopted models of lensing magnifications. (1): GLAFIC \citep{oguri10}, (2): Zitrin NFW \citep{zitrin13, zitrin15}, (-1): lensing models are not available (e.g., the sources are cluster member galaxies, or out of \hst\ footprint).
}
\vspace{-2mm}
\tablenotetext{e}{These quantities, derived with \textsc{magphys} \citep{dacunha08,dacunha15,battisti19}, are not corrected for lensing magnification.}
\vspace{-2mm}
\tablenotetext{f}{Modeled with MBB spectrum with fixed dust emissivity at $\beta=1.8$ (Section~\ref{ss:04a_method}).}
\label{tab:04_sed}
\end{deluxetable*}


\clearpage
\bibliography{00_main}{}
\bibliographystyle{aasjournal}


\end{document}